\documentclass[12pt,letterpaper]{article}
\usepackage[
left=1.25in,
right=1.25in,
top=1.25in,
bottom=1.25in
]{geometry}
\usepackage{setspace}
\usepackage{makeidx}
\usepackage{eurosym}
\usepackage{amsthm}
\usepackage{epsfig,lscape,rotate}
\usepackage{subcaption}
\usepackage{graphicx}
\usepackage{amsmath}
\usepackage{epstopdf}
\usepackage{lscape}
\usepackage{rotating}
\usepackage{amssymb}
\usepackage{amsfonts}
\usepackage{euscript}
\usepackage{mathrsfs}
\usepackage{natbib}
\usepackage{multirow}
\usepackage{algorithm}
\usepackage{booktabs}
\usepackage[usenames]{color}
\usepackage{xr}
\usepackage{tabularx}
\usepackage{threeparttable}
\usepackage{adjustbox}
\usepackage{tikz}
\usepackage{float}

\newtheorem{theorem}{Theorem}

\newtheorem{assumption}{Assumption}

\newtheorem{lemma}{Lemma}

\newtheorem{definition}{Definition}
\newtheorem{remark}{Remark}
\newtheoremstyle{ecta}
{\medskipamount}{\bigskipamount}{\normalfont}{1.4em}{\scshape}{:}{1em}{}
\theoremstyle{ecta}
\newtheorem*{example*}{Example}

\newcolumntype{Y}{>{\centering\arraybackslash}X}

\input{tcilatex}
\begin{document}

\title{Model Selection in Panel Data Models: A Generalization of the Vuong
Test}
\author{Jinyong Hahn\thanks{%
Department of Economics, UCLA, Los Angeles, CA 90095, USA. Email:
hahn@econ.ucla.edu.} \\
UCLA \and Zhipeng Liao\thanks{%
Department of Economics, UCLA, Los Angeles, CA 90095, USA. Email:
zhipeng.liao@econ.ucla.edu.} \\
UCLA \and Konrad Menzel\thanks{%
Department of Economics, NYU, New York, NY 10012, USA. Email: km125@nyu.edu.}
\\
NYU \and Quang Vuong\thanks{%
Department of Economics, NYU, New York, NY 10012, USA. Email: qvuong@nyu.edu.%
} \\
NYU}
\date{\today }
\maketitle

\begin{abstract}
This paper generalizes the classical \cite{Vuong1989} test to panel data
models by employing modified profile likelihoods and the Kullback--Leibler
information criterion. Unlike the standard likelihood function, the profile
likelihood lacks certain regular properties, making modification necessary.
We adopt a generalized panel data framework that incorporates group fixed
effects for time and individual pairs, rather than traditional individual
fixed effects. Applications of our approach include linear models with
non-nested specifications of individual-time effects.

\bigskip

\noindent JEL Classification: C14, C31, C32\bigskip

\noindent\textit{Keywords: } Panel data models, Vuong test, Bias correction,
Grouped heterogeneity.
\end{abstract}

\section{Introduction\label{sec:intro}}

With the advent of big data, modern empirical research increasingly adopts
more complex models. These models often involve high-dimensional nuisance
parameters, which can introduce substantial biases in standard estimators -
manifesting as variations of the incidental parameters problem. Classical
references include \cite{NeymanScott1948} and \cite{Nickell1981} among
others. Numerous studies have been conducted to address and mitigate these
issues. It is now well established that these issues represent a form of
higher-order bias, and a variety of methodological approaches have been
proposed to overcome them. See \cite{ArellanoHahn2007}, \cite%
{Arellano-Bonhomme}, \cite{BesterHansen2009}, \cite{CARRO2007503}, \cite%
{Dhaene-Jochmans}, \cite{FERNANDEZVAL200971}, \cite{Fernandez-Val-Weidner-1}%
, \cite{Fernandez-Val-Weidner-2}, \cite{HahnKuersteiner2002}, and \cite%
{HahnNewey2004}, for example.

Although the literature is well developed, its generalizations have only
recently found application in substantively meaningful but technically
complex contexts, such as network-type models. See \cite%
{Bonhomme-Lamadon-Manresa-1}, \cite{Jochmans-Weidner}, \cite%
{bonhomme2025neymanorthogonalizationapproachincidentalparameter}. Motivated
by these developments, we propose to investigate issues of model
specification in nonlinear panel data models. Like most econometric
frameworks, panel data models often rely on potentially restrictive
assumptions. However, to the best of our knowledge, there is currently no
literature that provides a systematic approach for comparing potentially
non-nested models in this context. We would like to close this gap in the
literature by proposing a panel generalization of the classical \cite%
{Vuong1989} test.

Specifically, we consider a model characterized by the joint log-likelihood
of the data $\{z_{i,t}\}_{i\leq n,t\leq T}$ specified as: 
\begin{equation}
L_{n,T}(\phi )\equiv \sum_{i\leq n}\sum_{t\leq T}\log f\left( z_{i,t};\theta
,\gamma _{g(i),m(t)}\right) ,  \label{general_spec}
\end{equation}%
where $f\left( \cdot \right) $ is a known function, $\theta $ represents
unknown parameters that are individual/time-invariant, $\gamma _{g(i),m(t)}$
denotes the individual and time fixed effect which has a grouping structure
determined by cluster/group assignment functions $g(\cdot )$ and $m(\cdot )$
with ranges $\mathcal{G}\equiv \{1,\ldots ,G\}$ and $\mathcal{M}\equiv
\{1,\ldots ,M\}$, respectively, and $\phi \equiv (\theta ^{\top },(\gamma
_{g,m}^{\top })_{g\in \mathcal{G},m\in \mathcal{M}})^{\top }$ include all
unknown parameters specified in the model. For ease of notations, we let $%
I_{g}$ for $g\in \mathcal{G}$ and $I_{m}$ for $m\in \mathcal{M}$ denote the
partitions of $\{1,\ldots ,n\}$ and $\{1,\ldots ,T\}$ according to the group
assignment functions $g(\cdot )$ and $m(\cdot )$, respectively. The pseduo
true value $\phi ^{\ast }$ is defined as the maximizer of $\mathbb{E}\left[
L_{n,T}(\phi )\right] $ over the parameter space, and it is often estimated
by the maximizer $\hat{\phi}$ of $L_{n,T}(\phi )$ over the same space. Our
group structure is inspired the analyses by \cite{Bonhomme-Manresa} or \cite%
{Bonhomme-Lamadon-Manresa-2}, yet it is general enough to include the
classical panel models with individual fixed effects.

In practice, different ways of specifying the likelihood function $f\left(
\cdot \right) $ and the group assignment functions $g(\cdot )$ and $m(\cdot
) $ lead to different modeling strategies in panel data models, and we often
have the situation to compare competing modeling strategies and hope to find
the one which is closer (or the closest) to the true data generating process
in terms of the Kullback--Leibler (KL) distance. It would be reasonable to
compare models through the \cite{Vuong1989} test, which tests the null
hypothesis:%
\begin{equation*}
H_{0}:\mathbb{E}\left[ L_{1,n,T}(\phi _{1}^{\ast })\right] =\mathbb{E}\left[
L_{2,n,T}(\phi _{2}^{\ast })\right] ,
\end{equation*}%
where $L_{j,n,T}(\cdot )$ represents the joint log-likelihood specified by
modeling strategy $j$ with $\phi _{j}^{\ast }$ denoting the pseudo true
value for $j=1,2$. The alternative hypothesis can be%
\begin{equation*}
H_{1}^{\text{2-sided}}:\mathbb{E}\left[ L_{1,n,T}(\phi _{1}^{\ast })\right]
\neq \mathbb{E}\left[ L_{2,n,T}(\phi _{2}^{\ast })\right] \text{ \ or \ }%
H_{1}^{\text{1-sided}}:\mathbb{E}\left[ L_{1,n,T}(\phi _{1}^{\ast })\right] >%
\mathbb{E}\left[ L_{2,n,T}(\phi _{2}^{\ast })\right] .
\end{equation*}%
The classical Vuong test employs the quasi-likelihood ratio (QLR), defined
as: 
\begin{equation}
QLR_{n,T}\equiv (nT)^{-1/2}(L_{1,n,T}(\hat{\phi}_{1})-L_{2,n,T}(\hat{\phi}%
_{2})),  \label{G_QLR}
\end{equation}%
to construct the test statistic, where $\hat{\phi}_{j}$ denotes the
estimator of $\phi _{j}^{\ast }$ in model $j$. It turns out that in panel
models, the log-likelihoods $L_{1,n,T}$ and $L_{2,n,T}$ are biased due to
the incidental parameter problems. Panel adaptation of the Vuong test thus
requires working with bias-corrected versions of the log likelihoods, which
we will call $LM_{1,n,T}^{\ast }$ and $LM_{2,n,T}^{\ast }$. We also show
that there is additional complication because the order of magnitude of the
standard error of the test statistic (after such modification) can depend on
the nature of the hypothesis. Our technical contribution consists of
estimation of $LM_{1,n,T}^{\ast }$ and $LM_{2,n,T}^{\ast }$ to make the
modification feasible as well as estimation of the standard error while
avoiding the uniformity issue.

We extend the classical \cite{Vuong1989} test to panel models, contributing
to the understanding of model selection complexities in high-dimensional
settings. The Vuong test originally compares two non-nested parametric
models and formally tests the hypothesis that their KL distances from the
true data distribution are equal. Originally designed for parsimoniously
parameterized models, this test has been expanded in various studies, see, 
\cite{RiversVuong2002}, \cite{ChenHongShum}, and \cite{Shi2015b}, for
example. In this paper, we demonstrate that the incidental parameters
problem in panel data contexts invalidates the classical Vuong test, which
compares maximized likelihood ratios between two models. We present a
modified test statistic that ensures validity for panel data analysis.

Our modification follows a precedent in \cite{LeePhillips2015} and \cite%
{Liao&Shi2020}. \cite{LeePhillips2015} explored model selection in panel
data settings using nested hypothesis testing. Our work complements theirs
by introducing approaches for testing non-nested hypotheses. \cite%
{Liao&Shi2020} examined the Vuong test for semiparametric models in a
cross-sectional setting, whereas we extend the analysis to a panel data
context with fixed effects. Our paper and \cite{LeePhillips2015} share the
common feature in the sense that both analyses utilize application of
profiling to the Kullback--Leibler information criterion (KLIC). It is
well-known that profile likelihood does not share the standard properties of
the genuine likelihood function. Modification of profile likelihood was
discussed a way of overcoming this problem. See \cite{ArellanoHahn2007}, 
\cite{ArellanoHahn2016}, \cite{CoxReid1987}, \cite{DiCiccioStern1993}, \cite%
{DiCiccioMartinSternYoung}, \cite{CoxFergusonReid}, and \cite{PaceSalvan},
for example. Following this tradition, our paper as well as \cite%
{LeePhillips2015} propose a modified profile likelihood that provides
approximations to the standard KLIC, and provided a valid method of model
selection in panel data models with individual fixed effects. Interestingly,
we find some close connection between the modification in \cite{Liao&Shi2020}
and the panel type modification following \cite{ArellanoHahn2007}, \cite%
{ArellanoHahn2016}, and \cite{LeePhillips2015}.

Recently developed methods in panel data allow applied researchers to choose
among a broad range of alternative models for separable, interactive, or
discretely supported unobserved heterogeneity across units and time. It is
widely understood that parameter estimates are often sensitive to the chosen
specification, however there is often no clear theoretical guidance on how
to choose among alternative models. Our paper makes further contribution in
this regard by considering the cases where the definition of ``individuals''
may not be agreed across different models. In empirical practice, the
group/cluster structure may not be obviously agreed among researchers. The
competing group/cluster structures may be nested in the sense that one
structure is ``finer'' than the other one, but they may be non-nested. In
this situation, extension of the insights in previous papers such as \cite%
{LeePhillips2015} or \cite{Liao&Shi2020} may not be straightforward. We
establish that the classical Vuong test is valid after some modification. In
doing so, we also make a technical contribution that may be of some
independent interest. \cite{FernandezValWeidener2016} analyzed the
asymptotic distribution of the (pseudo) MLE in nonlinear panel data models
with individual and time effects. Under the assumption that the objective
function is globally concave in all the parameters, they established the
asymptotic bias formula. Our paper contains a result\footnote{%
See Theorem \ref{Rate_gamma} in the Online Appendix and Theorem \ref%
{Rep_Theta} in Appendix \ref{Sec:AP2}, along with the related discussion
therein, for further details.} that relaxes this concavity assumption,
although it is achieved at the cost of restricting the group structure.
Therefore, our result can be argued to complement \cite%
{FernandezValWeidener2016}.

The remainder of the paper is organized as follows. In Section \ref{Sec:
Simple-Vuong},\ we introduce the Vuong test for comparing classical panel
data models that allow for potential group heterogeneity. Section \ref%
{sec:TWE} presents a test for comparing the two-way fixed effects (TWFE)
model against a heterogeneous time fixed effects model. Section \ref%
{sec:conclusion} concludes.\ The proofs of the main results, along with
auxiliary lemmas, are provided in the Appendix. Additionally, Appendix \ref%
{Sec:AP2} develops a general asymptotic theory for estimators in panel
models with individual and time fixed effects, as well as the asymptotic
distribution of the statistic $QLR_{n,T}$ in this broader setting -- results
that are of independent interest.\footnote{%
The technical proofs for this general theory, together with the proofs of
the auxiliary lemmas used to establish the main results in Sections \ref%
{Sec: Simple-Vuong} and \ref{sec:TWE}, are provided in the Online Appendix.}

The following notation will be adopted throughout the paper.\ We use $K$ to
denote a generic strictly positive constant that may vary from one instance
to another but remains independent of the panel dimensions $n$ {and }$T$.\
We adopt the convention that a summation over an empty set equals zero. We
use $a\equiv b$ to indicate that $a$ is defined as $b$. For real numbers $%
a_{1},\ldots ,a_{m}$, $(a_{j})_{j\leq m}\equiv (a_{1},\ldots ,a_{m})^{\top }$%
. For any matrix $A$, $A^{\top }$\ denotes the transpose of $A$,\ and $%
\left\Vert A\right\Vert $ denotes the Euclidean norm of $A$. For any doubly
indexed sequence $a_{i,t}$ (where $i=1,\ldots ,n$ and $t=1,\ldots ,T$), we
define\ $\bar{a}_{i}\equiv T^{-1}\sum_{t\leq T}a_{i,t}$, $\bar{a}_{t}\equiv
n^{-1}\sum_{i\leq n}a_{i,t}$ and $\bar{a}\equiv (nT)^{-1}\sum_{t\leq
T}\sum_{i\leq n}a_{i,t}$. The summation $\sum_{i^{\prime }\neq i}$ is taken
over all $i^{\prime }$ except $i$, which means $\sum_{i^{\prime }\neq
i}a_{i^{\prime },t}=\sum_{i^{\prime }=1}^{i-1}a_{i^{\prime
},t}+\sum_{i^{\prime }=i+1}^{n}a_{i^{\prime },t}$.\ For two sequences of
positive numbers $a_{n}$ and $b_{n}$, we write $a_{n}\succ b_{n}$ if $%
a_{n}\geq c_{n}b_{n}$ for some strictly positive sequence $c_{n}\rightarrow
\infty $. The notation $\left\Vert \cdot \right\Vert _{p}$ denotes the $%
L_{p} $-norm. Finally, for any function $g(\cdot )$, define $\mathbb{E}_{T}%
\left[ g\left( z_{i,t}\right) \right] \equiv T^{-1}\sum_{t\leq T}\mathbb{E}%
\left[ g\left( z_{i,t}\right) \right] $ and $\widehat{\mathbb{E}}_{T}\left[
g\left( z_{i,t}\right) \right] \equiv T^{-1}\sum_{t\leq T}g\left(
z_{i,t}\right) $.

\section{Vuong Test for Classical Panel Models\ \label{Sec: Simple-Vuong}}

In this section, we investigate the Vuong test for panel models with
time-invariant individual effects.\footnote{%
A simple version of the test appeared in \cite{hahn-liu}, which later became
a chapter in Liu's UCLA dissertation. Our current paper substantially
extends and generalizes the earlier simple result.} With a dataset\ $\left\{
z_{i,t}\right\} _{i\leq n,t\leq T}$, we compare two models, denoted as model
1 and model 2 respectively. Model 1 represents the classical panel model
with individual effects. In contrast, model 2 may specify a different joint
likelihood function and/or adopt a distinct clustering or grouping structure
for the individual effects.

For each model $j$ ($j=1,2$), the joint log-likelihood function simplifies
from the general specification in (\ref{general_spec}) to the following
form: 
\begin{equation*}
L_{j,n,T}(\phi _{j})\equiv \sum_{i\leq n}\sum_{t\leq T}\log
f_{j}(z_{i,t};\theta _{j},\gamma _{g_{j}(i)})=\sum_{g\in \mathcal{G}%
_{j}}\sum_{i\in I_{j,g}}\sum_{t\leq T}\psi _{j}\left( z_{i,t};\phi
_{j,g}\right) ,
\end{equation*}%
where\ $\phi _{j}\equiv (\theta _{j}^{\top },(\gamma _{j,g})_{g\in \mathcal{G%
}_{j}}^{\top })^{\top }$, $\psi _{j}\left( z_{i,t};\phi _{j,g}\right) \equiv
\log f_{j}(z_{i,t};\theta _{j},\gamma _{j,g})$, $\phi _{j,g}\equiv (\theta
_{j}^{\top },\gamma _{j,g})^{\top }$,\ $\mathcal{G}_{j}\equiv \{1,\ldots
,G_{j}\}$ defines the range of the grouping function $g_{j}(\cdot )$, and $%
I_{j,g}$ ($g\in \mathcal{G}_{j}$) represents the partition of $\{1,\ldots
,n\}$ induced by $g_{j}(\cdot )$. For model 1, it follows that $\mathcal{G}%
_{1}=\{1,\ldots ,n\}$, implying that each individual has its own group. For
model 2, however, the partitions $\{I_{2,g}\}_{g\in \mathcal{G}_{2}}$ may
form any arbitrary grouping of $\{1,\ldots ,n\}$\ allowing for greater
flexibility in the specification of individual effects.\footnote{%
This means that model 1 is nested within model 2 in terms of the grouping
structure, although the likelihood specifications may differ. In many
applications, the grouping structure is expected to be identical across
models.}

As this section is relatively long, we provide a roadmap to facilitate
reading. Sections \ref{Sec: Incid_Para} - \ref{section:lm-variance} provides
the theoretical foundation of the infeasible test, while Sections \ref%
{section-biasestimation} - \ref{section:varianceestimation} presents the
actual algorithm for implementation of the feasible version of the test. In
Section \ref{Sec: Incid_Para}, we provide an intuitive overview of the
adjustment to the maximized joint likelihood and discuss its importance for
model comparison using the Vuong test in the panel data models considered in
this paper. In Section \ref{section:log-likelihoods}, we present the log
likelihood of each model. In Section \ref{section:lm-bias}, we present an
infeasible adjustment that corrects for the bias in the likelihood, and in
Section \ref{section:lm-variance}, we present a theoretical analysis for the
asymptotic variance of the modified likelihood. Asymptotic properties of the
(infeasible) modified test statistic is presented in Theorem ~\ref{C_L0}. In
Sections \ref{section-biasestimation} and \ref{section:varianceestimation},
we present actual algorithms for estimation of the modified log likelihoods
and the asymptotic variances. Section \ref{section:feasiblestatistic}
presents Theorem~\ref{C_T1} that characterizes the asymptotic properties of
the feasible test statistic.

\subsection{Incidental Parameter Problem in Panel Data Models \label{Sec:
Incid_Para}}

To streamline notation and improve clarity, we focus here on a simplified
version of (\ref{general_spec}) that includes only individual fixed effects,
specified as:%
\begin{equation}
L_{n,T}(\phi )\equiv \sum_{i\leq n}\sum_{t\leq T}\log f\left( z_{i,t};\theta
,\gamma _{i}\right) =\sum_{i\leq n}\sum_{t\leq T}\psi \left( z_{i,t};\phi
_{i}\right) ,  \label{Fixed_Eff_Model}
\end{equation}%
where $\psi \left( z_{i,t};\phi _{i}\right) \equiv \log f\left(
z_{i,t};\theta ,\gamma _{i}\right) $, $\phi _{i}\equiv (\theta ^{\top
},\gamma _{i})^{\top }$\ and $\phi \equiv (\theta ^{\top },(\gamma
_{i})_{i\leq n}^{\top })^{\top }$.\ Since QLR statistic defined in (\ref%
{G_QLR}) involves the maximized joint likelihood\ $L_{n,T}(\hat{\phi})$,
understanding the QLR's properties requires investigating the estimation
error of $L_{n,T}(\hat{\phi})$.

Let $\phi ^{\ast }\equiv (\theta ^{\ast \top },(\gamma _{i}^{\ast })_{i\leq
n}^{\top })^{\top }$ denote the pseudo-true parameter which maximizes $%
\mathbb{E}\left[ L_{n,T}(\phi )\right] $.\footnote{%
Potential dependence of the pseudo-true parameters $\phi ^{\ast }$ on the
sample size is not made explicit for notational simplicity throughout the
paper.} Define:%
\begin{equation}
\psi \left( z_{i,t}\right) \equiv \psi \left( z_{i,t};\phi _{i}^{\ast
}\right) \text{, \ }\psi _{\gamma }\left( z_{i,t}\right) \equiv \partial
\psi \left( z_{i,t};\phi _{i}^{\ast }\right) /\partial \gamma _{i}\text{ \
and \ }\psi _{\gamma \gamma }\left( z_{i,t}\right) \equiv \partial ^{2}\psi
\left( z_{i,t};\phi _{i}^{\ast }\right) /\partial \gamma _{i}^{2},
\label{derv_1}
\end{equation}%
where $\phi _{i}^{\ast }\equiv (\theta ^{\ast \top },\gamma _{i}^{\ast
})^{\top }$. It can be shown\footnote{%
See Theorem \ref{Rep_L} in the Appendix, specialized to the case $\mathcal{G}%
=\{1,\ldots ,n\}$ and $\mathcal{M}=\{1\}$, so that $G=n$ and $M=1$}\ that
under the asymptotics with $nT^{-1}\rightarrow \rho $\ where $\rho \in
(0,\infty )$:%
\begin{equation}
L_{n,T}(\hat{\phi})-\mathbb{E}\left[ L_{n,T}(\phi ^{\ast })\right]
=L_{n,T}(\phi ^{\ast })-\mathbb{E}\left[ L_{n,T}(\phi ^{\ast })\right]
+2^{-1}\sum_{i\leq n}\tilde{\Psi}_{\gamma ,i}^{2}+o_{p}(n^{1/2}),
\label{L_nT_exp}
\end{equation}%
where $L_{n,T}(\phi ^{\ast })=\sum_{i\leq n}\sum_{t\leq T}\psi \left(
z_{i,t}\right) $, and for any $i\leq n$,%
\begin{equation}
\tilde{\Psi}_{\gamma ,i}\equiv T^{-1/2}\sum_{t\leq T}\tilde{\psi}_{\gamma
}^{\ast }\left( z_{i,t}\right) ,\text{ \ \ and \ \ }\tilde{\psi}_{\gamma
}^{\ast }\left( z_{i,t}\right) \equiv \frac{\psi _{\gamma }\left(
z_{i,t}\right) -\mathbb{E}\left[ \psi _{\gamma }\left( z_{i,t}\right) \right]
}{(\mathbb{E}_{T}\left[ -\psi _{\gamma \gamma }\left( z_{i,t}\right) \right]
)^{1/2}}.  \label{derv_2}
\end{equation}%
The expansion in (\ref{L_nT_exp}) shows that the estimation error of $%
L_{n,T}(\hat{\phi})$ has two leading components. The first component, $%
L_{n,T}(\phi ^{\ast })-\mathbb{E}\left[ L_{n,T}(\phi ^{\ast })\right] $,
results from estimating\ $\mathbb{E}\left[ L_{n,T}(\phi ^{\ast })\right] $
at the given pseudo-true parameter $\phi ^{\ast }$. When scaled by a factor
of $(nT)^{-1/2}$, this component converges in distribution to a normal
distribution under some regularity conditions. The second component, $%
2^{-1}\sum_{i\leq n}\tilde{\Psi}_{\gamma ,i}^{2}$, arises from estimating
the unknown pseudo-true parameter $\phi ^{\ast }$ and is determined by the
estimation error of the incidental parameters $(\gamma _{i}^{\ast })_{i\leq
n}$.\footnote{%
Since the dimension of $\theta ^{\ast }$ is fixed, the contribution of its
estimation error to $L_{n,T}(\hat{\phi})$ is asymptotically negligible
compared with the two leading components, and is therefore included in the
remainder term of (\ref{L_nT_exp}), represented as $o_{p}(n^{1/2})$.}

From the expansion in (\ref{L_nT_exp}), it is evident that in panel data
models, the $L_{n,T}(\hat{\phi})$ has a first-order bias, stemming from $%
2^{-1}\sum_{i\leq n}\tilde{\Psi}_{\gamma ,i}^{2}$. To evaluate this bias
explicitly, suppose that $\{\tilde{\psi}_{\gamma }\left( z_{i,t}\right)
\}_{i,t}$ are i.i.d. across $t\leq T$ for any $i$. Since $\mathbb{E}[\tilde{%
\psi}_{\gamma }^{\ast }\left( z_{i,t}\right) ]=0$, it follows that $\mathbb{E%
}[2^{-1}\sum_{i\leq n}\tilde{\Psi}_{\gamma ,i}^{2}]=2^{-1}\sum_{i\leq
n}\sigma _{\gamma ,i}^{2}$, where $\sigma _{\gamma ,i}^{2}\equiv \mathbb{E}[%
\tilde{\psi}_{\gamma }^{\ast }\left( z_{i,t}\right) ^{2}]$. Assuming that $%
\sigma _{\gamma ,i}^{2}$ is bounded above and below, the bias from the mean
of $2^{-1}\sum_{i\leq n}\tilde{\Psi}_{\gamma ,i}^{2}$ is of the same order
as $L_{n,T}(\phi ^{\ast })-\mathbb{E}\left[ L_{n,T}(\phi ^{\ast })\right] $.
This provides the explanation why the maximized joint likelihood $L_{n,T}(%
\hat{\phi})$ should be modified with a bias correction based on the mean of\ 
$2^{-1}\sum_{i\leq n}\tilde{\Psi}_{\gamma ,i}^{2}$.

The above discussion demonstrates that the maximized joint likelihood\ $%
L_{n,T}(\hat{\phi})$ requires a first-order bias correction in the form of a
consistent estimator of $2^{-1}\sum_{i\leq n}\mathbb{E}[\tilde{\Psi}%
_{\gamma,i}^{2}]$. As a result, the QLR statistic in the classical Vuong
test must also be adjusted with a bias correction to ensure valid inference.
This bias correction for the QLR statistic is given by the difference in the
bias of $L_{j,n,T}(\hat{\phi}_{j})$ ($j=1,2$) from the two competing models.

There is an additional complication. Depending on whether the models are
nested or overlapping, the first components, $L_{j,n,T}(\phi_{j}^{\ast })-%
\mathbb{E}[L_{j,n,T}(\phi_{j}^{\ast})]$, from the two models may cancel out.
See \cite{Vuong1989} for discussion in the cross section context. In such
cases, the second components, denoted as $2^{-1}\sum_{i\leq n}\tilde{\Psi }%
_{j,\gamma,i}^{2}$ ($j=1,2$),\ can become the primary drivers of the
asymptotic distribution of the bias-corrected QLR statistic.\footnote{%
In the classical nested case, where the functional form of the likelihood $f$
is identical across the two competing models, the second component of the
test statistic cancels out. In this setting, bias correction of the
estimator itself, along with further refinements of (\ref{L_nT_exp}), is
needed. This case was previously studied by \cite{LeePhillips2015}, and we
do not pursue it further in this paper.}\ Consequently, the estimation error
of the incidental parameters influences not only the bias but also the
variance of the QLR statistic.

\subsection{Quasi-likelihood Estimator\label{section:log-likelihoods}}

The pseudo true parameter $\phi _{j}^{\ast }$ for model $j$ is estimated as
the maximizer $\hat{\phi}_{j}$ of $L_{j,n,T}(\phi _{j})$ where $\phi
_{j}\equiv (\theta _{j}^{\top },(\gamma _{j,g})_{g\in \mathcal{G}_{j}}^{\top
})^{\top }$. The estimator\ $\hat{\phi}_{j}$\ satisfies the following
first-order conditions:%
\begin{equation}
\sum_{g\in \mathcal{G}_{j}}\sum_{i\in I_{j,g}}\sum_{t\leq T}\psi _{j,\theta
}(z_{i,t};\hat{\phi}_{j,g})=0_{d_{\theta }\times 1},  \label{C_foc_1}
\end{equation}%
and for any $g\in \mathcal{G}_{j}$,%
\begin{equation}
\sum_{i\in I_{j,g}}\sum_{t\leq T}\psi _{j,\gamma }(z_{i,t};\hat{\phi}%
_{j,g})=0,  \label{C_foc_2}
\end{equation}%
where $\phi _{j,g}\equiv (\theta _{j}^{\top },\gamma _{j,g})^{\top }$\ and $%
\psi _{j,a}(z_{i,t};\phi _{j,g})\equiv \partial \psi _{j}(z_{i,t};\phi
_{j,g})/\partial a$ for $a\in \{\theta _{j},\gamma _{j,g}\}$.\ The estimator 
$\hat{\phi}_{j}$ is subsequently used to compute the maximized joint
likelihood $L_{j,n,T}(\hat{\phi}_{j})$\ for model $j$.

Following the discussion in the previous subsection, the maximized joint
quasi-likelihood for each model must be adjusted to account for the
respective first-order bias, which can be derived from the asymptotic
expansion of $L_{j,n,T}(\hat{\phi}_{j})$. To present this expansion in a
more general setting than previously considered, we first update the
notations introduced earlier. Let $\psi _{j}\left( z_{i,t}\right) $, $\psi
_{j,\gamma }\left( z_{i,t}\right) $, $\psi _{j,\gamma \gamma }\left(
z_{i,t}\right) $ and $\tilde{\Psi}_{j,\gamma ,i}$ represent the counterparts
to their definitions in (\ref{derv_1}) and (\ref{derv_2}), with $\phi
_{j,i}\equiv \phi _{j,g}$,%
\begin{equation*}
\psi _{j,\gamma }^{\ast }\left( z_{i,t}\right) \equiv \psi _{j,\gamma
}\left( z_{i,t}\right) -\mathbb{E}[\psi _{j,\gamma }\left( z_{i,t}\right) ]%
\text{ \ and \ }\tilde{\psi}_{j,\gamma }^{\ast }\left( z_{i,t}\right) \equiv 
\frac{\psi _{j,\gamma }^{\ast }\left( z_{i,t}\right) }{(n_{j,g}^{-1}\sum_{i%
\in I_{j,g}}\mathbb{E}_{T}\left[ -\psi _{j,\gamma \gamma }\left(
z_{i,t}\right) \right] )^{1/2}},
\end{equation*}%
for any $i\in I_{j,g}$ and any $g\in \mathcal{G}_{j}$.\footnote{%
The demeaned version of $\psi _{j,\gamma }\left( z_{i,t}\right) $, i.e., $%
\psi _{j,\gamma }^{\ast }\left( z_{i,t}\right) $ is introduced here because
the first-order condition for $\phi _{j}^{\ast }$ only ensures that $%
\sum_{i\in I_{j,g}}\sum_{t\leq T}\mathbb{E}[\psi _{j,\gamma }\left(
z_{i,t}\right) ]=0$. This condition does not guarantee that $\mathbb{E}[\psi
_{j,\gamma }\left( z_{i,t}\right) ]=0$. If $\psi _{j,\gamma }\left(
z_{i,t}\right) $ is stationry across $t$ for each $i$, the first-order
condition for $\phi _{j}^{\ast }$ is simplified to $\sum_{i\in I_{j,g}}%
\mathbb{E}[\psi _{j,\gamma }\left( z_{i,t}\right) ]=0$. In this case, $%
\mathbb{E}[\psi _{1,\gamma }\left( z_{i,t}\right) ]=0$ holds for model 1,
but is still not guaranteed for model 2.}

\subsection{Infeasible Modified QLR Statistic - Bias\label{section:lm-bias}}

With the updated notations, we can obtain\footnote{%
See Theorem \ref{Rep_L} together with the decomposition in (\ref{Exp_2nd_1})
in the Appendix.} the following expansion:%
\begin{equation}
L_{j,n,T}(\hat{\phi}_{j})=T^{1/2}\sum_{i\leq n}(\tilde{\Psi}_{j,i}+\tilde{V}%
_{j,i}+\tilde{U}_{j,i})+o_{p}(G_{j}^{1/2}),  \label{C_Exp_1}
\end{equation}%
where the components are defined\footnote{%
The $\tilde{\Psi}_{j,\gamma ,i}$ is not the derivative of $\tilde{\Psi}%
_{j,i} $, as can be seen from its definition (\ref{derv_2}).} as:%
\begin{equation*}
\tilde{\Psi}_{j,i}\equiv T^{-1/2}\sum_{t\leq T}\psi _{j}(z_{i,t};\phi
_{j,i}^{\ast })\text{, \ \ }\tilde{V}_{j,i}\equiv \frac{\tilde{\Psi}%
_{j,\gamma ,i}^{2}}{2n_{j,g}T^{1/2}}\text{ \ \ and \ \ }\tilde{U}%
_{j,i}\equiv \sum_{\{i^{\prime }\in I_{j,g}:i^{\prime }<i\}}\frac{\tilde{\Psi%
}_{j,\gamma ,i}\tilde{\Psi}_{j,\gamma ,i^{\prime }}}{n_{j,g}T^{1/2}}.
\end{equation*}%
It is reasonable to expect $\mathbb{E}[\tilde{U}_{j,i}]=0$ for $j=1,2$ and
for any $i\leq n$,\footnote{%
For example, it is implied by Assumption \ref{A1}\ in the Appendix.}\textbf{%
\ }which implies that the first-order bias arises from the mean of $\tilde{V}%
_{j,i}$. To account for this bias, we define the (infeasible) modified
maximized joint likelihood as:%
\begin{equation}
LM_{j,n,T}^{\ast }(\hat{\phi}_{j})\equiv L_{j,n,T}(\hat{\phi}%
_{j})-T^{1/2}\sum_{i\leq n}\mathbb{E}[\tilde{V}_{j,i}].  \label{C_Exp_2}
\end{equation}%
Using this expression, we define the (infeasible) modified QLR statistic,
whose asymptotic expansion follows from (\ref{C_Exp_1}): 
\begin{equation}
MQLR_{n,T}^{\ast }\equiv (nT)^{-1/2}\left( LM_{1,n,T}^{\ast }(\hat{\phi}%
_{1})-LM_{2,n,T}^{\ast }(\hat{\phi}_{2})\right) =n^{-1/2}\sum_{i\leq n}(%
\tilde{\Psi}_{i}+\tilde{V}_{i}^{\ast }+\tilde{U}_{i})+o_{p}(T^{-1/2}),
\label{C_Exp_3}
\end{equation}%
where $\tilde{\Psi}_{i}\equiv \tilde{\Psi}_{1,i}-\tilde{\Psi}_{2,i}$, $%
\tilde{V}_{i}^{\ast }\equiv \tilde{V}_{1,i}-\tilde{V}_{2,i}-\mathbb{E}[%
\tilde{V}_{1,i}-\tilde{V}_{2,i}]$ and $U_{i}\equiv \tilde{U}_{1,i}-\tilde{U}%
_{2,i}$.

\subsection{Infeasible Modified QLR Statistic - Asymptotic Distribution\label%
{section:lm-variance}}

After bias correction, the main remaining components arising from the
estimation errors of the incidental parameters, i.e., $\tilde{V}_{i}^{\ast }+%
\tilde{U}_{i}$ primarily affect the variance of the modified QLR statistic\ $%
MQLR_{n,T}^{\ast}$.\ This effect becomes particularly significant when the
two models are overlapping. In such cases, the first component on the
right-hand side of (\ref{C_Exp_1}), i.e., $n^{-1/2}\sum_{i\leq n}\tilde{\Psi 
}_{i}$, can become arbitrarily small or even vanish, leaving the asymptotic
distribution of $MQLR_{n,T}^{\ast}$ predominantly determined by the
estimation errors of the incidental parameters, specifically $%
n^{-1/2}\sum_{i\leq n}(\tilde{V}_{i}^{\ast}+\tilde{U}_{i})$.

To facilitate the discussion of the variance and asymptotic distribution of $%
MQLR_{n,T}^{\ast}$, we follow the literature (see, e.g., \cite{Vuong1989}, 
\cite{Shi2015b} and \cite{Liao&Shi2020}) and formally define the
relationships between models in the panel setting below.

\begin{definition}
(i) Model 1 and model 2 are \textbf{strictly non-nested} if there do not
exist $\phi _{1,i}\in \Phi _{1}$ and $\phi _{2,i}\in \Phi _{12}$ such that%
\begin{equation}
\psi _{1}\left( z;\phi _{1,i}\right) =\psi _{2}\left( z;\phi _{2,i}\right) 
\text{,}  \label{Def_1}
\end{equation}%
for any $i$ and any $z$ in the support of $z_{i,t}$; (ii) the two models are 
\textbf{overlapping} if they are not strictly non-nested; (iii) model 1 and
model 2 are said to be \textbf{nested }if, for any $\phi _{j,i}\in \Phi _{j}$%
, there exists a $\phi _{j^{\prime },i}\in \Phi _{j^{\prime }}$ (where $%
j,j^{\prime }=1,2$ and $j\neq j^{\prime }$) such that (\ref{Def_1}) holds.
\end{definition}

When the two models are strictly non-nested, the variance of $%
n^{-1/2}\sum_{i\leq n}\tilde{\Psi}_{i}$ is bounded away from zero. In this
case, the component $n^{-1/2}\sum_{i\leq n}(\tilde{V}_{i}^{\ast }+\tilde{U}%
_{i})$ on the right-hand side of (\ref{C_Exp_1}) becomes asymptotically
negligible,\ as it is of order $O_{p}(T^{-1/2})$ under certain regularity
conditions.\footnote{%
This order follows from an approximation of\ $n^{-1}\sum_{i\leq n}(\mathrm{%
Var}(\tilde{V}_{i})+\mathrm{Var}(\tilde{U}_{i}))$ (see (\ref%
{P_C_L_Auxillary9_11}) in the Online Appendix, which appears in the proof of
Lemma \ref{C_L_Auxillary9}), together with an upper bound on the variance of 
$\tilde{\psi}_{j,\gamma }^{\ast }\left( z_{i,t}\right) $ for $j=1,2$.}\
Consequently, the asymptotic distribution of $MQLR_{n,T}^{\ast }$ is
primarily determined by $n^{-1/2}\sum_{i\leq n}\tilde{\Psi}_{i}$.
Conversely, when the two models are nested, $\tilde{\Psi}_{i}=0$ under the
null hypothesis. In this scenario, the asymptotic distribution of $%
MQLR_{n,T}^{\ast }$ is determined by\ $n^{-1/2}\sum_{i\leq n}(\tilde{V}%
_{i}^{\ast }+\tilde{U}_{i})$. As a result, deriving the asymptotic
distribution of $MQLR_{n,T}^{\ast }$ is relatively straightforward, when the
models are known to be either strictly non-nested or nested. In both cases,
the dominating term on the right-hand side of (\ref{C_Exp_1}) is
identifiable, and the asymptotic distribution of $MQLR_{n,T}^{\ast }$ can be
deduced from that of the dominating term.

The asymptotic distribution of $QLR_{n,T}^{\ast }$ is more challenging to
derive when the two models are overlapping. Because in this case, it is in
general unclear which components between $n^{-1/2}\sum_{i\leq n}\tilde{\Psi}%
_{i}$ and $n^{-1/2}\sum_{i\leq n}(\tilde{V}_{i}^{\ast }+\tilde{U}_{i})$
could be the dominating term, and both could be equally important. To
address this issue, we follow \cite{Liao&Shi2020} to consider both terms
together when deriving the asymptotic distribution of $MQLR_{n,T}^{\ast }$.

\begin{theorem}
\label{C_L0} Under Assumptions \ref{A1} and \ref{C_A1}\ in the Appendix, we
have%
\begin{equation}
\frac{MQLR_{n,T}^{\ast }-\overline{QLR}_{n,T}}{\omega _{n,T}}\rightarrow
_{d}N(0,1)\text{,}  \label{C_P1_1}
\end{equation}%
where 
\begin{equation}
\omega _{n,T}^{2}\equiv n^{-1}\sum_{i\leq n}\mathrm{Var}(\tilde{\Psi}_{i}+%
\tilde{V}_{i}+\tilde{U}_{i})\text{ \ and \ }\overline{QLR}_{n,T}\equiv
(nT)^{-1/2}\mathbb{E}[L_{1,n,T}(\phi _{1}^{\ast })-L_{1,n,T}(\phi _{2}^{\ast
})].  \label{def_omega}
\end{equation}
\end{theorem}

\begin{remark}
Theorem \ref{C_L0} provides the asymptotic distribution of $MQLR_{n,T}^{\ast
}$ under both the null and alternative hypotheses. Since $\overline{QLR}%
_{n,T}=0$ under the null hypothesis, we can construct a test for the null by
using a consistent estimator of $\omega _{n,T}$, along with a feasible
version of the modified QLR statistic. This feasible statistic replaces the
unknown bias correction terms in $MQLR_{n,T}^{\ast }$ with their consistent
estimators.\footnote{%
We impose Assumption \ref{C_A2} to simplify the estimation of the bias
correction and the variance of $MQLR_{n,T}^{\ast }$.\ All assumptions are
collected in the Appendix.}
\end{remark}

\subsection{Estimation of Bias \label{section-biasestimation}}

We first discuss the estimation of the bias correction terms. Under
Assumption \ref{C_A2}\ in the Appendix, the definition of $\tilde{V}_{j,i}$
implies that: 
\begin{equation}
T^{1/2}\sum_{i\leq n}\mathbb{E}[\tilde{V}_{j,i}]=\sum_{g\in \mathcal{G}%
_{j}}\sum_{i\in I_{j,g}}(2n_{j,g})^{-1}\sigma _{j,\gamma ,i}^{2}\text{, \ \
where }\sigma _{j,\gamma ,i}\equiv (\mathbb{E}_{T}[\tilde{\psi}_{j,\gamma
}^{\ast }\left( z_{i,t}\right) ^{2}])^{1/2}.  \label{C_Bias_1}
\end{equation}%
This motivates a straightforward plug-in estimator of the bias correction
term for model $j$, relying on a consistent estimator of $\sigma _{j,\gamma
,i}^{2}$\ for any $i\in I_{j,g}$ and any $g\in \mathcal{G}_{j}$: 
\begin{equation}
\hat{\sigma}_{j,\gamma ,i}^{2}\equiv \frac{\widehat{\mathbb{E}}_{T}[\hat{\psi%
}_{j,\gamma }(z_{i,t})^{2}]-(\widehat{\mathbb{E}}_{T}[\hat{\psi}_{j,\gamma
}(z_{i,t})])^{2}}{|\hat{\Psi}_{j,\gamma \gamma }(\hat{\phi}_{j,g})|},
\label{C_Var_1}
\end{equation}%
where%
\begin{equation}
\hat{\psi}_{j,\gamma }(z_{i,t})\equiv \psi _{j,\gamma }(z_{i,t};\hat{\phi}%
_{j,g})\text{ \ \ and \ \ }\hat{\Psi}_{j,\gamma \gamma }(\phi _{j,g})\equiv
n_{j,g}^{-1}\sum_{i\in I_{j,g}}\widehat{\mathbb{E}}_{T}[\psi _{j,\gamma
\gamma }(z_{i,t};\phi _{j,g})].  \label{C_Var_2}
\end{equation}%
The bias correction term for model $j$ is then estimated as%
\begin{equation}
R_{j,n,T}(\hat{\phi}_{j})\equiv \sum_{g\in \mathcal{G}_{j}}\sum_{i\in
I_{j,g}}(2n_{j,g})^{-1}\hat{\sigma}_{j,\gamma ,i}^{2}.  \label{C_B_est}
\end{equation}%
Using this, the feasible modified QLR statistic is defined as: 
\begin{equation}
MQLR_{n,T}\equiv (nT)^{-1/2}\left( LM_{1,n,T}(\hat{\phi}_{1})-LM_{2,n,T}(%
\hat{\phi}_{2})\right) ,  \label{C_MQLR}
\end{equation}%
where 
\begin{equation*}
LM_{j,n,T}(\hat{\phi}_{j})\equiv L_{j,n,T}(\hat{\phi}_{j})-R_{j,n,T}(\hat{%
\phi}_{j})
\end{equation*}%
denotes the modified maximized joint likelihood for model $j$.

\begin{theorem}
\textit{\label{C_L1} }Under Assumptions \ref{A1}, \ref{C_A1} and \ref{C_A2}
in the Appendix, we have%
\begin{equation*}
\frac{R_{j,n,T}(\hat{\phi}_{j})-T^{1/2}\sum_{i\leq n}\mathbb{E}[\tilde{V}%
_{j,i}]}{(nT)^{1/2}\omega _{n,T}}=O_{p}(T^{-1/2}).
\end{equation*}
\end{theorem}

Theorem \ref{C_L1} establishes the consistency of $R_{j,n,T}(\hat{\phi}_{j})$
as an estimator of the bias of the maximized joint likelihood $L_{j,n,T}(%
\hat{\phi}_{j})$ in model $j$. Combined with the result in (\ref{C_P1_1}),
this directly leads to the asymptotic normality of the modified QLR
statistic $QLR_{n,T}$:%
\begin{equation}
\frac{MQLR_{n,T}-\overline{QLR}_{n,T}}{\omega_{n,T}}\rightarrow_{d}N(0,1)%
\text{.}  \label{C_L1_2}
\end{equation}
To apply this result for inference under the null hypothesis, it remains to
construct a consistent estimator for $\omega_{n,T}^{2}$, which we now
proceed to address.

\subsection{Estimation of Variance\ \label{section:varianceestimation}}

A consistent estimator of $\omega _{n,T}^{2}$, which was defined in (\ref%
{def_omega}), can be constructed using the sample analog of the variance of $%
n^{-1/2}\sum_{i\leq n}\tilde{\Psi}_{i}$, and a variance correction term to
account for estimation errors of the incidental parameters. Define $\sigma
_{n,T}^{2}\equiv n^{-1}\sum_{i\leq n}\mathrm{Var}(\tilde{\Psi}_{i})$ and $%
\Delta \psi (z_{i,t},\phi _{i})\equiv \psi _{1}(z_{i,t};\phi _{1,i})-\psi
_{2}(z_{i,t};\phi _{2,i})$. Under Assumption \ref{C_A2}, a natural estimator
for $\sigma _{n,T}^{2}$ is given by: 
\begin{equation}
\hat{\sigma}_{n,T}^{2}\equiv (nT)^{-1}\sum_{i\leq n}\sum_{t\leq T}\Delta
\psi (z_{i,t};\hat{\phi}_{i})^{2}-(nT)^{-1}\left( MQLR_{n,T}\right) ^{2}.
\label{Var_Est_1}
\end{equation}%
The theoretical property of $\hat{\sigma}_{n,T}^{2}$ is provided in the
lemma below.\footnote{%
The sample variance\ $\hat{\sigma}_{n,T}^{2}$ will be used below as a
building block for variance estimation and therefore requires $\{\Delta \psi
(z_{i,t},\phi _{i}^{\ast })\}_{t}$ to be serially uncorrelated, as assumed
in Assumption \ref{C_A2}. If serial correlation is present, an
autocorrelation-consistent estimator is required. Since variance estimation
is mainly needed under the null for size control, Assumption \ref{C_A2} is
imposed only under the null; Lemma \ref{C_T2} in the Appendix shows that
consistency of the test does not rely on this assumption.}

\begin{theorem}
\textit{\label{C_L2}\ }Under Assumptions \ref{A1}, \ref{C_A1} and \ref{C_A2}
in the Appendix, we have:%
\begin{equation}
\frac{\hat{\sigma}_{n,T}^{2}-(\sigma _{n,T}^{2}+2\sigma _{S,n,T}^{2})}{%
\omega _{n,T}^{2}}=O_{p}(T^{-1/2}),  \label{C_L2_1}
\end{equation}%
where%
\begin{equation}
\sigma _{S,n,T}^{2}\equiv (2nT)^{-1}\sum_{g\in \mathcal{G}_{2}}\sum_{i\in
I_{2,g}}\left( \sigma _{1,\gamma ,i}^{4}+n_{2,g}^{-2}\sigma _{2,\gamma
,i}^{2}\sum_{i^{\prime }\in I_{2,g}}s_{2,\gamma ,i^{\prime
}}^{2}-2n_{2,g}^{-1}\sigma _{12,\gamma ,i}^{2}\right) ,  \label{C_Var_3}
\end{equation}%
$s_{2,\gamma ,i}^{2}\equiv \mathbb{E}_{T}[\tilde{\psi}_{2,\gamma }\left(
z_{i,t}\right) ^{2}]$ and $\sigma _{12,\gamma ,i}\equiv \mathbb{E}_{T}[%
\tilde{\psi}_{1,\gamma }^{\ast }\left( z_{i,t}\right) \tilde{\psi}_{2,\gamma
}^{\ast }\left( z_{i,t}\right) ]$ for any $i\in I_{2,g}$ and any $g\in 
\mathcal{G}_{2}$.\footnote{%
Since $\mathbb{E}_{T}[\tilde{\psi}_{1,\gamma }(z_{i,t})]=0$ by the
first-order condition for $\phi _{1}^{\ast }$, we have $\sigma _{12,\gamma
,i}=\mathbb{E}_{T}[\tilde{\psi}_{1,\gamma }\left( z_{i,t}\right) \tilde{\psi}%
_{2,\gamma }\left( z_{i,t}\right) ]$ for any $i\leq n$.}
\end{theorem}

Theorem \ref{C_L2} shows that $\hat{\sigma}_{n,T}^{2}$ tends to overestimate
the true variance\ $\sigma _{n,T}^{2}$ by $2\sigma _{S,n,T}^{2}$. It is
reasonable to expect the latter to be of order $O(T^{-1})$, which is
formalized by Assumptions \ref{C_A1}(iii) and \ref{A4}. On the other hand,\
it can be shown\footnote{%
See Lemma \ref{C_L_Auxillary9} in the Online Appendix.} that 
\begin{equation}
\frac{\omega _{n,T}^{2}-(\sigma _{n,T}^{2}+\sigma _{U,n,T}^{2})}{\omega
_{n,T}^{2}}=O(T^{-1/2}),  \label{C_Var_4}
\end{equation}%
where 
\begin{equation}
\sigma _{U,n,T}^{2}\equiv (2nT)^{-1}\sum_{g\in \mathcal{G}_{2}}\sum_{i\in
I_{2,g}}\left( \sigma _{1,\gamma ,i}^{4}+n_{2,g}^{-2}\sigma _{2,\gamma
,i}^{2}\sum_{i^{\prime }\in I_{2,g}}\sigma _{2,\gamma ,i^{\prime
}}^{2}-2n_{2,g}^{-1}\sigma _{12,\gamma ,i}^{2}\right) ,  \label{C_Var_5}
\end{equation}%
which represents the leading term in the variance of $n^{-1/2}\sum_{i\leq n}(%
\tilde{V}_{i}^{\ast }+\tilde{U}_{i})$.

Comparing their expressions in (\ref{C_Var_3}) and (\ref{C_Var_5}), it is
evident that $\sigma _{S,n,T}^{2}\geq \sigma _{U,n,T}^{2}$ in general, with
the inequality becoming strict when $\mathbb{E}[\tilde{\psi}_{j,\gamma
}\left( z_{i,t}\right) ]\neq 0$. From (\ref{C_L2_1}) and (\ref{C_Var_4}), it
follows that $\hat{\sigma}_{n,T}^{2}$ is a consistent estimator of $\omega
_{n,T}^{2}$ when $\sigma _{n,T}^{2}$ dominates $\sigma _{S,n,T}^{2}$. Since $%
\sigma _{S,n,T}^{2}=O(T^{-1})$, this dominance is guaranteed when the two
models are strictly non-nested, as $\sigma _{n,T}^{2}$ is bounded away from
zero in such cases. Conversely, if $\sigma _{n,T}^{2}$ is of the same or
even smaller order than\ $\sigma _{S,n,T}^{2}$, a scenario that arises when
the two models are (nearly) nested or overlapping, the result in (\ref%
{C_L2_1}) implies that $\hat{\sigma}_{n,T}^{2}$ overestimates $\omega
_{n,T}^{2}$ by $2\sigma _{S,n,T}^{2}-\sigma _{U,n,T}^{2}$, and becomes an
inconsistent estimator of $\omega _{n,T}^{2}$.

The over-estimation issue in $\hat{\sigma}_{n,T}^{2}$ can be adjusted
through consistent estimators of $\sigma _{U,n,T}^{2}$ and $\sigma
_{S,n,T}^{2}$. From its expression in (\ref{C_Var_5}), $\sigma _{U,n,T}^{2}$
can be estimated by replacing $\sigma _{j,\gamma ,i}^{2}$ with $\hat{\sigma}%
_{j,\gamma ,i}^{2}$ defined in (\ref{C_Var_1}), and replacing $\sigma
_{12,\gamma ,i}^{2}$\ by%
\begin{equation*}
\hat{\sigma}_{12,\gamma ,i}^{2}\equiv \frac{(\widehat{\mathbb{E}}_{T}[\hat{%
\psi}_{1,\gamma }(z_{i,t})\hat{\psi}_{2,\gamma }(z_{i,t})])^{2}}{|\hat{\Psi}%
_{1,\gamma \gamma }(\hat{\phi}_{1,g})\hat{\Psi}_{2,\gamma \gamma }(\hat{\phi}%
_{2,g})|}.
\end{equation*}%
Specifically, the estimator of $\sigma _{U,n,T}^{2}$ is defined as%
\begin{equation}
\hat{\sigma}_{U,n,T}^{2}\equiv (2nT)^{-1}\sum_{g\in \mathcal{G}%
_{2}}\sum_{i\in I_{2,g}}\left( \hat{\sigma}_{1,\gamma ,i}^{4}+n_{2,g}^{-2}%
\hat{\sigma}_{2,\gamma ,i}^{2}\sum_{i^{\prime }\in I_{2,g}}\hat{\sigma}%
_{2,\gamma ,i^{\prime }}^{2}-2n_{2,g}^{-1}\hat{\sigma}_{12,\gamma
,i}^{2}\right) .  \label{Var_Est_U}
\end{equation}%
Similarly, we define the estimator of $\sigma _{S,n,T}^{2}$ as%
\begin{equation}
\hat{\sigma}_{S,n,T}^{2}\equiv (2nT)^{-1}\sum_{g\in \mathcal{G}%
_{2}}\sum_{i\in I_{2,g}}\left( \hat{\sigma}_{1,\gamma ,i}^{4}+n_{2,g}^{-2}%
\hat{\sigma}_{2,\gamma ,i}^{2}\sum_{i^{\prime }\in I_{2,g}}\hat{s}_{2,\gamma
,i^{\prime }}^{2}-2n_{2,g}^{-1}\hat{\sigma}_{12,\gamma ,i}^{2}\right) .
\label{Var_Est_S}
\end{equation}%
Here, for any $i\in I_{2,g}$ and any $g\in \mathcal{G}_{2}$:%
\begin{equation}
\hat{s}_{2,\gamma ,i}^{2}\equiv \frac{\widehat{\mathbb{E}}_{T}[\hat{\psi}%
_{j,\gamma }(z_{i,t})^{2}]}{|\hat{\Psi}_{j,\gamma \gamma }(\hat{\phi}_{j,g})|%
}.  \label{Var_Est_3}
\end{equation}%
Thus, $\omega _{n,T}^{2}$ can be consistently estimated in a general setting
as $\hat{\sigma}_{n,T}^{2}+\hat{\sigma}_{U,n,T}^{2}-2\hat{\sigma}%
_{S,n,T}^{2} $, as indicated by Theorem \ref{C_L3} below.

\begin{theorem}
\textit{\label{C_L3}\ }Under Assumptions \ref{A1}, \ref{C_A1} and \ref{C_A2}
in the Appendix, we have:%
\begin{equation}
\frac{\hat{\sigma}_{U,n,T}^{2}-\sigma_{U,n,T}^{2}}{\omega_{n,T}^{2}}%
=O_{p}(n^{-1/2})\ \ \ \ \text{and \ \ }\frac{\hat{\sigma}_{S,n,T}^{2}-%
\sigma_{S,n,T}^{2}}{\omega_{n,T}^{2}}=O_{p}(n^{-1/2}).  \label{C_L3_1}
\end{equation}
\end{theorem}

One drawback of using $\hat{\sigma}_{n,T}^{2}-2\hat{\sigma}_{S,n,T}^{2}+\hat{%
\sigma}_{U,n,T}^{2}$ as a variance estimator is that it may take negative
values in finite samples. To address this issue, we follow \cite%
{Liao&Shi2020} and propose the following hybrid variance estimator: 
\begin{equation}
\hat{\omega}_{n,T}^{2}=\max \left\{ \hat{\sigma}_{n,T}^{2}+\hat{\sigma}%
_{U,n,T}^{2}-2\hat{\sigma}_{S,n,T}^{2},\hat{\sigma}_{U,n,T}^{2}\right\} .
\label{Var_Est_5}
\end{equation}

\begin{remark}
By definition, $\hat{\sigma}_{U,n,T}^{2}$ can be expressed as%
\begin{align}
\hat{\sigma}_{U,n,T}^{2}& =(2nT)^{-1}\sum_{g\in \mathcal{G}_{2}}\sum_{i\in
I_{2,g}}(\hat{\sigma}_{1,\gamma ,i}^{4}-2n_{2,g}^{-1}\hat{\sigma}_{12,\gamma
,i}^{2}+n_{2,g}^{-2}\hat{\sigma}_{2,\gamma ,i}^{4})  \notag \\
& +(2nT)^{-1}\sum_{g\in \mathcal{G}_{2}}n_{2,g}^{-2}\sum_{i\in
I_{2,g}}\sum_{i^{\prime }\in I_{2,g},i^{\prime }\neq i}\hat{\sigma}%
_{2,\gamma ,i}^{2}\hat{\sigma}_{2,\gamma ,i^{\prime }}^{2}.
\label{Var_Est_6}
\end{align}%
Since\ $\hat{\sigma}_{12,\gamma ,i}^{2}\leq \hat{\sigma}_{1,\gamma ,i}^{2}%
\hat{\sigma}_{2,\gamma ,i}^{2}$ for any $i$, the first term on the
right-hand side of (\ref{Var_Est_6}) is non-negative. Additionally, the
second term after the equality in (\ref{Var_Est_6}) is also non-negative by
definition. Therefore, $\hat{\sigma}_{U,n,T}^{2}$ is guaranteed to be
non-negative, and by construction, $\hat{\omega}_{n,T}^{2}$ is also
non-negative. This ensures the practical reliability of $\hat{\omega}%
_{n,T}^{2}$ as a variance estimator.
\end{remark}

\subsection{Feasible Modified QLR Statistic and Generalized Vuong Test\label%
{section:feasiblestatistic}}

Given the variance estimator $\hat{\omega}_{n,T}^{2}$, the two-sided and
one-sided Vuong tests are defined as:%
\begin{eqnarray}
\varphi _{n,T}^{2\text{-side}}(p) &\equiv &1\left\{ \left\vert
MQLR_{n,T}\right\vert >\hat{\omega}_{n,T}z_{1-p/2}\right\} \text{,}  \notag
\\
\varphi _{n,T}^{1\text{-side}}(p) &\equiv &1\left\{ MQLR_{n,T}>\hat{\omega}%
_{n,T}z_{1-p}\right\} ,  \label{C_Vuong_test}
\end{eqnarray}%
respectively, where $z_{1-p}$ is the $1-p$ quantile of the standard normal
distribution, and $p\in (0,1)$ denotes the significance level.

\begin{theorem}
\label{C_T1}\ Suppose that Assumptions \ref{A1}, \ref{C_A1} and \ref{C_A2}
in the Appendix hold. Further suppose that as $n,T\rightarrow \infty $,%
\begin{equation}
\frac{\overline{QLR}_{n,T}}{\omega _{n,T}}\rightarrow c,  \label{C_T1_1}
\end{equation}%
for some constant $c\ $in the extended real line. Then as $n,T\rightarrow
\infty $,%
\begin{equation}
\mathbb{E}[\varphi _{n,T}^{2\text{-side}}(p)]\rightarrow 2-\Phi
(z_{1-p/2}-c)-\Phi (z_{1-p/2}+c),  \label{C_T1_2}
\end{equation}%
and%
\begin{equation}
\mathbb{E}[\varphi _{n,T}^{1\text{-side}}(p)]\rightarrow 1-\Phi (z_{1-p}-c),
\label{C_T1_3}
\end{equation}%
where $\Phi (\cdot )$ denotes the cumulative distribution function of the
standard normal.
\end{theorem}

\begin{remark}
Theorem \ref{C_T1} establishes both the size and power properties of the
Vuong test defined in (\ref{C_Vuong_test}). Under the null hypothesis where $%
\overline{QLR}_{n,T}=0$, (\ref{C_T1_1}) holds with $c=0$. In this scenario, (%
\ref{C_T1_2}) and (\ref{C_T1_3}) imply that 
\begin{equation}
\mathbb{E}[\varphi _{n,T}^{2\text{-side}}(p)]\rightarrow 2(1-\Phi
(z_{1-p/2}))=p\text{ \ \ \ and \ \ \ }\mathbb{E}[\varphi _{n,T}^{1\text{-side%
}}(p)]\rightarrow p,  \label{C_Vuong_Size}
\end{equation}%
which establishes the size control over data generating processes in which
the two models being compared are either nested or overlapping. Furthermore,
(\ref{C_T1_2}) demonstrates that the two-sided Vuong test has power against
local alternatives with $\left\vert c\right\vert \in (0,\infty )$, and is
consistent against any fixed alternatives, while the one-sided Vuong test
share the similar power properties with $c\in (0,\infty )$.
\end{remark}

\section{Testing Heterogeneous Time Effects vs. TWFE\label{sec:TWE}}

The general result in the previous section was based on Theorems \ref%
{Rep_Theta} -- \ref{QLR_Stat} in Appendix \ref{Sec:AP2}. As discussed in the
introduction, the result established in Appendix \ref{Sec:AP2} is a
technical contribution that relaxes the concavity assumption in \cite%
{FernandezValWeidener2016}, but it was done at the cost of restricting the
group structure. In this section, we consider potentially more complicated
group structure but we do so in the context of the linear model, which
satisfies the concavity assumption. Because of substantive economic interest
associated with the linear model, in particular, the literature initiated by 
\cite{Bonhomme-Manresa}, we believe that the linear model may deserve a
special attention.

\cite{Bonhomme-Manresa} have recently proposed a new model that extends
beyond the scope of traditional panel data models. In light of its
computational demands, one may naturally question whether a standard TWFE
specification could offer a practical alternative. Obviously the two models
are not nested, so it is of interest to find a Vuong test for this
comparison.

We consider the linear panel model%
\begin{equation*}
y_{i,t}=x_{i,t}^{\top }\beta +\gamma _{g(i),m(t)}+\varepsilon _{i,t},
\end{equation*}%
where $y_{i,t}$ is the dependent variable, $x_{i,t}$ are the observed
regressors, and $\varepsilon _{i,t}$ is i.i.d. across $i$ and $t$ and with
mean zero and variance $\sigma ^{2}$. Here, the $g(\cdot )$ and $m(\cdot )$
are \emph{known} cluster/group assignment functions with range $\mathcal{G}%
=\{1,\ldots ,G\}$ and $\mathcal{M}=\{1,\ldots ,M\}$ respectively, and $%
\gamma _{g(i),m(t)}$ denotes the unknown fixed effect.\footnote{%
The $\gamma _{g(i),m(t)}$ takes different forms under various model
specifications. For example, in models involving time invariant individual
fixed effects only, we have $\gamma _{g(i),m(t)}=\gamma _{i}$, e.g., with $%
\mathcal{G}=\{1,\ldots ,n\}$, and $\mathcal{M}=\{1\}$. In models involving
only time fixed effects, we have $\gamma _{g(i),m(t)}=\gamma _{t}$, e.g.,
with $\mathcal{G}=\{1\}$, and $\mathcal{M}=\{1,\ldots ,T\}$. In the Section %
\ref{sec:linear-section} of the Online Appendix, we consider non-nested
hypothesis testing comparing several variants of the linear model.} In the
main text of the paper, we focus on the Vuong test comparing the
heterogeneous time effects model (as considered by \cite{Bonhomme-Manresa})
and the TWFE model, denoted as model 1 and model 2, respectively.

Specifically, model 1 defines the joint log-likelihood as%
\begin{equation}
\sum_{g\in \mathcal{G}_{1}}\sum_{i\in I_{g}}\sum_{t\leq T}\psi \left(
z_{i,t};\theta _{1},\gamma _{1,g,t}\right) ,  \label{M_1}
\end{equation}%
where $\mathcal{G}_{1}\equiv \{1,\ldots ,G_{1}\}$ and $G_{1}$ is a positive
integer. The function $\psi \left( z_{i,t};\theta ,\gamma \right) $ is given
by 
\begin{equation}
\psi \left( z_{i,t};\theta ,\gamma \right) =(-2)^{-1}(y_{i,t}-x_{i,t}^{\top
}\theta -\gamma )^{2}.  \label{M_Form}
\end{equation}%
In contrast, model 2 assumes that the joint log-likelihood is 
\begin{equation}
\sum_{i\leq n}\sum_{t\leq T}\psi \left( z_{i,t};\theta _{2},\gamma
_{2,i}+\gamma _{2,t}\right) ,  \label{M_2}
\end{equation}%
where $\gamma _{2,i}$ represents the time-invariant individual effect, and $%
\gamma _{2,t}$ captures the homogeneous time effect. We impose the following
normalization condition on model 2:%
\begin{equation}
\sum_{t\leq T}\gamma _{2,t}=0,  \label{Norm_C}
\end{equation}%
to ensure unique identification of the pseudo-true parameters.

In the following, we introduce the algorithm for calculating the modified
QLR statistic and explain its asymptotic properties. Even though the
technical details are more involved and use more complex notation, the
results in this section are presented in a way quite similar to those in the
previous one. To make things easier to follow, we have organized this
section in a way that mirrors the structure of the (second half of the)
previous section.

\subsection{Estimation under Group-Time Effects and TWFE}

The pseudo true parameters $\theta _{1}^{\ast }$ and $\gamma _{1,g,t}^{\ast
} $ (for $g\in \mathcal{G}_{1}$ and $t\leq T$) for model 1, defined as the
maximizers of the population version of the objective in (\ref{M_1}), take
the following forms: 
\begin{equation*}
\theta _{1}^{\ast }=\frac{\Sigma _{\dot{x}}^{-1}}{nT}\sum_{i\leq
n}\sum_{t\leq T}\mathbb{E}[\dot{x}_{i,t}^{\ast }y_{i,t}]\text{ \ \ \ \ and \
\ \ \ }\gamma _{1,g,t}^{\ast }=\mathbb{E}[\bar{y}_{g,t}]-\mathbb{E}[\bar{x}%
_{g,t}^{\top }]\theta _{1}^{\ast },
\end{equation*}%
where $\Sigma _{\dot{x}}\equiv (nT)^{-1}\sum_{i\leq n}\sum_{t\leq T}\mathbb{E%
}[\dot{x}_{i,t}^{\ast }\dot{x}_{i,t}^{\ast \top }]$, $\dot{x}_{i,t}^{\ast
}=x_{i,t}-\mathbb{E}[\bar{x}_{g,t}]$, and for any $i\in I_{g}$ and any $g\in 
\mathcal{G}_{1}$, $\bar{x}_{g,t}\equiv n_{g}^{-1}\sum_{i\in I_{g}}x_{i,t}$
and $\bar{y}_{g,t}\equiv n_{g}^{-1}\sum_{i\in I_{g}}y_{i,t}$.\ The
estimators of $\theta _{1}^{\ast }$ and $\gamma _{1,g,t}^{\ast }$ are their
sample analogs:%
\begin{equation}
\hat{\theta}_{1}=\frac{\hat{\Sigma}_{\dot{x}}^{-1}}{nT}\sum_{i\leq
n}\sum_{t\leq T}\dot{x}_{i,t}y_{i,t}\text{ \ \ \ \ and \ \ \ \ }\hat{\gamma}%
_{1,g,t}=\bar{y}_{g,t}-\bar{x}_{g,t}\hat{\theta}_{1},  \label{HTFE_1}
\end{equation}%
where $\dot{x}_{i,t}\equiv x_{i,t}-\bar{x}_{g,t}$ and $\hat{\Sigma}_{\dot{x}%
}\equiv (nT)^{-1}\sum_{i\leq n}\sum_{t\leq T}\dot{x}_{i,t}\dot{x}%
_{i,t}^{\top }$.

Applying (\ref{Norm_C}), we can solve the pseudo true parameters $\theta
_{2}^{\ast }$, $(\gamma _{2,i}^{\ast })_{i\leq n}$\ and\ $(\gamma
_{2,t}^{\ast })_{t\leq T}$ in model 2 as%
\begin{equation*}
\theta _{2}^{\ast }=\frac{\Sigma _{\ddot{x}^{\ast }}^{-1}}{nT}\sum_{i\leq
n}\sum_{t\leq T}\mathbb{E}[\ddot{x}_{i,t}^{\ast }y_{i,t}]\text{, \ }\gamma
_{2,i}^{\ast }=\mathbb{E}[\bar{y}_{i}]-\mathbb{E}[\bar{x}_{i}^{\top }]\theta
_{2}^{\ast }\text{ \ and \ }\gamma _{2,t}^{\ast }=\mathbb{E}[\bar{y}_{t}-%
\bar{y}]-\mathbb{E}[\bar{x}_{t}^{\top }-\bar{x}^{\top }]\theta _{2}^{\ast },
\end{equation*}%
where $\Sigma _{\ddot{x}}\equiv (nT)^{-1}\sum_{i\leq n}\sum_{t\leq T}\mathbb{%
E}[\ddot{x}_{i,t}^{\ast }\ddot{x}_{i,t}^{\ast \top }]$ and $\ddot{x}%
_{i,t}^{\ast }\equiv x_{i,t}-\mathbb{E}[\bar{x}_{i}-\bar{x}_{t}+\bar{x}]$.
The estimators of the pseudo true parameters are their empirical
counterparts:%
\begin{equation}
\hat{\theta}_{2}=\frac{\hat{\Sigma}_{\ddot{x}}^{-1}}{nT}\sum_{i\leq
n}\sum_{t\leq T}\ddot{x}_{i,t}y_{i,t}\text{, \ }\hat{\gamma}_{2,i}=\bar{y}%
_{i}-\bar{x}_{i}^{\top }\hat{\theta}_{2}\text{ \ and \ }\hat{\gamma}_{2,t}=(%
\bar{y}_{t}-\bar{y})-(\bar{x}_{t}-\bar{x})^{\top }\hat{\theta}_{2},
\label{TFE_2}
\end{equation}%
where\ $\ddot{x}_{i,t}\equiv x_{i,t}-\bar{x}_{i}-\bar{x}_{t}+\bar{x}$ and $%
\hat{\Sigma}_{\ddot{x}}\equiv (nT)^{-1}\sum_{i\leq n}\sum_{t\leq T}\ddot{x}%
_{i,t}\ddot{x}_{i,t}^{\top }$.

\subsection{Asymptotic Distribution of the QLR Statistic}

Using the estimators of the pseudo true parameters from both models, we
obtain the QLR statistic for model comparison as 
\begin{equation*}
QLR_{n,T}=(nT)^{-1/2}\left( \sum_{i\leq n}\sum_{t\leq T}\frac{\hat{%
\varepsilon}_{1,i,t}^{2}}{-2}-\sum_{i\leq n}\sum_{t\leq T}\frac{\hat{%
\varepsilon}_{2,i,t}^{2}}{-2}\right)
\end{equation*}%
where $\hat{\varepsilon}_{j,i,t}\equiv y_{i,t}-x_{i,t}^{\top }\hat{\theta}%
_{j}-\hat{\gamma}_{j,i,t}$\ for $j=1,2$, $\hat{\gamma}_{1,i,t}\equiv \hat{%
\gamma}_{1,g,t}$ for any $i\in I_{g}$ and any $g\in \mathcal{G}_{1}$, and $%
\hat{\gamma}_{2,i,t}\equiv \hat{\gamma}_{2,i}+\hat{\gamma}_{2,t}$ for any $%
i\leq n$ and any $t\leq T$.

We next derive the asymptotic distribution of $QLR_{n,T}$. Some notation are
needed. Let $\varepsilon _{j,i,t}\equiv y_{i,t}-x_{i,t}^{\top }\theta
_{j}^{\ast }-\gamma _{j,i,t}^{\ast }$ and $\varepsilon _{j,i,t}^{\ast
}\equiv \varepsilon _{j,i,t}-\mathbb{E}[\varepsilon _{j,i,t}]$ for $j=1,2$,
where $\gamma _{1,i,t}^{\ast }\equiv \gamma _{1,g,t}^{\ast }$ for any $i\in
I_{g}$ and any $g\in \mathcal{G}_{1}$, and $\gamma _{2,i,t}^{\ast }\equiv
\gamma _{2,i}^{\ast }+\gamma _{2,t}^{\ast }$ for any $i\leq n$ and any $%
t\leq T$. Without loss of generality,\ we suppose that for any $g\in G_{1}$, 
$I_{g}=\{N_{g}+1,\ldots ,N_{g}+n_{g}\}$ where $N_{g}\equiv \sum_{g^{\prime
}<g}n_{g^{\prime }}$.

For model 1, we define 
\begin{equation*}
\tilde{V}_{1,i}=\frac{\sum_{t\leq T}\varepsilon _{1,i,t}^{\ast 2}}{%
2n_{g}T^{1/2}}\text{ \ \ \ \ \ and \ \ \ \ \ }\tilde{U}_{1,i}=\frac{%
\sum_{i^{\prime }=N_{g}+1}^{i-1}\sum_{t\leq T}\varepsilon _{1,i,t}^{\ast
}\varepsilon _{1,i^{\prime }t}^{\ast }}{n_{g}T^{1/2}},
\end{equation*}%
for any $g\in \mathcal{G}_{1}$ and any $i\in I_{g}$. For model 2, we let%
\begin{equation*}
\tilde{V}_{2,i}=\frac{(\sum_{t\leq T}\varepsilon _{2,i,t}^{\ast })^{2}}{%
2T^{3/2}}+\frac{\sum_{t\leq T}\varepsilon _{2,i,t}^{\ast 2}}{2nT^{1/2}}\text{
\ \ \ \ and \ \ \ \ }\tilde{U}_{2,i}=\frac{\sum_{i^{\prime
}=1}^{i-1}\sum_{t\leq T}\varepsilon _{2,i,t}^{\ast }\varepsilon
_{2,i^{\prime }t}^{\ast }}{nT^{1/2}}.
\end{equation*}%
It can be shown\footnote{%
See Lemma \ref{TWFE_Est2} and Lemma \ref{GTFE_Est2} in the Online Appendix.}
that%
\begin{equation}
QLR_{n,T}=n^{-1/2}\sum_{i\leq n}(\tilde{\Psi}_{i}+\tilde{V}_{i}+\tilde{U}%
_{i})+o_{p}(n^{-1/2}),  \label{QLR-approx-linear}
\end{equation}%
where\ $\tilde{\Psi}_{i}=(2T^{1/2})^{-1}\sum_{t\leq T}(\varepsilon
_{2,i,t}^{2}-\varepsilon _{1,i,t}^{2})$, $\tilde{V}_{i}=\tilde{V}_{1,i}-%
\tilde{V}_{2,i}$ and $\tilde{U}_{i}=\tilde{U}_{1,i}-\tilde{U}_{2,i}$. The
variance $\omega _{n,T}^{2}$ of the $QLR_{n,T}$ is defined as in (\ref%
{def_omega}), with $\tilde{\Psi}_{i}$, $\tilde{V}_{i}$ and $\tilde{U}_{i}$
taking the specific forms given above. Below is the asymptotic property of
the infeasible test statistic that mirrors Theorem \ref{C_L0}:

\begin{theorem}
\textit{\label{TWFE_Vuong}\ Under\ Assumptions \ref{A1},\ }\ref{A8} and \ref%
{A9} in the Appendix, we have%
\begin{equation*}
\frac{QLR_{n,T}-\mathbb{E}[S_{n,T}]-\overline{QLR}_{n,T}}{\omega _{n,T}}%
\rightarrow _{d}N(0,1),
\end{equation*}%
where $\mathbb{E}[S_{n,T}]=(2nT)^{-1/2}\sum_{g\in \mathcal{G}_{1}}\sum_{i\in
I_{g}}\mathbb{E}\left[ \sum_{t\leq T}(n_{g}^{-1}\varepsilon _{1,i,t}^{\ast
2}-n^{-1}\varepsilon _{2,i,t}^{\ast 2})-T^{-1}(\sum_{t\leq T}\varepsilon
_{2,i,t}^{\ast })^{2}\right] $.
\end{theorem}

\subsection{Bias/Variance Estimation and the Feasible Test}

Following the discussion in Section \ref{Sec: Simple-Vuong}, we present a
feasible modified QLR procedure by estimating the bias $\mathbb{E}[S_{n,T}]$
and the variance $\omega _{n,T}^{2}$. Consistent estimation of these
quantities is required under the null hypothesis to ensure correct size
control of the test. For this purpose, we assume that $(\varepsilon
_{1,i,t},\varepsilon _{2,i,t})$ are independent across $t$ with
time-invariant first and second moments; this condition is imposed in
Assumption \ref{A10}.\footnote{%
The independence of $(\varepsilon _{1,i,t},\varepsilon _{2,i,t})$ over $t$
is imposed primarily to simplify the form of $\omega _{n,T}^{2}$ and its
consistent estimator. This condition is only required under the null
hypothesis; see Lemma \ref{TWFE_F_Vuong_Power} in the Appendix for
consistency of the test without this assumption.}

Under this condition, we have 
\begin{equation*}
\mathbb{E}[S_{n,T}]=(2nT)^{-1/2}\sum_{g\in \mathcal{G}_{1}}\sum_{i\in I_{g}}%
\left[ Tn_{g}^{-1}\sigma _{1,i}^{2}-(1+Tn^{-1})\sigma _{2,i}^{2}\right]
\end{equation*}%
where $\sigma _{j,i}^{2}\equiv \mathrm{Var}(\varepsilon _{j,i,t})$ can be
estimated by $\hat{\sigma}_{j,i}^{2}\equiv T^{-1}\sum_{t\leq T}\hat{%
\varepsilon}_{j,i,t}^{2}$. Therefore, the bias term can be estimated by%
\begin{equation}
\widehat{\mathbb{E}[S_{n,T}]}=(2nT)^{-1/2}\sum_{g\in \mathcal{G}%
_{1}}\sum_{i\in I_{g}}\left[ Tn_{g}^{-1}\hat{\sigma}_{1,i}^{2}-(1+Tn^{-1})%
\hat{\sigma}_{2,i}^{2}\right] ,  \label{TWE_Best}
\end{equation}%
which leads to the modified QLR statistic:%
\begin{equation}
MQLR_{n,T}=QLR_{n,T}-\widehat{\mathbb{E}[S_{n,T}]}.  \label{def_MQLR}
\end{equation}

It can be shown\footnote{%
See Lemma \ref{TWFE_Var} in the Appendix.}\ that $\omega _{n,T}^{2}$ can be
approximated by\ $\sigma _{n,T}^{2}+\sigma _{U,n,T}^{2}$, where 
\begin{equation}
\sigma _{n,T}^{2}\equiv (4n)^{-1}\sum_{i\leq n}\mathrm{Var}(\varepsilon
_{2,i,t}^{2}-\varepsilon _{1,i,t}^{2})  \label{def__Variance_1}
\end{equation}%
and%
\begin{align}
\sigma _{U,n,T}^{2}& \equiv (2nT)^{-1}\sum_{i\leq n}\sigma
_{2,i}^{4}+(2n)^{-1}\sum_{g\in \mathcal{G}_{1}}n_{g}^{-2}\left( \sum_{i\in
I_{g}}\sigma _{1,i}^{2}\right) ^{2}  \notag \\
& +(2n^{3})^{-1}\left( \sum_{i\leq n}\sigma _{2,i}^{2}\right)
^{2}-n^{-2}\sum_{g\in \mathcal{G}_{1}}n_{g}^{-1}\left( \sum_{i\in
I_{g}}\sigma _{1,2,i}\right) ^{2}.  \label{def__Variance_2}
\end{align}%
We consider the sample variance%
\begin{equation}
\hat{\sigma}_{n,T}^{2}\equiv (4nT)^{-1}\sum_{i\leq n}\sum_{t\leq T}(\hat{%
\varepsilon}_{2,i,t}^{2}-\hat{\varepsilon}_{1,i,t}^{2})^{2}-(nT)^{-1}\left(
MQLR_{n,T}\right) ^{2},  \label{def__Variance_Est_1}
\end{equation}%
as an estimator of $\sigma _{n,T}^{2}$. The variance $\sigma _{U,n,T}^{2}$
from estimating the incidental parameters is then estimated by%
\begin{align}
\hat{\sigma}_{U,n,T}^{2}& \equiv (2nT)^{-1}\sum_{i\leq n}\hat{\sigma}%
_{2,i}^{4}+(2n)^{-1}\sum_{g\in \mathcal{G}_{1}}n_{g}^{-2}\left( \sum_{i\in
I_{g}}\hat{\sigma}_{1,i}^{2}\right) ^{2}  \notag \\
& +(2n^{3})^{-1}\left( \sum_{i\leq n}\hat{\sigma}_{2,i}^{2}\right)
^{2}-n^{-2}\sum_{g\in \mathcal{G}_{1}}n_{g}^{-1}\left( \sum_{i\in I_{g}}\hat{%
\sigma}_{1,2,i}\right) ^{2},  \label{def__Variance_Est_2}
\end{align}%
where\ $\hat{\sigma}_{1,2,i}\equiv T^{-1}\sum_{t\leq T}\hat{\varepsilon}%
_{1,i,t}\hat{\varepsilon}_{2,i,t}$.

Proceeding as in the previous section, our estimate of $\omega _{n,T}^{2}$
is thus 
\begin{equation}
\hat{\omega}_{n,T}^{2}\equiv \max \left\{ \hat{\sigma}_{n,T}^{2}-\hat{\sigma}%
_{U,n,T}^{2},\hat{\sigma}_{U,n,T}^{2}\right\} ,  \label{def_omega-hat}
\end{equation}%
and the Vuong test follows (\ref{C_Vuong_test}) with $MQLR_{n,T}$ and $\hat{%
\omega}_{n,T}^{2}$ constructed in (\ref{def_MQLR}) and (\ref{def_omega-hat}%
), respectively.\footnote{%
In Lemma \ref{TWFE_Var_U} of the Appendix, we show that\ $\hat{\sigma}%
_{U,n,T}^{2}$ is positive in finite samples, which implies that the
estimator of $\omega _{n,T}^{2}$ constructed in (\ref{def_omega-hat}) below
is also positive.} Mirroring Theorem \ref{C_T1}, we have:

\begin{theorem}
\textit{\label{TWFE_F_Vuong} }Suppose that\ \textit{Assumptions \ref{A1},\ }%
\ref{A8}, \ref{A9} and \ref{A10} in the Appendix hold. Further suppose that
as $n,T\rightarrow \infty $, 
\begin{equation}
\frac{(nT)^{-1/2}\sum_{i\leq n}\sum_{t\leq T}\mathbb{E}[\varepsilon
_{2,i,t}^{2}-\varepsilon _{1,i,t}^{2}]}{\omega _{n,T}}\rightarrow c.
\label{TWFE_F_Vuong_1}
\end{equation}%
Then we have as $n,T\rightarrow \infty $,%
\begin{equation*}
\mathbb{E}[\varphi _{n,T}^{2\text{-side}}(p)]\rightarrow 2-\Phi
(z_{1-p/2}-c)-\Phi (z_{1-p/2}+c),
\end{equation*}%
and%
\begin{equation*}
\mathbb{E}[\varphi _{n,T}^{1\text{-side}}(p)]\rightarrow 1-\Phi (z_{1-p}-c),
\end{equation*}%
where $\varphi _{n,T}^{2\text{-side}}$ and $\varphi _{n,T}^{1\text{-side}}$
follow the identical form as (\ref{C_Vuong_test}).
\end{theorem}

\section{Conclusion\label{sec:conclusion}}

This paper extends the classical \cite{Vuong1989} test to panel data models
with fixed effects, addressing challenges that arise from high-dimensional
nuisance parameters. By modifying the profile likelihood and applying bias
correction techniques, we develop a valid procedure for comparing non-nested
panel data models. Our analysis emphasizes the need for variance adjustments
in the presence of incidental parameter problems and shows how these
modifications support more reliable model selection in high-dimensional
settings.

We contribute to the literature by generalizing the Vuong test to
accommodate grouped heterogeneity in both individual and time effects.
Furthermore, we propose a methodological framework for model selection when
competing specifications define cross-sectional units differently---a common
issue in empirical work. The theoretical results in this paper highlight the
critical role of bias correction in panel data analysis and complement
existing research on non-nested hypothesis testing.

\bigskip

\appendix

\setcounter{theorem}{0} \renewcommand{\thetheorem}{A\arabic{theorem}}

\begin{center}
{\Large \textbf{Appendix}}
\end{center}

Throughout the appendix, $K$ denotes a generic finite constant that may vary
from line to line but does not depend on $n$ and $T$. For brevity, proofs of
lemmas are omitted here and are provided in the Online Appendix.

\section{Assumptions and Proofs for Section\ \protect\ref{Sec: Simple-Vuong} 
\label{AP: S-Vuong-proof}}

We first state sufficient conditions for the main results in Section\ \ref%
{Sec: Simple-Vuong}, namely, Theorems \ref{C_L0}, \ref{C_L1}, \ref{C_L2}, %
\ref{C_L3} and \ref{C_T1}, and then provide their proofs. Auxiliary lemmas
used in these proofs are collected in Subsection \ref%
{sec:auxiliary-lemmas-main} below.

\begin{assumption}
\textit{\label{A1}\ (i) }$n,T\rightarrow \infty $ such that $n/T\rightarrow
\rho \text{, }$where $0<\rho <\infty $; (ii) for each $i$, $\left\{
z_{i,t}\right\} _{t\geq 1}$ is an alpha-mixing process; (iii) $\left\{
z_{i,t}\right\} _{t\geq 1}$ are independent across $i$; (iv) $%
\sup_{i}\left\vert \alpha _{i}(j)\right\vert \leq Ka^{j}$ for some $a$ such
that $a\in (0,1)$, where%
\begin{equation*}
\alpha _{i}(j)\equiv \sup_{t}\sup_{A\in \mathcal{A}_{t}^{i},\text{ }B\in 
\mathcal{B}_{t+j}^{i}}\left\vert \mathbb{P}\left( A\cap B\right) -\mathbb{P}%
\left( A\right) \mathbb{P}\left( B\right) \right\vert ,
\end{equation*}%
where $\mathcal{A}_{t}^{i}$ and $\mathcal{B}_{t}^{i}$ denote the
sigma-fields generated by $\left\{ z_{i,s}\right\} _{s\leq t}$ and $\left\{
z_{i,s^{\prime }}\right\} _{s^{\prime }\geq t}$, respectively.
\end{assumption}

\begin{assumption}
\textit{\label{C_A1}\ (i)\ Assumptions\ }\ref{A2}, \ref{A3} and \ref{A4} in
Section \ref{Sec:AP2} hold \textit{for both models 1 and 2};\ (ii) for some $%
\delta \in (0,1)$, $\mathbb{E}[(\Delta \psi (z_{i,t},\phi _{i}^{\ast
})/\omega _{n,T})^{4+\delta }]<K$;\ (iii) $\mathbb{E}\left[ \psi _{j,\gamma
}(z_{i,t})^{8}+M(z_{i,t})^{8}\right] <K$, where $M(z_{i,t})$ is defined in
Assumption \ref{A3}; (iv) $G_{2}^{-1}\max_{g\in \mathcal{G}%
_{2}}n_{g}^{-1}=o(1)$; (v) $n\omega _{n,T}^{2}\geq K^{-1}$.
\end{assumption}

\begin{assumption}
\textit{\label{C_A2} For each }$i=1,\ldots ,n$, the process $\{\Delta \psi
(z_{i,t},\phi _{i}^{\ast })\}_{t\geq 1}$\ is serially uncorrelated.
Moreover\ $\mathbb{E}[\psi _{j,\gamma }^{\ast }\left( z_{i,t+1}\right)
|\{(\psi _{j,\gamma }^{\ast }\left( z_{i,s}\right) ,\psi _{2,\gamma }^{\ast
}\left( z_{i,s}\right) \}_{s\leq t}\}]=0$\ for $j=1,2$ and any $t\geq 1$.
\end{assumption}

Assumption \ref{A1} specifies conditions governing the joint asymptotics and
the cross-sectional and time-series dependence structure of the data.
Together with Assumption \ref{A1}, Assumption\ \ref{C_A1}(i) is used to
obtain the expansion of $MQLR_{n,T}^{\ast }$ in (\ref{C_Exp_3}). Assumptions %
\ref{C_A1}(ii, iii) ensure the Lindeberg condition required for the
martingale central limit theorem. Assumption \ref{C_A1}(iv) requires that,
under model 2, the number of groups and/or the size of the smallest group
diverges. Assumption \ref{C_A1}(v) guarantees that the remainder term $%
o_{p}(n^{-1/2})$ in the expansion of the (infeasible) modified QLR statistic
in (\ref{C_Exp_3}) becomes asymptotically negligible after scaling by $%
\omega _{n,T}$. Finally, Assumption \ref{C_A2} is imposed to establish the
consistency of the bias and variance estimators for the QLR statistic.

\bigskip

\noindent \textsc{Proof of Theorem \ref{C_L0}}.\ To establish the claim, we
apply Theorem \ref{QLR_Stat} in Appendix \ref{Sec:AP2}. To this end, it
suffices to verify its sufficient conditions.

First, note that Assumption \ref{A1} and Assumptions \ref{A2}, \ref{A3}, \ref%
{A4} in Appendix \ref{Sec:AP2} are maintained for both models. Moreover,
Assumption \ref{A6}(i) in Appendix \ref{Sec:AP2} is implied by Assumption %
\ref{C_A1}. It therefore remains to verify that Assumptions \ref{A5} and \ref%
{A6}(ii) in Appendix \ref{Sec:AP2} also hold. Since the conditions in
Assumption \ref{A7} in Appendix \ref{Sec:AP2} are subsumed by Assumptions %
\ref{C_A1}(ii)--(iv), it follows from Theorem \ref{MGCLT_V} in Appendix \ref%
{Sec:AP2} that Assumption \ref{A6}(ii) in Appendix \ref{Sec:AP2} is
satisfied.

It remains to verify Assumption \ref{A5} in Appendix \ref{Sec:AP2}. For
model $j=1,2$, we have\ $\mathcal{G}_{j}\equiv \{1,\ldots ,G_{j}\}$ and $%
\mathcal{M}_{j}\equiv \{1\}$.\ Consequently, $G_{j}\leq n$, $M_{j}=1$ and $%
T_{m}=T$. Therefore, we can establish the following inequality:%
\begin{equation}
(G_{j}M_{j})^{2/p}\max_{g\in \mathcal{G}_{j}}(n_{j,g}T_{m})^{-1/2}\leq
n^{2/p}T^{-1/2}=o(1),  \label{P_C_L0_1}
\end{equation}%
where the final equality follows from Assumption \ref{A1}(i) and $p>4$.
Similarly,%
\begin{equation}
(G_{j}M_{j})(nT)^{-1/2}\leq n^{1/2}T^{-1/2}=O(1).  \label{P_C_L0_2}
\end{equation}%
Combining (\ref{P_C_L0_1}) with (\ref{P_C_L0_2}), we conclude that
Assumption \ref{A5} in Appendix \ref{Sec:AP2} holds for both models 1 and
2.\hfill $Q.E.D.$

\bigskip

\noindent \textsc{Proof of Theorem \ref{C_L1}}.\ For any $i\in I_{j,g}$ and
any $g\in \mathcal{G}_{j}$, Lemma \ref{AU_L4}(i) implies that, with
probability approaching 1\ (wpa1),%
\begin{equation*}
\hat{\sigma}_{j,\gamma ,i}^{2}-\sigma _{j,\gamma ,i}^{2}=\frac{\widehat{%
\mathbb{E}}_{T}[\hat{\psi}_{j,\gamma }^{2}(z_{i,t})]}{-\hat{\Psi}_{j,\gamma
\gamma }(\hat{\phi}_{j,g})}-\frac{\mathbb{E}_{T}[\psi _{j,\gamma }^{2}\left(
z_{i,t}\right) ]}{-\Psi _{j,\gamma \gamma ,g}}-\frac{(\widehat{\mathbb{E}}%
_{T}[\hat{\psi}_{j,\gamma }(z_{i,t})])^{2}}{-\hat{\Psi}_{j,\gamma \gamma }(%
\hat{\phi}_{j,g})}+\frac{(\mathbb{E}_{T}[\psi _{j,\gamma }\left(
z_{i,t}\right) ])^{2}}{-\Psi _{j,\gamma \gamma ,g}},
\end{equation*}%
where $\Psi _{j,\gamma \gamma ,g}\equiv n_{j,g}^{-1}\sum_{i\in I_{j,g}}%
\mathbb{E}_{T}\left[ \psi _{j,\gamma \gamma }\left( z_{i,t}\right) \right] $%
. Summing over $i$ and $g$ yields 
\begin{align}
\sum_{g\in \mathcal{G}_{j}}n_{j,g}^{-1}\sum_{i\in I_{j,g}}(\hat{\sigma}%
_{j,\gamma ,i}^{2}-\sigma _{j,\gamma ,i}^{2})& =-\sum_{g\in \mathcal{G}%
_{j}}\sum_{i\in I_{j,g}}\left( \frac{\widehat{\mathbb{E}}_{T}[\hat{\psi}%
_{j,\gamma }^{2}(z_{i,t})]}{n_{j,g}\hat{\Psi}_{j,\gamma \gamma }(\hat{\phi}%
_{j,g})}-\frac{\mathbb{E}_{T}[\psi _{j,\gamma }^{2}\left( z_{i,t}\right) ]}{%
n_{j,g}\Psi _{j,\gamma \gamma ,g}}\right)  \notag \\
& +\sum_{g\in \mathcal{G}_{j}}\sum_{i\in I_{j,g}}\left( \frac{(\widehat{%
\mathbb{E}}_{T}[\hat{\psi}_{j,\gamma }(z_{i,t})])^{2}}{n_{j,g}\hat{\Psi}%
_{j,\gamma \gamma }(\hat{\phi}_{j,g})}-\frac{(\mathbb{E}_{T}[\psi _{j,\gamma
}\left( z_{i,t}\right) ])^{2}}{n_{j,g}\Psi _{j,\gamma \gamma ,g}}\right) ,
\label{P_C_L_Bias_1}
\end{align}%
which together with (\ref{AU_L1_1}) and (\ref{AU_L1_2}) in Lemma \ref{AU_L1}
implies%
\begin{equation}
R_{j,n,T}(\hat{\phi}_{j})-\sum_{g\in \mathcal{G}_{j}}n_{j,g}^{-1}\sum_{i\in
I_{j,g}}\sigma _{j,\gamma ,i}^{2}=O_{p}(1).  \label{P_C_L_Bias_2}
\end{equation}%
Finally, under Assumption \ref{C_A2}, $T^{1/2}\sum_{i\leq n}\mathbb{E}[%
\tilde{V}_{j,i}]=2\sum_{g\in \mathcal{G}_{j}}n_{j,g}^{-1}\sum_{i\in
I_{j,g}}\sigma _{j,\gamma ,i}^{2}$. This identity,\ combined with Assumption %
\ref{C_A1}(v), establishes the claim of the lemma.\hfill $Q.E.D.$

\bigskip

\noindent \textsc{Proof of Theorem \ref{C_L2}}.\ From\ (\ref{AU_L1_3}) in
Lemma \ref{AU_L1}, (\ref{AU_L2_1}) in Lemma \ref{AU_L2} and Lemma \ref{AU_L4}%
(ii), we have%
\begin{align}
\hat{\sigma}_{n,T}^{2}& \equiv (nT)^{-1}\sum_{i\leq n}\sum_{t\leq T}\Delta
\psi (z_{i,t},\hat{\phi}_{i})^{2}-(nT)^{-1}\left( MQLR_{n,T}\right) ^{2} 
\notag \\
& =n^{-1}\sum_{i\leq n}\mathbb{E}_{T}[\Delta \psi (z_{i,t})^{2}]-(nT)^{-1}(%
\overline{QLR}_{n,T})^{2}+2\sigma _{S,n,T}^{2}+O_{p}(T^{-3/2}+\omega
_{n,T}(nT)^{-1/2})  \notag \\
& =\sigma _{n,T}^{2}+2\sigma _{S,n,T}^{2}+O_{p}(T^{-3/2}+\omega
_{n,T}(nT)^{-1/2}),  \label{P_C_L2_1}
\end{align}%
where the last equality follows from the definition of $\sigma _{n,T}^{2}$
and Assumption \ref{C_A1}(i). Thus,%
\begin{equation}
\frac{\hat{\sigma}_{n,T}^{2}-(\sigma _{n,T}^{2}+2\sigma _{S,n,T}^{2})}{%
\omega _{n,T}^{2}}=O_{p}(\omega _{n,T}^{-2}T^{-3/2}+\omega
_{n,T}^{-1}(nT)^{-1/2})=O_{p}(T^{-1/2}),  \label{P_C_L2_2}
\end{equation}%
where the second equality holds by Assumptions \ref{A1}(i) and \ref{C_A1}%
(v). The claim of the lemma follows from (\ref{P_C_L2_2}).\hfill $Q.E.D.$

\bigskip

\noindent \textsc{Proof of Theorem \ref{C_L3}}. From (\ref{AU_L1_4}) in
Lemma \ref{AU_L2} and (\ref{AU_L3_1}) in Lemma \ref{AU_L3}, we have:%
\begin{eqnarray}
\sum_{g\in \mathcal{G}_{j}}n_{j,g}^{-1}\sum_{i\in I_{j,g}}(\hat{\sigma}%
_{j,\gamma ,i}^{2}-\sigma _{j,\gamma ,i}^{2})^{2} &\leq &2\sum_{g\in 
\mathcal{G}_{j}}n_{j,g}^{-1}\sum_{i\in I_{j,g}}\left( \frac{(\widehat{%
\mathbb{E}}_{T}[\hat{\psi}_{j,\gamma }(z_{i,t})])^{2}}{\hat{\Psi}_{j,\gamma
\gamma }(\hat{\phi}_{j,g})}-\frac{(\mathbb{E}_{T}[\psi _{j,\gamma
}(z_{i,t})])^{2}}{\Psi _{j,\gamma \gamma ,g}}\right) ^{2}  \notag \\
&&+2\sum_{g\in \mathcal{G}_{j}}n_{j,g}^{-1}\sum_{i\in I_{j,g}}(\hat{s}%
_{j,\gamma ,i}^{2}-s_{j,\gamma ,i}^{2})^{2}\overset{}{=}O_{p}(1).
\label{P_C_L3_1}
\end{eqnarray}%
This along with the triangle inequality implies that%
\begin{eqnarray}
\left\vert \sum_{g\in \mathcal{G}_{j}}n_{j,g}^{-1}\sum_{i\in I_{j,g}}(\hat{%
\sigma}_{j,\gamma ,i}^{4}-\sigma _{j,\gamma ,i}^{4})\right\vert &\leq
&\sum_{g\in \mathcal{G}_{j}}n_{j,g}^{-1}\sum_{i\in I_{j,g}}(\hat{\sigma}%
_{j,\gamma ,i}^{2}-\sigma _{j,\gamma ,i}^{2})^{2}  \notag \\
&&+2\sum_{g\in \mathcal{G}_{j}}n_{j,g}^{-1}\sum_{i\in I_{j,g}}|\hat{\sigma}%
_{j,\gamma ,i}^{2}-\sigma _{j,\gamma ,i}^{2}|\sigma _{j,\gamma ,i}^{2} 
\notag \\
&=&2\sum_{g\in \mathcal{G}_{j}}n_{j,g}^{-1}\sum_{i\in I_{j,g}}|\hat{\sigma}%
_{j,\gamma ,i}^{2}-\sigma _{j,\gamma ,i}^{2}|\sigma _{j,\gamma
,i}^{2}+O_{p}(1).  \label{P_C_L3_2}
\end{eqnarray}%
From Assumptions \ref{C_A1}(iii) and \ref{A4}, it follows that%
\begin{equation}
\max_{i\leq n}s_{j,\gamma ,i}^{2}=\frac{\max_{i\leq n}T^{-1}\sum_{t\leq T}%
\mathbb{E}[\psi _{j,\gamma }\left( z_{i,t}\right) ^{2}]}{-\Psi _{j,\gamma
\gamma ,g}}\leq K.  \label{P_C_L3_2b}
\end{equation}%
Thus, by the Cauchy-Schwarz inequality and (\ref{P_C_L3_1}), 
\begin{equation*}
\sum_{g\in \mathcal{G}_{j}}n_{j,g}^{-1}\sum_{i\in I_{j,g}}|\hat{\sigma}%
_{j,\gamma ,i}^{2}-\sigma _{j,\gamma ,i}^{2}|\sigma _{j,\gamma ,i}^{2}\leq
KG_{j}^{1/2}\left( \sum_{g\in \mathcal{G}_{j}}n_{j,g}^{-1}\sum_{i\in
I_{j,g}}(\hat{\sigma}_{1,\gamma ,i}^{2}-\sigma _{1,\gamma
,i}^{2})^{2}\right) ^{1/2}=O_{p}(G_{j}^{1/2}).
\end{equation*}%
Together with (\ref{P_C_L3_2}), this implies 
\begin{equation}
n^{-1}\sum_{g\in \mathcal{G}_{j}}n_{j,g}^{-1}\sum_{i\in I_{j,g}}(\hat{\sigma}%
_{j,\gamma ,i}^{4}-\sigma _{j,\gamma ,i}^{4})=O_{p}(n^{-1/2}).
\label{P_C_L3_3}
\end{equation}%
Similarly, we can apply the Cauchy-Schwarz inequality and (\ref{AU_L3_2})
in\ Lemma \ref{AU_L3} to show that 
\begin{equation}
n^{-1}\sum_{g\in \mathcal{G}_{j}}n_{2,g}^{-1}\sum_{i\in I_{j,g}}(\hat{\sigma}%
_{12,\gamma ,i}^{2}-\sigma _{12,\gamma ,i}^{2})=O_{p}(n^{-1/2}).
\label{P_C_L3_4}
\end{equation}%
By the definitions of $\hat{\sigma}_{U,n,T}^{2}$ and $\sigma _{U,n,T}^{2}$,
we obtain the following expression:%
\begin{align}
\hat{\sigma}_{U,n,T}^{2}-\sigma _{U,n,T}^{2}& =(2nT)^{-1}\sum_{i=1}^{n}(\hat{%
\sigma}_{1,\gamma ,i}^{4}-\sigma _{1,\gamma ,i}^{4})-(nT)^{-1}\sum_{g\in 
\mathcal{G}_{j}}n_{2,g}^{-1}\sum_{i\in I_{j,g}}(\hat{\sigma}_{12,\gamma
,i}^{2}-\sigma _{12,\gamma ,i}^{2})  \notag \\
& +(2nT)^{-1}\sum_{g\in \mathcal{G}_{2}}\left( \left( n_{2,g}^{-1}\sum_{i\in
I_{2,g}}\hat{\sigma}_{2,\gamma ,i}^{2}\right) ^{2}-\left(
n_{2,g}^{-1}\sum_{i\in I_{2,g}}\sigma _{2,\gamma ,i}^{2}\right) ^{2}\right) .
\label{P_C_L3_5}
\end{align}%
From (\ref{P_C_L3_3}) and (\ref{P_C_L3_4}), and Assumption \ref{C_A1}(v), it
follows that 
\begin{equation}
\frac{(2nT)^{-1}\sum_{i=1}^{n}(\hat{\sigma}_{1,\gamma ,i}^{4}-\sigma
_{1,\gamma ,i}^{4})}{\omega _{n,T}^{2}}=O_{p}(n^{-1/2}),  \label{P_C_L3_6}
\end{equation}%
and 
\begin{equation}
\frac{(nT)^{-1}\sum_{g\in \mathcal{G}_{j}}n_{2,g}^{-1}\sum_{i\in I_{j,g}}(%
\hat{\sigma}_{12,\gamma ,i}^{2}-\sigma _{12,\gamma ,i}^{2})}{\omega
_{n,T}^{2}}=O_{p}(n^{-1/2}).  \label{P_C_L3_7}
\end{equation}%
The first claim in (\ref{C_L3_1}) follows from (\ref{P_C_L3_5}), (\ref%
{P_C_L3_6}), (\ref{P_C_L3_7}) and (\ref{AU_L3_3}) in Lemma \ref{AU_L3}.

Next, by the definitions of $\hat{\sigma}_{S,n,T}^{2}$ and $\sigma
_{S,n,T}^{2}$, we obtain%
\begin{align}
\hat{\sigma}_{S,n,T}^{2}-\sigma _{S,n,T}^{2}& =(2nT)^{-1}\sum_{i=1}^{n}(\hat{%
\sigma}_{1,\gamma ,i}^{4}-\sigma _{1,\gamma ,i}^{4})-(nT)^{-1}\sum_{g\in 
\mathcal{G}_{j}}n_{2,g}^{-1}\sum_{i\in I_{j,g}}(\hat{\sigma}_{12,\gamma
,i}^{2}-\sigma _{12,\gamma ,i}^{2})  \notag \\
& +(2nT)^{-1}\sum_{g\in \mathcal{G}_{2}}n_{2,g}^{-2}\sum_{i\in
I_{2,g}}\left( \hat{\sigma}_{2,\gamma ,i}^{2}\sum_{i^{\prime }\in I_{2,g}}%
\hat{s}_{2,\gamma ,i^{\prime }}^{2}-\sigma _{2,\gamma ,i}^{2}\sum_{i^{\prime
}\in I_{2,g}}s_{2,\gamma ,i^{\prime }}^{2}\right) .  \label{P_C_L3_8}
\end{align}%
Therefore, the second claim in (\ref{C_L3_1}) follows from (\ref{P_C_L3_6})
and (\ref{P_C_L3_7}), and (\ref{AU_L3_4}) in Lemma \ref{AU_L3}.\hfill $%
Q.E.D. $

\bigskip

\noindent \textsc{Proof of Theorem \ref{C_T1}}.\ (a) By the definition of $%
\hat{\omega}_{n,T}^{2}$, we have%
\begin{equation*}
\hat{\sigma}_{n,T}^{2}+\hat{\sigma}_{U,n,T}^{2}-2\hat{\sigma}%
_{S,n,T}^{2}\leq \hat{\omega}_{n,T}^{2}\leq \max \left\{ \hat{\sigma}%
_{n,T}^{2}+\hat{\sigma}_{U,n,T}^{2}-2\hat{\sigma}_{S,n,T}^{2},\hat{\sigma}%
_{U,n,T}^{2}+\sigma _{n,T}^{2}\right\} ,
\end{equation*}%
which implies:%
\begin{equation}
\frac{\left\vert \hat{\omega}_{n,T}^{2}-\omega _{n,T}^{2}\right\vert }{%
\omega _{n,T}^{2}}\leq \frac{\left\vert \hat{\sigma}_{n,T}^{2}+\hat{\sigma}%
_{U,n,T}^{2}-2\hat{\sigma}_{S,n,T}^{2}-\omega _{n,T}^{2}\right\vert }{\omega
_{n,T}^{2}}+\frac{\left\vert \hat{\sigma}_{U,n,T}^{2}+\sigma
_{n,T}^{2}-\omega _{n,T}^{2}\right\vert }{\omega _{n,T}^{2}}.
\label{P_C_T1_1}
\end{equation}%
Using Theorem \ref{C_L2}, Theorem \ref{C_L3}, (\ref{AU_L2_2}) in\ Lemma \ref%
{AU_L2},\ and the triangle inequality, the first term after the inequality
in (\ref{P_C_T1_1}) satisfies:%
\begin{align}
\frac{\left\vert \hat{\sigma}_{n,T}^{2}+\hat{\sigma}_{U,n,T}^{2}-2\hat{\sigma%
}_{S,n,T}^{2}-\omega _{n,T}^{2}\right\vert }{\omega _{n,T}^{2}}& \leq \frac{%
\left\vert \hat{\sigma}_{n,T}^{2}-(\sigma _{n,T}^{2}+2\sigma
_{S,n,T}^{2})\right\vert }{\omega _{n,T}^{2}}+\frac{\left\vert \hat{\sigma}%
_{U,n,T}^{2}-\sigma _{U,n,T}^{2}\right\vert }{\omega _{n,T}^{2}}  \notag \\
& +\frac{2\left\vert \hat{\sigma}_{S,n,T}^{2}-\sigma _{S,n,T}^{2}\right\vert 
}{\omega _{n,T}^{2}}+\frac{\left\vert \omega _{n,T}^{2}-(\sigma
_{n,T}^{2}+\sigma _{U,n,T}^{2})\right\vert }{\omega _{n,T}^{2}}  \notag \\
& =O_{p}(T^{-1/2}+n^{-1/2}+\omega _{n,T}^{-1}T^{-1})=O_{p}(T^{-1/2})
\label{P_C_T1_2}
\end{align}%
where the second equality holds by Assumptions \ref{A1}(i) and \ref{C_A1}%
(v). For the second term after the inequality in (\ref{P_C_T1_1}), we can
bound it as%
\begin{equation}
\frac{\left\vert \hat{\sigma}_{U,n,T}^{2}+\sigma _{n,T}^{2}-\omega
_{n,T}^{2}\right\vert }{\omega _{n,T}^{2}}\leq \frac{\left\vert \hat{\sigma}%
_{U,n,T}^{2}-\sigma _{U,n,T}^{2}\right\vert }{\omega _{n,T}^{2}}+\frac{%
\left\vert \omega _{n,T}^{2}-\sigma _{n,T}^{2}-\sigma
_{U,n,T}^{2}\right\vert }{\omega _{n,T}^{2}}=O_{p}(T^{-1/2}),
\label{P_C_T1_3}
\end{equation}%
where the equality follows from Theorem \ref{C_L2}, (\ref{AU_L2_2}) in\
Lemma \ref{AU_L2}, and Assumptions \ref{A1}(i) and \ref{C_A1}(v).\ Combining
the results of (\ref{P_C_T1_1}), (\ref{P_C_T1_2}) and (\ref{P_C_T1_3}), we
conclude:%
\begin{equation}
\frac{\hat{\omega}_{n,T}^{2}}{\omega _{n,T}^{2}}=1+O_{p}(T^{-1/2}).
\label{P_C_T1_4}
\end{equation}%
Using (\ref{C_L1_2}), (\ref{C_T1_1}) and (\ref{P_C_T1_4}), we apply
Slutsky's theorem to deduce: 
\begin{align}
\frac{\left\vert MQLR_{n,T}\right\vert -\hat{\omega}_{n,T}z_{1-p/2}}{\omega
_{n,T}}& =\left\vert \frac{MQLR_{n,T}-\overline{QLR}_{n,T}}{\omega _{n,T}}+%
\frac{\overline{QLR}_{n,T}}{\omega _{n,T}}\right\vert -\frac{\hat{\omega}%
_{n,T}}{\omega _{n,T}}z_{1-p/2}  \notag \\
& \rightarrow _{d}\left\vert Z+c\right\vert -z_{1-p/2},  \label{P_C_T1_5}
\end{align}%
where $Z$ denotes a standard normal random variable.\ Therefore, as $%
n,T\rightarrow \infty $,%
\begin{align}
\mathbb{E}[\varphi _{n,T}^{2\text{-side}}(p)]& =\mathbb{P}\left( \left\vert
MQLR_{n,T}\right\vert >\hat{\omega}_{n,T}z_{1-p/2}\right)  \notag \\
& =\mathbb{P}\left( \frac{\left\vert MQLR_{n,T}\right\vert -\hat{\omega}%
_{n,T}z_{1-p/2}}{\omega _{n,T}}>0\right) \rightarrow \mathbb{P}\left(
\left\vert Z+c\right\vert >z_{1-p/2}\right) ,  \label{P_C_T1_6}
\end{align}%
which shows (\ref{C_T1_2}). The second claim of the theorem, stated in (\ref%
{C_T1_3}), follows by the similar arguments and its proof is omitted.\hfill $%
Q.E.D.$

\subsection{Auxiliary Results and Lemmas\label{sec:auxiliary-lemmas-main}}

This subsection presents the power analysis of the proposed test without
Assumption \ref{C_A2}, and collects the auxiliary lemmas used in the proofs
of Theorems \ref{C_L0}-\ref{C_T1}.

\begin{lemma}
\label{C_T2}\ Suppose that Assumptions \ref{A1} and \ref{C_A1} hold. Then:

(a) If $|\overline{QLR}_{n,T}|\succ G_{1}+G_{2}$, we have $\mathbb{E}%
[\varphi _{n,T}^{2\text{-side}}(p)]\rightarrow 1$ as $n,T\rightarrow \infty $%
.

(b) If $\overline{QLR}_{n,T}\succ G_{1}+G_{2}$, we have $\mathbb{E}[\varphi
_{n,T}^{1\text{-side}}(p)]\rightarrow 1$\ as $n,T\rightarrow \infty $.
\end{lemma}

\begin{lemma}
\textit{\label{AU_L1}\ }Under Assumptions \ref{A1} and \ref{C_A1}, we have%
\begin{eqnarray}
\sum_{g\in \mathcal{G}_{j}}\sum_{i\in I_{j,g}}\left( \frac{\widehat{\mathbb{E%
}}_{T}[\hat{\psi}_{j,\gamma }(z_{i,t})^{2}]}{n_{j,g}\hat{\Psi}_{j,\gamma
\gamma }(\hat{\phi}_{j,g})}-\frac{\mathbb{E}_{T}[\psi _{j,\gamma }\left(
z_{i,t}\right) ^{2}]}{n_{j,g}\Psi _{j,\gamma \gamma ,g}}\right) &=&O_{p}(1),
\label{AU_L1_1} \\
\sum_{g\in \mathcal{G}_{j}}\sum_{i\in I_{j,g}}\left( \frac{(\widehat{\mathbb{%
E}}_{T}[\hat{\psi}_{j,\gamma }(z_{i,t})])^{2}}{n_{j,g}\hat{\Psi}_{j,\gamma
\gamma }(\hat{\phi}_{j,g})}-\frac{(\mathbb{E}_{T}[\psi _{j,\gamma }\left(
z_{i,t}\right) ])^{2}}{n_{j,g}\Psi _{j,\gamma \gamma ,g}}\right) &=&O_{p}(1),
\label{AU_L1_2} \\
\frac{\left( MQLR_{n,T}\right) ^{2}-(\overline{QLR}_{n,T})^{2}}{%
(nT)^{1/2}\omega _{n,T}^{2}} &=&O_{p}(1).  \label{AU_L1_3} \\
\sum_{g\in \mathcal{G}_{j}}n_{j,g}^{-1}\sum_{i\in I_{j,g}}\left( \frac{(%
\widehat{\mathbb{E}}_{T}[\hat{\psi}_{j,\gamma }(z_{i,t})])^{2}}{\hat{\Psi}%
_{j,\gamma \gamma }(\hat{\phi}_{j,g})}-\frac{(\mathbb{E}_{T}[\psi _{j,\gamma
}(z_{i,t})])^{2}}{\Psi _{j,\gamma \gamma ,g}}\right) ^{2} &=&O_{p}(1).
\label{AU_L1_4}
\end{eqnarray}
\end{lemma}

\begin{lemma}
\textit{\label{AU_L2} }Under Assumptions \ref{A1}, \ref{C_A1} and \ref{C_A2}%
,\ we have%
\begin{eqnarray}
\frac{n^{-1}\sum_{i\leq n}(\widehat{\mathbb{E}}_{T}[\Delta \psi (z_{i,t},%
\hat{\phi}_{i})^{2}]-\mathbb{E}_{T}[\Delta \psi (z_{i,t})^{2}])-2\sigma
_{S,n,T}^{2}}{\omega _{n,T}(nT)^{-1/2}+T^{-3/2}} &=&O_{p}(1),
\label{AU_L2_1} \\
\frac{T(\omega _{n,T}^{2}-\sigma _{n,T}^{2}-\sigma _{U,n,T}^{2})}{\omega
_{n,T}} &=&O(1).  \label{AU_L2_2}
\end{eqnarray}
\end{lemma}

\begin{lemma}
\textit{\label{AU_L3}\ }Under Assumptions \ref{A1}, \ref{C_A1} and \ref{C_A2}%
,\ we have%
\begin{eqnarray}
\sum_{g\in \mathcal{G}_{j}}n_{j,g}^{-1}\sum_{i\in I_{j,g}}(\hat{s}_{j,\gamma
,i}^{2}-s_{j,\gamma ,i}^{2})^{2} &=&O_{p}(1),  \label{AU_L3_1} \\
\sum_{g\in \mathcal{G}_{2}}n_{2,g}^{-1}\sum_{i\in I_{2,g}}(\hat{\sigma}%
_{12,\gamma ,i}-\sigma _{12,\gamma ,i})^{2} &=&O_{p}(1),  \label{AU_L3_2} \\
\frac{\sum_{g\in \mathcal{G}_{2}}\left( \left( n_{2,g}^{-1}\sum_{i\in
I_{2,g}}\hat{\sigma}_{2,\gamma ,i}^{2}\right) ^{2}-\left(
n_{2,g}^{-1}\sum_{i\in I_{2,g}}\sigma _{2,\gamma ,i}^{2}\right) ^{2}\right) 
}{(n^{1/2}T)\omega _{n,T}^{2}} &=&O_{p}(1),  \label{AU_L3_3} \\
\frac{\sum_{g\in \mathcal{G}_{2}}n_{2,g}^{-2}\sum_{i\in I_{2,g}}\left( \hat{%
\sigma}_{2,\gamma ,i}^{2}\sum_{i^{\prime }\in I_{2,g}}\hat{s}_{2,\gamma
,i^{\prime }}^{2}-\sigma _{2,\gamma ,i}^{2}\sum_{i^{\prime }\in
I_{2,g}}s_{2,\gamma ,i^{\prime }}^{2}\right) }{(n^{1/2}T)\omega _{n,T}^{2}}
&=&O_{p}(1).  \label{AU_L3_4}
\end{eqnarray}
\end{lemma}

\begin{lemma}
\textit{\label{AU_L4}\ }Under Assumptions \ref{A1} and \ref{C_A1}, we have:\
(i) $\min_{g\in \mathcal{G}_{j}}(-\hat{\Psi}_{j,\gamma \gamma }(\hat{\phi}%
_{j,g}))\geq K^{-1}$; and (ii) $\omega _{n,T}^{2}\leq K$.
\end{lemma}

\section{Assumptions and Proofs for Section \protect\ref{sec:TWE}\label%
{sec:proofTWE}}

\begin{assumption}
\textit{\label{A8}\ (i) }$\lambda _{\min }(\Sigma _{\ddot{x}})\geq K^{-1}$\
and $\lambda _{\min }(\Sigma _{\dot{x}})\geq K^{-1}$;\ (ii) $\max_{i,t}%
\mathbb{E}[y_{i,t}^{8}+\left\Vert x_{i,t}\right\Vert ^{8}]\leq K$.
\end{assumption}

\begin{assumption}
\textit{\label{A9}\ (i) }$n\omega _{n,T}^{2}\geq K^{-1}$; (ii) $\mathbb{E}%
[\omega _{n,T}^{-4}|\varepsilon _{2,i,t}^{2}-\varepsilon
_{1,i,t}^{2}|^{4}]<K $;\ (iii) $\max_{g\in \mathcal{G}_{1}}n_{g}^{-1}=o(1)$.
\end{assumption}

\begin{assumption}
\textit{\label{A10}\ (i) }The vectors $(\varepsilon _{1,i,t},\varepsilon
_{2,i,t})$ are independent across $t$, and $\mathbb{E}[\varepsilon _{j,i,t}]$%
, $\mathbb{E}[\varepsilon _{j,i,t}^{2}]$\ and $\mathbb{E}[\varepsilon
_{1,i,t}\varepsilon _{2,i,t}]$ are invariant in $t$; (ii) $%
G_{1}n^{-1/2}+n^{1/2}\sum_{g\in \mathcal{G}_{1}}n_{g}^{-1}=o(1)$.
\end{assumption}

Assumptions \ref{A8} and \ref{A9} are imposed to study the properties of the
estimators in (\ref{HTFE_1}) and (\ref{TFE_2}), and to derive the asymptotic
distribution of $QLR_{n,T}$ for comparing the heterogeneous time effects
model with the TWFE model. Assumption \ref{A8} ensures the pseudo true
parameters are uniquely identified (and well-defined). Assumptions \ref{A9}%
(i, ii) are the counterparts of Assumptions \ref{A6}(i) and \ref{A7}(i) in
Appendix \ref{Sec:AP2}, respectively, and are used to establish the
asymptotic distribution of $QLR_{n,T}$. Finally, Assumption \ref{A10} is
imposed primarily to simplify the variance of the QLR statistic and to
ensure the consistency of the proposed variance estimator.

\bigskip

\noindent \textsc{Proof of Theorem \ref{TWFE_Vuong}.} We begin by observing
that 
\begin{align}
\frac{\mathbb{E}\left[ \left\vert \sum_{i=1}^{n}(\tilde{V}_{1,i}-\mathbb{E}[%
\tilde{V}_{1,i}])\right\vert ^{2}\right] }{n\omega _{n,T}^{2}}& =\frac{%
\sum_{i=1}^{n}\mathbb{E}[(\tilde{V}_{1,i}-\mathbb{E}[\tilde{V}_{1,i}])^{2}]}{%
n\omega _{n,T}^{2}}  \notag \\
& \leq K\sum_{g\in \mathcal{G}_{1}}n_{g}^{-2}\sum_{i\in I_{g}}\mathbb{E}%
\left[ \left( T^{-1/2}\sum_{t\leq T}(\varepsilon _{1,i,t}^{\ast 2}-\mathbb{E}%
[\varepsilon _{1,i,t}^{\ast 2}])\right) ^{2}\right]  \notag \\
& \leq K\sum_{g\in \mathcal{G}_{1}}n_{g}^{-2}\sum_{i\in I_{g}}\max_{t\leq
T}\left\Vert \varepsilon _{1,i,t}^{\ast 2}\right\Vert _{2+\delta }^{2}\leq
KG_{1}\max_{g\in \mathcal{G}_{1}}n_{g}^{-1}=o(1),  \label{P_TWFE_Vuong_1}
\end{align}%
where the first inequality follows from Assumption \ref{A9}(i). The second
inequality holds for any $\delta \in (0,1)$ by the covariance inequality for
strong mixing processes together with Assumption \ref{A1}(iv). The third
inequality follows from $\mathbb{E}[\varepsilon _{1,i,t}^{8}]\leq K$, which
is implied by Assumptions \ref{A1} and \ref{A8} (see, Lemma \ref{TWFE_Para}%
(ii) in the Online Appendix). The final equality follows from Assumption \ref%
{A9} (iii). Similarly, we can establish that%
\begin{equation}
\frac{\mathbb{E}\left[ \left\vert (nT^{1/2})^{-1}\sum_{i=1}^{n}\sum_{t\leq
T}(\varepsilon _{2,i,t}^{\ast 2}-\mathbb{E}[\varepsilon _{2,i,t}^{\ast
2}])\right\vert ^{2}\right] }{n\omega _{n,T}^{2}}\leq Kn^{-1}=o(1),
\label{P_TWFE_Vuong_2}
\end{equation}%
which along with (\ref{QLR-approx-linear}), (\ref{P_TWFE_Vuong_1}) and
Markov's inequality implies that%
\begin{equation}
\frac{\widehat{QLR}_{n,T}-\mathbb{E}[S_{n,T}]-\overline{QLR}_{n,T}}{\omega
_{n,T}}=\sum_{i\leq n}\xi _{nT,i}+o_{p}(1),  \label{P_TWFE_Vuong_3}
\end{equation}%
where $\xi _{nT,i}=(n^{1/2}\omega _{n,T})^{-1}(\tilde{\Psi}_{i}^{\ast }+%
\tilde{V}_{i}^{\ast }+\tilde{U}_{i})$\ and $\tilde{\Psi}_{i}^{\ast }=\tilde{%
\Psi}_{i}-\mathbb{E}[\tilde{\Psi}_{i}]$. It can be shown\footnote{%
See Lemma \ref{TWFE_MGCLT} in the Online Appendix \ref{Proof_App_TWE}.} that
sufficient conditions of the martingale CLT for $\sum_{i\leq n}\xi _{nT,i}$
are satisfied, which establishes the asymptotic normality of (\ref%
{P_TWFE_Vuong_3}).\hfill $Q.E.D.$

\bigskip

\noindent \textsc{Proof of Theorem \ref{TWFE_F_Vuong}.} Under Assumptions %
\ref{A1},\ \ref{A8}, \ref{A9} and \ref{A10}, we have 
\begin{eqnarray}
\frac{MQLR_{n,T}}{\hat{\omega}_{n,T}} &=&\frac{\omega _{n,T}}{\hat{\omega}%
_{n,T}}\left( \frac{QLR_{n,T}-\mathbb{E}[S_{n,T}]-\overline{QLR}_{n,T}}{%
\omega _{n,T}}+\frac{\overline{QLR}_{n,T}}{\omega _{n,T}}-\frac{\widehat{%
\mathbb{E}[S_{n,T}]}-\mathbb{E}[S_{n,T}]}{\omega _{n,T}}\right)  \notag \\
&=&(1+o_{p}(1))\left( \frac{QLR_{n,T}-\mathbb{E}[S_{n,T}]-\overline{QLR}%
_{n,T}}{\omega _{n,T}}+\frac{\overline{QLR}_{n,T}}{\omega _{n,T}}%
+o_{p}(1)\right)  \notag \\
&\rightarrow &\hspace{-0.75em}_{d} {\text{ \ }}Z+c,  \label{P_TWFE_F_Vuong_1}
\end{eqnarray}%
where $\overline{QLR}_{n,T}=(nT)^{-1/2}\sum_{i\leq n}\sum_{t\leq T}\mathbb{E}%
[\varepsilon _{2,i,t}^{2}-\varepsilon _{1,i,t}^{2}]$. The first equality
follows from the definition of $MQLR_{n,T}$, while the second equality is
implied by Lemma \ref{TWE_Bias&Var_Est_Null}. The stated asymptotic
distribution follows from Theorem \ref{TWFE_Vuong}, (\ref{TWFE_F_Vuong_1})
and Slutsky's theorem. The claim of the theorem then follows directly from (%
\ref{P_TWFE_F_Vuong_1}) together with arguments analogous to those used in
the proof of Theorem \ref{C_T1}.\hfill $Q.E.D.$

\subsection{Auxiliary Results and Lemmas\label{sec:auxiliary-lemmas-twfe}}

\begin{lemma}
\textit{\label{TWFE_F_Vuong_Power} }Suppose that\ \textit{Assumptions \ref%
{A1},\ }\ref{A8} and \ref{A9} hold. Then:

(a) If $\left\vert (nT)^{-1/2}\sum_{i\leq n}\sum_{t\leq T}\mathbb{E}%
[\varepsilon _{2,i,t}^{2}-\varepsilon _{1,i,t}^{2}]\right\vert \succ G_{1}$,
then $\mathbb{E}[\varphi _{n,T}^{2\text{-side}}(p)]\rightarrow 1$, as $%
n,T\rightarrow \infty $.

(b) If $(nT)^{-1/2}\sum_{i\leq n}\sum_{t\leq T}\mathbb{E}[\varepsilon
_{2,i,t}^{2}-\varepsilon _{1,i,t}^{2}]\succ G_{1}$, then $\mathbb{E}[\varphi
_{n,T}^{1\text{-side}}(p)]\rightarrow 1$, as $n,T\rightarrow \infty $.
\end{lemma}

\begin{lemma}
\textit{\label{TWFE_Var}\ }Under\ \textit{Assumptions \ref{A1},\ }\ref{A8}, %
\ref{A9} and \ref{A10}, we have $\omega _{n,T}^{2}=\sigma _{n,T}^{2}+\sigma
_{U,n,T}^{2}+o(\omega _{n,T}^{2})$.
\end{lemma}

\begin{lemma}
\textit{\label{TWFE_Var_U}\ We have }$\hat{\sigma}_{U,n,T}^{2}\geq
(2nT)^{-1}\sum_{i\leq n}\hat{\sigma}_{2,i}^{4}+n^{-3}\sum_{g=2}^{G_{1}}%
\sum_{g^{\prime }=1}^{g-1}(\sum_{i\in I_{g}}\hat{\sigma}_{2,i}^{2})(\sum_{i%
\in I_{g^{\prime }}}\hat{\sigma}_{2,i}^{2})$.
\end{lemma}

\begin{lemma}
\textit{\label{TWE_Bias&Var_Est_Null}\ Suppose }Assumptions \textit{\ref{A1}%
,\ }\ref{A8}, \ref{A9} and \ref{A10} hold. Then, under the null hypothesis,%
\begin{equation}
\widehat{\mathbb{E}[S_{n,T}]}-\mathbb{E}[S_{n,T}]=o_{p}(\omega _{n,T})\text{
\ \ \ and \ \ }\hat{\omega}_{n,T}^{2}/\omega _{n,T}^{2}=1+o_{p}(1).
\label{TWE_Bias&Var_Est_Null_1}
\end{equation}
\end{lemma}

\begin{lemma}
\textit{\label{TWE_Bias&Var_Est_Alt}\ Suppose }Assumptions \textit{\ref{A1}%
,\ }\ref{A8} and \ref{A9} hold. Then $\hat{\omega}_{n,T}^{2}=O_{p}(1)$ and $%
\widehat{\mathbb{E}[S_{n,T}]}=O_{p}(G_{1})$. Moreover, $\left\vert \mathbb{E}%
[S_{n,T}]\right\vert \leq KG_{1}$ and $\omega _{n,T}^{2}\leq K$.
\end{lemma}

\section{General Theory for Nonlinear Panel Likelihood Models\label{Sec:AP2}}

In this section, we present general results on the properties of estimators
for nonlinear panel models and derive the asymptotic distribution of the QLR
statistic for comparing different panel models. The proofs of these results
are provided in the Online Appendix of the paper.

\subsection{General Framework\label{sec:main}}

We consider a general nonlinear panel model with joint likelihood specified
in (\ref{general_spec}). Recall that $I_{g}$ for $g\in \mathcal{G}$ and $%
I_{m}$ for $m\in \mathcal{M}$ denote the partitions of $\{1,\ldots ,n\}$ and 
$\{1,\ldots ,T\}$, respectively, induced by the group assignment functions $%
g(\cdot )$ and $m(\cdot )$. For notational convenience, we assume that $%
m(\cdot )$ is nondecreasing. This implies that for any two consecutive time
groups $m$ and $m+1$, we have $\min I_{m+1}=\max I_{m}+1$.

Under this notation, the joint likelihood in (\ref{general_spec}) can be
written as 
\begin{equation*}
\sum_{g\in \mathcal{G}}\sum_{m\in \mathcal{M}}\sum_{i\in I_{g}}\sum_{t\in
I_{m}}\psi \left( z_{i,t};\theta ,\gamma _{g,m}\right) ,
\end{equation*}%
where $\psi \left( z_{i,t};\theta ,\gamma _{g,m}\right) \equiv \log f\left(
z_{i,t};\theta ,\gamma _{g,m}\right) $. Let $(\theta ^{\ast \top },\gamma
_{g,m}^{\ast \top })^{\top }$ denote the pseudo true value defined as%
\begin{equation*}
(\theta ^{\ast \top },\gamma _{g,m}^{\ast \top })^{\top }=\underset{(\theta
^{\top },\gamma _{g,m}^{\top })^{\top }\in \Phi }{\arg \max }\Psi
_{g,m}(\theta ,\gamma _{g,m}),
\end{equation*}%
where\ $\Psi _{g,m}(\theta ,\gamma _{g,m})\equiv (n_{g}T_{m})^{-1}\sum_{i\in
I_{g}}\sum_{t\in I_{m}}\mathbb{E}\left[ \psi \left( z_{i,t};\theta ,\gamma
_{g,m}\right) \right] $,\ $\Phi $ denotes the parameter space including $%
(\theta ^{\ast \top },\gamma _{g,m}^{\ast \top })^{\top }$ in its interior,
and $n_{g}$ and $T_{m}$ denote the sizes of $I_{g}$ and $I_{m}$
respectively. The unknown pseudo true parameters are estimated through
M-estimation, i.e., 
\begin{equation*}
(\hat{\theta}^{\top },\hat{\gamma}_{\mathcal{G},1}^{\top },\ldots ,\hat{%
\gamma}_{\mathcal{G},M}^{\top })^{\top }\equiv \underset{\theta ,\{\gamma _{%
\mathcal{G},m}\}_{m\in \mathcal{M}}}{\arg \max }\sum_{g\in \mathcal{G}%
}\sum_{m\in \mathcal{M}}\sum_{i\in I_{g}}\sum_{t\in I_{m}}\psi \left(
z_{i,t};\theta ,\gamma _{g,m}\right) ,
\end{equation*}%
where $\gamma _{\mathcal{G},m}=(\gamma _{g,m}^{\top })_{g\in \mathcal{G}%
}^{\top }$ for $m=1,\ldots ,M$. The M-estimator $(\hat{\theta}^{\top },\hat{%
\gamma}_{\mathcal{G},1}^{\top },\ldots ,\hat{\gamma}_{\mathcal{G},M}^{\top
})^{\top }$ satisfies the first-order conditions%
\begin{eqnarray}
\sum_{g\in \mathcal{G}}\sum_{m\in \mathcal{M}}\sum_{i\in I_{g}}\sum_{t\in
I_{m}}\frac{\partial \psi (z_{i,t};\hat{\theta},\hat{\gamma}_{g,m})}{%
\partial \theta } &=&0_{d_{\theta }\times 1},  \label{Foc_theta} \\
\sum_{i\in I_{g}}\sum_{t\in I_{m}}\frac{\partial \psi (z_{i,t};\hat{\theta},%
\hat{\gamma}_{g,m})}{\partial \gamma } &=&0_{d_{\gamma }\times 1},
\label{Foc_gamma}
\end{eqnarray}%
and for any $g\in \mathcal{G}$ and$\ m\in \mathcal{M}$.

\subsection{Asymptotic Theory of the M-Estimator \label{Sec: M Theory}}

We establish the asymptotic properties of the M-estimator $(\hat{\theta}%
^{\top },\hat{\gamma}_{\mathcal{G},1}^{\top },\ldots ,\hat{\gamma}_{\mathcal{%
G},M}^{\top })^{\top }$,\ together with the asymptotic expansion of the
maximized joint likelihood in this subsection. We begin by stating
sufficient conditions under which these results hold.

\begin{assumption}
\textit{\label{A2} }$\min_{g\in \mathcal{G},m\in \mathcal{M}}\left[ \Psi
_{g,m}(\phi _{g,m}^{\ast })-\sup_{\left\{ \phi _{g,m}\in \Phi :\text{ }%
||\phi _{g,m}-\phi _{g,m}^{\ast }||>\eta \right\} }\Psi _{g,m}(\phi _{g,m})%
\right] >0$ \textit{for each }$\eta >0$, where $\Psi _{g,m}(\phi
_{g,m})\equiv \Psi _{g,m}(\theta ,\gamma _{g,m})$ and $\phi _{g,m}\equiv
(\theta ^{\top },\gamma _{g,m}^{\top })^{\top }$.
\end{assumption}

\begin{assumption}
\textit{\label{A3} }Let $\psi (z_{i,t};\tilde{\phi})$ be a function indexed
by the parameter $\tilde{\phi}=(\tilde{\phi}_{j})_{j\leq k}\in \limfunc{int}%
\Phi $, where $k\equiv \dim (\tilde{\phi})$ and $\Phi $ is a compact, convex
subset of $\mathbb{R}^{k}$. Let $\nu \equiv \left( \nu _{1},...,\nu
_{k}\right) ^{\top }$\ be a vector of non-negative integers, $\left\vert \nu
\right\vert =\sum_{j=1}^{k}\nu _{j}$ and $D^{\nu }\psi (z_{i,t},\tilde{\phi}%
)=\partial ^{\left\vert \nu \right\vert }\psi (z_{i,t},\tilde{\phi}%
)/(\partial \tilde{\phi}_{1}^{\nu _{1}}\cdots \partial \tilde{\phi}_{k}^{\nu
_{k}})$. Then: (i) there exists a function $M\left( z_{i,t}\right) $ such
that 
\begin{equation*}
\left\vert D^{\nu }\psi (z_{i,t},\tilde{\phi}_{1})-D^{\nu }\psi (z_{i,t},%
\tilde{\phi}_{2})\right\vert \leq M\left( z_{i,t}\right) ||\tilde{\phi}_{1}-%
\tilde{\phi}_{2}||
\end{equation*}%
for all $\tilde{\phi}_{1},\tilde{\phi}_{2}\in \Phi $ and $\left\vert
v\right\vert \leq 4$; (ii) the function $M(z_{i,t})$ satisfies $\sup_{i,t}%
\mathbb{E}[\left\vert M(z_{i,t})\right\vert ^{p+\delta }]<K$ for some
integer $p>\max \{4,k+\delta \}$ and for some $\delta >0$; (iii) for all
multi-indices $v$ with $\left\vert v\right\vert \leq 4$, $\max_{g\in 
\mathcal{G},m\in \mathcal{M}}\sup_{i,t}\mathbb{E}\left[ \left\vert D^{\nu
}\psi \left( z_{i,t},\phi _{g,m}^{\ast }\right) \right\vert ^{4+\delta }%
\right] <K$ where $\phi _{g,m}^{\ast }\equiv (\theta ^{\ast \top },\gamma
_{g,m}^{\ast \top })^{\top }$.
\end{assumption}

Let $\hat{\Psi}_{ab,g,m}\equiv (n_{g}T_{m})^{-1}\sum_{i\in I_{g}}\sum_{t\in
I_{m}}\psi _{ab}(z_{i,t};\phi _{g,m}^{\ast })$ and $\Psi _{ab,g,m}\equiv 
\mathbb{E}[\hat{\Psi}_{ab,g,m}]$\ for $a,b\in \{\gamma ,\theta \}$, where $%
\psi _{ab}(z_{i,t};\phi )\equiv \frac{\partial \psi (z_{i,t};\phi )}{%
\partial a\partial b^{\top }}$.

\begin{assumption}
\textit{\label{A4} }The smallest eigenvalues of $-\Psi _{\gamma \gamma ,g,m}$
and $\Psi _{\theta \gamma ,g,m}\Psi _{\gamma \gamma ,g,m}\Psi _{\gamma
\theta ,g,m}-\Psi _{\theta \theta ,g,m}$ are bounded below by $K^{-1}$
uniformly over $g$, $m$, $n$ and $T$.
\end{assumption}

Assumption \ref{A2} is the identification uniqueness condition to ensure the
consistency of the M-estimator. Assumption \ref{A3} imposes smoothness and
uniform moment bounds on the quasi-likelihood function and its derivatives.
Assumption \ref{A4} provides local identification conditions for $\theta
^{\ast }$ and $\gamma _{g,m}^{\ast }$.

\begin{assumption}
\textit{\label{A5} (i)\ }$(GM)^{2/p}\max_{g\in \mathcal{G},m\in \mathcal{M}%
}(n_{g}T_{m})^{-1/2}=o(1)$;\ (ii) $(GM)(nT)^{-1/2}=O(1)$.
\end{assumption}

Assumption \ref{A5}(i) is used to establish the consistency of the
M-estimator. It accommodates a wide range of grouping structures, including
Examples 1--4 in Section \ref{sec:linear-section} of the Online Appendix as
special cases. To better illustrate the restriction imposed by Assumption %
\ref{A5}(i) on the number of groups, let's consider a simple scenario where
the sizes of the groups $I_{g}$ ($g\in \mathcal{G}$) are proportional to
each other and the sizes of $I_{m}$ ($m\in \mathcal{G}$) are also
proportional to each other, i.e.,%
\begin{equation*}
K^{-1}\leq \min \left\{ \frac{\min_{g\in \mathcal{G}}n_{g}}{\max_{g\in 
\mathcal{G}}n_{g}},\frac{\min_{m\in \mathcal{M}}T_{m}}{\max_{m\in \mathcal{M}%
}T_{m}}\right\} \leq \max \left\{ \frac{\max_{g\in \mathcal{G}}n_{g}}{%
\min_{g\in \mathcal{G}}n_{g}},\frac{\max_{m\in \mathcal{M}}T_{m}}{\min_{m\in 
\mathcal{M}}T_{m}}\right\} \leq K,
\end{equation*}%
where $K$ denotes a generic finite constant. In this case 
\begin{equation}
K^{-1}\leq n(Gn_{g})^{-1}\leq K\text{ \ \ and \ \ }K^{-1}\leq
T(MT_{m})^{-1}\leq K  \label{gp_size_1}
\end{equation}%
for any $g\in \mathcal{G}$ and any $m\in \mathcal{M}$. Under (\ref{gp_size_1}%
), Assumption \ref{A5}(i) becomes 
\begin{equation}
(GM)^{1/2+2/p}(nT)^{-1/2}=o(1)  \label{gp_size_2}
\end{equation}%
which requires that the number of groups $GM$ is of a smaller order of $nT$.
Assumption \ref{A5}(ii) is used to establish the $(nT)^{-1/2}$-consistency
of $\hat{\theta}$.

Let $\hat{\Psi}_{\gamma ,g,m}\equiv (n_{g}T_{m})^{-1}\sum_{i\in
I_{g}}\sum_{t\in I_{m}}\psi _{\gamma }(z_{i,t})$, where $\psi _{\gamma
}(z_{i,t})\equiv \partial \psi (z_{i,t};\phi _{g,m}^{\ast })/\partial \gamma 
$. Define $\hat{U}_{\gamma ,g,m}\equiv \hat{\Psi}_{\theta \gamma ,g,m}-\Psi
_{\theta \gamma ,g,m}\Psi _{\gamma \gamma ,g,m}^{-1}\hat{\Psi}_{\gamma
\gamma ,g,m}$, and for $j=1,\ldots ,d_{\gamma }$, 
\begin{equation*}
\Psi _{\gamma _{j}\gamma \gamma ,g,m}\equiv (n_{g}T_{m})^{-1}\sum_{i\in
I_{g}}\sum_{t\in I_{m}}\mathbb{E}\left[ \frac{\partial ^{2}}{\partial \gamma
\partial \gamma ^{\top }}\left( \frac{\partial \psi (z_{i,t};\phi
_{g,m}^{\ast })}{\partial \gamma _{j}}\right) \right] .
\end{equation*}%
The matrix $\Psi _{\theta _{j}\gamma \gamma ,g,m}$ is defined in the similar
way for $j=1,\ldots ,d_{\theta }$. Theorem \ref{Rep_Theta} below provides
the second order expansion of the estimation error in $\hat{\theta}$.

\begin{theorem}
\label{Rep_Theta} Under Assumptions \ref{A1}, \ref{A2}, \ref{A3}, \ref{A4}
and \ref{A5}, we have%
\begin{align*}
\hat{\theta}-\theta ^{\ast }& =H_{\theta \theta }^{-1}(nT)^{-1}\sum_{g\in 
\mathcal{G}}\sum_{m\in \mathcal{M}}(n_{g}T_{m})(\hat{\Psi}_{\theta
,g,m}-\Psi _{\theta \gamma ,g,m}\Psi _{\gamma \gamma ,g,m}^{-1}\hat{\Psi}%
_{\gamma ,g,m}) \\
& \text{ \ \ \ }+H_{\theta \theta }^{-1}(nT)^{-1}\sum_{g\in \mathcal{G}%
}\sum_{m\in \mathcal{M}}(n_{g}T_{m})B_{\theta ,g,m}+o_{p}\left(
(nT)^{-1/2}\right) ,
\end{align*}%
where%
\begin{align*}
H_{\theta \theta }& \equiv -(nT)^{-1}\sum_{g\in \mathcal{G}}\sum_{m\in 
\mathcal{M}}(n_{g}T_{m})(\Psi _{\theta \theta ,g,m}-\Psi _{\theta \gamma
,g,m}\Psi _{\gamma \gamma ,g,m}^{-1}\Psi _{\gamma \theta ,g,m}), \\
B_{\theta ,g,m}& \equiv 2^{-1}\mathbb{E}\left[ \left( \hat{\Psi}_{\gamma
,g,m}^{\top }\Psi _{\gamma \gamma ,g,m}^{-1}\Psi _{\theta _{j}\gamma \gamma
,g,m}\Psi _{\gamma \gamma ,g,m}^{-1}\hat{\Psi}_{\gamma ,g,m}\right) _{j\leq
d_{\theta }}\right] -\mathbb{E}\left[ \hat{U}_{\gamma ,g,m}\Psi _{\gamma
\gamma ,g,m}^{-1}\hat{\Psi}_{\gamma ,g,m}\right] \\
& \text{ \ \ \ }-2^{-1}\Psi _{\theta \gamma ,g,m}\Psi _{\gamma \gamma
,g,m}^{-1}\mathbb{E}\left[ \left( \hat{\Psi}_{\gamma ,g,m}^{\top }\Psi
_{\gamma \gamma ,g,m}^{-1}\Psi _{\gamma _{j}\gamma \gamma ,g,m}\Psi _{\gamma
\gamma ,g,m}^{-1}\hat{\Psi}_{\gamma ,g,m}\right) _{j\leq d_{\gamma }}\right]
.
\end{align*}
\end{theorem}

Theorem \ref{Rep_L} establishes a second-order expansion of the likelihood
function evaluated at the M-estimator $\hat{\phi}\equiv (\hat{\theta}^{\top
},\hat{\gamma}_{\mathcal{G},1}^{\top },\ldots ,\hat{\gamma}_{\mathcal{G}%
,M}^{\top })^{\top }$.

\begin{theorem}
\textit{\ \label{Rep_L} }Under Assumptions \ref{A1}, \ref{A2}, \ref{A3}, \ref%
{A4} and \ref{A5}, we have%
\begin{equation*}
L_{n,T}(\hat{\phi})=L_{n,T}(\phi ^{\ast })+2^{-1}\sum_{g\in \mathcal{G}%
}\sum_{m\in \mathcal{M}}(n_{g}T_{m})\hat{\Psi}_{\gamma ,g,m}^{\top }(-\Psi
_{\gamma \gamma ,g,m})^{-1}\hat{\Psi}_{\gamma ,g,m}+o_{p}\left(
(GM)^{1/2}\right) ,
\end{equation*}%
where $L_{n,T}(\hat{\phi})=\sum_{g\in \mathcal{G}}\sum_{m\in \mathcal{M}%
}\sum_{i\in I_{g}}\sum_{t\in I_{m}}\psi (z_{i,t};\hat{\theta},\hat{\gamma}%
_{g,m})$, and $L_{n,T}(\phi ^{\ast })$ is defined analogously with $\hat{\phi%
}$ replaced by the pseudo true parameter $\phi ^{\ast }$.
\end{theorem}

Let $\hat{\Psi}_{g,m}(\phi _{g,m})\equiv (n_{g}T_{m})^{-1}\sum_{i\in
I_{g}}\sum_{t\in I_{m}}\psi (z_{i,t};\theta ,\gamma _{g,m})$. Then the joint
likelihood evaluated at the pseudo true parameter can be written as $%
L_{n,T}(\phi ^{\ast })=\sum_{g\in \mathcal{G}}\sum_{m\in \mathcal{M}%
}(n_{g}T_{m})\hat{\Psi}_{g,m}(\phi _{g,m}^{\ast })$. Define%
\begin{equation*}
R_{n,T}^{\ast }\equiv \frac{(nT)^{-1/2}}{2}\sum_{g\in \mathcal{G}}\sum_{m\in 
\mathcal{M}}(n_{g}T_{m})\mathbb{E}[\hat{\Psi}_{\gamma ,g,m}^{\top }(-\Psi
_{\gamma \gamma ,g,m}^{-1})\hat{\Psi}_{\gamma ,g,m}].
\end{equation*}%
With this notation, Theorem \ref{Rep_L} implies the following expansion for
the maximized joint likelihood:%
\begin{align}
& (nT)^{-1/2}L_{n,T}(\hat{\phi})-R_{n,T}^{\ast }  \notag \\
& =(nT)^{-1/2}\sum_{g\in \mathcal{G}}\sum_{m\in \mathcal{M}}(n_{g}T_{m})%
\mathbb{E}[\hat{\Psi}_{g,m}(\phi _{g,m}^{\ast })]  \label{R_L-1} \\
& \text{ \ }+(nT)^{-1/2}\sum_{g\in \mathcal{G}}\sum_{m\in \mathcal{M}%
}(n_{g}T_{m})(\hat{\Psi}_{g,m}(\phi _{g,m}^{\ast })-\mathbb{E}[\hat{\Psi}%
_{g,m}(\phi _{g,m}^{\ast })])  \label{R_L-2} \\
& \text{ \ }-(4nT)^{-1/2}\sum_{g\in \mathcal{G}}\sum_{m\in \mathcal{M}%
}(n_{g}T_{m})(\hat{\Psi}_{\gamma ,g,m}^{\top }\Psi _{\gamma \gamma ,g,m}^{-1}%
\hat{\Psi}_{\gamma ,g,m}-\mathbb{E}[\hat{\Psi}_{\gamma ,g,m}^{\top }\Psi
_{\gamma \gamma ,g,m}^{-1}\hat{\Psi}_{\gamma ,g,m}])  \label{R_L-3} \\
& \text{ \ }+o_{p}\left( (GM)^{1/2}(nT)^{-1/2}\right) ,  \label{R_L-4}
\end{align}%
where the first term (\ref{R_L-1}) is the population likelihood function
evaluated at the pseudo true parameter, the second term (\ref{R_L-2}) and
the third term (\ref{R_L-3}) are the first-order term and the second-order
term respectively, (\ref{R_L-4}) represents the higher order term dominated
by the term (\ref{R_L-3}). The second term (\ref{R_L-2}) plus the third term
(\ref{R_L-3})\ after self-normalization/studentization can be approximated
by a standard normal random variable, which as we shall see in the next
section, enables us to achieve valid model comparison inference uniformly
over different model structures.

\subsection{Generic QLR: Expansion and Limit Theory \label{V_Test}}

Suppose that, for the same data set $\left\{ z_{i,t}\right\} _{i\leq n,t\leq
T}$, we consider two models $j=1,2$. For model $j$, the joint log-likelihood
function is%
\begin{equation*}
\sum_{t\leq T}\sum_{i\leq n}\log f_{j}(z_{i,t};\theta _{j},\gamma _{j,i,t}),
\end{equation*}%
where $f_{j}\left( \cdot \right) $ is known, $\theta _{j}$ collects
individual- and time-invariant parameters, and $\gamma _{j,i,t}\equiv \gamma
_{g_{j}(i),m_{j}(t)}$ denotes fixed effects indexed by known group
assignment function $g_{j}(\cdot )$ and $m_{j}(\cdot )$ with ranges $%
\mathcal{G}_{j}=\{1,\ldots ,G_{j}\}$ and $\mathcal{M}_{j}=\{1,\ldots
,M_{j}\} $ respectively. For any $i$ and $t$ with $g_{j}(i)=g$ and $%
m_{j}(t)=m$, let%
\begin{equation*}
\psi _{j}\left( z_{i,t};\theta _{j},\gamma _{j,g,m}\right) \equiv \log
f_{j}(z_{i,t};\theta _{j},\gamma _{j,i,t}).
\end{equation*}%
Then the joint log-likelihood function can be written as%
\begin{equation*}
\sum_{g\in \mathcal{G}_{j}}\sum_{m\in \mathcal{M}_{j}}\sum_{i\in
I_{g}}\sum_{t\in I_{m}}\psi _{j}\left( z_{i,t};\theta _{j},\gamma
_{j,g,m}\right) \equiv \sum_{g\in \mathcal{G}_{j}}\sum_{m\in \mathcal{M}%
_{j}}(n_{g}T_{m})\hat{\Psi}_{j,g,m}(\phi _{j,g,m}),
\end{equation*}%
where $\hat{\Psi}_{j,g,m}(\phi _{j,g,m})\equiv (n_{g}T_{m})^{-1}\sum_{i\in
I_{g}}\sum_{t\in I_{m}}\psi _{j}\left( z_{i,t};\theta _{j},\gamma
_{j,g,m}\right) $ and $\phi _{j,g,m}\equiv (\theta _{j}^{\top },\gamma
_{j,g,m}^{\top })^{\top }$. The two models may differ in their
specifications of the likelihood functions $f_{j}(\cdot )$ and/or the group
assignment functions $g_{j}(\cdot )$ and/or $m_{j}(\cdot )$.

Let $\phi _{j,i,t}^{\ast }\equiv (\theta _{j}^{\ast \top },\gamma
_{j,i,t}^{\ast \top })^{\top }$\ denote the pseudo true parameters of model $%
j$. The Vuong test compares two models by testing 
\begin{equation}
H_{0}:\sum_{t\leq T}\sum_{i\leq n}\mathbb{E}\left[ \Delta \psi (z_{i,t})%
\right] =0,\text{ where }\Delta \psi (z_{i,t})\equiv \psi _{1}(z_{i,t};\phi
_{1,i,t}^{\ast })-\psi _{2}(z_{i,t};\phi _{2,i,t}^{\ast })  \label{H0}
\end{equation}%
using the QLR statistic 
\begin{equation}
QLR_{n,T}\equiv (nT)^{-1/2}\left( \sum_{g\in \mathcal{G}_{1}}\sum_{m\in 
\mathcal{M}_{1}}(n_{g}T_{m})\hat{\Psi}_{1,g,m}(\hat{\phi}_{1,g,m})-\sum_{g%
\in \mathcal{G}_{2}}\sum_{m\in \mathcal{M}_{2}}(n_{g}T_{m})\hat{\Psi}%
_{2,g,m}(\hat{\phi}_{2,g,m})\right) ,  \label{QLR_S}
\end{equation}%
where $\hat{\phi}_{j,g,m}\equiv (\hat{\theta}_{j}^{\top },\hat{\gamma}%
_{j,g,m}^{\top })^{\top }\ $denotes the estimator of $\phi _{j,g,m}^{\ast }$
in model $j$.

Let $\tilde{\Psi}_{j,i}\equiv T^{-1/2}\sum_{t\leq T}\psi _{j}(z_{i,t};\phi
_{j,i,t}^{\ast })$ and\ $\hat{\Psi}_{j,\gamma ,g,m}\equiv
(n_{g}T_{m})^{-1}\sum_{i\in I_{g}}\sum_{t\in I_{m}}\psi _{j,\gamma
}(z_{i,t};\phi _{j,g,m}^{\ast })$. To study the properties of the QLR
statistic, we suppose that Assumptions \ref{A1}, \ref{A2}, \ref{A3}, \ref{A4}
and \ref{A5} hold for model 1 and model 2, which enables us to apply the
expansion of the likelihood function listed in (\ref{R_L-1})-(\ref{R_L-4})
to both models to obtain%
\begin{equation}
QLR_{n,T}=n^{-1/2}\sum_{i\leq n}(\tilde{\Psi}_{1,i}-\tilde{\Psi}%
_{2,i})+S_{n,T}+o_{p}(n^{-1/2}),  \label{R_QLR}
\end{equation}%
where 
\begin{equation*}
S_{n,T}\equiv \sum_{j=1,2}\sum_{g\in \mathcal{G}_{j}}\sum_{m\in \mathcal{M}%
_{j}}\frac{(-1)^{j}(n_{g}T_{m})\hat{\Psi}_{j,\gamma ,g,m}^{\top }\Psi
_{j,\gamma \gamma ,g,m}^{-1}\hat{\Psi}_{j,\gamma ,g,m}}{2(nT)^{1/2}}.
\end{equation*}%
Since $z_{i,t}$ are independent across $i$ for all $t$, $\hat{\Psi}_{1,i}-%
\hat{\Psi}_{2,i}$ are independent across $i$. Consequently, the asymptotic
distribution of $n^{-1/2}\sum_{i\leq n}(\tilde{\Psi}_{1,i}-\tilde{\Psi}%
_{2,i})$ is straightforward to derive. However, when the two models are
nested or overlapping non-nested, its variance may be zero or arbitrarily
close to zero, invalidating inference based solely on this leading term. To
conduct valid inference for (\ref{R_L-1}) in a more general setting, the
second term $S_{n,T}$ on the right-hand side of (\ref{R_QLR}) must therefore
be taken into account when deriving the asymptotic distribution of the QLR
statistic.

For this purpose, we define%
\begin{equation*}
\tilde{\Psi}_{j,\gamma ,i,m}\equiv T_{m}^{-1/2}\sum_{t\in I_{m}}\tilde{\psi}%
_{j,\gamma }(z_{i,t};\phi _{j,g,m}^{\ast })\text{,}
\end{equation*}%
where $\tilde{\psi}_{j,\gamma }(z_{i,t};\phi _{j,g,m}^{\ast })\equiv (-\Psi
_{j,\gamma \gamma ,g,m})^{-1/2}\psi _{j,\gamma }(z_{i,t};\phi _{j,g,m}^{\ast
})$ for any $i\in I_{g}$ and $t\in I_{m}$, and for any $g\in \mathcal{G}_{j}$
and $m\in \mathcal{M}_{j}$. Then for any $g\in \mathcal{G}_{j}$ and $m\in 
\mathcal{M}_{j}$, we obtain\footnote{%
We adopt the convention that the summaion over an empty set is zero.
Therefore under this trandition, the summation $\sum_{\{i^{\prime }\in I_{g}:%
\text{ }i^{\prime }<i\}}\tilde{\Psi}_{j\gamma ,im}^{\top }\tilde{\Psi}%
_{j\gamma ,i^{\prime }m}=0$ if the set $\{i^{\prime }\in I_{g}:$ $i^{\prime
}<i\}$ is empty.} 
\begin{equation}
\frac{(n_{g}T_{m})\hat{\Psi}_{j,\gamma ,g,m}^{\top }\Psi _{j,\gamma \gamma
,g,m}^{-1}\hat{\Psi}_{j,\gamma ,g,m}}{2T^{1/2}}=\sum_{i\in I_{g}}\frac{%
\tilde{\Psi}_{j,\gamma ,i,m}^{\top }\tilde{\Psi}_{j,\gamma ,i,m}}{%
2n_{g}T^{1/2}}+\sum_{i\in I_{g}}\sum_{\{i^{\prime }\in I_{g}:\text{ }%
i^{\prime }<i\}}\frac{\tilde{\Psi}_{j,\gamma ,i,m}^{\top }\tilde{\Psi}%
_{j,\gamma ,i^{\prime },m}}{n_{g}T^{1/2}}.  \label{Exp_2nd_1}
\end{equation}%
The decomposition in (\ref{Exp_2nd_1}) is useful to study $QLR_{n,T}$ with $%
S_{n,T}$ taken into account, since by the independence of $z_{i,t}$ across $%
i $ for all $t$, $\tilde{\Psi}_{j,\gamma ,i,m}^{\top }\tilde{\Psi}_{j,\gamma
,i,m}$ are independent across $i$ for all $m$, and $\sum_{\{i^{\prime }\in
I_{g}:\text{ }i^{\prime }<i\}}\tilde{\Psi}_{j,\gamma ,i,m}^{\top }\tilde{\Psi%
}_{j,\gamma ,i^{\prime },m}$ are uncorrelated across $i$ for all $m$.

To employ the decomposition in (\ref{Exp_2nd_1}) in derivation of the
asymptotic distribution of $QLR_{n,T}$, we define 
\begin{equation}
\tilde{V}_{j,i}\equiv \frac{\sum_{m\in \mathcal{M}_{j}}\tilde{\Psi}%
_{j,\gamma ,i,m}^{\top }\tilde{\Psi}_{j,\gamma ,i,m}}{2n_{g_{j}(i)}T^{1/2}}%
\text{ \ \ \ and \ \ \ }\tilde{U}_{j,i}\equiv \frac{\sum_{\{i^{\prime }\in
I_{g_{j}(i)}:\text{ }i^{\prime }<i\}}\sum_{m\in \mathcal{M}_{j}}\tilde{\Psi}%
_{j,\gamma ,i,m}^{\top }\tilde{\Psi}_{j,\gamma ,i^{\prime },m}}{%
n_{g_{j}(i)}T^{1/2}}  \label{Exp_2nd_2}
\end{equation}%
for $j=1,2$ and for any $i=1,\ldots ,n$. Then 
\begin{equation*}
\sum_{g\in \mathcal{G}_{j}}\sum_{m\in \mathcal{M}_{j}}\frac{(n_{g}T_{m})\hat{%
\Psi}_{j,\gamma ,g,m}^{\top }\Psi _{j,\gamma \gamma ,g,m}^{-1}\hat{\Psi}%
_{j,\gamma ,g,m}}{2T^{1/2}}=\sum_{g\in \mathcal{G}_{j}}\sum_{i\in I_{g}}(%
\tilde{V}_{j,i}+\tilde{U}_{j,i})=\sum_{i\leq n}(\tilde{V}_{j,i}+\tilde{U}%
_{j,i}),
\end{equation*}%
and the expansion of the QLR statistic in (\ref{R_QLR}) can be written as%
\begin{equation}
QLR_{n,T}=n^{-1/2}\sum_{i\leq n}(\tilde{\Psi}_{i}+\tilde{V}_{i}+\tilde{U}%
_{i})+o_{p}(n^{-1/2})  \label{R_QLR_2}
\end{equation}%
where 
\begin{equation}
\tilde{\Psi}_{i}\equiv \tilde{\Psi}_{1,i}-\tilde{\Psi}_{2,i},\quad \tilde{V}%
_{i}\equiv \tilde{V}_{1,i}-\tilde{V}_{2,i},\quad \tilde{U}_{i}\equiv \tilde{U%
}_{1,i}-\tilde{U}_{2,i}.  \label{R_QLR_3}
\end{equation}

Let $\mathcal{F}_{nT,i}$ denote the sigma field generated by\ $\left\{
z_{i^{\prime }t}\right\} _{t\leq T,i^{\prime }\leq i}$.\ Then by the
independence of $\left\{ z_{i,t}\right\} _{t\leq T}$ across $i$, we have%
\begin{equation}
\mathbb{E}[\tilde{\Psi}_{i}^{\ast }+\tilde{V}_{i}^{\ast }+\tilde{U}_{i}|%
\mathcal{F}_{nT,i-1}]=0  \label{MDS_1}
\end{equation}%
for any $i=1,\ldots ,n$, where $\tilde{\Psi}_{i}^{\ast }\equiv \tilde{\Psi}%
_{i}-\mathbb{E}[\tilde{\Psi}_{i}]$ and\ $\tilde{V}_{i}^{\ast }\equiv \tilde{V%
}_{i}-\mathbb{E}[\tilde{V}_{i}]$. This shows that $\{\tilde{\Psi}_{i}^{\ast
}+\tilde{V}_{i}^{\ast }+\tilde{U}_{i}\}_{i\leq n}$ forms a martingale
difference array. In view of the expansion of the QLR statistic in (\ref%
{R_QLR_2}) and the martingale property in (\ref{MDS_1}), an inference
procedure which is valid uniformly over the model relationship can be
established by the martingale CLT.

Let $\omega _{n,T}\equiv (n^{-1}\sum_{i\leq n}\mathrm{Var}(\tilde{\Psi}_{i}+%
\tilde{V}_{i}+\tilde{U}_{i}))^{1/2}$. The following assumption is imposed to
derive the asymptotic distribution of $QLR_{n,T}$.

\begin{assumption}
\textit{\label{A6}\ (i) }$n\omega_{n,T}^{2}\geq K^{-1}$; (ii) $(n\omega
_{n,T}^{2})^{-1/2}\sum_{i=1}^{n}(\tilde{\Psi}_{i}^{\ast}+\tilde{V}_{i}^{\ast
}+\tilde{U}_{i})\rightarrow_{d}N(0,1)$ as $n,T\rightarrow\infty$.
\end{assumption}

Assumption \ref{A6}(i) ensures that the second term in the expansion of the
QLR statistic later in (\ref{R_QLR_2}) is asymptotically negligible, while
Assumption \ref{A6}(ii) requires that the leading term in the expansion
given in (\ref{C_Exp_3}) admits an asymptotic normal distribution.
Assumption \ref{A6}(ii) can be derived by verifying the sufficient
conditions of the martingale CLT\ (e.g., Corollary 3.1 of \cite%
{HallHeyde1980}). Specifically, let%
\begin{equation}
\xi _{nT,i}\equiv (n\omega _{n,T}^{2})^{-1/2}(\tilde{\Psi}_{i}^{\ast }+%
\tilde{V}_{i}^{\ast }+\tilde{U}_{i}).  \label{MDS}
\end{equation}%
Then Assumption \ref{A6}(ii) holds under the following conditions:%
\begin{equation}
\sum_{i\leq n}\mathbb{E}[\xi _{nT,i}^{2}|\mathcal{F}_{nT,i-1}]\rightarrow
_{p}1,  \label{C1_MGCLT}
\end{equation}%
and for any $\varepsilon >0$,%
\begin{equation}
\sum_{i\leq n}\mathbb{E}[\xi _{nT,i}^{2}I\{|\xi _{nT,i}|>\varepsilon \}|%
\mathcal{F}_{nT,i-1}]\rightarrow _{p}0.  \label{C2_MGCLT}
\end{equation}%
Details on verification of Assumption \ref{A6}(ii) under $d_{\gamma }=1$ and
some low level conditions are provided in Theorem \ref{MGCLT_V} below.

\begin{theorem}
\textit{\label{QLR_Stat}\ }Suppose that Assumptions \ref{A1}, \ref{A2}, \ref%
{A3}, \ref{A4} and \ref{A5} hold for model 1 and model 2. Then under
Assumption \ref{A6}, we have%
\begin{equation}
\frac{QLR_{n,T}-\mathbb{E}[S_{n,T}]-\overline{QLR}_{n,T}}{\omega _{n,T}}%
\rightarrow _{d}\text{ }N(0,\text{ }1)  \label{QLR_Stat_1}
\end{equation}%
where $\overline{QLR}_{n,T}\equiv n^{-1/2}\sum_{i\leq n}\mathbb{E}[\tilde{%
\Psi}_{i}]$.
\end{theorem}

\noindent \textsc{Proof of Theorem \ref{QLR_Stat}}. Since Assumptions \ref%
{A1}, \ref{A2}, \ref{A3}, \ref{A4} and \ref{A5} hold for both models, we can
invoke (\ref{R_QLR_2}) and then apply Assumption \ref{A6}(i) to get%
\begin{equation*}
\frac{QLR_{n,T}-\mathbb{E}[S_{n,T}]-QLR_{n,T}^{\ast }}{\omega _{n,T}}%
=(n\omega _{n,T}^{2})^{-1/2}\sum_{i\leq n}(\tilde{\Psi}_{i}^{\ast }+\tilde{V}%
_{i}^{\ast }+\tilde{U}_{i})+o_{p}\left( 1\right)
\end{equation*}%
which together with Assumption \ref{A6}(ii) and Slutsky's theorem shows (\ref%
{QLR_Stat_1}).\hfill $Q.E.D.$

We conclude this section by presenting low-level sufficient conditions for
verifying Assumption \ref{A6}(ii) in the case $d_{\gamma }=1$. Let $%
A_{j,i}\equiv (\tilde{\Psi}_{j,\gamma ,i,m})_{m\in \mathcal{M}_{j}}$ for $%
i=1,\ldots ,n$. Then%
\begin{equation*}
\tilde{V}_{j,i}=\frac{A_{j,i}^{\top }A_{j,i}}{2n_{g_{j}(i)}T^{1/2}}\text{ \
\ \ \ \ and \ \ \ \ }\tilde{U}_{j,i}=\sum_{\{i^{\prime }\in I_{g_{j}(i)}:%
\text{ }i^{\prime }<i\}}\frac{A_{j,i}^{\top }A_{j,i^{\prime }}}{%
n_{g_{j}(i)}T^{1/2}}.
\end{equation*}

\begin{assumption}
\textit{\label{A7}\ (i) }$\mathbb{E}\left[ (\Delta \psi (z_{i,t}))^{4}\omega
_{n,T}^{-4}\right] <K$; (ii) $\mathbb{E}\left[ |\psi _{j,\gamma
}(z_{i,t};\phi _{j,i,t}^{\ast })|^{4+\delta }\right] <K$; (iii) $%
G_{j}^{-1}\max_{g\in \mathcal{G}_{j}}n_{g}^{-1}=o(1)$.
\end{assumption}

Assumptions \ref{A7}(i, ii) are used to show the Lindeberg condition for the
martingale CLT and to bound the moments of $\tilde{\Psi}_{j,\gamma ,i,m}$.
Assumption \ref{A7}(iii) requires that under both model 1 and model 2, the
number of the groups and/or the size of the smallest group across $i$ should
go to infinity.

\begin{theorem}
\textit{\label{MGCLT_V}\ }Assumption \ref{A6}(ii) holds under Assumptions %
\ref{A1}, \ref{A5}(ii), \ref{A6}(i) and \ref{A7}.
\end{theorem}

The proof of this theorem is provided in the Online Appendix.

\bigskip

{\small 
\bibliographystyle{econometrica}
\bibliography{Panel_Voung}
}

\newpage

\clearpage
\newgeometry{
	left=0.80in,
	right=0.80in,
	top=1.00in,
	bottom=1.00in
}
\fontsize{11pt}{13.6pt}\selectfont

\setcounter{theorem}{0}
\setcounter{lemma}{0}
\setcounter{proposition}{0}
\setcounter{corollary}{0}
\setcounter{assumption}{0}
\setcounter{example}{0}
\setcounter{definition}{0}
\setcounter{remark}{0}
\setcounter{equation}{0}

\renewcommand{\thetheorem}{S\arabic{theorem}}
\renewcommand{\thelemma}{S\arabic{lemma}}
\renewcommand{\theproposition}{S\arabic{proposition}}
\renewcommand{\thecorollary}{S\arabic{corollary}}
\renewcommand{\theassumption}{S\arabic{assumption}}
\renewcommand{\theexample}{S\arabic{example}}
\renewcommand{\thedefinition}{S\arabic{definition}}
\renewcommand{\theremark}{S\arabic{remark}}
\renewcommand{\theequation}{S\arabic{equation}}

\begin{center}
	{\Large Online Appendix to ``Model Selection in}\\[0.3em]
	{\Large Panel Data Models: A Generalization of the Vuong Test''}
\end{center}

\vspace{0.4in}

\begin{abstract}
	\thispagestyle{empty}\noindent This supplement consists of four appendices.
	Appendix \ref{Proof_App_Main} collects the proofs of the auxiliary lemmas
	used to establish the main results in Section \ref{Sec: Simple-Vuong} of the
	paper. Appendix \ref{Proof_App_TWE} contains the proofs of the auxiliary
	lemmas used in deriving the main results in Section \ref{sec:TWE} of the
	paper. Appendix \ref{Proof_App_General_Model} provides the proofs of the
	main results presented in Section \ref{Sec:AP2} of the paper. Finally,
	Appendix \ref{sec:linear-section} presents several illustrative examples
	based on the linear panel model. \newline
	\newline
\end{abstract}

\setstretch{1}

\section{Proofs for Lemmas in Appendix \protect\ref%
	{sec:auxiliary-lemmas-main}\label{Proof_App_Main}}

\noindent \textsc{Proof of Lemma \ref{C_T2}}.\ (a) First, by Assumptions \ref%
{A1} and \ref{A3}(iii), we can write%
\begin{eqnarray}
	\mathbb{E}\left[ \left( T^{-1/2}\sum_{t\leq T}\psi _{j}(z_{i,t};\phi
	_{j,i}^{\ast })\right) ^{2}\right] &\leq &KT^{-1}\sum_{0\leq h\leq
		T}\sum_{t\leq T-h}\left\vert \mathrm{Cov}(\psi _{j}(z_{i,t};\phi
	_{j,i}^{\ast }),\psi _{j}(z_{i,t+h};\phi _{j,i}^{\ast }))\right\vert  \notag
	\\
	&\leq &KT^{-1}\sum_{0\leq h\leq T}\sum_{t\leq T-h}\max_{t\leq T}||\psi
	_{j}(z_{i,t};\phi _{j,i}^{\ast })||_{2+\delta }^{2}\alpha _{i}^{\delta
		/(2+\delta )}(h)  \notag \\
	&\leq &\max_{t\leq T}||\psi _{j}(z_{i,t};\phi _{j,i}^{\ast })||_{2+\delta
	}^{2}\sum_{0\leq h\leq T}(1-h/T)a^{\delta h/(2+\delta )}\leq K,
	\label{P_C_T2_0}
\end{eqnarray}%
where the second inequality follows from the covariance inequality for
strong mixing processes (see, e.g., Corollary 14.3 in \cite{davidsonbook}),
and the last inequality uses Assumption \ref{A1}(iv). Therefore, 
\begin{equation*}
	T^{1/2}\sum_{i\leq n}\mathbb{E}[\tilde{V}_{j,i}]=2\sum_{g\in \mathcal{G}%
		_{j}}n_{j,g}^{-1}\sum_{i\in I_{j,g}}\mathbb{E}\left[ \left(
	T^{-1/2}\sum_{t\leq T}\psi _{j}(z_{i,t};\phi _{j,i}^{\ast })\right) ^{2}%
	\right] \leq KG_{j}.
\end{equation*}%
It follows that 
\begin{equation}
	\left\vert T^{1/2}\sum_{i\leq n}\mathbb{E}[\tilde{V}_{i}]\right\vert
	=\left\vert T^{1/2}\sum_{i\leq n}(\mathbb{E}[\tilde{V}_{1,i}]-\mathbb{E}[%
	\tilde{V}_{2,i}])\right\vert \leq K(G_{1}+G_{2}).  \label{P_C_T2_1}
\end{equation}%
Moreover, by Assumptions \ref{A3}(iii) and \ref{A4}, and (\ref{P_C_L_Bias_2}%
) (which follows from Lemma\ \ref{AU_L1} under Assumptions \ref{A1} and \ref%
{C_A1} without imposing Assumption \ref{C_A2}), we obtain%
\begin{equation}
	R_{j,n,T}(\hat{\phi}_{j})=\sum_{g\in \mathcal{G}_{j}}n_{j,g}^{-1}\sum_{i\in
		I_{j,g}}\sigma _{j,\gamma ,i}^{2}+O_{p}(1)=O_{p}(G_{j}).  \label{P_C_T2_2}
\end{equation}%
Combining (\ref{P_C_T2_1}) and (\ref{P_C_T2_2}) with Theorem \ref{C_L0}, we
obtain 
\begin{eqnarray}
	MQLR_{n,T} &=&\overline{QLR}_{n,T}+(MQLR_{n,T}^{\ast }-\overline{QLR}_{n,T})
	\notag \\
	&&-(R_{1,n,T}(\hat{\phi}_{j})-R_{2,n,T}(\hat{\phi}_{j}))+T^{1/2}\sum_{i\leq
		n}\mathbb{E}[\tilde{V}_{i}]  \notag \\
	&=&\overline{QLR}_{n,T}+O_{p}(\omega _{n,T}+G_{1}+G_{2})=\overline{QLR}%
	_{n,T}+O_{p}(G_{1}+G_{2}),  \label{P_C_T2_3}
\end{eqnarray}%
where the last equality follows from Lemma \ref{AU_L4}(ii). From the
definition of $\hat{\omega}_{n,T}^{2}$ and Lemma \ref{C_L_Auxillary13}, 
\begin{equation}
	\hat{\omega}_{n,T}^{2}\leq \max \left\{ \hat{\sigma}_{n,T}^{2}+\hat{\sigma}%
	_{U,n,T}^{2},\hat{\sigma}_{U,n,T}^{2}\right\} \leq K,  \label{P_C_T2_4}
\end{equation}%
with probability approaching 1. It follows from (\ref{P_C_T2_3}) and (\ref%
{P_C_T2_4}) that 
\begin{equation*}
	\mathbb{E}[\varphi _{n,T}^{2\text{-side}}(p)]=\mathbb{P}\left(
	|MQLR_{n,T}|\geq \hat{\omega}_{n,T}z_{1-p/2}\right) \geq \mathbb{P}\left( |%
	\overline{QLR}_{n,T}|\geq Kz_{1-p/2}+O_{p}(G_{1}+G_{2})\right) -o(1).
\end{equation*}%
Since $|\overline{QLR}_{n,T}|\succ G_{1}+G_{2}$, the right-hand side
converges to 1. This establishes $\mathbb{E}[\varphi _{n,T}^{2\text{-side}%
}(p)]\rightarrow 1$, as $n,T\rightarrow \infty $.

(b) The proof for the one-sided test follows by analogous arguments and is
therefore omitted.\hfill $Q.E.D.$

\bigskip

\noindent \textsc{Proof of lemma \ref{AU_L1}.} The result in (\ref{AU_L1_1})
follows from Assumption \ref{A1}(i) together with Lemma \ref{C_L_Auxillary2}%
, Lemma \ref{C_L_Auxillary3} and Lemma \ref{C_L_Auxillary3c}. The claim in (%
\ref{AU_L1_2}) is obtained by applying Assumption \ref{A1}(i), Lemma \ref%
{C_L_Auxillary2b}, Lemma \ref{C_L_Auxillary3b} and Lemma \ref%
{C_L_Auxillary3c}. The assertion in (\ref{AU_L1_3}) is established in Lemma %
\ref{C_L_Auxillary4}. Finally, the last claim of the lemma follows from
Lemma \ref{C_L_Auxillary7c}.\hfill $Q.E.D.$

\bigskip

\noindent \textsc{Proof of lemma \ref{AU_L2}.} To prove the claim in (\ref%
{AU_L2_1}), we first note that, by Lemma \ref{C_L_Auxillary7a}, 
\begin{eqnarray}
	&&n^{-1}\sum_{i\leq n}(\widehat{\mathbb{E}}_{T}[\Delta \psi (z_{i,t},\hat{%
		\phi}_{i})^{2}]-\mathbb{E}_{T}[\Delta \psi (z_{i,t})^{2}])  \notag \\
	&=&2(nT)^{-1}\sum_{i\leq n}\sum_{t\leq T}\Delta \psi (z_{i,t})\Delta \psi
	_{\phi }(z_{i,t};\phi _{i}^{\ast })(\hat{\phi}_{i}-\phi _{i}^{\ast })  \notag
	\\
	&&+(nT)^{-1}\sum_{i\leq n}\sum_{t\leq T}(\hat{\phi}_{i}-\phi _{i}^{\ast
	})^{\top }\Delta \psi _{\phi }(z_{i,t};\phi _{i}^{\ast })^{\top }\Delta \psi
	_{\phi }(z_{i,t};\phi _{i}^{\ast })(\hat{\phi}_{i}-\phi _{i}^{\ast })  \notag
	\\
	&&+O_{p}(T^{-3/2}+\omega _{n,T}T^{-1}+\omega _{n,T}^{2}(nT)^{-1/2}).
	\label{P_AU_L2_1}
\end{eqnarray}%
Together with Lemma \ref{C_L_Auxillary5}, Lemma \ref{C_L_Auxillary10} and
Assumption \ref{A1}(i), this implies%
\begin{eqnarray}
	&&n^{-1}\sum_{i\leq n}(\widehat{\mathbb{E}}_{T}[\Delta \psi (z_{i,t},\hat{%
		\phi}_{i})^{2}]-\mathbb{E}_{T}[\Delta \psi (z_{i,t})^{2}])  \notag \\
	&=&(nT)^{-1}\sum_{i\leq n}\sum_{t\leq T}(\hat{\phi}_{i}-\phi _{i}^{\ast
	})^{\top }\Delta \psi _{\phi }(z_{i,t};\phi _{i}^{\ast })^{\top }\Delta \psi
	_{\phi }(z_{i,t};\phi _{i}^{\ast })(\hat{\phi}_{i}-\phi _{i}^{\ast })  \notag
	\\
	&&+O_{p}(T^{-3/2}+\omega _{n,T}(nT)^{-1/2}).  \label{P_AU_L2_2}
\end{eqnarray}%
Under Assumption \ref{C_A2}, we have 
\begin{equation*}
	\mathbb{E}\left[ \left( T^{-1/2}\sum_{t\leq T}\tilde{\psi}_{j,\gamma }^{\ast
	}\left( z_{i,t}\right) \right) ^{2}\right] =T^{-1}\sum_{t\leq T}\mathbb{E}[%
	\tilde{\psi}_{j,\gamma }^{\ast }\left( z_{i,t}\right) ^{2}]=\sigma
	_{j,\gamma ,i}^{2}
\end{equation*}%
and 
\begin{equation*}
	T^{-1}\sum_{t,t^{\prime }\leq T}\mathbb{E}[\tilde{\psi}_{1,\gamma }^{\ast
	}(z_{i,t})\tilde{\psi}_{2,\gamma }^{\ast }(z_{i,t^{\prime
	}})]=T^{-1}\sum_{t\leq T}\mathbb{E}[\tilde{\psi}_{1,\gamma }^{\ast }\left(
	z_{i,t}\right) \tilde{\psi}_{1,\gamma }^{\ast }\left( z_{i,t}\right)
	]=\sigma _{12,\gamma ,i},
\end{equation*}%
where $\tilde{\psi}_{j,\gamma }^{\ast }(z_{i,t})\equiv \psi _{j,\gamma
}^{\ast }(z_{i,t})/(-\Psi _{j,\gamma \gamma ,g})^{1/2}$ ($j=1,2$), for any $%
i\in I_{2,g}$ and any $g\in \mathcal{G}_{2}$. Combining these results with
Assumption \ref{A1}(i), (\ref{P_AU_L2_2}) and Lemma \ref{C_L_Auxillary6}
yields%
\begin{eqnarray}
	&&n^{-1}\sum_{i\leq n}(\widehat{\mathbb{E}}_{T}[\Delta \psi (z_{i,t},\hat{%
		\phi}_{i})^{2}]-\mathbb{E}_{T}[\Delta \psi (z_{i,t})^{2}])  \notag \\
	&=&(nT)^{-1}\sum_{g\in \mathcal{G}_{2}}\left( \sum_{i\in I_{2,g}}\sigma
	_{1,\gamma ,i}^{4}+n_{2,g}^{-2}\sum_{i\in I_{j,g}}s_{j,\gamma
		,i}^{2}\sum_{i\in I_{j,g}}\sigma _{j,\gamma ,i}^{2}-2n_{2,g}^{-1}\sum_{i\in
		I_{2,g}}\sigma _{12,\gamma ,i}^{2}\right)   \notag \\
	&&+O_{p}(T^{-3/2}+\omega _{n,T}(nT)^{-1/2}).  \label{P_AU_L2_3}
\end{eqnarray}%
Combining (\ref{P_AU_L2_3}) with the definition of $\sigma _{S,n,T}^{2}$
establishes the claim in (\ref{AU_L2_1}). The result in (\ref{AU_L2_2}) is
proved in Lemma \ref{C_L_Auxillary9}.\hfill $Q.E.D.$

\bigskip

\noindent \textsc{Proof of lemma \ref{AU_L3}.} The claims in (\ref{AU_L3_1}%
), (\ref{AU_L3_2}), (\ref{AU_L3_3}) and (\ref{AU_L3_4}) follow from Lemmas %
\ref{C_L_Auxillary7b}, \ref{C_L_Auxillary7d}, \ref{C_L_Auxillary11} and \ref%
{C_L_Auxillary12}, respectively.\hfill $Q.E.D.$

\bigskip

\noindent \textsc{Proof of Lemma \ref{AU_L4}.} The claims in part (i) and
part (ii) follow directly from Lemma \ref{C_L_Auxillary1}(ii) and Lemma \ref%
{C_L_Auxillary10}, respectively.\hfill $Q.E.D.$

\subsection{Auxiliary Lemmas}

This section provides the lemmas used in the proof of Lemmas \ref{AU_L1}-\ref%
{AU_L3}. We begin by recalling notation introduced in the main text and then
introduce additional notation. For $j=1,2$, and any $g\in \mathcal{G}_{j}$,%
\begin{eqnarray*}
	\hat{\Psi}_{j,\gamma \gamma }(\hat{\phi}_{j,g}) &\equiv
	&n_{j,g}^{-1}\sum_{i\in I_{j,g}}\widehat{\mathbb{E}}_{T}\left[ \hat{\psi}%
	_{j,\gamma \gamma }\left( z_{i,t}\right) \right] \text{,} \\
	\Psi _{j,\gamma \gamma ,g} &\equiv &n_{j,g}^{-1}\sum_{i\in I_{j,g}}\mathbb{E}%
	_{T}\left[ \psi _{j,\gamma \gamma }\left( z_{i,t}\right) \right] \text{, } \\
	\hat{\Psi}_{j,\gamma ,g} &\equiv &n_{j,g}^{-1}\sum_{i\in I_{j,g}}\widehat{%
		\mathbb{E}}_{T}[\psi _{j,\gamma }^{\ast }(z_{i,t})],
\end{eqnarray*}%
where\ $\hat{\psi}_{j,\gamma \gamma }\left( z_{i,t}\right) \equiv \psi
_{j,\gamma \gamma }(z_{i,t};\hat{\phi}_{j,g})$, $\psi _{j,\gamma }^{\ast
}(z_{i,t})\equiv \psi _{j,\gamma }(z_{i,t})-\mathbb{E}\left[ \psi _{j,\gamma
}(z_{i,t})\right] $,\ $\psi _{j,\gamma \gamma }\left( z_{i,t}\right) \equiv
\psi _{j,\gamma \gamma }(z_{i,t};\phi _{j,g}^{\ast })$ and $\psi _{j,\gamma
}(z_{i,t})\equiv \psi _{j,\gamma }(z_{i,t};\phi _{j,g}^{\ast })$.

\begin{lemma}
	\textit{\label{C_L_Auxillary1}\ Suppose }Assumptions \ref{A1} and \ref{C_A1}
	hold. For model\ $j$ ($j=1,2$), we have:
	
	(i) $\max_{g\in \mathcal{G}_{j}}|\hat{\gamma}_{j,g}-\gamma _{j,g}^{\ast
	}|=o_{p}(1)$;
	
	(ii) $\min_{g\in\mathcal{G}_{j}}(-\hat{\Psi}_{j,\gamma\gamma}(\hat{\phi}%
	_{j,g}))\geq K^{-1}$ wpa1;
	
	(iii)\ $\hat{\theta}_{j}=\theta_{j}^{\ast}+O_{p}((nT)^{-1/2})$;
	
	(iv)\ $\sum_{g\in\mathcal{G}_{j}}|(n_{j,g}T)^{1/2}(\hat{\gamma}%
	_{j,g}-\gamma_{j,g}^{\ast})|^{s}=O_{p}\left( G_{j}\right) $ for $s\in
	\{1,\ldots,4\}$;
	
	(v) $\sum_{g\in\mathcal{G}_{j}}n_{j,g}(\hat{\Psi}_{j,\gamma\gamma}(\hat{\phi 
	}_{j,g})-\Psi_{j,\gamma\gamma,g})^{2}=O_{p}(G_{j}T^{-1});$
	
	(vi) $\sum_{g\in\mathcal{G}_{j}}n_{j,g}|\hat{\gamma}_{j,g}-\gamma_{j,g}^{%
		\ast }+\Psi_{j,\gamma\gamma,g}^{-1}\hat{\Psi}_{j,%
		\gamma,g}|=O_{p}((nT^{-1})^{1/2})$;
	
	(vii) $\sum_{g\in\mathcal{G}_{j}}n_{j,g}(\hat{\gamma}_{j,g}-\gamma_{j,g}^{%
		\ast}+\Psi_{j,\gamma\gamma,g}^{-1}\hat{\Psi}_{j,\gamma,g})^{2}=O_{p}(T^{-1})$%
	;
	
	(viii) $\max_{g\in\mathcal{G}_{j}}n_{j,g}^{2}\mathbb{E}[\hat{\Psi}%
	_{j,\gamma,g}^{4}]=O_{p}(T^{-2})$.
\end{lemma}

\noindent \textsc{Proof of Lemma \ref{C_L_Auxillary1}}. We begin the proof
by noting that Assumptions \ref{A2}, \ref{A3}, \ref{A4} hold for both models
1 and 2 under Assumption \ref{C_A1}(i). Additionally, Assumption \ref{A5}
has been verified for both models under Assumption \ref{A1}(i) in the proof
of Theorem \ref{C_L0}. Under these assumptions, the claims in parts (i),
(ii), (iii), and (iv) of the lemma are established using Lemma \ref%
{Consistency}, Lemma \ref{L3AB}(ii), Lemma \ref{Rate_theta}, and Lemma \ref%
{Rep_L_A} from Section \ref{Sec:AP2}, respectively.

To establish the claim in part (v) of the lemma, we begin by applying the
Cauchy-Schwarz inequality to get%
\begin{equation}
	\sum_{g\in \mathcal{G}_{j}}n_{j,g}(\hat{\Psi}_{j,\gamma \gamma }(\hat{\phi}%
	_{j,g})-\Psi _{j,\gamma \gamma ,g})^{2}\leq 2\sum_{g\in \mathcal{G}%
		_{j}}n_{j,g}(\hat{\Psi}_{j,\gamma \gamma }(\hat{\phi}_{j,g})-\hat{\Psi}%
	_{j,\gamma \gamma ,g})^{2}+2\sum_{g\in \mathcal{G}_{j}}n_{j,g}(\hat{\Psi}%
	_{j,\gamma \gamma ,g}-\Psi _{j,\gamma \gamma ,g})^{2},
	\label{P_C_L_Auxillary1_3}
\end{equation}%
where $\hat{\Psi}_{j,\gamma \gamma ,g}\equiv \hat{\Psi}_{j,\gamma \gamma
}(\phi _{j,g}^{\ast })$ and $\hat{\Psi}_{j,\gamma \gamma }(\phi _{j,g}^{\ast
})$ is defined analogously to $\hat{\Psi}_{j,\gamma \gamma }(\hat{\phi}%
_{j,g})$ with $\hat{\phi}_{j,g}$ replaced by $\phi _{j,g}^{\ast }$. Under
Assumptions \ref{A1} and \ref{A3}(iii), we obtain:%
\begin{equation*}
	\mathbb{E}\left[ \sum_{g\in \mathcal{G}_{j}}n_{j,g}(\hat{\Psi}_{j,\gamma
		\gamma ,g}-\Psi _{j,\gamma \gamma ,g})^{2}\right] =\sum_{g\in \mathcal{G}%
		_{j}}n_{j,g}^{-1}\sum_{i\in I_{j,g}}\mathbb{E}\left[ \left\vert
	T^{-1}\sum_{t\leq T}(\psi _{j,\gamma \gamma }(z_{i,t})-\mathbb{E}[\psi
	_{j,\gamma \gamma }(z_{i,t})]\right\vert ^{2}\right] \leq KG_{j}T^{-1}.
\end{equation*}%
By Markov's inequality, it follows that:%
\begin{equation}
	\sum_{g\in \mathcal{G}_{j}}n_{j,g}(\hat{\Psi}_{j,\gamma \gamma ,g}-\Psi
	_{j,\gamma \gamma ,g})^{2}=O_{p}\left( G_{j}T^{-1}\right) .
	\label{P_C_L_Auxillary1_4}
\end{equation}%
Next, using Assumption \ref{A3}(i) and the Cauchy-Schwarz inequality, we
have:%
\begin{equation}
	\sum_{g\in \mathcal{G}_{j}}n_{j,g}(\hat{\Psi}_{j,\gamma \gamma }(\hat{\phi}%
	_{j,g})-\hat{\Psi}_{j,\gamma \gamma }(\phi _{j,g}^{\ast }))^{2}\leq \left(
	\sum_{g\in \mathcal{G}_{j}}n_{j,g}^{2}||\hat{\phi}_{j,g}-\phi _{j,g}^{\ast
	}||^{4}\right) ^{1/2}\times \left( \sum_{g\in \mathcal{G}%
		_{j}}(n_{j,g}T)^{-1}\sum_{i\in I_{j,g}}\sum_{t\leq T}M(z_{i,t})^{4}\right)
	^{1/2}.  \label{P_C_L_Auxillary1_5}
\end{equation}%
The second term in the product satisfies:%
\begin{equation}
	\sum_{g\in \mathcal{G}_{j}}(n_{j,g}T)^{-1}\sum_{i\in I_{j,g}}\sum_{t\leq
		T}M(z_{i,t})^{4}=O_{p}(G_{j})  \label{P_C_L_Auxillary1_6}
\end{equation}%
by Assumption \ref{A3}(ii) and Markov's inequality. From parts (i) and (iii)
of the lemma, we know: 
\begin{equation}
	\sum_{g\in \mathcal{G}_{j}}n_{j,g}^{2}||\hat{\phi}_{j,g}-\phi _{j,g}^{\ast
	}||^{4}=K\left( \sum_{g\in \mathcal{G}_{j}}n_{j,g}^{2}|\hat{\gamma}%
	_{g,m}-\gamma _{g,m}^{\ast }|^{4}+n^{2}||\hat{\theta}_{j}-\theta _{j}^{\ast
	}||^{4}\right) =O_{p}(G_{j}T^{-2}),  \label{P_C_L_Auxillary1_7}
\end{equation}%
which, together with (\ref{P_C_L_Auxillary1_5})-(\ref{P_C_L_Auxillary1_7}),
implies that 
\begin{equation}
	\sum_{g\in \mathcal{G}_{j}}n_{j,g}(\hat{\Psi}_{j,\gamma \gamma ,g}(\hat{\phi}%
	_{j,g})-\hat{\Psi}_{j,\gamma \gamma ,g})^{2}=O_{p}\left( G_{j}T^{-1}\right) .
	\label{P_C_L_Auxillary1_8}
\end{equation}%
The claim in part (v) of the lemma follows from (\ref{P_C_L_Auxillary1_3}), (%
\ref{P_C_L_Auxillary1_4}) and (\ref{P_C_L_Auxillary1_8}).

To establish the claim in part (vi) of the lemma, observe that from (\ref%
{P_Rate_theta_1}) of Lemma \ref{FOCs}: 
\begin{equation}
	(\hat{\gamma}_{j,g}-\gamma _{j,g}^{\ast })+\frac{\hat{\Psi}_{j,\gamma ,g}}{%
		\hat{\Psi}_{j,\gamma \gamma ,g}}=-\frac{\hat{\Psi}_{j,\gamma \theta ,g}}{%
		\hat{\Psi}_{j,\gamma \gamma ,g}}(\hat{\theta}_{j}-\theta _{j}^{\ast })-\frac{%
		(\hat{\phi}_{j,g}-\phi _{j,g}^{\ast })^{\top }\tilde{\Psi}_{j,\gamma \phi
			\phi ,g}(\hat{\phi}_{j,g}-\phi _{j,g}^{\ast })}{2\hat{\Psi}_{j,\gamma \gamma
			,g}},  \label{P_C_L_Auxillary1_9}
\end{equation}%
where%
\begin{equation*}
	\hat{\Psi}_{j,\gamma \theta ,g}\equiv (n_{j,g}T)^{-1}\sum_{i\in
		I_{j,g}}\sum_{t\leq T}\frac{\partial ^{2}\psi _{j}(z_{i,t};\phi _{j,g}^{\ast
		})}{\partial \gamma _{j,g}\partial \theta _{j}^{\top }}\text{ and }\tilde{%
		\Psi}_{j,\gamma \phi \phi ,g}\equiv (n_{j,g}T)^{-1}\sum_{i\in
		I_{j,g}}\sum_{t\leq T}\frac{\partial ^{2}\psi _{j,\gamma }(z_{i,t};\tilde{%
			\phi}_{j,g})}{\partial \phi _{j,g}\partial \phi _{j,g}^{\top }}.
\end{equation*}%
Here, $\tilde{\phi}_{j,g}$ lies between $\hat{\phi}_{j,g}$ and $\phi
_{j,g}^{\ast }$. By Assumption \ref{A3}(ii), the triangle inequality, and
Markov's inequality, we have 
\begin{equation}
	n^{-1}\sum_{g\in \mathcal{G}_{j}}n_{j,g}||\hat{\Psi}_{j,\gamma \theta
		,g}||\leq (nT)^{-1}\sum_{g\in \mathcal{G}_{j}}\sum_{i\in I_{j,g}}\sum_{t\leq
		T}\left\Vert \frac{\partial ^{2}\psi _{j}(z_{i,t};\phi _{j,g}^{\ast })}{%
		\partial \gamma _{j,g}\partial \theta _{j}^{\top }}\right\Vert =O_{p}(1).
	\label{P_C_L_Auxillary1_10a}
\end{equation}%
This, combined with parts (ii) and (iii) of the lemma, shows:%
\begin{equation}
	\left\vert \sum_{g\in \mathcal{G}_{j}}n_{j,g}\frac{\hat{\Psi}_{j,\gamma
			\theta ,g}}{\hat{\Psi}_{j,\gamma \gamma ,g}}(\hat{\theta}_{j}-\theta
	_{j}^{\ast })\right\vert \leq \frac{||\hat{\theta}_{j}-\theta _{j}^{\ast
		}||\sum_{g\in \mathcal{G}_{j}}n_{j,g}||\hat{\Psi}_{j,\gamma \theta ,g}||}{%
		\min_{g\in \mathcal{G}_{j}}|\hat{\Psi}_{j,\gamma \gamma ,g}|}%
	=O_{P}((nT^{-1})^{1/2}).  \label{P_C_L_Auxillary1_10}
\end{equation}%
By the triangle inequality and the Cauchy-Schwarz inequality:%
\begin{equation}
	\left\vert \sum_{g\in \mathcal{G}_{j}}n_{j,g}\frac{(\hat{\phi}_{j,g}-\phi
		_{j,g}^{\ast })^{\top }\tilde{\Psi}_{j,\gamma \phi \phi ,g}(\hat{\phi}%
		_{j,g}-\phi _{j,g}^{\ast })}{\hat{\Psi}_{j,\gamma \gamma ,g}}\right\vert
	\leq \frac{\max_{g\in \mathcal{G}_{j}}||\tilde{\Psi}_{j,\gamma \phi \phi
			,g}||}{\min_{g\in \mathcal{G}_{j}}|\hat{\Psi}_{j,\gamma \gamma ,g}|}%
	\sum_{g\in \mathcal{G}_{j}}n_{j,g}||\hat{\phi}_{j,g}-\phi _{j,g}^{\ast
	}||^{2}.  \label{P_C_L_Auxillary1_11}
\end{equation}%
In (\ref{P_Rate_theta_2}) of the Supplementary Appendix, we have shown that
under Assumptions \ref{A1} and \ref{A2}-\ref{A5}: 
\begin{equation}
	\max_{g\in \mathcal{G}_{j}}||\tilde{\Psi}_{j,\gamma \phi \phi ,g}||=O_{p}(1).
	\label{P_C_L_Auxillary1_12}
\end{equation}%
Using parts (iii, iv) of the lemma, we have%
\begin{equation}
	\sum_{g\in \mathcal{G}_{j}}n_{j,g}||\hat{\phi}_{j,g}-\phi _{j,g}^{\ast
	}||^{2}\leq K\left( \sum_{g\in \mathcal{G}_{j}}n_{j,g}|\hat{\gamma}%
	_{j,g}-\gamma _{j,g}^{\ast }|^{2}+n||\hat{\theta}_{j}-\theta _{j}^{\ast
	}||^{2}\right) =O_{p}(T^{-1}).  \label{P_C_L_Auxillary1_13}
\end{equation}%
From part (ii) of the lemma and (\ref{P_C_L_Auxillary1_11})-(\ref%
{P_C_L_Auxillary1_13}), it follows that:%
\begin{equation}
	\sum_{g\in \mathcal{G}_{j}}n_{j,g}\frac{(\hat{\phi}_{j,g}-\phi _{j,g}^{\ast
		})^{\top }\tilde{\Psi}_{j,\gamma \phi \phi ,g}(\hat{\phi}_{j,g}-\phi
		_{j,g}^{\ast })}{\hat{\Psi}_{j,\gamma \gamma ,g}}=O_{p}(T^{-1}).
	\label{P_C_L_Auxillary1_14}
\end{equation}%
Combining this result with (\ref{P_C_L_Auxillary1_9}) and (\ref%
{P_C_L_Auxillary1_10}) establishes the claim in this part.

To establish the claim in part (vii), similar arguments as in (\ref%
{P_C_L_Auxillary1_10a}) yield:%
\begin{equation*}
	n^{-1}\sum_{g\in\mathcal{G}_{j}}n_{j,g}||\hat{\Psi}_{j,\gamma\theta,g}||^{2}%
	\leq(nT)^{-1}\sum_{g\in\mathcal{G}_{j}}\sum_{i\in I_{j,g}}\sum_{t\leq
		T}\left\Vert \frac{\partial^{2}\psi_{j}(z_{i,t};\phi_{j,g}^{\ast})}{%
		\partial\gamma_{j,g}\partial\theta_{j}^{\top}}\right\Vert ^{2}=O_{p}(1).
\end{equation*}
Combining this with parts (ii) and (iii) of the lemma, we obtain:%
\begin{equation}
	\sum_{g\in\mathcal{G}_{j}}n_{j,g}\left\vert \frac{\hat{\Psi}_{j,\gamma
			\theta,g}}{\hat{\Psi}_{j,\gamma\gamma,g}}(\hat{\theta}_{j}-\theta_{j}^{\ast
	})\right\vert ^{2}\leq\frac{||\hat{\theta}_{j}-\theta_{j}^{\ast}||^{2}%
		\sum_{g\in\mathcal{G}_{j}}n_{j,g}||\hat{\Psi}_{j,\gamma\theta,g}||^{2}}{%
		\min_{g\in\mathcal{G}_{j}}\hat{\Psi}_{j,\gamma\gamma,g}^{2}}=O_{P}(T^{-1}).
	\label{P_C_L_Auxillary1_15}
\end{equation}
The Cauchy-Schwarz inequality, part (ii) of the lemma, and earlier bounds in
(\ref{P_C_L_Auxillary1_7}) and (\ref{P_C_L_Auxillary1_12}) further imply:%
\begin{align}
	& \sum_{g\in\mathcal{G}_{j}}n_{j,g}\left\vert \frac{(\hat{\phi}%
		_{j,g}-\phi_{j,g}^{\ast})^{\top}\tilde{\Psi}_{j,\gamma\phi\phi,g}(\hat{\phi}%
		_{j,g}-\phi_{j,g}^{\ast})}{\hat{\Psi}_{j,\gamma\gamma,g}}\right\vert ^{2} 
	\notag \\
	& \leq\frac{\max_{g\in\mathcal{G}_{j}}\left\Vert \tilde{\Psi}_{j,\gamma
			\phi\phi,g}\right\Vert ^{2}\sum_{g\in\mathcal{G}_{j}}n_{j,g}||\hat{\phi}%
		_{j,g}-\phi_{j,g}^{\ast}||^{4}}{\min_{g\in\mathcal{G}_{j}}\hat{\Psi}%
		_{j,\gamma\gamma,g}^{2}}=O_{p}(G_{j}T^{-2}).  \label{P_C_L_Auxillary1_16}
\end{align}
The claim in part (vii) of the lemma now follows from (\ref%
{P_C_L_Auxillary1_9}), (\ref{P_C_L_Auxillary1_15}) and (\ref%
{P_C_L_Auxillary1_16}).

To establish the claim in part (viii), observe that: 
\begin{equation}
	n_{j,g}^{2}\hat{\Psi}_{j,\gamma ,g}^{4}=T^{-2}\left(
	(n_{j,g}T)^{-1/2}\sum_{i\in I_{j,g}}\sum_{t\leq T}\psi _{j,\gamma }^{\ast
	}(z_{i,t})\right) ^{4}.  \label{P_C_L_Auxillary1_17}
\end{equation}%
Using Rosenthal's inequality (see, e.g., (\ref{L0-1}) in Lemma \ref{L0}) and
Assumptions \ref{C_A1}(iii), we have%
\begin{equation}
	\mathbb{E}\left[ \left( (n_{j,g}T)^{-1/2}\sum_{i\in I_{j,g}}\sum_{t\leq
		T}\psi _{j,\gamma }^{\ast }(z_{i,t})\right) ^{4}\right] \leq K.
	\label{P_C_L_Auxillary1_18}
\end{equation}%
The claim follows from (\ref{P_C_L_Auxillary1_17}) and (\ref%
{P_C_L_Auxillary1_18}). \hfill $Q.E.D.$

\bigskip

\begin{lemma}
	\textit{\label{C_L_Auxillary2} }Under Assumptions \ref{A1} and \ref{C_A1},
	we have%
	\begin{equation}
		\sum_{g\in \mathcal{G}_{j}}\sum_{i\in I_{j,g}}\frac{\widehat{\mathbb{E}}_{T}[%
			\hat{\psi}_{j,\gamma }^{2}(z_{i,t})]-\widehat{\mathbb{E}}_{T}[\psi
			_{j,\gamma }^{2}(z_{i,t})]}{n_{j,g}\hat{\Psi}_{j,\gamma \gamma }(\hat{\phi}%
			_{j,g})}=O_{p}\left( (nT^{-1})^{1/2}+G_{j}T^{-1}\right) .
		\label{C_L_Auxillary2_1}
	\end{equation}
\end{lemma}

\noindent {\textsc{Proof of Lemma \ref{C_L_Auxillary2}}}.\ Define:%
\begin{equation}
	\psi _{j,\gamma \phi }(z_{i,t};\phi _{j,g})\equiv \frac{\partial \psi
		_{j,\gamma }(z_{i,t};\phi _{j,g})}{\partial \phi _{j,g}}\text{ \ \ and \ \ }%
	\psi _{j,\gamma \phi \phi }(z_{i,t};\phi _{j,g})\equiv \frac{\partial
		^{2}\psi _{j,\gamma }(z_{i,t};\phi _{j,g})}{\partial \phi _{j,g}\partial
		\phi _{j,g}^{\top }}.  \label{P_C_L_Auxillary2_0}
\end{equation}%
Using the Taylor expansion, we obtain:%
\begin{align}
	& \sum_{i\in I_{j,g}}\sum_{t\leq T}(\hat{\psi}_{j,\gamma }^{2}(z_{i,t})-\psi
	_{j,\gamma }^{2}(z_{i,t}))  \notag \\
	& =2\sum_{i\in I_{j,g}}\sum_{t\leq T}\psi _{j,\gamma }(z_{i,t};\phi
	_{j,g}^{\ast })\psi _{j,\gamma \phi }^{\top }(z_{i,t};\phi _{j,g}^{\ast })(%
	\hat{\phi}_{j,g}-\phi _{j,g}^{\ast })  \notag \\
	& +(\hat{\phi}_{j,g}-\phi _{j,g}^{\ast })^{\top }\sum_{i\in
		I_{j,g}}\sum_{t\leq T}\psi _{j,\gamma \phi }(z_{i,t};\tilde{\phi}_{j,g})\psi
	_{j,\gamma \phi }^{\top }(z_{i,t};\tilde{\phi}_{j,g})(\hat{\phi}_{j,g}-\phi
	_{j,g}^{\ast })  \notag \\
	& +(\hat{\phi}_{j,g}-\phi _{j,g}^{\ast })^{\top }\sum_{i\in
		I_{j,g}}\sum_{t\leq T}\psi _{j,\gamma }(z_{i,t};\tilde{\phi}_{j,g})\psi
	_{j,\gamma \phi \phi }(z_{i,t};\tilde{\phi}_{j,g})(\hat{\phi}_{j,g}-\phi
	_{j,g}^{\ast }),  \label{P_C_L_Auxillary2_1}
\end{align}%
where $\tilde{\phi}_{j,g}$ lies between $\hat{\phi}_{j,g}$ and $\phi
_{j,g}^{\ast }$ for any $g\in \mathcal{G}_{j}$. From this expression, it is
evident that (\ref{C_L_Auxillary2_1}) follows if the following conditions
are satisfied: 
\begin{gather}
	\sum_{g\in \mathcal{G}_{j}}\sum_{i\in I_{j,g}}\sum_{t\leq T}\frac{\psi
		_{j,\gamma }(z_{i,t};\phi _{j,g}^{\ast })\psi _{j,\gamma \phi }^{\top
		}(z_{i,t};\phi _{j,g}^{\ast })(\hat{\phi}_{j,g}-\phi _{j,g}^{\ast })}{%
		(n_{j,g}T)\hat{\Psi}_{j,\gamma \gamma }(\hat{\phi}_{j,g})}%
	=O_{p}((nT^{-1})^{1/2}+G_{j}T^{-1}),  \label{P_C_L_Auxillary2_2} \\
	\sum_{g\in \mathcal{G}_{j}}\frac{(\hat{\phi}_{j,g}-\phi _{j,g}^{\ast
		})^{\top }\sum_{i\in I_{j,g}}\sum_{t\leq T}\psi _{j,\gamma \phi }(z_{i,t};%
		\tilde{\phi}_{j,g})\psi _{j,\gamma \phi }^{\top }(z_{i,t};\tilde{\phi}%
		_{j,g})(\hat{\phi}_{j,g}-\phi _{j,g}^{\ast })}{(n_{j,g}T)\hat{\Psi}%
		_{j,\gamma \gamma }(\hat{\phi}_{j,g})}=O_{p}(G_{j}T^{-1}),
	\label{P_C_L_Auxillary2_3} \\
	\sum_{g\in \mathcal{G}_{j}}\frac{(\hat{\phi}_{j,g}-\phi _{j,g}^{\ast
		})^{\top }\sum_{i\in I_{j,g}}\sum_{t\leq T}\psi _{j,\gamma }(z_{i,t};\tilde{%
			\phi}_{j,g})\psi _{j,\gamma \phi \phi }(z_{i,t};\tilde{\phi}_{j,g})(\hat{\phi%
		}_{j,g}-\phi _{j,g}^{\ast })}{(n_{j,g}T)\hat{\Psi}_{j,\gamma \gamma }(\hat{%
			\phi}_{j,g})}=O_{p}(G_{j}T^{-1}).  \label{P_C_L_Auxillary2_4}
\end{gather}%
We now proceed to verify (\ref{P_C_L_Auxillary2_2}), (\ref%
{P_C_L_Auxillary2_3}) and (\ref{P_C_L_Auxillary2_4}).

Let\ $C_{1,j,g}(z_{i,t})\equiv \psi _{j,\gamma }(z_{i,t})\psi _{j,\gamma
	\phi }^{\top }(z_{i,t},\phi _{j,g}^{\ast })$\ and\ $C_{1,j,g}^{\ast
}(z_{i,t})\equiv C_{1,j,g}(z_{i,t})-\mathbb{E}[C_{1,j,g}(z_{i,t})]$.
Applying Assumptions \ref{A1}, \ref{A3},\ and the covariance inequality for
strong mixing processes, we obtain:%
\begin{equation}
	\mathbb{E}\left[ \left\Vert T^{-1/2}\sum_{t\leq T}C_{1,j,g}^{\ast
	}(z_{i,t})\right\Vert ^{2}\right] \leq K.  \label{P_C_L_Auxillary2_5a}
\end{equation}%
Using the triangle inequality and the Cauchy-Schwarz inequality, it follows
that%
\begin{align}
	& \left\vert \sum_{g\in \mathcal{G}_{j}}\frac{(n_{j,g}T)^{-1}\sum_{i\in
			I_{j,g}}\sum_{t\leq T}C_{1,j,g}^{\ast }(z_{i,t})}{\hat{\Psi}_{j,\gamma
			\gamma }(\hat{\phi}_{j,g})}(\hat{\phi}_{j,g}-\phi _{j,g}^{\ast })\right\vert
	\notag \\
	& \leq \frac{\sum_{g\in \mathcal{G}_{j}}\left\vert (n_{j,g}T)^{-1}\sum_{i\in
			I_{j,g}}\sum_{t\leq T}C_{1,j,g}^{\ast }(z_{i,t})(\hat{\phi}_{j,g}-\phi
		_{j,g}^{\ast })\right\vert }{\min_{g\in \mathcal{G}_{j}}(|\hat{\Psi}%
		_{j,\gamma \gamma }(\hat{\phi}_{j,g})|)}  \notag \\
	& \leq \frac{\left( \sum_{g\in \mathcal{G}_{j}}\left\Vert
		(n_{j,g}T)^{-1}\sum_{i\in I_{j,g}}\sum_{t\leq T}C_{1,j,g}^{\ast
		}(z_{i,t})\right\Vert ^{2}\right) ^{1/2}\left( \sum_{g\in \mathcal{G}_{j}}||%
		\hat{\phi}_{j,g}-\phi _{j,g}^{\ast }||^{2}\right) ^{1/2}}{\min_{g\in 
			\mathcal{G}_{j}}(|\hat{\Psi}_{j,\gamma \gamma }(\hat{\phi}_{j,g})|)}.
	\label{P_C_L_Auxillary2_5}
\end{align}%
Under Assumption \ref{A1} and (\ref{P_C_L_Auxillary2_5a}), we have%
\begin{align*}
	\mathbb{E}\left[ \sum_{g\in \mathcal{G}_{j}}\left\Vert
	(n_{j,g}T)^{-1}\sum_{i\in I_{j,g}}\sum_{t\leq T}C_{1,j,g}^{\ast
	}(z_{i,t})\right\Vert ^{2}\right] & =\sum_{g\in \mathcal{G}%
		_{j}}(n_{j,g}T)^{-2}\sum_{i\in I_{j,g}}\mathbb{E}\left[ \left\Vert
	\sum_{t\leq T}C_{1,j,g}^{\ast }(z_{i,t})\right\Vert ^{2}\right] \\
	& \leq \sum_{g\in \mathcal{G}_{j}}(n_{j,g}T)^{-2}\sum_{i\in I_{j,g}}KT\leq
	KG_{j}T^{-1},
\end{align*}%
which, combined with Markov's inequality, implies that%
\begin{equation}
	\sum_{g\in \mathcal{G}_{j}}\left\Vert (n_{j,g}T)^{-1}\sum_{i\in
		I_{j,g}}\sum_{t\leq T}C_{1,j,g}^{\ast }(z_{i,t})\right\Vert
	^{2}=O_{p}(G_{j}T^{-1}).  \label{P_C_L_Auxillary2_6}
\end{equation}%
Thus, by collecting the results in Lemma \ref{C_L_Auxillary1}(ii), (\ref%
{P_C_L_Auxillary1_13}), (\ref{P_C_L_Auxillary2_5}) and (\ref%
{P_C_L_Auxillary2_6}), we conclude:%
\begin{equation}
	\sum_{g\in \mathcal{G}_{j}}\left( \frac{\sum_{i\in I_{j,g}}\sum_{t\leq
			T}C_{1,j,g}^{\ast }(z_{i,t})}{(n_{j,g}T)\hat{\Psi}_{j,\gamma \gamma }(\hat{%
			\phi}_{j,g})}\right) (\hat{\phi}_{j,g}-\phi _{j,g}^{\ast })=O_{p}\left(
	G_{j}T^{-1}\right) .  \label{P_C_L_Auxillary2_7}
\end{equation}

By the triangle inequality and the Cauchy-Schwarz inequality, we obtain%
\begin{align}
	& \left\Vert \sum_{g\in\mathcal{G}_{j}}\frac{\hat{\Psi}_{j,\gamma\gamma}(%
		\hat{\phi}_{j,g})-\Psi_{j,\gamma\gamma,g}}{(n_{j,g}T)\hat{\Psi}%
		_{j,\gamma\gamma}(\hat{\phi}_{j,g})\Psi_{j,\gamma\gamma,g}}\sum_{i\in
		I_{j,g}}\sum_{t\leq T}\mathbb{E}[C_{1,j,g}(z_{i,t})](\hat{\phi}_{j,g}-\phi
	_{j,g}^{\ast})\right\Vert  \notag \\
	& \leq\frac{\left( \sum_{g\in\mathcal{G}_{j}}(\hat{\Psi}_{j,\gamma\gamma }(%
		\hat{\phi}_{j,g})-\Psi_{j,\gamma\gamma,g})^{2}\right) ^{1/2}}{T\min _{g\in%
			\mathcal{G}_{j}}(|\hat{\Psi}_{j,\gamma\gamma}(\hat{\phi}_{j,g})\Psi_{j,%
			\gamma\gamma,g}|)}  \notag \\
	& \times\left( \sum_{g\in\mathcal{G}_{j}}\left\Vert
	(n_{j,g}T)^{-1}\sum_{i\in I_{j,g}}\sum_{t\leq T}\mathbb{E}%
	[C_{1,j,g}(z_{i,t})]\right\Vert ^{2}||\hat{\phi}_{j,g}-\phi_{j,g}^{%
		\ast}||^{2}\right) ^{1/2}.  \label{P_C_L_Auxillary2_8}
\end{align}
Since\ $\left\Vert \mathbb{E}[C_{1,j,g}(z_{i,t})]\right\Vert \leq K$ under
Assumption \ref{A3}(iii), by (\ref{P_C_L_Auxillary1_13}) and the
Cauchy-Schwarz inequality, we have 
\begin{equation}
	\sum_{g\in\mathcal{G}_{j}}\left\Vert (n_{j,g}T)^{-1}\sum_{i\in
		I_{j,g}}\sum_{t\leq T}\mathbb{E}[C_{1t,j,g}(z_{i,t})]\right\Vert ^{2}||\hat{%
		\phi }_{j,g}-\phi_{j,g}^{\ast}||^{2}\leq K\sum_{g\in\mathcal{G}_{j}}||\hat{%
		\phi }_{j,g}-\phi_{j,g}^{\ast}||^{2}=O_{p}(G_{j}T^{-1}).
	\label{P_C_L_Auxillary2_9}
\end{equation}
Combining this with Assumption \ref{A4}, which is upheld in Assumption \ref%
{C_A1}(i), Lemma \ref{C_L_Auxillary1}(ii, v) and (\ref{P_C_L_Auxillary2_8}),
it follows that%
\begin{equation}
	\sum_{g\in\mathcal{G}_{j}}\frac{\hat{\Psi}_{j,\gamma\gamma}(\hat{\phi}%
		_{j,g})-\Psi_{j,\gamma\gamma,g}}{(n_{j,g}T)\hat{\Psi}_{j,\gamma\gamma}(\hat{%
			\phi}_{j,g})\Psi_{j,\gamma\gamma,g}}\sum_{i\in I_{j,g}}\sum_{t\leq T}\mathbb{%
		E}[C_{1,j,g}(z_{i,t})](\hat{\phi}_{j,g}-\phi_{j,g}^{%
		\ast})=O_{p}(G_{j}T^{-1}).  \label{P_C_L_Auxillary2_10}
\end{equation}

By the triangle inequality, we have 
\begin{align}
	& \left\Vert \sum_{g\in\mathcal{G}_{j}}\frac{\sum_{i\in I_{j,g}}\sum_{t\leq
			T}\mathbb{E}[C_{1,j,g}(z_{i,t})]}{(n_{j,g}T)\Psi_{j,\gamma\gamma,g}}(\hat {%
		\phi}_{j,g}-\phi_{j,g}^{\ast})\right\Vert  \notag \\
	& \leq\left\Vert \sum_{g\in\mathcal{G}_{j}}\frac{\sum_{i\in
			I_{j,g}}\sum_{t\leq T}\mathbb{E}[\psi_{j,\gamma}(z_{i,t};\phi_{j,g}^{\ast})%
		\psi_{j,\gamma\theta}^{\top}(z_{i,t};\phi_{j,g}^{\ast})]}{%
		(n_{j,g}T)\Psi_{j,\gamma\gamma,g}}(\hat{\theta}_{j}-\theta_{j}^{\ast})\right%
	\Vert  \notag \\
	& +\left\Vert \sum_{g\in\mathcal{G}_{j}}\frac{\sum_{i\in I_{j,g}}\sum_{t\leq
			T}\mathbb{E}[\psi_{j,\gamma}(z_{i,t};\phi_{j,g}^{\ast})\psi_{j,\gamma\gamma
		}(z_{i,t};\phi_{j,g}^{\ast})]}{(n_{j,g}T)\Psi_{j,\gamma\gamma,g}}(\hat{%
		\gamma }_{j,g}-\gamma_{j,g}^{\ast})\right\Vert ,  \label{P_C_L_Auxillary2_11}
\end{align}
where $\psi_{j,\gamma a}(z_{i,t};\phi_{j,g})\equiv\partial\psi_{j,\gamma
}(z_{i,t};\phi_{j,g})/\partial a$ for $a\in\{ \theta_{j},\gamma_{j,g}\}$.
The first term in the inequality above can be bounded as follows: 
\begin{align}
	& \left\Vert \sum_{g\in\mathcal{G}_{j}}\frac{\sum_{i\in I_{j,g}}\sum_{t\leq
			T}\mathbb{E}[\psi_{j,\gamma}(z_{i,t};\phi_{j,g}^{\ast})\psi_{j,\gamma\theta
		}^{\top}(z_{i,t};\phi_{j,g}^{\ast})]}{(n_{j,g}T)\Psi_{j,\gamma\gamma,g}}(%
	\hat{\theta}_{j}-\theta_{j}^{\ast})\right\Vert  \notag \\
	& \leq\frac{\sum_{g\in\mathcal{G}_{j}}(n_{j,g}T)^{-1}\sum_{i\in
			I_{j,g}}\sum_{t\leq T}\left\Vert \mathbb{E}[\psi_{j,\gamma}(z_{i,t};%
		\phi_{j,g}^{\ast
		})\psi_{j,\gamma\theta}^{\top}(z_{i,t};\phi_{j,g}^{\ast})]\right\Vert }{%
		\min_{g\in\mathcal{G}_{j}}(-\Psi_{j,\gamma\gamma,g})}||\hat{\theta}%
	_{j}-\theta_{j}^{\ast}||  \notag \\
	& \leq KG_{j}\left\Vert \hat{\theta}_{j}-\theta_{j}^{\ast}\right\Vert
	=O_{p}((nT^{-1})^{1/2}).  \label{P_C_L_Auxillary2_12}
\end{align}
where the first inequality follows from the triangle and Cauchy-Schwarz
inequalities, the second inequality follows from Assumptions \ref{A3}(iii)
and \ref{A4}, and the equality holds due to Lemma \ref{C_L_Auxillary1}(iii).
For the second term after the inequality in (\ref{P_C_L_Auxillary2_11}), we
have:%
\begin{align}
	& \left\vert \sum_{g\in\mathcal{G}_{j}}\frac{\sum_{i\in I_{j,g}}\sum_{t\leq
			T}\mathbb{E}[\psi_{j,\gamma}(z_{i,t};\phi_{j,g}^{\ast})\psi_{j,\gamma\gamma
		}(z_{i,t};\phi_{j,g}^{\ast})]}{(n_{j,g}T)\Psi_{j,\gamma\gamma,g}}(\hat{%
		\gamma }_{j,g}-\gamma_{j,g}^{\ast})\right\vert  \notag \\
	& \leq\left\vert \sum_{g\in\mathcal{G}_{j}}\frac{\sum_{i\in
			I_{j,g}}\sum_{t\leq T}\mathbb{E}[\psi_{j,\gamma}(z_{i,t};\phi_{j,g}^{\ast})%
		\psi_{j,\gamma\gamma}(z_{i,t};\phi_{j,g}^{\ast})]}{(n_{j,g}T)\Psi
		_{j,\gamma\gamma,g}}\Psi_{j,\gamma\gamma,g}^{-1}\hat{\Psi}_{j,\gamma
		,g}\right\vert  \notag \\
	& +\left\vert \sum_{g\in\mathcal{G}_{j}}\frac{\sum_{i\in I_{j,g}}\sum_{t\leq
			T}\mathbb{E}[\psi_{j,\gamma}(z_{i,t};\phi_{j,g}^{\ast})\psi_{j,\gamma\gamma
		}(z_{i,t};\phi_{j,g}^{\ast})]}{(n_{j,g}T)\Psi_{j,\gamma\gamma,g}}(\hat{%
		\gamma }_{j,g}-\gamma_{j,g}^{\ast}+\Psi_{j,\gamma\gamma,g}^{-1}\hat{\Psi}%
	_{j,\gamma,g})\right\vert .  \label{P_C_L_Auxillary2_13}
\end{align}
Using Assumptions \ref{A3}(iii) and \ref{A4}, and Lemma\ \ref{C_L_Auxillary1}%
(vi), we get%
\begin{align}
	& \left\vert \sum_{g\in\mathcal{G}_{j}}\frac{\sum_{i\in I_{j,g}}\sum_{t\leq
			T}\mathbb{E}[\psi_{j,\gamma}(z_{i,t};\phi_{j,g}^{\ast})\psi_{j,\gamma\gamma
		}(z_{i,t};\phi_{j,g}^{\ast})]}{(n_{j,g}T)\Psi_{j,\gamma\gamma,g}}(\hat{%
		\gamma }_{j,g}-\gamma_{j,g}^{\ast}+\Psi_{j,\gamma\gamma,g}^{-1}\hat{\Psi}%
	_{j,\gamma,g})\right\vert  \notag \\
	& \leq K\sum_{g\in\mathcal{G}_{j}}|\hat{\gamma}_{j,g}-\gamma_{j,g}^{\ast
	}+\Psi_{j,\gamma\gamma,g}^{-1}\hat{\Psi}_{j,%
		\gamma,g}|=O_{p}((nT^{-1})^{1/2}).  \label{P_C_L_Auxillary2_14}
\end{align}
By Assumptions \ref{A1}, \ref{A3} and \ref{A4}, and Lemma \ref%
{C_L_Auxillary1}(viii), it follows that%
\begin{equation*}
	\mathbb{E}\left[ \left\vert \sum_{g\in\mathcal{G}_{j}}\frac{\sum_{i\in
			I_{j,g}}\sum_{t\leq T}\mathbb{E}[\psi_{j,\gamma}(z_{i,t};\phi_{j,g}^{\ast
		})\psi_{j,\gamma\gamma}(z_{i,t};\phi_{j,g}^{\ast})]}{(n_{j,g}T)\Psi
		_{j,\gamma\gamma,g}}\Psi_{j,\gamma\gamma,g}^{-1}\hat{\Psi}_{j,\gamma
		,g}\right\vert ^{2}\right] \leq K\sum_{g\in\mathcal{G}_{j}}\mathbb{E}[\hat{%
		\Psi}_{j,\gamma,g}^{2}]\leq KnT^{-1}.
\end{equation*}
This, together with Markov's inequality, (\ref{P_C_L_Auxillary2_11})-(\ref%
{P_C_L_Auxillary2_14}) shows that 
\begin{equation}
	\sum_{g\in\mathcal{G}_{j}}\frac{\sum_{i\in I_{j,g}}\sum_{t\leq T}\mathbb{E}%
		[C_{1,j,g}(z_{i,t})]}{(n_{j,g}T)\Psi_{j,\gamma\gamma,g}}(\hat{\phi }%
	_{j,g}-\phi_{j,g}^{\ast})=O_{p}((nT^{-1})^{1/2}).
	\label{P_C_L_Auxillary2_14b}
\end{equation}
Collecting the results in (\ref{P_C_L_Auxillary2_7}), (\ref%
{P_C_L_Auxillary2_10}) and (\ref{P_C_L_Auxillary2_14b}), we obtain%
\begin{equation*}
	T^{-1}\sum_{g\in\mathcal{G}_{j}}\sum_{i\in I_{j,g}}\sum_{t\leq T}\frac {%
		\psi_{j,\gamma}(z_{i,t};\phi_{j,g}^{\ast})\psi_{j,\gamma\phi}^{%
			\top}(z_{i,t};\phi_{j,g}^{\ast})(\hat{\phi}_{j,g}-\phi_{j,g}^{\ast})}{n_{j,g}%
		\hat{\Psi}_{j,\gamma\gamma}(\hat{\phi}_{j,g})}=O_{p}\left(
	(nT^{-1})^{1/2}+G_{j}T^{-1}\right) ,
\end{equation*}
which establishes\ (\ref{P_C_L_Auxillary2_2}).

We now proceed to verify\ (\ref{P_C_L_Auxillary2_3}). Using the triangle
inequality and the Cauchy-Schwarz inequality, we obtain:%
\begin{align}
	& \left\vert \sum_{g\in\mathcal{G}_{j}}\frac{(\hat{\phi}_{j,g}-\phi
		_{j,g}^{\ast})^{\top}\sum_{i\in I_{j,g}}\sum_{t\leq T}\psi_{j,\gamma\phi
		}(z_{i,t};\tilde{\phi}_{j,g})\psi_{j,\gamma\phi}^{\top}(z_{i,t};\tilde{\phi }%
		_{j,g})(\hat{\phi}_{j,g}-\phi_{j,g}^{\ast})}{(n_{j,g}T)\hat{\Psi}%
		_{j,\gamma\gamma}(\hat{\phi}_{j,g})}\right\vert  \notag \\
	& \leq\frac{\sum_{g\in\mathcal{G}_{j}}||\hat{\phi}_{j,g}-\phi_{j,g}^{\ast
		}||^{2}(n_{j,g}T)^{-1}\sum_{i\in I_{j,g}}\sum_{t\leq T}||\psi_{j,\gamma\phi
		}(z_{i,t};\tilde{\phi}_{j,g})||^{2}}{\min_{g\in\mathcal{G}_{j}}(|\hat{\Psi }%
		_{j,\gamma\gamma,g}(\hat{\phi}_{j,g})|)}.  \label{P_C_L_Auxillary2_15}
\end{align}
By Assumption \ref{A3} and the Cauchy-Schwarz inequality, we have:%
\begin{equation*}
	\sum_{i\in I_{j,g}}\sum_{t\leq T}||\psi_{j,\gamma\phi}(z_{i,t};\tilde{\phi }%
	_{j,g})||^{2}\leq K\sum_{i\in I_{j,g}}\sum_{t\leq T}\left( M(z_{i,t})^{2}||%
	\hat{\phi}_{j,g}-\phi_{j,g}^{\ast}||^{2}+||\psi_{j,\gamma\phi}(z_{i,t};%
	\phi_{j,g}^{\ast})||^{2}\right) .
\end{equation*}
This implies that%
\begin{align}
	& \sum_{g\in\mathcal{G}_{j}}||\hat{\phi}_{j,g}-\phi_{j,g}^{%
		\ast}||^{2}(n_{j,g}T)^{-1}\sum_{i\in I_{j,g}}\sum_{t\leq
		T}||\psi_{j,\gamma\phi }(z_{i,t};\tilde{\phi}_{j,g})||^{2}  \notag \\
	& \leq K\sum_{g\in\mathcal{G}_{j}}||\hat{\phi}_{j,g}-\phi_{j,g}^{%
		\ast}||^{4}(n_{j,g}T)^{-1}\sum_{i\in I_{j,g}}\sum_{t\leq T}M(z_{i,t})^{2} 
	\notag \\
	& +K\sum_{g\in\mathcal{G}_{j}}||\hat{\phi}_{j,g}-\phi_{j,g}^{%
		\ast}||^{2}(n_{j,g}T)^{-1}\sum_{i\in I_{j,g}}\sum_{t\leq
		T}||\psi_{j,\gamma\phi }(z_{i,t};\phi_{j,g}^{\ast})||^{2}  \notag \\
	& \leq K\left( 1+\max_{g\in\mathcal{G}_{j}}||\hat{\phi}_{j,g}-\phi
	_{j,g}^{\ast}||^{2}\right) \sum_{g\in\mathcal{G}_{j}}||\hat{\phi}%
	_{j,g}-\phi_{j,g}^{\ast}||^{2}(n_{j,g}T)^{-1}\sum_{i\in I_{j,g}}\sum_{t\leq
		T}\tilde{M}_{1}(z_{i,t})^{2},  \label{P_C_L_Auxillary2_16}
\end{align}
where $\tilde{M}_{1}(z_{i,t})\equiv
M(z_{i,t})+||\psi_{j,\gamma\phi}(z_{i,t};\phi_{j,g}^{\ast})||$. By the
Cauchy-Schwarz inequality, 
\begin{align}
	& \sum_{g\in\mathcal{G}_{j}}||\hat{\phi}_{j,g}-\phi_{j,g}^{%
		\ast}||^{2}(n_{j,g}T)^{-1}\sum_{i\in I_{j,g}}\sum_{t\leq T}\tilde{M}%
	_{1}(z_{i,t})^{2}  \notag \\
	& \leq\left( \sum_{g\in\mathcal{G}_{j}}||\hat{\phi}_{j,g}-\phi_{j,g}^{\ast
	}||^{4}\right) ^{1/2}\left( \sum_{g\in\mathcal{G}_{j}}(n_{j,g}T)^{-1}\sum_{i%
		\in I_{j,g}}\sum_{t\leq T}\tilde{M}_{1}(z_{i,t})^{4}\right) ^{1/2}.
	\label{P_C_L_Auxillary2_17}
\end{align}
From Assumptions \ref{A3}(ii, iii) and Markov's inequality, it follows that 
\begin{equation}
	\sum_{g\in\mathcal{G}_{j}}(n_{j,g}T)^{-1}\sum_{i\in I_{j,g}}\sum_{t\leq T}%
	\mathbb{E}[\tilde{M}_{1}(z_{i,t})^{4}]\leq O_{p}(G_{j}),
	\label{P_C_L_Auxillary2_17b}
\end{equation}
which together with (\ref{P_C_L_Auxillary1_7}) and (\ref{P_C_L_Auxillary2_17}%
) establishes that 
\begin{equation*}
	\sum_{g\in\mathcal{G}_{j}}||\hat{\phi}_{j,g}-\phi_{j,g}^{%
		\ast}||^{2}(n_{j,g}T)^{-1}\sum_{i\in I_{j,g}}\sum_{t\leq T}\tilde{M}%
	_{1}(z_{i,t})^{2}=O_{p}(G_{j}T^{-1}).
\end{equation*}
This along with Lemma\ \ref{C_L_Auxillary1}(i, ii) and (\ref%
{P_C_L_Auxillary2_16}) shows that%
\begin{equation}
	\sum_{g\in\mathcal{G}_{j}}||\hat{\phi}_{j,g}-\phi_{j,g}^{%
		\ast}||^{2}(n_{j,g}T)^{-1}\sum_{i\in I_{j,g}}\sum_{t\leq
		T}||\psi_{j,\gamma\phi}(z_{i,t};\tilde{\phi}_{j,g})||^{2}=O_{p}(G_{j}T^{-1}).
	\label{P_C_L_Auxillary2_18}
\end{equation}
Combining Lemma \ref{C_L_Auxillary1}(ii), and results (\ref%
{P_C_L_Auxillary2_15}) and (\ref{P_C_L_Auxillary2_18}), we verify that: 
\begin{equation*}
	\sum_{g\in\mathcal{G}_{j}}\frac{(\hat{\phi}_{j,g}-\phi_{j,g}^{\ast})^{\top
		}\sum_{i\in I_{j,g}}\sum_{t\leq T}\psi_{j,\gamma\phi}(z_{i,t};\tilde{\phi }%
		_{j,g})\psi_{j,\gamma\phi}^{\top}(z_{i,t};\tilde{\phi}_{j,g})(\hat{\phi }%
		_{j,g}-\phi_{j,g}^{\ast})}{(n_{j,g}T)\hat{\Psi}_{j,\gamma\gamma}(\hat{\phi }%
		_{j,g})}=O_{p}(G_{j}T^{-1}),
\end{equation*}
which shows that (\ref{P_C_L_Auxillary2_3}) holds.

To verify (\ref{P_C_L_Auxillary2_4}), we first note that, by the triangle
inequality and the Cauchy-Schwarz inequality:%
\begin{align}
	& \left\vert \sum_{g\in\mathcal{G}_{j}}\frac{(\hat{\phi}_{j,g}-\phi
		_{j,g}^{\ast})^{\top}\sum_{i\in I_{j,g}}\sum_{t\leq
			T}\psi_{j,\gamma}(z_{i,t};\tilde{\phi}_{j,g})\psi_{j,\gamma\phi\phi}(z_{i,t};%
		\tilde{\phi}_{j,g})(\hat{\phi}_{j,g}-\phi_{j,g}^{\ast})}{(n_{j,g}T)\hat{\Psi}%
		_{j,\gamma\gamma}(\hat{\phi}_{j,g})}\right\vert  \notag \\
	& \leq\frac{\sum_{g\in\mathcal{G}_{j}}||\hat{\phi}_{j,g}-\phi_{j,g}^{\ast
		}||^{2}(n_{j,g}T)^{-1}\sum_{i\in I_{j,g}}\sum_{t\leq T}\left\Vert
		\psi_{j,\gamma\phi}(z_{i,t};\tilde{\phi}_{j,g})\right\Vert \left\Vert
		\psi_{j,\gamma\phi\phi}(z_{i,t};\tilde{\phi}_{j,g})\right\Vert }{\min _{g\in%
			\mathcal{G}_{j}}(|\hat{\Psi}_{j,\gamma\gamma,g}(\hat{\phi}_{j,g})|)}.
	\label{P_C_L_Auxillary2_19}
\end{align}
Using Assumption \ref{A3} and the Cauchy-Schwarz inequality, it follows that%
\begin{align*}
	& \sum_{i\in I_{j,g}}\sum_{t\leq T}\left\Vert \psi_{j,\gamma\phi}(z_{i,t};%
	\tilde{\phi}_{j,g})\right\Vert \left\Vert \psi_{j,\gamma\phi\phi }(z_{i,t};%
	\tilde{\phi}_{j,g})\right\Vert \\
	& \leq||\hat{\phi}_{j,g}-\phi_{j,g}^{\ast}||^{2}\sum_{i\in
		I_{j,g}}\sum_{t\leq T}M(z_{i,t})^{2}+\sum_{i\in I_{j,g}}\sum_{t\leq
		T}\left\Vert \psi_{j,\gamma\phi}(z_{i,t};\phi_{j,g}^{\ast})\right\Vert
	\left\Vert \psi_{j,\gamma\phi\phi}(z_{i,t};\phi_{j,g}^{\ast})\right\Vert \\
	& \text{ \ \ \ }+||\hat{\phi}_{j,g}-\phi_{j,g}^{\ast}||\sum_{i\in
		I_{j,g}}\sum_{t\leq T}M(z_{i,t})(\left\Vert \psi_{j,\gamma\phi}(z_{i,t};\phi
	_{j,g}^{\ast})\right\Vert +\left\Vert
	\psi_{j,\gamma\phi\phi}(z_{i,t};\phi_{j,g}^{\ast})\right\Vert ) \\
	& \leq K(1+\max_{g\in\mathcal{G}_{j}}(||\hat{\phi}_{j,g}-\phi_{j,g}^{\ast
	}||^{2}+||\hat{\phi}_{j,g}-\phi_{j,g}^{\ast}||))\sum_{i\in
		I_{j,g}}\sum_{t\leq T}\tilde{M}_{2}(z_{i,t})^{2},
\end{align*}
where $\tilde{M}_{2}(z_{i,t})\equiv\tilde{M}_{1}(z_{i,t})+||\psi_{j,\gamma
	\phi\phi}(z_{i,t};\phi_{j,g}^{\ast})||$. Therefore,%
\begin{align}
	& \left\vert \sum_{g\in\mathcal{G}_{j}}\frac{(\hat{\phi}_{j,g}-\phi
		_{j,g}^{\ast})^{\top}\sum_{i\in I_{j,g}}\sum_{t\leq
			T}\psi_{j,\gamma}(z_{i,t};\tilde{\phi}_{j,g})\psi_{j,\gamma\phi\phi}(z_{i,t};%
		\tilde{\phi}_{j,g})(\hat{\phi}_{j,g}-\phi_{j,g}^{\ast})}{(n_{j,g}T)\hat{\Psi}%
		_{j,\gamma\gamma}(\hat{\phi}_{j,g})}\right\vert  \notag \\
	& \leq\frac{K(1+\max_{g\in\mathcal{G}_{j}}(||\hat{\phi}_{j,g}-\phi
		_{j,g}^{\ast}||^{2}+||\hat{\phi}_{j,g}-\phi_{j,g}^{\ast}||))}{\min _{g\in%
			\mathcal{G}_{j}}(|\hat{\Psi}_{j,\gamma\gamma,g}(\hat{\phi}_{j,g})|)}  \notag
	\\
	& \text{ \ \ \ }\times\sum_{g\in\mathcal{G}_{j}}||\hat{\phi}_{j,g}-\phi
	_{j,g}^{\ast}||^{2}(n_{j,g}T)^{-1}\sum_{i\in I_{j,g}}\sum_{t\leq T}\tilde {M}%
	_{2}(z_{i,t})^{2}.  \label{P_C_L_Auxillary2_20}
\end{align}
Following similar arguments to those used to establish (\ref%
{P_C_L_Auxillary2_18}), we can show that 
\begin{equation*}
	\sum_{g\in\mathcal{G}_{j}}||\hat{\phi}_{j,g}-\phi_{j,g}^{%
		\ast}||^{2}(n_{j,g}T)^{-1}\sum_{i\in I_{j,g}}\sum_{t\leq T}\tilde{M}%
	_{2}(z_{i,t})^{2}=O_{p}(G_{j}T^{-1}),
\end{equation*}
which, combined with (\ref{P_C_L_Auxillary2_20}) and Lemma \ref%
{C_L_Auxillary1}(i, ii, iii), verifies (\ref{P_C_L_Auxillary2_4}).\hfill$%
Q.E.D.$

\bigskip

\begin{lemma}
	\textit{\label{C_L_Auxillary2b}\ }Under Assumptions \ref{A1} and \ref{C_A1},
	we have%
	\begin{equation}
		\sum_{g\in \mathcal{G}_{2}}\sum_{i\in I_{2,g}}\frac{(\widehat{\mathbb{E}}%
			_{T}[\hat{\psi}_{2,\gamma }(z_{i,t})])^{2}-(\widehat{\mathbb{E}}_{T}[\psi
			_{2,\gamma }(z_{i,t})])^{2}}{n_{2,g}\hat{\Psi}_{2,\gamma \gamma }(\hat{\phi}%
			_{2,g})}=O_{p}((nT^{-1})^{1/2}+G_{2}T^{-1}).  \label{C_L_Auxillary2b_1}
	\end{equation}
\end{lemma}

\noindent {\textsc{Proof of Lemma \ref{C_L_Auxillary2b}}}. First, we notice
that the left-hand side (LHS) of (\ref{C_L_Auxillary2b_1}) can be written as 
\begin{align}
	& \sum_{g\in \mathcal{G}_{2}}\sum_{i\in I_{2,g}}\frac{(\sum_{t\leq T}\hat{%
			\psi}_{2,\gamma }(z_{i,t}))^{2}-(\sum_{t\leq T}\psi _{2,\gamma
		}(z_{i,t}))^{2}}{n_{2,g}T^{2}\hat{\Psi}_{2,\gamma \gamma }(\hat{\phi}_{2,g})}
	\notag \\
	& =\sum_{g\in \mathcal{G}_{2}}\sum_{i\in I_{2,g}}\frac{(\sum_{t\leq T}(\hat{%
			\psi}_{2,\gamma }(z_{i,t})-\psi _{2,\gamma }(z_{i,t})))^{2}}{n_{2,g}T^{2}%
		\hat{\Psi}_{2,\gamma \gamma }(\hat{\phi}_{2,g})}  \notag \\
	& +2\sum_{g\in \mathcal{G}_{2}}\sum_{i\in I_{2,g}}\frac{\sum_{t\leq T}(\hat{%
			\psi}_{2,\gamma }(z_{i,t})-\psi _{2,\gamma }(z_{i,t}))}{n_{2,g}T^{2}\hat{\Psi%
		}_{2,\gamma \gamma }(\hat{\phi}_{2,g})}\sum_{t\leq T}\psi _{2,\gamma
	}(z_{i,t}).  \label{P_C_L_Auxillary2b_1}
\end{align}%
By Assumption \ref{A3}(i) and\ the Cauchy-Schwarz inequality, the first term
on the right-hand side (RHS) of the above equation can be bounded as:%
\begin{align}
	& \sum_{g\in \mathcal{G}_{2}}\sum_{i\in I_{2,g}}\frac{(\sum_{t\leq T}(\hat{%
			\psi}_{2,\gamma }(z_{i,t})-\psi _{2,\gamma }(z_{i,t})))^{2}}{n_{2,g}T^{2}|%
		\hat{\Psi}_{2,\gamma \gamma }(\hat{\phi}_{2,g})|}  \notag \\
	& \leq \frac{\sum_{g\in \mathcal{G}_{2}}n_{2,g}^{-1}\sum_{i\in
			I_{2,g}}(T^{-1}\sum_{t\leq T}M(z_{i,t})||\hat{\phi}_{2,g}-\phi _{2,g}^{\ast
		}||)^{2}}{\min_{g\in \mathcal{G}_{2}}|\hat{\Psi}_{2,\gamma \gamma }(\hat{\phi%
		}_{2,g})|}  \notag \\
	& \leq \frac{\left( \sum_{g\in \mathcal{G}_{2}}||\hat{\phi}_{2,g}-\phi
		_{2,g}^{\ast }||^{4}\right) ^{1/2}\left( \sum_{g\in \mathcal{G}%
			_{2}}(n_{2,g}T)^{-1}\sum_{i\in I_{2,g}}\sum_{t\leq T}M(z_{i,t})^{4}\right)
		^{1/2}}{\min_{g\in \mathcal{G}_{2}}|\hat{\Psi}_{2,\gamma \gamma }(\hat{\phi}%
		_{2,g})|}.  \label{P_C_L_Auxillary2b_2}
\end{align}%
From Assumption \ref{A3} and Markov's inequality, it follows that 
\begin{equation}
	\sum_{g\in \mathcal{G}_{2}}(n_{2,g}T)^{-1}\sum_{i\in I_{2,g}}\sum_{t\leq
		T}M(z_{i,t})^{4}=O_{p}(G_{2}),  \label{P_C_L_Auxillary2b_3}
\end{equation}%
which along with (\ref{P_C_L_Auxillary1_7}) and (\ref{P_C_L_Auxillary2b_2})
implies that 
\begin{equation}
	\sum_{g\in \mathcal{G}_{2}}\sum_{i\in I_{2,g}}\frac{(\sum_{t\leq T}(\hat{\psi%
		}_{2,\gamma }(z_{i,t})-\psi _{2,\gamma }(z_{i,t})))^{2}}{n_{2,g}T^{2}\hat{%
			\Psi}_{2,\gamma \gamma }(\hat{\phi}_{2,g})}=O_{p}(G_{2}T^{-1}).
	\label{P_C_L_Auxillary2b_3b}
\end{equation}

To investigate the second term in the RHS of (\ref{P_C_L_Auxillary2b_1}), we
first apply the Taylor expansion to obtain: 
\begin{align}
	& \sum_{g\in \mathcal{G}_{2}}\sum_{i\in I_{2,g}}\frac{T^{-1}\sum_{t\leq T}(%
		\hat{\psi}_{2,\gamma }(z_{i,t})-\psi _{2,\gamma }(z_{i,t}))}{n_{2,g}\hat{\Psi%
		}_{2,\gamma \gamma }(\hat{\phi}_{2,g})}T^{-1}\sum_{t\leq T}\psi _{2,\gamma
	}(z_{i,t})  \notag \\
	& =\sum_{g\in \mathcal{G}_{2}}\sum_{i\in I_{2,g}}\frac{T^{-1}\sum_{t\leq
			T}\psi _{2,\gamma \phi }(z_{i,t},\tilde{\phi}_{2,g})(\hat{\phi}_{2,g}-\phi
		_{2,g}^{\ast })}{n_{2,g}\hat{\Psi}_{2,\gamma \gamma }(\hat{\phi}_{2,g})}%
	T^{-1}\sum_{t\leq T}\psi _{2,\gamma }(z_{i,t}),  \label{P_C_L_Auxillary2b_4}
\end{align}%
where $\tilde{\phi}_{j,g}$ lies between $\hat{\phi}_{j,g}$ and $\phi
_{j,g}^{\ast }$. By Assumption \ref{A3} and the Cauchy-Schwarz inequality,
it follows that%
\begin{align}
	& \left\vert \sum_{g\in \mathcal{G}_{2}}\sum_{i\in I_{2,g}}\frac{%
		T^{-1}\sum_{t\leq T}(\psi _{2,\gamma \phi }(z_{i,t},\tilde{\phi}_{2,g})-\psi
		_{2,\gamma \phi }(z_{i,t},\phi _{2,g}^{\ast }))(\tilde{\phi}_{2,g}-\phi
		_{2,g}^{\ast })}{n_{2,g}\hat{\Psi}_{2,\gamma \gamma }(\hat{\phi}_{2,g})}%
	T^{-1}\sum_{t\leq T}\psi _{2,\gamma }(z_{i,t})\right\vert  \notag \\
	& \leq \max_{i\leq n}\left\vert T^{-1}\sum_{t\leq T}\psi _{2,\gamma
	}(z_{i,t})\right\vert \frac{\sum_{g\in \mathcal{G}_{2}}||\tilde{\phi}%
		_{2,g}-\phi _{2,g}^{\ast }||^{2}(n_{2,g}T)^{-1}\sum_{i\in
			I_{2,g}}\sum_{t\leq T}M(z_{i,t})}{\min_{g\in \mathcal{G}_{2}}\hat{\Psi}%
		_{2,\gamma \gamma }(\hat{\phi}_{2,g})}.  \label{P_C_L_Auxillary2b_5}
\end{align}%
Note that by a maximal inequality under the $L_{p}$-norm,%
\begin{equation}
	\left\Vert \max_{i\leq n}\left\vert T^{-1}\sum_{t\leq T}(\psi _{2,\gamma
	}(z_{i,t})-\mathbb{E}[\psi _{2,\gamma }(z_{i,t})])\right\vert \right\Vert
	_{p}\leq Kn^{1/p}\max_{i\leq n}\left\Vert T^{-1}\sum_{t\leq T}(\psi
	_{2,\gamma }(z_{i,t})-\mathbb{E}[\psi _{2,\gamma }(z_{i,t})])\right\Vert
	_{p}.  \label{P_C_L_Auxillary2b_6}
\end{equation}%
By Assumption \ref{A3}, we have 
\begin{equation}
	\mathbb{E}[|\psi _{2,\gamma }(z_{i,t})|^{p}]\leq K,
	\label{P_C_L_Auxillary2b_7}
\end{equation}%
which implies that $\left\Vert \psi _{2,\gamma }(z_{i,t})-\mathbb{E}[\psi
_{2,\gamma }(z_{i,t})]\right\Vert _{p}\leq K$. Thus, applying Rosenthal's
inequality, we get%
\begin{equation*}
	\mathbb{E}\left[ \left\vert \sum_{t\leq T}(\psi _{2,\gamma }(z_{i,t})-%
	\mathbb{E}[\psi _{2,\gamma }(z_{i,t})])\right\vert ^{p}\right] \leq KT^{p/2}
\end{equation*}%
which along with (\ref{P_C_L_Auxillary2b_6}) shows that 
\begin{equation*}
	\left\Vert \max_{i\leq n}\left\vert T^{-1}\sum_{t\leq T}(\psi _{2,\gamma
	}(z_{i,t})-\mathbb{E}[\psi _{2,\gamma }(z_{i,t})])\right\vert \right\Vert
	_{p}\leq Kn^{1/p}T^{-1/2}.
\end{equation*}%
Combining this result with Markov's inequality and Assumption \ref{A1}(i),
we establish that%
\begin{equation}
	\max_{i\leq n}\left\vert T^{-1}\sum_{t\leq T}(\psi _{2,\gamma }(z_{i,t})-%
	\mathbb{E}[\psi _{2,\gamma }(z_{i,t})])\right\vert =O_{p}(T^{-1/2+1/p}).
	\label{P_C_L_Auxillary2b_8}
\end{equation}%
Therefore, from (\ref{P_C_L_Auxillary2b_7}) and (\ref{P_C_L_Auxillary2b_8}),
it follows that 
\begin{align}
	\max_{i\leq n}\left\vert T^{-1}\sum_{t\leq T}\psi _{2,\gamma
	}(z_{i,t})\right\vert & \leq \max_{i\leq n}\left\vert T^{-1}\sum_{t\leq
		T}(\psi _{2,\gamma }(z_{i,t})-\mathbb{E}[\psi _{2,\gamma
	}(z_{i,t})])\right\vert  \notag \\
	& +\max_{i\leq n}\left\vert T^{-1}\sum_{t\leq T}\mathbb{E}[\psi _{2,\gamma
	}(z_{i,t})]\right\vert \overset{}{=}O_{p}(1).  \label{P_C_L_Auxillary2b_9}
\end{align}%
By the Cauchy-Schwarz inequality,\ Assumption \ref{A3}(ii) and (\ref%
{P_C_L_Auxillary1_7}), we have 
\begin{align}
	& \frac{\sum_{g\in \mathcal{G}_{2}}||\hat{\phi}_{2,g}-\phi _{2,g}^{\ast
		}||^{2}(n_{2,g}T)^{-1}\sum_{i\in I_{2,g}}\sum_{t\leq T}M(z_{i,t}).}{%
		\min_{g\in \mathcal{G}_{2}}|\hat{\Psi}_{2,\gamma \gamma }(\hat{\phi}_{2,g})|}
	\notag \\
	& \leq \frac{\left( \sum_{g\in \mathcal{G}_{2}}||\hat{\phi}_{2,g}-\phi
		_{2,g}^{\ast }||^{4}\right) ^{1/2}\left( \sum_{g\in \mathcal{G}%
			_{2}}(n_{2,g}T)^{-1}\sum_{i\in I_{2,g}}\sum_{t\leq T}M(z_{i,t})^{2}\right)
		^{1/2}}{\min_{g\in \mathcal{G}_{2}}|\hat{\Psi}_{2,\gamma \gamma }(\hat{\phi}%
		_{2,g})|}  \notag \\
	& =O_{p}(G_{2}T^{-1}).  \label{P_C_L_Auxillary2b_10}
\end{align}%
From (\ref{P_C_L_Auxillary2b_4}), (\ref{P_C_L_Auxillary2b_5}), (\ref%
{P_C_L_Auxillary2b_9}) and (\ref{P_C_L_Auxillary2b_10}), we establish: 
\begin{align}
	& \sum_{g\in \mathcal{G}_{2}}\sum_{i\in I_{2,g}}\frac{T^{-1}\sum_{t\leq T}(%
		\hat{\psi}_{2,\gamma }(z_{i,t})-\psi _{2,\gamma }(z_{i,t}))}{n_{2,g}\hat{\Psi%
		}_{2,\gamma \gamma }(\hat{\phi}_{2,g})}T^{-1}\sum_{t\leq T}\psi _{2,\gamma
	}(z_{i,t})  \notag \\
	& =\sum_{g\in \mathcal{G}_{2}}\sum_{i\in I_{2,g}}\frac{T^{-1}\sum_{t\leq
			T}\psi _{2,\gamma \theta }(z_{i,t})(\hat{\theta}_{2}-\theta _{2}^{\ast })}{%
		n_{2,g}\hat{\Psi}_{2,\gamma \gamma }(\hat{\phi}_{2,g})}T^{-1}\sum_{t\leq
		T}\psi _{2,\gamma }(z_{i,t})  \notag \\
	& +\sum_{g\in \mathcal{G}_{2}}(\hat{\gamma}_{2,g}-\gamma _{2,g}^{\ast
	})\sum_{i\in I_{2,g}}\frac{T^{-1}\sum_{t\leq T}\psi _{2,\gamma \gamma
		}(z_{i,t})}{n_{2,g}\hat{\Psi}_{2,\gamma \gamma }(\hat{\phi}_{2,g})}%
	T^{-1}\sum_{t\leq T}\psi _{2,\gamma }(z_{i,t})+O_{p}(G_{2}T^{-1}).
	\label{P_C_L_Auxillary2b_11}
\end{align}%
The first term on the RHS of (\ref{P_C_L_Auxillary2b_11}) can be bounded as 
\begin{align}
	& \left\vert \sum_{g\in \mathcal{G}_{2}}\sum_{i\in I_{2,g}}\frac{%
		T^{-1}\sum_{t\leq T}\psi _{2,\gamma \theta }(z_{i,t})(\hat{\theta}%
		_{2}-\theta _{2}^{\ast })}{n_{2,g}\hat{\Psi}_{2,\gamma \gamma }(\hat{\phi}%
		_{2,g})}T^{-1}\sum_{t\leq T}\psi _{2,\gamma }(z_{i,t})\right\vert  \notag \\
	& \leq \frac{||\hat{\theta}_{2}-\theta _{2}^{\ast }||\max_{i\leq
			n}\left\vert T^{-1}\sum_{t\leq T}\psi _{2,\gamma }(z_{i,t})\right\vert }{%
		\min_{g\in \mathcal{G}_{2}}|\hat{\Psi}_{2,\gamma \gamma }(\hat{\phi}_{2,g})|}%
	\sum_{g\in \mathcal{G}_{2}}(n_{2,g}T)^{-1}\sum_{i\in I_{2,g}}\sum_{t\leq
		T}||\psi _{2,\gamma \theta }(z_{i,t})||.  \label{P_C_L_Auxillary2b_12}
\end{align}%
By Assumption \ref{A3} and Markov's inequality, it follows that 
\begin{equation*}
	\sum_{g\in \mathcal{G}_{2}}(n_{2,g}T)^{-1}\sum_{i\in I_{2,g}}\sum_{t\leq
		T}||\psi _{2,\gamma \theta }(z_{i,t})||=O_{p}(G_{2}),
\end{equation*}%
which together with Lemma \ref{C_L_Auxillary1}(ii, iii) and (\ref%
{P_C_L_Auxillary2b_9}) implies that%
\begin{equation}
	\sum_{g\in \mathcal{G}_{2}}\sum_{i\in I_{2,g}}\frac{T^{-1}\sum_{t\leq T}\psi
		_{2,\gamma \theta }(z_{i,t})(\hat{\theta}_{2}-\theta _{2}^{\ast })}{n_{2,g}%
		\hat{\Psi}_{2,\gamma \gamma }(\hat{\phi}_{2,g})}T^{-1}\sum_{t\leq T}\psi
	_{2,\gamma }(z_{i,t})=O_{p}(G_{2}(nT)^{-1/2}).  \label{P_C_L_Auxillary2b_13}
\end{equation}%
To study the second term on the RHS of (\ref{P_C_L_Auxillary2b_11}), we
first write it as%
\begin{align}
	& \sum_{g\in \mathcal{G}_{2}}(\hat{\gamma}_{2,g}-\gamma _{2,g}^{\ast
	})\sum_{i\in I_{2,g}}\frac{T^{-1}\sum_{t\leq T}\psi _{2,\gamma \gamma
		}(z_{i,t})}{n_{2,g}\hat{\Psi}_{2,\gamma \gamma }(\hat{\phi}_{2,g})}%
	T^{-1}\sum_{t\leq T}\psi _{2,\gamma }(z_{i,t})  \notag \\
	& =\sum_{g\in \mathcal{G}_{2}}(\hat{\gamma}_{2,g}-\gamma _{2,g}^{\ast
	})\sum_{i\in I_{2,g}}\frac{T^{-1}\sum_{t\leq T}(\psi _{2,\gamma \gamma
		}(z_{i,t})-\mathbb{E}[\psi _{2,\gamma \gamma }(z_{i,t})])}{n_{2,g}\hat{\Psi}%
		_{2,\gamma \gamma }(\hat{\phi}_{2,g})}T^{-1}\sum_{t\leq T}\psi _{2,\gamma
	}(z_{i,t})  \notag \\
	& +\sum_{g\in \mathcal{G}_{2}}(\hat{\gamma}_{2,g}-\gamma _{2,g}^{\ast
	})\sum_{i\in I_{2,g}}\frac{\mathbb{E}[\psi _{2,\gamma \gamma }(z_{i,t})]}{%
		n_{2,g}\hat{\Psi}_{2,\gamma \gamma }(\hat{\phi}_{2,g})}T^{-1}\sum_{t\leq
		T}(\psi _{2,\gamma }(z_{i,t})-\mathbb{E}[\psi _{2,\gamma }(z_{i,t})])  \notag
	\\
	& -\sum_{g\in \mathcal{G}_{2}}(\hat{\gamma}_{2,g}-\gamma _{2,g}^{\ast
	})\sum_{i\in I_{2,g}}\frac{\mathbb{E}[\psi _{2,\gamma \gamma }(z_{i,t})]%
		\mathbb{E}[\psi _{2,\gamma }(z_{i,t})]}{n_{2,g}\hat{\Psi}_{2,\gamma \gamma }(%
		\hat{\phi}_{2,g})\Psi _{2,\gamma \gamma ,g}}(\hat{\Psi}_{2,\gamma \gamma }(%
	\hat{\phi}_{2,g})-\Psi _{2,\gamma \gamma ,g})  \notag \\
	& +\sum_{g\in \mathcal{G}_{2}}(\hat{\gamma}_{2,g}-\gamma _{2,g}^{\ast })%
	\frac{\sum_{i\in I_{2,g}}\mathbb{E}[\psi _{2,\gamma \gamma }(z_{i,t})]%
		\mathbb{E}[\psi _{2,\gamma }(z_{i,t})]}{n_{2,g}\Psi _{2,\gamma \gamma ,g}}.
	\label{P_C_L_Auxillary2b_14}
\end{align}%
By the Cauchy-Schwarz inequality, we can bound the first term on the RHS of (%
\ref{P_C_L_Auxillary2b_14}) as 
\begin{align}
	& \left\vert \sum_{g\in \mathcal{G}_{2}}(\hat{\gamma}_{j,g}-\gamma
	_{j,g}^{\ast })\sum_{i\in I_{2,g}}\frac{\sum_{t\leq T}(\psi _{2,\gamma
			\gamma }(z_{i,t})-\mathbb{E}[\psi _{2,\gamma \gamma }(z_{i,t})])}{(n_{2,g}T)%
		\hat{\Psi}_{2,\gamma \gamma }(\hat{\phi}_{j,g})}T^{-1}\sum_{t\leq T}\psi
	_{2,\gamma }(z_{i,t})\right\vert  \notag \\
	& \leq \frac{\max_{i\leq n}\left\vert T^{-1}\sum_{t\leq T}\psi _{j,\gamma
		}(z_{i,t})\right\vert }{\min_{g\in \mathcal{G}_{2}}|\hat{\Psi}_{2,\gamma
			\gamma }(\hat{\phi}_{j,g})|}\left( \sum_{g\in \mathcal{G}_{2}}(\hat{\gamma}%
	_{j,g}-\gamma _{j,g}^{\ast })^{2}\right) ^{1/2}  \notag \\
	& \times \left( \sum_{g\in \mathcal{G}_{2}}n_{2,g}^{-1}\sum_{i\in
		I_{2,g}}\left( T^{-1}\sum_{t\leq T}(\psi _{2,\gamma \gamma }(z_{i,t})-%
	\mathbb{E}[\psi _{2,\gamma \gamma }(z_{i,t})])\right) ^{2}\right) ^{1/2}.
	\label{P_C_L_Auxillary2b_15a}
\end{align}%
Under Assumptions \ref{A1} and \ref{A3}, we can apply the covariance
inequality for mixing processes to obtain%
\begin{equation*}
	\mathbb{E}\left[ \sum_{g\in \mathcal{G}_{2}}n_{2,g}^{-1}\sum_{i\in
		I_{2,g}}\left( T^{-1}\sum_{t\leq T}(\psi _{2,\gamma \gamma }(z_{i,t})-%
	\mathbb{E}[\psi _{2,\gamma \gamma }(z_{i,t})])\right) ^{2}\right] \leq
	KG_{2}T^{-1},
\end{equation*}%
which together with Markov's inequality, Lemma \ref{C_L_Auxillary1}(ii, iv),
(\ref{P_C_L_Auxillary2b_9}) and (\ref{P_C_L_Auxillary2b_15a}) implies that
we can bound the first term on the RHS of (\ref{P_C_L_Auxillary2b_14}) by%
\begin{equation}
	\sum_{g\in \mathcal{G}_{2}}(\hat{\gamma}_{2,g}-\gamma _{2,g}^{\ast
	})\sum_{i\in I_{2,g}}\frac{\sum_{t\leq T}(\psi _{2,\gamma \gamma }(z_{i,t})-%
		\mathbb{E}[\psi _{2,\gamma \gamma }(z_{i,t})])}{(n_{2,g}T)\hat{\Psi}%
		_{2,\gamma \gamma }(\hat{\phi}_{2,g})}T^{-1}\sum_{t\leq T}\psi _{2,\gamma
	}(z_{i,t})=O_{p}(G_{2}T^{-1}).  \label{P_C_L_Auxillary2b_15}
\end{equation}%
As for the second term on the RHS of (\ref{P_C_L_Auxillary2b_14}), we have
by the Cauchy-Schwarz inequality and Assumption \ref{A3}%
\begin{align}
	& \left\vert \sum_{g\in \mathcal{G}_{2}}(\hat{\gamma}_{2,g}-\gamma
	_{2,g}^{\ast })\sum_{i\in I_{2,g}}\frac{\mathbb{E}[\psi _{2,\gamma \gamma
		}(z_{i,t})]}{n_{2,g}\hat{\Psi}_{2,\gamma \gamma }(\hat{\phi}_{2,g})}%
	T^{-1}\sum_{t\leq T}(\psi _{2,\gamma }(z_{i,t})-\mathbb{E}[\psi _{2,\gamma
	}(z_{i,t})])\right\vert  \notag \\
	& \leq \frac{K\left( \sum_{g\in \mathcal{G}_{2}}(\hat{\gamma}_{2,g}-\gamma
		_{2,g}^{\ast })^{2}\right) ^{1/2}}{\min_{g\in \mathcal{G}_{2}}|\hat{\Psi}%
		_{2,\gamma \gamma }(\hat{\phi}_{2,g})|}\left( \sum_{g\in \mathcal{G}%
		_{2}}n_{2,g}^{-1}\sum_{i\in I_{2,g}}\left( T^{-1}\sum_{t\leq T}(\psi
	_{2,\gamma }(z_{i,t})-\mathbb{E}[\psi _{2,\gamma }(z_{i,t})])\right)
	^{2}\right) ^{1/2}.  \label{P_C_L_Auxillary2b_16}
\end{align}%
Under Assumptions \ref{C_A2} and \ref{A3}, it follows that%
\begin{equation*}
	\mathbb{E}\left[ \sum_{g\in \mathcal{G}_{2}}n_{2,g}^{-1}\sum_{i\in
		I_{2,g}}\left( T^{-1}\sum_{t\leq T}(\psi _{2,\gamma }(z_{i,t})-\mathbb{E}%
	[\psi _{2,\gamma }(z_{i,t})])\right) ^{2}\right] \leq KG_{2}T^{-1},
\end{equation*}%
which together with Markov's inequality, Lemma \ref{C_L_Auxillary1}(ii, iv)
and (\ref{P_C_L_Auxillary2b_16}) implies that we can bound the second term
on the RHS of (\ref{P_C_L_Auxillary2b_14}) by%
\begin{equation}
	\sum_{g\in \mathcal{G}_{2}}\sum_{g\in \mathcal{G}_{2}}(\hat{\gamma}%
	_{2,g}-\gamma _{2,g}^{\ast })\sum_{i\in I_{2,g}}\frac{\mathbb{E}[\psi
		_{2,\gamma \gamma }(z_{i,t})]}{n_{2,g}\hat{\Psi}_{2,\gamma \gamma }(\hat{\phi%
		}_{2,g})}T^{-1}\sum_{t\leq T}(\psi _{2,\gamma }(z_{i,t})-\mathbb{E}[\psi
	_{2,\gamma }(z_{i,t})])=O_{p}(G_{2}T^{-1}).  \label{P_C_L_Auxillary2b_17}
\end{equation}%
As for the third term on the RHS of (\ref{P_C_L_Auxillary2b_14}), we have by
the Cauchy-Schwarz inequality%
\begin{align}
	& \left\vert \sum_{g\in \mathcal{G}_{2}}(\hat{\gamma}_{2,g}-\gamma
	_{2,g}^{\ast })\sum_{i\in I_{2,g}}\frac{\mathbb{E}[\psi _{2,\gamma \gamma
		}(z_{i,t})]\mathbb{E}[\psi _{2,\gamma }(z_{i,t})]}{n_{2,g}\hat{\Psi}%
		_{2,\gamma \gamma }(\hat{\phi}_{2,g})\Psi _{2,\gamma \gamma ,g}}(\hat{\Psi}%
	_{2,\gamma \gamma }(\hat{\phi}_{2,g})-\Psi _{2,\gamma \gamma ,g})\right\vert
	\notag \\
	& \leq \frac{K\left( \sum_{g\in \mathcal{G}_{2}}(\hat{\gamma}_{2,g}-\gamma
		_{2,g}^{\ast })^{2}\right) ^{1/2}\left( \sum_{g\in \mathcal{G}_{2}}(\hat{\Psi%
		}_{2,\gamma \gamma }(\hat{\phi}_{2,g})-\Psi _{2,\gamma \gamma
			,g})^{2}\right) ^{1/2}}{\min_{g\in \mathcal{G}_{2}}|\hat{\Psi}_{2,\gamma
			\gamma }(\hat{\phi}_{2,g})\Psi _{2,\gamma \gamma ,g})|}=O_{p}(G_{2}T^{-1})
	\label{P_C_L_Auxillary2b_18}
\end{align}%
where the equality follows from Assumption \ref{A4} and Lemma \ref%
{C_L_Auxillary1}(iv, v).

Combining the results from\ (\ref{P_C_L_Auxillary2b_11}),\ (\ref%
{P_C_L_Auxillary2b_13}), (\ref{P_C_L_Auxillary2b_14}), (\ref%
{P_C_L_Auxillary2b_15}), (\ref{P_C_L_Auxillary2b_17}) and (\ref%
{P_C_L_Auxillary2b_18}), we obtain%
\begin{align}
	& \sum_{g\in\mathcal{G}_{2}}\sum_{i\in I_{2,g}}\frac{T^{-1}\sum_{t\leq T}(%
		\hat{\psi}_{2,\gamma}(z_{i,t})-\psi_{2,\gamma}(z_{i,t}))}{n_{2,g}\hat{\Psi }%
		_{2,\gamma\gamma}(\hat{\phi}_{2,g})}T^{-1}\sum_{t\leq T}\psi_{2,\gamma
	}(z_{i,t})  \notag \\
	& =\sum_{g\in\mathcal{G}_{2}}(\hat{\gamma}_{2,g}-\gamma_{2,g}^{\ast})\frac{%
		\sum_{i\in I_{2,g}}\mathbb{E}[\psi_{2,\gamma\gamma}(z_{i,t})]\mathbb{E}%
		[\psi_{2,\gamma}(z_{i,t})]}{n_{2,g}\Psi_{2,\gamma\gamma,g}}%
	+O_{p}(G_{2}T^{-1}).  \label{P_C_L_Auxillary2b_19}
\end{align}
In view of the results established in (\ref{P_C_L_Auxillary2b_1}), (\ref%
{P_C_L_Auxillary2b_3b}) and (\ref{P_C_L_Auxillary2b_19}), it is evident that
the claim of the lemma follows if 
\begin{equation}
	\sum_{g\in\mathcal{G}_{2}}(\hat{\gamma}_{2,g}-\gamma_{2,g}^{\ast})\frac {%
		\sum_{i\in I_{2,g}}\mathbb{E}[\psi_{2,\gamma\gamma}(z_{i,t})]\mathbb{E}%
		[\psi_{2,\gamma}(z_{i,t})]}{n_{2,g}\Psi_{2,\gamma\gamma,g}}%
	=O_{p}(G_{2}^{1/2}T^{-1/2}).  \label{P_C_L_Auxillary2b_20}
\end{equation}

We next turn to verify (\ref{P_C_L_Auxillary2b_20}). Using the expression in
(\ref{P_C_L_Auxillary1_9}), we can write 
\begin{align}
	& \sum_{g\in\mathcal{G}_{2}}(\hat{\gamma}_{2,g}-\gamma_{2,g}^{\ast})\frac{%
		\sum_{i\in I_{2,g}}\mathbb{E}[\psi_{2,\gamma\gamma}(z_{i,t})]\mathbb{E}%
		[\psi_{2,\gamma}(z_{i,t})]}{n_{2,g}\Psi_{2,\gamma\gamma,g}}  \notag \\
	& =-\sum_{g\in\mathcal{G}_{2}}\frac{\hat{\Psi}_{2,\gamma,g}}{\Psi
		_{2,\gamma\gamma,g}}\frac{\sum_{i\in I_{2,g}}\mathbb{E}[\psi_{2,\gamma\gamma
		}(z_{i,t})]\mathbb{E}[\psi_{2,\gamma}(z_{i,t})]}{n_{2,g}\Psi_{2,\gamma
			\gamma,g}}  \notag \\
	& -\sum_{g\in\mathcal{G}_{2}}(\hat{\Psi}_{2,\gamma\gamma,g}^{-1}-\Psi_{2,%
		\gamma\gamma,g}^{-1})\hat{\Psi}_{2,\gamma,g}\frac{\sum_{i\in I_{2,g}}\mathbb{%
			E}[\psi_{2,\gamma\gamma}(z_{i,t})]\mathbb{E}[\psi_{2,\gamma}(z_{i,t})]}{%
		n_{2,g}\Psi_{2,\gamma\gamma,g}}  \notag \\
	& -\sum_{g\in\mathcal{G}_{2}}\frac{\hat{\Psi}_{2,\gamma\theta,g}(\hat{\theta 
		}_{2}-\theta_{2}^{\ast})}{\hat{\Psi}_{2,\gamma\gamma,g}}\frac{\sum_{i\in
			I_{2,g}}\mathbb{E}[\psi_{2,\gamma\gamma}(z_{i,t})]\mathbb{E}[\psi_{2,\gamma
		}(z_{i,t})]}{n_{2,g}\Psi_{2,\gamma\gamma,g}}  \notag \\
	& -\sum_{g\in\mathcal{G}_{2}}\frac{(\hat{\phi}_{2,g}-\phi_{2,g}^{\ast})^{%
			\top}\tilde{\Psi}_{2,\gamma\phi\phi,g}(\hat{\phi}_{2,g}-\phi_{2,g}^{\ast})}{2%
		\hat{\Psi}_{2,\gamma\gamma,g}}\frac{\sum_{i\in I_{2,g}}\mathbb{E}%
		[\psi_{2,\gamma\gamma}(z_{i,t})]\mathbb{E}[\psi_{2,\gamma}(z_{i,t})]}{%
		n_{2,g}\Psi_{2,\gamma\gamma,g}}.  \label{P_C_L_Auxillary2b_21}
\end{align}
By Assumptions \ref{A1}, \ref{C_A1}(iii) and \ref{A3}, and Lemma \ref%
{C_L_Auxillary1}(viii) 
\begin{align*}
	& \mathbb{E}\left[ \left( \sum_{g\in\mathcal{G}_{2}}\frac{\hat{\Psi }%
		_{2,\gamma,g}}{\Psi_{2,\gamma\gamma,g}}\frac{\sum_{i\in I_{2,g}}\mathbb{E}%
		[\psi_{2,\gamma\gamma}(z_{i,t})]\mathbb{E}[\psi_{2,\gamma}(z_{i,t})]}{%
		n_{2,g}\Psi_{2,\gamma\gamma,g}}\right) ^{2}\right] \\
	& =\sum_{g\in\mathcal{G}_{2}}\mathbb{E}[\hat{\Psi}_{2,\gamma,g}^{2}]\left( 
	\frac{\sum_{i\in I_{2,g}}\mathbb{E}[\psi_{2,\gamma\gamma}(z_{i,t})]\mathbb{E}%
		[\psi_{2,\gamma}(z_{i,t})]}{n_{2,g}\Psi_{2,\gamma\gamma,g}^{2}}\right)
	^{2}\leq K\sum_{g\in\mathcal{G}_{2}}\mathbb{E}[\hat{\Psi}_{2,\gamma,g}^{2}]%
	\leq KG_{2}T^{-1},
\end{align*}
where the second inequality follows from (\ref{P_C_L_Auxillary1_18}).\textbf{%
	\ }Thus, by Markov's inequality, 
\begin{equation}
	\sum_{g\in\mathcal{G}_{2}}\frac{\hat{\Psi}_{2,\gamma,g}}{\Psi_{2,\gamma
			\gamma,g}}\frac{\sum_{i\in I_{2,g}}\mathbb{E}[\psi_{2,\gamma\gamma}(z_{i,t})]%
		\mathbb{E}[\psi_{2,\gamma}(z_{i,t})]}{n_{2,g}\Psi_{2,\gamma\gamma ,g}}%
	=O_{p}(G_{2}^{1/2}T^{-1/2}).  \label{P_C_L_Auxillary2b_22}
\end{equation}
By the Cauchy-Schwarz inequality, Assumptions \ref{A3} and \ref{A4}, we have%
\begin{align*}
	& \left\vert \sum_{g\in\mathcal{G}_{2}}(\hat{\Psi}_{2,\gamma\gamma,g}^{-1}-%
	\Psi_{2,\gamma\gamma,g}^{-1})\hat{\Psi}_{2,\gamma,g}\frac{\sum_{i\in I_{2,g}}%
		\mathbb{E}[\psi_{2,\gamma\gamma}(z_{i,t})]\mathbb{E}[\psi_{2,\gamma
		}(z_{i,t})]}{n_{2,g}\Psi_{2,\gamma\gamma,g}}\right\vert \\
	& \leq\frac{K\left( \sum_{g\in\mathcal{G}_{2}}(\hat{\Psi}_{2,\gamma\gamma
			,g}-\Psi_{2,\gamma\gamma,g})^{2}\right) ^{1/2}\left( \sum_{g\in \mathcal{G}%
			_{2}}\hat{\Psi}_{2,\gamma,g}^{2}\right) ^{1/2}}{\min _{g\in\mathcal{G}_{2}}|%
		\hat{\Psi}_{2,\gamma\gamma}(\hat{\phi}_{2,g})|},
\end{align*}
which along with Lemma \ref{C_L_Auxillary1}(ii, v), (\ref%
{P_C_L_Auxillary1_18}) and Markov's inequality shows that 
\begin{equation}
	\sum_{g\in\mathcal{G}_{2}}(\hat{\Psi}_{2,\gamma\gamma,g}^{-1}-\Psi
	_{2,\gamma\gamma,g}^{-1})\hat{\Psi}_{2,\gamma,g}\frac{\sum_{i\in I_{2,g}}%
		\mathbb{E}[\psi_{2,\gamma\gamma}(z_{i,t})]\mathbb{E}[\psi_{2,%
			\gamma}(z_{i,t})]}{n_{2,g}\Psi_{2,\gamma\gamma,g}}=O_{p}(G_{2}T^{-1}).
	\label{P_C_L_Auxillary2b_23}
\end{equation}
By the Cauchy-Schwarz inequality and Assumption \ref{A4}, we have%
\begin{align}
	& \left\vert \sum_{g\in\mathcal{G}_{2}}\frac{\hat{\Psi}_{2,\gamma\theta ,g}(%
		\hat{\theta}_{2}-\theta_{2}^{\ast})}{\hat{\Psi}_{2,\gamma\gamma,g}}\frac{%
		\sum_{i\in I_{2,g}}\mathbb{E}[\psi_{2,\gamma\gamma}(z_{i,t})]\mathbb{E}%
		[\psi_{2,\gamma}(z_{i,t})]}{n_{2,g}\Psi_{2,\gamma\gamma,g}}\right\vert 
	\notag \\
	& \leq\frac{K||\hat{\theta}_{2}-\theta_{2}^{\ast}||}{\min_{g\in \mathcal{G}%
			_{2}}|\hat{\Psi}_{2,\gamma\gamma}(\hat{\phi}_{2,g})|}\sum _{g\in\mathcal{G}%
		_{2}}(||\hat{\Psi}_{2,\gamma\theta,g}-\mathbb{E}[\hat{\Psi }%
	_{2,\gamma\theta,g}]||+||\mathbb{E}[\hat{\Psi}_{2,\gamma\theta,g}]||).
	\label{P_C_L_Auxillary2b_24}
\end{align}
Under Assumptions \ref{A1} and \ref{A3}, we can apply the covariance
inequality for mixing processes to obtain%
\begin{equation*}
	\mathbb{E}\left[ ||\hat{\Psi}_{2,\gamma\theta,g}-\mathbb{E}[\hat{\Psi }%
	_{2,\gamma\theta,g}]||^{2}\right] \leq K(n_{2,g}T)^{-1},
\end{equation*}
which together with Markov's inequality, Assumption \ref{A3}, Lemma \ref%
{C_L_Auxillary1}(ii, iii) and (\ref{P_C_L_Auxillary2b_24}) implies that 
\begin{equation}
	\sum_{g\in\mathcal{G}_{2}}\frac{\hat{\Psi}_{2,\gamma\theta,g}(\hat{\theta}%
		_{2}-\theta_{2}^{\ast})}{\hat{\Psi}_{2,\gamma\gamma,g}}\frac{\sum_{i\in
			I_{2,g}}\mathbb{E}[\psi_{2,\gamma\gamma}(z_{i,t})]\mathbb{E}[\psi_{2,\gamma
		}(z_{i,t})]}{n_{2,g}\Psi_{2,\gamma\gamma,g}}=O_{p}(G_{2}T^{-1}).
	\label{P_C_L_Auxillary2b_25}
\end{equation}
By the Cauchy-Schwarz inequality, Assumptions \ref{A3} and \ref{A4}, we have%
\begin{align}
	& \left\vert \sum_{g\in\mathcal{G}_{2}}\frac{(\hat{\phi}_{2,g}-\phi
		_{2,g}^{\ast})^{\top}\tilde{\Psi}_{2,\gamma\phi\phi,g}(\hat{\phi}%
		_{2,g}-\phi_{2,g}^{\ast})}{2\hat{\Psi}_{2,\gamma\gamma,g}}\frac{\sum_{i\in
			I_{2,g}}\mathbb{E}[\psi_{2,\gamma\gamma}(z_{i,t})]\mathbb{E}%
		[\psi_{2,\gamma}(z_{i,t})]}{n_{2,g}\Psi_{2,\gamma\gamma,g}}\right\vert 
	\notag \\
	& \leq\frac{K\max_{g\in\mathcal{G}_{2}}||\tilde{\Psi}_{2,\gamma\phi\phi,g}||%
	}{\min_{g\in\mathcal{G}_{2}}|\hat{\Psi}_{2,\gamma\gamma}(\hat{\phi}_{j,g})|}%
	\sum_{g\in\mathcal{G}_{2}}||\hat{\phi}_{2,g}-\phi_{2,g}^{%
		\ast}||^{2}=O_{p}(G_{j}T^{-1})  \label{P_C_L_Auxillary2b_26}
\end{align}
where the equality follows from Lemma \ref{C_L_Auxillary1}(ii), (\ref%
{P_C_L_Auxillary1_12}) and (\ref{P_C_L_Auxillary1_13}). The claim in (\ref%
{P_C_L_Auxillary2b_20}) follows directly from (\ref{P_C_L_Auxillary2b_21}), (%
\ref{P_C_L_Auxillary2b_22}), (\ref{P_C_L_Auxillary2b_23}), (\ref%
{P_C_L_Auxillary2b_25}) and (\ref{P_C_L_Auxillary2b_26}).\hfill$Q.E.D.$

\bigskip

\begin{lemma}
	\textit{\label{C_L_Auxillary3}\ }Under Assumptions \ref{A1} and \ref{C_A1},
	we have%
	\begin{equation*}
		\sum_{g\in \mathcal{G}_{j}}\sum_{i\in I_{j,g}}\left( \frac{\widehat{\mathbb{E%
			}}_{T}[\psi _{j,\gamma }^{2}(z_{i,t})]}{n_{j,g}\hat{\Psi}_{j,\gamma \gamma }(%
			\hat{\phi}_{j,g})}-\frac{\widehat{\mathbb{E}}_{T}[\psi _{j,\gamma
			}^{2}(z_{i,t})]}{n_{j,g}\Psi _{j,\gamma \gamma ,g}}\right)
		=O_{p}((nT^{-1})^{1/2}+G_{j}T^{-1}).
	\end{equation*}
\end{lemma}

\noindent {\textsc{Proof of Lemma \ref{C_L_Auxillary3}}}. Let $%
C_{2,j,g}(z_{i,t})\equiv \psi _{j,\gamma }^{2}(z_{i,t})$ and $%
C_{2,j,g}^{\ast }(z_{i,t})\equiv C_{2,j,g}(z_{i,t})-\mathbb{E}%
[C_{2,j,g}(z_{i,t})]$. Using these notations, the following expression is
obtained:%
\begin{align}
	& T^{-1}\sum_{g\in \mathcal{G}_{j}}\sum_{i\in I_{j,g}}\sum_{t\leq T}\left( 
	\frac{\psi _{j,\gamma }^{2}(z_{i,t})}{n_{j,g}\hat{\Psi}_{j,\gamma \gamma }(%
		\hat{\phi}_{j,g})}-\frac{\psi _{j,\gamma }^{2}(z_{i,t})}{n_{j,g}\Psi
		_{j,\gamma \gamma ,g}}\right)  \notag \\
	& =\sum_{g\in \mathcal{G}_{j}}\left( \frac{1}{\hat{\Psi}_{j,\gamma \gamma }(%
		\hat{\phi}_{j,g})}-\frac{1}{\Psi _{j,\gamma \gamma ,g}}\right)
	(n_{j,g}T)^{-1}\sum_{i\in I_{j,g}}\sum_{t\leq T}\mathbb{E}%
	[C_{2,j,g}(z_{i,t})]  \notag \\
	& +\sum_{g\in \mathcal{G}_{j}}\left( \frac{1}{\hat{\Psi}_{j,\gamma \gamma }(%
		\hat{\phi}_{j,g})}-\frac{1}{\Psi _{j,\gamma \gamma ,g}}\right)
	(n_{j,g}T)^{-1}\sum_{i\in I_{j,g}}\sum_{t\leq T}C_{2,j,g}^{\ast }(z_{i,t}).
	\label{P_C_L_Auxillary3_1}
\end{align}%
Applying the triangle inequality and the Cauchy-Schwarz inequality, we can
bound the second term on the right-hand side of (\ref{P_C_L_Auxillary3_1})
as follows:%
\begin{align}
	& \left\vert \sum_{g\in \mathcal{G}_{j}}\left( \frac{1}{\hat{\Psi}_{j,\gamma
			\gamma }(\hat{\phi}_{j,g})}-\frac{1}{\Psi _{j,\gamma \gamma ,g}}\right)
	(n_{j,g}T)^{-1}\sum_{i\in I_{j,g}}\sum_{t\leq T}C_{2,j,g}^{\ast
	}(z_{i,t})\right\vert  \notag \\
	& \leq \frac{\sum_{g\in \mathcal{G}_{j}}\left\vert (\hat{\Psi}_{j,\gamma
			\gamma }(\hat{\phi}_{j,g})-\Psi _{j,\gamma \gamma
			,g})(n_{j,g}T)^{-1}\sum_{i\in I_{j,g}}\sum_{t\leq T}C_{2,j,g}^{\ast
		}(z_{i,t})\right\vert }{\min_{g\in \mathcal{G}_{j}}(|\hat{\Psi}_{j,\gamma
			\gamma }(\hat{\phi}_{j,g})\Psi _{j,\gamma \gamma ,g}|)}  \notag \\
	& \leq \frac{\left( \sum_{g\in \mathcal{G}_{j}}(\hat{\Psi}_{j,\gamma \gamma
		}(\hat{\phi}_{j,g})-\Psi _{j,\gamma \gamma ,g})^{2}\right) ^{1/2}\left(
		\sum_{g\in \mathcal{G}_{j}}((n_{j,g}T)^{-1}\sum_{i\in I_{j,g}}\sum_{t\leq
			T}C_{2,j,g}^{\ast }(z_{i,t}))^{2}\right) ^{1/2}}{T\min_{g\in \mathcal{G}%
			_{j}}(|\hat{\Psi}_{j,\gamma \gamma }(\hat{\phi}_{j,g})\Psi _{j,\gamma \gamma
			,g}|)}.  \label{P_C_L_Auxillary3_2}
\end{align}%
Using similar arguments to those leading up to (\ref{P_C_L_Auxillary2_6}),
we have%
\begin{equation*}
	\sum_{g\in \mathcal{G}_{j}}\left( (n_{j,g}T)^{-1}\sum_{i\in
		I_{j,g}}\sum_{t\leq T}C_{2,j,g}^{\ast }(z_{i,t})\right)
	^{2}=O_{p}(G_{j}T^{-1}),
\end{equation*}%
which together with Assumption \ref{A4}, Lemma \ref{C_L_Auxillary1}(ii, v)
and (\ref{P_C_L_Auxillary3_2}) implies that 
\begin{equation}
	\sum_{g\in \mathcal{G}_{j}}\left( \frac{1}{\hat{\Psi}_{j,\gamma \gamma }(%
		\hat{\phi}_{j,g})}-\frac{1}{\Psi _{j,\gamma \gamma ,g}}\right)
	(n_{j,g}T)^{-1}\sum_{i\in I_{j,g}}\sum_{t\leq T}C_{2,j,g}^{\ast
	}(z_{i,t})=O_{p}(G_{j}T^{-1}).  \label{P_C_L_Auxillary3_3}
\end{equation}

Next, observe that the first term on the right-hand side of (\ref%
{P_C_L_Auxillary3_1}) can be rewritten as%
\begin{align}
	& \sum_{g\in\mathcal{G}_{j}}\left( \frac{1}{\hat{\Psi}_{j,\gamma\gamma}(\hat{%
			\phi}_{j,g})}-\frac{1}{\Psi_{j,\gamma\gamma,g}}\right)
	(n_{j,g}T)^{-1}\sum_{i\in I_{j,g}}\sum_{t\leq T}\mathbb{E}%
	[C_{2,j,g}(z_{i,t})]  \notag \\
	& =\sum_{g\in\mathcal{G}_{j}}\frac{(\hat{\Psi}_{j,\gamma\gamma}(\hat{\phi }%
		_{j,g})-\Psi_{j,\gamma\gamma,g})^{2}}{\hat{\Psi}_{j,\gamma\gamma}(\hat{\phi }%
		_{j,g})\Psi_{j,\gamma\gamma,g}^{2}}(n_{j,g}T)^{-1}\sum_{i\in
		I_{j,g}}\sum_{t\leq T}\mathbb{E}[C_{2,j,g}(z_{i,t})]  \notag \\
	& -\sum_{g\in\mathcal{G}_{j}}\frac{(\hat{\Psi}_{j,\gamma\gamma}(\hat{\phi }%
		_{j,g})-\Psi_{j,\gamma\gamma,g})}{\Psi_{j,\gamma\gamma,g}^{2}}%
	(n_{j,g}T)^{-1}\sum_{i\in I_{j,g}}\sum_{t\leq T}\mathbb{E}%
	[C_{2,j,g}(z_{i,t})].  \label{P_C_L_Auxillary3_4a}
\end{align}
Under Assumption \ref{A4}, the first term on the RHS of (\ref%
{P_C_L_Auxillary3_4a}) can be bounded as follows: 
\begin{equation*}
	\left\vert \sum_{g\in\mathcal{G}_{j}}\frac{(\hat{\Psi}_{j,\gamma\gamma}(\hat{%
			\phi}_{j,g})-\Psi_{j,\gamma\gamma,g})^{2}}{\Psi_{j,\gamma\gamma,g}\hat{\Psi}%
		_{j,\gamma\gamma}^{2}(\hat{\phi}_{j,g})}(n_{j,g}T)^{-1}\sum_{i\in
		I_{j,g}}\sum_{t\leq T}\mathbb{E}[C_{2,j,g}(z_{i,t})]\right\vert \leq \frac{%
		K\sum_{g\in\mathcal{G}_{j}}(\hat{\Psi}_{j,\gamma\gamma}(\hat{\phi}%
		_{j,g})-\Psi_{j,\gamma\gamma,g})^{2}}{\min_{g\in\mathcal{G}%
			_{j}}(|\Psi_{j,\gamma\gamma,g}|\hat{\Psi}_{j,\gamma\gamma}^{2}(\hat{\phi}%
		_{j,g}))}.
\end{equation*}
Thus, by Lemma \ref{C_L_Auxillary1}(ii, v), it follows that 
\begin{equation}
	\sum_{g\in\mathcal{G}_{j}}\frac{(\hat{\Psi}_{j,\gamma\gamma}(\hat{\phi}%
		_{j,g})-\Psi_{j,\gamma\gamma,g})^{2}}{\Psi_{j,\gamma\gamma,g}\hat{\Psi }%
		_{j,\gamma\gamma}^{2}(\hat{\phi}_{j,g})}(n_{j,g}T)^{-1}\sum_{i\in
		I_{j,g}}\sum_{t\leq T}\mathbb{E}[C_{2,j,g}(z_{i,t})]=O_{p}(G_{j}T^{-1}).
	\label{P_C_L_Auxillary3_4}
\end{equation}
Combining this result with (\ref{P_C_L_Auxillary3_1}), (\ref%
{P_C_L_Auxillary3_3}) and (\ref{P_C_L_Auxillary3_4a}), we obtain:%
\begin{align}
	& T^{-1}\sum_{g\in\mathcal{G}_{j}}\sum_{i\in I_{j,g}}\sum_{t\leq T}\left( 
	\frac{\psi_{j,\gamma}^{2}(z_{i,t};\phi_{j,g}^{\ast})}{n_{j,g}\hat{\Psi }%
		_{j,\gamma\gamma}(\hat{\phi}_{j,g})}-\frac{\psi_{j,\gamma}^{2}(z_{i,t};%
		\phi_{j,g}^{\ast})}{n_{j,g}\Psi_{j,\gamma\gamma,g}}\right)  \notag \\
	& =\sum_{g\in\mathcal{G}_{j}}\frac{\sum_{i\in I_{j,g}}\sum_{t\leq T}\mathbb{E%
		}[C_{2,j,g}(z_{i,t})]}{(n_{j,g}T)\Psi_{j,\gamma\gamma,g}^{2}}%
	(\Psi_{j,\gamma\gamma,g}-\hat{\Psi}_{j,\gamma\gamma}(\hat{\phi}%
	_{j,g}))+O_{p}(G_{j}T^{-1}).  \label{P_C_L_Auxillary3_5}
\end{align}

We now take care of the first term on the RHS of (\ref{P_C_L_Auxillary3_5}).
Under Assumptions\ \ref{A1}, \ref{A3} and \ref{A4}, we have the following:%
\begin{align*}
	& \mathbb{E}\left[ \left( \sum_{g\in\mathcal{G}_{j}}\frac{\sum_{i\in
			I_{j,g}}\sum_{t\leq T}\mathbb{E}[C_{2,j,g}(z_{i,t})]}{(n_{j,g}T)\Psi
		_{j,\gamma\gamma,g}^{2}}(\Psi_{j,\gamma\gamma,g}-\hat{\Psi}_{j,\gamma\gamma
	}(\phi_{j,g}^{\ast}))\right) ^{2}\right] \\
	& =\sum_{g\in\mathcal{G}_{j}}\left( \frac{\sum_{i\in I_{j,g}}\sum_{t\leq T}%
		\mathbb{E}[C_{2,j,g}(z_{i,t})]}{(n_{j,g}T)\Psi_{j,\gamma\gamma,g}^{2}}%
	\right) ^{2}\mathbb{E}\left[ (\Psi_{j,\gamma\gamma,g}-\hat{\Psi}%
	_{j,\gamma\gamma}(\phi_{j,g}^{\ast}))^{2}\right] \\
	& \leq K\sum_{g\in\mathcal{G}_{j}}\mathbb{E}\left[ (\Psi_{j,\gamma\gamma ,g}-%
	\hat{\Psi}_{j,\gamma\gamma}(\phi_{j,g}^{\ast}))^{2}\right] \leq KG_{j}T^{-1}.
\end{align*}
Thus, by Markov's inequality, it follows that:%
\begin{equation}
	\sum_{g\in\mathcal{G}_{j}}\frac{\sum_{i\in I_{j,g}}\sum_{t\leq T}\mathbb{E}%
		[C_{2,j,g}(z_{i,t})]}{(n_{j,g}T)\Psi_{j,\gamma\gamma,g}^{2}}%
	(\Psi_{j,\gamma\gamma,g}-\hat{\Psi}_{j,\gamma\gamma,g}(\phi_{j,g}^{\ast
	}))=O_{p}((nT^{-1})^{1/2}).  \label{P_C_L_Auxillary3_6}
\end{equation}
Applying a Taylor expansion, we obtain:%
\begin{align}
	& \sum_{g\in\mathcal{G}_{j}}\frac{\sum_{i\in I_{j,g}}\sum_{t\leq T}\mathbb{E}%
		[C_{2,j,g}(z_{i,t})]}{(n_{j,g}T)\Psi_{j,\gamma\gamma,g}^{2}}(\hat{\Psi}%
	_{j,\gamma\gamma,g}(\hat{\phi}_{j,g})-\hat{\Psi}_{j,\gamma\gamma
		,g}(\phi_{j,g}^{\ast}))  \notag \\
	& =\sum_{g\in\mathcal{G}_{j}}\frac{\sum_{i\in I_{j,g}}\sum_{t\leq T}\mathbb{E%
		}[C_{2,j,g}(z_{i,t})]}{(n_{j,g}T)\Psi_{j,\gamma\gamma,g}^{2}}\hat{\Psi}%
	_{j,\gamma\gamma\phi,g}(\tilde{\phi}_{j,g})(\hat{\phi}_{j,g}-\phi_{j,g}^{%
		\ast}),  \label{P_C_L_Auxillary3_7}
\end{align}
where $\tilde{\phi}_{j,g}$ lies between $\hat{\phi}_{j,g}$ and $\phi
_{j,g}^{\ast}$, and%
\begin{equation*}
	\hat{\Psi}_{j,\gamma\gamma\phi}(\phi_{j,g})\equiv(n_{j,g}T)^{-1}\sum_{i\in
		I_{j,g}}\sum_{t\leq T}\psi_{j,\gamma\gamma\phi}(z_{i,t};\phi_{j,g})\text{ \
		and \ }\psi_{j,\gamma\gamma\phi}(z_{i,t};\phi_{j,g})\equiv\frac{\partial
		\psi_{j,\gamma\gamma}(z_{i,t};\phi_{j,g})}{\partial\phi_{j,g}}.
\end{equation*}
Using Assumptions \ref{A3} and \ref{A4}, along with the triangle and
Cauchy-Schwarz inequalities, we have 
\begin{align}
	& \left\vert \sum_{g\in\mathcal{G}_{j}}\frac{\sum_{i\in I_{j,g}}\sum_{t\leq
			T}\mathbb{E}[C_{2,j,g}(z_{i,t})]}{(n_{j,g}T)\Psi_{j,\gamma\gamma,g}^{2}}(%
	\hat{\Psi}_{j,\gamma\gamma\phi}(\tilde{\phi}_{j,g})-\hat{\Psi}_{j,\gamma
		\gamma\phi}(\phi_{j,g}^{\ast}))(\hat{\phi}_{j,g}-\phi_{j,g}^{\ast})\right%
	\vert  \notag \\
	& \leq\sum_{g\in\mathcal{G}_{j}}\frac{\sum_{i\in I_{j,g}}\sum_{t\leq T}%
		\mathbb{E}[C_{2,j,g}(z_{i,t})]}{(n_{j,g}T)\Psi_{j,\gamma\gamma,g}^{2}}%
	\left\Vert \hat{\Psi}_{j,\gamma\gamma\phi}(\tilde{\phi}_{j,g})-\hat{\Psi }%
	_{j,\gamma\gamma\phi}(\phi_{j,g}^{\ast})\right\Vert ||\hat{\phi}%
	_{j,g}-\phi_{j,g}^{\ast}||  \notag \\
	& \leq\sum_{g\in\mathcal{G}_{j}}\frac{\sum_{i\in I_{j,g}}\sum_{t\leq T}%
		\mathbb{E}[C_{2,j,g}(z_{i,t})]}{(n_{j,g}T)\Psi_{j,\gamma\gamma,g}^{2}}\frac{%
		\sum_{i\in I_{j,g}}\sum_{t\leq T}M(z_{i,t})}{(n_{j,g}T)}||\hat{\phi }%
	_{j,g}-\phi_{j,g}^{\ast}||^{2}  \notag \\
	& \leq K\left( \sum_{g\in\mathcal{G}_{j}}(n_{j,g}T)^{-1}\sum_{i\in
		I_{j,g}}\sum_{t\leq T}M(z_{i,t})^{2}\right) ^{1/2}\left( \sum_{g\in\mathcal{G%
		}_{j}}||\hat{\phi}_{j,g}-\phi_{j,g}^{\ast}||^{4}\right)
	^{1/2}=O_{p}(G_{j}T^{-1}),  \label{P_C_L_Auxillary3_8}
\end{align}
where the last equality uses (\ref{P_C_L_Auxillary1_7}) and Markov's
inequality.

Let $\Psi_{j,\gamma\gamma\phi,g}\equiv\mathbb{E}[\hat{\Psi}_{j,\gamma
	\gamma\phi}(\phi_{j,g}^{\ast})]$. By Assumptions \ref{A3} and \ref{A4},
along with the triangle inequality and the Cauchy-Schwarz inequality: 
\begin{align}
	& \left\vert \sum_{g\in\mathcal{G}_{j}}\frac{\sum_{i\in I_{j,g}}\sum_{t\leq
			T}\mathbb{E}[C_{2,j,g}(z_{i,t})]}{(n_{j,g}T)\Psi_{j,\gamma\gamma,g}^{2}}(%
	\hat{\Psi}_{j,\gamma\gamma\phi}(\phi_{j,g}^{\ast})-\Psi_{j,\gamma\gamma
		\phi,g})(\hat{\phi}_{j,g}-\phi_{j,g}^{\ast})\right\vert  \notag \\
	& \leq K\sum_{g\in\mathcal{G}_{j}}\left\vert (\hat{\Psi}_{j,\gamma\gamma\phi
	}(\phi_{j,g}^{\ast})-\Psi_{j,\gamma\gamma\phi,g})(\hat{\phi}_{j,g}-\phi
	_{j,g}^{\ast})\right\vert  \notag \\
	& \leq K\left( \sum_{g\in\mathcal{G}_{j}}\left\Vert \hat{\Psi}%
	_{j,\gamma\gamma\phi}(\phi_{j,g}^{\ast})-\Psi_{j,\gamma\gamma\phi
		,g}\right\Vert ^{2}\right) ^{1/2}\left( \sum_{g\in\mathcal{G}_{j}}||\hat{\phi%
	}_{j,g}-\phi_{j,g}^{\ast}||^{2}\right) ^{1/2}.  \label{P_C_L_Auxillary3_9}
\end{align}
Under Assumptions \ref{A1} and \ref{A3}, the covariance inequality for
mixing processes implies: 
\begin{equation*}
	\mathbb{E}\left[ \sum_{g\in\mathcal{G}_{j}}\left\Vert \hat{\Psi}%
	_{j,\gamma\gamma\phi}(\phi_{j,g}^{\ast})-\Psi_{j,\gamma\gamma\phi
		,g}\right\Vert ^{2}\right] \leq KG_{j}T^{-1},
\end{equation*}
which, combined with Markov's inequality, (\ref{P_C_L_Auxillary1_13}) and (%
\ref{P_C_L_Auxillary3_9}), establishes 
\begin{equation}
	\sum_{g\in\mathcal{G}_{j}}\frac{\sum_{i\in I_{j,g}}\sum_{t\leq T}\mathbb{E}%
		[C_{2,j,g}(z_{i,t})]}{(n_{j,g}T)\Psi_{j,\gamma\gamma,g}^{2}}(\hat{\Psi}%
	_{j,\gamma\gamma\phi}(\phi_{j,g}^{\ast})-\Psi_{j,\gamma\gamma \phi,g})(\hat{%
		\phi}_{j,g}-\phi_{j,g}^{\ast})=O_{p}(G_{j}T^{-1}).
	\label{P_C_L_Auxillary3_10}
\end{equation}
Combining the results in (\ref{P_C_L_Auxillary3_7}), (\ref%
{P_C_L_Auxillary3_8}) and (\ref{P_C_L_Auxillary3_10}), we have 
\begin{align}
	& \sum_{g\in\mathcal{G}_{j}}\frac{\sum_{i\in I_{j,g}}\sum_{t\leq T}\mathbb{E}%
		[C_{2,j,g}(z_{i,t})]}{(n_{j,g}T)\Psi_{j,\gamma\gamma,g}^{2}}(\hat{\Psi}%
	_{j,\gamma\gamma}(\hat{\phi}_{j,g})-\hat{\Psi}_{j,\gamma\gamma
	}(\phi_{j,g}^{\ast}))  \notag \\
	& =\sum_{g\in\mathcal{G}_{j}}\frac{\sum_{i\in I_{j,g}}\sum_{t\leq T}\mathbb{E%
		}[C_{2,j,g}(z_{i,t})]}{(n_{j,g}T)\Psi_{j,\gamma\gamma,g}^{2}}%
	\Psi_{j,\gamma\gamma\phi,g}(\hat{\phi}_{j,g}-\phi_{j,g}^{%
		\ast})+O_{p}(G_{j}T^{-1}).  \label{P_C_L_Auxillary3_10b}
\end{align}

Using the triangle inequality, we can write:%
\begin{align}
	& \left\vert \sum_{g\in\mathcal{G}_{j}}\frac{\sum_{i\in I_{j,g}}\sum_{t\leq
			T}\mathbb{E}[C_{2,j,g}(z_{i,t})]}{(n_{j,g}T)\Psi_{j,\gamma\gamma,g}^{2}}%
	\Psi_{j,\gamma\gamma\phi,g}(\hat{\phi}_{j,g}-\phi_{j,g}^{\ast})\right\vert 
	\notag \\
	& \leq\left\vert \sum_{g\in\mathcal{G}_{j}}\frac{\sum_{i\in
			I_{j,g}}\sum_{t\leq T}\mathbb{E}[C_{2,j,g}(z_{i,t})]}{(n_{j,g}T)\Psi_{j,%
			\gamma \gamma,g}^{2}}\Psi_{j,\gamma\gamma\theta,g}(\hat{\theta}_{j,g}-\theta
	_{j,g}^{\ast})\right\vert  \notag \\
	& +\left\vert \sum_{g\in\mathcal{G}_{j}}\frac{\sum_{i\in I_{j,g}}\sum_{t\leq
			T}\mathbb{E}[C_{2,j,g}(z_{i,t})]}{(n_{j,g}T)\Psi_{j,\gamma\gamma,g}^{2}}%
	\Psi_{j,\gamma\gamma\gamma,g}(\hat{\gamma}_{j,g}-\gamma_{j,g}^{\ast}+%
	\Psi_{j,\gamma\gamma,g}^{-1}\hat{\Psi}_{j,\gamma,g})\right\vert  \notag \\
	& +\left\vert \sum_{g\in\mathcal{G}_{j}}\frac{\sum_{i\in I_{j,g}}\sum_{t\leq
			T}\mathbb{E}[C_{2,j,g}(z_{i,t})]}{(n_{j,g}T)\Psi_{j,\gamma\gamma,g}^{2}}%
	\Psi_{j,\gamma\gamma\gamma,g}\Psi_{j,\gamma\gamma,g}^{-1}\hat{\Psi}%
	_{j,\gamma,g}\right\vert ,  \label{P_C_L_Auxillary3_11}
\end{align}
where $\psi_{j,\gamma\gamma a}(z_{i,t};\phi_{j,g})\equiv\partial\psi
_{j,\gamma\gamma}(z_{i,t};\phi_{j,g})/\partial a$ for $a\in\{\theta,\gamma\}$%
. By Assumptions \ref{A3} and \ref{A4}, and Lemma \ref{C_L_Auxillary1}(iii),%
\begin{align}
	& \left\vert \sum_{g\in\mathcal{G}_{j}}\frac{\sum_{i\in I_{j,g}}\sum_{t\leq
			T}\mathbb{E}[C_{2,j,g}(z_{i,t})]}{(n_{j,g}T)\Psi_{j,\gamma\gamma,g}^{2}}%
	\Psi_{j,\gamma\gamma\theta,g}(\hat{\theta}_{j,g}-\theta_{j,g}^{\ast
	})\right\vert  \notag \\
	& \leq\frac{K\sum_{g\in\mathcal{G}_{j}}(n_{j,g}T)^{-1}\sum_{i\in
			I_{j,g}}\sum_{t\leq T}\mathbb{E}[C_{2,j,g}(z_{i,t})]}{\min_{g\in\mathcal{G}%
			_{j}}\Psi_{j,\gamma\gamma,g}^{2}}||\hat{\theta}_{j,g}-\theta_{j,g}^{\ast }||
	\notag \\
	& \leq Kn||\hat{\theta}_{j,g}-\theta_{j,g}^{\ast}||=O_{p}((nT^{-1})^{1/2}).
	\label{P_C_L_Auxillary3_12}
\end{align}
By Assumptions \ref{A3} and \ref{A4}, and Lemma \ref{C_L_Auxillary1}(vi),%
\begin{equation}
	\left\vert \sum_{g\in\mathcal{G}_{j}}\frac{\sum_{i\in I_{j,g}}\sum_{t\leq T}%
		\mathbb{E}[C_{2,j,g}(z_{i,t})]}{(n_{j,g}T)\Psi_{j,\gamma\gamma,g}^{2}}%
	\Psi_{j,\gamma\gamma\gamma,g}(\hat{\gamma}_{j,g}-\gamma_{j,g}^{\ast}+%
	\Psi_{j,\gamma\gamma,g}^{-1}\hat{\Psi}_{j,\gamma,g})\right\vert
	=O_{p}((nT^{-1})^{1/2}).  \label{P_C_L_Auxillary3_13}
\end{equation}
Moreover, under Assumptions \ref{A1}, \ref{A3} and \ref{A4},\ and Lemma \ref%
{C_L_Auxillary1}(viii), it follows that 
\begin{align*}
	& \mathbb{E}\left[ \left( \sum_{g\in\mathcal{G}_{j}}\frac{\sum_{i\in
			I_{j,g}}\sum_{t\leq T}\mathbb{E}[C_{2,j,g}(z_{i,t})]}{(n_{j,g}T)\Psi
		_{j,\gamma\gamma,g}^{2}}\Psi_{j,\gamma\gamma\gamma,g}\Psi_{j,\gamma\gamma
		,g}^{-1}\hat{\Psi}_{j,\gamma,g}\right) ^{2}\right] \\
	& =\sum_{g\in\mathcal{G}_{j}}\left( \frac{\sum_{i\in I_{j,g}}\sum_{t\leq T}%
		\mathbb{E}[C_{2,j,g}(z_{i,t})]}{(n_{j,g}T)\Psi_{j,\gamma\gamma,g}^{2}}%
	\Psi_{j,\gamma\gamma\gamma,g}\Psi_{j,\gamma\gamma,g}^{-1}\right) ^{2}\mathbb{%
		E}[\hat{\Psi}_{j,\gamma,g}^{2}]\leq K\sum_{g\in\mathcal{G}_{j}}\mathbb{E}[%
	\hat{\Psi}_{j,\gamma,g}^{2}]\leq G_{j}T^{-1},
\end{align*}
which together with Markov's inequality implies that 
\begin{equation}
	\sum_{g\in\mathcal{G}_{j}}\frac{\sum_{i\in I_{j,g}}\sum_{t\leq T}\mathbb{E}%
		[C_{2,j,g}(z_{i,t})]}{(n_{j,g}T)\Psi_{j,\gamma\gamma,g}^{2}}%
	\Psi_{j,\gamma\gamma\gamma,g}\Psi_{j,\gamma\gamma,g}^{-1}\hat{\Psi}%
	_{j,\gamma,g}=O_{p}((nT^{-1})^{1/2}).  \label{P_C_L_Auxillary3_14}
\end{equation}
From (\ref{P_C_L_Auxillary3_11})-(\ref{P_C_L_Auxillary3_14}), it follows
that 
\begin{equation}
	\sum_{g\in\mathcal{G}_{j}}\frac{\sum_{i\in I_{j,g}}\sum_{t\leq T}\mathbb{E}%
		[C_{2,j,g}(z_{i,t})]}{(n_{j,g}T)\Psi_{j,\gamma\gamma,g}^{2}}%
	\Psi_{j,\gamma\gamma\gamma,g}(\hat{\phi}_{j,g}-\phi_{j,g}^{%
		\ast})=O_{p}((nT^{-1})^{1/2}).  \label{P_C_L_Auxillary3_15}
\end{equation}
Finally, combining (\ref{P_C_L_Auxillary3_10b}) and (\ref%
{P_C_L_Auxillary3_15}), we obtain 
\begin{equation}
	\sum_{g\in\mathcal{G}_{j}}\frac{\sum_{i\in I_{j,g}}\sum_{t\leq T}\mathbb{E}%
		[C_{2,j,g}(z_{i,t})]}{(n_{j,g}T)\Psi_{j,\gamma\gamma,g}^{2}}(\hat{\Psi}%
	_{j,\gamma\gamma}(\hat{\phi}_{j,g})-\hat{\Psi}_{j,\gamma\gamma
	}(\phi_{j,g}^{\ast}))=O_{p}((nT^{-1})^{1/2}+G_{j}T^{-1}),
	\label{P_C_L_Auxillary3_16}
\end{equation}
which, along with (\ref{P_C_L_Auxillary3_5}) and (\ref{P_C_L_Auxillary3_6}),
establishes the lemma's claim.\hfill$Q.E.D.$

\bigskip

\begin{lemma}
	\textit{\label{C_L_Auxillary3b} }Under Assumptions \ref{A1} and \ref{C_A1},
	we have%
	\begin{equation}
		\sum_{g\in \mathcal{G}_{j}}\sum_{i\in I_{j,g}}\left( \frac{(\widehat{\mathbb{%
					E}}_{T}[\psi _{j,\gamma }(z_{i,t})])^{2}}{n_{j,g}\hat{\Psi}_{j,\gamma \gamma
			}(\hat{\phi}_{j,g})\Psi _{j,\gamma \gamma ,g}}(\hat{\Psi}_{j,\gamma \gamma }(%
		\hat{\phi}_{j,g})-\Psi _{j,\gamma \gamma ,g})\right)
		=O_{p}((G_{j}T^{-1})^{1/2}).  \label{C_L_Auxillary3b_1}
	\end{equation}
\end{lemma}

\noindent{\textsc{Proof of Lemma \ref{C_L_Auxillary3b}}}. The term on the
LHS of (\ref{C_L_Auxillary3b_1}) can be written as 
\begin{align}
	& \sum_{g\in\mathcal{G}_{j}}\sum_{i\in I_{j,g}}\left( \frac{%
		(T^{-1}\sum_{t\leq T}\psi_{j,\gamma}(z_{i,t}))^{2}}{n_{j,g}\hat{\Psi}%
		_{j,\gamma \gamma}(\hat{\phi}_{j,g})\Psi_{j,\gamma\gamma,g}}(\hat{\Psi}%
	_{j,\gamma\gamma }(\hat{\phi}_{j,g})-\Psi_{j,\gamma\gamma,g})\right)  \notag
	\\
	& =\sum_{g\in\mathcal{G}_{j}}\sum_{i\in I_{j,g}}\left( \frac{(\mathbb{E}%
		[\psi_{j,\gamma}(z_{i,t})])^{2}}{n_{j,g}\hat{\Psi}_{j,\gamma\gamma}(\hat{%
			\phi }_{j,g})\Psi_{j,\gamma\gamma,g}}(\hat{\Psi}_{j,\gamma\gamma}(\hat{\phi}%
	_{j,g})-\Psi_{j,\gamma\gamma,g})\right)  \notag \\
	& +\sum_{g\in\mathcal{G}_{j}}\sum_{i\in I_{j,g}}\left( \frac{%
		(T^{-1}\sum_{t\leq T}\psi_{j,\gamma}(z_{i,t}))^{2}-(\mathbb{E}%
		[\psi_{j,\gamma }(z_{i,t})])^{2}}{n_{j,g}\hat{\Psi}_{j,\gamma\gamma}(\hat{%
			\phi}_{j,g})\Psi_{j,\gamma\gamma,g}}(\hat{\Psi}_{j,\gamma\gamma}(\hat{\phi}%
	_{j,g})-\Psi_{j,\gamma\gamma,g})\right) .  \label{P_C_L_Auxillary3b_1}
\end{align}

The first term on the RHS of (\ref{P_C_L_Auxillary3b_1}) can be expressed as 
\begin{align}
	& \sum_{g\in\mathcal{G}_{j}}\sum_{i\in I_{j,g}}\left( \frac{(\mathbb{E}%
		[\psi_{j,\gamma}(z_{i,t})])^{2}}{n_{j,g}\hat{\Psi}_{j,\gamma\gamma}(\hat{%
			\phi }_{j,g})\Psi_{j,\gamma\gamma,g}}(\hat{\Psi}_{j,\gamma\gamma}(\hat{\phi}%
	_{j,g})-\Psi_{j,\gamma\gamma,g})\right)  \notag \\
	& =\sum_{g\in\mathcal{G}_{j}}\sum_{i\in I_{j,g}}\left( \frac{(\mathbb{E}%
		[\psi_{j,\gamma}(z_{i,t})])^{2}}{n_{j,g}\Psi_{j,\gamma\gamma,g}^{2}}(\hat {%
		\Psi}_{j,\gamma\gamma}(\hat{\phi}_{j,g})-\Psi_{j,\gamma\gamma,g})\right) 
	\notag \\
	& -\sum_{g\in\mathcal{G}_{j}}\sum_{i\in I_{j,g}}\frac{(\mathbb{E}%
		[\psi_{j,\gamma}(z_{i,t})])^{2}(\hat{\Psi}_{j,\gamma\gamma}(\hat{\phi}%
		_{j,g})-\Psi_{j,\gamma\gamma,g})^{2}}{n_{j,g}\hat{\Psi}_{j,\gamma\gamma}(%
		\hat{\phi}_{j,g})\Psi_{j,\gamma\gamma,g}^{2}}.  \label{P_C_L_Auxillary3b_2}
\end{align}
By Assumption \ref{A4}, Lemma \ref{C_L_Auxillary1}(ii, v), we have 
\begin{align}
	& \left\vert \sum_{g\in\mathcal{G}_{j}}\sum_{i\in I_{j,g}}\frac {(\mathbb{E}%
		[\psi_{j,\gamma}(z_{i,t})])^{2}(\hat{\Psi}_{j,\gamma\gamma}(\hat{\phi}%
		_{j,g})-\Psi_{j,\gamma\gamma,g})^{2}}{n_{j,g}\hat{\Psi}_{j,\gamma\gamma}(%
		\hat{\phi}_{j,g})\Psi_{j,\gamma\gamma,g}^{2}}\right\vert  \notag \\
	& \leq\frac{K}{\min_{g\in\mathcal{G}_{j}}|\hat{\Psi}_{j,\gamma\gamma}(\hat{%
			\phi}_{j,g})\Psi_{j,\gamma\gamma,g}^{2}|}\sum_{g\in\mathcal{G}_{j}}(\hat{\Psi%
	}_{j,\gamma\gamma}(\hat{\phi}_{j,g})-\Psi_{j,\gamma%
		\gamma,g})^{2}=O_{p}(G_{j}T^{-1}).  \label{P_C_L_Auxillary3b_3}
\end{align}
By the similar arguments for deriving (\ref{P_C_L_Auxillary3_16}), we can
also show that 
\begin{equation*}
	\sum_{g\in\mathcal{G}_{j}}\sum_{i\in I_{j,g}}\left( \frac{(\mathbb{E}%
		[\psi_{j,\gamma}(z_{i,t})])^{2}}{n_{j,g}\Psi_{j,\gamma\gamma,g}^{2}}(\hat {%
		\Psi}_{j,\gamma\gamma}(\hat{\phi}_{j,g})-\Psi_{j,\gamma\gamma,g})\right)
	=O_{p}((G_{j}T^{-1})^{1/2}),
\end{equation*}
which along with (\ref{P_C_L_Auxillary3b_2}) and (\ref{P_C_L_Auxillary3b_3})
implies that 
\begin{equation}
	\sum_{g\in\mathcal{G}_{j}}\sum_{i\in I_{j,g}}\left( \frac{(\mathbb{E}%
		[\psi_{j,\gamma}(z_{i,t})])^{2}}{n_{j,g}\hat{\Psi}_{j,\gamma\gamma}(\hat{%
			\phi }_{j,g})\Psi_{j,\gamma\gamma,g}}(\hat{\Psi}_{j,\gamma\gamma}(\hat{\phi}%
	_{j,g})-\Psi_{j,\gamma\gamma,g})\right) =O_{p}((G_{j}T^{-1})^{1/2}).
	\label{P_C_L_Auxillary3b_4}
\end{equation}

The second term on the RHS of (\ref{P_C_L_Auxillary3b_1}) can be expressed as%
\begin{align}
	& \left\vert \sum_{g\in\mathcal{G}_{j}}\sum_{i\in I_{j,g}}\left( \frac{%
		(T^{-1}\sum_{t\leq T}\psi_{j,\gamma}(z_{i,t}))^{2}-(\mathbb{E}%
		[\psi_{j,\gamma}(z_{i,t})])^{2}}{n_{j,g}\hat{\Psi}_{j,\gamma\gamma}(\hat{%
			\phi }_{j,g})\Psi_{j,\gamma\gamma,g}}(\hat{\Psi}_{j,\gamma\gamma}(\hat{\phi}%
	_{j,g})-\Psi_{j,\gamma\gamma,g})\right) \right\vert  \notag \\
	& \leq\left\vert \sum_{g\in\mathcal{G}_{j}}\sum_{i\in I_{j,g}}\left( \frac{%
		(T^{-1}\sum_{t\leq T}\tilde{\psi}_{j,\gamma}^{\ast}(z_{i,t}))^{2}}{n_{j,g}|%
		\hat{\Psi}_{j,\gamma\gamma}(\hat{\phi}_{j,g})|}(\hat{\Psi}_{j,\gamma\gamma}(%
	\hat{\phi}_{j,g})-\Psi_{j,\gamma\gamma,g})\right) \right\vert  \notag \\
	& +2\left\vert \sum_{g\in\mathcal{G}_{j}}\sum_{i\in I_{j,g}}\left( \frac{%
		\mathbb{E}[\psi_{j,\gamma}(z_{i,t})]T^{-1}\sum_{t\leq T}\tilde{\psi }%
		_{j,\gamma}^{\ast}(z_{i,t})}{n_{j,g}|\hat{\Psi}_{j,\gamma\gamma}(\hat{\phi }%
		_{j,g})||\Psi_{j,\gamma\gamma,g}|^{1/2}}(\hat{\Psi}_{j,\gamma\gamma}(\hat{%
		\phi}_{j,g})-\Psi_{j,\gamma\gamma,g})\right) \right\vert
	\label{P_C_L_Auxillary3b_5}
\end{align}
with probability approaching 1. By the Cauchy-Schwarz inequality,%
\begin{align}
	& \left\vert \sum_{g\in\mathcal{G}_{j}}\sum_{i\in I_{j,g}}\left( \frac{%
		(T^{-1}\sum_{t\leq T}\tilde{\psi}_{j,\gamma}^{\ast}(z_{i,t}))^{2}}{n_{j,g}%
		\hat{\Psi}_{j,\gamma\gamma}(\hat{\phi}_{j,g})}(\hat{\Psi}_{j,\gamma\gamma}(%
	\hat{\phi}_{j,g})-\Psi_{j,\gamma\gamma,g})\right) \right\vert  \notag \\
	& \leq\frac{\left( \sum_{g\in\mathcal{G}_{j}}n_{j,g}^{-1}\sum_{i\in
			I_{j,g}}(T^{-1}\sum_{t\leq T}\tilde{\psi}_{j,\gamma}^{\ast}(z_{i,t}))^{4}%
		\right) ^{1/2}}{\min_{g\in\mathcal{G}_{j}}|\hat{\Psi}_{j,\gamma\gamma}(\hat{%
			\phi }_{j,g})\Psi_{j,\gamma\gamma,g}^{2}|}\left( \sum_{g\in\mathcal{G}_{j}}(%
	\hat{\Psi}_{j,\gamma\gamma}(\hat{\phi}_{j,g})-\Psi_{j,\gamma\gamma,g})^{2}%
	\right) ^{1/2}.  \label{P_C_L_Auxillary3b_6}
\end{align}
Under Assumptions\ \ref{A1}, \ref{C_A1}(iii), \ref{A3} and \ref{A4}, we can
apply Rosenthal's inequality to obtain%
\begin{equation}
	\mathbb{E}\left[ \sum_{g\in\mathcal{G}_{j}}n_{j,g}^{-1}\sum_{i\in
		I_{j,g}}(T^{-1}\sum_{t\leq T}\tilde{\psi}_{j,\gamma}^{\ast}(z_{i,t}))^{4}%
	\right] \leq KG_{j}T^{-1},  \label{P_C_L_Auxillary3b_7}
\end{equation}
which together with Markov's inequality, and Lemma \ref{C_L_Auxillary1}(ii,
v) shows that 
\begin{equation}
	\sum_{g\in\mathcal{G}_{j}}\sum_{i\in I_{j,g}}\left( \frac{(T^{-1}\sum_{t\leq
			T}\tilde{\psi}_{j,\gamma}^{\ast}(z_{i,t}))^{2}}{n_{j,g}\hat{\Psi}%
		_{j,\gamma\gamma}(\hat{\phi}_{j,g})}(\hat{\Psi}_{j,\gamma\gamma}(\hat{\phi }%
	_{j,g})-\Psi_{j,\gamma\gamma,g})\right) =O_{p}(G_{j}T^{-1}).
	\label{P_C_L_Auxillary3b_8}
\end{equation}
Similarly, we can show that 
\begin{align*}
	& \left\vert \sum_{g\in\mathcal{G}_{j}}\sum_{i\in I_{j,g}}\left( \frac{%
		\mathbb{E}[\psi_{j,\gamma}(z_{i,t})]T^{-1}\sum_{t\leq T}\tilde{\psi }%
		_{j,\gamma}^{\ast}(z_{i,t})}{n_{j,g}\hat{\Psi}_{j,\gamma\gamma}(\hat{\phi }%
		_{j,g})(-\Psi_{j,\gamma\gamma,g})^{1/2}}(\hat{\Psi}_{j,\gamma\gamma}(\hat{%
		\phi}_{j,g})-\Psi_{j,\gamma\gamma,g})\right) \right\vert \\
	& \leq\frac{\max_{g\in\mathcal{G}_{j}}|\mathbb{E}[\psi_{j,\gamma}(z_{i,t})]|%
		\left( \sum_{g\in\mathcal{G}_{j}}(\hat{\Psi}_{j,\gamma\gamma}(\hat{\phi}%
		_{j,g})-\Psi_{j,\gamma\gamma,g})^{2}\right) ^{1/2}}{\min _{g\in\mathcal{G}%
			_{j}}|\hat{\Psi}_{j,\gamma\gamma}(\hat{\phi}_{j,g})(-\Psi_{j,\gamma%
			\gamma,g})^{1/2}|} \\
	& \times\left( \sum_{g\in\mathcal{G}_{j}}n_{j,g}^{-1}\sum_{i\in
		I_{j,g}}(T^{-1}\sum_{t\leq T}\tilde{\psi}_{j,\gamma}^{\ast}(z_{i,t}))^{2}%
	\right) ^{1/2}\overset{}{=}O_{p}(G_{j}T^{-1}),
\end{align*}
which along with (\ref{P_C_L_Auxillary3b_5}) and (\ref{P_C_L_Auxillary3b_8})
implies that 
\begin{equation}
	\sum_{g\in\mathcal{G}_{j}}\sum_{i\in I_{j,g}}\left( \frac{(T^{-1}\sum_{t\leq
			T}\psi_{j,\gamma}(z_{i,t}))^{2}-(\mathbb{E}[\psi_{j,\gamma}(z_{i,t})])^{2}}{%
		n_{j,g}\hat{\Psi}_{j,\gamma\gamma}(\hat{\phi}_{j,g})\Psi_{j,\gamma\gamma,g}}(%
	\hat{\Psi}_{j,\gamma\gamma}(\hat{\phi}_{j,g})-\Psi_{j,\gamma\gamma
		,g})\right) =O_{p}(G_{j}T^{-1}).  \label{P_C_L_Auxillary3b_9}
\end{equation}
The claim of the lemma now follows from (\ref{P_C_L_Auxillary3b_1}), (\ref%
{P_C_L_Auxillary3b_4}) and (\ref{P_C_L_Auxillary3b_9}).\hfill$Q.E.D.$

\bigskip

\begin{lemma}
	\textit{\label{C_L_Auxillary3c}\ }Under Assumptions \ref{A1} and \ref{C_A1},
	we have%
	\begin{equation}
		\sum_{g\in \mathcal{G}_{j}}\sum_{i\in I_{j,g}}\frac{\widehat{\mathbb{E}}%
			_{T}[\psi _{j,\gamma }^{2}(z_{i,t})]-\mathbb{E}_{T}[\psi _{j,\gamma
			}^{2}(z_{i,t})]}{n_{j,g}\Psi _{j,\gamma \gamma ,g}}%
		=O_{p}((G_{j}T^{-1})^{1/2})  \label{C_L_Auxillary3c_1}
	\end{equation}%
	and%
	\begin{equation}
		\sum_{g\in \mathcal{G}_{j}}\sum_{i\in I_{j,g}}\frac{(\widehat{\mathbb{E}}%
			_{T}[\psi _{j,\gamma }(z_{i,t})])^{2}-(\mathbb{E}_{T}[\psi _{j,\gamma
			}\left( z_{i,t}\right) ])^{2}}{n_{j,g}\Psi _{j,\gamma \gamma ,g}}%
		=O_{p}((G_{j}T^{-1})^{1/2}).  \label{C_L_Auxillary3c_2}
	\end{equation}
\end{lemma}

\noindent {\textsc{Proof of Lemma \ref{C_L_Auxillary3c}}}. Recall that $\psi
_{j,\gamma }^{\ast }(z_{i,t})\equiv \psi _{j,\gamma }(z_{i,t})-\mathbb{E}%
[\psi _{j,\gamma }(z_{i,t})]$. Using this expression, we have 
\begin{equation*}
	\sum_{t\leq T}\psi _{j,\gamma }(z_{i,t})^{2}=\sum_{t\leq T}\psi _{j,\gamma
	}^{\ast }(z_{i,t})^{2}+2\sum_{t\leq T}\psi _{j,\gamma }^{\ast }(z_{i,t})%
	\mathbb{E}[\psi _{j,\gamma }(z_{i,t})]+\sum_{t\leq T}(\mathbb{E}[\psi
	_{j,\gamma }(z_{i,t})])^{2}
\end{equation*}%
and 
\begin{equation*}
	\sum_{t\leq T}\mathbb{E}[\psi _{j,\gamma }(z_{i,t})^{2}]=\sum_{t\leq T}%
	\mathbb{E}[\psi _{j,\gamma }^{\ast }(z_{i,t})^{2}]+\sum_{t\leq T}(\mathbb{E}%
	[\psi _{j,\gamma }(z_{i,t})])^{2}.
\end{equation*}%
Therefore,%
\begin{align}
	& \sum_{g\in \mathcal{G}_{j}}\sum_{i\in I_{j,g}}\frac{T^{-1}\sum_{t\leq
			T}(\psi _{j,\gamma }(z_{i,t})^{2}-\mathbb{E}[\psi _{j,\gamma }(z_{i,t})^{2}])%
	}{n_{j,g}\Psi _{j,\gamma \gamma ,g}}  \notag \\
	& =\sum_{g\in \mathcal{G}_{j}}\sum_{i\in I_{j,g}}\frac{T^{-1}\sum_{t\leq
			T}(\psi _{j,\gamma }^{\ast }(z_{i,t})^{2}-\mathbb{E}[\psi _{j,\gamma }^{\ast
		}(z_{i,t})^{2}])}{n_{j,g}\Psi _{j,\gamma \gamma ,g}}+2\sum_{g\in \mathcal{G}%
		_{j}}\sum_{i\in I_{j,g}}\frac{T^{-1}\sum_{t\leq T}\psi _{j,\gamma }^{\ast
		}(z_{i,t})\mathbb{E}[\psi _{j,\gamma }(z_{i,t})]}{n_{j,g}\Psi _{j,\gamma
			\gamma ,g}}.  \label{P_L_Auxillary3c_0a}
\end{align}%
By Assumptions \ref{A1}, \ref{C_A1}, \ref{A3}(iii) and \ref{A4}, it follows
that%
\begin{align}
	& \mathrm{Var}\left( \sum_{g\in \mathcal{G}_{j}}\sum_{i\in I_{j,g}}\frac{%
		T^{-1}\sum_{t\leq T}(\psi _{j,\gamma }^{\ast }(z_{i,t})^{2}-\mathbb{E}[\psi
		_{j,\gamma }^{\ast }(z_{i,t})^{2}])}{n_{j,g}\Psi _{j,\gamma \gamma ,g}}%
	\right)  \notag \\
	& =\sum_{g\in \mathcal{G}_{j}}\frac{\mathrm{Var}\left( \sum_{i\in
			I_{j,g}}\sum_{t\leq T}(\psi _{j,\gamma }^{\ast }(z_{i,t})^{2}-\mathbb{E}%
		[\psi _{j,\gamma }^{\ast }(z_{i,t})^{2}])\right) }{T^{2}n_{j,g}^{2}\Psi
		_{j,\gamma \gamma ,g}^{2}}  \notag \\
	& \leq K\sum_{g\in \mathcal{G}_{j}}\frac{\mathbb{E}\left[ \left\vert
		(n_{j,g}T)^{-1/2}\sum_{i\in I_{j,g}}\sum_{t\leq T}(\psi _{j,\gamma }^{\ast
		}(z_{i,t})^{2}-\mathbb{E}[\psi _{j,\gamma }^{\ast
		}(z_{i,t})^{2}])\right\vert ^{2}\right] }{Tn_{j,g}}\leq KG_{j}T^{-1},
	\label{P_L_Auxillary3c_0b}
\end{align}%
where the second inequality follows by (\ref{L0-1}) in Lemma \ref{L0}.
Similarly, 
\begin{align}
	& \mathrm{Var}\left( \sum_{g\in \mathcal{G}_{j}}\sum_{i\in I_{j,g}}\frac{%
		T^{-1}\sum_{t\leq T}\psi _{j,\gamma }^{\ast }(z_{i,t})\mathbb{E}[\psi
		_{j,\gamma }(z_{i,t})]}{n_{j,g}\Psi _{j,\gamma \gamma ,g}}\right)  \notag \\
	& =\sum_{g\in \mathcal{G}_{j}}\frac{\mathrm{Var}\left( \sum_{i\in
			I_{j,g}}\sum_{t\leq T}\psi _{j,\gamma }^{\ast }(z_{i,t})\mathbb{E}[\psi
		_{j,\gamma }(z_{i,t})]\right) }{T^{2}n_{j,g}^{2}\Psi _{j,\gamma \gamma
			,g}^{2}}  \notag \\
	& \leq K\sum_{g\in \mathcal{G}_{j}}\frac{\mathbb{E}\left[ \left\vert
		(n_{j,g}T)^{-1/2}\sum_{i\in I_{j,g}}\sum_{t\leq T}\psi _{j,\gamma }^{\ast
		}(z_{i,t})\mathbb{E}[\psi _{j,\gamma }(z_{i,t})]\right\vert ^{2}\right] }{%
		Tn_{j,g}}\leq KG_{j}T^{-1}  \label{P_L_Auxillary3c_0c}
\end{align}%
which, combined with (\ref{P_L_Auxillary3c_0a}), (\ref{P_L_Auxillary3c_0b})
and Markov's inequality, shows the claim in (\ref{C_L_Auxillary3c_1}).

To show the claim in (\ref{C_L_Auxillary3c_2}), we first express the term on
its LHS as%
\begin{align}
	& \sum_{g\in \mathcal{G}_{j}}\sum_{i\in I_{j,g}}\frac{(T^{-1}\sum_{t\leq
			T}\psi _{j,\gamma }(z_{i,t}))^{2}-T^{-1}\sum_{t\leq T}(\mathbb{E}[\psi
		_{j,\gamma }(z_{i,t})])^{2}}{n_{j,g}\Psi _{j,\gamma \gamma ,g}}  \notag \\
	& =\sum_{g\in \mathcal{G}_{j}}\sum_{i\in I_{j,g}}\frac{(T^{-1}\sum_{t\leq T}%
		\tilde{\psi}_{j,\gamma }^{\ast }(z_{i,t}))^{2}}{n_{j,g}}+2\sum_{g\in 
		\mathcal{G}_{j}}\sum_{i\in I_{j,g}}\frac{T^{-1}\sum_{t\leq T}\tilde{\psi}%
		_{j,\gamma }^{\ast }(z_{i,t})\mathbb{E}[\psi _{j,\gamma }(z_{i,t})]}{%
		n_{j,g}(-\Psi _{j,\gamma \gamma ,g})^{1/2}}.  \label{P_L_Auxillary3c_1}
\end{align}%
Under Assumptions \ref{A1}, \ref{C_A1}, \ref{A3} and \ref{A4}, we can apply
Markov's inequality to obtain:%
\begin{equation}
	\sum_{g\in \mathcal{G}_{j}}n_{j,g}^{-1}\sum_{i\in I_{j,g}}\left(
	T^{-1}\sum_{t\leq T}\tilde{\psi}_{j,\gamma }^{\ast }(z_{i,t})\right)
	^{2}=O_{p}(G_{j}T^{-1}),  \label{P_L_Auxillary3c_2}
\end{equation}%
and%
\begin{equation}
	\sum_{g\in \mathcal{G}_{j}}\sum_{i\in I_{j,g}}\frac{T^{-1}\sum_{t\leq T}%
		\tilde{\psi}_{j,\gamma }^{\ast }(z_{i,t})\mathbb{E}[\psi _{j,\gamma
		}(z_{i,t})]}{n_{j,g}(-\Psi _{j,\gamma \gamma ,g})^{1/2}}%
	=O_{p}((G_{j}T^{-1})^{1/2}).  \label{P_L_Auxillary3c_3}
\end{equation}%
The claim in (\ref{C_L_Auxillary3c_2}) now follows from (\ref%
{P_L_Auxillary3c_1}), (\ref{P_L_Auxillary3c_2}) and (\ref{P_L_Auxillary3c_3}%
).\hfill $Q.E.D.$

\bigskip

\begin{lemma}
	\textit{\label{C_L_Auxillary4}\ }Under Assumptions \ref{A1} and \ref{C_A1},
	we have%
	\begin{equation*}
		\frac{\left( MQLR_{n,T}\right) ^{2}-(\overline{QLR}_{n,T})^{2}}{%
			(nT)^{1/2}\omega _{n,T}^{2}}=O_{p}(1).
	\end{equation*}
\end{lemma}

\noindent{\textsc{Proof of Lemma \ref{C_L_Auxillary4}}}. First, observe that
by the definitions of $MQLR_{n,T}$ and $\overline{QLR}_{n,T}$, and the
triangle inequality, 
\begin{align}
	\frac{\left\vert MQLR_{n,T}-\overline{QLR}_{n,T}\right\vert }{\omega_{n,T}}
	& \leq\frac{\left\vert MQLR_{n,T}-MQLR_{n,T}^{\ast}\right\vert }{\omega
		_{n,T}}+\frac{\left\vert MQLR_{n,T}^{\ast}-\overline{QLR}_{n,T}\right\vert }{%
		\omega_{n,T}}  \notag \\
	& =\frac{\left\vert R_{j,n,T}(\hat{\phi}_{j})-T^{1/2}\sum_{i\leq n}\mathbb{E}%
		[\tilde{V}_{j,i}]\right\vert }{(nT)^{1/2}\omega_{n,T}}+O_{p}(1)=O_{p}(1),
	\label{P_C_L_Auxillary4_1}
\end{align}
where the first equality follows from Theorem \ref{C_L0}, and the second
equality follows from\ Theorem \ref{C_L1}.\ By the definition of $\omega
_{n,T}$, Assumption \ref{C_A1}(ii), the triangle inequality and H\"{o}lder's
inequality, we have%
\begin{equation}
	\frac{\left\vert \overline{QLR}_{n,T}\right\vert }{(nT)^{1/2}\omega_{n,T}}%
	\leq(nT)^{-1}\sum_{i\leq n}\sum_{t\leq T}\frac{\mathbb{E}[|\Delta\psi
		(z_{i,t})|]}{\omega_{n,T}}\leq K.  \label{P_C_L_Auxillary4_2}
\end{equation}
Combining the results in (\ref{P_C_L_Auxillary4_1}) and (\ref%
{P_C_L_Auxillary4_2}), we conclude that%
\begin{align}
	\frac{\left\vert \left( MQLR_{n,T}\right) ^{2}-(\overline{QLR}%
		_{n,T})^{2}\right\vert }{(nT)^{1/2}\omega_{n,T}^{2}} & \leq\frac{(MQLR_{n,T}-%
		\overline{QLR}_{n,T})^{2}}{(nT)^{1/2}\omega_{n,T}^{2}}  \notag \\
	& +\frac{2\overline{QLR}_{n,T}|MQLR_{n,T}-\overline{QLR}_{n,T}|}{%
		(nT)^{1/2}\omega_{n,T}^{2}}\overset{}{=}O_{p}(1),  \label{P_C_L_Auxillary4_3}
\end{align}
which shows the claim of the lemma.\hfill$Q.E.D.$

\bigskip

\begin{lemma}
	\textit{\label{C_L_Auxillary5} }Under Assumptions\ \ref{A1} and \ref{C_A1},
	we have%
	\begin{equation*}
		(nT)^{-1}\sum_{i\leq n}\sum_{t\leq T}\Delta \psi (z_{i,t})\psi _{j,\phi
		}(z_{i,t})(\hat{\phi}_{j,i}-\phi _{j,i}^{\ast })=O_{p}(\omega
		_{n,T}(nT)^{-1/2}),
	\end{equation*}%
	where\ $\psi _{j,\phi }(z_{i,t})\equiv \psi _{j,\phi }(z_{i,t},\phi
	_{j,i}^{\ast })$ and $\psi _{j,\phi }(z_{i,t},\phi _{j,i})\equiv \partial
	\psi _{j}(z_{i,t},\phi _{j,i})/\partial \phi _{j,i}$.
\end{lemma}

\noindent\textsc{Proof of Lemma \ref{C_L_Auxillary5}}. By the definition of $%
\psi_{j,\phi}(z_{i,t})$, we have 
\begin{equation}
	\psi_{j,\phi}(z_{i,t})(\hat{\phi}_{j,i}-\phi_{j,i}^{\ast})=\psi_{j,\theta
	}(z_{i,t})(\hat{\theta}_{j}-\theta_{j}^{\ast})+\psi_{j,\gamma}(z_{i,t})(\hat{%
		\gamma}_{j,i}-\gamma_{j,i}^{\ast}),  \label{P_C_L_Auxillary5_00}
\end{equation}
which leads to the expression:%
\begin{align}
	(nT)^{-1}\sum_{i\leq n}\sum_{t\leq
		T}\Delta\psi(z_{i,t})\psi_{j,\phi}(z_{i,t})(\hat{\phi}_{j,i}-\phi_{j,i}^{%
		\ast}) & =(nT)^{-1}\sum_{i\leq n}\sum_{t\leq
		T}\Delta\psi(z_{i,t})\psi_{j,\theta}(z_{i,t})(\hat{\theta}%
	_{j}-\theta_{j}^{\ast})  \notag \\
	& +(nT)^{-1}\sum_{i\leq n}\sum_{t\leq T}\Delta\psi(z_{i,t})\psi_{j,\gamma
	}(z_{i,t})(\hat{\gamma}_{j,i}-\gamma_{j,i}^{\ast}),
	\label{P_C_L_Auxillary5_0}
\end{align}
where $\psi_{j,a}(z_{i,t})\equiv\psi_{j,a}(z_{i,t},\phi_{j,i}^{\ast})$ and $%
\psi_{j,a}(z_{i,t},\phi_{j,i})\equiv\partial\psi_{j}(z_{i,t},\phi
_{j,i})/\partial a$ for $a\in\{\theta,\gamma\}$.

By the triangle inequality and the Cauchy-Schwarz inequality, the first term
after the equality in\ (\ref{P_C_L_Auxillary5_0}) can be bounded as: 
\begin{equation}
	\left\vert (nT)^{-1}\sum_{i\leq n}\sum_{t\leq
		T}\Delta\psi(z_{i,t})\psi_{j,\theta}(z_{i,t})(\hat{\theta}%
	_{j}-\theta_{j}^{\ast})\right\vert \leq\frac{||\hat{\theta}%
		_{j}-\theta_{j}^{\ast}||}{nT}\sum_{i\leq n}\sum_{t\leq T}\left\vert
	\Delta\psi(z_{i,t})\right\vert \left\Vert
	\psi_{j,\theta}(z_{i,t})\right\Vert .  \label{P_C_L_Auxillary5_1}
\end{equation}
Under Assumptions \ref{C_A1}(ii) and \ref{A3}, 
\begin{equation*}
	\sum_{i\leq n}\sum_{t\leq T}\frac{\mathbb{E}\left[ \left\vert \Delta
		\psi(z_{i,t})\right\vert \left\Vert \psi_{j,\theta}(z_{i,t})\right\Vert %
		\right] }{nT}\leq\sum_{i\leq n}\sum_{t\leq T}\frac{\left( \mathbb{E}\left[
		\Delta\psi(z_{i,t})^{2}\right] \mathbb{E}[\left\Vert
		\psi_{j,\theta}(z_{i,t})\right\Vert ^{2}]\right) ^{1/2}}{nT}\leq
	K\omega_{n,T},
\end{equation*}
which, along with Markov's inequality, Lemma \ref{C_L_Auxillary1}(iii) and (%
\ref{P_C_L_Auxillary5_1}), establishes%
\begin{equation}
	(nT)^{-1}\sum_{i\leq n}\sum_{t\leq T}\Delta\psi(z_{i,t})\psi_{j,\theta
	}(z_{i,t})(\hat{\theta}_{j}-\theta_{j}^{\ast})=O_{p}(%
	\omega_{n,T}(nT)^{-1/2}).  \label{P_C_L_Auxillary5_2}
\end{equation}

Next, we rewrite the second term after the equality in (\ref%
{P_C_L_Auxillary5_0}) as:%
\begin{align}
	& (nT)^{-1}\sum_{i\leq n}\sum_{t\leq T}\Delta \psi (z_{i,t})\psi _{j,\gamma
	}(z_{i,t})(\hat{\gamma}_{j,i}-\gamma _{j,i}^{\ast })  \notag \\
	& =(nT)^{-1}\sum_{g\in \mathcal{G}_{j}}\sum_{i\in I_{j,g}}\sum_{t\leq T}%
	\frac{\mathbb{E}\left[ \Delta \psi (z_{i,t})\psi _{j,\gamma }(z_{i,t})\right]
	}{-\Psi _{j,\gamma \gamma ,g}}\hat{\Psi}_{j,\gamma ,g}  \notag \\
	& +(nT)^{-1}\sum_{g\in \mathcal{G}_{j}}\sum_{i\in I_{j,g}}\sum_{t\leq T}%
	\frac{\Delta \psi (z_{i,t})\psi _{j,\gamma }(z_{i,t})-\mathbb{E}\left[
		\Delta \psi (z_{i,t})\psi _{j,\gamma }(z_{i,t})\right] }{-\Psi _{j,\gamma
			\gamma ,g}}\hat{\Psi}_{j,\gamma ,g}  \notag \\
	& +(nT)^{-1}\sum_{g\in \mathcal{G}_{j}}\sum_{i\in I_{j,g}}\sum_{t\leq
		T}\Delta \psi (z_{i,t})\psi _{j,\gamma }(z_{i,t})(\hat{\gamma}_{j,g}-\gamma
	_{j,g}^{\ast }+\Psi _{j,\gamma \gamma ,g}^{-1}\hat{\Psi}_{j,\gamma ,g}).
	\label{P_C_L_Auxillary5_3}
\end{align}%
Using Assumptions \ref{C_A1}(ii), \ref{A3}\ and \ref{A4}, we bound the
second moment of the first term:%
\begin{align*}
	& \mathbb{E}\left[ \left( (nT)^{-1}\sum_{g\in \mathcal{G}_{j}}\sum_{i\in
		I_{j,g}}\sum_{t\leq T}\frac{\mathbb{E}\left[ \Delta \psi (z_{i,t})\psi
		_{j,\gamma }(z_{i,t})\right] }{-\Psi _{j,\gamma \gamma ,g}}\hat{\Psi}%
	_{j,\gamma ,g}\right) ^{2}\right] \\
	& =\omega _{n,T}^{2}(nT)^{-2}\sum_{g\in \mathcal{G}_{j}}\mathbb{E}[\hat{\Psi}%
	_{j,\gamma ,g}^{2}]\left( \sum_{i\in I_{j,g}}\sum_{t\leq T}\frac{\mathbb{E}%
		\left[ \Delta \psi (z_{i,t})\psi _{j,\gamma }(z_{i,t})\right] }{-\Psi
		_{j,\gamma \gamma ,g}\omega _{n,T}}\right) ^{2} \\
	& \leq K\omega _{n,T}^{2}(nT)^{-2}\sum_{g\in \mathcal{G}_{j}}(n_{j,g}T)^{2}%
	\mathbb{E}[\hat{\Psi}_{j,\gamma ,g}^{2}]\leq K\omega _{n,T}^{2}(nT)^{-1},
\end{align*}%
where the second inequality follows from Lemma \ref{C_L_Auxillary1}(viii).\
Hence by Markov's inequality and Assumption \ref{A1}(i), the first term on
the right of (\ref{P_C_L_Auxillary5_3}) satisfies:%
\begin{equation}
	(nT)^{-1}\sum_{g\in \mathcal{G}_{j}}\sum_{i\in I_{j,g}}\sum_{t\leq T}\frac{%
		\mathbb{E}\left[ \Delta \psi (z_{i,t})\psi _{j,\gamma }(z_{i,t})\right] }{%
		-\Psi _{j,\gamma \gamma ,g}}\hat{\Psi}_{j,\gamma ,g}=O_{p}(\omega
	_{n,T}(nT)^{-1/2}).  \label{P_C_L_Auxillary5_3b}
\end{equation}%
For the second term in the RHS of (\ref{P_C_L_Auxillary5_3}), we can apply
the triangle inequality and Holder's inequality to obtain:%
\begin{align}
	& \mathbb{E}\left[ \left\vert (nT)^{-1}\sum_{g\in \mathcal{G}_{j}}\sum_{i\in
		I_{j,g}}\sum_{t\leq T}\frac{\Delta \psi (z_{i,t})\psi _{j,\gamma }(z_{i,t})-%
		\mathbb{E}\left[ \Delta \psi (z_{i,t})\psi _{j,\gamma }(z_{i,t})\right] }{%
		-\Psi _{j,\gamma \gamma ,g}}\hat{\Psi}_{j,\gamma ,g}\right\vert \right] 
	\notag \\
	& \leq K(nT)^{-1}\sum_{g\in \mathcal{G}_{j}}\sqrt{\mathbb{E}\left[ |\hat{\Psi%
		}_{j,\gamma ,g}|^{2}\right] }\sqrt{\mathbb{E}\left[ \left\vert \sum_{i\in
			I_{j,g}}\sum_{t\leq T}\Delta \psi (z_{i,t})\psi _{j,\gamma }(z_{i,t})-%
		\mathbb{E}\left[ \Delta \psi (z_{i,t})\psi _{j,\gamma }(z_{i,t})\right]
		\right\vert ^{2}\right] }.  \label{P_C_L_Auxillary5_3c}
\end{align}%
Under Assumptions\ \ref{A1} and \ref{C_A1}(ii, iii), it follows that%
\begin{eqnarray*}
	&&\mathbb{E}\left[ \left\vert \sum_{i\in I_{j,g}}\sum_{t\leq T}\frac{\Delta
		\psi (z_{i,t})\psi _{j,\gamma }(z_{i,t})-\mathbb{E}\left[ \Delta \psi
		(z_{i,t})\psi _{j,\gamma }(z_{i,t})\right] }{\omega _{n,T}}\right\vert ^{2}%
	\right] \\
	&=&\sum_{i\in I_{j,g}}\mathbb{E}\left[ \left\vert \sum_{t\leq T}\frac{\Delta
		\psi (z_{i,t})\psi _{j,\gamma }(z_{i,t})-\mathbb{E}\left[ \Delta \psi
		(z_{i,t})\psi _{j,\gamma }(z_{i,t})\right] }{\omega _{n,T}}\right\vert ^{2}%
	\right] \\
	&\leq &KT\sum_{i\in I_{j,g}}\mathbb{E}\left[ \left\vert \frac{\Delta \psi
		(z_{i,t})}{\omega _{n,T}}\psi _{j,\gamma }(z_{i,t})\right\vert ^{2+\delta /2}%
	\right] \\
	&\leq &KT\sum_{i\in I_{j,g}}\left( \mathbb{E}\left[ \left\vert \frac{\Delta
		\psi (z_{i,t})}{\omega _{n,T}}\right\vert ^{4+\delta }\right] \mathbb{E}%
	\left[ \left\vert \psi _{j,\gamma }(z_{i,t})\right\vert ^{4+\delta }\right]
	\right) ^{1/2}\leq K(n_{j,g}T),
\end{eqnarray*}%
where $\delta >0$ is defined in Assumption \ref{C_A1}(ii). The first
inequality follows from Assumption \ref{A1} together with a covariance
inequality for strongly mixing processes, and the second inequality follows
from H\"{o}lder's inequality. This together with Assumption \ref{A1}(i),
Lemma \ref{C_L_Auxillary1}(viii) and (\ref{P_C_L_Auxillary5_3c}), and
Markov's inequality implies that the first term on the right of (\ref%
{P_C_L_Auxillary5_3}) satisfies 
\begin{equation}
	(nT)^{-1}\sum_{g\in \mathcal{G}_{j}}\sum_{i\in I_{j,g}}\sum_{t\leq T}\frac{%
		\Delta \psi (z_{i,t})\psi _{j,\gamma }(z_{i,t})-\mathbb{E}\left[ \Delta \psi
		(z_{i,t})\psi _{j,\gamma }(z_{i,t})\right] }{-\Psi _{j,\gamma \gamma ,g}}%
	\hat{\Psi}_{j,\gamma ,g}=O_{p}(\omega _{n,T}(nT)^{-1/2}).
	\label{P_C_L_Auxillary5_4}
\end{equation}%
By the Cauchy-Schwarz inequality, the third term after the equality in (\ref%
{P_C_L_Auxillary5_3}) can be bounded as 
\begin{align}
	& \left\vert (nT)^{-1}\sum_{g\in \mathcal{G}_{j}}\sum_{i\in
		I_{j,g}}\sum_{t\leq T}\Delta \psi (z_{i,t})\psi _{j,\gamma }(z_{i,t})(\hat{%
		\gamma}_{j,g}-\gamma _{j,g}^{\ast }+\Psi _{j,\gamma \gamma ,g}^{-1}\hat{\Psi}%
	_{j,\gamma ,g})\right\vert  \notag \\
	& \leq n^{-1}\left( \sum_{g\in \mathcal{G}_{j}}n_{j,g}(\hat{\gamma}%
	_{j,g}-\gamma _{j,g}^{\ast }+\Psi _{j,\gamma \gamma ,g}^{-1}\hat{\Psi}%
	_{j,\gamma ,g})^{2}\right) ^{1/2}\left( \sum_{g\in \mathcal{G}%
		_{j}}T^{-1}\sum_{i\in I_{j,g}}\sum_{t\leq T}(\Delta \psi (z_{i,t})\psi
	_{j,\gamma }(z_{i,t}))^{2}\right) ^{1/2}.  \label{P_C_L_Auxillary5_5}
\end{align}%
By Assumptions \ref{C_A1}(iii) and \ref{A3}, and Markov's inequality,%
\begin{equation}
	\sum_{g\in \mathcal{G}_{j}}T^{-1}\sum_{i\in I_{j,g}}\sum_{t\leq T}(\Delta
	\psi (z_{i,t})\psi _{j,\gamma }(z_{i,t}))^{2}=O_{p}(\omega _{n,T}^{2}n).
	\label{P_C_L_Auxillary5_6}
\end{equation}%
Combining this with Assumption \ref{A1}(i), (\ref{P_C_L_Auxillary5_5}) and
Lemma \ref{C_L_Auxillary1}(vii), we establish a bound of the third term on
the right of (\ref{P_C_L_Auxillary5_3}): 
\begin{equation}
	(nT)^{-1}\sum_{g\in \mathcal{G}_{j}}\sum_{i\in I_{j,g}}\sum_{t\leq T}\Delta
	\psi (z_{i,t})\psi _{j,\gamma }(z_{i,t})(\hat{\gamma}_{j,g}-\gamma
	_{j,g}^{\ast }+\Psi _{j,\gamma \gamma ,g}^{-1}\hat{\Psi}_{j,\gamma
		,g})=O_{p}(\omega _{n,T}(nT)^{-1/2}).  \label{P_C_L_Auxillary5_7}
\end{equation}%
Collecting results from (\ref{P_C_L_Auxillary5_3}), (\ref%
{P_C_L_Auxillary5_3b}), (\ref{P_C_L_Auxillary5_4}) and (\ref%
{P_C_L_Auxillary5_7}), we obtain:%
\begin{equation}
	(nT)^{-1}\sum_{i\leq n}\sum_{t\leq T}\Delta \psi (z_{i,t})\psi _{j,\gamma
	}(z_{i,t})(\hat{\gamma}_{j,i}-\gamma _{j,i}^{\ast })=O_{p}(\omega
	_{n,T}(nT)^{-1/2}).  \label{P_C_L_Auxillary5_8}
\end{equation}%
The lemma follows from (\ref{P_C_L_Auxillary5_0}), (\ref{P_C_L_Auxillary5_2}%
) and (\ref{P_C_L_Auxillary5_8}).\hfill $Q.E.D.$

\bigskip

\begin{lemma}
	\textit{\label{C_L_Auxillary6a}\ }Under Assumptions\ \ref{A1} and \ref{C_A1}%
	, we have%
	\begin{eqnarray*}
		&&(nT)^{-1}\sum_{i\leq n}\sum_{t\leq T}(\psi _{j,\gamma }(z_{i,t})(\hat{%
			\gamma}_{j,i}-\gamma _{j,i}^{\ast }))^{2} \\
		&=&(nT)^{-1}\sum_{g\in \mathcal{G}_{j}}n_{j,g}^{-2}\left( \sum_{i\in
			I_{j,g}}s_{j,\gamma ,i}^{2}\right) \sum_{i\in I_{j,g}}\mathbb{E}\left[
		\left( T^{-1/2}\sum_{t\leq T}\tilde{\psi}_{j,\gamma }^{\ast }\left(
		z_{i,t}\right) \right) ^{2}\right] +O_{p}(G_{j}^{1/2}(nT)^{-1}).
	\end{eqnarray*}
\end{lemma}

\noindent \textsc{Proof of Lemma \ref{C_L_Auxillary6a}}. We start with the
following characterization:%
\begin{align}
	& (nT)^{-1}\sum_{i\leq n}\sum_{t\leq T}(\psi _{j,\gamma }(z_{i,t})(\hat{%
		\gamma}_{j,i}-\gamma _{j,i}^{\ast }))^{2}  \notag \\
	& =(nT)^{-1}\sum_{i\leq n}(\hat{\gamma}_{j,i}-\gamma _{j,i}^{\ast
	})^{2}\sum_{t\leq T}(\psi _{j,\gamma }^{2}(z_{i,t})-\mathbb{E}[\psi
	_{j,\gamma }^{2}(z_{i,t})])+n^{-1}\sum_{i\leq n}\mathbb{E}_{T}[\psi
	_{j,\gamma }^{2}(z_{i,t})](\hat{\gamma}_{j,i}-\gamma _{j,i}^{\ast })^{2}.
	\label{P_C_L_Auxillary6a_1}
\end{align}%
By the Cauchy-Schwarz inequality, the first term after the equality in (\ref%
{P_C_L_Auxillary6a_1}) can be bounded as:%
\begin{align}
	& \left\vert \sum_{i\leq n}(\hat{\gamma}_{j,i}-\gamma _{j,i}^{\ast
	})^{2}\sum_{t\leq T}(\psi _{j,\gamma }^{2}(z_{i,t})-\mathbb{E}[\psi
	_{j,\gamma }^{2}(z_{i,t})])\right\vert  \notag \\
	& =\left\vert \sum_{g\in \mathcal{G}_{j}}(\hat{\gamma}_{j,g}-\gamma
	_{j,g}^{\ast })^{2}\sum_{i\in I_{j,g}}\sum_{t\leq T}(\psi _{j,\gamma
	}^{2}(z_{i,t})-\mathbb{E}[\psi _{j,\gamma }^{2}(z_{i,t})])\right\vert  \notag
	\\
	& \leq \left( \sum_{g\in \mathcal{G}_{j}}n_{j,g}^{2}(\hat{\gamma}%
	_{j,g}-\gamma _{j,g}^{\ast })^{4}\right) ^{1/2}\left( \sum_{g\in \mathcal{G}%
		_{j}}\left( n_{j,g}^{-1}\sum_{i\in I_{j,g}}\sum_{t\leq T}\left( \psi
	_{j,\gamma }^{2}(z_{i,t})-\mathbb{E}[\psi _{j,\gamma }^{2}(z_{i,t})]\right)
	\right) ^{2}\right) ^{1/2}.  \label{P_C_L_Auxillary6a_2}
\end{align}%
Using Assumptions \ref{A1}, \ref{C_A1}(iii) and the covariance inequality
for strongly mixing processes, we can derive:%
\begin{eqnarray*}
	&&\mathbb{E}\left[ \sum_{g\in \mathcal{G}_{j}}\left( n_{j,g}^{-1}\sum_{i\in
		I_{j,g}}\sum_{t\leq T}(\psi _{j,\gamma }^{2}(z_{i,t})-\mathbb{E}[\psi
	_{j,\gamma }^{2}(z_{i,t})])\right) ^{2}\right] \\
	&=&\sum_{g\in \mathcal{G}_{j}}n_{j,g}^{-2}\sum_{i\in I_{j,g}}\mathbb{E}\left[
	\left( \sum_{t\leq T}(\psi _{j,\gamma }^{2}(z_{i,t})-\mathbb{E}[\psi
	_{j,\gamma }^{2}(z_{i,t})])\right) ^{2}\right] \\
	&\leq &KT\sum_{g\in \mathcal{G}_{j}}n_{j,g}^{-2}\sum_{i\in I_{j,g}}\mathbb{E}%
	\left[ |\psi _{j,\gamma }(z_{i,t})|^{4+\delta }\right] \leq KG_{j}T.
\end{eqnarray*}%
Applying Markov's inequality, we obtain: 
\begin{equation*}
	\sum_{g\in \mathcal{G}_{j}}\left( n_{j,g}^{-1}\sum_{i\in I_{j,g}}\sum_{t\leq
		T}(\psi _{j,\gamma }^{2}(z_{i,t})-\mathbb{E}[\psi _{j,\gamma
	}^{2}(z_{i,t})])\right) ^{2}=O_{p}(G_{j}T).
\end{equation*}%
Combining this result with (\ref{P_C_L_Auxillary6a_2}) and Lemma \ref%
{C_L_Auxillary1}(iv), we establish the bound for the first term on the right
of (\ref{P_C_L_Auxillary6a_1}) 
\begin{equation}
	(nT)^{-1}\sum_{i\leq n}(\hat{\gamma}_{j,i}-\gamma _{j,i}^{\ast
	})^{2}\sum_{t\leq T}(\psi _{j,\gamma }^{2}(z_{i,t})-\mathbb{E}[\psi
	_{j,\gamma }^{2}(z_{i,t})])=O_{p}(G_{j}(nT^{3/2})^{-1}).
	\label{P_C_L_Auxillary6a_3}
\end{equation}

Collecting the results from (\ref{P_C_L_Auxillary6a_1}) and (\ref%
{P_C_L_Auxillary6a_3}), we obtain:%
\begin{equation*}
	(nT)^{-1}\sum_{i\leq n}\sum_{t\leq T}(\psi _{j,\gamma }(z_{i,t})(\hat{\gamma}%
	_{j,i}-\gamma _{j,i}^{\ast }))^{2}=n^{-1}\sum_{i\leq n}\mathbb{E}_{T}[\psi
	_{j,\gamma }^{2}(z_{i,t})](\hat{\gamma}_{j,i}-\gamma _{j,i}^{\ast
	})^{2}+O_{p}(G_{j}(nT^{3/2})^{-1}),
\end{equation*}%
which we rewrite as:%
\begin{align}
	& (nT)^{-1}\sum_{i\leq n}\sum_{t\leq T}(\psi _{j,\gamma }(z_{i,t})(\hat{%
		\gamma}_{j,i}-\gamma _{j,i}^{\ast }))^{2}  \notag \\
	& =n^{-1}\sum_{g\in \mathcal{G}_{j}}\sum_{i\in I_{j,g}}\frac{\mathbb{E}%
		_{T}[\psi _{j,\gamma }^{2}(z_{i,t})]}{\Psi _{j,\gamma \gamma ,g}^{2}}\hat{%
		\Psi}_{j,\gamma ,g}^{2}-2n^{-1}\sum_{g\in \mathcal{G}_{j}}\sum_{i\in I_{j,g}}%
	\frac{\mathbb{E}_{T}[\psi _{j,\gamma }^{2}(z_{i,t})]}{\Psi _{j,\gamma \gamma
			,g}}\hat{\Psi}_{j,\gamma ,g}(\hat{\gamma}_{j,g}-\gamma _{j,g}^{\ast }+\Psi
	_{j,\gamma \gamma ,g}^{-1}\hat{\Psi}_{j,\gamma ,g})  \notag \\
	& +n^{-1}\sum_{g\in \mathcal{G}_{j}}\sum_{i\in I_{j,g}}\mathbb{E}_{T}[\psi
	_{j,\gamma }^{2}(z_{i,t})](\hat{\gamma}_{j,g}-\gamma _{j,g}^{\ast }+\Psi
	_{j,\gamma \gamma ,g}^{-1}\hat{\Psi}_{j,\gamma
		,g})^{2}+O_{p}(G_{j}(nT^{3/2})^{-1}).  \label{P_C_L_Auxillary6a_4}
\end{align}%
Recall that $s_{j,\gamma ,i}^{2}\equiv \mathbb{E}_{T}[\tilde{\psi}_{j,\gamma
}^{2}(z_{i,t})]$\ and $\tilde{\psi}_{j,\gamma }(z_{i,t})\equiv \psi
_{j,\gamma }\left( z_{i,t}\right) (-\Psi _{j,\gamma \gamma ,g})^{-1/2}$. The
first term after the second equality in (\ref{P_C_L_Auxillary6a_4}) can be
written as%
\begin{align}
	n^{-1}\sum_{g\in \mathcal{G}_{j}}\sum_{i\in I_{j,g}}\mathbb{E}_{T}[\psi
	_{j,\gamma }^{2}(z_{i,t})]\frac{\hat{\Psi}_{j,\gamma ,g}^{2}}{\Psi
		_{j,\gamma \gamma ,g}^{2}}& =n^{-1}\sum_{g\in \mathcal{G}_{j}}\sum_{i\in
		I_{j,g}}s_{j,\gamma ,i}^{2}\mathbb{E}\left[ \frac{\hat{\Psi}_{j,\gamma
			,g}^{2}}{-\Psi _{j,\gamma \gamma ,g}}\right]  \notag \\
	& +n^{-1}\sum_{g\in \mathcal{G}_{j}}\sum_{i\in I_{j,g}}s_{j,\gamma
		,i}^{2}\left( \frac{\hat{\Psi}_{j,\gamma ,g}^{2}}{-\Psi _{j,\gamma \gamma ,g}%
	}-\mathbb{E}\left[ \frac{\hat{\Psi}_{j,\gamma ,g}^{2}}{-\Psi _{j,\gamma
			\gamma ,g}}\right] \right) .  \label{P_C_L_Auxillary6a_5}
\end{align}%
Under Assumption \ref{A1}, we know that%
\begin{equation*}
	\mathbb{E}\left[ \frac{\hat{\Psi}_{j,\gamma ,g}^{2}}{-\Psi _{j,\gamma \gamma
			,g}}\right] =\mathbb{E}\left[ \left( (n_{j,g}T)^{-1}\sum_{i\in
		I_{j,g}}\sum_{t\leq T}\tilde{\psi}_{j,\gamma }^{\ast }\left( z_{i,t}\right)
	\right) ^{2}\right] =(n_{j,g}^{2}T)^{-1}\sum_{i\in I_{j,g}}\mathbb{E}\left[
	\left( T^{-1/2}\sum_{t\leq T}\tilde{\psi}_{j,\gamma }^{\ast }\left(
	z_{i,t}\right) \right) ^{2}\right] .
\end{equation*}%
Thus, we have:%
\begin{equation}
	n^{-1}\sum_{g\in \mathcal{G}_{j}}\sum_{i\in I_{j,g}}s_{j,\gamma ,i}^{2}%
	\mathbb{E}\left[ \frac{\hat{\Psi}_{j,\gamma ,g}^{2}}{-\Psi _{j,\gamma \gamma
			,g}}\right] =(nT)^{-1}\sum_{g\in \mathcal{G}_{j}}n_{j,g}^{-2}\left(
	\sum_{i\in I_{j,g}}s_{j,\gamma ,i}^{2}\right) \sum_{i\in I_{j,g}}\mathbb{E}%
	\left[ \left( T^{-1/2}\sum_{t\leq T}\tilde{\psi}_{j,\gamma }^{\ast }\left(
	z_{i,t}\right) \right) ^{2}\right] ,  \label{P_C_L_Auxillary6a_6}
\end{equation}%
where 
\begin{equation}
	\mathbb{E}\left[ \left( T^{-1/2}\sum_{t\leq T}\tilde{\psi}_{j,\gamma }^{\ast
	}\left( z_{i,t}\right) \right) ^{2}\right] \leq K
	\label{P_C_L_Auxillary6a_6b}
\end{equation}%
by arguments analogous to those used to establish (\ref{P_C_T2_0}). Using
Assumptions \ref{A1}, \ref{A3} and \ref{A4}, and Lemma \ref{C_L_Auxillary1}%
(viii), we also derive: 
\begin{align*}
	& \mathbb{E}\left[ \left\vert n^{-1}\sum_{g\in \mathcal{G}_{j}}\sum_{i\in
		I_{j,g}}s_{j,\gamma ,i}^{2}\left( \frac{\hat{\Psi}_{j,\gamma ,g}^{2}}{-\Psi
		_{j,\gamma \gamma ,g}}-\mathbb{E}\left[ \frac{\hat{\Psi}_{j,\gamma ,g}^{2}}{%
		-\Psi _{j,\gamma \gamma ,g}}\right] \right) \right\vert ^{2}\right] \\
	& \leq n^{-2}\sum_{g\in \mathcal{G}_{j}}\left( \sum_{i\in
		I_{j,g}}s_{j,\gamma ,i}^{2}\right) ^{2}\mathbb{E}\left[ \frac{\hat{\Psi}%
		_{j,\gamma ,g}^{4}}{\Psi _{j,\gamma \gamma ,g}^{2}}\right] \leq
	Kn^{-2}\sum_{g\in \mathcal{G}_{j}}n_{j,g}^{2}\mathbb{E}[\hat{\Psi}_{j,\gamma
		,g}^{4}]\leq KG_{j}(nT)^{-2},
\end{align*}%
which along with Markov's inequality implies that 
\begin{equation}
	n^{-1}\sum_{g\in \mathcal{G}_{j}}\sum_{i\in I_{j,g}}s_{j,\gamma
		,i}^{2}\left( \frac{\hat{\Psi}_{j,\gamma ,g}^{2}}{-\Psi _{j,\gamma \gamma ,g}%
	}-\mathbb{E}\left[ \frac{\hat{\Psi}_{j,\gamma ,g}^{2}}{-\Psi _{j,\gamma
			\gamma ,g}}\right] \right) =O_{p}(G_{j}^{1/2}(nT)^{-1}).
	\label{P_C_L_Auxillary6a_7}
\end{equation}%
Combining the results from (\ref{P_C_L_Auxillary6a_5}), (\ref%
{P_C_L_Auxillary6a_6}) and (\ref{P_C_L_Auxillary6a_7}), we establish a bound
for the first term on the right of (\ref{P_C_L_Auxillary6a_4}):%
\begin{eqnarray}
	n^{-1}\sum_{g\in \mathcal{G}_{j}}\sum_{i\in I_{j,g}}\mathbb{E}_{T}[\psi
	_{j,\gamma }^{2}(z_{i,t})]\frac{\hat{\Psi}_{j,\gamma ,g}^{2}}{\Psi
		_{j,\gamma \gamma ,g}^{2}} &=&(nT)^{-1}\sum_{g\in \mathcal{G}%
		_{j}}n_{j,g}^{-2}\left( \sum_{i\in I_{j,g}}s_{j,\gamma ,i}^{2}\right)
	\sum_{i\in I_{j,g}}\mathbb{E}\left[ \left( T^{-1/2}\sum_{t\leq T}\tilde{\psi}%
	_{j,\gamma }^{\ast }\left( z_{i,t}\right) \right) ^{2}\right]  \notag \\
	&&+O_{p}(G_{j}^{1/2}(nT)^{-1})\overset{}{=}O_{p}(G_{j}(nT)^{-1}),
	\label{P_C_L_Auxillary6a_8}
\end{eqnarray}%
where the second equality follows from Assumption \ref{A3} and (\ref%
{P_C_L_Auxillary6a_6b}).

By Assumption \ref{A3} and Lemma \ref{C_L_Auxillary1}(vii), we obtain a
bound for the third term on the right of (\ref{P_C_L_Auxillary6a_4}):%
\begin{align}
	& n^{-1}\sum_{g\in \mathcal{G}_{j}}\sum_{i\in I_{j,g}}\mathbb{E}_{T}[\psi
	_{j,\gamma }^{2}(z_{i,t})](\hat{\gamma}_{j,g}-\gamma _{j,g}^{\ast }+\Psi
	_{j,\gamma \gamma ,g}^{-1}\hat{\Psi}_{j,\gamma ,g})^{2}  \notag \\
	& \leq n^{-1}\sum_{g\in \mathcal{G}_{j}}n_{j,g}(\hat{\gamma}_{j,g}-\gamma
	_{j,g}^{\ast }+\Psi _{j,\gamma \gamma ,g}^{-1}\hat{\Psi}_{j,\gamma
		,g})^{2}=O_{p}((nT)^{-1}).  \label{P_C_L_Auxillary6a_9}
\end{align}%
From (\ref{P_C_L_Auxillary6a_8}) and (\ref{P_C_L_Auxillary6a_9}), and the
Cauchy-Schwarz inequality, we establish a bound for the second term on the
right of (\ref{P_C_L_Auxillary6a_4}): 
\begin{equation*}
	n^{-1}\sum_{g\in \mathcal{G}_{j}}\sum_{i\in I_{j,g}}\frac{\mathbb{E}%
		_{T}[\psi _{j,\gamma }^{2}(z_{i,t})]}{\Psi _{j,\gamma \gamma ,g}}\hat{\Psi}%
	_{j,\gamma ,g}(\hat{\gamma}_{j,g}-\gamma _{j,g}^{\ast }+\Psi _{j,\gamma
		\gamma ,g}^{-1}\hat{\Psi}_{j,\gamma ,g})=O_{p}(G_{j}^{1/2}(nT)^{-1}),
\end{equation*}%
which along with (\ref{P_C_L_Auxillary6a_4}), (\ref{P_C_L_Auxillary6a_8})
and (\ref{P_C_L_Auxillary6a_9}), completes the proof of the lemma..\hfill $%
Q.E.D.$

\bigskip

\begin{lemma}
	\textit{\label{C_L_Auxillary6b} }Under Assumptions \ref{A1} and \ref{C_A1}\
	we have%
	\begin{align}
		& (nT)^{-1}\sum_{i\leq n}(\hat{\gamma}_{1,i}-\gamma _{1,i}^{\ast })(\hat{%
			\gamma}_{2,i}-\gamma _{2,i}^{\ast })\sum_{t\leq T}\psi _{1,\gamma
		}(z_{i,t})\psi _{2,\gamma }(z_{i,t})  \notag \\
		& =(nT)^{-1}\sum_{g\in \mathcal{G}_{2}}n_{2,g}^{-1}\sum_{i\in I_{2,g}}\sigma
		_{12,\gamma ,i}\left( T^{-1}\sum_{t,t^{\prime }\leq T}\mathbb{E}[\tilde{\psi}%
		_{1,\gamma }^{\ast }(z_{i,t})\tilde{\psi}_{2,\gamma }^{\ast }(z_{i,t^{\prime
		}})]\right) +O_{p}(G_{2}^{1/2}(nT)^{-1}).  \label{C_L_Auxillary6b_1}
	\end{align}
\end{lemma}

\noindent \textsc{Proof of Lemma \ref{C_L_Auxillary6b}}. First, we observe
that the term on the LHS of (\ref{C_L_Auxillary6b_1}) can be written as:%
\begin{align}
	& (nT)^{-1}\sum_{i\leq n}(\hat{\gamma}_{1,i}-\gamma _{1,i}^{\ast })(\hat{%
		\gamma}_{2,i}-\gamma _{2,i}^{\ast })\sum_{t\leq T}\psi _{1,\gamma
	}(z_{i,t})\psi _{2,\gamma }(z_{i,t})  \notag \\
	& =n^{-1}\sum_{i\leq n}(\hat{\gamma}_{1,i}-\gamma _{1,i}^{\ast })(\hat{\gamma%
	}_{2,i}-\gamma _{2,i}^{\ast })\mathbb{E}_{T}[\psi _{1,\gamma }(z_{i,t})\psi
	_{2,\gamma }(z_{i,t})]  \notag \\
	& +(nT)^{-1}\sum_{i\leq n}(\hat{\gamma}_{1,i}-\gamma _{1,i}^{\ast })(\hat{%
		\gamma}_{2,i}-\gamma _{2,i}^{\ast })\sum_{t\leq T}(\psi _{1,\gamma
	}(z_{i,t})\psi _{2,\gamma }(z_{i,t})-\mathbb{E}[\psi _{1,\gamma
	}(z_{i,t})\psi _{2,\gamma }(z_{i,t})]).  \label{P_C_L_Auxillary6b_1}
\end{align}%
By the Cauchy-Schwarz inequality, we can bound the second term after the
equality in (\ref{P_C_L_Auxillary6b_1}) as:%
\begin{align}
	& \left\vert (nT)^{-1}\sum_{i\leq n}(\hat{\gamma}_{1,i}-\gamma _{1,i}^{\ast
	})(\hat{\gamma}_{2,i}-\gamma _{2,i}^{\ast })\sum_{t\leq T}(\psi _{1,\gamma
	}(z_{i,t})\psi _{2,\gamma }(z_{i,t})-\mathbb{E}[\psi _{1,\gamma
	}(z_{i,t})\psi _{2,\gamma }(z_{i,t})])\right\vert  \notag \\
	& \leq (nT)^{-1}\left( \sum_{i\leq n}(\hat{\gamma}_{1,i}-\gamma _{1,i}^{\ast
	})^{4}\right) ^{1/4}\times \left( \sum_{i\leq n}(\hat{\gamma}_{2,i}-\gamma
	_{2,i}^{\ast })^{4}\right) ^{1/4}  \notag \\
	& \times \left( \sum_{i\leq n}\left( \sum_{t\leq T}(\psi _{1,\gamma
	}(z_{i,t})\psi _{2,\gamma }(z_{i,t})-\mathbb{E}[\psi _{1,\gamma
	}(z_{i,t})\psi _{2,\gamma }(z_{i,t})])\right) ^{2}\right) ^{1/2}.
	\label{P_C_L_Auxillary6b_2}
\end{align}%
By Assumptions \ref{A1}, \ref{C_A1}(iii), the covariance inequality for
strongly mixing processes and H\"{o}lder's inequality,%
\begin{eqnarray*}
	&&\mathbb{E}\left[ \left( \sum_{t\leq T}(\psi _{1,\gamma }(z_{i,t})\psi
	_{2,\gamma }(z_{i,t})-\mathbb{E}[\psi _{1,\gamma }(z_{i,t})\psi _{2,\gamma
	}(z_{i,t})])\right) ^{2}\right] \\
	&\leq &KT\max_{t\leq T}\left\Vert \psi _{1,\gamma }(z_{i,t})\psi _{2,\gamma
	}(z_{i,t})\right\Vert _{2+\delta /2}^{2}\leq KT\max_{t\leq T}\left\Vert \psi
	_{1,\gamma }(z_{i,t})\right\Vert _{4+\delta }^{2}\left\Vert \psi _{2,\gamma
	}(z_{i,t})\right\Vert _{4+\delta }^{2}\leq KT.
\end{eqnarray*}%
Hence by Markov's inequality, 
\begin{equation*}
	\sum_{i\leq n}\left( \sum_{t\leq T}(\psi _{1,\gamma }(z_{i,t})\psi
	_{2,\gamma }(z_{i,t})-\mathbb{E}[\psi _{1,\gamma }(z_{i,t})\psi _{2,\gamma
	}(z_{i,t})])\right) ^{2}=O_{p}(nT),
\end{equation*}%
which together with (\ref{P_C_L_Auxillary6b_2}) and Lemma \ref%
{C_L_Auxillary1}(iv) leads to:%
\begin{equation}
	(nT)^{-1}\sum_{i\leq n}(\hat{\gamma}_{1,i}-\gamma _{1,i}^{\ast })(\hat{\gamma%
	}_{2,i}-\gamma _{2,i}^{\ast })\sum_{t\leq T}(\psi _{1,\gamma }(z_{i,t})\psi
	_{2,\gamma }(z_{i,t})-\mathbb{E}[\psi _{1,\gamma }(z_{i,t})\psi _{2,\gamma
	}(z_{i,t})])=O_{p}(T^{-3/2}).  \label{P_C_L_Auxillary6b_3}
\end{equation}

The first order condition for $\gamma _{1,i}^{\ast }$ implies that $\mathbb{E%
}_{T}[\psi _{1,\gamma }(z_{i,t})]=0$ for any $i$. Consequently, 
\begin{equation*}
	\mathbb{E}_{T}[\psi _{1,\gamma }(z_{i,t})\psi _{2,\gamma }(z_{i,t})]=\mathbb{%
		E}_{T}[\psi _{1,\gamma }(z_{i,t})\psi _{2,\gamma }(z_{i,t})]-\mathbb{E}%
	_{T}[\psi _{1,\gamma }(z_{i,t})]\mathbb{E}_{T}[\psi _{2,\gamma }(z_{i,t})]=%
	\mathbb{E}_{T}[\psi _{1,\gamma }^{\ast }(z_{i,t})\psi _{2,\gamma }^{\ast
	}(z_{i,t})]
\end{equation*}%
for any $i$. Using this expression, we can rewrite the first term after the
equality in (\ref{P_C_L_Auxillary6b_1}) as:%
\begin{align}
	& n^{-1}\sum_{g\in \mathcal{G}_{2}}(\hat{\gamma}_{2,g}-\gamma _{2,g}^{\ast
	})\sum_{i\in I_{2,g}}(\hat{\gamma}_{1,i}-\gamma _{1,i}^{\ast })\mathbb{E}%
	_{T}[\psi _{1,\gamma }^{\ast }(z_{i,t})\psi _{2,\gamma }^{\ast }(z_{i,t})] 
	\notag \\
	& =n^{-1}\sum_{g\in \mathcal{G}_{2}}\frac{\hat{\Psi}_{2,\gamma ,g}}{\hat{\Psi%
		}_{2,\gamma \gamma ,g}}\sum_{i\in I_{2,g}}\frac{\hat{\Psi}_{1,\gamma ,i}}{%
		\hat{\Psi}_{1,\gamma \gamma ,i}}\mathbb{E}_{T}[\psi _{1,\gamma }^{\ast
	}(z_{i,t})\psi _{2,\gamma }^{\ast }(z_{i,t})]  \notag \\
	& +n^{-1}\sum_{g\in \mathcal{G}_{2}}(\hat{\gamma}_{2,g}-\gamma _{2,g}^{\ast
	}+\hat{\Psi}_{2,\gamma \gamma ,g}^{-1}\hat{\Psi}_{2,\gamma ,g})\sum_{i\in
		I_{2,g}}(\hat{\gamma}_{1,i}-\gamma _{1,i}^{\ast }+\hat{\Psi}_{1,\gamma
		\gamma ,i}^{-1}\hat{\Psi}_{1,\gamma ,i})\mathbb{E}_{T}[\psi _{1,\gamma
	}^{\ast }(z_{i,t})\psi _{2,\gamma }^{\ast }(z_{i,t})]  \notag \\
	& -n^{-1}\sum_{g\in \mathcal{G}_{2}}\frac{\hat{\Psi}_{2,\gamma ,g}}{\hat{\Psi%
		}_{2,\gamma \gamma ,g}}\sum_{i\in I_{2,g}}(\hat{\gamma}_{1,i}-\gamma
	_{1,i}^{\ast }+\hat{\Psi}_{1,\gamma \gamma ,i}^{-1}\hat{\Psi}_{1,\gamma ,i})%
	\mathbb{E}_{T}[\psi _{1,\gamma }^{\ast }(z_{i,t})\psi _{2,\gamma }^{\ast
	}(z_{i,t})]  \notag \\
	& -n^{-1}\sum_{g\in \mathcal{G}_{2}}(\hat{\gamma}_{2,g}-\gamma _{2,g}^{\ast
	}+\hat{\Psi}_{2,\gamma \gamma ,g}^{-1}\hat{\Psi}_{2,\gamma ,g})\sum_{i\in
		I_{2,g}}\frac{\hat{\Psi}_{1,\gamma ,i}}{\hat{\Psi}_{1,\gamma \gamma ,i}}%
	\mathbb{E}_{T}[\psi _{1,\gamma }^{\ast }(z_{i,t})\psi _{2,\gamma }^{\ast
	}(z_{i,t})],  \label{P_C_L_Auxillary6b_4}
\end{align}%
where%
\begin{equation*}
	\hat{\Psi}_{j,\gamma \gamma ,g}\equiv (n_{j,g}T)^{-1}\sum_{i\in
		I_{j,g}}\sum_{t\leq T}\psi _{j,\gamma \gamma }(z_{i,t};\phi _{j,g}^{\ast }),
\end{equation*}%
for $j=1,2$, and any $g\in \mathcal{G}_{j}$. By the Cauchy-Schwarz
inequality,%
\begin{align}
	& \sum_{g\in \mathcal{G}_{2}}n_{2,g}^{2}\left( \frac{\hat{\Psi}_{2,\gamma ,g}%
	}{\hat{\Psi}_{2,\gamma \gamma ,g}}-\frac{\hat{\Psi}_{2,\gamma ,g}}{\mathbb{E}%
		[\hat{\Psi}_{2,\gamma \gamma ,g}]}\right) ^{2}  \notag \\
	& =\sum_{g\in \mathcal{G}_{2}}\frac{n_{2,g}^{2}\hat{\Psi}_{2,\gamma ,g}^{2}}{%
		\hat{\Psi}_{2,\gamma \gamma ,g}^{2}(\mathbb{E}[\hat{\Psi}_{2,\gamma \gamma
			,g}])^{2}}(\hat{\Psi}_{2,\gamma \gamma ,g}-\mathbb{E}[\hat{\Psi}_{2,\gamma
		\gamma ,g}])^{2}  \notag \\
	& \leq \frac{\left( \sum_{g\in \mathcal{G}_{2}}n_{2,g}^{2}\hat{\Psi}%
		_{2,\gamma ,g}^{4}\right) ^{1/2}\left( \sum_{g\in \mathcal{G}%
			_{2}}n_{2,g}^{2}(\hat{\Psi}_{2,\gamma \gamma ,g}-\mathbb{E}[\hat{\Psi}%
		_{2,\gamma \gamma ,g}])^{4}\right) ^{1/2}}{\min_{g\in \mathcal{G}_{2}}\hat{%
			\Psi}_{2,\gamma \gamma ,g}^{2}(\mathbb{E}[\hat{\Psi}_{2,\gamma \gamma
			,g}])^{2}},  \label{P_C_L_Auxillary6b_5}
\end{align}%
By Lemma \ref{C_L_Auxillary1}(viii) and Markov's inequality, we have%
\begin{equation}
	\sum_{g\in \mathcal{G}_{2}}n_{2,g}^{2}\hat{\Psi}_{2,\gamma
		,g}^{4}=O_{p}(G_{2}T^{-2}).  \label{P_C_L_Auxillary6b_5a}
\end{equation}%
Similarly, we can show that 
\begin{equation}
	\sum_{g\in \mathcal{G}_{2}}n_{2,g}^{2}(\hat{\Psi}_{2,\gamma \gamma ,g}-%
	\mathbb{E}[\hat{\Psi}_{2,\gamma \gamma ,g}])^{4}=O_{p}(G_{2}T^{-2}).
	\label{P_C_L_Auxillary6b_5b}
\end{equation}%
Combining the results from Assumption \ref{A4} and Lemma \ref{C_L_Auxillary1}%
(ii), (\ref{P_C_L_Auxillary6b_5}), (\ref{P_C_L_Auxillary6b_5a}) and (\ref%
{P_C_L_Auxillary6b_5b}), we establish:%
\begin{equation}
	\sum_{g\in \mathcal{G}_{2}}n_{2,g}^{2}\left( \frac{\hat{\Psi}_{2,\gamma ,g}}{%
		\hat{\Psi}_{2,\gamma \gamma ,g}}-\frac{\hat{\Psi}_{2,\gamma ,g}}{\mathbb{E}[%
		\hat{\Psi}_{2,\gamma \gamma ,g}]}\right) ^{2}=O_{p}(G_{2}T^{-2}).
	\label{P_C_L_Auxillary6b_6}
\end{equation}%
Similarly, we can show that%
\begin{align}
	& \sum_{g\in \mathcal{G}_{2}}\left( n_{2,g}^{-1}\sum_{i\in I_{2,g}}\left( 
	\frac{\hat{\Psi}_{1,\gamma ,i}}{\hat{\Psi}_{1,\gamma \gamma ,i}}-\frac{\hat{%
			\Psi}_{1,\gamma ,i}}{\mathbb{E}[\hat{\Psi}_{1,\gamma \gamma ,i}]}\right) 
	\mathbb{E}_{T}[\psi _{1,\gamma }^{\ast }(z_{i,t})\psi _{2,\gamma }^{\ast
	}(z_{i,t})]\right) ^{2}  \notag \\
	& \leq \sum_{g\in \mathcal{G}_{2}}n_{2,g}^{-1}\sum_{i\in I_{2,g}}\left( 
	\frac{\hat{\Psi}_{1,\gamma ,i}}{\hat{\Psi}_{1,\gamma \gamma ,i}}-\frac{\hat{%
			\Psi}_{1,\gamma ,i}}{\mathbb{E}[\hat{\Psi}_{1,\gamma \gamma ,i}]}\right)
	^{2}(\mathbb{E}_{T}[\psi _{1,\gamma }^{\ast }(z_{i,t})\psi _{2,\gamma
	}^{\ast }(z_{i,t})])^{2}  \notag \\
	& \leq \frac{K\sum_{g\in \mathcal{G}_{2}}n_{2,g}^{-1}\sum_{i\in I_{2,g}}\hat{%
			\Psi}_{1,\gamma ,i}^{2}(\hat{\Psi}_{1,\gamma \gamma ,i}-\mathbb{E}[\hat{\Psi}%
		_{1,\gamma \gamma ,i}])^{2}}{\min_{i\leq n}\hat{\Psi}_{1,\gamma \gamma
			,i}^{2}(\mathbb{E}[\hat{\Psi}_{1,\gamma \gamma ,i}])^{2}}=O_{p}(G_{2}T^{-2}).
	\label{P_C_L_Auxillary6b_7}
\end{align}%
By Assumptions\ \ref{A1}, \ref{C_A1}(iii), \ref{A4}, and Markov's
inequality, we have%
\begin{equation}
	\sum_{g\in \mathcal{G}_{2}}n_{2,g}^{2}\frac{\hat{\Psi}_{2,\gamma ,g}^{2}}{(%
		\mathbb{E}[\hat{\Psi}_{2,\gamma \gamma ,g}])^{2}}=O_{p}(nT^{-1})
	\label{P_C_L_Auxillary6b_8a}
\end{equation}%
and 
\begin{equation}
	\sum_{g\in \mathcal{G}_{2}}n_{2,g}^{-1}\left( \sum_{i\in I_{2,g}}\frac{%
		\mathbb{E}_{T}[\psi _{1,\gamma }^{\ast }(z_{i,t})\psi _{2,\gamma }^{\ast
		}(z_{i,t})]\hat{\Psi}_{1,\gamma ,i}}{\mathbb{E}[\hat{\Psi}_{1,\gamma \gamma
			,i}]}\right) ^{2}=O_{p}(G_{2}T^{-1}).  \label{P_C_L_Auxillary6b_8b}
\end{equation}%
Collecting the results in (\ref{P_C_L_Auxillary6b_6}), (\ref%
{P_C_L_Auxillary6b_7}), (\ref{P_C_L_Auxillary6b_8a}) and (\ref%
{P_C_L_Auxillary6b_8b}), and applying the Cauchy-Schwarz inequality, we
obtain an approximation to the first term on the right of (\ref%
{P_C_L_Auxillary6b_4}):%
\begin{align}
	& n^{-1}\sum_{g\in \mathcal{G}_{2}}\frac{\hat{\Psi}_{2,\gamma ,g}}{\hat{\Psi}%
		_{2,\gamma \gamma ,g}}\sum_{i\in I_{2,g}}\frac{\hat{\Psi}_{1,\gamma ,i}}{%
		\hat{\Psi}_{1,\gamma \gamma ,i}}\mathbb{E}_{T}[\psi _{1,\gamma }^{\ast
	}(z_{i,t})\psi _{2,\gamma }^{\ast }(z_{i,t})]  \notag \\
	& =n^{-1}\sum_{g\in \mathcal{G}_{2}}\frac{\hat{\Psi}_{2,\gamma ,g}}{\mathbb{E%
		}[\hat{\Psi}_{2,\gamma \gamma ,g}]}\sum_{i\in I_{2,g}}\frac{\hat{\Psi}%
		_{1,\gamma ,i}}{\mathbb{E}[\hat{\Psi}_{1,\gamma \gamma ,i}]}\mathbb{E}%
	_{T}[\psi _{1,\gamma }^{\ast }(z_{i,t})\psi _{2,\gamma }^{\ast }(z_{i,t})] 
	\notag \\
	& +n^{-1}\sum_{g\in \mathcal{G}_{2}}\frac{\hat{\Psi}_{2,\gamma ,g}}{\mathbb{E%
		}[\hat{\Psi}_{2,\gamma \gamma ,g}]}\sum_{i\in I_{2,g}}\left( \frac{\hat{\Psi}%
		_{1,\gamma ,i}}{\hat{\Psi}_{1,\gamma \gamma ,i}}-\frac{\hat{\Psi}_{1,\gamma
			,i}}{\mathbb{E}[\hat{\Psi}_{1,\gamma \gamma ,i}]}\right) \mathbb{E}_{T}[\psi
	_{1,\gamma }^{\ast }(z_{i,t})\psi _{2,\gamma }^{\ast }(z_{i,t})]  \notag \\
	& +n^{-1}\sum_{g\in \mathcal{G}_{2}}\left( \frac{\hat{\Psi}_{2,\gamma ,g}}{%
		\hat{\Psi}_{2,\gamma \gamma ,g}}-\frac{\hat{\Psi}_{2,\gamma ,g}}{\mathbb{E}[%
		\hat{\Psi}_{2,\gamma \gamma ,g}]}\right) \sum_{i\in I_{2,g}}\frac{\mathbb{E}%
		_{T}[\psi _{1,\gamma }^{\ast }(z_{i,t})\psi _{2,\gamma }^{\ast }(z_{i,t})]%
		\hat{\Psi}_{1,\gamma ,i}}{\mathbb{E}[\hat{\Psi}_{1,\gamma \gamma ,i}]} 
	\notag \\
	& +n^{-1}\sum_{g\in \mathcal{G}_{2}}\left( \frac{\hat{\Psi}_{2,\gamma ,g}}{%
		\hat{\Psi}_{2,\gamma \gamma ,g}}-\frac{\hat{\Psi}_{2,\gamma ,g}}{\mathbb{E}[%
		\hat{\Psi}_{2,\gamma \gamma ,g}]}\right) \sum_{i\in I_{2,g}}\left( \frac{%
		\hat{\Psi}_{1,\gamma ,i}}{\hat{\Psi}_{1,\gamma \gamma ,i}}-\frac{\hat{\Psi}%
		_{1,\gamma ,i}}{\mathbb{E}[\hat{\Psi}_{1,\gamma \gamma ,i}]}\right) \mathbb{E%
	}_{T}[\psi _{1,\gamma }^{\ast }(z_{i,t})\psi _{2,\gamma }^{\ast }(z_{i,t})] 
	\notag \\
	& =n^{-1}\sum_{g\in \mathcal{G}_{2}}\frac{\hat{\Psi}_{2,\gamma ,g}}{\mathbb{E%
		}[\hat{\Psi}_{2,\gamma \gamma ,g}]}\sum_{i\in I_{2,g}}\frac{\mathbb{E}%
		_{T}[\psi _{1,\gamma }^{\ast }(z_{i,t})\psi _{2,\gamma }^{\ast }(z_{i,t})]%
		\hat{\Psi}_{1,\gamma ,i}}{\mathbb{E}[\hat{\Psi}_{1,\gamma \gamma ,i}]}%
	+O_{p}(G_{2}^{1/2}(nT)^{-1}).  \label{P_C_L_Auxillary6b_9}
\end{align}%
Since $\mathbb{E}[\hat{\Psi}_{2,\gamma \gamma ,g}]=\Psi _{2,\gamma \gamma ,g}
$ and $\mathbb{E}[\hat{\Psi}_{1,\gamma \gamma ,i}]=\Psi _{2,\gamma \gamma ,i}
$, by the definition of $\sigma _{12,\gamma ,i}$, we have%
\begin{equation*}
	\sigma _{12,\gamma ,i}=\frac{\mathbb{E}_{T}[\psi _{1,\gamma }^{\ast
		}(z_{i,t})\psi _{2,\gamma }^{\ast }(z_{i,t})]}{(\mathbb{E}[\hat{\Psi}%
		_{2,\gamma \gamma ,g}]\mathbb{E}[\hat{\Psi}_{1,\gamma \gamma ,i}])^{1/2}}.
\end{equation*}%
Let $\tilde{\Psi}_{j,\gamma ,i}^{\ast }\equiv T^{-1/2}\sum_{t\leq T}\tilde{%
	\psi}_{j,\gamma }^{\ast }(z_{i,t})$. Then the leading term on the far RHS of
(\ref{P_C_L_Auxillary6b_9}) can be expressed as%
\begin{align}
	& n^{-1}\sum_{g\in \mathcal{G}_{2}}\frac{\hat{\Psi}_{2,\gamma ,g}}{\mathbb{E}%
		[\hat{\Psi}_{2,\gamma \gamma ,g}]}\sum_{i\in I_{2,g}}\frac{\mathbb{E}%
		_{T}[\psi _{1,\gamma }^{\ast }(z_{i,t})\psi _{2,\gamma }^{\ast }(z_{i,t})]%
		\hat{\Psi}_{1,\gamma ,i}}{\mathbb{E}[\hat{\Psi}_{1,\gamma \gamma ,i}]} 
	\notag \\
	& =n^{-1}\sum_{g\in \mathcal{G}_{2}}\frac{\hat{\Psi}_{2,\gamma ,g}-\mathbb{E}%
		[\hat{\Psi}_{2,\gamma ,g}]}{(\mathbb{E}[-\hat{\Psi}_{2,\gamma \gamma
			,g}])^{1/2}}\sum_{i\in I_{2,g}}\frac{\hat{\Psi}_{1,\gamma ,i}-\mathbb{E}[%
		\hat{\Psi}_{1,\gamma ,i}]}{(\mathbb{E}[-\hat{\Psi}_{1,\gamma \gamma
			,i}])^{1/2}}\sigma _{12,\gamma ,i}  \notag \\
	& =(nT)^{-1}\sum_{g\in \mathcal{G}_{2}}n_{2,g}^{-1}\sum_{i\in I_{2,g}}\tilde{%
		\Psi}_{2,\gamma ,i}^{\ast }\sum_{i^{\prime }\in I_{2,g}}\tilde{\Psi}%
	_{1,\gamma ,i^{\prime }}^{\ast }\sigma _{12,\gamma ,i^{\prime }}  \notag \\
	& =(nT)^{-1}\sum_{g\in \mathcal{G}_{2}}\sum_{i\in I_{2,g}}\frac{\sigma
		_{12,\gamma ,i}\tilde{\Psi}_{1,\gamma ,i}^{\ast }\tilde{\Psi}_{2,\gamma
			,i}^{\ast }}{n_{2,g}}+(nT)^{-1}\sum_{g\in \mathcal{G}_{2}}\sum_{i\in
		I_{2,g}}\sum_{i^{\prime }\in I_{2,g},i^{\prime }\neq i}\frac{\sigma
		_{12,\gamma ,i}\tilde{\Psi}_{1,\gamma ,i^{\prime }}^{\ast }\tilde{\Psi}%
		_{2,\gamma ,i^{\prime }}^{\ast }}{n_{2,g}}.  \label{P_C_L_Auxillary6b_10}
\end{align}%
By Assumption \ref{A1}, H\"{o}lder's inequality and Lemma \ref%
{C_L_Auxillary1}(viii), it follows that 
\begin{align*}
	& \mathbb{E}\left[ \left( (nT)^{-1}\sum_{g\in \mathcal{G}_{2}}\sum_{i\in
		I_{2,g}}\frac{\sigma _{12,\gamma ,i}}{n_{2,g}}(\tilde{\Psi}_{1,\gamma
		,i}^{\ast }\tilde{\Psi}_{2,\gamma ,i}^{\ast }-\mathbb{E}[\tilde{\Psi}%
	_{1,\gamma ,i}^{\ast }\tilde{\Psi}_{2,\gamma ,i}^{\ast }])\right) ^{2}\right]
	\\
	& \leq (nT)^{-2}\sum_{g\in \mathcal{G}_{2}}\sum_{i\in I_{2,g}}\frac{\sigma
		_{12,\gamma ,i}^{2}}{n_{2,g}^{2}}\mathbb{E}[(\tilde{\Psi}_{1,\gamma
		,i}^{\ast }\tilde{\Psi}_{2,\gamma ,i}^{\ast })^{2}]\leq KG_{2}(nT)^{-2},
\end{align*}%
where we write $\tilde{\Psi}_{2,\gamma ,i}^{\ast }\equiv \tilde{\Psi}%
_{2,\gamma ,g}^{\ast }$ for any $i\in I_{2,g}$ and any $g\in \mathcal{G}_{2}$%
. This along with Markov's inequality implies that 
\begin{equation}
	(nT)^{-1}\sum_{g\in \mathcal{G}_{2}}\sum_{i\in I_{2,g}}\frac{\sigma
		_{12,\gamma ,i}}{n_{2,g}}(\tilde{\Psi}_{1,\gamma ,i}^{\ast }\tilde{\Psi}%
	_{2,\gamma ,i}^{\ast }-\mathbb{E}[\tilde{\Psi}_{1,\gamma ,i}^{\ast }\tilde{%
		\Psi}_{2,\gamma ,i}^{\ast }])=O_{p}(G_{2}^{1/2}(nT)^{-1}).
	\label{P_C_L_Auxillary6b_11}
\end{equation}%
Since\ $\mathbb{E}[\tilde{\Psi}_{1,\gamma ,i}^{\ast }\tilde{\Psi}_{2,\gamma
	,i}^{\ast }]=T^{-1}\sum_{t,t^{\prime }\leq T}\mathbb{E}[\tilde{\psi}%
_{1,\gamma }^{\ast }(z_{i,t})\tilde{\psi}_{2,\gamma }^{\ast }(z_{i,t^{\prime
}})]$, from (\ref{P_C_L_Auxillary6b_11}) it follows that the first term on
the far RHS\ of (\ref{P_C_L_Auxillary6b_10}) can be alternatively written
as: 
\begin{eqnarray}
	&&(nT)^{-1}\sum_{g\in \mathcal{G}_{2}}\sum_{i\in I_{2,g}}\frac{\sigma
		_{12,\gamma ,i}\tilde{\Psi}_{1,\gamma ,i}^{\ast }\tilde{\Psi}_{2,\gamma
			,i}^{\ast }}{n_{2,g}}  \notag \\
	&=&(nT^{2})^{-1}\sum_{g\in \mathcal{G}_{2}}n_{2,g}^{-1}\sum_{i\in
		I_{2,g}}\sigma _{12,\gamma ,i}\left( \sum_{t,t^{\prime }\leq T}\mathbb{E}[%
	\tilde{\psi}_{1,\gamma }^{\ast }(z_{i,t})\tilde{\psi}_{2,\gamma }^{\ast
	}(z_{i,t^{\prime }})]\right) +O_{p}(G_{2}^{1/2}(nT)^{-1}).
	\label{P_C_L_Auxillary6b_12}
\end{eqnarray}%
For the second term on the RHS of the second equality in (\ref%
{P_C_L_Auxillary6b_10}), we can write it as 
\begin{align}
	& (nT)^{-1}\sum_{g\in \mathcal{G}_{2}}\sum_{i\in I_{2,g}}\sum_{i^{\prime
		}\in I_{2,g},i^{\prime }\neq i}\frac{\sigma _{12,\gamma ,i}\tilde{\Psi}%
		_{1,\gamma ,i}^{\ast }\tilde{\Psi}_{2,\gamma ,i^{\prime }}^{\ast }}{n_{2,g}}
	\notag \\
	& =(nT)^{-1}\sum_{g\in \mathcal{G}_{2}}\sum_{i\in I_{2,g}}\sum_{i^{\prime
		}\in I_{2,g},i^{\prime }<i}\left( \frac{\sigma _{12,\gamma ,i}\tilde{\Psi}%
		_{1,\gamma ,i}^{\ast }\tilde{\Psi}_{2,\gamma ,i^{\prime }}^{\ast }}{n_{2,g}}+%
	\frac{\sigma _{12,\gamma ,i^{\prime }}\tilde{\Psi}_{1,\gamma ,i^{\prime
		}}^{\ast }\tilde{\Psi}_{2,\gamma ,i}^{\ast }}{n_{2,g}}\right) .
	\label{P_C_L_Auxillary6b_13}
\end{align}%
By Assumption \ref{A1}, Lemma \ref{C_L_Auxillary1}(viii) and H\"{o}lder's
inequality, we have%
\begin{align*}
	& \mathbb{E}\left[ \left( (nT)^{-1}\sum_{g\in \mathcal{G}_{2}}\sum_{i\in
		I_{2,g}}\sum_{i^{\prime }\in I_{2,g},i^{\prime }<i}\frac{\sigma _{12,\gamma
			,i}\tilde{\Psi}_{1,\gamma ,i}^{\ast }\tilde{\Psi}_{2,\gamma ,i^{\prime
		}}^{\ast }}{n_{2,g}}\right) ^{2}\right]  \\
	& =(nT)^{-2}\sum_{g\in \mathcal{G}_{2}}\sum_{i\in I_{2,g}}\sum_{i^{\prime
		}\in I_{2,g},i^{\prime }<i}\frac{\sigma _{12,\gamma ,i}^{2}\mathbb{E}[(%
		\tilde{\Psi}_{1,\gamma ,i}^{\ast }\tilde{\Psi}_{2,\gamma ,i^{\prime }}^{\ast
		})^{2}]}{n_{2,g}^{2}}\leq G_{2}(nT)^{-2},
\end{align*}%
which together with Markov's inequality implies that 
\begin{equation}
	(nT)^{-1}\sum_{g\in \mathcal{G}_{2}}\sum_{i\in I_{2,g}}\sum_{i^{\prime }\in
		I_{2,g},i^{\prime }<i}\frac{\sigma _{12,\gamma ,i}\tilde{\Psi}_{1,\gamma
			,i}^{\ast }\tilde{\Psi}_{2,\gamma ,i^{\prime }}^{\ast }}{n_{2,g}}%
	=O_{p}(G_{2}^{1/2}(nT)^{-1}).  \label{P_C_L_Auxillary6b_14}
\end{equation}%
Similarly, we can show that 
\begin{equation}
	(nT)^{-1}\sum_{g\in \mathcal{G}_{2}}\sum_{i\in I_{2,g}}\sum_{i^{\prime }\in
		I_{2,g},i^{\prime }<i}\frac{\sigma _{12,\gamma ,i^{\prime }}\tilde{\Psi}%
		_{1,\gamma ,i^{\prime }}^{\ast }\tilde{\Psi}_{2,\gamma ,i}^{\ast }}{n_{2,g}}%
	=O_{p}(G_{2}^{1/2}(nT)^{-1}).  \label{P_C_L_Auxillary6b_15}
\end{equation}%
Collecting the results from (\ref{P_C_L_Auxillary6b_10}), and (\ref%
{P_C_L_Auxillary6b_12})-(\ref{P_C_L_Auxillary6b_15}) yields an approximation
to the first term on the right of (\ref{P_C_L_Auxillary6b_4}):%
\begin{eqnarray}
	&&n^{-1}\sum_{g\in \mathcal{G}_{2}}\frac{\hat{\Psi}_{2,\gamma ,g}}{\hat{\Psi}%
		_{2,\gamma \gamma ,g}}\sum_{i\in I_{2,g}}\frac{\hat{\Psi}_{1,\gamma ,i}}{%
		\hat{\Psi}_{1,\gamma \gamma ,i}}\mathbb{E}_{T}[\psi _{1,\gamma }^{\ast
	}(z_{i,t})\psi _{2,\gamma }^{\ast }(z_{i,t})]  \notag \\
	&=&(nT^{2})^{-1}\sum_{g\in \mathcal{G}_{2}}n_{2,g}^{-1}\sum_{i\in
		I_{2,g}}\sigma _{12,\gamma ,i}\left( \sum_{t,t^{\prime }\leq T}\mathbb{E}[%
	\tilde{\psi}_{1,\gamma }^{\ast }(z_{i,t})\tilde{\psi}_{2,\gamma }^{\ast
	}(z_{i,t^{\prime }})]\right) +O_{p}(G_{2}^{1/2}(nT)^{-1}).
	\label{P_C_L_Auxillary6b_16}
\end{eqnarray}

We now consider the second term on the right of (\ref{P_C_L_Auxillary6b_4}).
By the Cauchy-Schwarz inequality, Assumption \ref{A3} and Lemma \ref%
{C_L_Auxillary1}(vii), we have 
\begin{align}
	& \left\vert n^{-1}\sum_{g\in \mathcal{G}_{2}}(\hat{\gamma}_{2,g}-\gamma
	_{2,g}^{\ast }+\hat{\Psi}_{2,\gamma \gamma ,g}^{-1}\hat{\Psi}_{2,\gamma
		,g})\sum_{i\in I_{2,g}}(\hat{\gamma}_{1,i}-\gamma _{1,i}^{\ast }+\hat{\Psi}%
	_{1,\gamma \gamma ,i}^{-1}\hat{\Psi}_{1,\gamma ,i})\mathbb{E}_{T}[\psi
	_{1,\gamma }^{\ast }(z_{i,t})\psi _{2,\gamma }^{\ast }(z_{i,t})]\right\vert 
	\notag \\
	& \leq Kn^{-1}\left( \sum_{g\in \mathcal{G}_{2}}n_{2,g}(\hat{\gamma}%
	_{2,g}-\gamma _{2,g}^{\ast }+\hat{\Psi}_{2,\gamma \gamma ,g}^{-1}\hat{\Psi}%
	_{2,\gamma ,g})^{2}\right) ^{1/2}  \notag \\
	& \times \left( \sum_{i\leq n}(\hat{\gamma}_{1,i}-\gamma _{1,i}^{\ast }+\hat{%
		\Psi}_{1,\gamma \gamma ,i}^{-1}\hat{\Psi}_{1,\gamma ,i})^{2}\right) ^{1/2}%
	\overset{}{=}O_{p}((nT)^{-1}).  \label{P_C_L_Auxillary6b_17}
\end{align}

As for the third term on the right of (\ref{P_C_L_Auxillary6b_4}), we have
by Assumption \ref{A3}, Lemma \ref{C_L_Auxillary1}(ii, vi, viii), and
Markov's inequality:%
\begin{align}
	& \left\vert n^{-1}\sum_{g\in\mathcal{G}_{2}}\frac{\hat{\Psi}_{2,\gamma,g}}{%
		\hat{\Psi}_{2,\gamma\gamma,g}}\sum_{i\in I_{2,g}}(\hat{\gamma}%
	_{1,i}-\gamma_{1,i}^{\ast}+\hat{\Psi}_{1,\gamma\gamma,i}^{-1}\hat{\Psi}%
	_{1,\gamma ,i})\sigma_{12,\gamma,i}\right\vert  \notag \\
	& \leq n^{-1}\left( \sum_{g\in\mathcal{G}_{2}}\frac{n_{2,g}\hat{\Psi }%
		_{2,\gamma,g}^{2}}{\hat{\Psi}_{2,\gamma\gamma,g}^{2}}\right) ^{1/2}\left(
	\sum_{i\leq n}(\hat{\gamma}_{1,i}-\gamma_{1,i}^{\ast}+\hat{\Psi}%
	_{1,\gamma\gamma,i}^{-1}\hat{\Psi}_{1,\gamma,i})^{2}\sigma_{12,\gamma,i}^{2}%
	\right) ^{1/2}  \notag \\
	& =O_{p}(G_{2}^{1/2}(nT)^{-1}).  \label{P_C_L_Auxillary6b_18}
\end{align}
Finally, the last term on the right of (\ref{P_C_L_Auxillary6b_4}) can be
bounded by%
\begin{align}
	& \left\vert n^{-1}\sum_{g\in\mathcal{G}_{2}}(\hat{\gamma}_{2,g}-\gamma
	_{2,g}^{\ast}+\hat{\Psi}_{2,\gamma\gamma,g}^{-1}\hat{\Psi}%
	_{2,\gamma,g})\sum_{i\in I_{2,g}}\frac{\hat{\Psi}_{1,\gamma,i}}{\hat{\Psi}%
		_{1,\gamma \gamma,i}}\sigma_{12,\gamma,i}\right\vert  \notag \\
	& \leq n^{-1}\left( \sum_{g\in\mathcal{G}_{2}}n_{2,g}(\hat{\gamma}%
	_{2,g}-\gamma_{2,g}^{\ast}+\hat{\Psi}_{2,\gamma\gamma,g}^{-1}\hat{\Psi }%
	_{2,\gamma,g})^{2}\right) ^{1/2}\left( \sum_{g\in\mathcal{G}_{2}}\sum_{i\in
		I_{2,g}}\frac{\hat{\Psi}_{1,\gamma,i}^{2}}{\hat{\Psi}_{1,\gamma\gamma,i}^{2}}%
	\sigma_{12,\gamma,i}^{2}\right) ^{1/2}  \notag \\
	& =O_{p}(n^{-1}T^{-1/2}).  \label{P_C_L_Auxillary6b_19}
\end{align}
The claim of the lemma now follows from the results in (\ref%
{P_C_L_Auxillary6b_4}), (\ref{P_C_L_Auxillary6b_16}), (\ref%
{P_C_L_Auxillary6b_17}), (\ref{P_C_L_Auxillary6b_18}) and (\ref%
{P_C_L_Auxillary6b_19}).\hfill$Q.E.D.$

\bigskip

\begin{lemma}
	\textit{\label{C_L_Auxillary6}\ }Under\ Assumptions \ref{A1} and \ref{C_A1},
	we have%
	\begin{align*}
		& (nT)^{-1}\sum_{i\leq n}\sum_{t\leq T}(\psi _{1,\phi }(z_{i,t})(\hat{\phi}%
		_{1,i}-\phi _{1,i}^{\ast })-\psi _{2,\phi }(z_{i,t})(\hat{\phi}_{2,i}-\phi
		_{2,i}^{\ast }))^{2} \\
		& =(nT)^{-1}\sum_{g\in \mathcal{G}_{2}}\left( 
		\begin{array}{c}
			\sum_{i\in I_{2,g}}\sigma _{1,\gamma ,i}^{4}+n_{2,g}^{-2}\sum_{i\in
				I_{j,g}}s_{j,\gamma ,i}^{2}\sum_{i\in I_{j,g}}\mathbb{E}\left[ \left(
			T^{-1/2}\sum_{t\leq T}\tilde{\psi}_{j,\gamma }^{\ast }\left( z_{i,t}\right)
			\right) ^{2}\right]  \\ 
			-2n_{2,g}^{-1}\sum_{i\in I_{2,g}}\sigma _{12,\gamma ,i}\left(
			T^{-1}\sum_{t,t^{\prime }\leq T}\mathbb{E}[\tilde{\psi}_{1,\gamma }^{\ast
			}(z_{i,t})\tilde{\psi}_{2,\gamma }^{\ast }(z_{i,t^{\prime }})]\right) 
		\end{array}%
		\right) +O_{p}(T^{-3/2}).
	\end{align*}
\end{lemma}

\noindent\textsc{Proof of Lemma \ref{C_L_Auxillary6}}. From the expression
of $\psi_{j,\phi}(z_{i,t})(\hat{\phi}_{j,i}-\phi_{j,i}^{\ast})$ in (\ref%
{P_C_L_Auxillary5_00}), we have%
\begin{align}
	& (nT)^{-1}\sum_{i\leq n}\sum_{t\leq T}(\psi_{1,\phi}(z_{i,t})(\hat{\phi }%
	_{1,i}-\phi_{1,i}^{\ast})-\psi_{2,\phi}(z_{i,t})(\hat{\phi}_{2,i}-\phi
	_{2,i}^{\ast}))^{2}  \notag \\
	& =(nT)^{-1}\sum_{i\leq n}\sum_{t\leq T}(\Delta\psi_{\theta}(z_{i,t})(\hat{%
		\theta}-\theta^{\ast}))^{2}+(nT)^{-1}\sum_{i\leq n}\sum_{t\leq
		T}(\Delta\psi_{\gamma}(z_{i,t})(\hat{\gamma}_{i}-\gamma_{i}^{\ast}))^{2} 
	\notag \\
	& +2(nT)^{-1}\sum_{i\leq n}\sum_{t\leq T}(\hat{\theta}-\theta^{\ast})^{\top
	}\Delta\psi_{\theta}(z_{i,t})^{\top}\Delta\psi_{\gamma}(z_{i,t})(\hat{\gamma 
	}_{i}-\gamma_{i}^{\ast}),  \label{P_C_L_Auxillary6_1}
\end{align}
where%
\begin{align*}
	\Delta\psi_{\theta}(z_{i,t})(\hat{\theta}-\theta^{\ast}) & \equiv
	\psi_{1,\theta}(z_{i,t})(\hat{\theta}_{1}-\theta_{1}^{\ast})-\psi_{2,\theta
	}(z_{i,t})(\hat{\theta}_{2}-\theta_{2}^{\ast}), \\
	\Delta\psi_{\gamma}(z_{i,t})(\hat{\gamma}_{i}-\gamma_{i}^{\ast}) &
	\equiv\psi_{1,\gamma}(z_{i,t})(\hat{\gamma}_{1,i}-\gamma_{1,i}^{\ast})-%
	\psi_{2,\gamma}(z_{i,t})(\hat{\gamma}_{2,i}-\gamma_{2,i}^{\ast}).
\end{align*}
By the Cauchy-Schwarz inequality,%
\begin{equation}
	(nT)^{-1}\sum_{i\leq n}\sum_{t\leq T}(\psi_{j,\theta}(z_{i,t})(\hat{\theta }%
	_{j}-\theta_{j}^{\ast}))^{2}\leq||\hat{\theta}_{j}-\theta_{j}^{%
		\ast}||^{2}(nT)^{-1}\sum_{i\leq n}\sum_{t\leq T}\left\Vert
	\psi_{j,\theta}(z_{i,t})\right\Vert ^{2}.  \label{P_C_L_Auxillary6_2}
\end{equation}
Since $(nT)^{-1}\sum_{i\leq n}\sum_{t\leq T}\left\Vert
\psi_{j,\theta}(z_{i,t})\right\Vert ^{2}=O_{p}(1)$ by Assumption \ref{A3}
and Markov's inequality, it follows from Lemma \ref{C_L_Auxillary1}(iii) and
(\ref{P_C_L_Auxillary6_2}) that for both $j=1,2$:%
\begin{equation*}
	(nT)^{-1}\sum_{i\leq n}\sum_{t\leq T}(\psi_{j,\theta}(z_{i,t})(\hat{\theta }%
	_{j}-\theta_{j}^{\ast}))^{2}=O_{p}((nT)^{-1}),
\end{equation*}
which establishes:%
\begin{equation}
	(nT)^{-1}\sum_{t\leq T}(\Delta\psi_{\theta}(z_{i,t})(\hat{\theta}-\theta
	^{\ast}))^{2}=O_{p}((nT)^{-1}).  \label{P_C_L_Auxillary6_3}
\end{equation}

For $j,j^{\prime }=1,2$, we can apply the triangle inequality and
Cauchy-Schwarz inequality to get%
\begin{align}
	& \left\vert (nT)^{-1}\sum_{i\leq n}\sum_{t\leq T}\psi _{j,\theta }(z_{i,t})(%
	\hat{\theta}_{j}-\theta _{j}^{\ast })\psi _{j^{\prime },\gamma }(z_{i,t})(%
	\hat{\gamma}_{j^{\prime },i}-\gamma _{j^{\prime },i}^{\ast })\right\vert 
	\notag \\
	& \leq ||\hat{\theta}_{j}-\theta _{j}^{\ast }||(nT)^{-1}\sum_{i\leq n}|\hat{%
		\gamma}_{j^{\prime },i}-\gamma _{j^{\prime },i}^{\ast }|\sum_{t\leq
		T}\left\Vert \psi _{j,\theta }(z_{i,t})\psi _{j^{\prime },\gamma
	}(z_{i,t})\right\Vert  \notag \\
	& \leq ||\hat{\theta}_{j}-\theta _{j}^{\ast }||\left( n^{-1}\sum_{i\leq n}|%
	\hat{\gamma}_{j^{\prime },i}-\gamma _{j^{\prime },i}^{\ast }|^{2}\right)
	^{1/2}\left( (nT)^{-1}\sum_{i\leq n}\sum_{t\leq T}\left\Vert \psi _{j,\theta
	}(z_{i,t})\psi _{j^{\prime },\gamma }(z_{i,t})\right\Vert ^{2}\right) ^{1/2}.
	\label{P_C_L_Auxillary6_4}
\end{align}%
Since by Assumption \ref{A3} and Markov's inequality $(nT)^{-1}\sum_{i\leq
	n}\sum_{t\leq T}||\psi _{j,\theta }(z_{i,t})\psi _{j^{\prime },\gamma
}(z_{i,t})||^{2}=O_{p}(1)$, it follows from (\ref{P_C_L_Auxillary6_4}),
Lemma \ref{C_L_Auxillary1}(iii, iv) that%
\begin{equation}
	(nT)^{-1}\sum_{i\leq n}\sum_{t\leq T}\psi _{j,\theta }(z_{i,t})(\hat{\theta}%
	_{j}-\theta _{j}^{\ast })\psi _{j^{\prime },\gamma }(z_{i,t})(\hat{\gamma}%
	_{j^{\prime },i}-\gamma _{j^{\prime },i}^{\ast })=O_{p}((nT^{2})^{-1/2}).
	\label{P_C_L_Auxillary6_5}
\end{equation}%
Combining the results in (\ref{P_C_L_Auxillary6_1}), (\ref%
{P_C_L_Auxillary6_3}) and (\ref{P_C_L_Auxillary6_5}), we obtain%
\begin{align}
	& (nT)^{-1}\sum_{i\leq n}\sum_{t\leq T}(\psi _{1,\phi }(z_{i,t})(\hat{\phi}%
	_{1,i}-\phi _{1,i}^{\ast })-\psi _{2,\phi }(z_{i,t})(\hat{\phi}_{2,i}-\phi
	_{2,i}^{\ast }))^{2}  \notag \\
	& =(nT)^{-1}\sum_{i\leq n}(\hat{\gamma}_{1,i}-\gamma _{1,i}^{\ast
	})^{2}\sum_{t\leq T}\psi _{1,\gamma }^{2}(z_{i,t})+\sum_{i\leq n}(\hat{\gamma%
	}_{2,i}-\gamma _{2,i}^{\ast })^{2}\sum_{t\leq T}\psi _{2,\gamma
	}^{2}(z_{i,t})  \notag \\
	& -2(nT)^{-1}\sum_{i\leq n}(\hat{\gamma}_{1,i}-\gamma _{1,i}^{\ast })(\hat{%
		\gamma}_{2,i}-\gamma _{2,i}^{\ast })\sum_{t\leq T}\psi _{1,\gamma
	}(z_{i,t})\psi _{2,\gamma }(z_{i,t})+O_{p}((nT^{2})^{-1/2})
	\label{P_C_L_Auxillary6_6}
\end{align}%
which together with Lemma \ref{C_L_Auxillary6a} and Lemma \ref%
{C_L_Auxillary6b}, and Assumption \ref{A1}(i) establishes the claim of the
lemma.\hfill $Q.E.D.$

\bigskip

\begin{lemma}
	\textit{\label{C_L_Auxillary7a}\ }Under Assumptions \ref{A1} and \ref{C_A1}%
	,\ we have%
	\begin{eqnarray}
		(nT)^{-1}\sum_{i\leq n}\sum_{t\leq T}\Delta \psi (z_{i,t},\hat{\phi}%
		_{i})^{2} &=&(nT)^{-1}\sum_{i\leq n}\sum_{t\leq T}\mathbb{E}[\Delta \psi
		(z_{i,t})^{2}]  \notag \\
		&&+2(nT)^{-1}\sum_{i\leq n}\sum_{t\leq T}\Delta \psi (z_{i,t})\Delta \psi
		_{\phi }(z_{i,t};\phi _{i}^{\ast })(\hat{\phi}_{i}-\phi _{i}^{\ast })  \notag
		\\
		&&+(nT)^{-1}\sum_{i\leq n}\sum_{t\leq T}(\hat{\phi}_{i}-\phi _{i}^{\ast
		})^{\top }\Delta \psi _{\phi }(z_{i,t};\phi _{i}^{\ast })^{\top }\Delta \psi
		_{\phi }(z_{i,t};\phi _{i}^{\ast })(\hat{\phi}_{i}-\phi _{i}^{\ast })  \notag
		\\
		&&+O_{p}(T^{-3/2}+\omega _{n,T}T^{-1}+\omega _{n,T}^{2}(nT)^{-1/2})  \notag
		\\
		&=&(nT)^{-1}\sum_{i\leq n}\sum_{t\leq T}\mathbb{E}[\Delta \psi
		(z_{i,t})^{2}]+O_{p}(T^{-1/2}+\omega _{n,T}T^{-1}+\omega
		_{n,T}^{2}(nT)^{-1/2}).  \notag \\
		&&  \label{C_L_Auxillary7a_1}
	\end{eqnarray}
\end{lemma}

\noindent {\textsc{Proof of Lemma \ref{C_L_Auxillary7a}}}. Recall that $%
\Delta \psi (z_{i,t};\phi _{i})\equiv \psi _{1}(z_{i,t};\phi _{1,i})-\psi
_{2}(z_{i,t};\phi _{2,i})$ where $\phi _{i}\equiv (\phi _{1,i}^{\top },\phi
_{2,i}^{\top })^{\top }$. Define 
\begin{equation*}
	\Delta \psi _{\phi }(z_{i,t};\phi _{i})\equiv \frac{\partial \Delta \psi
		_{\phi }(z_{i,t};\phi _{i})}{\partial \phi _{i}}\text{ \ and \ }\Delta \psi
	_{\phi \phi }(z_{i,t};\phi _{i})\equiv \frac{\partial ^{2}\Delta \psi _{\phi
		}(z_{i,t};\phi _{i})}{\partial \phi _{i}\partial \phi _{i}^{^{\top }}}.
\end{equation*}%
Applying the Taylor expansion, we obtain%
\begin{align}
	& (nT)^{-1}\sum_{i\leq n}\sum_{t\leq T}(\psi _{1}(z_{i,t};\hat{\phi}%
	_{1,i})-\psi _{2}(z_{i,t};\hat{\phi}_{2,i}))^{2}  \notag \\
	& =(nT)^{-1}\sum_{i\leq n}\sum_{t\leq T}\Delta \psi
	(z_{i,t})^{2}+2(nT)^{-1}\sum_{i\leq n}\sum_{t\leq T}\Delta \psi
	(z_{i,t})\Delta \psi _{\phi }(z_{i,t};\phi _{i}^{\ast })(\hat{\phi}_{i}-\phi
	_{i}^{\ast })  \notag \\
	& +(nT)^{-1}\sum_{i\leq n}\sum_{t\leq T}(\hat{\phi}_{i}-\phi _{i}^{\ast
	})^{\top }\Delta \psi _{\phi }(z_{i,t};\tilde{\phi}_{i})^{\top }\Delta \psi
	_{\phi }(z_{i,t};\tilde{\phi}_{i})(\hat{\phi}_{i}-\phi _{i}^{\ast })  \notag
	\\
	& +(nT)^{-1}\sum_{i\leq n}\sum_{t\leq T}\Delta \psi (z_{i,t},\tilde{\phi}%
	_{i})(\hat{\phi}_{i}-\phi _{i}^{\ast })^{\top }\Delta \psi _{\phi \phi
	}(z_{i,t};\tilde{\phi}_{i})(\hat{\phi}_{i}-\phi _{i}^{\ast }),
	\label{P_C_L_Auxillary7_1}
\end{align}%
where $\tilde{\phi}_{i}$ is a mean value in between $\phi _{i}^{\ast }$ and $%
\hat{\phi}_{i}$. By Assumptions \ref{A1} and \ref{C_A1}(ii), it follows that%
\begin{eqnarray*}
	&&\mathbb{E}\left[ \left( \frac{(nT)^{-1}\sum_{i\leq n}\sum_{t\leq T}(\Delta
		\psi (z_{i,t})^{2}-\mathbb{E}[\Delta \psi (z_{i,t})^{2}])}{\omega _{n,T}^{2}}%
	\right) ^{2}\right] \\
	&=&(nT)^{-2}\sum_{i\leq n}\mathbb{E}\left[ \left( \sum_{t\leq T}\frac{\Delta
		\psi (z_{i,t})^{2}-\mathbb{E}[\Delta \psi (z_{i,t})^{2}]}{\omega _{n,T}^{2}}%
	\right) ^{2}\right] \\
	&\leq &K(n^{2}T)^{-1}\sum_{i\leq n}\max_{t\leq T}\left\Vert (\Delta \psi
	(z_{i,t})/\omega _{n,T})^{2}\right\Vert _{2+\delta /2}\leq K(nT)^{-1},
\end{eqnarray*}%
which together with Markov's inequality implies that 
\begin{equation}
	(nT)^{-1}\sum_{i\leq n}\sum_{t\leq T}\Delta \psi
	(z_{i,t})^{2}=(nT)^{-1}\sum_{i\leq n}\sum_{t\leq T}\mathbb{E}[\Delta \psi
	(z_{i,t})^{2}]+O_{p}(\omega _{n,T}^{2}(nT)^{-1/2}).
	\label{P_C_L_Auxillary7_2}
\end{equation}%
In view of (\ref{P_C_L_Auxillary7_1}) and (\ref{P_C_L_Auxillary7_2}), the
first equality in\ (\ref{C_L_Auxillary7a_1}) follows if%
\begin{align}
	& (nT)^{-1}\sum_{i\leq n}\sum_{t\leq T}(\hat{\phi}_{i}-\phi _{i}^{\ast
	})^{\top }\Delta \psi _{\phi }(z_{i,t};\tilde{\phi}_{i})^{\top }\Delta \psi
	_{\phi }(z_{i,t};\tilde{\phi}_{i})(\hat{\phi}_{i}-\phi _{i}^{\ast })  \notag
	\\
	& =(nT)^{-1}\sum_{i\leq n}\sum_{t\leq T}(\hat{\phi}_{i}-\phi _{i}^{\ast
	})^{\top }\Delta \psi _{\phi }(z_{i,t};\phi _{i}^{\ast })^{\top }\Delta \psi
	_{\phi }(z_{i,t};\phi _{i}^{\ast })(\hat{\phi}_{i}-\phi _{i}^{\ast
	})+O_{p}(T^{-3/2}),  \label{P_C_L_Auxillary7_3}
\end{align}%
and%
\begin{equation}
	(nT)^{-1}\sum_{i\leq n}\sum_{t\leq T}\Delta \psi (z_{i,t};\tilde{\phi}_{i})(%
	\hat{\phi}_{i}-\phi _{i}^{\ast })^{\top }\Delta \psi _{\phi \phi }(z_{i,t};%
	\tilde{\phi}_{i})(\hat{\phi}_{i}-\phi _{i}^{\ast })=O_{p}(T^{-3/2}+\omega
	_{n,T}T^{-1}).  \label{P_C_L_Auxillary7_4}
\end{equation}

By the definition of $\Delta \psi _{\phi }(z_{i,t},\phi _{i})$, we can
write: 
\begin{equation*}
	\Delta \psi _{\phi }(z_{i,t};\tilde{\phi}_{i})(\hat{\phi}_{i}-\phi
	_{i}^{\ast })=\psi _{1,\phi }(z_{i,t};\tilde{\phi}_{1,i})(\hat{\phi}%
	_{1,i}-\phi _{1,i}^{\ast })-\psi _{2,\phi }(z_{i,t};\tilde{\phi}_{2,i})(\hat{%
		\phi}_{2,i}-\phi _{2,i}^{\ast }).
\end{equation*}%
Therefore, (\ref{P_C_L_Auxillary7_3}) follows if for $j,j^{\prime }=1,2$ the
following hold: 
\begin{equation}
	(nT)^{-1}\sum_{i\leq n}\sum_{t\leq T}(\hat{\phi}_{j,i}-\phi _{j,i}^{\ast
	})^{\top }\left( 
	\begin{array}{c}
		\psi _{j,\phi }(z_{i,t};\tilde{\phi}_{j,i})^{\top }\psi _{j^{\prime },\phi
		}(z_{i,t};\tilde{\phi}_{j^{\prime },i}) \\ 
		-\psi _{j,\phi }(z_{i,t})\psi _{j^{\prime },\phi }(z_{i,t})%
	\end{array}%
	\right) (\hat{\phi}_{j^{\prime },i}-\phi _{j^{\prime },i}^{\ast
	})=O_{p}(T^{-3/2}).  \label{P_C_L_Auxillary7_5}
\end{equation}%
We next verify (\ref{P_C_L_Auxillary7_5}). By Lemma \ref{C_L_Auxillary1}%
(iii, iv), 
\begin{equation}
	n^{-1}\sum_{i\leq n}||\hat{\phi}_{j,i}-\phi _{j,i}^{\ast }||^{4}\leq K\left(
	||\hat{\theta}_{j}-\theta _{j}^{\ast }||^{4}+n^{-1}\sum_{g\in \mathcal{G}%
		_{j}}\sum_{i\in I_{g}}||\hat{\gamma}_{j,g}-\gamma _{j,g}^{\ast
	}||^{4}\right) =O_{p}(T^{-2}).  \label{P_C_L_Auxillary7_5b}
\end{equation}%
By the triangle inequality and Assumption \ref{A3}(i), we have%
\begin{align}
	& \left\vert (nT)^{-1}\sum_{i\leq n}\sum_{t\leq T}(\hat{\phi}_{j,i}-\phi
	_{j,i}^{\ast })^{\top }\left( 
	\begin{array}{c}
		\psi _{j,\phi }(z_{i,t};\tilde{\phi}_{j,i})^{\top }\psi _{j^{\prime },\phi
		}(z_{i,t};\tilde{\phi}_{j^{\prime },i}) \\ 
		-\psi _{j,\phi }(z_{i,t})\psi _{j^{\prime },\phi }(z_{i,t})%
	\end{array}%
	\right) (\hat{\phi}_{j^{\prime },i}-\phi _{j^{\prime },i}^{\ast })\right\vert
	\notag \\
	& \leq (nT)^{-1}\sum_{i\leq n}\sum_{t\leq T}\tilde{M}_{3}(z_{i,t})||\hat{\phi%
	}_{j,i}-\phi _{j,i}^{\ast }||^{2}||\hat{\phi}_{j^{\prime },i}-\phi
	_{j^{\prime },i}^{\ast }||^{2}  \notag \\
	& +(nT)^{-1}\sum_{i\leq n}\sum_{t\leq T}\tilde{M}_{3}(z_{i,t})||\hat{\phi}%
	_{j^{\prime },i}-\phi _{j^{\prime },i}^{\ast }||^{2}||\hat{\phi}_{j,i}-\phi
	_{j,i}^{\ast }||  \notag \\
	& +(nT)^{-1}\sum_{i\leq n}\sum_{t\leq T}\tilde{M}_{3}(z_{i,t})||\hat{\phi}%
	_{j,i}-\phi _{j,i}^{\ast }||^{2}||\hat{\phi}_{j^{\prime },i}-\phi
	_{j^{\prime },i}^{\ast }||,  \label{P_C_L_Auxillary7_6}
\end{align}%
where\ $\tilde{M}_{3}(z_{i,t})\equiv M(z_{i,t})(M(z_{i,t})+||\psi _{j,\phi
}(z_{i,t})||+||\psi _{j^{\prime },\phi }(z_{i,t})||)$. By Assumption \ref%
{C_A1}(iii) and Markov's inequality%
\begin{equation}
	(nT)^{-1}\sum_{i\leq n}\sum_{t\leq T}\tilde{M}_{3}(z_{i,t})^{4}=O_{p}(1).
	\label{P_C_L_Auxillary7_7}
\end{equation}%
From the Cauchy-Schwarz inequality, (\ref{P_C_L_Auxillary7_5b}) and (\ref%
{P_C_L_Auxillary7_7}), we have 
\begin{eqnarray}
	&&(nT)^{-1}\sum_{i\leq n}\sum_{t\leq T}\tilde{M}_{3}(z_{i,t})||\hat{\phi}%
	_{j^{\prime },i}-\phi _{j^{\prime },i}^{\ast }||^{2}||\hat{\phi}_{j,i}-\phi
	_{j,i}^{\ast }||  \notag \\
	&\leq &\left( (nT)^{-1}\sum_{i\leq n}\sum_{t\leq T}||\hat{\phi}_{j^{\prime
		},i}-\phi _{j^{\prime },i}^{\ast }||^{4}\right) ^{1/2}\times \left(
	(nT)^{-1}\sum_{i\leq n}\sum_{t\leq T}\tilde{M}_{3}(z_{i,t})^{4}\right) ^{1/4}
	\notag \\
	&&\times \left( (nT)^{-1}\sum_{i\leq n}\sum_{t\leq T}||\hat{\phi}_{j,i}-\phi
	_{j,i}^{\ast }||^{4}\right) ^{1/4}\overset{}{=}O_{p}(T^{-3/2}).
	\label{P_C_L_Auxillary7_8}
\end{eqnarray}%
Similarly,%
\begin{equation}
	(nT)^{-1}\sum_{i\leq n}\sum_{t\leq T}\tilde{M}_{3}(z_{i,t})||\hat{\phi}%
	_{j,i}-\phi _{j,i}^{\ast }||^{2}||\hat{\phi}_{j^{\prime },i}-\phi
	_{j^{\prime },i}^{\ast }||=O_{p}(T^{-3/2}).  \label{P_C_L_Auxillary7_9}
\end{equation}%
Moreover, by Lemma \ref{C_L_Auxillary1}(i, iii) and (\ref{P_C_L_Auxillary7_9}%
), 
\begin{eqnarray}
	&&(nT)^{-1}\sum_{i\leq n}\sum_{t\leq T}\tilde{M}_{3}(z_{i,t})||\hat{\phi}%
	_{j,i}-\phi _{j,i}^{\ast }||^{2}||\hat{\phi}_{j^{\prime },i}-\phi
	_{j^{\prime },i}^{\ast }||^{2}  \notag \\
	&\leq &\max_{i\leq n}||\hat{\phi}_{j^{\prime },i}-\phi _{j^{\prime
		},i}^{\ast }||(nT)^{-1}\sum_{i\leq n}\sum_{t\leq T}\tilde{M}_{3}(z_{i,t})||%
	\hat{\phi}_{j^{\prime },i}-\phi _{j^{\prime },i}^{\ast }||^{2}||\hat{\phi}%
	_{j,i}-\phi _{j,i}^{\ast }||=O_{p}(T^{-3/2}),  \label{P_C_L_Auxillary7_10}
\end{eqnarray}%
which, together with (\ref{P_C_L_Auxillary7_6}),\ (\ref{P_C_L_Auxillary7_8})
and (\ref{P_C_L_Auxillary7_9}) shows (\ref{P_C_L_Auxillary7_5}).

We next turn to verify (\ref{P_C_L_Auxillary7_4}). For $j,j^{\prime }=1,2$,
by the triangle inequality and the Cauchy-Schwarz inequality, 
\begin{align}
	& \left\vert (nT)^{-1}\sum_{i\leq n}\sum_{t\leq T}(\psi _{j}(z_{i,t};\tilde{%
		\phi}_{j,i})-\psi _{j}(z_{i,t};\phi _{j,i}^{\ast }))(\hat{\phi}_{j^{\prime
		},i}-\phi _{j^{\prime },i}^{\ast })^{\top }\psi _{j^{\prime },\phi \phi
	}(z_{i,t};\tilde{\phi}_{j^{\prime },i})(\hat{\phi}_{j^{\prime },i}-\phi
	_{j^{\prime },i}^{\ast })\right\vert  \notag \\
	& \leq (nT)^{-1}\sum_{i\leq n}\sum_{t\leq T}M(z_{i,t})||\psi _{j^{\prime
		},\phi \phi }(z_{i,t};\tilde{\phi}_{j^{\prime },i})||\times ||\hat{\phi}%
	_{j^{\prime },i}-\phi _{j^{\prime },i}^{\ast }||^{2}||\hat{\phi}_{j,i}-\phi
	_{j,i}^{\ast }||  \notag \\
	& \leq K(nT)^{-1}\sum_{i\leq n}\sum_{t\leq T}M^{2}(z_{i,t})||\hat{\phi}%
	_{j^{\prime },i}-\phi _{j^{\prime },i}^{\ast }||^{3}||\hat{\phi}_{j,i}-\phi
	_{j,i}^{\ast }||  \notag \\
	& +K(nT)^{-1}\sum_{i\leq n}\sum_{t\leq T}M(z_{i,t})||\psi _{j^{\prime },\phi
		\phi }(z_{i,t};\phi _{j^{\prime },i}^{\ast })||\times ||\hat{\phi}%
	_{j^{\prime },i}-\phi _{j^{\prime },i}^{\ast }||^{2}||\hat{\phi}_{j,i}-\phi
	_{j,i}^{\ast }||  \notag \\
	& \leq K\left( 1+\max_{i\leq n}||\hat{\phi}_{j^{\prime },i}-\phi _{j^{\prime
		},i}^{\ast }||\right) (nT)^{-1}\sum_{i\leq n}\sum_{t\leq T}\tilde{M}%
	_{4}(z_{i,t})||\hat{\phi}_{j^{\prime },i}-\phi _{j^{\prime },i}^{\ast
	}||^{2}||\hat{\phi}_{j,i}-\phi _{j,i}^{\ast }||,  \label{P_C_L_Auxillary7_12}
\end{align}%
where $\tilde{M}_{4}(z_{i,t})\equiv M(z_{i,t})(M(z_{i,t})+||\psi _{j^{\prime
	},\phi \phi }(z_{i,t};\phi _{j^{\prime },i}^{\ast })||)$. By the similar
arguments for deriving (\ref{P_C_L_Auxillary7_8}), we can show that 
\begin{equation*}
	(nT)^{-1}\sum_{i\leq n}\sum_{t\leq T}\tilde{M}_{4}(z_{i,t})||\hat{\phi}%
	_{j^{\prime },i}-\phi _{j^{\prime },i}^{\ast }||^{2}||\hat{\phi}_{j,i}-\phi
	_{j,i}^{\ast }||=O_{p}(T^{-3/2})
\end{equation*}%
which along with (\ref{P_C_L_Auxillary7_12}) and Lemma \ref{C_L_Auxillary1}%
(i, iii) establishes 
\begin{equation}
	(nT)^{-1}\sum_{i\leq n}\sum_{t\leq T}(\psi _{j}(z_{i,t};\tilde{\phi}%
	_{j,i})-\psi _{j}(z_{i,t};\phi _{j,i}^{\ast }))(\hat{\phi}_{j^{\prime
		},i}-\phi _{j^{\prime },i}^{\ast })^{\top }\psi _{j^{\prime },\phi \phi
	}(z_{i,t};\tilde{\phi}_{j^{\prime },i})(\hat{\phi}_{j^{\prime },i}-\phi
	_{j^{\prime },i}^{\ast })=O_{p}(T^{-3/2}).  \label{P_C_L_Auxillary7_13}
\end{equation}%
By the triangle inequality and the Cauchy-Schwarz inequality,%
\begin{align}
	& \left\vert (nT)^{-1}\sum_{i\leq n}\sum_{t\leq T}\Delta \psi (z_{i,t})(\hat{%
		\phi}_{j,i}-\phi _{j,i}^{\ast })^{\top }\psi _{j,\phi \phi }(z_{i,t};\tilde{%
		\phi}_{j,i})(\hat{\phi}_{j,i}-\phi _{j,i}^{\ast })\right\vert  \notag \\
	& \leq K(nT)^{-1}\sum_{i\leq n}\sum_{t\leq T}\left\Vert \Delta \psi
	(z_{i,t})\psi _{j,\phi \phi }(z_{i,t};\phi _{j,i}^{\ast })\right\Vert ||\hat{%
		\phi}_{j,i}-\phi _{j,i}^{\ast }||^{2}  \notag \\
	& +K\max_{i\leq n}||\hat{\phi}_{j,i}-\phi _{j,i}^{\ast
	}||(nT)^{-1}\sum_{i\leq n}\sum_{t\leq T}\left\vert \Delta \psi
	(z_{i,t})M(z_{i,t})\right\vert ||\hat{\phi}_{j,i}-\phi _{j,i}^{\ast }||^{2}.
	\label{P_C_L_Auxillary7_14}
\end{align}%
By Assumption \ref{C_A1}(ii) and \ref{A3}, and Markov's inequality, 
\begin{equation}
	(nT)^{-1}\sum_{i\leq n}\sum_{t\leq T}\left\Vert \Delta \psi (z_{i,t},\phi
	_{i}^{\ast })\psi _{j,\phi \phi }(z_{i,t};\phi _{j,i}^{\ast })\right\Vert
	^{2}=O_{p}(\omega _{n,T}^{2})  \label{P_C_L_Auxillary7_15}
\end{equation}%
and 
\begin{equation}
	(nT)^{-1}\sum_{i\leq n}\sum_{t\leq T}\left\vert \Delta \psi (z_{i,t};\phi
	_{i}^{\ast })M(z_{i,t})\right\vert ^{2}=O_{p}(\omega _{n,T}^{2}),
	\label{P_C_L_Auxillary7_16}
\end{equation}%
which together with Lemma \ref{C_L_Auxillary1}(i, iii), (\ref%
{P_C_L_Auxillary7_5b}) and (\ref{P_C_L_Auxillary7_14}) shows that 
\begin{equation}
	(nT)^{-1}\sum_{i\leq n}\sum_{t\leq T}\Delta \psi (z_{i,t})(\hat{\phi}%
	_{j,i}-\phi _{j,i}^{\ast })^{\top }\psi _{j,\phi \phi }(z_{i,t};\tilde{\phi}%
	_{j,i})(\hat{\phi}_{j,i}-\phi _{j,i}^{\ast })=O_{p}(\omega _{n,T}T^{-1}).
	\label{P_C_L_Auxillary7_17}
\end{equation}%
The desired result in (\ref{P_C_L_Auxillary7_4}) follows from (\ref%
{P_C_L_Auxillary7_13}) and (\ref{P_C_L_Auxillary7_17}).

The second equality in (\ref{C_L_Auxillary7a_1}) holds if%
\begin{equation}
	(nT)^{-1}\sum_{i\leq n}\sum_{t\leq T}\Delta \psi (z_{i,t})\Delta \psi _{\phi
	}(z_{i,t};\phi _{i}^{\ast })(\hat{\phi}_{i}-\phi _{i}^{\ast
	})=O_{p}(T^{-1/2}),  \label{P_C_L_Auxillary7_18}
\end{equation}%
and 
\begin{equation}
	(nT)^{-1}\sum_{i\leq n}\sum_{t\leq T}(\hat{\phi}_{j,i}-\phi _{j,i}^{\ast
	})^{\top }\psi _{j,\phi }(z_{i,t})\psi _{j^{\prime },\phi }(z_{i,t})(\hat{%
		\phi}_{j^{\prime },i}-\phi _{j^{\prime },i}^{\ast })=O_{p}(T^{-1}).
	\label{P_C_L_Auxillary7_19}
\end{equation}%
To show (\ref{P_C_L_Auxillary7_18}), we first observe that by the
Cauchy-Schwarz inequality, 
\begin{eqnarray}
	&&\left\vert (nT)^{-1}\sum_{i\leq n}\sum_{t\leq T}\Delta \psi
	(z_{i,t})\Delta \psi _{\phi }(z_{i,t};\phi _{i}^{\ast })(\hat{\phi}_{i}-\phi
	_{i}^{\ast })\right\vert ^{2}  \notag \\
	&\leq &K(nT)^{-1}\sum_{i\leq n}\sum_{t\leq T}\Delta \psi (z_{i,t})^{2} 
	\notag \\
	&&\times \sum_{j=1,2}\left( n^{-1}\sum_{i\leq n}||\hat{\phi}_{j,i}-\phi
	_{j,i}^{\ast }||^{4}\right) ^{1/2}\left( (nT)^{-1}\sum_{i\leq n}\sum_{t\leq
		T}\left\Vert \psi _{j,\phi }(z_{i,t})\right\Vert ^{4}\right) ^{1/2}.
	\label{P_C_L_Auxillary7_20}
\end{eqnarray}%
By Assumptions \ref{C_A1}(ii) and \ref{A3}(iii), and Markov's inequality, 
\begin{equation}
	(nT)^{-1}\sum_{i\leq n}\sum_{t\leq T}\Delta \psi (z_{i,t})^{2}=O_{p}(1)\text{
		\ and \ }(nT)^{-1}\sum_{i\leq n}\sum_{t\leq T}\left\Vert \psi _{j,\phi
	}(z_{i,t})\right\Vert ^{4}=O_{p}(1),  \label{P_C_L_Auxillary7_21}
\end{equation}%
which together with (\ref{P_C_L_Auxillary7_5b}) and (\ref%
{P_C_L_Auxillary7_20}) establishes\ (\ref{P_C_L_Auxillary7_18}).

Finally, we verify\ (\ref{P_C_L_Auxillary7_19}).\ From Assumption \ref{C_A1}%
(iii) and Markov's inequality, it follows that%
\begin{equation}
	(nT)^{-1}\sum_{i\leq n}\sum_{t\leq T}\left\Vert \psi _{j,\phi }(z_{i,t})\psi
	_{j^{\prime },\phi }(z_{i,t})\right\Vert ^{2}=O_{p}(1).
	\label{P_C_L_Auxillary7_22}
\end{equation}%
Therefore, by the triangle inequality and Cauchy-Schwarz inequality, 
\begin{eqnarray}
	&&\left\vert (nT)^{-1}\sum_{i\leq n}\sum_{t\leq T}(\hat{\phi}_{j,i}-\phi
	_{j,i}^{\ast })^{\top }\psi _{j,\phi }(z_{i,t})\psi _{j^{\prime },\phi
	}(z_{i,t})(\hat{\phi}_{j^{\prime },i}-\phi _{j^{\prime },i}^{\ast
	})\right\vert  \notag \\
	&\leq &(nT)^{-1}\sum_{i\leq n}\sum_{t\leq T}||\hat{\phi}_{j,i}-\phi
	_{j,i}^{\ast }||\times ||\hat{\phi}_{j^{\prime },i}-\phi _{j^{\prime
		},i}^{\ast }||\times \left\Vert \psi _{j,\phi }(z_{i,t})\psi _{j^{\prime
		},\phi }(z_{i,t})\right\Vert  \notag \\
	&\leq &\left( n^{-1}\sum_{i\leq n}||\hat{\phi}_{j,i}-\phi _{j,i}^{\ast
	}||^{2}||\hat{\phi}_{j^{\prime },i}-\phi _{j^{\prime },i}^{\ast
	}||^{2}\right) ^{1/2}  \notag \\
	&&\times \left( (nT)^{-1}\sum_{i\leq n}\sum_{t\leq T}\left\Vert \psi
	_{j,\phi }(z_{i,t})\psi _{j^{\prime },\phi }(z_{i,t})\right\Vert ^{2}\right)
	^{1/2}\overset{}{=}O_{p}(T^{-1}),  \label{P_C_L_Auxillary7_23}
\end{eqnarray}%
which establishes\ (\ref{P_C_L_Auxillary7_19}).\hfill $Q.E.D.$

\bigskip

\begin{lemma}
	\textit{\label{C_L_Auxillary7b}} Under Assumptions \ref{A1} and \ref{C_A1},\
	we have $\sum_{g\in \mathcal{G}_{j}}n_{j,g}^{-1}\sum_{i\in I_{j,g}}(\hat{s}%
	_{j,\gamma ,i}^{2}-s_{j,\gamma ,i}^{2})^{2}=O_{p}(1)$.
\end{lemma}

\noindent {\textsc{Proof of Lemma \ref{C_L_Auxillary7b}}}. By the
definitions of $\hat{s}_{j,\gamma ,i}^{2}$ and $s_{j,\gamma ,i}^{2}$, and
Lemma \ref{C_L_Auxillary1}(ii), we obtain the following expression: 
\begin{align*}
	\hat{s}_{j,\gamma ,i}^{2}-s_{j,\gamma ,i}^{2}& =\frac{T^{-1}\sum_{t\leq
			T}\psi _{j,\gamma }^{2}(z_{i,t};\hat{\phi}_{j,g})}{-\hat{\Psi}_{j,\gamma
			\gamma }(\hat{\phi}_{j,g})}-\frac{\mathbb{E}[\psi _{j,\gamma }^{2}(z_{i,t})]%
	}{-\Psi _{j,\gamma \gamma ,g}} \\
	& =\frac{T^{-1}\sum_{t\leq T}(\psi _{j,\gamma }^{2}(z_{i,t};\hat{\phi}%
		_{j,g})-\psi _{j,\gamma }^{2}(z_{i,t};\phi _{j,g}^{\ast }))}{-\hat{\Psi}%
		_{j,\gamma \gamma }(\hat{\phi}_{j,g})} \\
	& +\frac{T^{-1}\sum_{t\leq T}\psi _{j,\gamma }^{2}(z_{i,t};\phi _{j,g}^{\ast
		})}{\hat{\Psi}_{j,\gamma \gamma }(\hat{\phi}_{j,g})\Psi _{j,\gamma \gamma ,g}%
	}(\hat{\Psi}_{j,\gamma \gamma }(\hat{\phi}_{j,g})-\Psi _{j,\gamma \gamma ,g})
	\\
	& +\frac{T^{-1}\sum_{t\leq T}(\psi _{j,\gamma }^{2}(z_{i,t};\phi
		_{j,g}^{\ast })-\mathbb{E}[\psi _{j,\gamma }^{2}(z_{i,t})])}{-\Psi
		_{j,\gamma \gamma ,g}},
\end{align*}%
wpa1. From the expression above, it is evident that the claim of the lemma
follows if the following results hold:%
\begin{align}
	\sum_{g\in \mathcal{G}_{j}}\sum_{i\in I_{j,g}}\frac{(T^{-1}\sum_{t\leq
			T}(\psi _{j,\gamma }^{2}(z_{i,t};\hat{\phi}_{j,g})-\psi _{j,\gamma
		}^{2}(z_{i,t})))^{2}}{n_{j,g}\hat{\Psi}_{j,\gamma \gamma }^{2}(\hat{\phi}%
		_{j,g})}& =O_{p}(1),  \label{P_L_Auxillary7b_1} \\
	\sum_{g\in \mathcal{G}_{j}}\sum_{i\in I_{j,g}}\frac{(T^{-1}\sum_{t\leq
			T}\psi _{j,\gamma }^{2}(z_{i,t}))^{2}}{n_{j,g}\hat{\Psi}_{j,\gamma \gamma
		}^{2}(\hat{\phi}_{j,g})\Psi _{j,\gamma \gamma ,g}^{2}}(\hat{\Psi}_{j,\gamma
		\gamma }(\hat{\phi}_{j,g})-\Psi _{j,\gamma \gamma ,g})^{2}& =O_{p}(1),
	\label{P_L_Auxillary7b_2} \\
	\sum_{g\in \mathcal{G}_{j}}\sum_{i\in I_{j,g}}\frac{(T^{-1}\sum_{t\leq
			T}(\psi _{j,\gamma }^{2}(z_{i,t})-\mathbb{E}[\psi _{j,\gamma
		}^{2}(z_{i,t};\phi _{j,g}^{\ast })]))^{2}}{n_{j,g}\Psi _{j,\gamma \gamma
			,g}^{2}}& =O_{p}(1).  \label{P_L_Auxillary7b_3}
\end{align}%
We now proceed to verify (\ref{P_L_Auxillary7b_1}), (\ref{P_L_Auxillary7b_2}%
) and (\ref{P_L_Auxillary7b_3}).

By the Cauchy-Schwarz inequality, the term before the equality in (\ref%
{P_L_Auxillary7b_1}) can be bounded as follows: 
\begin{align}
	& \sum_{g\in\mathcal{G}_{j}}\sum_{i\in I_{j,g}}\frac{(T^{-1}\sum_{t\leq
			T}(\psi_{j,\gamma}^{2}(z_{i,t};\hat{\phi}_{j,g})-\psi_{j,%
			\gamma}^{2}(z_{i,t})))^{2}}{n_{j,g}\hat{\Psi}_{j,\gamma\gamma}^{2}(\hat{\phi}%
		_{j,g})}  \notag \\
	& \leq KT^{-1}\sum_{g\in\mathcal{G}_{j}}\sum_{i\in I_{j,g}}\frac{\left(
		T^{-1}\sum_{t\leq T}(\psi_{j,\gamma}(z_{i,t};\hat{\phi}_{j,g})-\psi_{j,%
			\gamma }(z_{i,t}))^{2}\right) ^{2}}{n_{j,g}\hat{\Psi}_{j,\gamma\gamma}^{2}(%
		\hat {\phi}_{j,g})}  \notag \\
	& +KT^{-1}\sum_{g\in\mathcal{G}_{j}}\sum_{i\in I_{j,g}}\frac{\left(
		T^{-1}\sum_{t\leq T}(\psi_{j,\gamma}(z_{i,t};\hat{\phi}_{j,g})-\psi_{j,%
			\gamma }(z_{i,t}))\psi_{j,\gamma}(z_{i,t})\right) ^{2}}{n_{j,g}\hat{\Psi}%
		_{j,\gamma\gamma}^{2}(\hat{\phi}_{j,g})}.  \label{P_L_Auxillary7b_4}
\end{align}
By Assumption \ref{A3} and the Cauchy-Schwarz inequality,\ we have%
\begin{align}
	& \sum_{g\in\mathcal{G}_{j}}\sum_{i\in I_{j,g}}\frac{\left(
		T^{-1}\sum_{t\leq T}(\psi_{j,\gamma}(z_{i,t};\hat{\phi}_{j,g})-\psi_{j,%
			\gamma }(z_{i,t}))^{2}\right) ^{2}}{n_{j,g}\hat{\Psi}_{j,\gamma\gamma}^{2}(%
		\hat {\phi}_{j,g})}  \notag \\
	& \leq\frac{\sum_{g\in\mathcal{G}_{j}}n_{j,g}^{-1}\sum_{i\in I_{j,g}}\left(
		T^{-1}\sum_{t\leq T}M(z_{i,t})^{2}||\hat{\phi}_{j,g}-\phi_{j,g}^{\ast}||^{2}%
		\right) ^{2}}{\min_{g\in\mathcal{G}_{j}}\hat{\Psi}_{j,\gamma\gamma}^{2}(\hat{%
			\phi}_{j,g})}  \notag \\
	& \leq K\frac{\max_{g\in\mathcal{G}_{j}}||\hat{\phi}_{j,g}-\phi_{j,g}^{\ast
		}||^{4}}{\min_{g\in\mathcal{G}_{j}}\hat{\Psi}_{j,\gamma\gamma}^{2}(\hat{\phi 
		}_{j,g})}\sum_{g\in\mathcal{G}_{j}}n_{j,g}^{-1}\sum_{i\in I_{j,g}}\left(
	T^{-1}\sum_{t\leq T}(M(z_{i,t})^{2}-\mathbb{E}[M(z_{i,t})^{2}])\right) ^{2} 
	\notag \\
	& +K\frac{\sum_{g\in\mathcal{G}_{j}}||\hat{\phi}_{j,g}-\phi_{j,g}^{\ast
		}||^{4}n_{j,g}^{-1}\sum_{i\in I_{j,g}}\left( T^{-1}\sum_{t\leq T}(\mathbb{E}%
		[M(z_{i,t})^{2}])\right) ^{2}}{\min_{g\in\mathcal{G}_{j}}\hat {\Psi}%
		_{j,\gamma\gamma}^{2}(\hat{\phi}_{j,g})}.  \label{P_L_Auxillary7b_5}
\end{align}
Applying Assumptions \ref{A1}, \ref{A3},\ and the covariance inequality for
strong mixing processes, we obtain:%
\begin{equation}
	\mathbb{E}\left[ \left( T^{-1/2}\sum_{t\leq T}(M(z_{i,t})^{2}-\mathbb{E}%
	[M(z_{i,t})^{2}])\right) ^{2}\right] \leq K.  \label{P_L_Auxillary7b_6}
\end{equation}
Combining this with Lemma \ref{C_L_Auxillary1}(i, ii, iii) and Markov's
inequality, we establish%
\begin{equation}
	\frac{\max_{g\in\mathcal{G}_{j}}||\hat{\phi}_{j,g}-\phi_{j,g}^{\ast}||^{4}}{%
		\min_{g\in\mathcal{G}_{j}}\hat{\Psi}_{j,\gamma\gamma}^{2}(\hat{\phi}_{j,g})}%
	\sum_{g\in\mathcal{G}_{j}}n_{j,g}^{-1}\sum_{i\in I_{j,g}}\left(
	T^{-1}\sum_{t\leq T}(M(z_{i,t})^{2}-\mathbb{E}[M(z_{i,t})^{2}])\right)
	^{2}=o_{p}(G_{j}T^{-1}).  \label{P_L_Auxillary7b_7}
\end{equation}
Since $T^{-1}\sum_{t\leq T}(\mathbb{E}[M(z_{i,t})^{2}])\leq K$, from Lemma %
\ref{C_L_Auxillary1}(ii) and (\ref{P_C_L_Auxillary1_7}) it follows that%
\begin{equation}
	\frac{\sum_{g\in\mathcal{G}_{j}}||\hat{\phi}_{j,g}-\phi_{j,g}^{%
			\ast}||^{4}n_{j,g}^{-1}\sum_{i\in I_{j,g}}\left( T^{-1}\sum_{t\leq T}(%
		\mathbb{E}[M(z_{i,t})^{2}])\right) ^{2}}{\min_{g\in\mathcal{G}_{j}}\hat {\Psi%
		}_{j,\gamma\gamma}^{2}(\hat{\phi}_{j,g})}=O_{p}(G_{j}T^{-2}),
	\label{P_L_Auxillary7b_8}
\end{equation}
which along with (\ref{P_L_Auxillary7b_5}) and (\ref{P_L_Auxillary7b_7})
verifies (\ref{P_L_Auxillary7b_1}).

To verify (\ref{P_L_Auxillary7b_2}), we first observe that by Lemma \ref%
{C_L_Auxillary1}(ii), the term on its LHS satisfies:%
\begin{align}
	& \sum_{g\in \mathcal{G}_{j}}\sum_{i\in I_{j,g}}\frac{(T^{-1}\sum_{t\leq
			T}\psi _{j,\gamma }^{2}(z_{i,t}))^{2}}{n_{j,g}\hat{\Psi}_{j,\gamma \gamma
		}^{2}(\hat{\phi}_{j,g})\Psi _{j,\gamma \gamma ,g}^{2}}(\hat{\Psi}_{j,\gamma
		\gamma }(\hat{\phi}_{j,g})-\Psi _{j,\gamma \gamma ,g})^{2}  \notag \\
	& \leq \frac{\sum_{g\in \mathcal{G}_{j}}(\hat{\Psi}_{j,\gamma \gamma }(\hat{%
			\phi}_{j,g})-\Psi _{j,\gamma \gamma ,g})^{2}}{\min_{g\in \mathcal{G}_{j}}%
		\hat{\Psi}_{j,\gamma \gamma }^{2}(\hat{\phi}_{j,g})\Psi _{j,\gamma \gamma
			,g}^{2}}\max_{i\leq n}\left( T^{-1}\sum_{t\leq T}\psi _{j,\gamma
	}^{2}(z_{i,t})\right) ^{2}.  \label{P_L_Auxillary7b_9}
\end{align}%
By the similar arguments for showing (\ref{P_C_L_Auxillary2b_9}), we can
establish 
\begin{equation*}
	\max_{i\leq n}T^{-1}\sum_{t\leq T}\psi _{j,\gamma }^{2}(z_{i,t})=O_{p}(1).
\end{equation*}%
Combining this with\ (\ref{P_L_Auxillary7b_9}), Assumption \ref{A4} and
Lemma \ref{C_L_Auxillary1}(ii, v), we obtain:%
\begin{equation}
	\sum_{g\in \mathcal{G}_{j}}\sum_{i\in I_{j,g}}\frac{(T^{-1}\sum_{t\leq
			T}\psi _{j,\gamma }^{2}(z_{i,t}))^{2}}{n_{j,g}\hat{\Psi}_{j,\gamma \gamma
		}^{2}(\hat{\phi}_{j,g})\Psi _{j,\gamma \gamma ,g}^{2}}(\hat{\Psi}_{j,\gamma
		\gamma }(\hat{\phi}_{j,g})-\Psi _{j,\gamma \gamma
		,g})^{2}=O_{p}(G_{j}T^{-1}).  \label{P_L_Auxillary7b_10}
\end{equation}

To verify (\ref{P_L_Auxillary7b_3}), we observe that by\ Assumptions \ref{A1}%
, \ref{A3} and \ref{A4}, we have%
\begin{eqnarray}
	&&\sum_{g\in \mathcal{G}_{j}}\sum_{i\in I_{j,g}}\frac{\mathbb{E}\left[
		(T^{-1}\sum_{t\leq T}(\psi _{j,\gamma }^{2}(z_{i,t})-\mathbb{E}[\psi
		_{j,\gamma }^{2}(z_{i,t})]))^{2}\right] }{n_{j,g}\Psi _{j,\gamma \gamma
			,g}^{2}}  \notag \\
	&\leq &K\sum_{g\in \mathcal{G}_{j}}\sum_{i\in I_{j,g}}\frac{\max_{t\leq
			T}\left\Vert \psi _{j,\gamma }^{2}(z_{i,t})\right\Vert _{2+\delta /2}^{2}}{%
		n_{j,g}T\Psi _{j,\gamma \gamma ,g}^{2}}\leq KG_{j}T^{-1}.
	\label{P_L_Auxillary7b_11}
\end{eqnarray}
The condition in (\ref{P_L_Auxillary7b_3}) now follows from Assumptions \ref%
{A1}(i), (\ref{P_L_Auxillary7b_11}) and Markov's inequality.\hfill $Q.E.D.$

\bigskip

\begin{lemma}
	\textit{\label{C_L_Auxillary7c}}\ Under Assumptions \ref{A1} and \ref{C_A1}%
	,\ we have%
	\begin{equation}
		\sum_{g\in \mathcal{G}_{j}}n_{j,g}^{-1}\sum_{i\in I_{j,g}}\left( \frac{(%
			\widehat{\mathbb{E}}_{T}[\hat{\psi}_{j,\gamma }(z_{i,t})])^{2}}{\hat{\Psi}%
			_{j,\gamma \gamma }(\hat{\phi}_{j,g})}-\frac{(\mathbb{E}_{T}[\psi _{j,\gamma
			}(z_{i,t})])^{2}}{\Psi _{j,\gamma \gamma ,g}}\right) ^{2}=O_{p}(1).
		\label{C_L_Auxillary7c_1}
	\end{equation}
\end{lemma}

\noindent {\textsc{Proof of Lemma \ref{C_L_Auxillary7c}}}. Since we can write%
\begin{align*}
	& \frac{(T^{-1}\sum_{t\leq T}\psi _{j,\gamma }(z_{i,t};\hat{\phi}_{j,g}))^{2}%
	}{\hat{\Psi}_{j,\gamma \gamma }(\hat{\phi}_{j,g})}-\frac{(\mathbb{E}[\psi
		_{j,\gamma }(z_{i,t})])^{2}}{\Psi _{j,\gamma \gamma ,g}} \\
	& =\frac{(T^{-1}\sum_{t\leq T}\psi _{j,\gamma }(z_{i,t};\hat{\phi}%
		_{j,g}))^{2}-(T^{-1}\sum_{t\leq T}\psi _{j,\gamma }(z_{i,t}))^{2}}{\hat{\Psi}%
		_{j,\gamma \gamma }(\hat{\phi}_{j,g})} \\
	& -\frac{(T^{-1}\sum_{t\leq T}\psi _{j,\gamma }(z_{i,t}))^{2}}{\hat{\Psi}%
		_{j,\gamma \gamma }(\hat{\phi}_{j,g})\Psi _{j,\gamma \gamma ,g}}(\hat{\Psi}%
	_{j,\gamma \gamma }(\hat{\phi}_{j,g})-\Psi _{j,\gamma \gamma ,g}) \\
	& +\frac{(T^{-1}\sum_{t\leq T}\psi _{j,\gamma }(z_{i,t}))^{2}-(\mathbb{E}%
		[\psi _{j,\gamma }(z_{i,t})])^{2}}{\Psi _{j,\gamma \gamma ,g}},
\end{align*}%
the claim in (\ref{C_L_Auxillary7c_1}) follows if the following conditions
hold:%
\begin{align}
	\sum_{g\in \mathcal{G}_{j}}\sum_{i\in I_{j,g}}\frac{((T^{-1}\sum_{t\leq
			T}\psi _{j,\gamma }(z_{i,t};\hat{\phi}_{j,g}))^{2}-(T^{-1}\sum_{t\leq T}\psi
		_{j,\gamma }(z_{i,t}))^{2})^{2}}{n_{j,g}\hat{\Psi}_{j,\gamma \gamma }^{2}(%
		\hat{\phi}_{j,g})}& =O_{p}(1),  \label{P_C_L_Auxillary7c_1} \\
	\sum_{g\in \mathcal{G}_{j}}\sum_{i\in I_{j,g}}\frac{(T^{-1}\sum_{t\leq
			T}\psi _{j,\gamma }(z_{i,t}))^{4}}{n_{j,g}\hat{\Psi}_{j,\gamma \gamma }^{2}(%
		\hat{\phi}_{j,g})\Psi _{j,\gamma \gamma ,g}^{2}}(\hat{\Psi}_{j,\gamma \gamma
	}(\hat{\phi}_{j,g})-\Psi _{j,\gamma \gamma ,g})^{2}& =O_{p}(1),
	\label{P_C_L_Auxillary7c_2} \\
	\sum_{g\in \mathcal{G}_{j}}\sum_{i\in I_{j,g}}\frac{((T^{-1}\sum_{t\leq
			T}\psi _{j,\gamma }(z_{i,t}))^{2}-(\mathbb{E}[\psi _{j,\gamma }(z_{i,t};\phi
		_{j,g}^{\ast })])^{2})^{2}}{n_{j,g}\Psi _{j,\gamma \gamma ,g}^{2}}&
	=O_{p}(1).  \label{P_C_L_Auxillary7c_3}
\end{align}%
We now proceed to verify (\ref{P_C_L_Auxillary7c_1}), (\ref%
{P_C_L_Auxillary7c_2}) and (\ref{P_C_L_Auxillary7c_3}).

By the Cauchy-Schwarz inequality, the term before the equality in (\ref%
{P_C_L_Auxillary7c_1}) can be bounded as follows: 
\begin{align}
	& \sum_{g\in \mathcal{G}_{j}}\sum_{i\in I_{j,g}}\frac{((T^{-1}\sum_{t\leq
			T}\psi _{j,\gamma }(z_{i,t};\hat{\phi}_{j,g}))^{2}-(T^{-1}\sum_{t\leq T}\psi
		_{j,\gamma }(z_{i,t}))^{2})^{2}}{n_{j,g}\hat{\Psi}_{j,\gamma \gamma }^{2}(%
		\hat{\phi}_{j,g})}  \notag \\
	& \leq K\sum_{g\in \mathcal{G}_{j}}\sum_{i\in I_{j,g}}\frac{%
		(T^{-1}\sum_{t\leq T}(\psi _{j,\gamma }(z_{i,t};\hat{\phi}_{j,g})-\psi
		_{j,\gamma }(z_{i,t})))^{4}}{n_{j,g}\hat{\Psi}_{j,\gamma \gamma }^{2}(\hat{%
			\phi}_{j,g})}  \notag \\
	& +K\sum_{g\in \mathcal{G}_{j}}\sum_{i\in I_{j,g}}\frac{(T^{-1}\sum_{t\leq
			T}(\psi _{j,\gamma }(z_{i,t};\hat{\phi}_{j,g})-\psi _{j,\gamma
		}(z_{i,t})))^{2}(T^{-1}\sum_{t\leq T}\psi _{j,\gamma }(z_{i,t}))^{2}}{n_{j,g}%
		\hat{\Psi}_{j,\gamma \gamma }^{2}(\hat{\phi}_{j,g})}  \notag \\
	& \leq \frac{K\max_{i\leq n}(T^{-1}\sum_{t\leq T}M(z_{i,t}))^{4}\sum_{g\in 
			\mathcal{G}_{j}}||\hat{\phi}_{j,g}-\phi _{j,g}^{\ast }||^{4}}{\min_{g\in 
			\mathcal{G}_{j}}\hat{\Psi}_{j,\gamma \gamma }^{2}(\hat{\phi}_{j,g})}  \notag
	\\
	& +\frac{K\max_{i\leq n}(T^{-1}\sum_{t\leq T}\psi _{j,\gamma
		}(z_{i,t}))^{2}(T^{-1}\sum_{t\leq T}M(z_{i,t}))^{2}\sum_{g\in \mathcal{G}%
			_{j}}||\hat{\phi}_{j,g}-\phi _{j,g}^{\ast }||^{2}}{\min_{g\in \mathcal{G}%
			_{j}}\hat{\Psi}_{j,\gamma \gamma }^{2}(\hat{\phi}_{j,g})}.
	\label{P_C_L_Auxillary7c_4}
\end{align}%
By the similar arguments for showing (\ref{P_C_L_Auxillary2b_9}),\ we can
show that%
\begin{equation}
	\max_{i\leq n}T^{-1}\sum_{t\leq T}M(z_{i,t})=O_{p}(1),
	\label{P_C_L_Auxillary7c_5}
\end{equation}%
which together with Markov's inequality, Assumption \ref{A3} and Lemma \ref%
{C_L_Auxillary1}(i, ii, iii) implies that 
\begin{equation}
	\frac{\max_{i\leq n}(T^{-1}\sum_{t\leq T}M(z_{i,t}))^{4}\sum_{g\in \mathcal{G%
			}_{j}}||\hat{\phi}_{j,g}-\phi _{j,g}^{\ast }||^{4}}{\min_{g\in \mathcal{G}%
			_{j}}\hat{\Psi}_{j,\gamma \gamma }^{2}(\hat{\phi}_{j,g})}=O_{p}(G_{j}T^{-2}).
	\label{P_C_L_Auxillary7c_6}
\end{equation}%
From Assumption \ref{A3}, Lemma \ref{C_L_Auxillary1}(ii), (\ref%
{P_C_L_Auxillary2b_9}) and (\ref{P_C_L_Auxillary7c_5}), we also have%
\begin{equation*}
	\frac{\max_{i\leq n}(T^{-1}\sum_{t\leq T}\psi _{j,\gamma
		}(z_{i,t}))^{2}(T^{-1}\sum_{t\leq T}M(z_{i,t}))^{2}\sum_{g\in \mathcal{G}%
			_{j}}||\hat{\phi}_{j,g}-\phi _{j,g}^{\ast }||^{2}}{\min_{g\in \mathcal{G}%
			_{j}}\hat{\Psi}_{j,\gamma \gamma }^{2}(\hat{\phi}_{j,g})}=O_{p}(G_{j}T^{-1}),
\end{equation*}%
which together with Assumption \ref{A1}(i), (\ref{P_C_L_Auxillary7c_4}) and (%
\ref{P_C_L_Auxillary7c_6}) verifies (\ref{P_C_L_Auxillary7c_1}).

To verify (\ref{P_C_L_Auxillary7c_2}), we first observe that the term on its
LHS satisfies:%
\begin{align*}
	& \sum_{g\in \mathcal{G}_{j}}\sum_{i\in I_{j,g}}\frac{(T^{-1}\sum_{t\leq
			T}\psi _{j,\gamma }(z_{i,t}))^{4}}{n_{j,g}\hat{\Psi}_{j,\gamma \gamma }^{2}(%
		\hat{\phi}_{j,g})\Psi _{j,\gamma \gamma ,g}^{2}}(\hat{\Psi}_{j,\gamma \gamma
	}(\hat{\phi}_{j,g})-\Psi _{j,\gamma \gamma ,g})^{2} \\
	& \leq \frac{\max_{i\leq n}(T^{-1}\sum_{t\leq T}\psi _{j,\gamma
		}(z_{i,t}))^{4}}{\min_{g\in \mathcal{G}_{j}}\hat{\Psi}_{j,\gamma \gamma
		}^{2}(\hat{\phi}_{j,g})\Psi _{j,\gamma \gamma ,g}^{2}}\sum_{g\in \mathcal{G}%
		_{j}}(\hat{\Psi}_{j,\gamma \gamma }(\hat{\phi}_{j,g})-\Psi _{j,\gamma \gamma
		,g})^{2}.
\end{align*}%
This combined with Assumption \ref{A4}, Lemma \ref{C_L_Auxillary1}(ii, v)
and (\ref{P_C_L_Auxillary2b_9}) implies that 
\begin{equation}
	\sum_{g\in \mathcal{G}_{j}}\sum_{i\in I_{j,g}}\frac{(T^{-1}\sum_{t\leq
			T}\psi _{j,\gamma }(z_{i,t}))^{4}}{n_{j,g}\hat{\Psi}_{j,\gamma \gamma }^{2}(%
		\hat{\phi}_{j,g})\Psi _{j,\gamma \gamma ,g}^{2}}(\hat{\Psi}_{j,\gamma \gamma
	}(\hat{\phi}_{j,g})-\Psi _{j,\gamma \gamma ,g})^{2}=O_{p}(G_{j}T^{-1}).
	\label{P_C_L_Auxillary7c_7}
\end{equation}%
The claim in (\ref{P_C_L_Auxillary7c_2}) follows from Assumption \ref{A1}(i)
and (\ref{P_C_L_Auxillary7c_7}).

To verify (\ref{P_C_L_Auxillary7c_3}), we first observe that by Assumption %
\ref{A4}, the term on its LHS satisfies:%
\begin{align}
	& \sum_{g\in \mathcal{G}_{j}}\sum_{i\in I_{j,g}}\frac{((T^{-1}\sum_{t\leq
			T}\psi _{j,\gamma }(z_{i,t}))^{2}-(\mathbb{E}[\psi _{j,\gamma
		}(z_{i,t})])^{2})^{2}}{n_{j,g}\Psi _{j,\gamma \gamma ,g}^{2}}  \notag \\
	& \leq K\sum_{g\in \mathcal{G}_{j}}n_{j,g}^{-1}\sum_{i\in I_{j,g}}\left(
	T^{-1}\sum_{t\leq T}(\psi _{j,\gamma }(z_{i,t})-\mathbb{E}[\psi _{j,\gamma
	}(z_{i,t})])\right) ^{4}  \notag \\
	& +K\sum_{g\in \mathcal{G}_{j}}n_{j,g}^{-1}\sum_{i\in I_{j,g}}\left(
	T^{-1}\sum_{t\leq T}(\psi _{j,\gamma }(z_{i,t})-\mathbb{E}[\psi _{j,\gamma
	}(z_{i,t})])\right) ^{2}(\mathbb{E}[\psi _{j,\gamma }(z_{i,t})])^{2}.
	\label{P_C_L_Auxillary7c_8}
\end{align}%
Applying Rosenthal's inequality and Assumption \ref{A3}, we get%
\begin{equation*}
	\mathbb{E}\left[ \left( T^{-1}\sum_{t\leq T}(\psi _{j,\gamma }(z_{i,t})-%
	\mathbb{E}[\psi _{j,\gamma }(z_{i,t})])\right) ^{4}\right] \leq KT^{-2},
\end{equation*}%
which together with Markov's inequality implies that 
\begin{equation}
	\sum_{g\in \mathcal{G}_{j}}n_{j,g}^{-1}\sum_{i\in I_{j,g}}\left(
	T^{-1}\sum_{t\leq T}(\psi _{j,\gamma }(z_{i,t})-\mathbb{E}[\psi _{j,\gamma
	}(z_{i,t})])\right) ^{4}=O_{p}(T^{-1}).  \label{P_C_L_Auxillary7c_9}
\end{equation}%
Similarly, we can show that 
\begin{equation*}
	\sum_{g\in \mathcal{G}_{j}}n_{j,g}^{-1}\sum_{i\in I_{j,g}}\left(
	T^{-1}\sum_{t\leq T}(\psi _{j,\gamma }(z_{i,t})-\mathbb{E}[\psi _{j,\gamma
	}(z_{i,t})])\right) ^{2}(\mathbb{E}[\psi _{j,\gamma
	}(z_{i,t})])^{2}=O_{p}(1),
\end{equation*}%
which together with (\ref{P_C_L_Auxillary7c_8}) and (\ref%
{P_C_L_Auxillary7c_9}) verifies (\ref{P_C_L_Auxillary7c_3}).\hfill $Q.E.D.$

\bigskip

\begin{lemma}
	\textit{\label{C_L_Auxillary7d}} Under Assumptions \ref{A1} and \ref{C_A1},\
	we have%
	\begin{equation}
		\sum_{g\in \mathcal{G}_{2}}n_{2,g}^{-1}\sum_{i\in I_{2,g}}(\hat{\sigma}%
		_{12,\gamma ,i}-\sigma _{12,\gamma ,i})^{2}=O_{p}(1).
		\label{C_L_Auxillary7d_1}
	\end{equation}
\end{lemma}

\noindent {\textsc{Proof of Lemma \ref{C_L_Auxillary7d}}}.\ By the
definitions of $\hat{\sigma}_{12,\gamma ,i}^{2}$ and $\sigma _{12,\gamma
	,i}^{2}$, we can write 
\begin{align}
	\hat{\sigma}_{12,\gamma ,i}-\sigma _{12,\gamma ,i}& =\frac{\widehat{\mathbb{E%
		}}_{T}[\psi _{1,\gamma }(z_{i,t};\hat{\phi}_{1,i})\psi _{2,\gamma }(z_{i,t};%
		\hat{\phi}_{2,i})]}{|\hat{\Psi}_{1,\gamma \gamma }(\hat{\phi}_{1,i})\hat{\Psi%
		}_{2,\gamma \gamma }(\hat{\phi}_{2,i})|^{1/2}}-\frac{\mathbb{E}_{T}[\psi
		_{1,\gamma }(z_{i,t})\psi _{2,\gamma }(z_{i,t})]}{(\Psi _{1,\gamma \gamma
			,i}\Psi _{1,\gamma \gamma ,i})^{1/2}}  \notag \\
	& =\frac{T^{-1}\sum_{t\leq T}(\psi _{1,\gamma }(z_{i,t};\hat{\phi}%
		_{1,i})\psi _{2,\gamma }(z_{i,t};\hat{\phi}_{2,i})-\mathbb{E}[\psi
		_{1,\gamma }(z_{i,t})\psi _{2,\gamma }(z_{i,t})])}{|\hat{\Psi}_{1,\gamma
			\gamma }(\hat{\phi}_{1,i})\hat{\Psi}_{2,\gamma \gamma }(\hat{\phi}%
		_{2,i})|^{1/2}}  \notag \\
	& +\frac{\mathbb{E}_{T}[\tilde{\psi}_{1,\gamma }(z_{i,t})\tilde{\psi}%
		_{2,\gamma }(z_{i,t})]\left( |\hat{\Psi}_{1,\gamma \gamma }(\hat{\phi}_{1,i})%
		\hat{\Psi}_{2,\gamma \gamma }(\hat{\phi}_{2,i})|^{1/2}-(\Psi _{1,\gamma
			\gamma ,i}\Psi _{2,\gamma \gamma ,i})^{1/2}\right) }{|\hat{\Psi}_{1,\gamma
			\gamma }(\hat{\phi}_{1,i})\hat{\Psi}_{2,\gamma \gamma }(\hat{\phi}%
		_{2,i})|^{1/2}}.  \label{P_C_L_Auxillary7d_1}
\end{align}%
From the expression above, the claim of the lemma follows if 
\begin{equation}
	\sum_{g\in \mathcal{G}_{2}}\sum_{i\in I_{2,g}}\frac{(T^{-1}\sum_{t\leq
			T}(\psi _{1,\gamma }(z_{i,t};\hat{\phi}_{1,i})\psi _{2,\gamma }(z_{i,t};\hat{%
			\phi}_{2,i})-\mathbb{E}[\psi _{1,\gamma }(z_{i,t})\psi _{2,\gamma
		}(z_{i,t})]))^{2}}{n_{2,g}\hat{\Psi}_{1,\gamma \gamma }(\hat{\phi}_{1,i})%
		\hat{\Psi}_{2,\gamma \gamma }(\hat{\phi}_{2,i})}=O_{p}(1)
	\label{P_C_L_Auxillary7d_2}
\end{equation}%
and%
\begin{equation}
	\sum_{g\in \mathcal{G}_{2}}\sum_{i\in I_{2,g}}\frac{(\mathbb{E}[\tilde{\psi}%
		_{1,\gamma }(z_{i,t})\tilde{\psi}_{2,\gamma }(z_{i,t})](|\hat{\Psi}%
		_{1,\gamma \gamma }(\hat{\phi}_{1,i})\hat{\Psi}_{2,\gamma \gamma }(\hat{\phi}%
		_{2,i})|^{1/2}-(\Psi _{1,\gamma \gamma ,i}\Psi _{2,\gamma \gamma
			,i})^{1/2}))^{2}}{n_{2,g}\hat{\Psi}_{1,\gamma \gamma }(\hat{\phi}_{1,i})\hat{%
			\Psi}_{2,\gamma \gamma }(\hat{\phi}_{2,i})}=O_{p}(1).
	\label{P_C_L_Auxillary7d_3}
\end{equation}

To verify (\ref{P_C_L_Auxillary7d_2}), we first observe that by Assumption %
\ref{A3} and the triangle inequality:%
\begin{align}
	& \left\vert T^{-1}\sum_{t\leq T}(\psi _{1,\gamma }(z_{i,t};\hat{\phi}%
	_{1,i})\psi _{2,\gamma }(z_{i,t};\hat{\phi}_{2,i})-\psi _{1,\gamma
	}(z_{i,t})\psi _{2,\gamma }(z_{i,t}))\right\vert  \notag \\
	& \leq \left\vert T^{-1}\sum_{t\leq T}(\psi _{1,\gamma }(z_{i,t};\hat{\phi}%
	_{1,i})-\psi _{1,\gamma }(z_{i,t}))(\psi _{2,\gamma }(z_{i,t};\hat{\phi}%
	_{2,i})-\psi _{2,\gamma }(z_{i,t}))\right\vert  \notag \\
	& +\left\vert T^{-1}\sum_{t\leq T}(\psi _{1,\gamma }(z_{i,t};\hat{\phi}%
	_{1,i})-\psi _{1,\gamma }(z_{i,t}))\psi _{2,\gamma }(z_{i,t})\right\vert 
	\notag \\
	& +\left\vert T^{-1}\sum_{t\leq T}(\psi _{2,\gamma }(z_{i,t};\hat{\phi}%
	_{2,i})-\psi _{2,\gamma }(z_{i,t}))\psi _{1,\gamma }(z_{i,t})\right\vert 
	\notag \\
	& \leq K(1+\max_{i\leq n}||\hat{\phi}_{1,i}-\phi _{1,i}^{\ast }||)(||\hat{%
		\phi}_{1,i}-\phi _{1,g}^{\ast }||+||\hat{\phi}_{2,i}-\phi _{2,i}^{\ast
	}||)T^{-1}\sum_{t\leq T}M_{5}(z_{i,t})^{2},  \label{P_C_L_Auxillary7d_4}
\end{align}%
where\ $M_{5}(z_{i,t})\equiv M(z_{i,t})+\psi _{1,\gamma }(z_{i,t})+\psi
_{2,\gamma }(z_{i,t})$. Thus, by the Cauchy-Schwarz inequality, 
\begin{align}
	& \left\vert \sum_{g\in \mathcal{G}_{2}}\sum_{i\in I_{2,g}}\frac{%
		(T^{-1}\sum_{t\leq T}(\psi _{1,\gamma }(z_{i,t};\hat{\phi}_{1,i})\psi
		_{2,\gamma }(z_{i,t};\hat{\phi}_{2,i})-\psi _{1,\gamma }(z_{i,t})\psi
		_{2,\gamma }(z_{i,t})))^{2}}{n_{2,g}\hat{\Psi}_{1,\gamma \gamma }(\hat{\phi}%
		_{1,i})\hat{\Psi}_{2,\gamma \gamma }(\hat{\phi}_{2,i})}\right\vert  \notag \\
	& \leq \frac{K(1+\max_{i\leq n}||\hat{\phi}_{1,i}-\phi _{1,i}^{\ast }||^{2})%
	}{\min_{i\leq n}|\hat{\Psi}_{1,\gamma \gamma }(\hat{\phi}_{1,i})\hat{\Psi}%
		_{2,\gamma \gamma }(\hat{\phi}_{2,i})|}  \notag \\
	& \times \sum_{g\in \mathcal{G}_{2}}n_{2,g}^{-1}\sum_{i\in I_{2,g}}(||\hat{%
		\phi}_{1,i}-\phi _{1,g}^{\ast }||+||\hat{\phi}_{2,i}-\phi _{2,i}^{\ast
	}||)^{2}\left( T^{-1}\sum_{t\leq T}M_{1}(z_{i,t})^{2}\right) ^{2}  \notag \\
	& \leq \frac{K(1+\max_{i\leq n}||\hat{\phi}_{1,i}-\phi _{1,i}^{\ast
		}||)\max_{i\leq n}\left( T^{-1}\sum_{t\leq T}M_{1}(z_{i,t})^{2}\right) ^{2}}{%
		\min_{i\leq n}|\hat{\Psi}_{1,\gamma \gamma }(\hat{\phi}_{1,i})\hat{\Psi}%
		_{2,\gamma \gamma }(\hat{\phi}_{2,i})|}  \notag \\
	& \times \left( \sum_{g\in \mathcal{G}_{2}}n_{2,g}^{-1}\sum_{i\in I_{2,g}}(||%
	\hat{\phi}_{1,i}-\phi _{1,g}^{\ast }||^{2}+||\hat{\phi}_{2,i}-\phi
	_{2,i}^{\ast }||^{2})\right) .  \label{P_C_L_Auxillary7d_5}
\end{align}%
By the similar arguments for showing (\ref{P_C_L_Auxillary2b_9}),\ we can
show that%
\begin{equation*}
	\max_{i\leq n}T^{-1}\sum_{t\leq T}M_{1}(z_{i,t})^{2}=O_{p}(1),
\end{equation*}%
which, along with Assumptions \ref{A1}(i) and \ref{A3}, Lemma \ref%
{C_L_Auxillary1}(ii, iii, iv) and (\ref{P_C_L_Auxillary7d_5}) implies that 
\begin{equation}
	\sum_{g\in \mathcal{G}_{2}}\sum_{i\in I_{2,g}}\frac{(T^{-1}\sum_{t\leq
			T}(\psi _{1,\gamma }(z_{i,t};\hat{\phi}_{1,i})\psi _{2,\gamma }(z_{i,t};\hat{%
			\phi}_{2,i})-\psi _{1,\gamma }(z_{i,t})\psi _{2,\gamma }(z_{i,t})))^{2}}{%
		n_{2,g}\hat{\Psi}_{1,\gamma \gamma }(\hat{\phi}_{1,i})\hat{\Psi}_{2,\gamma
			\gamma }(\hat{\phi}_{2,i})}=O_{p}(1).  \label{P_C_L_Auxillary7d_6}
\end{equation}%
Next, by Assumptions \ref{A1} and \ref{A3}, 
\begin{eqnarray*}
	&&\mathbb{E}\left[ \sum_{g\in \mathcal{G}_{2}}\sum_{i\in I_{2,g}}\left(
	T^{-1}\sum_{t\leq T}(\psi _{1,\gamma }(z_{i,t})\psi _{2,\gamma }(z_{i,t})-%
	\mathbb{E}[\psi _{1,\gamma }(z_{i,t})\psi _{2,\gamma }(z_{i,t})])\right) ^{2}%
	\right] \\
	&=&\sum_{g\in \mathcal{G}_{2}}\sum_{i\in I_{2,g}}\mathbb{E}\left[ \left(
	T^{-1}\sum_{t\leq T}(\psi _{1,\gamma }(z_{i,t})\psi _{2,\gamma }(z_{i,t})-%
	\mathbb{E}[\psi _{1,\gamma }(z_{i,t})\psi _{2,\gamma }(z_{i,t})])\right) ^{2}%
	\right] \\
	&\leq &KT^{-1}\sum_{g\in \mathcal{G}_{2}}\sum_{i\in I_{2,g}}\max_{t\leq
		T}\left\Vert \psi _{1,\gamma }(z_{i,t})\psi _{2,\gamma }(z_{i,t})\right\Vert
	_{2+\delta /2}^{2}\leq K,
\end{eqnarray*}%
which along with Markov's inequality, Assumptions \ref{A1}(i) and Lemma \ref%
{C_L_Auxillary1}(ii) shows that 
\begin{equation}
	\sum_{g\in \mathcal{G}_{2}}\sum_{i\in I_{2,g}}\frac{(T^{-1}\sum_{t\leq
			T}(\psi _{1,\gamma }(z_{i,t})\psi _{2,\gamma }(z_{i,t})-\mathbb{E}[\psi
		_{1,\gamma }(z_{i,t})\psi _{2,\gamma }(z_{i,t})]))^{2}}{n_{2,g}\hat{\Psi}%
		_{1,\gamma \gamma }(\hat{\phi}_{1,i})\hat{\Psi}_{2,\gamma \gamma }(\hat{\phi}%
		_{2,i})}=O_{p}(1).  \label{P_C_L_Auxillary7d_7}
\end{equation}%
The desired result in (\ref{P_C_L_Auxillary7d_2}) follows from (\ref%
{P_C_L_Auxillary7d_6}) and (\ref{P_C_L_Auxillary7d_7}).

To verify (\ref{P_C_L_Auxillary7d_3}), we first notice that the term on its
LHS satisfies:%
\begin{align*}
	& \sum_{g\in \mathcal{G}_{2}}\sum_{i\in I_{2,g}}\frac{(\mathbb{E}[\tilde{\psi%
		}_{1,\gamma }(z_{i,t})\tilde{\psi}_{2,\gamma }(z_{i,t})](|\hat{\Psi}%
		_{1,\gamma \gamma }(\hat{\phi}_{1,i})\hat{\Psi}_{2,\gamma \gamma }(\hat{\phi}%
		_{2,i})|^{1/2}-(\Psi _{1,\gamma \gamma ,i}\Psi _{2,\gamma \gamma
			,i})^{1/2}))^{2}}{n_{2,g}|\hat{\Psi}_{1,\gamma \gamma }(\hat{\phi}_{1,i})%
		\hat{\Psi}_{2,\gamma \gamma }(\hat{\phi}_{2,i})|} \\
	& \leq \frac{K\sum_{g\in \mathcal{G}_{2}}n_{2,g}^{-1}\sum_{i\in I_{2,g}}(|%
		\hat{\Psi}_{1,\gamma \gamma }(\hat{\phi}_{1,i})\hat{\Psi}_{2,\gamma \gamma }(%
		\hat{\phi}_{2,i})|^{1/2}-(\Psi _{1,\gamma \gamma ,i}\Psi _{2,\gamma \gamma
			,i})^{1/2}))^{2}}{\min_{i\leq n}|\hat{\Psi}_{1,\gamma \gamma }(\hat{\phi}%
		_{1,i})\hat{\Psi}_{2,\gamma \gamma }(\hat{\phi}_{2,i})|} \\
	& \leq \frac{K\sum_{g\in \mathcal{G}_{2}}n_{2,g}^{-1}\sum_{i\in I_{2,g}}(%
		\hat{\Psi}_{1,\gamma \gamma }(\hat{\phi}_{1,i})\hat{\Psi}_{2,\gamma \gamma }(%
		\hat{\phi}_{2,i})-\Psi _{1,\gamma \gamma ,i}\Psi _{2,\gamma \gamma ,i}))^{2}%
	}{\min_{i\leq n}|\hat{\Psi}_{1,\gamma \gamma }(\hat{\phi}_{1,i})\hat{\Psi}%
		_{2,\gamma \gamma }(\hat{\phi}_{2,i})|^{2}} \\
	& \leq K\max_{i\leq n}(\hat{\Psi}_{1,\gamma \gamma }(\hat{\phi}_{2,i})-\Psi
	_{1,\gamma \gamma ,i})^{2}\frac{\sum_{g\in \mathcal{G}_{2}}(\hat{\Psi}%
		_{2,\gamma \gamma }(\hat{\phi}_{2,i})-\Psi _{2,\gamma \gamma ,i})^{2}}{%
		\min_{i\leq n}|\hat{\Psi}_{1,\gamma \gamma }(\hat{\phi}_{1,i})\hat{\Psi}%
		_{2,\gamma \gamma }(\hat{\phi}_{2,i})|^{2}} \\
	& +\frac{K\sum_{g\in \mathcal{G}_{2}}n_{2,g}^{-1}\sum_{i\in I_{2,g}}(\hat{%
			\Psi}_{1,\gamma \gamma }(\hat{\phi}_{1,i})-\Psi _{1,\gamma \gamma ,i})^{2}}{%
		\min_{i\leq n}|\hat{\Psi}_{1,\gamma \gamma }(\hat{\phi}_{1,i})\hat{\Psi}%
		_{2,\gamma \gamma }(\hat{\phi}_{2,i})|^{2}}+\frac{K\sum_{g\in \mathcal{G}%
			_{2}}(\hat{\Psi}_{2,\gamma \gamma }(\hat{\phi}_{1,i})-\Psi _{2,\gamma \gamma
			,i})^{2}}{\min_{i\leq n}|\hat{\Psi}_{1,\gamma \gamma }(\hat{\phi}_{1,i})\hat{%
			\Psi}_{2,\gamma \gamma }(\hat{\phi}_{2,i})|^{2}},
\end{align*}%
which together with Lemma \ref{C_L_Auxillary1}(ii, v) shows (\ref%
{P_C_L_Auxillary7d_3}).\hfill $Q.E.D.$

\bigskip

\begin{lemma}
	\textit{\label{C_L_Auxillary9}\ }Under Assumptions \ref{A1}, \ref{C_A1} and %
	\ref{C_A2},\ we have:%
	\begin{equation*}
		\omega _{n,T}^{2}=\sigma _{n,T}^{2}+(2nT)^{-1}\sum_{g\in \mathcal{G}%
			_{2}}\sum_{i\in I_{2,g}}\left( \sigma _{1,\gamma ,i}^{4}+n_{2,g}^{-2}\sigma
		_{2,\gamma ,i}^{2}\sum_{i^{\prime }\in I_{2,g}}\sigma _{2,\gamma ,i^{\prime
		}}^{2}-2n_{2,g}^{-1}\sigma _{12,\gamma ,i}^{2}\right) +O(\omega
		_{n,T}T^{-1}).
	\end{equation*}
\end{lemma}

\noindent {\textsc{Proof of Lemma \ref{C_L_Auxillary9}}}.\ To prove the
claim in the lemma, we first observe that by Assumption \ref{A1}(iii),%
\begin{equation*}
	\mathrm{Cov}(\tilde{\Psi}_{i},\tilde{U}_{i})=0\text{ \ \ and \ \ }\mathrm{Cov%
	}(\tilde{V}_{i},\tilde{U}_{i})=0.
\end{equation*}%
Therefore, 
\begin{equation*}
	\mathrm{Var}(\tilde{\Psi}_{i}+\tilde{V}_{i}+\tilde{U}_{i})=\mathrm{Var}(%
	\tilde{\Psi}_{i}+\tilde{V}_{i})+\mathrm{Var}(\tilde{U}_{i}).
\end{equation*}%
Together with the definition of $\omega _{n,T}^{2}$, this yields%
\begin{align}
	\omega _{n,T}^{2}& =n^{-1}\sum_{i\leq n}(\mathrm{Var}(\tilde{\Psi}_{i}+%
	\tilde{V}_{i})+\mathrm{Var}(\tilde{U}_{i}))  \notag \\
	& =n^{-1}\sum_{i\leq n}(\mathrm{Var}(\tilde{\Psi}_{i})+\mathrm{Var}(\tilde{V}%
	_{i})+2\mathrm{Cov}(\tilde{\Psi}_{i},\tilde{V}_{i})+\mathrm{Var}(\tilde{U}%
	_{i}))  \notag \\
	& =\sigma _{n,T}^{2}+n^{-1}\sum_{i\leq n}(\mathrm{Var}(\tilde{V}_{i})+%
	\mathrm{Var}(\tilde{U}_{i}))+2n^{-1}\sum_{i\leq n}\mathrm{Cov}(\tilde{\Psi}%
	_{i},\tilde{V}_{i}).  \label{P_C_L_Auxillary9_1}
\end{align}%
We next study the second and the third terms in the far RHS of (\ref%
{P_C_L_Auxillary9_1}).

For model 1,\ $\tilde{U}_{1,i}=0$ and%
\begin{align}
	\tilde{V}_{1,i}& =(2T^{1/2})^{-1}\tilde{\Psi}_{1,\gamma
		,i}^{2}=(2T^{1/2})^{-1}\left( T^{-1/2}\sum_{t\leq T}\tilde{\psi}_{1,\gamma
	}^{\ast }\left( z_{i,t}\right) \right) ^{2}  \notag \\
	& =(2T^{3/2})^{-1}\sum_{t\leq T}\tilde{\psi}_{1,\gamma }^{\ast }\left(
	z_{i,t}\right) ^{2}+T^{-3/2}\sum_{t=2}^{T}\sum_{t^{\prime }=1}^{t-1}\tilde{%
		\psi}_{1,\gamma }^{\ast }\left( z_{i,t}\right) \tilde{\psi}_{1,\gamma
	}^{\ast }\left( z_{i,t^{\prime }}\right) ,  \label{P_C_L_Auxillary9_2}
\end{align}%
while for model 2, we have%
\begin{align}
	\tilde{V}_{2,i}& =(2n_{2,g}T^{1/2})^{-1}\tilde{\Psi}_{2,\gamma
		,i}^{2}=(2n_{2,g}T^{1/2})^{-1}\left( T^{-1/2}\sum_{t\leq T}\tilde{\psi}%
	_{2,\gamma }^{\ast }\left( z_{i,t}\right) \right) ^{2}  \notag \\
	& =(2T^{3/2})^{-1}\sum_{t\leq T}n_{2,g}^{-1}\tilde{\psi}_{2,\gamma }^{\ast
	}\left( z_{i,t}\right) ^{2}+T^{-3/2}\sum_{t=2}^{T}\sum_{t^{\prime
		}=1}^{t-1}n_{2,g}^{-1}\tilde{\psi}_{2,\gamma }^{\ast }\left( z_{i,t}\right) 
	\tilde{\psi}_{2,\gamma }^{\ast }\left( z_{i,t^{\prime }}\right) ,
	\label{P_C_L_Auxillary9_3}
\end{align}%
and \ 
\begin{equation}
	\tilde{U}_{2,i}=(n_{2,g}T^{3/2})^{-1}\sum_{\{i^{\prime }\in
		I_{2,g}:i^{\prime }<i\}}\left( \sum_{t\leq T}\tilde{\psi}_{2,\gamma }^{\ast
	}\left( z_{i,t}\right) \right) \left( \sum_{t^{\prime }\leq T}\tilde{\psi}%
	_{2,\gamma }^{\ast }\left( z_{i^{\prime },t^{\prime }}\right) \right) .
	\label{P_C_L_Auxillary9_4}
\end{equation}%
Subtracting (\ref{P_C_L_Auxillary9_3}) from (\ref{P_C_L_Auxillary9_2})
yields 
\begin{eqnarray}
	\tilde{V}_{1,i}-\tilde{V}_{2,i} &=&(2T^{3/2})^{-1}\sum_{t\leq T}(\tilde{\psi}%
	_{1,\gamma }^{\ast }\left( z_{i,t}\right) ^{2}-n_{2,g}^{-1}\tilde{\psi}%
	_{2,\gamma }^{\ast }\left( z_{i,t}\right) ^{2})  \notag \\
	&&+T^{-3/2}\sum_{t=2}^{T}\sum_{t^{\prime }=1}^{t-1}(\tilde{\psi}_{1,\gamma
	}^{\ast }\left( z_{i,t}\right) \tilde{\psi}_{1,\gamma }^{\ast }\left(
	z_{i,t^{\prime }}\right) -n_{2,g}^{-1}\tilde{\psi}_{2,\gamma }^{\ast }\left(
	z_{i,t}\right) \tilde{\psi}_{2,\gamma }^{\ast }\left( z_{i,t^{\prime
	}}\right) ).  \label{P_C_L_Auxillary9_5a}
\end{eqnarray}%
For any $h\geq 1$ with $t-h\geq 1$ and\ $j,j^{\prime }\in \{1,2\}$,%
\begin{equation}
	\mathrm{Cov}\left( \tilde{\psi}_{j,\gamma }^{\ast }\left( z_{i,t-h}\right)
	^{2},\tilde{\psi}_{j^{\prime },\gamma }^{\ast }\left( z_{i,t}\right)
	\sum_{t^{\prime }=1}^{t-1}\tilde{\psi}_{j^{\prime },\gamma }^{\ast }\left(
	z_{i,t^{\prime }}\right) \right) =\sum_{t^{\prime }=1}^{t-1}\mathbb{E}\left[ 
	\tilde{\psi}_{j,\gamma }^{\ast }\left( z_{i,t-h}\right) ^{2}\tilde{\psi}%
	_{j^{\prime },\gamma }^{\ast }\left( z_{i,t^{\prime }}\right) \tilde{\psi}%
	_{j^{\prime },\gamma }^{\ast }\left( z_{i,t}\right) \right] =0,
	\label{P_C_L_Auxillary9_5b}
\end{equation}%
where the second equality is by Assumption \ref{C_A2}. For any $h\geq 0$
with $t+h\leq T$ and $j,j^{\prime }\in \{1,2\}$, we have%
\begin{eqnarray}
	&&\left\vert \mathrm{Cov}\left( \tilde{\psi}_{j,\gamma }^{\ast }\left(
	z_{i,t+h}\right) ^{2},\tilde{\psi}_{j^{\prime },\gamma }^{\ast }\left(
	z_{i,t}\right) \sum_{t^{\prime }=1}^{t-1}\tilde{\psi}_{j^{\prime },\gamma
	}^{\ast }\left( z_{i,t^{\prime }}\right) \right) \right\vert  \notag \\
	&\leq &K\alpha _{i}^{1/4}\left\Vert \tilde{\psi}_{j,\gamma }^{\ast }\left(
	z_{i,t+h}\right) ^{2}\right\Vert _{4}\left\Vert \tilde{\psi}_{j^{\prime
		},\gamma }^{\ast }\left( z_{i,t}\right) \sum_{t^{\prime }=1}^{t-1}\tilde{\psi%
	}_{j^{\prime },\gamma }^{\ast }\left( z_{i,t^{\prime }}\right) \right\Vert
	_{2}  \notag \\
	&\leq &K\alpha _{i}^{1/4}(h)\left\Vert \tilde{\psi}_{j,\gamma }^{\ast
	}\left( z_{i,t+h}\right) ^{2}\right\Vert _{4}\left\Vert \tilde{\psi}%
	_{j^{\prime },\gamma }^{\ast }\left( z_{i,t}\right) \right\Vert
	_{4}\left\Vert \sum_{t^{\prime }=1}^{t-1}\tilde{\psi}_{j^{\prime },\gamma
	}^{\ast }\left( z_{i,t^{\prime }}\right) \right\Vert _{4}\leq K\alpha
	_{i}^{1/4}(h)t^{1/2},  \label{P_C_L_Auxillary9_5c}
\end{eqnarray}%
where we use the covariance inequality for strong mixing processes, H\"{o}%
lder's inequality, Assumption \ref{C_A1}(iii), and a Rosenthal-type
inequality $||\sum_{t^{\prime }=1}^{t-1}\tilde{\psi}_{j^{\prime },\gamma
}^{\ast }\left( z_{i,t^{\prime }}\right) ||_{4}\leq Kt^{1/2}$ (cf. (\ref%
{P_L0_2}) in the proof of Lemma \ref{L0}). Combining (\ref%
{P_C_L_Auxillary9_5b}) and (\ref{P_C_L_Auxillary9_5c}) yields%
\begin{eqnarray}
	&&\left\vert \mathrm{Cov}\left( \sum_{t\leq T}(\tilde{\psi}_{1,\gamma
	}^{\ast }\left( z_{i,t}\right) ^{2}-n_{2,g}^{-1}\tilde{\psi}_{2,\gamma
	}^{\ast }\left( z_{i,t}\right) ^{2}),\sum_{t=2}^{T}\sum_{t^{\prime }=1}^{t-1}%
	\tilde{\psi}_{1,\gamma }^{\ast }\left( z_{i,t}\right) \tilde{\psi}_{1,\gamma
	}^{\ast }\left( z_{i,t^{\prime }}\right) -n_{2,g}^{-1}\tilde{\psi}_{2,\gamma
	}^{\ast }\left( z_{i,t}\right) \tilde{\psi}_{2,\gamma }^{\ast }\left(
	z_{i,t^{\prime }}\right) \right) \right\vert  \notag \\
	&\leq &K\sum_{j\leq 2}\sum_{j^{\prime }\leq 2}\sum_{0\leq h\leq
		T-1}\sum_{t\leq T-h}\mathrm{Cov}\left( \tilde{\psi}_{j,\gamma }^{\ast
	}\left( z_{i,t+h}\right) ^{2},\sum_{t^{\prime }=1}^{t-1}\tilde{\psi}%
	_{j^{\prime },\gamma }^{\ast }\left( z_{i,t}\right) \tilde{\psi}_{j^{\prime
		},\gamma }^{\ast }\left( z_{i,t^{\prime }}\right) \right)  \notag \\
	&\leq &K\sum_{j\leq 2}\sum_{j^{\prime }\leq 2}\sum_{0\leq h\leq
		T-1}\sum_{t\leq T-h}\alpha _{i}^{1/4}(h)t^{1/2}\leq K\sum_{j\leq
		2}\sum_{j^{\prime }\leq 2}\sum_{0\leq h\leq T}\alpha
	_{i}^{1/4}(h)(T-h)^{3/2}\leq KT^{3/2}.  \label{P_C_L_Auxillary9_5d}
\end{eqnarray}%
The last inequality follows since 
\begin{equation}
	\sum_{0\leq h\leq T}\alpha _{i}^{1/4}(h)(T-h)^{3/2}=T^{3/2}\sum_{0\leq h\leq
		T}\alpha _{i}^{1/4}(h)(1-h/T)^{3/2}\leq T^{3/2}\sum_{0\leq h\leq T}\alpha
	_{i}^{1/4}(h)\leq KT^{3/2},  \label{P_C_L_Auxillary9_5da}
\end{equation}%
where the final inequality is implied by Assumption \ref{A1}(iv). Using (\ref%
{P_C_L_Auxillary9_5a}) and (\ref{P_C_L_Auxillary9_5d}),%
\begin{align}
	n^{-1}\sum_{i\leq n}\mathrm{Var}(\tilde{V}_{i})& =n^{-1}\sum_{i\leq n}%
	\mathrm{Var}(\tilde{V}_{1,i}-\tilde{V}_{2,i})  \notag \\
	& =n^{-1}\sum_{i\leq n}\mathrm{Var}\left(
	T^{-3/2}\sum_{t=2}^{T}\sum_{t^{\prime }=1}^{t-1}(\tilde{\psi}_{1,\gamma
	}^{\ast }\left( z_{i,t}\right) \tilde{\psi}_{1,\gamma }^{\ast }\left(
	z_{i,t^{\prime }}\right) -n_{2,g}^{-1}\tilde{\psi}_{2,\gamma }^{\ast }\left(
	z_{i,t}\right) \tilde{\psi}_{2,\gamma }^{\ast }\left( z_{i,t^{\prime
	}}\right) )\right)  \notag \\
	& +n^{-1}\sum_{i\leq n}\mathrm{Var}\left( (2T^{3/2})^{-1}\sum_{t\leq T}(%
	\tilde{\psi}_{1,\gamma }^{\ast }\left( z_{i,t}\right) ^{2}-n_{2,g}^{-1}%
	\tilde{\psi}_{2,\gamma }^{\ast }\left( z_{i,t}\right) ^{2})\right)
	+O_{p}(T^{-3/2}).  \label{P_C_L_Auxillary9_5e}
\end{align}%
From Assumptions \ref{A1} and \ref{C_A1}(iii) and the covariance inequality
of strong mixing process, it follows that 
\begin{eqnarray}
	&&n^{-1}\sum_{i\leq n}\mathrm{Var}\left( (2T^{3/2})^{-1}\sum_{t\leq T}(%
	\tilde{\psi}_{1,\gamma }^{\ast }\left( z_{i,t}\right) ^{2}-n_{2,g}^{-1}%
	\tilde{\psi}_{2,\gamma }^{\ast }\left( z_{i,t}\right) ^{2})\right)  \notag \\
	&\leq &K(nT^{3})^{-1}\sum_{j\leq 2}\sum_{i\leq n}\mathrm{Var}\left(
	\sum_{t\leq T}\tilde{\psi}_{j,\gamma }^{\ast }\left( z_{i,t}\right)
	^{2}\right) \leq KT^{-2}.  \label{P_C_L_Auxillary9_5f}
\end{eqnarray}%
Let $\sigma _{j,\gamma ,i,t}\equiv (\mathbb{E}[\tilde{\psi}_{j,\gamma
}^{\ast }\left( z_{i,t}\right) ^{2}])^{1/2}$ and $\sigma _{12,\gamma
	,i,t}\equiv \mathbb{E}[\tilde{\psi}_{1,\gamma }^{\ast }\left( z_{i,t}\right) 
\tilde{\psi}_{2,\gamma }^{\ast }\left( z_{i,t}\right) ]$. A direct
calculation using Assumption \ref{C_A2} yields%
\begin{eqnarray}
	&&\mathrm{Var}\left( T^{-3/2}\sum_{t=2}^{T}\sum_{t^{\prime }=1}^{t-1}(\tilde{%
		\psi}_{1,\gamma }^{\ast }\left( z_{i,t}\right) \tilde{\psi}_{1,\gamma
	}^{\ast }\left( z_{i,t^{\prime }}\right) -n_{2,g}^{-1}\tilde{\psi}_{2,\gamma
	}^{\ast }\left( z_{i,t}\right) \tilde{\psi}_{2,\gamma }^{\ast }\left(
	z_{i,t^{\prime }}\right) )\right)  \notag \\
	&=&T^{-3}\sum_{t=2}^{T}\sum_{t^{\prime }=1}^{t-1}\mathrm{Var}(\tilde{\psi}%
	_{1,\gamma }^{\ast }\left( z_{i,t}\right) \tilde{\psi}_{1,\gamma }^{\ast
	}\left( z_{i,t^{\prime }}\right) -n_{2,g}^{-1}\tilde{\psi}_{2,\gamma }^{\ast
	}\left( z_{i,t}\right) \tilde{\psi}_{2,\gamma }^{\ast }\left( z_{i,t^{\prime
	}}\right) )  \notag \\
	&=&T^{-3}\sum_{t=2}^{T}\sum_{t^{\prime }=1}^{t-1}(\mathrm{Var}(\tilde{\psi}%
	_{1,\gamma }^{\ast }\left( z_{i,t}\right) \tilde{\psi}_{1,\gamma }^{\ast
	}\left( z_{i,t^{\prime }}\right) )+n_{2,g}^{-2}\mathrm{Var}(\tilde{\psi}%
	_{2,\gamma }^{\ast }\left( z_{i,t}\right) \tilde{\psi}_{2,\gamma }^{\ast
	}\left( z_{i,t^{\prime }}\right) ))  \notag \\
	&&-2T^{-3}\sum_{t=2}^{T}\sum_{t^{\prime }=1}^{t-1}\mathrm{Cov}(\tilde{\psi}%
	_{1,\gamma }^{\ast }\left( z_{i,t}\right) \tilde{\psi}_{1,\gamma }^{\ast
	}\left( z_{i,t^{\prime }}\right) ,n_{2,g}^{-1}\tilde{\psi}_{2,\gamma }^{\ast
	}\left( z_{i,t}\right) \tilde{\psi}_{2,\gamma }^{\ast }\left( z_{i,t^{\prime
	}}\right) )  \notag \\
	&=&T^{-3}\sum_{t=2}^{T}\sum_{t^{\prime }=1}^{t-1}(\sigma _{1,\gamma
		,i,t}^{2}\sigma _{1,\gamma ,i,t^{\prime }}^{2}+n_{2,g}^{-2}\sigma _{2,\gamma
		,i,t}^{2}\sigma _{2,\gamma ,i,t^{\prime }}^{2}-2n_{2,g}^{-1}\sigma
	_{12,\gamma ,i,t}\sigma _{12,\gamma ,i,t^{\prime }})  \notag \\
	&=&(2T^{3})^{-1}\left[ \left( \sum_{t\leq T}\sigma _{1,\gamma
		,i,t}^{2}\right) ^{2}+n_{2,g}^{-2}\left( \sum_{t\leq T}\sigma _{2,\gamma
		,i,t}^{2}\right) ^{2}-2n_{2,g}^{-1}\left( \sum_{t\leq T}\sigma _{12,\gamma
		,i,t}\right) ^{2}\right]  \notag \\
	&&-T^{-3}\left( \sum_{t\leq T}\sigma _{1,\gamma
		,i,t}^{4}+n_{2,g}^{-2}\sum_{t\leq T}\sigma _{2,\gamma
		,i,t}^{4}-2n_{2,g}^{-1}\sum_{t\leq T}\sigma _{12,\gamma ,i,t}^{2}\right)
	\label{P_C_L_Auxillary9_5g}
\end{eqnarray}%
Under Assumption \ref{C_A1}(iii),%
\begin{equation*}
	T^{-3}\left\vert \sum_{t\leq T}\sigma _{1,\gamma
		,i,t}^{4}+n_{2,g}^{-2}\sum_{t\leq T}\sigma _{2,\gamma
		,i,t}^{4}-2n_{2,g}^{-1}\sum_{t\leq T}\sigma _{12,\gamma ,i,t}^{2}\right\vert
	\leq KT^{-2},
\end{equation*}%
which along with (\ref{P_C_L_Auxillary9_5g}) and the definitions of $\sigma
_{j,\gamma ,i}^{2}$ and $\sigma _{12,\gamma ,i}$ implies that 
\begin{eqnarray}
	&&n^{-1}\sum_{i\leq n}\mathrm{Var}\left(
	T^{-3/2}\sum_{t=2}^{T}\sum_{t^{\prime }=1}^{t-1}(\tilde{\psi}_{1,\gamma
	}^{\ast }\left( z_{i,t}\right) \tilde{\psi}_{1,\gamma }^{\ast }\left(
	z_{i,t^{\prime }}\right) -n_{2,g}^{-1}\tilde{\psi}_{2,\gamma }^{\ast }\left(
	z_{i,t}\right) \tilde{\psi}_{2,\gamma }^{\ast }\left( z_{i,t^{\prime
	}}\right) )\right)  \notag \\
	&=&(2nT)^{-1}\sum_{i\leq n}(\sigma _{1,\gamma ,i}^{4}+n_{2,g}^{-2}\sigma
	_{2,\gamma ,i}^{4}-2n_{2,g}^{-1}\sigma _{12,\gamma ,i}^{2})+O(T^{-2}).
	\label{P_C_L_Auxillary9_5h}
\end{eqnarray}%
Using (\ref{P_C_L_Auxillary9_5e}), (\ref{P_C_L_Auxillary9_5f})\ and (\ref%
{P_C_L_Auxillary9_5h}), we obtain%
\begin{equation}
	n^{-1}\sum_{i\leq n}\mathrm{Var}(\tilde{V}_{i})=(2nT)^{-1}\sum_{i\leq
		n}(\sigma _{1,\gamma ,i}^{4}+n_{2,g}^{-2}\sigma _{2,\gamma
		,i}^{4}-2n_{2,g}^{-1}\sigma _{12,\gamma ,i}^{2})+O(T^{-3/2}).
	\label{P_C_L_Auxillary9_6}
\end{equation}%
Similarly, by Assumption \ref{C_A2} we have 
\begin{align}
	n^{-1}\sum_{i\leq n}\mathrm{Var}(\tilde{U}_{i})& =n^{-1}\sum_{i\leq n}%
	\mathrm{Var}(\tilde{U}_{2,i})=(nT)^{-1}\sum_{g\in \mathcal{G}%
		_{2}}n_{2,g}^{-2}\sum_{i\in I_{2,g}}\sigma _{2,\gamma
		,i}^{2}\sum_{\{i^{\prime }\in I_{2,g}:i^{\prime }<i\}}\sigma _{2,\gamma
		,i^{\prime }}^{2}  \notag \\
	& =(2nT)^{-1}\sum_{g\in \mathcal{G}_{2}}n_{2,g}^{-2}\left[ \left( \sum_{i\in
		I_{2,g}}\sigma _{2,\gamma ,i}^{2}\right) ^{2}-\sum_{i\in I_{2,g}}\sigma
	_{2,\gamma ,i}^{4}\right] ,  \label{P_C_L_Auxillary9_10}
\end{align}%
which, along with (\ref{P_C_L_Auxillary9_6}) shows that 
\begin{align}
	& n^{-1}\sum_{i\leq n}(\mathrm{Var}(\tilde{V}_{i})+\mathrm{Var}(\tilde{U}%
	_{i}))  \notag \\
	& =(2nT)^{-1}\sum_{g\in \mathcal{G}_{2}}\sum_{i\in I_{2,g}}\left( \sigma
	_{1,\gamma ,i}^{4}+n_{2,g}^{-2}\sigma _{2,\gamma ,i}^{2}\sum_{i^{\prime }\in
		I_{2,g}}\sigma _{2,\gamma ,i^{\prime }}^{2}-2n_{2,g}^{-1}\sigma _{12,\gamma
		,i}^{2}\right) +O(T^{-2}).  \label{P_C_L_Auxillary9_11}
\end{align}

To bound the third term in the far RHS of (\ref{P_C_L_Auxillary9_1}), we
observe that by Assumptions \ref{A1}, \ref{C_A1}(ii, iii) and the covariance
inequality of strong mixing process,%
\begin{eqnarray*}
	&&\left\vert \mathrm{Cov}\left( \sum_{t\leq T}\Delta \psi \left(
	z_{i,t}\right) ,\sum_{t\leq T}\tilde{\psi}_{j,\gamma }^{\ast
	}(z_{i,t})^{2}\right) \right\vert \\
	&\leq &\sum_{h\leq T}\sum_{t\leq T-h}\left( \left\vert \mathrm{Cov}\left(
	\Delta \psi \left( z_{i,t+h}\right) ,\tilde{\psi}_{j,\gamma }^{\ast
	}(z_{i,t})^{2}\right) \right\vert +\left\vert \mathrm{Cov}\left( \Delta \psi
	\left( z_{i,t}\right) ,\tilde{\psi}_{j,\gamma }^{\ast
	}(z_{i,t+h})^{2}\right) \right\vert \right) \\
	&\leq &\sum_{h\leq T}\sum_{t\leq T-h}\alpha _{i}^{\delta /(2+\delta
		)}(h)\left( \left\Vert \Delta \psi \left( z_{i,t+h}\right) \right\Vert
	_{2+\delta }\left\Vert \tilde{\psi}_{j,\gamma }^{\ast
	}(z_{i,t})^{2}\right\Vert _{2+\delta }+\left\Vert \Delta \psi \left(
	z_{i,t}\right) \right\Vert _{2+\delta }\left\Vert \tilde{\psi}_{j,\gamma
	}^{\ast }(z_{i,t+h})^{2}\right\Vert _{2+\delta }\right) \\
	&\leq &K\omega _{n,T}\sum_{h\leq T}(T-h)\alpha _{i}^{\delta /(2+\delta
		)}(h)\leq K\omega _{n,T}T,
\end{eqnarray*}%
where the final inequality follows by arguments analogous to those used to
establish (\ref{P_C_L_Auxillary9_5da}). This implies%
\begin{eqnarray}
	\left\vert n^{-1}\sum_{i\leq n}\mathrm{Cov}(\tilde{\Psi}_{i},\tilde{V}%
	_{i})\right\vert &\leq &n^{-1}\sum_{i\leq n}\left\vert \mathrm{Cov}\left( 
	\tilde{\Psi}_{i},(2T^{3/2})^{-1}\sum_{t\leq T}(\tilde{\psi}_{1,\gamma
	}^{\ast }(z_{i,t})^{2}-n_{2,g}^{-1}\tilde{\psi}_{2,\gamma }^{\ast
	}(z_{i,t})^{2})\right) \right\vert  \notag \\
	&=&(2nT^{2})^{-1}\sum_{i\leq n}\left\vert \mathrm{Cov}\left( \sum_{t\leq
		T}\Delta \psi \left( z_{i,t}\right) ,\sum_{t\leq T}(\tilde{\psi}_{1,\gamma
	}^{\ast }(z_{i,t})^{2}-n_{2,g}^{-1}\tilde{\psi}_{2,\gamma }^{\ast
	}(z_{i,t})^{2})\right) \right\vert  \notag \\
	&\leq &K\omega _{n,T}T^{-1}.  \label{P_C_L_Auxillary9_12}
\end{eqnarray}%
The claim of the lemma follows from Assumption \ref{C_A1}(v), (\ref%
{P_C_L_Auxillary9_1}), (\ref{P_C_L_Auxillary9_11}) and (\ref%
{P_C_L_Auxillary9_12}).\hfill $Q.E.D.$

\bigskip

\begin{lemma}
	\textit{\label{C_L_Auxillary10}\ }Under Assumptions \ref{A1} and \ref{C_A1}%
	,\ we have $\omega _{n,T}^{2}\leq K$.
\end{lemma}

\noindent {\textsc{Proof of Lemma \ref{C_L_Auxillary10}}}. From the
definition of $\omega _{n,T}^{2}$, we have 
\begin{equation}
	\omega _{n,T}^{2}=n^{-1}\sum_{i\leq n}\mathrm{Var}(\tilde{\Psi}_{i}+\tilde{V}%
	_{i}+\tilde{U}_{i})\leq Kn^{-1}\sum_{i\leq n}(\mathrm{Var}(\tilde{\Psi}_{i})+%
	\mathrm{Var}(\tilde{V}_{i})+\mathrm{Var}(\tilde{U}_{i})).
	\label{P_C_L_Auxillary10_1}
\end{equation}%
Under Assumptions \ref{A1} and \ref{A3}(iii), we \ can use covariance
inequality for strong mixing processes to obtain%
\begin{equation}
	n^{-1}\sum_{i\leq n}\mathrm{Var}(\tilde{\Psi}_{i})=n^{-1}\sum_{i\leq n}%
	\mathrm{Var}\left( T^{-1/2}\sum_{t\leq T}\Delta \psi \left( z_{i,t}\right)
	\right) \leq Kn^{-1}\sum_{i\leq n}\max_{t\leq T}\left\Vert \Delta \psi
	\left( z_{i,t}\right) \right\Vert _{2+\delta }^{2}\leq K,
	\label{P_C_L_Auxillary10_2}
\end{equation}%
where the first inequality follows from arguments analogous to those used in
establishing (\ref{P_C_T2_0}). From the definition of $\tilde{V}_{i}$, we
have 
\begin{eqnarray}
	n^{-1}\sum_{i\leq n}\mathrm{Var}(\tilde{V}_{i}) &\leq &n^{-1}\sum_{i\leq n}%
	\mathrm{Var}\left( (2T^{1/2})^{-1}\tilde{\Psi}_{1,\gamma
		,i}^{2}-(2n_{2,g}T^{1/2})^{-1}\tilde{\Psi}_{2,\gamma ,i}^{2}\right)  \notag
	\\
	&\leq &(2T)^{-1}\max_{j=1,2}n^{-1}\sum_{i\leq n}\mathbb{E}\left[ \left(
	T^{-1/2}\sum_{t\leq T}\tilde{\psi}_{j,\gamma }^{\ast }\left( z_{i,t}\right)
	\right) ^{4}\right] \leq KT^{-1},  \label{P_C_L_Auxillary10_3}
\end{eqnarray}%
where the last inequality is by Assumptions \ref{A1} and \ref{A3}(iii), and
Rosenthal's inequality. Similarly, 
\begin{eqnarray*}
	\mathrm{Var}(\tilde{U}_{i}) &=&n_{2,g}^{-2}T^{-1}\sum_{\{i^{\prime }\in
		I_{2,g}:i^{\prime }<i\}}\mathrm{Var}\left( \left( T^{-1/2}\sum_{t\leq T}%
	\tilde{\psi}_{2,\gamma }^{\ast }\left( z_{i,t}\right) \right) \left(
	T^{-1/2}\sum_{t^{\prime }\leq T}\tilde{\psi}_{2,\gamma }^{\ast }\left(
	z_{i^{\prime },t^{\prime }}\right) \right) \right) \\
	&\leq &n_{2,g}^{-2}T^{-1}\sum_{\{i^{\prime }\in I_{2,g}:i^{\prime }<i\}}%
	\mathbb{E}\left[ \left( T^{-1/2}\sum_{t\leq T}\tilde{\psi}_{2,\gamma }^{\ast
	}\left( z_{i,t}\right) \right) ^{2}\right] \mathbb{E}\left[ \left(
	T^{-1/2}\sum_{t^{\prime }\leq T}\tilde{\psi}_{2,\gamma }^{\ast }\left(
	z_{i^{\prime },t^{\prime }}\right) \right) ^{2}\right] \leq
	Kn_{2,g}^{-1}T^{-1}
\end{eqnarray*}%
which implies that 
\begin{equation}
	n^{-1}\sum_{i\leq n}\mathrm{Var}(\tilde{U}_{i})\leq n^{-1}\sum_{g\in 
		\mathcal{G}_{2}}\sum_{i\in I_{g}}Kn_{2,g}^{-1}T^{-1}\leq KT^{-1}.
	\label{P_C_L_Auxillary10_4}
\end{equation}%
The claim of the lemma follows from (\ref{P_C_L_Auxillary10_1})-(\ref%
{P_C_L_Auxillary10_4}).\hfill $Q.E.D.$

\bigskip

\begin{lemma}
	\textit{\label{C_L_Auxillary11}\ }Under Assumptions \ref{A1} and \ref{C_A1}%
	,\ we have: 
	\begin{equation*}
		\frac{(2nT)^{-1}\sum_{g\in \mathcal{G}_{2}}\left( \left(
			n_{2,g}^{-1}\sum_{i\in I_{2,g}}\hat{\sigma}_{2,\gamma ,i}^{2}\right)
			^{2}-\left( n_{2,g}^{-1}\sum_{i\in I_{2,g}}\sigma _{2,\gamma ,i}^{2}\right)
			^{2}\right) }{\omega _{n,T}^{2}}=O_{p}(n^{-1/2}).
	\end{equation*}
\end{lemma}

\noindent {\textsc{Proof of Lemma \ref{C_L_Auxillary11}}}. Next note that%
\begin{align}
	& \frac{\left\vert (2nT)^{-1}\sum_{g\in \mathcal{G}_{2}}\left( \left(
		n_{2,g}^{-1}\sum_{i\in I_{2,g}}\hat{\sigma}_{2,\gamma ,i}^{2}\right)
		^{2}-\left( n_{2,g}^{-1}\sum_{i\in I_{2,g}}\sigma _{2,\gamma ,i}^{2}\right)
		^{2}\right) \right\vert }{\omega _{n,T}^{2}}  \notag \\
	& =\frac{\left\vert (2nT)^{-1}\sum_{g\in \mathcal{G}_{2}}\left(
		n_{2,g}^{-1}\sum_{i\in I_{2,g}}(\hat{\sigma}_{2,\gamma ,i}^{2}-\sigma
		_{2,\gamma ,i}^{2})\right) \left( n_{2,g}^{-1}\sum_{i\in I_{2,g}}(\hat{\sigma%
		}_{2,\gamma ,i}^{2}+\sigma _{2,\gamma ,i}^{2})\right) \right\vert }{\omega
		_{n,T}^{2}}  \notag \\
	& \leq \frac{(2nT)^{-1}\sum_{g\in \mathcal{G}_{2}}\left(
		n_{2,g}^{-1}\sum_{i\in I_{2,g}}(\hat{\sigma}_{2,\gamma ,i}^{2}-\sigma
		_{2,\gamma ,i}^{2})\right) ^{2}}{\omega _{n,T}^{2}}  \notag \\
	& +\frac{\left\vert (nT)^{-1}\sum_{g\in \mathcal{G}_{2}}\left(
		n_{2,g}^{-1}\sum_{i\in I_{2,g}}(\hat{\sigma}_{2,\gamma ,i}^{2}-\sigma
		_{2,\gamma ,i}^{2})\right) \left( n_{2,g}^{-1}\sum_{i\in I_{2,g}}\sigma
		_{2,\gamma ,i}^{2}\right) \right\vert }{\omega _{n,T}^{2}}
	\label{P_C_L_Auxillary11_1}
\end{align}%
From Lemma \ref{C_L_Auxillary7b} and (\ref{C_L_Auxillary7c_1}) in Lemma \ref%
{C_L_Auxillary7c}, we have%
\begin{eqnarray}
	\sum_{g\in \mathcal{G}_{j}}n_{j,g}^{-1}\sum_{i\in I_{j,g}}(\hat{\sigma}%
	_{j,\gamma ,i}^{2}-\sigma _{j,\gamma ,i}^{2})^{2} &\leq &2\sum_{g\in 
		\mathcal{G}_{j}}n_{j,g}^{-1}\sum_{i\in I_{j,g}}\left( \frac{(\widehat{%
			\mathbb{E}}_{T}[\hat{\psi}_{j,\gamma }(z_{i,t})])^{2}}{\hat{\Psi}_{j,\gamma
			\gamma }(\hat{\phi}_{j,g})}-\frac{(\mathbb{E}_{T}[\psi _{j,\gamma
		}(z_{i,t})])^{2}}{\Psi _{j,\gamma \gamma ,g}}\right) ^{2}  \notag \\
	&&+2\sum_{g\in \mathcal{G}_{j}}n_{j,g}^{-1}\sum_{i\in I_{j,g}}(\hat{s}%
	_{j,\gamma ,i}^{2}-s_{j,\gamma ,i}^{2})^{2}\overset{}{=}O_{p}(1).
	\label{P_C_L_Auxillary11_2}
\end{eqnarray}%
Therefore,%
\begin{align}
	& \frac{(2nT)^{-1}\sum_{g\in \mathcal{G}_{2}}\left( n_{2,g}^{-1}\sum_{i\in
			I_{2,g}}(\hat{\sigma}_{2,\gamma ,i}^{2}-\sigma _{2,\gamma ,i}^{2})\right)
		^{2}}{\omega _{n,T}^{2}}  \notag \\
	& \leq \frac{(2nT)^{-1}\sum_{g\in \mathcal{G}_{2}}n_{2,g}^{-1}\sum_{i\in
			I_{2,g}}(\hat{\sigma}_{2,\gamma ,i}^{2}-\sigma _{2,\gamma ,i}^{2})^{2}}{%
		\omega _{n,T}^{2}}=O_{p}(n^{-1}).  \label{P_C_L_Auxillary11_3}
\end{align}%
By the Cauchy-Schwarz inequality,%
\begin{align}
	& \frac{\left\vert (nT)^{-1}\sum_{g\in \mathcal{G}_{2}}\left(
		n_{2,g}^{-1}\sum_{i\in I_{2,g}}(\hat{\sigma}_{2,\gamma ,i}^{2}-\sigma
		_{2,\gamma ,i}^{2})\right) \left( n_{2,g}^{-1}\sum_{i\in I_{2,g}}\sigma
		_{2,\gamma ,i}^{2}\right) \right\vert }{\omega _{n,T}^{2}}  \notag \\
	& \leq \frac{(nT)^{-1}\left( \sum_{g\in \mathcal{G}_{2}}\left(
		n_{2,g}^{-1}\sum_{i\in I_{2,g}}(\hat{\sigma}_{2,\gamma ,i}^{2}-\sigma
		_{2,\gamma ,i}^{2})\right) ^{2}\right) ^{1/2}\left( \sum_{g\in \mathcal{G}%
			_{2}}\left( n_{2,g}^{-1}\sum_{i\in I_{2,g}}\sigma _{2,\gamma ,i}^{2}\right)
		^{2}\right) ^{1/2}}{\omega _{n,T}^{2}}  \notag \\
	& \leq \frac{(nT)^{-1}\left( \sum_{g\in \mathcal{G}_{2}}n_{2,g}^{-1}\sum_{i%
			\in I_{2,g}}(\hat{\sigma}_{2,\gamma ,i}^{2}-\sigma _{2,\gamma
			,i}^{2})^{2}\right) ^{1/2}\left( \sum_{g\in \mathcal{G}_{2}}n_{2,g}^{-1}%
		\sum_{i\in I_{2,g}}\sigma _{2,\gamma ,i}^{4}\right) ^{1/2}}{\omega _{n,T}^{2}%
	}  \notag \\
	& \leq \frac{KG_{2}^{1/2}\left( \sum_{g\in \mathcal{G}_{2}}n_{2,g}^{-1}%
		\sum_{i\in I_{2,g}}(\hat{\sigma}_{2,\gamma ,i}^{2}-\sigma _{2,\gamma
			,i}^{2})^{2}\right) ^{1/2}}{(nT)\omega _{n,T}^{2}}  \notag \\
	& =O_{p}(n^{-1/2}(G_{2}n)^{-1/2}(T\omega _{n,T}^{2})^{-1})=O_{p}(n^{-1/2}),
	\label{P_C_L_Auxillary11_4}
\end{align}%
where the third inequality follows by Assumption \ref{A3}(iii), and the
equality follows from (\ref{P_C_L_Auxillary11_2}) and Assumptions \ref{A1}%
(i) and \ref{C_A1}(v). Combining the results in (\ref{P_C_L_Auxillary11_1}),
(\ref{P_C_L_Auxillary11_3}) and (\ref{P_C_L_Auxillary11_4}), we get the
desired conclusion.\hfill $Q.E.D.$

\bigskip

\begin{lemma}
	\label{C_L_Auxillary12}\textit{\ }Under Assumptions \ref{A1} and \ref{C_A1}%
	,\ we have: 
	\begin{equation*}
		\frac{(nT)^{-1}\sum_{g\in \mathcal{G}_{2}}n_{2,g}^{-2}\sum_{i\in
				I_{2,g}}\left( \hat{\sigma}_{2,\gamma ,i}^{2}\sum_{i^{\prime }\in I_{2,g}}%
			\hat{s}_{2,\gamma ,i^{\prime }}^{2}-\sigma _{2,\gamma ,i}^{2}\sum_{i^{\prime
				}\in I_{2,g}}s_{2,\gamma ,i^{\prime }}^{2}\right) }{\omega _{n,T}^{2}}%
		=O_{p}(n^{-1/2}).
	\end{equation*}
\end{lemma}

\noindent {\textsc{Proof of Lemma \ref{C_L_Auxillary12}}}. We apply the
triangle inequality along with the Cauchy-Schwarz inequality, and obtain%
\begin{align*}
	& \left\vert (nT)^{-1}\sum_{g\in \mathcal{G}_{2}}n_{2,g}^{-2}\left(
	\sum_{i\in I_{2,g}}\hat{\sigma}_{2,\gamma ,i}^{2}\sum_{i^{\prime }\in
		I_{2,g}}\hat{s}_{2,\gamma ,i^{\prime }}^{2}-\sum_{i\in I_{2,g}}\sigma
	_{2,\gamma ,i}^{2}\sum_{i^{\prime }\in I_{2,g}}s_{2,\gamma ,i^{\prime
	}}^{2}\right) \right\vert \\
	& \leq \left\vert (nT)^{-1}\sum_{g\in \mathcal{G}_{2}}n_{2,g}^{-2}%
	\sum_{i,i^{\prime }\in I_{2,g}}(\hat{\sigma}_{2,\gamma ,i}^{2}-\sigma
	_{2,\gamma ,i}^{2})(\hat{s}_{2,\gamma ,i^{\prime }}^{2}-s_{2,\gamma
		,i^{\prime }}^{2})\right\vert \\
	& +\left\vert (nT)^{-1}\sum_{g\in \mathcal{G}_{2}}n_{2,g}^{-2}\sum_{i,i^{%
			\prime }\in I_{2,g}}(\hat{\sigma}_{2,\gamma ,i}^{2}-\sigma _{2,\gamma
		,i}^{2})s_{2,\gamma ,i^{\prime }}^{2}\right\vert +\left\vert
	(nT)^{-1}\sum_{g\in \mathcal{G}_{2}}n_{2,g}^{-2}\sum_{i,i^{\prime }\in
		I_{2,g}}\sigma _{2,\gamma ,i}^{2}(\hat{s}_{2,\gamma ,i^{\prime
	}}^{2}-s_{2,\gamma ,i^{\prime }}^{2})\right\vert \\
	& \leq (nT)^{-1}\left( \sum_{g\in \mathcal{G}_{2}}n_{2,g}^{-1}\sum_{i\in
		I_{2,g}}(\hat{\sigma}_{2,\gamma ,i}^{2}-\sigma _{2,\gamma
		,i}^{2})^{2}\right) ^{1/2}\left( \sum_{g\in \mathcal{G}_{2}}n_{2,g}^{-1}%
	\sum_{i^{\prime }\in I_{2,g}}(\hat{s}_{2,\gamma ,i^{\prime
	}}^{2}-s_{2,\gamma ,i^{\prime }}^{2})^{2}\right) ^{1/2} \\
	& +(nT)^{-1}\left( \sum_{g\in \mathcal{G}_{2}}n_{2,g}^{-1}\sum_{i\in
		I_{2,g}}(\hat{\sigma}_{2,\gamma ,i}^{2}-\sigma _{2,\gamma
		,i}^{2})^{2}\right) ^{1/2}\left( \sum_{g\in \mathcal{G}_{2}}n_{2,g}^{-1}%
	\sum_{i^{\prime }\in I_{2,g}}s_{2,\gamma ,i^{\prime }}^{4}\right) ^{1/2} \\
	& +(nT)^{-1}\left( \sum_{g\in \mathcal{G}_{2}}n_{2,g}^{-1}\sum_{i\in
		I_{2,g}}\sigma _{2,\gamma ,i}^{4}\right) ^{1/2}\left( \sum_{g\in \mathcal{G}%
		_{2}}n_{2,g}^{-1}\sum_{i^{\prime }\in I_{2,g}}(\hat{s}_{2,\gamma ,i^{\prime
	}}^{2}-s_{2,\gamma ,i^{\prime }}^{2})^{2}\right) ^{1/2},
\end{align*}%
which along with Assumptions \ref{A3} and \ref{A4}, (\ref%
{P_C_L_Auxillary11_2}) and Lemma \ref{C_L_Auxillary7b} implies that 
\begin{equation}
	(nT)^{-1}\sum_{g\in \mathcal{G}_{2}}n_{2,g}^{-2}\left( \sum_{i\in I_{2,g}}%
	\hat{\sigma}_{2,\gamma ,i}^{2}\sum_{i^{\prime }\in I_{2,g}}\hat{s}_{2,\gamma
		,i^{\prime }}^{2}-\sum_{i\in I_{2,g}}\sigma _{2,\gamma
		,i}^{2}\sum_{i^{\prime }\in I_{2,g}}s_{2,\gamma ,i^{\prime }}^{2}\right)
	=O_{p}((n^{1/2}T)^{-1}).  \label{P_C_L_Auxillary12_1}
\end{equation}%
The claim of the lemma follows from (\ref{P_C_L_Auxillary12_1}) and
Assumption \ref{C_A1}(v).\hfill $Q.E.D.$

\bigskip

\begin{lemma}
	\textit{\label{C_L_Auxillary13}\ }Under Assumptions \ref{A1} and \ref{C_A1},
	we have: (i) $\hat{\sigma}_{U,n,T}^{2}\leq K$ and $\hat{\sigma}%
	_{n,T}^{2}\leq K$ wpa1;\ and (ii) $\hat{\sigma}_{S,n,T}^{2}\geq 0$.
\end{lemma}

\noindent \textsc{Proof of Lemma \ref{C_L_Auxillary13}.} (i) Since $|\hat{%
	\sigma}_{12,\gamma ,i}|\leq \hat{\sigma}_{1,\gamma ,i}^{2}\hat{\sigma}%
_{2,\gamma ,i}^{2}$ for all $i$, from (\ref{Var_Est_6}), it follows that%
\begin{eqnarray}
	\hat{\sigma}_{U,n,T}^{2} &\leq &(2nT)^{-1}\sum_{g\in \mathcal{G}%
		_{2}}\sum_{i\in I_{2,g}}(\hat{\sigma}_{1,\gamma ,i}^{4}+2n_{2,g}^{-1}\hat{%
		\sigma}_{1,\gamma ,i}^{2}\hat{\sigma}_{2,\gamma ,i}^{2}+n_{2,g}^{-2}\hat{%
		\sigma}_{2,\gamma ,i}^{4})  \notag \\
	&&+(2nT)^{-1}\sum_{g\in \mathcal{G}_{2}}n_{2,g}^{-2}\sum_{i\in
		I_{2,g}}\sum_{i^{\prime }\in I_{2,g},i^{\prime }\neq i}\hat{\sigma}%
	_{2,\gamma ,i}^{2}\hat{\sigma}_{2,\gamma ,i^{\prime }}^{2}  \notag \\
	&\leq &(nT)^{-1}\sum_{g\in \mathcal{G}_{2}}\sum_{i\in I_{2,g}}\left( \hat{%
		\sigma}_{1,\gamma ,i}^{4}+n_{2,g}^{-2}\hat{\sigma}_{2,\gamma
		,i}^{2}\sum_{i^{\prime }\in I_{2,g}}\hat{\sigma}_{2,\gamma ,i^{\prime
	}}^{2}\right)  \notag \\
	&\leq &(nT)^{-1}\sum_{i\leq n}\hat{s}_{1,\gamma ,i}^{4}+(nT)^{-1}\sum_{g\in 
		\mathcal{G}_{2}}\left( n_{2,g}^{-1}\sum_{i\in I_{2,g}}\hat{s}_{2,\gamma
		,i}^{2}\right) ^{2}.  \label{P_C_L_Auxillary13_1}
\end{eqnarray}%
By Assumptions \ref{A3} and \ref{A4}, $s_{j,\gamma ,i}^{2}\leq K$. Together
with Lemma \ref{C_L_Auxillary7b}, this implies 
\begin{eqnarray}
	(nT)^{-1}\sum_{i\leq n}\hat{s}_{1,\gamma ,i}^{4} &\leq &2\left(
	(nT)^{-1}\sum_{i\leq n}(\hat{s}_{1,\gamma ,i}^{2}-s_{1,\gamma
		,i}^{2})^{2}+(nT)^{-1}\sum_{i\leq n}s_{1,\gamma ,i}^{4}\right)  \notag \\
	&=&2(nT)^{-1}\sum_{i\leq n}s_{1,\gamma ,i}^{4}+O_{p}((nT)^{-1})\leq K,
	\label{P_C_L_Auxillary13_2}
\end{eqnarray}%
wpa1. Similarly, 
\begin{eqnarray}
	(nT)^{-1}\sum_{g\in \mathcal{G}_{2}}\left( n_{2,g}^{-1}\sum_{i\in I_{2,g}}%
	\hat{s}_{2,\gamma ,i}^{2}\right) ^{2} &\leq &2(nT)^{-1}\sum_{g\in \mathcal{G}%
		_{2}}n_{2,g}^{-1}\sum_{i\in I_{2,g}}s_{2,\gamma ,i}^{4}  \notag \\
	&&+2(nT)^{-1}\sum_{g\in \mathcal{G}_{2}}n_{2,g}^{-1}\sum_{i\in I_{2,g}}(\hat{%
		s}_{2,\gamma ,i}^{2}-s_{2,\gamma ,i}^{2})^{2}  \notag \\
	&\leq &2(nT)^{-1}\sum_{g\in \mathcal{G}_{2}}n_{2,g}^{-1}\sum_{i\in
		I_{2,g}}s_{2,\gamma ,i}^{4}+O_{p}((nT)^{-1})\leq K,
	\label{P_C_L_Auxillary13_3}
\end{eqnarray}%
with probability approaching 1. Combining (\ref{P_C_L_Auxillary13_1}), (\ref%
{P_C_L_Auxillary13_2}) and (\ref{P_C_L_Auxillary13_3}) yields $\hat{\sigma}%
_{U,n,T}^{2}\leq K$ with probability approaching 1.

Next, by Lemma \ref{C_L_Auxillary7a} and Lemma \ref{C_L_Auxillary10}, 
\begin{equation}
	\hat{\sigma}_{n,T}^{2}\leq (nT)^{-1}\sum_{i\leq n}\sum_{t\leq T}\Delta \psi
	(z_{i,t},\hat{\phi}_{i})^{2}=(nT)^{-1}\sum_{i\leq n}\sum_{t\leq T}\mathbb{E}%
	[\Delta \psi (z_{i,t})^{2}]+O_{p}(T^{-1/2}).  \label{P_C_L_Auxillary13_4}
\end{equation}%
Since $\mathbb{E}[\Delta \psi (z_{i,t})^{2}]\leq K$ by Assumption \ref{A3}%
(iii), from (\ref{P_C_L_Auxillary13_4}), we obtain $\hat{\sigma}%
_{n,T}^{2}\leq K$ wpa1.

Finally, since $\hat{s}_{2,\gamma ,i}^{2}\geq \hat{\sigma}_{2,\gamma
	,i}^{2}\geq 0$ for all $i\leq n$, the definition of $\hat{\sigma}%
_{S,n,T}^{2} $ implies%
\begin{eqnarray}
	\hat{\sigma}_{S,n,T}^{2} &=&(2nT)^{-1}\sum_{g\in \mathcal{G}_{2}}\sum_{i\in
		I_{2,g}}(\hat{\sigma}_{1,\gamma ,i}^{4}+n_{2,g}^{-2}\hat{\sigma}_{2,\gamma
		,i}^{2}\hat{s}_{2,\gamma ,i}^{2}-2n_{2,g}^{-1}\hat{\sigma}_{12,\gamma
		,i}^{2})  \notag \\
	&&+(2nT)^{-1}\sum_{g\in \mathcal{G}_{2}}\sum_{i\in I_{2,g}}\left(
	n_{2,g}^{-2}\hat{\sigma}_{2,\gamma ,i}^{2}\sum_{i^{\prime }\in
		I_{2,g},i^{\prime }\neq i}\hat{s}_{2,\gamma ,i^{\prime }}^{2}\right)  \notag
	\\
	&\geq &(2nT)^{-1}\sum_{g\in \mathcal{G}_{2}}\sum_{i\in I_{2,g}}(\hat{\sigma}%
	_{1,\gamma ,i}^{4}+n_{2,g}^{-2}\hat{\sigma}_{2,\gamma ,i}^{4}-2n_{2,g}^{-1}%
	\hat{\sigma}_{12,\gamma ,i}^{2}).  \label{P_C_L_Auxillary13_5}
\end{eqnarray}%
Next, note that 
\begin{equation}
	\hat{\sigma}_{1,\gamma ,i}^{4}+n_{2,g}^{-2}\hat{\sigma}_{2,\gamma
		,i}^{4}-2n_{2,g}^{-1}\hat{\sigma}_{12,\gamma ,i}^{2}=(\hat{\sigma}_{1,\gamma
		,i}^{2}-n_{2,g}^{-1}\hat{\sigma}_{2,\gamma ,i}^{2})^{2}+2n_{2,g}^{-1}(\hat{%
		\sigma}_{1,\gamma ,i}^{2}\hat{\sigma}_{2,\gamma ,i}^{2}-\hat{\sigma}%
	_{12,\gamma ,i}^{2}).  \label{P_C_L_Auxillary13_6}
\end{equation}%
Moreover, since $\widehat{\mathbb{E}}_{T}[\hat{\psi}_{1,\gamma }(z_{i,t})]=0$
by the first-order condition of $(\hat{\gamma}_{j,i})_{i\leq n}$, we may
write 
\begin{equation*}
	\hat{\sigma}_{12,\gamma ,i}^{2}=\frac{(\widehat{\mathbb{E}}_{T}[(\hat{\psi}%
		_{1,\gamma }(z_{i,t})-\widehat{\mathbb{E}}_{T}[\hat{\psi}_{1,\gamma
		}(z_{i,t})])(\hat{\psi}_{2,\gamma }(z_{i,t})-\widehat{\mathbb{E}}_{T}[\hat{%
			\psi}_{2,\gamma }(z_{i,t})])])^{2}}{|\hat{\Psi}_{1,\gamma \gamma }(\hat{\phi}%
		_{1,g})\hat{\Psi}_{2,\gamma \gamma }(\hat{\phi}_{2,g})|}.
\end{equation*}%
Together with the definitions of $\hat{\sigma}_{j,\gamma ,i}^{2}$ and the
Cauchy-Schwarz inequality, this yields 
\begin{equation}
	\hat{\sigma}_{12,\gamma ,i}^{2}\leq \hat{\sigma}_{1,\gamma ,i}^{2}\hat{\sigma%
	}_{2,\gamma ,i}^{2}.  \label{P_C_L_Auxillary13_7}
\end{equation}%
Combining (\ref{P_C_L_Auxillary13_5}), (\ref{P_C_L_Auxillary13_6}) and (\ref%
{P_C_L_Auxillary13_7}) therefore establishes that $\hat{\sigma}%
_{S,n,T}^{2}\geq 0$.\hfill $Q.E.D.$

\section{Proofs for Lemmas in Appendix \protect\ref{sec:proofTWE}\label%
	{Proof_App_TWE}}

Throughout the proof, we let $x_{j,i,t}$, $\bar{x}_{j,i}$, $\bar{x}_{j,t}$
and $\bar{x}_{j}$ denote the $j$th entries of $x_{i,t}$, $\bar{x}_{i}$, $%
\bar{x}_{t}$ and $\bar{x}$ for $j=1,\ldots ,d_{x}$\ respectively.

\begin{lemma}
	\textit{\label{TWFE_MGCLT} Under the conditions of Theorem \ref{TWFE_Vuong},
		we have }$\sum_{i\leq n}\xi _{nT,i}\rightarrow _{d}N(0,1)$\textit{, }where $%
	\xi _{nT,i}$ is defined in (\textit{\ref{P_TWFE_Vuong_3}}) in the proof of 
	\textit{Theorem \ref{TWFE_Vuong}.}
\end{lemma}

\noindent{\textsc{Proof of Lemma \ref{TWFE_MGCLT}}}. The claim of the lemma
follows by arguments analogous to those used in the proof of Theorem \ref%
{MGCLT_V}, provided that Lemma \ref{C1_MGCLT_L1}, Lemma \ref{C1_MGCLT_L2}
and Lemma \ref{C2_MGCLT_L1} still hold. We next sketch the verification of
these lemmas.

First, we observe that $\psi_{j}(z_{i,t};\phi_{j,i,t}^{\ast})=\varepsilon
_{j,i,t}^{2}$ and $\psi_{j\gamma}(z_{i,t};\phi_{j,i,t}^{\ast})=\varepsilon
_{j,i,t}$ for $j=1,2$. This shows that Assumptions \ref{A6}(i) and \ref{A7}%
(i) are maintained in Assumptions \ref{A9}(ii, iii), respectively. Moreover,
Assumption \ref{A7}(ii) holds due to Lemma \ref{TWFE_Para}(ii) and Lemma \ref%
{GTFE_1}(iii). Additionally, Assumptions \ref{A5}(ii) and \ref{A7}(iii) hold
for model 1. The only hindering factors for the direct application of
Theorem \ref{MGCLT_V} are Assumptions \ref{A5}(ii) and \ref{A7}(iii) for
model 2. This is because the two-way fixed effect model involves $n$ and $T$
groups across $i$ and $T$ separately.

However as there is only one group across $i$ in the definition of $\tilde{U}%
_{2,i}$, Assumptions \ref{A5}(ii) and \ref{A7}(iii) hold for model 2 with $%
\mathcal{G}_{2}=\{1\}$ and $\mathcal{M}_{2}=\{1\ldots ,T\}$. Consequently,
we can apply Lemma \ref{C1_MGCLT_L2} and Lemma \ref{C2_MGCLT_L1} to obtain 
\begin{equation}
	(n\omega _{n,T}^{2})^{-1}\sum_{i\leq n}\left( \mathbb{E}[\tilde{U}_{i}^{2}|%
	\mathcal{F}_{nT,i-1}]-\mathbb{E}[\tilde{U}_{i}^{2}]\right) =O_{p}\left(
	T^{-1/2}+\max_{g\in \mathcal{G}_{1}}n_{g}^{-1/2}\right) ,
	\label{P_TWFE_Vuong_4}
\end{equation}%
and%
\begin{equation}
	\max_{j=1,2}\max_{g\in \mathcal{G}_{j}}\max_{i\in I_{g}}\mathbb{E}\left[
	\left\vert n_{g}^{1/2}\tilde{U}_{j,i}\right\vert ^{4}\right] \leq K,
	\label{P_TWFE_Vuong_5}
\end{equation}%
respectively. Applying similar reasoning as presented in the proof of Lemma %
\ref{C1_MGCLT_L1}, 
\begin{equation}
	(n\omega _{n,T}^{2})^{-1}\sum_{i\leq n}\mathbb{E}\left[ (\tilde{\Psi}_{i}+%
	\tilde{V}_{i}^{\ast })\tilde{U}_{j,i}|\mathcal{F}_{nT,i-1}\right]
	=O_{p}\left( T^{-1/2}+\max_{g\in \mathcal{G}_{1}}n_{g}^{-1}\right) .
	\label{P_TWFE_Vuong_6}
\end{equation}%
This expression serves as the counterpart of (\ref{C1_MGCLT_L1_1}).\ The
detailed proof for (\ref{P_TWFE_Vuong_6}) is provided at the end of this
proof. In view of (\ref{P_TWFE_Vuong_4}), (\ref{P_TWFE_Vuong_5}) and (\ref%
{P_TWFE_Vuong_6}), we can verify (\ref{C1_MGCLT}) and (\ref{C2_MGCLT}) using
the same arguments in the proof of Theorem \ref{MGCLT_V}.

We proceed to verify (\ref{P_TWFE_Vuong_6}). Applying Lemma\ \ref%
{C1_MGCLT_L1} with $\mathcal{G}_{1}=\{1,\ldots ,G_{1}\}$, $\mathcal{M}%
_{1}=\{1\ldots ,T\}$, $\mathcal{G}_{2}=\{1,\ldots ,n\}$ and $\mathcal{M}%
_{2}=\{1\}$\ obtains%
\begin{equation}
	(n\omega _{n,T}^{2})^{-1}\sum_{i\leq n}\mathbb{E}\left[ (\tilde{\Psi}_{i}+%
	\tilde{V}_{i}^{\ast })\tilde{U}_{1,i}|\mathcal{F}_{nT,i-1}\right]
	=O_{p}\left( T^{-1/2}+\max_{g\in \mathcal{G}_{1}}n_{g}^{-1}\right) .
	\label{P_TWFE_Vuong_7}
\end{equation}%
By the independence of $\{z_{i,t}\}_{t\leq T}$ across $i$ we have%
\begin{equation}
	\frac{\sum_{i\leq n}\mathbb{E}\left[ (\tilde{\Psi}_{i}+\tilde{V}_{i}^{\ast })%
		\tilde{U}_{2,i}|\mathcal{F}_{nT,i-1}\right] }{n\omega _{n,T}^{2}}%
	=\sum_{i=1}^{n-1}\frac{C_{2,i}^{\top }e_{2,i}}{n^{2}T^{1/2}\omega _{n,T}}
	\label{P_TWFE_Vuong_8}
\end{equation}%
where $C_{2,i}\equiv \sum_{i^{\prime }=i+1}^{n}c_{j,i^{\prime }}$,\ $%
c_{j,i^{\prime }}\equiv \mathbb{E}[(\tilde{\Psi}_{i^{\prime }}+\tilde{V}%
_{i^{\prime }}^{\ast })e_{2,i^{\prime }}/\omega _{n,T}]$ and $%
e_{2,i}=(\varepsilon _{2,i,t})_{t\leq T}$. Moreover, 
\begin{equation}
	\mathbb{E}\left[ \left\vert \sum_{i\leq n}\frac{C_{2,i}^{\top }e_{2,i}}{%
		n^{2}T^{1/2}\omega _{n,T}}\right\vert ^{2}\right] \leq \frac{K\max_{i\leq
			n}\lambda _{\max }(\mathrm{Cov}(e_{2,i}))\sum_{i\leq n}C_{2,i}^{\top }C_{2,i}%
	}{n^{4}T\omega _{n,T}^{2}}\leq \frac{K\sum_{i\leq n}C_{2,i}^{\top }C_{2,i}}{%
		n^{3}T},  \label{P_TWFE_Vuong_9}
\end{equation}%
where the second inequality is by Assumption \ref{A9}(ii) and (\ref%
{P_TWFE_Xe_5f}). By the definition of $\tilde{V}_{i}^{\ast }$,%
\begin{equation}
	\mathbb{E}[|\tilde{V}_{i}^{\ast }|^{2}]=\frac{\mathbb{E}\left[ \left(
		(\sum_{t\leq T}\varepsilon _{2,i,t})^{2}-\mathbb{E}[(\sum_{t\leq
			T}\varepsilon _{2,i,t})^{2}]\right) ^{2}\right] }{(2T^{3/2})^{2}}\leq \frac{%
		\mathbb{E}\left[ (T^{-1/2}\sum_{t\leq T}\varepsilon _{2,i,t})^{4}\right] }{4T%
	}\leq KT^{-1},  \label{P_TWFE_Vuong_10}
\end{equation}%
where the second inequality is by Assumption \ref{A1}(ii, iv) and
Rosenthal's inequality of strong mixing processes. In view of Assumptions %
\ref{A9}(ii, iii) and (\ref{P_TWFE_Vuong_10}), we can use the same arguments
for proving (\ref{P_C1_MGCLT_L1_10}) to show that%
\begin{equation}
	\max_{i\leq n}|c_{j,i}|^{2}\leq K\left( 1+M_{2}\max_{g\in \mathcal{G}%
		_{2}}n_{g}^{-2}\right) \leq K.  \label{P_TWFE_Vuong_11}
\end{equation}%
This, along with the definition of $C_{2,i}$, implies that 
\begin{equation}
	\max_{i\leq n}\frac{C_{2,i}^{\top }C_{2,i}}{n^{2}}\leq \max_{i\leq n}\left( 
	\frac{\sum_{i^{\prime }=i+1}^{n}|c_{j,i^{\prime }}|}{n}\right) ^{2}\leq K.
	\label{P_TWFE_Vuong_12}
\end{equation}%
Therefore by Assumption \ref{A9}(ii), (\ref{P_TWFE_Vuong_9}) and (\ref%
{P_TWFE_Vuong_12}), 
\begin{equation*}
	\mathbb{E}\left[ \left\vert \sum_{i\leq n}\frac{C_{2,i}^{\top }e_{2,i}}{%
		T^{1/2}n^{2}\omega _{n,T}}\right\vert ^{2}\right] \leq KT^{-1}
\end{equation*}%
which, together with the Markov inequality and (\ref{P_TWFE_Vuong_8}),
implies that 
\begin{equation}
	\frac{\sum_{i\leq n}\mathbb{E}\left[ (\tilde{\Psi}_{i}+\tilde{V}_{i}^{\ast })%
		\tilde{U}_{2,i}|\mathcal{F}_{nT,i-1}\right] }{n\omega _{n,T}^{2}}%
	=O_{p}(T^{-1/2})  \label{P_TWFE_Vuong_13}
\end{equation}%
The claim in (\ref{P_TWFE_Vuong_6}) now follows from (\ref{P_TWFE_Vuong_7})
and the above result.\hfill $Q.E.D.$

\bigskip

\noindent \textsc{Proof of Lemma \ref{TWFE_F_Vuong_Power}.} (a) We first
observe that 
\begin{eqnarray}
	\mathbb{E}[\varphi _{n,T}^{2\text{-side}}(p)] &=&\mathbb{P}\left( \frac{%
		\left\vert MQLR_{n,T}\right\vert }{\hat{\omega}_{n,T}}>z_{1-p/2}\right) 
	\notag \\
	&\geq &\mathbb{P}\left( \left\vert \overline{QLR}_{n,T}\right\vert >\hat{%
		\omega}_{n,T}z_{1-p/2}+\left\vert QLR_{n,T}-|\widehat{\mathbb{E}[S_{n,T}]}%
	-QLR_{n,T}^{\ast }\right\vert \right) .  \label{P_TWFE_F_Vuong_2}
\end{eqnarray}%
From Lemma \ref{TWE_Bias&Var_Est_Alt}, it follows that 
\begin{equation}
	\hat{\omega}_{n,T}z_{1-p/2}+|\widehat{\mathbb{E}[S_{n,T}]}-\mathbb{E}%
	[S_{n,T}]|=O_{p}(G_{1}).  \label{P_TWFE_F_Vuong_3}
\end{equation}%
By Theorem \ref{TWFE_Vuong} and Lemma \ref{TWE_Bias&Var_Est_Alt}, 
\begin{equation}
	QLR_{n,T}-\mathbb{E}[S_{n,T}]-\overline{QLR}_{n,T}=O_{p}(\omega
	_{n,T})=O_{p}(1).  \label{P_TWFE_F_Vuong_4}
\end{equation}%
Therefore, by the triangle inequality 
\begin{equation*}
	\hat{\omega}_{n,T}z_{1-p/2}+\left\vert QLR_{n,T}-|\widehat{\mathbb{E}%
		[S_{n,T}]}-\overline{QLR}_{n,T}\right\vert =O_{p}(1).
\end{equation*}%
This along with $\left\vert \overline{QLR}_{n,T}\right\vert \succ G_{1}$ and
(\ref{P_TWFE_F_Vuong_2}) implies that 
\begin{equation*}
	\mathbb{E}[\varphi _{n,T}^{2\text{-side}}(p)]\geq \mathbb{P}\left(
	\left\vert QLR_{n,T}^{\ast }\right\vert >O_{p}(G_{1})\right) \rightarrow 1,
\end{equation*}%
as $n,T\rightarrow \infty $. This establishes the claim in this part of the
lemma.

(b) The proof for the one-sided test follows by analogous arguments and is
therefore omitted.\hfill $Q.E.D.$

\bigskip

\noindent \textsc{Proof of Lemma }{\textsc{\ref{TWFE_Var}}}. From the
definitions of $\tilde{V}_{j,i}$ and $\tilde{U}_{j,i}$, we have 
\begin{equation}
	\tilde{V}_{i}=\tilde{V}_{1,i}-\tilde{V}_{2,i}=(2T^{1/2})^{-1}\sum_{t\leq
		T}(n_{g}^{-1}\varepsilon _{1,i,t}^{\ast 2}-(n^{-1}+T^{-1})\varepsilon
	_{2,i,t}^{\ast 2})-T^{-3/2}\sum_{t=2}^{T}\sum_{t^{\prime
		}=1}^{t-1}\varepsilon _{2,i,t}^{\ast }\varepsilon _{2,i,t^{\prime }}^{\ast },
	\label{P_TWFE_Variance_1}
\end{equation}%
and 
\begin{equation}
	\tilde{U}_{i}=\tilde{U}_{1,i}-\tilde{U}_{2,i}=\sum_{i^{\prime
		}=N_{g}+1}^{i-1}T^{-1/2}\sum_{t\leq T}(n_{g}^{-1}\varepsilon _{1,i,t}^{\ast
	}\varepsilon _{1,i^{\prime },t}^{\ast }-n^{-1}\varepsilon _{2,i,t}^{\ast
	}\varepsilon _{2,i^{\prime },t}^{\ast })-\sum_{i^{\prime
		}=1}^{N_{g}}T^{-1/2}\sum_{t\leq T}n^{-1}\varepsilon _{2,i,t}^{\ast
	}\varepsilon _{2,i^{\prime }t}^{\ast }.  \label{P_TWFE_Variance_2}
\end{equation}%
Recall that $\tilde{\Psi}_{i}=(2T^{1/2})^{-1}\sum_{t\leq T}(\varepsilon
_{2,i,t}^{2}-\varepsilon _{1,i,t}^{2})$. From Assumption \ref{A10}(i), it
follows that%
\begin{eqnarray}
	\mathrm{Cov}(\tilde{\Psi}_{i},\tilde{V}_{i}) &=&(4T)^{-1}\mathrm{Cov}\left(
	\sum_{t\leq T}(\varepsilon _{2,i,t}^{2}-\varepsilon
	_{1,i,t}^{2}),\sum_{t\leq T}(n_{g}^{-1}\varepsilon _{1,i,t}^{\ast
		2}-(n^{-1}+T^{-1})\varepsilon _{2,i,t}^{\ast 2})\right)  \notag \\
	&&-(2T^{2})^{-1}\mathrm{Cov}\left( \sum_{t\leq T}(\varepsilon
	_{2,i,t}^{2}-\varepsilon _{1,i,t}^{2}),\sum_{t=2}^{T}\sum_{t^{\prime
		}=1}^{t-1}\varepsilon _{2,i,t}^{\ast }\varepsilon _{2,i,t^{\prime }}^{\ast
	}\right)  \notag \\
	&=&(4n_{g}T)^{-1}\mathrm{Cov}\left( \sum_{t\leq T}(\varepsilon
	_{2,i,t}^{2}-\varepsilon _{1,i,t}^{2}),\sum_{t\leq T}\varepsilon
	_{1,i,t}^{\ast 2}\right)  \notag \\
	&&-(4T)^{-1}(n^{-1}+T^{-1})\mathrm{Cov}\left( \sum_{t\leq T}(\varepsilon
	_{2,i,t}^{2}-\varepsilon _{1,i,t}^{2}),\sum_{t\leq T}\varepsilon
	_{2,i,t}^{\ast 2}\right) .  \label{P_TWFE_Variance_3}
\end{eqnarray}%
Using Assumptions \ref{A1} and \ref{A9}(ii), Lemma \ref{TWFE_Para}(ii),
Lemma \ref{GTFE_1}(iii), and the covariance inequality of strong mixing
process,%
\begin{equation*}
	\left\vert \mathrm{Cov}\left( \sum_{t\leq T}(\varepsilon
	_{2,i,t}^{2}-\varepsilon _{1,i,t}^{2}),\sum_{t\leq T}\varepsilon
	_{1,i,t}^{\ast 2}\right) \right\vert +\left\vert \mathrm{Cov}\left(
	\sum_{t\leq T}(\varepsilon _{2,i,t}^{2}-\varepsilon
	_{1,i,t}^{2}),\sum_{t\leq T}\varepsilon _{2,i,t}^{\ast 2}\right) \right\vert
	<K\omega _{n,T}T\text{,}
\end{equation*}%
which together with Assumption \ref{A1}(i) and (\ref{P_TWFE_Variance_3})
implies that%
\begin{eqnarray}
	\left\vert n^{-1}\sum_{i\leq n}\mathrm{Cov}(\tilde{\Psi}_{i},\tilde{V}%
	_{i})\right\vert &\leq &K\omega _{n,T}n^{-1}\sum_{g\in \mathcal{G}%
		_{1}}\sum_{i\in I_{g}}(n_{g}^{-1}+n^{-1}+T^{-1})  \notag \\
	&=&K\omega _{n,T}n^{-1}\sum_{g\in \mathcal{G}_{1}}(1+n_{g}(n^{-1}+T^{-1}))%
	\leq KG_{1}\omega _{n,T}n^{-1}.  \label{P_TWFE_Variance_4}
\end{eqnarray}%
Here the last inequality holds by $n_{g}(n^{-1}+T^{-1})\leq K$.\ Under
Assumption \ref{A10}(i) and some direct calculation, we can show that 
\begin{equation}
	\mathrm{Cov}(\tilde{\Psi}_{i},\tilde{U}_{i})=0\text{ \ \ and \ \ }\mathrm{Cov%
	}(\tilde{V}_{i},\tilde{U}_{i})=0.  \label{P_TWFE_Variance_5}
\end{equation}%
Since $\sigma _{n,T}^{2}=n^{-1}\sum_{i\leq n}\mathrm{Var}(\tilde{\Psi}_{i})$%
, combining\ (\ref{P_TWFE_Variance_4}) and (\ref{P_TWFE_Variance_5}) yields%
\begin{eqnarray}
	\omega _{n,T}^{2} &=&n^{-1}\sum_{i\leq n}\mathrm{Var}(\tilde{\Psi}_{i}+%
	\tilde{V}_{i}+\tilde{U}_{i})  \notag \\
	&=&n^{-1}\sum_{i\leq n}(\mathrm{Var}(\tilde{\Psi}_{i})+\mathrm{Var}(\tilde{V}%
	_{i})+\mathrm{Var}(\tilde{U}_{i})+2\mathrm{Cov}(\tilde{\Psi}_{i},\tilde{V}%
	_{i}))  \notag \\
	&=&\sigma _{n,T}^{2}+n^{-1}\sum_{i\leq n}\mathrm{Var}(\tilde{V}%
	_{i})+n^{-1}\sum_{i\leq n}\mathrm{Var}(\tilde{U}_{i})+O(G_{1}\omega
	_{n,T}n^{-1}).  \label{P_TWFE_Variance_6}
\end{eqnarray}%
To finish the proof, it remains to derive the approximations for $%
n^{-1}\sum_{i\leq n}\mathrm{Var}(\tilde{V}_{i})$ and $n^{-1}\sum_{i\leq n}%
\mathrm{Var}(\tilde{U}_{i})$.

Using Assumption \ref{A10}(i), we have%
\begin{eqnarray*}
	\mathrm{Var}(\tilde{U}_{i}) &=&\sum_{i^{\prime }=N_{g}+1}^{i-1}T^{-1}\mathrm{%
		Var}\left( \sum_{t\leq T}(n_{g}^{-1}\varepsilon _{1,i,t}^{\ast }\varepsilon
	_{1,i^{\prime },t}^{\ast }-n^{-1}\varepsilon _{2,i,t}^{\ast }\varepsilon
	_{2,i^{\prime },t}^{\ast })\right) +\sum_{i^{\prime }=1}^{N_{g}}T^{-1}%
	\mathrm{Var}\left( \sum_{t\leq T}n^{-1}\varepsilon _{2,i,t}^{\ast
	}\varepsilon _{2,i^{\prime }t}^{\ast }\right) \\
	&=&\sum_{i^{\prime }=N_{g}+1}^{i-1}T^{-1}\sum_{t\leq T}\mathrm{Var}\left(
	n_{g}^{-1}\varepsilon _{1,i,t}^{\ast }\varepsilon _{1,i^{\prime },t}^{\ast
	}-n^{-1}\varepsilon _{2,i,t}^{\ast }\varepsilon _{2,i^{\prime },t}^{\ast
	}\right) +\sum_{i^{\prime }=1}^{N_{g}}T^{-1}\sum_{t\leq T}\mathrm{Var}\left(
	n^{-1}\varepsilon _{2,i,t}^{\ast }\varepsilon _{2,i^{\prime }t}^{\ast
	}\right) \\
	&=&\sum_{i^{\prime }=N_{g}+1}^{i-1}T^{-1}\sum_{t\leq T}(n_{g}^{-2}\sigma
	_{1,i}^{2}\sigma _{1,i^{\prime }}^{2}+n^{-2}\sigma _{2,i}^{2}\sigma
	_{2,i^{\prime }}^{2}-2n_{g}^{-1}n^{-1}\sigma _{1,2,i}\sigma _{1,2,i^{\prime
	}})+\sum_{i^{\prime }=1}^{N_{g}}T^{-1}\sum_{t\leq T}n^{-2}\sigma
	_{2,i}^{2}\sigma _{2,i^{\prime }}^{2} \\
	&=&n_{g}^{-2}\sum_{i^{\prime }=N_{g}+1}^{i-1}\sigma _{1,i}^{2}\sigma
	_{1,i^{\prime }}^{2}+n^{-2}\sum_{i^{\prime }=1}^{i-1}\sigma _{2,i}^{2}\sigma
	_{2,i^{\prime }}^{2}-2(nn_{g})^{-1}\sum_{i^{\prime }=N_{g}+1}^{i-1}\sigma
	_{1,2,i}\sigma _{1,2,i^{\prime }}.
\end{eqnarray*}%
This implies that%
\begin{eqnarray}
	n^{-1}\sum_{i\leq n}\mathrm{Var}(\tilde{U}_{i}) &=&n^{-1}\sum_{g\in \mathcal{%
			G}_{1}}n_{g}^{-2}\sum_{i\in I_{g}}\sum_{i^{\prime }=N_{g}+1}^{i-1}\sigma
	_{1,i}^{2}\sigma _{1,i^{\prime }}^{2}+n^{-3}\sum_{i\leq n}\sum_{i^{\prime
		}=1}^{i-1}\sigma _{2,i}^{2}\sigma _{2,i^{\prime }}^{2}  \notag \\
	&&-2n^{-2}\sum_{g\in \mathcal{G}_{1}}n_{g}^{-1}\sum_{i\in
		I_{g}}\sum_{i^{\prime }=N_{g}+1}^{i-1}\sigma _{1,2,i}\sigma _{1,2,i^{\prime
	}}.  \notag \\
	&=&(2n)^{-1}\sum_{g\in \mathcal{G}_{1}}\left( n_{g}^{-1}\sum_{i\in
		I_{g}}\sigma _{1,i}^{2}\right) ^{2}+(2n)^{-1}\left( n^{-1}\sum_{i\leq
		n}\sigma _{2,i}^{2}\right) ^{2}-n^{-2}\sum_{g\in \mathcal{G}%
		_{1}}n_{g}^{-1}\left( \sum_{i\in I_{g}}\sigma _{1,2,i}\right) ^{2}  \notag \\
	&&-(2n)^{-1}\sum_{g\in \mathcal{G}_{1}}n_{g}^{-2}\sum_{i\in I_{g}}\sigma
	_{1,i}^{4}-(2n^{3})^{-1}\sum_{i\leq n}\sigma _{2,i}^{4}+n^{-2}\sum_{g\in 
		\mathcal{G}_{1}}n_{g}^{-1}\sum_{i\in I_{g}}\sigma _{1,2,i}^{2}.
	\label{P_TWFE_Variance_7}
\end{eqnarray}%
Since $\sigma _{1,i}^{4}+\sigma _{2,i}^{4}+\sigma _{1,2,i}^{2}<K$ by Lemma %
\ref{TWFE_Para}(ii), Lemma \ref{GTFE_1}(iii), it follows that 
\begin{equation}
	(2n)^{-1}\sum_{g\in \mathcal{G}_{1}}n_{g}^{-2}\sum_{i\in I_{g}}\sigma
	_{1,i}^{4}+(2n^{3})^{-1}\sum_{i\leq n}\sigma _{2,i}^{4}+n^{-2}\sum_{g\in 
		\mathcal{G}_{1}}n_{g}^{-1}\sum_{i\in I_{g}}\sigma _{1,2,i}^{2}=O\left(
	n^{-1}\sum_{g\in \mathcal{G}_{1}}n_{g}^{-1}\right) .
	\label{P_TWFE_Variance_8}
\end{equation}%
Similarly,%
\begin{eqnarray}
	\mathrm{Var}(\tilde{V}_{i}) &=&(2T^{1/2})^{-2}\sum_{t\leq T}\mathrm{Var}%
	(n_{g}^{-1}\varepsilon _{1,i,t}^{\ast 2}-(n^{-1}+T^{-1})\varepsilon
	_{2,i,t}^{\ast 2})+T^{-3}\sum_{t=2}^{T}\sum_{t^{\prime }=1}^{t-1}\mathrm{Var}%
	(\varepsilon _{2,i,t}^{\ast }\varepsilon _{2,i,t^{\prime }}^{\ast })  \notag
	\\
	&=&T^{-3}\sum_{t=2}^{T}\sum_{t^{\prime }=1}^{t-1}\sigma
	_{2,i}^{4}+(2T^{1/2})^{-2}\sum_{t\leq T}\mathrm{Var}(n_{g}^{-1}\varepsilon
	_{1,i,t}^{\ast 2}-(n^{-1}+T^{-1})\varepsilon _{2,i,t}^{\ast 2}),  \notag \\
	&=&(2T)^{-1}\sigma _{2,i}^{4}+(2T^{1/2})^{-2}\sum_{t\leq T}\mathrm{Var}%
	(n_{g}^{-1}\varepsilon _{1,i,t}^{\ast 2}-(n^{-1}+T^{-1})\varepsilon
	_{2,i,t}^{\ast 2})-(2T^{2})^{-1}\sigma _{2,i}^{4}.  \label{P_TWFE_Variance_9}
\end{eqnarray}%
Since 
\begin{equation*}
	\mathrm{Var}(n_{g}^{-1}\varepsilon _{1,i,t}^{\ast
		2}-(n^{-1}+T^{-1})\varepsilon _{2,i,t}^{\ast 2})\leq Kn_{g}^{-2}\text{ \ \
		and \ \ }\sigma _{2,i}^{4}\leq K,
\end{equation*}%
from (\ref{P_TWFE_Variance_9}) and Assumption \ref{A1}(i), it follows that 
\begin{equation}
	n^{-1}\sum_{i\leq n}\mathrm{Var}(\tilde{V}_{i})=(2nT)^{-1}\sum_{i\leq
		n}\sigma _{2,i}^{4}+O\left( n^{-1}\sum_{g\in \mathcal{G}_{1}}n_{g}^{-1}%
	\right) .  \label{P_TWFE_Variance_10}
\end{equation}%
Collecting the results in (\ref{P_TWFE_Variance_6}), (\ref{P_TWFE_Variance_7}%
), (\ref{P_TWFE_Variance_8}) and (\ref{P_TWFE_Variance_10}), we have%
\begin{eqnarray}
	\omega _{n,T}^{2} &=&\sigma _{n,T}^{2}+(2nT)^{-1}\sum_{i\leq n}\sigma
	_{2,i}^{4}+(2n)^{-1}\left( n^{-1}\sum_{i\leq n}\sigma _{2,i}^{2}\right)
	^{2}+(2n)^{-1}\sum_{g\in \mathcal{G}_{1}}\left( n_{g}^{-1}\sum_{i\in
		I_{g}}\sigma _{1,i}^{2}\right) ^{2}  \notag \\
	&&-n^{-2}\sum_{g\in \mathcal{G}_{1}}n_{g}^{-1}\left( \sum_{i\in I_{g}}\sigma
	_{1,2,i}\right) ^{2}+O\left( G_{1}\omega _{n,T}n^{-1}+n^{-1}\sum_{g\in 
		\mathcal{G}_{1}}n_{g}^{-1}\right) .  \label{P_TWFE_Variance_11}
\end{eqnarray}%
The claim of the lemma follows from (\ref{P_TWFE_Variance_11}), the
definition of $\sigma _{U,n,T}^{2}$ and\ Assumption \ref{A10}(ii).\hfill $%
Q.E.D.$

\bigskip

\noindent \textsc{Proof of Theorem \ref{TWFE_Var_U}.} From the definition of 
$\hat{\sigma}_{U,n,T}^{2}$, we can write%
\begin{eqnarray}
	\hat{\sigma}_{U,n,T}^{2} &=&(2nT)^{-1}\sum_{i\leq n}\hat{\sigma}%
	_{2,i}^{4}+(2n)^{-1}\sum_{g\in \mathcal{G}_{1}}n_{g}^{-2}\left( \sum_{i\in
		I_{g}}\hat{\sigma}_{1,i}^{2}\right) ^{2}  \notag \\
	&&+(2n^{3})^{-1}\left( \sum_{g\in \mathcal{G}_{1}}\sum_{i\in I_{g}}\hat{%
		\sigma}_{2,i}^{2}\right) ^{2}-n^{-2}\sum_{g\in \mathcal{G}%
		_{1}}n_{g}^{-1}\left( \sum_{i\in I_{g}}\hat{\sigma}_{1,2,i}\right) ^{2} 
	\notag \\
	&=&(2nT)^{-1}\sum_{i\leq n}\hat{\sigma}_{2,i}^{4}+n^{-3}\sum_{g=2}^{G_{1}}%
	\sum_{g^{\prime }=1}^{g-1}\left( \sum_{i\in I_{g}}\hat{\sigma}%
	_{2,i}^{2}\right) \left( \sum_{i\in I_{g^{\prime }}}\hat{\sigma}%
	_{2,i}^{2}\right)  \notag \\
	&&+n^{-1}\sum_{g\in \mathcal{G}_{1}}\left( n_{g}^{-2}\left( \sum_{i\in I_{g}}%
	\hat{\sigma}_{1,i}^{2}\right) ^{2}+n^{-2}\left( \sum_{i\in I_{g}}\hat{\sigma}%
	_{2,i}^{2}\right) ^{2}-2(n_{g}n)^{-1}\left( \sum_{i\in I_{g}}\hat{\sigma}%
	_{1,2,i}\right) ^{2}\right) .  \notag \\
	&&  \label{P_TWFE_Var_U_1}
\end{eqnarray}%
Thus it suffices to show that, for each $g\in \mathcal{G}_{1}$, 
\begin{equation}
	n_{g}^{-2}\left( \sum_{i\in I_{g}}\hat{\sigma}_{1,i}^{2}\right)
	^{2}+n^{-2}\left( \sum_{i\in I_{g}}\hat{\sigma}_{2,i}^{2}\right)
	^{2}-2n^{-1}n_{g}^{-1}\left( \sum_{i\in I_{g}}\hat{\sigma}_{1,2,i}\right)
	^{2}\geq 0.  \label{P_TWFE_Var_U_2}
\end{equation}%
By the triangle inequality and the Cauchy-Schwarz inequality, 
\begin{equation*}
	\left( \sum_{i\in I_{g}}\hat{\sigma}_{1,2,i}\right) ^{2}\leq \left(
	\sum_{i\in I_{g}}\hat{\sigma}_{1,i}\hat{\sigma}_{2,i}\right) ^{2}\leq \left(
	\sum_{i\in I_{g}}\hat{\sigma}_{1,i}^{2}\right) \left( \sum_{i\in I_{g}}\hat{%
		\sigma}_{2,i}^{2}\right) .
\end{equation*}%
Therefore for any $g\in \mathcal{G}_{1}$ 
\begin{eqnarray*}
	&&n_{g}^{-2}\left( \sum_{i\in I_{g}}\hat{\sigma}_{1,i}^{2}\right)
	^{2}+n^{-2}\left( \sum_{i\in I_{g}}\hat{\sigma}_{2,i}^{2}\right)
	^{2}-2n^{-1}n_{g}^{-1}\left( \sum_{i\in I_{g}}\hat{\sigma}_{1,2,i}\right)
	^{2} \\
	&\geq &n_{g}^{-2}\left( \sum_{i\in I_{g}}\hat{\sigma}_{1,i}^{2}\right)
	^{2}+n^{-2}\left( \sum_{i\in I_{g}}\hat{\sigma}_{2,i}^{2}\right)
	^{2}-2n^{-1}n_{g}^{-1}\left( \sum_{i\in I_{g}}\hat{\sigma}_{1,i}^{2}\right)
	\left( \sum_{i\in I_{g}}\hat{\sigma}_{2,i}^{2}\right) \\
	&=&\left( n_{g}^{-1}\sum_{i\in I_{g}}\hat{\sigma}_{1,i}^{2}-n^{-1}\sum_{i\in
		I_{g}}\hat{\sigma}_{2,i}^{2}\right) ^{2}\geq 0,
\end{eqnarray*}%
which establishes (\ref{P_TWFE_Var_U_2}) and hence the lemma.\hfill $Q.E.D.$

\bigskip

\noindent \textsc{Proof of Lemma \ref{TWE_Bias&Var_Est_Null}}. The first
claim of the lemma follows directly from Lemma \ref{TWE_Var_Est_6a}. We now
establish the second claim. By the definition of $\hat{\omega}_{n,T}^{2}$,
we have the deterministic bounds%
\begin{equation*}
	\hat{\sigma}_{n,T}^{2}-\hat{\sigma}_{U,n,T}^{2}\leq \hat{\omega}%
	_{n,T}^{2}\leq \max \left\{ \hat{\sigma}_{n,T}^{2}-\hat{\sigma}_{U,n,T}^{2},%
	\hat{\sigma}_{U,n,T}^{2}+\sigma _{n,T}^{2}\right\} .
\end{equation*}%
Therefore,%
\begin{equation}
	\frac{\left\vert \hat{\omega}_{n,T}^{2}-\omega _{n,T}^{2}\right\vert }{%
		\omega _{n,T}^{2}}\leq \frac{\left\vert \hat{\sigma}_{n,T}^{2}-\hat{\sigma}%
		_{U,n,T}^{2}-\omega _{n,T}^{2}\right\vert }{\omega _{n,T}^{2}}+\frac{%
		\left\vert \hat{\sigma}_{U,n,T}^{2}+\sigma _{n,T}^{2}-\omega
		_{n,T}^{2}\right\vert }{\omega _{n,T}^{2}}.
	\label{P_TWE_Bias_Var_Est_Null_3}
\end{equation}%
For the first term, by the triangle inequality,%
\begin{eqnarray}
	\frac{\left\vert \hat{\sigma}_{n,T}^{2}-\hat{\sigma}_{U,n,T}^{2}-\omega
		_{n,T}^{2}\right\vert }{\omega _{n,T}^{2}} &\leq &\frac{\left\vert \hat{%
			\sigma}_{n,T}^{2}-\hat{\sigma}_{U,n,T}^{2}-(\sigma _{n,T}^{2}+\sigma
		_{U,n,T}^{2})\right\vert }{\omega _{n,T}^{2}}+\frac{\left\vert \omega
		_{n,T}^{2}-(\sigma _{n,T}^{2}+\sigma _{U,n,T}^{2})\right\vert }{\omega
		_{n,T}^{2}}  \notag \\
	&=&\frac{\left\vert \hat{\sigma}_{n,T}^{2}-\sigma _{n,T}^{2}-2\sigma
		_{U,n,T}^{2}-(\hat{\sigma}_{U,n,T}^{2}-\sigma _{U,n,T}^{2})\right\vert }{%
		\omega _{n,T}^{2}}+o_{p}(1)  \notag \\
	&\leq &\frac{\left\vert \hat{\sigma}_{n,T}^{2}-\sigma _{n,T}^{2}-2\sigma
		_{U,n,T}^{2}\right\vert }{\omega _{n,T}^{2}}+\frac{\left\vert \hat{\sigma}%
		_{U,n,T}^{2}-\sigma _{U,n,T}^{2}\right\vert }{\omega _{n,T}^{2}}+o_{p}(1),
	\label{P_TWE_Bias_Var_Est_Null_4}
\end{eqnarray}%
where the equality is by Lemma \ref{TWFE_Var}. Under Assumption \ref{A10},
we have $\sigma _{U,n,T}^{2}=s_{U,n,T}^{2}$, where $s_{U,n,T}^{2}$ is
defined in Lemma \ref{TWE_Var_Est_8}. It therefore follows from Lemma \ref%
{TWE_Var_Est_8} that%
\begin{equation}
	\hat{\sigma}_{U,n,T}^{2}-\sigma _{U,n,T}^{2}=O_{p}\left( n^{-1}\left(
	\sum_{g\in \mathcal{G}_{1}}n_{g}^{-1}\right) ^{1/2}\right) =O_{p}\left(
	\omega _{n,T}^{2}\left( \sum_{g\in \mathcal{G}_{1}}n_{g}^{-1}\right)
	^{1/2}\right) =o_{p}(\omega _{n,T}^{2}).  \label{P_TWE_Bias_Var_Est_Null_5}
\end{equation}%
Here, the second equality follows from Assumption \ref{A9}(i), while the
third equality is implied by Assumption \ref{A10}(ii).\ Combining (\ref%
{P_TWE_Bias_Var_Est_Null_4}), (\ref{P_TWE_Bias_Var_Est_Null_5}) and Lemma %
\ref{TWE_Var_Est_7} yields: 
\begin{equation}
	\left\vert \hat{\sigma}_{n,T}^{2}-\hat{\sigma}_{U,n,T}^{2}-\omega
	_{n,T}^{2}\right\vert =o_{p}(\omega _{n,T}^{2}).
	\label{P_TWE_Bias_Var_Est_Null_6}
\end{equation}%
For the second term in (\ref{P_TWE_Bias_Var_Est_Null_3}), Lemma \ref%
{TWFE_Var} and (\ref{P_TWE_Bias_Var_Est_Null_5}) imply 
\begin{equation*}
	\frac{\left\vert \hat{\sigma}_{U,n,T}^{2}+\sigma _{n,T}^{2}-\omega
		_{n,T}^{2}\right\vert }{\omega _{n,T}^{2}}\leq \frac{\left\vert \hat{\sigma}%
		_{U,n,T}^{2}-\sigma _{U,n,T}^{2}\right\vert }{\omega _{n,T}^{2}}+\frac{%
		\left\vert \omega _{n,T}^{2}-\sigma _{n,T}^{2}-\sigma
		_{U,n,T}^{2}\right\vert }{\omega _{n,T}^{2}}=o_{p}(1).
\end{equation*}%
Substituting these bounds into (\ref{P_TWE_Bias_Var_Est_Null_3}) gives%
\begin{equation*}
	\left\vert \hat{\omega}_{n,T}^{2}-\omega _{n,T}^{2}\right\vert =o_{p}(\omega
	_{n,T}^{2}),
\end{equation*}%
which is equivalent to the second claim in (\ref{TWE_Bias&Var_Est_Null_1}%
).\hfill $Q.E.D.$

\bigskip

\noindent \textsc{Proof of Lemma \ref{TWE_Bias&Var_Est_Alt}}. By Lemma \ref%
{TWE_Var_Est_0},%
\begin{eqnarray*}
	\hat{\sigma}_{n,T}^{2} &\leq &(nT)^{-1}\sum_{i\leq n}\sum_{t\leq T}(\hat{%
		\varepsilon}_{2,i,t}^{2}-\hat{\varepsilon}_{1,i,t}^{2})^{2} \\
	&=&(nT)^{-1}\sum_{i\leq n}\sum_{t\leq T}(\ddot{\varepsilon}_{2,i,t}^{2}-\dot{%
		\varepsilon}_{1,i,t}^{2})^{2}+O_{p}(\omega _{n,T}n^{-1}+G_{1}^{1/2}n^{-3/2})
	\\
	&=&(nT)^{-1}\sum_{i\leq n}\sum_{t\leq T}(\ddot{\varepsilon}_{2,i,t}^{2}-\dot{%
		\varepsilon}_{1,i,t}^{2})^{2}+O_{p}(n^{-1}) \\
	&\leq &2(nT)^{-1}\sum_{i\leq n}\sum_{t\leq T}(\ddot{\varepsilon}_{2,i,t}^{4}+%
	\dot{\varepsilon}_{1,i,t}^{4})+o_{p}(1),
\end{eqnarray*}%
where the third line uses the facts that $\omega _{n,T}\leq K$ (established
below), and $G_{1}^{1/2}n^{-3/2}=(G_{1}n^{-1})^{1/2}n^{-1}\leq Kn^{-1}$. By
Lemma \ref{TWFE_Para}(ii) and Lemma \ref{GTFE_1}(iii), (\ref%
{P_TWE_Var_Est_0_14}), and Markov's inequality, 
\begin{equation*}
	(nT)^{-1}\sum_{i\leq n}\sum_{t\leq T}(\ddot{\varepsilon}_{2,i,t}^{4}+\dot{%
		\varepsilon}_{1,i,t}^{4})=O_{p}(1)
\end{equation*}%
and thus, 
\begin{equation}
	\hat{\sigma}_{n,T}^{2}=O_{p}(1).  \label{P_TWE_Bias&Var_Est_Alt_1b}
\end{equation}%
Next by Assumptions \ref{A1}(i) and \ref{A9}(iii), and Lemma \ref%
{TWE_Var_Est_8},%
\begin{equation}
	\hat{\sigma}_{U,n,T}^{2}-s_{U,n,T}^{2}=o_{p}(1).
	\label{P_TWE_Bias&Var_Est_Alt_1c}
\end{equation}%
Moreover, by Assumption \ref{A1}(i), Lemma \ref{TWFE_Para}(ii) and Lemma \ref%
{GTFE_1}(iii),%
\begin{eqnarray*}
	s_{U,n,T}^{2} &\leq &(2nT)^{-1}\sum_{i\leq n}\sigma
	_{2,i}^{2}s_{2,i}^{2}+(2n)^{-1}\sum_{g\in \mathcal{G}_{1}}n_{g}^{-2}\left(
	\sum_{i\in I_{g}}s_{1,i}^{2}\right) ^{2} \\
	&&+(2n^{3})^{-1}\left( \sum_{i\leq n}s_{2,i}^{2}\right)
	^{2}+n^{-2}\sum_{g\in \mathcal{G}_{1}}n_{g}^{-1}\left( \sum_{i\in
		I_{g}}s_{1,2,i}\right) ^{2}\overset{}{\leq }K,
\end{eqnarray*}%
and hence, by (\ref{P_TWE_Bias&Var_Est_Alt_1c}), 
\begin{equation}
	\hat{\sigma}_{U,n,T}^{2}=O_{p}(1).  \label{P_TWE_Bias&Var_Est_Alt_1d}
\end{equation}%
Combining (\ref{P_TWE_Bias&Var_Est_Alt_1b}) and (\ref%
{P_TWE_Bias&Var_Est_Alt_1d}) gives%
\begin{equation*}
	\hat{\omega}_{n,T}^{2}\leq \hat{\sigma}_{n,T}^{2}+\hat{\sigma}%
	_{U,n,T}^{2}=O_{p}(1)
\end{equation*}%
which proves the first claim of the lemma.

Next note that%
\begin{eqnarray}
	\widehat{\mathbb{E}[S_{n,T}]} &=&(2nT)^{-1/2}\left( \sum_{g\in \mathcal{G}%
		_{1}}\sum_{i\in I_{g}}Tn_{g}^{-1}\hat{\sigma}_{1,i}^{2}-(1+Tn^{-1})\sum_{g%
		\in \mathcal{G}_{1}}\sum_{i\in I_{g}}\hat{\sigma}_{2,i}^{2}\right)  \notag \\
	&=&(2nT)^{-1/2}\left( \sum_{g\in \mathcal{G}_{1}}\sum_{i\in
		I_{g}}Tn_{g}^{-1}s_{1,i}^{2}-(1+Tn^{-1})\sum_{g\in \mathcal{G}%
		_{1}}\sum_{i\in I_{g}}s_{2,i}^{2}\right)  \notag \\
	&&+(2nT)^{-1/2}\left( \sum_{g\in \mathcal{G}_{1}}\sum_{i\in
		I_{g}}Tn_{g}^{-1}(\hat{\sigma}_{1,i}^{2}-s_{1,i}^{2})-(1+Tn^{-1})\sum_{g\in 
		\mathcal{G}_{1}}\sum_{i\in I_{g}}(\hat{\sigma}_{2,i}^{2}-s_{2,i}^{2})\right)
	\notag \\
	&=&(2nT)^{-1/2}\left( \sum_{g\in \mathcal{G}_{1}}\sum_{i\in
		I_{g}}Tn_{g}^{-1}s_{1,i}^{2}-(1+Tn^{-1})\sum_{g\in \mathcal{G}%
		_{1}}\sum_{i\in I_{g}}s_{2,i}^{2}\right) +O_{p}\left( \sum_{g\in \mathcal{G}%
		_{1}}n_{g}^{-1}\right) ,  \notag \\
	&&  \label{P_TWE_Bias&Var_Est_Alt_2}
\end{eqnarray}%
where the last step follows from Lemma \ref{TWFE_Est4} and Lemma \ref%
{GTFE_Est3}. By Assumption \ref{A1}(i) and Lemma \ref{TWFE_Para}(ii),%
\begin{equation}
	(2nT)^{-1/2}(1+Tn^{-1})\sum_{g\in \mathcal{G}_{1}}\sum_{i\in
		I_{g}}s_{2,i}^{2}\leq K,  \label{P_TWE_Bias&Var_Est_Alt_3}
\end{equation}%
and by Assumption \ref{A1}(i) and Lemma \ref{GTFE_1}(iii),%
\begin{equation}
	(2nT)^{-1/2}\sum_{g\in \mathcal{G}_{1}}\sum_{i\in
		I_{g}}Tn_{g}^{-1}s_{1,i}^{2}\leq KG_{1}.  \label{P_TWE_Bias&Var_Est_Alt_4}
\end{equation}%
Combining (\ref{P_TWE_Bias&Var_Est_Alt_2}), (\ref{P_TWE_Bias&Var_Est_Alt_3})
and (\ref{P_TWE_Bias&Var_Est_Alt_4}) yields 
\begin{equation}
	\left\vert \widehat{\mathbb{E}[S_{n,T}]}\right\vert \leq KG_{1}+O_{p}\left(
	n^{-1}+\sum_{g\in \mathcal{G}_{1}}n_{g}^{-1}\right) =O_{p}(G_{1}),
	\label{P_TWE_Bias&Var_Est_Alt_5}
\end{equation}%
where the final equality uses Assumptions \ref{A1}(i) and \ref{A9}(iii).
This proves the second claim of the lemma.\ In addition, since $\sigma
_{2,i}^{2}\leq s_{2,i}^{2}$ and $\sigma _{1,i}^{2}\leq s_{1,i}^{2}$ for any $%
i$, the bound $\left\vert \mathbb{E}[S_{n,T}]\right\vert \leq KG_{1}$
follows from the definition of $\mathbb{E}[S_{n,T}]$, (\ref%
{P_TWE_Bias&Var_Est_Alt_3}), (\ref{P_TWE_Bias&Var_Est_Alt_4}) and the
triangle inequality.

Finally, to show $\omega _{n,T}^{2}\leq K$, note that 
\begin{equation}
	\omega _{n,T}^{2}=n^{-1}\sum_{i\leq n}\mathrm{Var}(\tilde{\Psi}_{i}+\tilde{V}%
	_{i}+\tilde{U}_{i})\leq Kn^{-1}\sum_{i\leq n}(\mathrm{Var}(\tilde{\Psi}_{i})+%
	\mathrm{Var}(\tilde{V}_{i})+\mathrm{Var}(\tilde{U}_{i})).
	\label{P_TWE_Bias&Var_Est_Alt_6}
\end{equation}%
By Assumption \ref{A1}, Lemma \ref{TWFE_Para}(ii) and Lemma \ref{GTFE_1}%
(iii), 
\begin{equation}
	\mathrm{Var}(\tilde{\Psi}_{i})=\mathrm{Var}\left( (2T^{1/2})^{-1}\sum_{t\leq
		T}(\varepsilon _{2,i,t}^{2}-\varepsilon _{1,i,t}^{2})\right) \leq
	K\max_{t\leq T}\left\Vert \varepsilon _{2,i,t}^{2}-\varepsilon
	_{1,i,t}^{2}\right\Vert _{2+\delta }^{2}\leq K.
	\label{P_TWE_Bias&Var_Est_Alt_7}
\end{equation}%
Similarly,%
\begin{eqnarray}
	\mathrm{Var}(\tilde{V}_{i}) &\leq &2(\mathrm{Var}(\tilde{V}_{1,i})+\mathrm{%
		Var}(\tilde{V}_{2,i}))  \notag \\
	&\leq &K\left( n_{g}^{-2}\mathrm{Var}\left( T^{-1/2}\sum_{t\leq
		T}\varepsilon _{1,i,t}^{\ast 2}\right) +\mathrm{Var}\left( T^{-3/2}\left(
	\sum_{t\leq T}\varepsilon _{2,i,t}^{\ast }\right) ^{2}\right) +n^{-2}\mathrm{%
		Var}\left( T^{-1/2}\sum_{t\leq T}\varepsilon _{2,i,t}^{\ast 2}\right) \right)
	\notag \\
	&\leq &K\left( n_{g}^{-2}+\mathbb{E}\left[ T^{-3}\left( \sum_{t\leq
		T}\varepsilon _{2,i,t}^{\ast }\right) ^{4}\right] \right) \leq
	K(n_{g}^{-2}+T^{-1}),  \label{P_TWE_Bias&Var_Est_Alt_8}
\end{eqnarray}%
and%
\begin{eqnarray}
	\mathrm{Var}(\tilde{U}_{i}) &\leq &2(\mathrm{Var}(\tilde{U}_{1,i})+\mathrm{%
		Var}(\tilde{U}_{2,i}))  \notag \\
	&=&2\left( n_{g}^{-2}\sum_{i^{\prime }=N_{g}+1}^{i-1}\mathrm{Var}\left(
	T^{-1/2}\sum_{t\leq T}\varepsilon _{1,i,t}^{\ast }\varepsilon _{1,i^{\prime
		}t}^{\ast }\right) +n^{-2}\sum_{i^{\prime }=1}^{i-1}\mathrm{Var}\left(
	T^{-1/2}\sum_{t\leq T}\varepsilon _{2,i,t}^{\ast }\varepsilon _{2,i^{\prime
		}t}^{\ast }\right) \right)  \notag \\
	&\leq &K(n_{g}^{-1}+n^{-1}).  \label{P_TWE_Bias&Var_Est_Alt_9}
\end{eqnarray}%
Substituting (\ref{P_TWE_Bias&Var_Est_Alt_7}), (\ref%
{P_TWE_Bias&Var_Est_Alt_8}) and (\ref{P_TWE_Bias&Var_Est_Alt_9}) into (\ref%
{P_TWE_Bias&Var_Est_Alt_6}) yields $\omega _{n,T}^{2}\leq
K(1+n_{g}^{-1}+n^{-1}+T^{-1})\leq K$, which proves the final claim.\hfill $%
Q.E.D.$

\subsection{Auxiliary Lemmas}

\begin{lemma}
	\textit{\label{TWE_Var_Est_0}\ Suppose }Assumptions \textit{\ref{A1},\ }\ref%
	{A8} and \ref{A9} hold. Then%
	\begin{equation*}
		(nT)^{-1}\sum_{i\leq n}\sum_{t\leq T}(\hat{\varepsilon}_{2,i,t}^{2}-\hat{%
			\varepsilon}_{1,i,t}^{2})^{2}=(nT)^{-1}\sum_{i\leq n}\sum_{t\leq T}(\ddot{%
			\varepsilon}_{2,i,t}^{2}-\dot{\varepsilon}_{1,i,t}^{2})^{2}+O_{p}(\omega
		_{n,T}n^{-1}+G_{1}^{1/2}n^{-3/2}).
	\end{equation*}
\end{lemma}

\noindent \textsc{Proof of Lemma \ref{TWE_Var_Est_0}}. By definition $\hat{%
	\varepsilon}_{2,i,t}=\ddot{\varepsilon}_{2,i,t}-\ddot{x}_{i,t}^{\top }(\hat{%
	\theta}_{2}-\theta _{2}^{\ast })$ where $\ddot{\varepsilon}_{2,i,t}\equiv
\varepsilon _{2,i,t}-\bar{\varepsilon}_{2,i}-\bar{\varepsilon}_{2,t}+\bar{%
	\varepsilon}_{2}$, which implies%
\begin{equation}
	\hat{\varepsilon}_{2,i,t}^{2}=\ddot{\varepsilon}_{2,i,t}^{2}-2\ddot{%
		\varepsilon}_{2,i,t}\ddot{x}_{i,t}^{\top }(\hat{\theta}_{2}-\theta
	_{2}^{\ast })+(\hat{\theta}_{2}-\theta _{2}^{\ast })^{\top }\ddot{x}_{i,t}%
	\ddot{x}_{i,t}^{\top }(\hat{\theta}_{2}-\theta _{2}^{\ast }).
	\label{P_TWE_Var_Est_0_1}
\end{equation}%
Similarly, from the definition of $\hat{\varepsilon}_{1,i,t}$, 
\begin{equation}
	\hat{\varepsilon}_{1,i,t}^{2}=\dot{\varepsilon}_{1,i,t}^{2}-2\dot{\varepsilon%
	}_{1,i,t}\dot{x}_{i,t}^{\top }(\hat{\theta}_{1}-\theta _{1}^{\ast })+(\hat{%
		\theta}_{1}-\theta _{1}^{\ast })^{\top }\dot{x}_{i,t}\dot{x}_{i,t}^{\top }(%
	\hat{\theta}_{1}-\theta _{1}^{\ast }).  \label{P_TWE_Var_Est_0_2}
\end{equation}%
Subtracting the population quantity and using (\ref{P_TWE_Var_Est_0_1}) and (%
\ref{P_TWE_Var_Est_0_2}), we obtain:%
\begin{eqnarray}
	&&(nT)^{-1}\sum_{i\leq n}\sum_{t\leq T}(\hat{\varepsilon}_{2,i,t}^{2}-\hat{%
		\varepsilon}_{1,i,t}^{2})^{2}-(nT)^{-1}\sum_{i\leq n}\sum_{t\leq T}(\ddot{%
		\varepsilon}_{2,i,t}^{2}-\dot{\varepsilon}_{1,i,t}^{2})^{2}  \notag \\
	&=&4(nT)^{-1}\sum_{i\leq n}\sum_{t\leq T}(\ddot{\varepsilon}_{2,i,t}\ddot{x}%
	_{i,t}^{\top }(\hat{\theta}_{2}-\theta _{2}^{\ast })-\dot{\varepsilon}%
	_{1,i,t}\dot{x}_{i,t}^{\top }(\hat{\theta}_{1}-\theta _{1}^{\ast }))^{2} 
	\notag \\
	&&+(nT)^{-1}\sum_{i\leq n}\sum_{t\leq T}((\hat{\theta}_{2}-\theta _{2}^{\ast
	})^{\top }\ddot{x}_{i,t}\ddot{x}_{i,t}^{\top }(\hat{\theta}_{2}-\theta
	_{2}^{\ast })-(\hat{\theta}_{1}-\theta _{1}^{\ast })^{\top }\dot{x}_{i,t}%
	\dot{x}_{i,t}^{\top }(\hat{\theta}_{1}-\theta _{1}^{\ast }))^{2}  \notag \\
	&&-4(nT)^{-1}\sum_{i\leq n}\sum_{t\leq T}(\ddot{\varepsilon}_{2,i,t}^{2}-%
	\dot{\varepsilon}_{1,i,t}^{2})(\ddot{\varepsilon}_{2,i,t}\ddot{x}%
	_{i,t}^{\top }(\hat{\theta}_{2}-\theta _{2}^{\ast })-\dot{\varepsilon}%
	_{1,i,t}\dot{x}_{i,t}^{\top }(\hat{\theta}_{1}-\theta _{1}^{\ast }))  \notag
	\\
	&&+2(nT)^{-1}\sum_{i\leq n}\sum_{t\leq T}(\ddot{\varepsilon}_{2,i,t}^{2}-%
	\dot{\varepsilon}_{1,i,t}^{2})((\hat{\theta}_{2}-\theta _{2}^{\ast })^{\top }%
	\ddot{x}_{i,t}\ddot{x}_{i,t}^{\top }(\hat{\theta}_{2}-\theta _{2}^{\ast })-(%
	\hat{\theta}_{1}-\theta _{1}^{\ast })^{\top }\dot{x}_{i,t}\dot{x}%
	_{i,t}^{\top }(\hat{\theta}_{1}-\theta _{1}^{\ast }))  \notag \\
	&&-4(nT)^{-1}\sum_{i\leq n}\sum_{t\leq T}\left( 
	\begin{array}{c}
		\ddot{\varepsilon}_{2,i,t}\ddot{x}_{i,t}^{\top }(\hat{\theta}_{2}-\theta
		_{2}^{\ast }) \\ 
		-\dot{\varepsilon}_{1,i,t}\dot{x}_{i,t}^{\top }(\hat{\theta}_{1}-\theta
		_{1}^{\ast })%
	\end{array}%
	\right) \left( 
	\begin{array}{c}
		(\hat{\theta}_{2}-\theta _{2}^{\ast })^{\top }\ddot{x}_{i,t}\ddot{x}%
		_{i,t}^{\top }(\hat{\theta}_{2}-\theta _{2}^{\ast }) \\ 
		-(\hat{\theta}_{1}-\theta _{1}^{\ast })^{\top }\dot{x}_{i,t}\dot{x}%
		_{i,t}^{\top }(\hat{\theta}_{1}-\theta _{1}^{\ast })%
	\end{array}%
	\right) .  \label{P_TWE_Var_Est_0_3}
\end{eqnarray}%
Under Assumptions \ref{A1} and \ref{A8}(ii), Lemma \ref{TWFE_Para}(ii) and
Lemma \ref{GTFE_1}(iii), and Markov's inequality,%
\begin{equation}
	(nT)^{-1}\sum_{i\leq n}\sum_{t\leq T}(\dot{\varepsilon}_{1,i,t}^{4}+\ddot{%
		\varepsilon}_{2,i,t}^{4}+||\dot{x}_{i,t}||^{4}+||\ddot{x}%
	_{i,t}||^{4})=O_{p}(1).  \label{P_TWE_Var_Est_0_4}
\end{equation}%
Combining this with Lemmas \ref{TWFE_Est1} and \ref{GTFE_Est1}, and applying
the Cauchy--Schwarz inequality yields 
\begin{eqnarray}
	\sum_{i\leq n}\sum_{t\leq T}(\ddot{\varepsilon}_{2,i,t}\ddot{x}_{i,t}^{\top
	}(\hat{\theta}_{2}-\theta _{2}^{\ast })-\dot{\varepsilon}_{1,i,t}\dot{x}%
	_{i,t}^{\top }(\hat{\theta}_{1}-\theta _{1}^{\ast }))^{2} &=&O_{p}(1),
	\label{P_TWE_Var_Est_0_5} \\
	\sum_{i\leq n}\sum_{t\leq T}((\hat{\theta}_{2}-\theta _{2}^{\ast })^{\top }%
	\ddot{x}_{i,t}\ddot{x}_{i,t}^{\top }(\hat{\theta}_{2}-\theta _{2}^{\ast })-(%
	\hat{\theta}_{1}-\theta _{1}^{\ast })^{\top }\dot{x}_{i,t}\dot{x}%
	_{i,t}^{\top }(\hat{\theta}_{1}-\theta _{1}^{\ast }))^{2}
	&=&O_{p}((nT)^{-1}),  \label{P_TWE_Var_Est_0_6} \\
	\sum_{i\leq n}\sum_{t\leq T}(\ddot{\varepsilon}_{2,i,t}^{2}-\dot{\varepsilon}%
	_{1,i,t}^{2})\left( 
	\begin{array}{c}
		(\hat{\theta}_{2}-\theta _{2}^{\ast })^{\top }\ddot{x}_{i,t}\ddot{x}%
		_{i,t}^{\top }(\hat{\theta}_{2}-\theta _{2}^{\ast }) \\ 
		-(\hat{\theta}_{1}-\theta _{1}^{\ast })^{\top }\dot{x}_{i,t}\dot{x}%
		_{i,t}^{\top }(\hat{\theta}_{1}-\theta _{1}^{\ast })%
	\end{array}%
	\right) &=&O_{p}(1),  \label{P_TWE_Var_Est_0_7} \\
	\sum_{i\leq n}\sum_{t\leq T}\left( 
	\begin{array}{c}
		\ddot{\varepsilon}_{2,i,t}\ddot{x}_{i,t}^{\top }(\hat{\theta}_{2}-\theta
		_{2}^{\ast }) \\ 
		-\dot{\varepsilon}_{1,i,t}\dot{x}_{i,t}^{\top }(\hat{\theta}_{1}-\theta
		_{1}^{\ast })%
	\end{array}%
	\right) \left( 
	\begin{array}{c}
		(\hat{\theta}_{2}-\theta _{2}^{\ast })^{\top }\ddot{x}_{i,t}\ddot{x}%
		_{i,t}^{\top }(\hat{\theta}_{2}-\theta _{2}^{\ast }) \\ 
		-(\hat{\theta}_{1}-\theta _{1}^{\ast })^{\top }\dot{x}_{i,t}\dot{x}%
		_{i,t}^{\top }(\hat{\theta}_{1}-\theta _{1}^{\ast })%
	\end{array}%
	\right) &=&O_{p}((nT)^{-1/2}).  \label{P_TWE_Var_Est_0_8}
\end{eqnarray}%
Therefore, using Assumption \ref{A1}(i) and (\ref{P_TWE_Var_Est_0_3}), (\ref%
{P_TWE_Var_Est_0_5})-(\ref{P_TWE_Var_Est_0_8}), we obtain 
\begin{eqnarray}
	&&(nT)^{-1}\sum_{i\leq n}\sum_{t\leq T}(\hat{\varepsilon}_{2,i,t}^{2}-\hat{%
		\varepsilon}_{1,i,t}^{2})^{2}-(nT)^{-1}\sum_{i\leq n}\sum_{t\leq T}(\ddot{%
		\varepsilon}_{2,i,t}^{2}-\dot{\varepsilon}_{1,i,t}^{2})^{2}  \notag \\
	&=&-4(nT)^{-1}\sum_{i\leq n}\sum_{t\leq T}(\ddot{\varepsilon}_{2,i,t}^{2}-%
	\dot{\varepsilon}_{1,i,t}^{2})(\ddot{\varepsilon}_{2,i,t}\ddot{x}%
	_{i,t}^{\top }(\hat{\theta}_{2}-\theta _{2}^{\ast })-\dot{\varepsilon}%
	_{1,i,t}\dot{x}_{i,t}^{\top }(\hat{\theta}_{1}-\theta _{1}^{\ast
	}))+O_{p}(n^{-2}).  \label{P_TWE_Var_Est_0_9}
\end{eqnarray}%
It therefore suffices to show that%
\begin{equation}
	(nT)^{-1}\sum_{i\leq n}\sum_{t\leq T}(\ddot{\varepsilon}_{2,i,t}^{2}-\dot{%
		\varepsilon}_{1,i,t}^{2})(\ddot{\varepsilon}_{2,i,t}\ddot{x}_{i,t}^{\top }(%
	\hat{\theta}_{2}-\theta _{2}^{\ast })-\dot{\varepsilon}_{1,i,t}\dot{x}%
	_{i,t}^{\top }(\hat{\theta}_{1}-\theta _{1}^{\ast }))=O_{p}(\omega
	_{n,T}n^{-1}+G_{1}^{1/2}n^{-3/2}),  \label{P_TWE_Var_Est_0_10}
\end{equation}%
which together with (\ref{P_TWE_Var_Est_0_9}) completes the proof.

From the definitions of $\ddot{\varepsilon}_{2,i,t}$ and $\dot{\varepsilon}%
_{1,i,t}$, we have 
\begin{equation}
	\ddot{\varepsilon}_{2,i,t}^{2}=(\varepsilon _{2,i,t}-\bar{\varepsilon}_{2,i}-%
	\bar{\varepsilon}_{2,t}+\bar{\varepsilon}_{2})^{2}=\varepsilon
	_{2,i,t}^{2}-2\varepsilon _{2,i,t}(\bar{\varepsilon}_{2,i}^{\ast }+\bar{%
		\varepsilon}_{2,t}^{\ast }-\bar{\varepsilon}_{2}^{\ast })+(\bar{\varepsilon}%
	_{2,i}^{\ast }+\bar{\varepsilon}_{2,t}^{\ast }-\bar{\varepsilon}_{2}^{\ast
	})^{2}  \label{P_TWE_Var_Est_0_11}
\end{equation}%
and%
\begin{equation}
	\dot{\varepsilon}_{1,i,t}^{2}=(\varepsilon _{1,i,t}-\bar{\varepsilon}%
	_{1,g,t})^{2}=(\varepsilon _{1,i,t}-\bar{\varepsilon}_{1,g,t}^{\ast
	})^{2}=\varepsilon _{1,i,t}^{2}-2\varepsilon _{1,i,t}\bar{\varepsilon}%
	_{1,g,t}^{\ast }+\bar{\varepsilon}_{1,g,t}^{\ast 2}.
	\label{P_TWE_Var_Est_0_12}
\end{equation}%
Substituting these expansions yields 
\begin{eqnarray}
	&&(nT)^{-1}\sum_{i\leq n}\sum_{t\leq T}(\ddot{\varepsilon}_{2,i,t}^{2}-\dot{%
		\varepsilon}_{1,i,t}^{2})(\ddot{\varepsilon}_{2,i,t}\ddot{x}_{i,t}^{\top }(%
	\hat{\theta}_{2}-\theta _{2}^{\ast })-\dot{\varepsilon}_{1,i,t}\dot{x}%
	_{i,t}^{\top }(\hat{\theta}_{1}-\theta _{1}^{\ast }))  \notag \\
	&=&(nT)^{-1}\sum_{i\leq n}\sum_{t\leq T}(\varepsilon
	_{2,i,t}^{2}-\varepsilon _{1,i,t}^{2})(\ddot{\varepsilon}_{2,i,t}\ddot{x}%
	_{i,t}^{\top }(\hat{\theta}_{2}-\theta _{2}^{\ast })-\dot{\varepsilon}%
	_{1,i,t}\dot{x}_{i,t}^{\top }(\hat{\theta}_{1}-\theta _{1}^{\ast }))  \notag
	\\
	&&-2(nT)^{-1}\sum_{g\in \mathcal{G}_{1}}\sum_{i\in I_{g}}\sum_{t\leq
		T}(\varepsilon _{2,i,t}(\bar{\varepsilon}_{2,i}^{\ast }+\bar{\varepsilon}%
	_{2,t}^{\ast }-\bar{\varepsilon}_{2}^{\ast })-\varepsilon _{1,i,t}\bar{%
		\varepsilon}_{1,g,t}^{\ast })(\ddot{\varepsilon}_{2,i,t}\ddot{x}_{i,t}^{\top
	}(\hat{\theta}_{2}-\theta _{2}^{\ast })-\dot{\varepsilon}_{1,i,t}\dot{x}%
	_{i,t}^{\top }(\hat{\theta}_{1}-\theta _{1}^{\ast }))  \notag \\
	&&+(nT)^{-1}\sum_{g\in \mathcal{G}_{1}}\sum_{i\in I_{g}}\sum_{t\leq T}((\bar{%
		\varepsilon}_{2,i}^{\ast }+\bar{\varepsilon}_{2,t}^{\ast }-\bar{\varepsilon}%
	_{2}^{\ast })^{2}-\bar{\varepsilon}_{1,g,t}^{\ast 2})(\ddot{\varepsilon}%
	_{2,i,t}\ddot{x}_{i,t}^{\top }(\hat{\theta}_{2}-\theta _{2}^{\ast })-\dot{%
		\varepsilon}_{1,i,t}\dot{x}_{i,t}^{\top }(\hat{\theta}_{1}-\theta _{1}^{\ast
	})).  \label{P_TWE_Var_Est_0_13}
\end{eqnarray}%
By Assumption \ref{A1}, Lemma \ref{TWFE_Para}(ii) and Lemma \ref{GTFE_1}%
(iii),%
\begin{equation}
	\mathbb{E}[\bar{\varepsilon}_{2,i}^{\ast 4}+\bar{\varepsilon}_{2,t}^{\ast
		4}]\leq Kn^{-2}\text{, \ \ }\mathbb{E}[\bar{\varepsilon}_{2}^{\ast 4}]\leq
	Kn^{-4}\text{ \ \ and \ \ }\mathbb{E}[\bar{\varepsilon}_{1,g,t}^{\ast
		4}]\leq Kn_{g}^{-2}.  \label{P_TWE_Var_Est_0_14}
\end{equation}%
Combining this with Assumption \ref{A1}(i), (\ref{P_TWE_Var_Est_0_5}), the
Cauchy--Schwarz inequality, and Markov's inequality leads to 
\begin{eqnarray}
	&&\left\vert (nT)^{-1}\sum_{g\in \mathcal{G}_{1}}\sum_{i\in
		I_{g}}\sum_{t\leq T}((\bar{\varepsilon}_{2,i}^{\ast }+\bar{\varepsilon}%
	_{2,t}^{\ast }-\bar{\varepsilon}_{2}^{\ast })^{2}-\bar{\varepsilon}%
	_{1,g,t}^{\ast 2})(\ddot{\varepsilon}_{2,i,t}\ddot{x}_{i,t}^{\top }(\hat{%
		\theta}_{2}-\theta _{2}^{\ast })-\dot{\varepsilon}_{1,i,t}\dot{x}%
	_{i,t}^{\top }(\hat{\theta}_{1}-\theta _{1}^{\ast }))\right\vert ^{2}  \notag
	\\
	&\leq &(nT)^{-1}\sum_{g\in \mathcal{G}_{1}}\sum_{i\in I_{g}}\sum_{t\leq T}((%
	\bar{\varepsilon}_{2,i}^{\ast }+\bar{\varepsilon}_{2,t}^{\ast }-\bar{%
		\varepsilon}_{2}^{\ast })^{2}-\bar{\varepsilon}_{1,g,t}^{\ast 2})^{2}  \notag
	\\
	&&\times (nT)^{-1}\sum_{i\leq n}\sum_{t\leq T}(\ddot{\varepsilon}_{2,i,t}%
	\ddot{x}_{i,t}^{\top }(\hat{\theta}_{2}-\theta _{2}^{\ast })-\dot{\varepsilon%
	}_{1,i,t}\dot{x}_{i,t}^{\top }(\hat{\theta}_{1}-\theta _{1}^{\ast }))^{2}%
	\overset{}{=}O_{p}\left( n^{-3}\sum_{g\in \mathcal{G}_{1}}n_{g}^{-1}\right)
	\label{P_TWE_Var_Est_0_15}
\end{eqnarray}%
and%
\begin{eqnarray}
	&&\left\vert (nT)^{-1}\sum_{g\in \mathcal{G}_{1}}\sum_{i\in
		I_{g}}\sum_{t\leq T}(\varepsilon _{2,i,t}(\bar{\varepsilon}_{2,i}^{\ast }+%
	\bar{\varepsilon}_{2,t}^{\ast }-\bar{\varepsilon}_{2}^{\ast })-\varepsilon
	_{1,i,t}\bar{\varepsilon}_{1,g,t}^{\ast })(\ddot{\varepsilon}_{2,i,t}\ddot{x}%
	_{i,t}^{\top }(\hat{\theta}_{2}-\theta _{2}^{\ast })-\dot{\varepsilon}%
	_{1,i,t}\dot{x}_{i,t}^{\top }(\hat{\theta}_{1}-\theta _{1}^{\ast
	}))\right\vert ^{2}  \notag \\
	&\leq &K(nT)^{-1}\sum_{g\in \mathcal{G}_{1}}\sum_{i\in I_{g}}\sum_{t\leq
		T}(\varepsilon _{2,i,t}^{2}(\bar{\varepsilon}_{2,i}^{\ast }+\bar{\varepsilon}%
	_{2,t}^{\ast }-\bar{\varepsilon}_{2}^{\ast })^{2}+\varepsilon _{1,i,t}^{2}%
	\bar{\varepsilon}_{1,g,t}^{\ast 2})  \notag \\
	&&\times (nT)^{-1}\sum_{i\leq n}\sum_{t\leq T}(\ddot{\varepsilon}_{2,i,t}%
	\ddot{x}_{i,t}^{\top }(\hat{\theta}_{2}-\theta _{2}^{\ast })-\dot{\varepsilon%
	}_{1,i,t}\dot{x}_{i,t}^{\top }(\hat{\theta}_{1}-\theta _{1}^{\ast }))^{2}%
	\overset{}{=}O_{p}\left( G_{1}n^{-3}\right) .  \label{P_TWE_Var_Est_0_16}
\end{eqnarray}%
Moreover, by Assumption \ref{A9}(ii), (\ref{P_TWE_Var_Est_0_5}), and
Markov's inequality,%
\begin{eqnarray}
	&&\left\vert (nT)^{-1}\sum_{i\leq n}\sum_{t\leq T}(\varepsilon
	_{2,i,t}^{2}-\varepsilon _{1,i,t}^{2})(\ddot{\varepsilon}_{2,i,t}\ddot{x}%
	_{i,t}^{\top }(\hat{\theta}_{2}-\theta _{2}^{\ast })-\dot{\varepsilon}%
	_{1,i,t}\dot{x}_{i,t}^{\top }(\hat{\theta}_{1}-\theta _{1}^{\ast
	}))\right\vert ^{2}  \notag \\
	&\leq &\omega _{n,T}^{2}\times (nT)^{-1}\sum_{i\leq n}\sum_{t\leq T}\frac{%
		(\varepsilon _{2,i,t}^{2}-\varepsilon _{1,i,t}^{2})^{2}}{\omega _{n,T}^{2}} 
	\notag \\
	&&\times (nT)^{-1}\sum_{i\leq n}\sum_{t\leq T}(\ddot{\varepsilon}_{2,i,t}%
	\ddot{x}_{i,t}^{\top }(\hat{\theta}_{2}-\theta _{2}^{\ast })-\dot{\varepsilon%
	}_{1,i,t}\dot{x}_{i,t}^{\top }(\hat{\theta}_{1}-\theta _{1}^{\ast }))^{2}%
	\overset{}{=}O_{p}(\omega _{n,T}^{2}n^{-2}).  \label{P_TWE_Var_Est_0_17}
\end{eqnarray}%
Combining (\ref{P_TWE_Var_Est_0_13}), (\ref{P_TWE_Var_Est_0_15}), (\ref%
{P_TWE_Var_Est_0_16}) and (\ref{P_TWE_Var_Est_0_17}), we conclude that%
\begin{eqnarray*}
	&&(nT)^{-1}\sum_{i\leq n}\sum_{t\leq T}(\ddot{\varepsilon}_{2,i,t}^{2}-\dot{%
		\varepsilon}_{1,i,t}^{2})(\ddot{\varepsilon}_{2,i,t}\ddot{x}_{i,t}^{\top }(%
	\hat{\theta}_{2}-\theta _{2}^{\ast })-\dot{\varepsilon}_{1,i,t}\dot{x}%
	_{i,t}^{\top }(\hat{\theta}_{1}-\theta _{1}^{\ast })) \\
	&=&O_{p}\left( \omega _{n,T}n^{-1}+n^{-3/2}\left( \sum_{g\in \mathcal{G}%
		_{1}}n_{g}^{-1}\right) ^{1/2}+G_{1}^{1/2}n^{-3/2}\right) ,
\end{eqnarray*}%
which, together with $\sum_{g\in \mathcal{G}_{1}}n_{g}^{-1}\leq
G_{1}\max_{g\in \mathcal{G}_{1}}n_{g}^{-1}$, establishes the claim in (\ref%
{P_TWE_Var_Est_0_10}).\hfill $Q.E.D.$

\bigskip

\begin{lemma}
	\textit{\label{TWE_Var_Est_1}\ Suppose }Assumptions \textit{\ref{A1},\ }\ref%
	{A8} and \ref{A9} hold. Then%
	\begin{eqnarray*}
		&&(nT)^{-1}\sum_{i\leq n}\sum_{t\leq T}(\ddot{\varepsilon}_{2,i,t}^{2}-\dot{%
			\varepsilon}_{1,i,t}^{2})^{2}-(nT)^{-1}\sum_{i\leq n}\sum_{t\leq
			T}(\varepsilon _{2,i,t}^{2}-\varepsilon _{1,i,t}^{2})^{2} \\
		&=&4(nT)^{-1}\sum_{g\in \mathcal{G}_{1}}\sum_{i\in I_{g}}\sum_{t\leq
			T}(\varepsilon _{2,i,t}(\bar{\varepsilon}_{2,i}^{\ast }+\bar{\varepsilon}%
		_{2,t}^{\ast }-\bar{\varepsilon}_{2}^{\ast })-\varepsilon _{1,i,t}\bar{%
			\varepsilon}_{1,g,t}^{\ast })^{2}+O_{p}\left( \omega
		_{n,T}G_{1}n^{-1}+n^{-1}\sum_{g\in \mathcal{G}_{1}}n_{g}^{-1/2}\right) .
	\end{eqnarray*}
\end{lemma}

\noindent \textsc{Proof of Lemma \ref{TWE_Var_Est_1}}. Using the
decompositions in (\ref{P_TWE_Var_Est_0_11}) and (\ref{P_TWE_Var_Est_0_12}),
we obtain%
\begin{eqnarray}
	&&(nT)^{-1}\sum_{i\leq n}\sum_{t\leq T}(\ddot{\varepsilon}_{2,i,t}^{2}-\dot{%
		\varepsilon}_{1,i,t}^{2})^{2}-(nT)^{-1}\sum_{i\leq n}\sum_{t\leq
		T}(\varepsilon _{2,i,t}^{2}-\varepsilon _{1,i,t}^{2})^{2}  \notag \\
	&=&4(nT)^{-1}\sum_{g\in \mathcal{G}_{1}}\sum_{i\in I_{g}}\sum_{t\leq
		T}(\varepsilon _{2,i,t}(\bar{\varepsilon}_{2,i}^{\ast }+\bar{\varepsilon}%
	_{2,t}^{\ast }-\bar{\varepsilon}_{2}^{\ast })-\varepsilon _{1,i,t}\bar{%
		\varepsilon}_{1,g,t}^{\ast })^{2}  \notag \\
	&&+(nT)^{-1}\sum_{g\in \mathcal{G}_{1}}\sum_{i\in I_{g}}\sum_{t\leq T}((\bar{%
		\varepsilon}_{2,i}^{\ast }+\bar{\varepsilon}_{2,t}^{\ast }-\bar{\varepsilon}%
	_{2}^{\ast })^{2}-\bar{\varepsilon}_{1,g,t}^{\ast 2})^{2}  \notag \\
	&&-4(nT)^{-1}\sum_{g\in \mathcal{G}_{1}}\sum_{i\in I_{g}}\sum_{t\leq
		T}(\varepsilon _{2,i,t}(\bar{\varepsilon}_{2,i}^{\ast }+\bar{\varepsilon}%
	_{2,t}^{\ast }-\bar{\varepsilon}_{2}^{\ast })-\varepsilon _{1,i,t}\bar{%
		\varepsilon}_{1,g,t}^{\ast })((\bar{\varepsilon}_{2,i}^{\ast }+\bar{%
		\varepsilon}_{2,t}^{\ast }-\bar{\varepsilon}_{2}^{\ast })^{2}-\bar{%
		\varepsilon}_{1,g,t}^{\ast 2})  \notag \\
	&&-4(nT)^{-1}\sum_{g\in \mathcal{G}_{1}}\sum_{i\in I_{g}}\sum_{t\leq
		T}(\varepsilon _{2,i,t}^{2}-\varepsilon _{1,i,t}^{2})(\varepsilon _{2,i,t}(%
	\bar{\varepsilon}_{2,i}^{\ast }+\bar{\varepsilon}_{2,t}^{\ast }-\bar{%
		\varepsilon}_{2}^{\ast })-\varepsilon _{1,i,t}\bar{\varepsilon}%
	_{1,g,t}^{\ast })  \notag \\
	&&+2(nT)^{-1}\sum_{g\in \mathcal{G}_{1}}\sum_{i\in I_{g}}\sum_{t\leq
		T}(\varepsilon _{2,i,t}^{2}-\varepsilon _{1,i,t}^{2})((\bar{\varepsilon}%
	_{2,i}^{\ast }+\bar{\varepsilon}_{2,t}^{\ast }-\bar{\varepsilon}_{2}^{\ast
	})^{2}-\bar{\varepsilon}_{1,g,t}^{\ast 2}).  \label{P_TWE_Var_Est_1_1}
\end{eqnarray}%
By (\ref{P_TWE_Var_Est_0_14}), together with the Cauchy--Schwarz and
Markov's inequalities, 
\begin{equation}
	(nT)^{-1}\sum_{g\in \mathcal{G}_{1}}\sum_{i\in I_{g}}\sum_{t\leq T}((\bar{%
		\varepsilon}_{2,i}^{\ast }+\bar{\varepsilon}_{2,t}^{\ast }-\bar{\varepsilon}%
	_{2}^{\ast })^{2}-\bar{\varepsilon}_{1,g,t}^{\ast 2})^{2}=O_{p}\left(
	n^{-1}\sum_{g\in \mathcal{G}_{1}}n_{g}^{-1}\right) .
	\label{P_TWE_Var_Est_1_2}
\end{equation}%
Next, note that%
\begin{eqnarray}
	&&(nT)^{-1}\sum_{g\in \mathcal{G}_{1}}\sum_{i\in I_{g}}\sum_{t\leq
		T}(\varepsilon _{2,i,t}(\bar{\varepsilon}_{2,i}^{\ast }+\bar{\varepsilon}%
	_{2,t}^{\ast }-\bar{\varepsilon}_{2}^{\ast })-\varepsilon _{1,i,t}\bar{%
		\varepsilon}_{1,g,t}^{\ast })((\bar{\varepsilon}_{2,i}^{\ast }+\bar{%
		\varepsilon}_{2,t}^{\ast }-\bar{\varepsilon}_{2}^{\ast })^{2}-\bar{%
		\varepsilon}_{1,g,t}^{\ast 2})  \notag \\
	&=&(nT)^{-1}\sum_{g\in \mathcal{G}_{1}}\sum_{i\in I_{g}}\sum_{t\leq
		T}\varepsilon _{2,i,t}(\bar{\varepsilon}_{2,i}^{\ast }+\bar{\varepsilon}%
	_{2,t}^{\ast }-\bar{\varepsilon}_{2}^{\ast })^{3}-(nT)^{-1}\sum_{g\in 
		\mathcal{G}_{1}}\sum_{i\in I_{g}}\sum_{t\leq T}\varepsilon _{2,i,t}(\bar{%
		\varepsilon}_{2,i}^{\ast }+\bar{\varepsilon}_{2,t}^{\ast }-\bar{\varepsilon}%
	_{2}^{\ast })\bar{\varepsilon}_{1,g,t}^{\ast 2}  \notag \\
	&&-(nT)^{-1}\sum_{g\in \mathcal{G}_{1}}\sum_{i\in I_{g}}\sum_{t\leq
		T}\varepsilon _{1,i,t}\bar{\varepsilon}_{1,g,t}^{\ast }(\bar{\varepsilon}%
	_{2,i}^{\ast }+\bar{\varepsilon}_{2,t}^{\ast }-\bar{\varepsilon}_{2}^{\ast
	})^{2}+(nT)^{-1}\sum_{g\in \mathcal{G}_{1}}\sum_{i\in I_{g}}\sum_{t\leq
		T}\varepsilon _{1,i,t}\bar{\varepsilon}_{1,g,t}^{\ast 3}.
	\label{P_TWE_Var_Est_1_3}
\end{eqnarray}%
From Assumption \ref{A1}(i), Lemma \ref{TWFE_Para}(ii), Lemma \ref{GTFE_1}%
(iii), and (\ref{P_TWE_Var_Est_0_14}),%
\begin{equation*}
	\mathbb{E}[|\varepsilon _{2,i,t}(\bar{\varepsilon}_{2,i}^{\ast }+\bar{%
		\varepsilon}_{2,t}^{\ast }-\bar{\varepsilon}_{2}^{\ast })^{3}|]\leq
	\left\Vert \varepsilon _{2,i,t}\right\Vert _{4}\left\Vert (\bar{\varepsilon}%
	_{2,i}^{\ast }+\bar{\varepsilon}_{2,t}^{\ast }-\bar{\varepsilon}_{2}^{\ast
	})^{3}\right\Vert _{4/3}\leq Kn^{-3/2}.
\end{equation*}%
Hence, by Markov's inequality, 
\begin{equation}
	(nT)^{-1}\sum_{g\in \mathcal{G}_{1}}\sum_{i\in I_{g}}\sum_{t\leq
		T}\varepsilon _{2,i,t}(\bar{\varepsilon}_{2,i}^{\ast }+\bar{\varepsilon}%
	_{2,t}^{\ast }-\bar{\varepsilon}_{2}^{\ast })^{3}=O_{p}(n^{-3/2}).
	\label{P_TWE_Var_Est_1_4}
\end{equation}%
Similarly, the remaining three terms in (\ref{P_TWE_Var_Est_1_3}) satisfy:%
\begin{eqnarray}
	(nT)^{-1}\sum_{g\in \mathcal{G}_{1}}\sum_{i\in I_{g}}\sum_{t\leq
		T}\varepsilon _{2,i,t}(\bar{\varepsilon}_{2,i}^{\ast }+\bar{\varepsilon}%
	_{2,t}^{\ast }-\bar{\varepsilon}_{2}^{\ast })\bar{\varepsilon}_{1,g,t}^{\ast
		2} &=&O_{p}\left( G_{1}n^{-3/2}\right) ,  \label{P_TWE_Var_Est_1_5} \\
	(nT)^{-1}\sum_{g\in \mathcal{G}_{1}}\sum_{i\in I_{g}}\sum_{t\leq
		T}\varepsilon _{1,i,t}\bar{\varepsilon}_{1,g,t}^{\ast }(\bar{\varepsilon}%
	_{2,i}^{\ast }+\bar{\varepsilon}_{2,t}^{\ast }-\bar{\varepsilon}_{2}^{\ast
	})^{2} &=&O_{p}\left( n^{-2}\sum_{g\in \mathcal{G}_{1}}n_{g}^{1/2}\right) ,
	\label{P_TWE_Var_Est_1_6} \\
	(nT)^{-1}\sum_{g\in \mathcal{G}_{1}}\sum_{i\in I_{g}}\sum_{t\leq
		T}\varepsilon _{1,i,t}\bar{\varepsilon}_{1,g,t}^{\ast 3} &=&O_{p}\left(
	n^{-1}\sum_{g\in \mathcal{G}_{1}}n_{g}^{-1/2}\right) .
	\label{P_TWE_Var_Est_1_7}
\end{eqnarray}%
Combining (\ref{P_TWE_Var_Est_1_3})-(\ref{P_TWE_Var_Est_1_7}) yields%
\begin{equation}
	(nT)^{-1}\sum_{g\in \mathcal{G}_{1}}\sum_{i\in I_{g}}\sum_{t\leq
		T}(\varepsilon _{2,i,t}(\bar{\varepsilon}_{2,i}^{\ast }+\bar{\varepsilon}%
	_{2,t}^{\ast }-\bar{\varepsilon}_{2}^{\ast })-\varepsilon _{1,i,t}\bar{%
		\varepsilon}_{1,g,t}^{\ast })((\bar{\varepsilon}_{2,i}^{\ast }+\bar{%
		\varepsilon}_{2,t}^{\ast }-\bar{\varepsilon}_{2}^{\ast })^{2}-\bar{%
		\varepsilon}_{1,g,t}^{\ast 2})=O_{p}\left( n^{-1}\sum_{g\in \mathcal{G}%
		_{1}}n_{g}^{-1/2}\right) .  \label{P_TWE_Var_Est_1_8}
\end{equation}%
Let $\bar{u}_{1,i}\equiv T^{-1}\sum_{t\leq T}(\varepsilon
_{2,i,t}^{2}-\varepsilon _{1,i,t}^{2})\varepsilon _{2,i,t}$ and $\bar{u}%
_{1,i}^{\ast }\equiv \bar{u}_{1,i}-\mathbb{E}[\bar{u}_{1,i}]$. By Assumption %
\ref{A9}(ii), Lemma \ref{TWFE_Para}(ii) and H\"{o}lder's inequality%
\begin{eqnarray}
	\left\vert \mathbb{E}[\bar{u}_{1,i}]\right\vert &\leq &T^{-1}\sum_{t\leq T}%
	\mathbb{E}[\left\vert (\varepsilon _{2,i,t}^{2}-\varepsilon
	_{1,i,t}^{2})\varepsilon _{2,i,t}]\right\vert  \notag \\
	&\leq &\omega _{n,T}T^{-1}\sum_{t\leq T}||(\varepsilon
	_{2,i,t}^{2}-\varepsilon _{1,i,t}^{2})/\omega _{n,T}||_{2}||\varepsilon
	_{2,i,t}||_{2}\leq K\omega _{n,T}.  \label{P_TWE_Var_Est_1_9a}
\end{eqnarray}%
Together with Assumption\ \ref{A1} and (\ref{P_TWE_Var_Est_0_14}), this
implies 
\begin{equation*}
	\mathbb{E}\left[ \left\vert n^{-1}\sum_{i\leq n}\mathbb{E}[\bar{u}_{1,i}]%
	\bar{\varepsilon}_{2,i}^{\ast }\right\vert ^{2}\right] =n^{-2}\sum_{i\leq n}(%
	\mathbb{E}[\bar{u}_{1,i}])^{2}\mathbb{E}[\bar{\varepsilon}_{2,i}^{\ast
		2}]\leq K\omega _{n,T}^{2}n^{-2}.
\end{equation*}%
Therefore, by Markov's inequality, 
\begin{equation}
	n^{-1}\sum_{i\leq n}\mathbb{E}[\bar{u}_{1,i}]\bar{\varepsilon}_{2,i}^{\ast
	}=O_{p}(\omega _{n,T}n^{-1}).  \label{P_TWE_Var_Est_1_9b}
\end{equation}%
Next, by Assumptions \ref{A1} and \ref{A9}(ii), together with the covariance
inequality for strong mixing processes,%
\begin{eqnarray}
	\mathbb{E}[\bar{u}_{1,i}^{\ast 2}] &=&\mathrm{Var}\left( T^{-1}\sum_{t\leq
		T}(\varepsilon _{2,i,t}^{2}-\varepsilon _{1,i,t}^{2})\varepsilon
	_{2,i,t}\right)  \notag \\
	&\leq &2T^{-2}\sum_{0\leq h\leq T}\sum_{t\leq T-h}\left\vert \mathrm{Cov}%
	((\varepsilon _{2,i,t}^{2}-\varepsilon _{1,i,t}^{2})\varepsilon
	_{2,i,t},(\varepsilon _{2,i,t+h}^{2}-\varepsilon _{1,i,t+h}^{2})\varepsilon
	_{2,i,t+h}\right\vert  \notag \\
	&\leq &2T^{-2}\max_{t\leq T}\left\Vert (\varepsilon _{2,i,t}^{2}-\varepsilon
	_{1,i,t}^{2})\varepsilon _{2,i,t}\right\Vert _{2+\delta /3}^{2}\sum_{0\leq
		h\leq T}\sum_{t\leq T-h}\alpha _{i}^{\delta /(6+\delta )}(h)  \notag \\
	&\leq &KT^{-1}\max_{t\leq T}\left\Vert (\varepsilon _{2,i,t}^{2}-\varepsilon
	_{1,i,t}^{2})\varepsilon _{2,i,t}\right\Vert _{2+\delta /3}^{2}.
	\label{P_TWE_Var_Est_1_9c}
\end{eqnarray}%
Here the last inequality follows since, by Assumption \ref{A1}(iv), 
\begin{equation*}
	T^{-1}\sum_{0\leq h\leq T}\sum_{t\leq T-h}\alpha _{i}^{\delta /(6+\delta
		)}(h)\leq K\sum_{0\leq h\leq T}(1-h/T)a^{\delta h/(6+\delta )}\leq K.
\end{equation*}%
By H\"{o}lder's inequality%
\begin{eqnarray}
	\left\Vert (\varepsilon _{2,i,t}^{2}-\varepsilon _{1,i,t}^{2})\varepsilon
	_{2,i,t}\right\Vert _{2+\delta /3}^{2+\delta /3} &=&\mathbb{E}[|(\varepsilon
	_{2,i,t}^{2}-\varepsilon _{1,i,t}^{2})\varepsilon _{2,i,t}|^{2+\delta /3}] 
	\notag \\
	&\leq &(\mathbb{E}[|\varepsilon _{2,i,t}^{2}-\varepsilon
	_{1,i,t}^{2}|^{3+\delta /2}])^{2/3}(E[|\varepsilon _{2,i,t}|^{6+\delta
	}])^{1/3}  \notag \\
	&\leq &K\omega _{n,T}^{2+\delta /3}(\mathbb{E}[|(\varepsilon
	_{2,i,t}^{2}-\varepsilon _{1,i,t}^{2})/\omega _{n,T}|^{3+\delta
		/2}])^{2/3}\leq K\omega _{n,T}^{2+\delta /3},  \label{P_TWE_Var_Est_1_9d}
\end{eqnarray}%
where the second inequality follows from Lemma \ref{TWFE_Para}(ii) and the
last from Assumption \ref{A9}(ii). Combining\ (\ref{P_TWE_Var_Est_1_9b}) and
(\ref{P_TWE_Var_Est_1_9c}) yields%
\begin{equation}
	\mathbb{E}[\bar{u}_{1,i}^{\ast 2}]\leq K\omega _{n,T}^{2}T^{-1}.
	\label{P_TWE_Var_Est_1_9e}
\end{equation}%
Finally, together with (\ref{P_TWE_Var_Est_0_14}), the triangle inequality,
and H\"{o}lder's inequality 
\begin{equation*}
	\mathbb{E}\left[ \left\vert n^{-1}\sum_{i\leq n}\bar{u}_{1,i}^{\ast }\bar{%
		\varepsilon}_{2,i}^{\ast }\right\vert \right] \leq n^{-1}\sum_{i\leq n}%
	\mathbb{E}[|\bar{u}_{1,i}^{\ast }\bar{\varepsilon}_{2,i}^{\ast }|]\leq
	n^{-1}\sum_{i\leq n}||\bar{u}_{1,i}^{\ast }||_{2}||\bar{\varepsilon}%
	_{2,i}^{\ast }||_{2}\leq K\omega _{n,T}(nT)^{-1/2}.
\end{equation*}%
Therefore, by Assumption\ \ref{A1}(i) and Markov's inequality,%
\begin{equation}
	n^{-1}\sum_{i\leq n}\bar{u}_{1,i}^{\ast }\bar{\varepsilon}_{2,i}^{\ast
	}=O_{p}(\omega _{n,T}n^{-1}).  \label{P_TWE_Var_Est_1_9f}
\end{equation}%
Combining\ (\ref{P_TWE_Var_Est_1_9b})\ and\ (\ref{P_TWE_Var_Est_1_9f}),\ we
obtain%
\begin{equation}
	(nT)^{-1}\sum_{i\leq n}\sum_{t\leq T}(\varepsilon _{2,i,t}^{2}-\varepsilon
	_{1,i,t}^{2})\varepsilon _{2,i,t}\bar{\varepsilon}_{2,i}^{\ast
	}=n^{-1}\sum_{i\leq n}\mathbb{E}[\bar{u}_{1,i}]\bar{\varepsilon}_{2,i}^{\ast
	}+n^{-1}\sum_{i\leq n}\bar{u}_{1,i}^{\ast }\bar{\varepsilon}_{2,i}^{\ast
	}=O_{p}(\omega _{n,T}n^{-1}).  \label{P_TWE_Var_Est_1_9}
\end{equation}%
Similarly,%
\begin{equation*}
	(nT)^{-1}\sum_{i\leq n}\sum_{t\leq T}(\varepsilon _{2,i,t}^{2}-\varepsilon
	_{1,i,t}^{2})\varepsilon _{2,i,t}(\bar{\varepsilon}_{2,t}^{\ast }-\bar{%
		\varepsilon}_{2}^{\ast })=O_{p}(\omega _{n,T}n^{-1})
\end{equation*}%
which, along with (\ref{P_TWE_Var_Est_1_9}), yields: 
\begin{equation}
	(nT)^{-1}\sum_{g\in \mathcal{G}_{1}}\sum_{i\in I_{g}}\sum_{t\leq
		T}(\varepsilon _{2,i,t}^{2}-\varepsilon _{1,i,t}^{2})\varepsilon _{2,i,t}(%
	\bar{\varepsilon}_{2,i}^{\ast }+\bar{\varepsilon}_{2,t}^{\ast }-\bar{%
		\varepsilon}_{2}^{\ast })=O_{p}(\omega _{n,T}n^{-1}).
	\label{P_TWE_Var_Est_1_10}
\end{equation}%
Next, define $\bar{u}_{2,g,t}\equiv n_{g}^{-1}\sum_{i\in I_{g}}(\varepsilon
_{2,i,t}^{2}-\varepsilon _{1,i,t}^{2})\varepsilon _{1,i,t}$ and $\bar{u}%
_{2,g,t}^{\ast }\equiv \bar{u}_{2,g,t}-\mathbb{E}[\bar{u}_{2,g,t}]$. By
arguments analogous to those used to establish (\ref{P_TWE_Var_Est_1_9a}), 
\begin{equation*}
	\left\vert \mathbb{E}[\bar{u}_{2,g,t}]\right\vert \leq K\omega _{n,T}\text{
		\ \ \ and \ \ \ }\mathbb{E}[\bar{u}_{2,g,t}^{\ast 2}]\leq K\omega
	_{n,T}^{2}n_{g}^{-1}.
\end{equation*}%
Hence by (\ref{P_TWE_Var_Est_0_14}), Holder's inequality and Markov's
inequality,%
\begin{eqnarray}
	&&(nT)^{-1}\sum_{g\in \mathcal{G}_{1}}\sum_{i\in I_{g}}\sum_{t\leq
		T}(\varepsilon _{2,i,t}^{2}-\varepsilon _{1,i,t}^{2})\varepsilon _{1,i,t}%
	\bar{\varepsilon}_{1,g,t}^{\ast }  \notag \\
	&=&(nT)^{-1}\sum_{g\in \mathcal{G}_{1}}\sum_{t\leq T}n_{g}\mathbb{E}[\bar{u}%
	_{2,g,t}]\bar{\varepsilon}_{1,g,t}^{\ast }+(nT)^{-1}\sum_{g\in \mathcal{G}%
		_{1}}\sum_{t\leq T}n_{g}\bar{\varepsilon}_{1,g,t}^{\ast }\bar{u}%
	_{2,g,t}^{\ast }\overset{}{=}O_{p}(\omega _{n,T}G_{1}n^{-1}).
	\label{P_TWE_Var_Est_1_11}
\end{eqnarray}%
By (\ref{P_TWE_Var_Est_1_10}) and (\ref{P_TWE_Var_Est_1_11}), the fourth
term in\ (\ref{P_TWE_Var_Est_1_1}) satisfies%
\begin{equation}
	(nT)^{-1}\sum_{g\in \mathcal{G}_{1}}\sum_{i\in I_{g}}\sum_{t\leq
		T}(\varepsilon _{2,i,t}^{2}-\varepsilon _{1,i,t}^{2})(\varepsilon _{2,i,t}(%
	\bar{\varepsilon}_{2,i}^{\ast }+\bar{\varepsilon}_{2,t}^{\ast }-\bar{%
		\varepsilon}_{2}^{\ast })-\varepsilon _{1,i,t}\bar{\varepsilon}%
	_{1,g,t}^{\ast })=O_{p}(\omega _{n,T}G_{1}n^{-1}).
	\label{P_TWE_Var_Est_1_12}
\end{equation}%
Moreover, by Assumption \ref{A9}(i), (\ref{P_TWE_Var_Est_0_14}), and
Holder's inequality%
\begin{eqnarray*}
	&&(nT)^{-1}\sum_{g\in \mathcal{G}_{1}}\sum_{i\in I_{g}}\sum_{t\leq T}\mathbb{%
		E}[|(\varepsilon _{2,i,t}^{2}-\varepsilon _{1,i,t}^{2})(\bar{\varepsilon}%
	_{2,i}^{\ast }+\bar{\varepsilon}_{2,t}^{\ast }-\bar{\varepsilon}_{2}^{\ast
	})^{2}|] \\
	&\leq &\omega _{n,T}(nT)^{-1}\sum_{g\in \mathcal{G}_{1}}\sum_{i\in
		I_{g}}\sum_{t\leq T}\left\Vert (\varepsilon _{2,i,t}^{2}-\varepsilon
	_{1,i,t}^{2})/\omega _{n,T}\right\Vert _{2}\left\Vert (\bar{\varepsilon}%
	_{2,i}^{\ast }+\bar{\varepsilon}_{2,t}^{\ast }-\bar{\varepsilon}_{2}^{\ast
	})^{2}\right\Vert _{2}\leq K\omega _{n,T}n^{-1}.
\end{eqnarray*}%
Together with Markov's inequality, this implies

\begin{equation}
	(nT)^{-1}\sum_{g\in \mathcal{G}_{1}}\sum_{i\in I_{g}}\sum_{t\leq
		T}(\varepsilon _{2,i,t}^{2}-\varepsilon _{1,i,t}^{2})(\bar{\varepsilon}%
	_{2,i}^{\ast }+\bar{\varepsilon}_{2,t}^{\ast }-\bar{\varepsilon}_{2}^{\ast
	})^{2}=O_{p}(\omega _{n,T}n^{-1}).  \label{P_TWE_Var_Est_1_13}
\end{equation}%
Similarly, 
\begin{equation}
	(nT)^{-1}\sum_{g\in \mathcal{G}_{1}}\sum_{i\in I_{g}}\sum_{t\leq
		T}(\varepsilon _{2,i,t}^{2}-\varepsilon _{1,i,t}^{2})\bar{\varepsilon}%
	_{1,g,t}^{\ast 2}=O_{p}(\omega _{n,T}G_{1}n^{-1}).
	\label{P_TWE_Var_Est_1_14}
\end{equation}%
Combining (\ref{P_TWE_Var_Est_1_13}) and (\ref{P_TWE_Var_Est_1_14}) yields%
\begin{equation}
	(nT)^{-1}\sum_{g\in \mathcal{G}_{1}}\sum_{i\in I_{g}}\sum_{t\leq
		T}(\varepsilon _{2,i,t}^{2}-\varepsilon _{1,i,t}^{2})((\bar{\varepsilon}%
	_{2,i}^{\ast }+\bar{\varepsilon}_{2,t}^{\ast }-\bar{\varepsilon}_{2}^{\ast
	})^{2}-\bar{\varepsilon}_{1,g,t}^{\ast 2})=O_{p}(\omega _{n,T}G_{1}n^{-1}).
	\label{P_TWE_Var_Est_1_15}
\end{equation}%
The claim of the lemma follows immediately from (\ref{P_TWE_Var_Est_1_1}), (%
\ref{P_TWE_Var_Est_1_2}), (\ref{P_TWE_Var_Est_1_8}), (\ref%
{P_TWE_Var_Est_1_12}) and (\ref{P_TWE_Var_Est_1_15}). \hfill $Q.E.D.$

\bigskip

\begin{lemma}
	\textit{\label{TWE_Var_Est_2}\ Suppose }Assumptions \textit{\ref{A1},\ }\ref%
	{A8} and \ref{A9} hold. Then%
	\begin{eqnarray*}
		&&(nT)^{-1}\sum_{g\in \mathcal{G}_{1}}\sum_{i\in I_{g}}\sum_{t\leq
			T}(\varepsilon _{2,i,t}(\bar{\varepsilon}_{2,i}^{\ast }+\bar{\varepsilon}%
		_{2,t}^{\ast }-\bar{\varepsilon}_{2}^{\ast })-\varepsilon _{1,i,t}\bar{%
			\varepsilon}_{1,g,t}^{\ast })^{2} \\
		&=&n^{-1}\sum_{i\leq n}s_{2,i}^{2}\mathbb{E}[\bar{\varepsilon}_{2,i}^{\ast
			2}]+n^{-3}\sum_{i\leq n}\sum_{i^{\prime }\leq n}\left( T^{-1}\sum_{t\leq
			T}s_{2,i,t}^{2}\sigma _{2,i^{\prime },t}^{2}\right) \\
		&&+(nT)^{-1}\sum_{t\leq T}\sum_{g\in \mathcal{G}_{1}}\left(
		n_{g}^{-1}\sum_{i\in I_{g}}s_{1,i,t}^{2}\right) \left( n_{g}^{-1}\sum_{i\in
			I_{g}}\sigma _{1,i,t}^{2}\right) \\
		&&-2(n^{2}T)^{-1}\sum_{t\leq T}\sum_{g\in \mathcal{G}_{1}}n_{g}s_{1,2,g,t}%
		\sigma _{1,2,g,t}+O_{p}\left( n^{-1}\sum_{g\in \mathcal{G}%
			_{1}}n_{g}^{-1/2}\right) ,
	\end{eqnarray*}%
	where%
	\begin{equation*}
		s_{j,i,t}^{2}\equiv \mathbb{E}[\varepsilon _{j,i,t}^{2}]\text{, }%
		s_{j,i}^{2}=T^{-1}\sum_{t\leq T}s_{j,i,t}^{2}\text{, }s_{1,2,g,t}\equiv
		n_{g}^{-1}\sum_{i\in I_{g}}\mathbb{E}[\varepsilon _{1,i,t}\varepsilon
		_{2,i,t}]\text{ and }\sigma _{1,2,g,t}\equiv n_{g}^{-1}\sum_{i\in I_{g}}%
		\mathbb{E}[\varepsilon _{1,i,t}^{\ast }\varepsilon _{2,i,t}^{\ast }].
	\end{equation*}
\end{lemma}

\noindent \textsc{Proof of Lemma \ref{TWE_Var_Est_2}}. By Lemma \ref%
{TWFE_Para}(ii) and Lemma \ref{GTFE_1}(iii), (\ref{P_TWE_Var_Est_0_14}) and H%
\"{o}lder's inequality,%
\begin{equation*}
	\mathbb{E}[|(\varepsilon _{2,i,t}(\bar{\varepsilon}_{2,i}^{\ast }+\bar{%
		\varepsilon}_{2,t}^{\ast })-\varepsilon _{1,i,t}\bar{\varepsilon}%
	_{1,g,t}^{\ast })\varepsilon _{2,i,t}|]\leq \left\Vert \varepsilon
	_{2,i,t}^{2}\right\Vert _{2}\left\Vert \bar{\varepsilon}_{2,i}^{\ast }+\bar{%
		\varepsilon}_{2,t}^{\ast }\right\Vert _{2}+\left\Vert \varepsilon
	_{1,i,t}\varepsilon _{2,i,t}\right\Vert _{2}||\bar{\varepsilon}%
	_{1,g,t}^{\ast }||_{2}\leq Kn_{g}^{-1/2}.
\end{equation*}%
Therefore, by Markov's inequality, 
\begin{equation}
	(nT)^{-1}\sum_{g\in \mathcal{G}_{1}}\sum_{i\in I_{g}}\sum_{t\leq
		T}(\varepsilon _{2,i,t}(\bar{\varepsilon}_{2,i}^{\ast }+\bar{\varepsilon}%
	_{2,t}^{\ast })-\varepsilon _{1,i,t}\bar{\varepsilon}_{1,g,t}^{\ast
	})\varepsilon _{2,i,t}=O_{p}\left( n^{-1}\sum_{g\in \mathcal{G}%
		_{1}}n_{g}^{1/2}\right) .  \label{P_TWE_Var_Est_2_0}
\end{equation}%
By Assumption \ref{A1}(i), (\ref{P_TWE_Var_Est_0_14}) and Markov's
inequality,%
\begin{equation}
	\bar{\varepsilon}_{2}^{\ast }=O_{p}(n^{-1}),  \label{P_TWE_Var_Est_2_1}
\end{equation}%
which together with Assumptions \ref{A9}(i) and (\ref{P_TWE_Var_Est_2_0})
implies that%
\begin{equation}
	(nT)^{-1}\sum_{g\in \mathcal{G}_{1}}\sum_{i\in I_{g}}\sum_{t\leq
		T}(\varepsilon _{2,i,t}(\bar{\varepsilon}_{2,i}^{\ast }+\bar{\varepsilon}%
	_{2,t}^{\ast })-\varepsilon _{1,i,t}\bar{\varepsilon}_{1,g,t}^{\ast
	})\varepsilon _{2,i,t}\bar{\varepsilon}_{2}^{\ast }=O_{p}\left(
	n^{-2}\sum_{g\in \mathcal{G}_{1}}n_{g}^{1/2}\right) .
	\label{P_TWE_Var_Est_2_2}
\end{equation}%
Therefore,%
\begin{eqnarray}
	&&(nT)^{-1}\sum_{g\in \mathcal{G}_{1}}\sum_{i\in I_{g}}\sum_{t\leq
		T}(\varepsilon _{2,i,t}(\bar{\varepsilon}_{2,i}^{\ast }+\bar{\varepsilon}%
	_{2,t}^{\ast }-\bar{\varepsilon}_{2}^{\ast })-\varepsilon _{1,i,t}\bar{%
		\varepsilon}_{1,g,t}^{\ast })^{2}  \notag \\
	&=&(nT)^{-1}\sum_{g\in \mathcal{G}_{1}}\sum_{i\in I_{g}}\sum_{t\leq
		T}(\varepsilon _{2,i,t}(\bar{\varepsilon}_{2,i}^{\ast }+\bar{\varepsilon}%
	_{2,t}^{\ast })-\varepsilon _{1,i,t}\bar{\varepsilon}_{1,g,t}^{\ast
	})^{2}+O_{p}\left( n^{-2}\sum_{g\in \mathcal{G}_{1}}n_{g}^{1/2}\right) 
	\notag \\
	&=&(nT)^{-1}\sum_{g\in \mathcal{G}_{1}}\sum_{i\in I_{g}}\sum_{t\leq
		T}\varepsilon _{2,i,t}^{2}(\bar{\varepsilon}_{2,i}^{\ast }+\bar{\varepsilon}%
	_{2,t}^{\ast })^{2}+(nT)^{-1}\sum_{i\leq n}\sum_{t\leq T}\varepsilon
	_{1,i,t}^{2}\bar{\varepsilon}_{1,g,t}^{\ast 2}  \notag \\
	&&-2(nT)^{-1}\sum_{g\in \mathcal{G}_{1}}\sum_{i\in I_{g}}\sum_{t\leq
		T}\varepsilon _{1,i,t}\varepsilon _{2,i,t}(\bar{\varepsilon}_{2,i}^{\ast }+%
	\bar{\varepsilon}_{2,t}^{\ast })\bar{\varepsilon}_{1,g,t}^{\ast
	}+O_{p}\left( n^{-2}\sum_{g\in \mathcal{G}_{1}}n_{g}^{1/2}\right) .
	\label{P_TWE_Var_Est_2_3}
\end{eqnarray}%
The claim of the lemma follows from (\ref{P_TWE_Var_Est_2_3}), Lemmas \ref%
{TWE_Var_Est_3}-\ref{TWE_Var_Est_5} below.\hfill $Q.E.D.$

\bigskip

\begin{lemma}
	\textit{\label{TWE_Var_Est_3}\ Suppose }Assumptions \textit{\ref{A1},\ }\ref%
	{A8} and \ref{A9} hold. Then%
	\begin{eqnarray*}
		(nT)^{-1}\sum_{i\leq n}\sum_{t\leq T}\varepsilon _{2,i,t}^{2}(\bar{%
			\varepsilon}_{2,i}^{\ast }+\bar{\varepsilon}_{2,t}^{\ast })^{2}
		&=&n^{-1}\sum_{i\leq n}s_{2,i}^{2}\mathbb{E}[\bar{\varepsilon}_{2,i}^{\ast
			2}] \\
		&&+n^{-3}\sum_{i\leq n}\sum_{i^{\prime }\leq n}\left( T^{-1}\sum_{t\leq
			T}s_{2,i,t}^{2}\sigma _{2,i^{\prime },t}^{2}\right) +O_{p}\left(
		n^{-3/2}\right) .
	\end{eqnarray*}
\end{lemma}

\noindent \textsc{Proof of Lemma \ref{TWE_Var_Est_3}}. We begin by
decomposing 
\begin{eqnarray}
	(nT)^{-1}\sum_{i\leq n}\sum_{t\leq T}\varepsilon _{2,i,t}^{2}(\bar{%
		\varepsilon}_{2,i}^{\ast }+\bar{\varepsilon}_{2,t}^{\ast })^{2}
	&=&(nT)^{-1}\sum_{i\leq n}\sum_{t\leq T}\varepsilon _{2,i,t}^{2}\bar{%
		\varepsilon}_{2,i}^{\ast 2}+(nT)^{-1}\sum_{i\leq n}\sum_{t\leq T}\varepsilon
	_{2,i,t}^{2}\bar{\varepsilon}_{2,t}^{\ast 2}  \notag \\
	&&+2(nT)^{-1}\sum_{i\leq n}\sum_{t\leq T}\varepsilon _{2,i,t}^{2}\bar{%
		\varepsilon}_{2,i}^{\ast }\bar{\varepsilon}_{2,t}^{\ast }.
	\label{P_TWE_Var_Est_3_0}
\end{eqnarray}%
We analyze the three terms on the right-hand side in turn.

Using Assumption \ref{A1}, Lemma \ref{TWFE_Para}(ii), (\ref%
{P_TWE_Var_Est_0_14}), the Cauchy-Schwarz inequality and Markov's
inequality, we obtain%
\begin{eqnarray}
	&&n^{-1}\sum_{i\leq n}\mathbb{E}\left[ \left\vert T^{-1}\sum_{t\leq
		T}(\varepsilon _{2,i,t}^{2}-\mathbb{E}[\varepsilon _{2,i,t}^{2}])\right\vert 
	\bar{\varepsilon}_{2,i}^{\ast 2}\right]  \notag \\
	&\leq &n^{-1}\sum_{i\leq n}\left( \mathbb{E}\left[ \left( T^{-1}\sum_{t\leq
		T}(\varepsilon _{2,i,t}^{2}-\mathbb{E}[\varepsilon _{2,i,t}^{2}])\right) ^{2}%
	\right] \mathbb{E}\left[ \bar{\varepsilon}_{2,i}^{\ast 4}\right] \right)
	^{1/2}\leq Kn^{-3/2}.  \label{P_TWE_Var_Est_3_4}
\end{eqnarray}%
This, together with Markov's inequality, implies that 
\begin{eqnarray}
	(nT)^{-1}\sum_{i\leq n}\sum_{t\leq T}\varepsilon _{2,i,t}^{2}\bar{\varepsilon%
	}_{2,i}^{\ast 2} &=&(nT)^{-1}\sum_{i\leq n}\sum_{t\leq T}\mathbb{E}%
	[\varepsilon _{2,i,t}^{2}]\bar{\varepsilon}_{2,i}^{\ast
		2}+(nT)^{-1}\sum_{i\leq n}\sum_{t\leq T}(\varepsilon _{2,i,t}^{2}-\mathbb{E}%
	[\varepsilon _{2,i,t}^{2}])\bar{\varepsilon}_{2,i}^{\ast 2}  \notag \\
	&=&(nT)^{-1}\sum_{i\leq n}\sum_{t\leq T}s_{2,i,t}^{2}\bar{\varepsilon}%
	_{2,i}^{\ast 2}+O_{p}(n^{-3/2})  \notag \\
	&=&n^{-1}\sum_{i\leq n}s_{2,i}^{2}\bar{\varepsilon}_{2,i}^{\ast
		2}+O_{p}(n^{-3/2}),  \label{P_TWE_Var_Est_3_5}
\end{eqnarray}%
where $s_{2,i}^{2}=T^{-1}\sum_{t\leq T}s_{2,i,t}^{2}$. Using Assumption \ref%
{A1}, Lemma \ref{TWFE_Para}(ii), and (\ref{P_TWE_Var_Est_0_14}), we further
have 
\begin{equation*}
	\mathbb{E}\left[ \left( n^{-1}\sum_{i\leq n}s_{2,i}^{2}(\bar{\varepsilon}%
	_{2,i}^{\ast 2}-\mathbb{E}[\bar{\varepsilon}_{2,i}^{\ast 2}])\right) ^{2}%
	\right] =n^{-2}\sum_{i\leq n}\mathbb{E}[s_{2,i}^{4}(\bar{\varepsilon}%
	_{2,i}^{\ast 2}-\mathbb{E}[\bar{\varepsilon}_{2,i}^{\ast 2}])^{2}]\leq
	Kn^{-2}\sum_{i\leq n}\mathbb{E}[\bar{\varepsilon}_{2,i}^{\ast 4}]\leq
	Kn^{-3},
\end{equation*}%
which implies 
\begin{equation}
	n^{-1}\sum_{i\leq n}s_{2,i}^{2}\bar{\varepsilon}_{2,i}^{\ast
		2}=n^{-1}\sum_{i\leq n}s_{2,i}^{2}\mathbb{E}[\bar{\varepsilon}_{2,i}^{\ast
		2}]+O_{p}(n^{-3/2}).  \label{P_TWE_Var_Est_3_6}
\end{equation}%
Combining (\ref{P_TWE_Var_Est_3_5}) and (\ref{P_TWE_Var_Est_3_6}) yields 
\begin{equation}
	(nT)^{-1}\sum_{i\leq n}\sum_{t\leq T}\varepsilon _{2,i,t}^{2}\bar{\varepsilon%
	}_{2,i}^{\ast 2}=n^{-1}\sum_{i\leq n}s_{2,i}^{2}\mathbb{E}[\bar{\varepsilon}%
	_{2,i}^{\ast 2}]+O_{p}(n^{-3/2}).  \label{P_TWE_Var_Est_3_7}
\end{equation}

Using Assumption \ref{A1}, (\ref{P_TWE_Var_Est_0_14}) and Lemma \ref%
{TWFE_Para}(ii), we have 
\begin{eqnarray*}
	&&T^{-1}\sum_{t\leq T}\mathbb{E}\left[ \left\vert n^{-1}\sum_{i\leq
		n}(\varepsilon _{2,i,t}^{2}-\mathbb{E}[\varepsilon _{2,i,t}^{2}])\right\vert 
	\bar{\varepsilon}_{2,t}^{\ast 2}\right] \\
	&\leq &T^{-1}\sum_{t\leq T}\left( \mathbb{E}\left[ \left( n^{-1}\sum_{i\leq
		n}(\varepsilon _{2,i,t}^{2}-\mathbb{E}[\varepsilon _{2,i,t}^{2}])\right) ^{2}%
	\right] \mathbb{E}\left[ \bar{\varepsilon}_{2,t}^{\ast 4}\right] \right)
	^{1/2}\leq Kn^{-3/2}.
\end{eqnarray*}%
Together with Markov's inequality, this implies that%
\begin{eqnarray}
	(nT)^{-1}\sum_{i\leq n}\sum_{t\leq T}\varepsilon _{2,i,t}^{2}\bar{\varepsilon%
	}_{2,t}^{\ast 2} &=&T^{-1}\sum_{t\leq T}\left( n^{-1}\sum_{i\leq
		n}(\varepsilon _{2,i,t}^{2}-\mathbb{E}[\varepsilon _{2,i,t}^{2}])\right) 
	\bar{\varepsilon}_{2,t}^{\ast 2}  \notag \\
	&&+(nT)^{-1}\sum_{i\leq n}\sum_{t\leq T}\mathbb{E}[\varepsilon _{2,i,t}^{2}]%
	\bar{\varepsilon}_{2,t}^{\ast 2}  \notag \\
	&=&T^{-1}\sum_{t\leq T}s_{2,t}^{2}\bar{\varepsilon}_{2,t}^{\ast
		2}+O_{p}(n^{-3/2}),  \label{P_TWE_Var_Est_3_8}
\end{eqnarray}%
where $s_{2,t}^{2}\equiv n^{-1}\sum_{i\leq n}s_{2,i,t}^{2}$. By the
definition of $\ \bar{\varepsilon}_{2,t}^{\ast }$, we can write%
\begin{equation}
	\bar{\varepsilon}_{2,t}^{\ast 2}=n^{-2}\sum_{i\leq n}\varepsilon
	_{2,i,t}^{\ast 2}+2n^{-2}\sum_{i=2}^{n}\sum_{i^{\prime }=1}^{i-1}\varepsilon
	_{2,i,t}^{\ast }\varepsilon _{2,i^{\prime },t}^{\ast }.
	\label{P_TWE_Var_Est_3_9}
\end{equation}%
Using this expression, we may write 
\begin{equation}
	T^{-1}\sum_{t\leq T}s_{2,t}^{2}\bar{\varepsilon}_{2,t}^{\ast
		2}=(n^{2}T)^{-1}\sum_{t\leq T}\sum_{i\leq n}s_{2,t}^{2}\varepsilon
	_{2,i,t}^{\ast 2}+2(n^{2}T)^{-1}\sum_{t\leq T}\sum_{i=2}^{n}\sum_{i^{\prime
		}=1}^{i-1}s_{2,t}^{2}\varepsilon _{2,i,t}^{\ast }\varepsilon _{2,i^{\prime
		},t}^{\ast }.  \label{P_TWE_Var_Est_3_10}
\end{equation}%
Under Assumption \ref{A1} and Lemma \ref{TWFE_Para}(ii), we have 
\begin{eqnarray}
	&&\mathbb{E}\left[ \left( (n^{2}T)^{-1}\sum_{i\leq n}\sum_{t\leq
		T}s_{2,t}^{2}(\varepsilon _{2,i,t}^{\ast 2}-\sigma _{2,i,t}^{2})\right) ^{2}%
	\right]  \notag \\
	&=&(n^{2}T)^{-2}\sum_{i\leq n}\mathbb{E}\left[ \left( \sum_{t\leq
		T}s_{2,t}^{2}(\varepsilon _{2,i,t}^{\ast 2}-\sigma _{2,i,t}^{2})\right) ^{2}%
	\right]  \notag \\
	&\leq &K(n^{4}T)^{-1}\sum_{i\leq n}\max_{t\leq T}\left\Vert
	s_{2,t}^{2}(\varepsilon _{2,i,t}^{\ast 2}-\sigma _{2,i,t}^{2})\right\Vert
	_{2+\delta }^{2}\leq Kn^{-4}  \label{P_TWE_Var_Est_3_11}
\end{eqnarray}%
and%
\begin{eqnarray}
	&&\mathbb{E}\left[ \left( (n^{2}T)^{-1}\sum_{i=2}^{n}\sum_{i^{\prime
		}=1}^{i-1}\sum_{t\leq T}s_{2,t}^{2}\varepsilon _{2,i,t}^{\ast }\varepsilon
	_{2,i^{\prime },t}^{\ast }\right) ^{2}\right]  \notag \\
	&=&(n^{2}T)^{-2}\sum_{i=2}^{n}\sum_{i^{\prime }=1}^{i-1}\mathbb{E}\left[
	\left( \sum_{t\leq T}s_{2,t}^{2}\varepsilon _{2,i,t}^{\ast }\varepsilon
	_{2,i^{\prime },t}^{\ast }\right) ^{2}\right]  \notag \\
	&\leq &K(n^{4}T)^{-1}\sum_{i=2}^{n}\sum_{i^{\prime }=1}^{i-1}\max_{t\leq
		T}||\varepsilon _{2,i,t}^{\ast }\varepsilon _{2,i^{\prime },t}^{\ast
	}||_{2+\delta }^{2}  \notag \\
	&\leq &K(n^{4}T)^{-1}\sum_{i=2}^{n}\sum_{i^{\prime }=1}^{i-1}\max_{t\leq
		T}||\varepsilon _{2,i,t}^{\ast }||_{4+2\delta }^{2}||\varepsilon
	_{2,i^{\prime },t}^{\ast }||_{4+2\delta }^{2}\leq Kn^{-3}.
	\label{P_TWE_Var_Est_3_12}
\end{eqnarray}%
Therefore, by (\ref{P_TWE_Var_Est_3_10})-(\ref{P_TWE_Var_Est_3_12}), and
Markov's inequality,%
\begin{equation}
	T^{-1}\sum_{t\leq T}s_{2,t}^{2}\bar{\varepsilon}_{2,t}^{\ast
		2}=(n^{2}T)^{-1}\sum_{t\leq T}\sum_{i\leq n}s_{2,t}^{2}\sigma
	_{2,i,t}^{2}+O_{p}(n^{-3/2}).  \label{P_TWE_Var_Est_3_13}
\end{equation}%
Combining (\ref{P_TWE_Var_Est_3_8}), (\ref{P_TWE_Var_Est_3_9}) and (\ref%
{P_TWE_Var_Est_3_13}) yields%
\begin{equation}
	(nT)^{-1}\sum_{i\leq n}\sum_{t\leq T}\varepsilon _{2,i,t}^{2}\bar{\varepsilon%
	}_{2,t}^{\ast 2}=n^{-3}\sum_{i\leq n}\sum_{i^{\prime }\leq n}\left(
	T^{-1}\sum_{t\leq T}s_{2,i,t}^{2}\sigma _{2,i^{\prime },t}^{2}\right)
	+O_{p}(n^{-3/2}).  \label{P_TWE_Var_Est_3_14}
\end{equation}

The third term in (\ref{P_TWE_Var_Est_3_0}) can be written as 
\begin{eqnarray}
	(nT)^{-1}\sum_{i\leq n}\sum_{t\leq T}\varepsilon _{2,i,t}^{2}\bar{\varepsilon%
	}_{2,i}^{\ast }\bar{\varepsilon}_{2,t}^{\ast } &=&(nT)^{-1}\sum_{t\leq T}%
	\bar{\varepsilon}_{2,t}^{\ast }\sum_{i\leq n}\mathbb{E}[\varepsilon
	_{2,i,t}^{2}\bar{\varepsilon}_{2,i}^{\ast }]  \notag \\
	&&+(nT)^{-1}\sum_{t\leq T}\bar{\varepsilon}_{2,t}^{\ast }\sum_{i\leq
		n}(\varepsilon _{2,i,t}^{2}\bar{\varepsilon}_{2,i}^{\ast }-\mathbb{E}%
	[\varepsilon _{2,i,t}^{2}\bar{\varepsilon}_{2,i}^{\ast }]).
	\label{P_TWE_Var_Est_3_15}
\end{eqnarray}%
By the Cauchy-Schwarz inequality,%
\begin{eqnarray}
	&&\left( (nT)^{-1}\sum_{t\leq T}\bar{\varepsilon}_{2,t}^{\ast }\sum_{i\leq
		n}(\varepsilon _{2,i,t}^{2}\bar{\varepsilon}_{2,i}^{\ast }-\mathbb{E}%
	[\varepsilon _{2,i,t}^{2}\bar{\varepsilon}_{2,i}^{\ast }])\right) ^{2} 
	\notag \\
	&\leq &T^{-1}\sum_{t\leq T}\bar{\varepsilon}_{2,t}^{\ast 2}\times
	T^{-1}\sum_{t\leq T}\left( n^{-1}\sum_{i\leq n}(\varepsilon _{2,i,t}^{2}\bar{%
		\varepsilon}_{2,i}^{\ast }-\mathbb{E}[\varepsilon _{2,i,t}^{2}\bar{%
		\varepsilon}_{2,i}^{\ast }])\right) ^{2}.  \label{P_TWE_Var_Est_3_16}
\end{eqnarray}%
From Assumption \ref{A1}, Lemma \ref{TWFE_Para}(ii) and (\ref%
{P_TWE_Var_Est_0_14}), it follows that 
\begin{eqnarray*}
	T^{-1}\sum_{t\leq T}\mathbb{E}\left[ \left( n^{-1}\sum_{i\leq n}(\varepsilon
	_{2,i,t}^{2}\bar{\varepsilon}_{2,i}^{\ast }-\mathbb{E}[\varepsilon
	_{2,i,t}^{2}\bar{\varepsilon}_{2,i}^{\ast }])\right) ^{2}\right] &\leq
	&(n^{2}T)^{-1}\sum_{t\leq T}\sum_{i\leq n}\mathbb{E}[\varepsilon _{2,i,t}^{4}%
	\bar{\varepsilon}_{2,i}^{\ast 2}] \\
	&\leq &(n^{2}T)^{-1}\sum_{t\leq T}\sum_{i\leq n}(\mathbb{E}[\varepsilon
	_{2,i,t}^{8}]\mathbb{E}[\bar{\varepsilon}_{2,i}^{\ast 4}])^{1/2}\leq Kn^{-2}.
\end{eqnarray*}%
Hence by Markov's inequality, 
\begin{equation}
	T^{-1}\sum_{t\leq T}\left( n^{-1}\sum_{i\leq n}(\varepsilon _{2,i,t}^{2}\bar{%
		\varepsilon}_{2,i}^{\ast }-\mathbb{E}[\varepsilon _{2,i,t}^{2}\bar{%
		\varepsilon}_{2,i}^{\ast }])\right) ^{2}=O_{p}(n^{-2}).
	\label{P_TWE_Var_Est_3_17}
\end{equation}%
Since $T^{-1}\sum_{t\leq T}\bar{\varepsilon}_{2,t}^{\ast 2}=O_{p}(n^{-1})$
by (\ref{P_TWE_Var_Est_0_14}), from (\ref{P_TWE_Var_Est_3_16}) and (\ref%
{P_TWE_Var_Est_3_17}) it follows that%
\begin{equation}
	(nT)^{-1}\sum_{t\leq T}\bar{\varepsilon}_{2,t}^{\ast }\sum_{i\leq
		n}(\varepsilon _{2,i,t}^{2}\bar{\varepsilon}_{2,i}^{\ast }-\mathbb{E}%
	[\varepsilon _{2,i,t}^{2}\bar{\varepsilon}_{2,i}^{\ast }])=O_{p}(n^{-3/2}).
	\label{P_TWE_Var_Est_3_18}
\end{equation}%
Let $v_{2,t}\equiv \sum_{i\leq n}\mathbb{E}[\varepsilon _{2,i,t}^{2}\bar{%
	\varepsilon}_{2,i}^{\ast }]$. Then we may write: 
\begin{equation*}
	(nT)^{-1}\sum_{t\leq T}\bar{\varepsilon}_{2,t}^{\ast }\sum_{i\leq n}\mathbb{E%
	}[\varepsilon _{2,i,t}^{2}\bar{\varepsilon}_{2,i}^{\ast
	}]=(nT)^{-1}\sum_{t\leq T}v_{2,t}\bar{\varepsilon}_{2,t}^{\ast
	}=(n^{2}T)^{-1}\sum_{i\leq n}\sum_{t\leq T}v_{2,t}\varepsilon _{2,i,t}^{\ast
	}.
\end{equation*}%
Note that $\left\vert \sum_{i\leq n}\mathbb{E}[\varepsilon _{2,i,t}^{2}\bar{%
	\varepsilon}_{2,i}^{\ast }]\right\vert \leq K$ by Assumption \ref{A1}(i), (%
\ref{P_TWE_Var_Est_0_14}) and Lemma \ref{TWFE_Para}(ii). Therefore, 
\begin{eqnarray*}
	\mathbb{E}\left[ \left( (nT)^{-1}\sum_{t\leq T}\bar{\varepsilon}_{2,t}^{\ast
	}\sum_{i\leq n}\mathbb{E}[\varepsilon _{2,i,t}^{2}\bar{\varepsilon}%
	_{2,i}^{\ast }]\right) ^{2}\right] &=&(n^{2}T)^{-2}\mathbb{E}\left[ \left(
	\sum_{i\leq n}\sum_{t\leq T}v_{2,t}\varepsilon _{2,i,t}^{\ast }\right) ^{2}%
	\right] \\
	&=&(n^{2}T)^{-2}\sum_{i\leq n}\mathbb{E}\left[ \left( \sum_{t\leq
		T}v_{2,t}\varepsilon _{2,i,t}^{\ast }\right) ^{2}\right] \\
	&\leq &K(n^{4}T)^{-1}\sum_{i\leq n}\max_{t\leq T}\left\Vert
	v_{2,t}\varepsilon _{2,i,t}^{\ast }\right\Vert _{2+\delta }^{2}\leq Kn^{-4},
\end{eqnarray*}%
which together with Markov's inequality implies that 
\begin{equation}
	(nT)^{-1}\sum_{t\leq T}\bar{\varepsilon}_{2,t}^{\ast }\sum_{i\leq n}\mathbb{E%
	}[\varepsilon _{2,i,t}^{2}\bar{\varepsilon}_{2,i}^{\ast }]=O_{p}(n^{-2}).
	\label{P_TWE_Var_Est_3_19}
\end{equation}%
From (\ref{P_TWE_Var_Est_3_15}), (\ref{P_TWE_Var_Est_3_18}) and (\ref%
{P_TWE_Var_Est_3_19}), we obtain 
\begin{equation}
	(nT)^{-1}\sum_{t\leq T}\sum_{i\leq n}\varepsilon _{2,i,t}^{2}\bar{\varepsilon%
	}_{2,i}^{\ast }\bar{\varepsilon}_{2,t}^{\ast }=O_{p}(n^{-3/2}).
	\label{P_TWE_Var_Est_3_20}
\end{equation}%
The claim of the lemma now follows from (\ref{P_TWE_Var_Est_3_0}), (\ref%
{P_TWE_Var_Est_3_7}), (\ref{P_TWE_Var_Est_3_14}) and (\ref%
{P_TWE_Var_Est_3_19}).\hfill $Q.E.D.$

\bigskip

\begin{lemma}
	\textit{\label{TWE_Var_Est_4}\ Suppose }Assumptions \textit{\ref{A1},\ }\ref%
	{A8} and \ref{A9}. Then%
	\begin{eqnarray*}
		(nT)^{-1}\sum_{g\in \mathcal{G}_{1}}\sum_{i\in I_{g}}\sum_{t\leq
			T}\varepsilon _{1,i,t}^{2}\bar{\varepsilon}_{1,g,t}^{\ast 2}
		&=&(nT)^{-1}\sum_{t\leq T}\sum_{g\in \mathcal{G}_{1}}\left(
		n_{g}^{-1}\sum_{i\in I_{g}}s_{1,i,t}^{2}\right) \left( n_{g}^{-1}\sum_{i\in
			I_{g}}\sigma _{1,i,t}^{2}\right) \\
		&&+O_{p}\left( n^{-1}\sum_{g\in \mathcal{G}_{1}}n_{g}^{-1/2}\right) .
	\end{eqnarray*}
\end{lemma}

\noindent \textsc{Proof of Lemma \ref{TWE_Var_Est_4}}. We begin the proof by
writing 
\begin{eqnarray}
	(nT)^{-1}\sum_{g\in \mathcal{G}_{1}}\sum_{i\in I_{g}}\sum_{t\leq
		T}\varepsilon _{1,i,t}^{2}\bar{\varepsilon}_{1,g,t}^{\ast 2}
	&=&(nT)^{-1}\sum_{g\in \mathcal{G}_{1}}\sum_{i\in I_{g}}\sum_{t\leq T}%
	\mathbb{E}[\varepsilon _{1,i,t}^{2}]\bar{\varepsilon}_{1,g,t}^{\ast 2} 
	\notag \\
	&&+(nT)^{-1}\sum_{g\in \mathcal{G}_{1}}\sum_{i\in I_{g}}\sum_{t\leq
		T}(\varepsilon _{1,i,t}^{2}-\mathbb{E}[\varepsilon _{1,i,t}^{2}])\bar{%
		\varepsilon}_{1,g,t}^{\ast 2}.  \label{P_TWE_Var_Est_4_1}
\end{eqnarray}%
By H\"{o}lder's inequality, the second term in (\ref{P_TWE_Var_Est_4_1})
satisfies%
\begin{eqnarray}
	\mathbb{E}\left[ \left\vert \sum_{g\in \mathcal{G}_{1}}\sum_{i\in
		I_{g}}\sum_{t\leq T}(\varepsilon _{1,i,t}^{2}-\mathbb{E}[\varepsilon
	_{1,i,t}^{2}])\bar{\varepsilon}_{1,g,t}^{\ast 2}\right\vert \right] &=&%
	\mathbb{E}\left[ \left\vert \sum_{t\leq T}\sum_{g\in \mathcal{G}_{1}}\bar{%
		\varepsilon}_{1,g,t}^{\ast 2}\sum_{i\in I_{g}}(\varepsilon _{1,i,t}^{2}-%
	\mathbb{E}[\varepsilon _{1,i,t}^{2}])\right\vert \right]  \notag \\
	&\leq &\sum_{t\leq T}\sum_{g\in \mathcal{G}_{1}}\mathbb{E}\left[ \left\vert 
	\bar{\varepsilon}_{1,g,t}^{\ast 2}\sum_{i\in I_{g}}(\varepsilon _{1,i,t}^{2}-%
	\mathbb{E}[\varepsilon _{1,i,t}^{2}])\right\vert \right]  \notag \\
	&\leq &\sum_{t\leq T}\sum_{g\in \mathcal{G}_{1}}||\bar{\varepsilon}%
	_{1,g,t}^{\ast 2}||_{2}\left\Vert \sum_{i\in I_{g}}(\varepsilon _{1,i,t}^{2}-%
	\mathbb{E}[\varepsilon _{1,i,t}^{2}])\right\Vert _{2}  \notag \\
	&\leq &KT\sum_{g\in \mathcal{G}_{1}}n_{g}^{-1/2},  \label{P_TWE_Var_Est_4_2}
\end{eqnarray}%
where the last inequality holds by Assumption \ref{A1}, Lemma \ref{TWFE_Para}%
(ii) and (\ref{P_TWE_Var_Est_0_14}). Therefore, 
\begin{equation}
	(nT)^{-1}\sum_{g\in \mathcal{G}_{1}}\sum_{i\in I_{g}}\sum_{t\leq
		T}(\varepsilon _{1,i,t}^{2}-\mathbb{E}[\varepsilon _{1,i,t}^{2}])\bar{%
		\varepsilon}_{1,g,t}^{\ast 2}=O_{p}\left( n^{-1}\sum_{g\in \mathcal{G}%
		_{1}}n_{g}^{-1/2}\right) .  \label{P_TWE_Var_Est_4_3}
\end{equation}%
The first term in (\ref{P_TWE_Var_Est_4_1}) satisfies 
\begin{eqnarray*}
	\mathbb{E}\left[ \left( \sum_{g\in \mathcal{G}_{1}}\sum_{i\in
		I_{g}}\sum_{t\leq T}\mathbb{E}[\varepsilon _{1,i,t}^{2}](\bar{\varepsilon}%
	_{1,g,t}^{\ast 2}-\mathbb{E}[\bar{\varepsilon}_{1,g,t}^{\ast 2}])\right) ^{2}%
	\right] &=&\mathbb{E}\left[ \left( \sum_{g\in \mathcal{G}_{1}}\sum_{t\leq T}(%
	\bar{\varepsilon}_{1,g,t}^{\ast 2}-\mathbb{E}[\bar{\varepsilon}%
	_{1,g,t}^{\ast 2}])\sum_{i\in I_{g}}\mathbb{E}[\varepsilon
	_{1,i,t}^{2}]\right) ^{2}\right] \\
	&=&\sum_{g\in \mathcal{G}_{1}}\mathbb{E}\left[ \left( \sum_{t\leq T}(\bar{%
		\varepsilon}_{1,g,t}^{\ast 2}-\mathbb{E}[\bar{\varepsilon}_{1,g,t}^{\ast
		2}])\sum_{i\in I_{g}}\mathbb{E}[\varepsilon _{1,i,t}^{2}]\right) ^{2}\right]
	\\
	&\leq &TK\sum_{g\in \mathcal{G}_{1}}\max_{t\leq T}\left\Vert (\bar{%
		\varepsilon}_{1,g,t}^{\ast 2}-\mathbb{E}[\bar{\varepsilon}_{1,g,t}^{\ast
		2}])\sum_{i\in I_{g}}\mathbb{E}[\varepsilon _{1,i,t}^{2}]\right\Vert
	_{2+\delta }^{2} \\
	&\leq &KT\sum_{g\in \mathcal{G}_{1}}n_{g}^{2}\left\Vert \bar{\varepsilon}%
	_{1,g,t}^{\ast 2}\right\Vert _{2+\delta }^{2}\leq KG_{1}T,
\end{eqnarray*}%
which, together with Assumption \ref{A1}(i) and Markov's inequality, implies 
\begin{equation}
	(nT)^{-1}\sum_{g\in \mathcal{G}_{1}}\sum_{i\in I_{g}}\sum_{t\leq T}\mathbb{E}%
	[\varepsilon _{1,i,t}^{2}](\bar{\varepsilon}_{1,g,t}^{\ast 2}-\mathbb{E}[%
	\bar{\varepsilon}_{1,g,t}^{\ast 2}])=O_{p}(G_{1}^{1/2}n^{-3/2}).
	\label{P_TWE_Var_Est_4_4}
\end{equation}%
It is clear that $\mathbb{E}[\bar{\varepsilon}_{1,g,t}^{\ast
	2}]=n_{g}^{-2}\sum_{i\in I_{g}}\sigma _{1,i,t}^{2}$. Hence 
\begin{eqnarray}
	(nT)^{-1}\sum_{i\leq n}\sum_{t\leq T}\mathbb{E}[\varepsilon _{1,i,t}^{2}]%
	\mathbb{E}[\bar{\varepsilon}_{1,g,t}^{\ast 2}] &=&(nT)^{-1}\sum_{t\leq
		T}\sum_{g\in \mathcal{G}_{1}}\mathbb{E}[\bar{\varepsilon}_{1,g,t}^{\ast
		2}]\sum_{i\in I_{g}}\mathbb{E}[\varepsilon _{1,i,t}^{2}]  \notag \\
	&=&(nT)^{-1}\sum_{t\leq T}\sum_{g\in \mathcal{G}_{1}}\left(
	n_{g}^{-1}\sum_{i\in I_{g}}s_{1,i,t}^{2}\right) \left( n_{g}^{-1}\sum_{i\in
		I_{g}}\sigma _{1,i,t}^{2}\right) .  \label{P_TWE_Var_Est_4_5}
\end{eqnarray}%
The claim of the lemma follows from (\ref{P_TWE_Var_Est_4_1}), (\ref%
{P_TWE_Var_Est_4_3}), (\ref{P_TWE_Var_Est_4_4}) and (\ref{P_TWE_Var_Est_4_5}%
).\hfill $Q.E.D.$

\bigskip

\begin{lemma}
	\textit{\label{TWE_Var_Est_5}\ Suppose }Assumptions \textit{\ref{A1},\ }\ref%
	{A8} and \ref{A9} hold. Then%
	\begin{equation*}
		(nT)^{-1}\sum_{g\in \mathcal{G}_{1}}\sum_{i\in I_{g}}\sum_{t\leq
			T}\varepsilon _{1,i,t}\varepsilon _{2,i,t}(\bar{\varepsilon}_{2,i}^{\ast }+%
		\bar{\varepsilon}_{2,t}^{\ast })\bar{\varepsilon}_{1,g,t}^{\ast
		}=(n^{2}T)^{-1}\sum_{t\leq T}\sum_{g\in \mathcal{G}_{1}}n_{g}s_{1,2,g,t}%
		\sigma _{1,2,g,t}+O_{p}(G_{1}n^{-3/2}).
	\end{equation*}
\end{lemma}

\noindent \textsc{Proof of Lemma \ref{TWE_Var_Est_5}}. Let $\bar{v}%
_{1,g,t}\equiv n_{g}^{-1}\sum_{i\in I_{g}}\varepsilon _{1,i,t}\varepsilon
_{2,i,t}\bar{\varepsilon}_{2,i}^{\ast }$ and $\bar{v}_{1,g,t}^{\ast }\equiv 
\bar{v}_{1,g,t}-\mathbb{E}[\bar{v}_{1,g,t}]$. Then 
\begin{eqnarray}
	\sum_{g\in \mathcal{G}_{1}}\sum_{t\leq T}\sum_{i\in I_{g}}\varepsilon
	_{1,i,t}\varepsilon _{2,i,t}\bar{\varepsilon}_{2,i}^{\ast }\bar{\varepsilon}%
	_{1,g,t}^{\ast } &=&\sum_{g\in \mathcal{G}_{1}}\sum_{t\leq T}n_{g}\bar{v}%
	_{1,g,t}\bar{\varepsilon}_{1,g,t}^{\ast }  \notag \\
	&=&\sum_{g\in \mathcal{G}_{1}}\sum_{t\leq T}n_{g}\mathbb{E}[\bar{v}_{1,g,t}]%
	\bar{\varepsilon}_{1,g,t}^{\ast }+\sum_{g\in \mathcal{G}_{1}}\sum_{t\leq
		T}n_{g}\bar{v}_{1,g,t}^{\ast }\bar{\varepsilon}_{1,g,t}^{\ast }.
	\label{P_TWE_Var_Est_5_1}
\end{eqnarray}%
From Assumption \ref{A1}, Lemma \ref{TWFE_Para}(ii) and Lemma \ref{GTFE_1}%
(iii),%
\begin{equation}
	\mathbb{E}[\bar{v}_{1,g,t}^{\ast 2}]\leq n_{g}^{-2}\sum_{i\in I_{g}}\mathbb{E%
	}[\varepsilon _{1,i,t}^{2}\varepsilon _{2,i,t}^{2}\bar{\varepsilon}%
	_{2,i}^{\ast 2}]\leq Kn_{g}^{-2}\sum_{i\in I_{g}}\left\Vert \bar{\varepsilon}%
	_{2,i}^{\ast 2}\right\Vert _{2}\leq Kn_{g}^{-1}T^{-1}.
	\label{P_TWE_Var_Est_5_2}
\end{equation}%
This, together with (\ref{P_TWE_Var_Est_0_14}) and H\"{o}lder's inequality,
implies that 
\begin{equation*}
	\sum_{g\in \mathcal{G}_{1}}\sum_{t\leq T}n_{g}\mathbb{E}[|\bar{v}%
	_{1,g,t}^{\ast }\bar{\varepsilon}_{1,g,t}^{\ast }|]\leq \sum_{g\in \mathcal{G%
		}_{1}}\sum_{t\leq T}n_{g}\left\Vert \bar{v}_{1,g,t}^{\ast }\right\Vert
	_{2}\left\Vert \bar{\varepsilon}_{1,g,t}^{\ast }\right\Vert _{2}\leq
	KG_{1}T^{1/2}.
\end{equation*}%
Therefore, by Assumption \ref{A1}(i) and Markov's inequality, 
\begin{equation}
	(nT)^{-1}\sum_{g\in \mathcal{G}_{1}}\sum_{t\leq T}n_{g}\bar{v}_{1,g,t}^{\ast
	}\bar{\varepsilon}_{1,g,t}^{\ast }=O_{p}(G_{1}n^{-3/2}).
	\label{P_TWE_Var_Est_5_3}
\end{equation}%
From Assumption \ref{A1}, Lemma \ref{TWFE_Para}(ii) and Lemma \ref{GTFE_1}%
(iii),%
\begin{equation*}
	\left\vert \mathbb{E}[\bar{v}_{1,g,t}]\right\vert \leq n_{g}^{-1}\sum_{i\in
		I_{g}}\mathbb{E}[|\varepsilon _{1,i,t}\varepsilon _{2,i,t}\bar{\varepsilon}%
	_{2,i}^{\ast }|]\leq n_{g}^{-1}\sum_{i\in I_{g}}\left\Vert \varepsilon
	_{1,i,t}\varepsilon _{2,i,t}\right\Vert _{2}\left\Vert \bar{\varepsilon}%
	_{2,i}^{\ast }\right\Vert _{2}\leq KT^{-1/2},
\end{equation*}%
which, together with Assumption \ref{A1} and (\ref{P_TWE_Var_Est_0_14}),
implies that 
\begin{eqnarray*}
	\mathbb{E}\left[ \left( \sum_{g\in \mathcal{G}_{1}}\sum_{t\leq T}n_{g}%
	\mathbb{E}[\bar{v}_{1,g,t}]\bar{\varepsilon}_{1,g,t}^{\ast }\right) ^{2}%
	\right] &=&\sum_{g\in \mathcal{G}_{1}}n_{g}^{2}\mathbb{E}\left[ \left(
	\sum_{t\leq T}\mathbb{E}[\bar{v}_{1,g,t}]\bar{\varepsilon}_{1,g,t}^{\ast
	}\right) ^{2}\right] \\
	&=&\sum_{g\in \mathcal{G}_{1}}\sum_{i\in I_{g}}\mathbb{E}\left[ \left(
	\sum_{t\leq T}\mathbb{E}[\bar{v}_{1,g,t}]\varepsilon _{1,i,t}^{\ast }\right)
	^{2}\right] \\
	&\leq &KT\sum_{g\in \mathcal{G}_{1}}\sum_{i\in I_{g}}\max_{t\leq
		T}\left\Vert \mathbb{E}[\bar{v}_{1,g,t}]\varepsilon _{1,i,t}^{\ast
	}\right\Vert _{2+\delta }^{2}\leq Kn.
\end{eqnarray*}%
Therefore, by Assumption \ref{A1}(i) and Markov's inequality,%
\begin{equation}
	(nT)^{-1}\sum_{g\in \mathcal{G}_{1}}\sum_{t\leq T}n_{g}\mathbb{E}[\bar{v}%
	_{1,g,t}]\bar{\varepsilon}_{1,g,t}^{\ast }=O_{p}(n^{-3/2}),
	\label{P_TWE_Var_Est_5_4}
\end{equation}%
From (\ref{P_TWE_Var_Est_5_1}), (\ref{P_TWE_Var_Est_5_3}) and (\ref%
{P_TWE_Var_Est_5_4}), it follows that 
\begin{equation}
	(nT)^{-1}\sum_{g\in \mathcal{G}_{1}}\sum_{t\leq T}\sum_{i\in
		I_{g}}\varepsilon _{1,i,t}\varepsilon _{2,i,t}\bar{\varepsilon}_{2,i}^{\ast }%
	\bar{\varepsilon}_{1,g,t}^{\ast }=O_{p}(G_{1}n^{-3/2}).
	\label{P_TWE_Var_Est_5_5}
\end{equation}

Let $\bar{v}_{2,g,t}\equiv n_{g}^{-1}\sum_{i\in I_{g}}\varepsilon
_{1,i,t}\varepsilon _{2,i,t}$ and $\bar{v}_{2,g,t}^{\ast }\equiv \bar{v}%
_{2,g,t}-\mathbb{E}[\bar{v}_{2,g,t}]$. Then 
\begin{eqnarray}
	\sum_{t\leq T}\sum_{g\in \mathcal{G}_{1}}\sum_{i\in I_{g}}\varepsilon
	_{1,i,t}\varepsilon _{2,i,t}\bar{\varepsilon}_{2,t}^{\ast }\bar{\varepsilon}%
	_{1,g,t}^{\ast } &=&\sum_{t\leq T}\bar{\varepsilon}_{2,t}^{\ast }\sum_{g\in 
		\mathcal{G}_{1}}n_{g}\bar{\varepsilon}_{1,g,t}^{\ast }\bar{v}_{2,g,t}  \notag
	\\
	&=&\sum_{t\leq T}\bar{\varepsilon}_{2,t}^{\ast }\sum_{g\in \mathcal{G}%
		_{1}}n_{g}\bar{\varepsilon}_{1,g,t}^{\ast }\mathbb{E}[\bar{v}%
	_{2,g,t}]+\sum_{t\leq T}\bar{\varepsilon}_{2,t}^{\ast }\sum_{g\in \mathcal{G}%
		_{1}}n_{g}\bar{\varepsilon}_{1,g,t}^{\ast }\bar{v}_{2,g,t}^{\ast }.
	\label{P_TWE_Var_Est_5_6}
\end{eqnarray}%
Using Assumption \ref{A1}, Lemma \ref{TWFE_Para}(ii) and Lemma \ref{GTFE_1}%
(iii),%
\begin{equation}
	\mathbb{E}[\bar{v}_{2,g,t}^{\ast 2}]\leq n_{g}^{-2}\sum_{i\in I_{g}}\mathbb{E%
	}[|\varepsilon _{1,i,t}\varepsilon _{2,i,t}|^{2}]\leq Kn_{{}g}^{-1}.
	\label{P_TWE_Var_Est_5_7}
\end{equation}%
This, together with (\ref{P_TWE_Var_Est_0_14}) and H\"{o}lder's inequality,
implies that%
\begin{equation*}
	\sum_{t\leq T}\sum_{g\in \mathcal{G}_{1}}n_{g}\mathbb{E}[|\bar{\varepsilon}%
	_{2,t}^{\ast }\bar{\varepsilon}_{1,g,t}^{\ast }\bar{v}_{2,g,t}^{\ast }|]\leq
	\sum_{t\leq T}\sum_{g\in \mathcal{G}_{1}}n_{g}\left\Vert \bar{v}%
	_{2,g,t}^{\ast }\right\Vert _{2}\left\Vert \bar{\varepsilon}_{2,t}^{\ast
	}\right\Vert _{4}\left\Vert \bar{\varepsilon}_{1,g,t}^{\ast }\right\Vert
	_{4}\leq KG_{1}Tn^{-1/2}.
\end{equation*}%
Therefore, by Assumption \ref{A1}(i) and Markov's inequality, the second
term in (\ref{P_TWE_Var_Est_5_6}) satisfies 
\begin{equation}
	(nT)^{-1}\sum_{t\leq T}\bar{\varepsilon}_{2,t}^{\ast }\sum_{g\in \mathcal{G}%
		_{1}}n_{g}\bar{\varepsilon}_{1,g,t}^{\ast }\bar{v}_{2,g,t}^{\ast
	}=O_{p}(G_{1}n^{-3/2}).  \label{P_TWE_Var_Est_5_8}
\end{equation}%
Note that $\mathbb{E}[\bar{v}_{2,g,t}]=s_{1,2,g,t}$. Let $\tilde{\varepsilon}%
_{1,i,t}^{\ast }\equiv s_{1,2,g,t}\varepsilon _{1,i,t}^{\ast }$. Then the
first term in (\ref{P_TWE_Var_Est_5_6}) can be written as%
\begin{eqnarray}
	\sum_{t\leq T}\bar{\varepsilon}_{2,t}^{\ast }\sum_{g\in \mathcal{G}_{1}}n_{g}%
	\bar{\varepsilon}_{1,g,t}^{\ast }\mathbb{E}[\bar{v}_{2,g,t}]
	&=&n^{-1}\sum_{t\leq T}\left( \sum_{i\leq n}\tilde{\varepsilon}%
	_{1,i,t}^{\ast }\right) \left( \sum_{i\leq n}\varepsilon _{2,i,t}^{\ast
	}\right)  \notag \\
	&=&n^{-1}\sum_{i\leq n}\sum_{t\leq T}\tilde{\varepsilon}_{1,i,t}^{\ast
	}\varepsilon _{2,i,t}^{\ast }  \notag \\
	&&+n^{-1}\sum_{i^{\prime }=2}^{n}\sum_{i=1}^{i-1}\sum_{t\leq T}(\tilde{%
		\varepsilon}_{1,i^{\prime },t}^{\ast }\varepsilon _{2,i,t}^{\ast }+\tilde{%
		\varepsilon}_{1,i,t}^{\ast }\varepsilon _{2,i^{\prime },t}^{\ast }).
	\label{P_TWE_Var_Est_5_9}
\end{eqnarray}%
Under Assumption \ref{A1}, Lemma \ref{TWFE_Para}(ii) and Lemma \ref{GTFE_1}%
(iii), we have 
\begin{eqnarray*}
	\mathbb{E}\left[ \left( \sum_{i\leq n}\sum_{t\leq T}(\tilde{\varepsilon}%
	_{1,i,t}^{\ast }\varepsilon _{2,i,t}^{\ast }-\mathbb{E}[\tilde{\varepsilon}%
	_{1,i,t}^{\ast }\varepsilon _{2,i,t}^{\ast }])\right) ^{2}\right]
	&=&\sum_{i\leq n}\mathbb{E}\left[ \left( \sum_{t\leq T}(\tilde{\varepsilon}%
	_{1,i,t}^{\ast }\varepsilon _{2,i,t}^{\ast }-\mathbb{E}[\tilde{\varepsilon}%
	_{1,i,t}^{\ast }\varepsilon _{2,i,t}^{\ast }])\right) ^{2}\right] \\
	&\leq &KT\sum_{i\leq n}\max_{t\leq T}\left\Vert \tilde{\varepsilon}%
	_{1,i,t}^{\ast }\varepsilon _{2,i,t}^{\ast }\right\Vert _{2+\delta }^{2}\leq
	KnT,
\end{eqnarray*}%
which together with Assumption \ref{A1}(i) and Markov's inequality implies
that 
\begin{eqnarray}
	(n^{2}T)^{-1}\sum_{i\leq n}\sum_{t\leq T}\tilde{\varepsilon}_{1,i,t}^{\ast
	}\varepsilon _{2,i,t}^{\ast } &=&(n^{2}T)^{-1}\sum_{i\leq n}\sum_{t\leq T}%
	\mathbb{E}[\tilde{\varepsilon}_{1,i,t}^{\ast }\varepsilon _{2,i,t}^{\ast
	}]+O_{p}(n^{-2})  \notag \\
	&=&(n^{2}T)^{-1}\sum_{t\leq T}\sum_{g\in \mathcal{G}_{1}}n_{g}s_{1,2,g,t}%
	\sigma _{1,2,g,t}+O_{p}(n^{-2}).  \label{P_TWE_Var_Est_5_10}
\end{eqnarray}%
Under Assumption \ref{A1}, Lemma \ref{TWFE_Para}(ii) and Lemma \ref{GTFE_1}%
(iii), we have%
\begin{equation}
	\mathbb{E}\left[ \left( \sum_{i^{\prime }=2}^{n}\sum_{i=1}^{i-1}\sum_{t\leq
		T}\tilde{\varepsilon}_{1,i^{\prime },t}^{\ast }\varepsilon _{2,i,t}^{\ast
	}\right) ^{2}\right] =\sum_{i^{\prime }=2}^{n}\sum_{i=1}^{i-1}\mathbb{E}%
	\left[ \left( \sum_{t\leq T}\tilde{\varepsilon}_{1,i^{\prime },t}^{\ast
	}\varepsilon _{2,i,t}^{\ast }\right) ^{2}\right] \leq Kn^{2}T.
	\label{P_TWE_Var_Est_5_11}
\end{equation}%
Similarly, 
\begin{equation*}
	\mathbb{E}\left[ \left( \sum_{i^{\prime }=2}^{n}\sum_{i=1}^{i-1}\sum_{t\leq
		T}\tilde{\varepsilon}_{1,i,t}^{\ast }\varepsilon _{2,i^{\prime },t}^{\ast
	}\right) ^{2}\right] \leq Kn^{2}T
\end{equation*}%
which together with Assumption \ref{A1}(i), (\ref{P_TWE_Var_Est_5_11}) and
Markov's inequality implies that%
\begin{equation}
	(n^{2}T)^{-1}\sum_{i^{\prime }=2}^{n}\sum_{i=1}^{i-1}\sum_{t\leq T}(\tilde{%
		\varepsilon}_{1,i^{\prime },t}^{\ast }\varepsilon _{2,i,t}^{\ast }+\tilde{%
		\varepsilon}_{1,i,t}^{\ast }\varepsilon _{2,i^{\prime },t}^{\ast
	})=O_{p}(n^{-3/2}).  \label{P_TWE_Var_Est_5_12}
\end{equation}%
From (\ref{P_TWE_Var_Est_5_9}), (\ref{P_TWE_Var_Est_5_10}) and (\ref%
{P_TWE_Var_Est_5_12}), it follows that 
\begin{equation*}
	(nT)^{-1}\sum_{t\leq T}\bar{\varepsilon}_{2,t}^{\ast }\sum_{g\in \mathcal{G}%
		_{1}}n_{g}\bar{\varepsilon}_{1,g,t}^{\ast }\mathbb{E}[\bar{v}%
	_{2,g,t}]=(n^{2}T)^{-1}\sum_{t\leq T}\sum_{g\in \mathcal{G}%
		_{1}}n_{g}s_{1,2,g,t}\sigma _{1,2,g,t}+O_{p}(n^{-3/2}).
\end{equation*}%
This, along with (\ref{P_TWE_Var_Est_5_8}), establishes the claim of the
lemma.\hfill $Q.E.D.$

\bigskip

\begin{lemma}
	\textit{\label{TWE_Var_Est_6a}\ Suppose }Assumptions \textit{\ref{A1},\ }\ref%
	{A8}, \ref{A9} and \ref{A10} hold. Then 
	\begin{equation}
		\frac{\widehat{\mathbb{E}[S_{n,T}]}-\mathbb{E}[S_{n,T}]}{\omega _{n,T}}%
		=o_{p}(1).  \label{TWE_Var_Est_6a_1}
	\end{equation}
\end{lemma}

\noindent \textsc{Proof of Lemma \ref{TWE_Bias&Var_Est_Null}}. Under
Assumption \ref{A10}(i),%
\begin{eqnarray*}
	\mathbb{E}[S_{n,T}] &=&(2nT)^{-1/2}\sum_{g\in \mathcal{G}_{1}}\sum_{i\in
		I_{g}}\mathbb{E}\left[ \sum_{t\leq T}(n_{g}^{-1}\varepsilon _{1,i,t}^{\ast
		2}-n^{-1}\varepsilon _{2,i,t}^{\ast 2})-T^{-1}\left( \sum_{t\leq
		T}\varepsilon _{2,i,t}^{\ast }\right) ^{2}\right] \\
	&=&(2nT)^{-1/2}\sum_{g\in \mathcal{G}_{1}}\sum_{i\in I_{g}}\left[
	\sum_{t\leq T}(n_{g}^{-1}\sigma _{1,i}^{2}-n^{-1}\sigma _{2,i}^{2})-\sigma
	_{2,i}^{2}\right] \\
	&=&(2nT)^{-1/2}\sum_{g\in \mathcal{G}_{1}}\sum_{i\in
		I_{g}}(Tn_{g}^{-1}\sigma _{1,i}^{2}-(1+Tn^{-1})\sigma _{2,i}^{2}).
\end{eqnarray*}%
Therefore, 
\begin{equation}
	\widehat{\mathbb{E}[S_{n,T}]}-\mathbb{E}[S_{n,T}]=(2nT)^{-1/2}\left(
	\sum_{g\in \mathcal{G}_{1}}\sum_{i\in I_{g}}Tn_{g}^{-1}(\hat{\sigma}%
	_{1,i}^{2}-\sigma _{1,i}^{2})-(1+Tn^{-1})\sum_{g\in \mathcal{G}%
		_{1}}\sum_{i\in I_{g}}(\hat{\sigma}_{2,i}^{2}-\sigma _{2,i}^{2})\right) .
	\label{P_TWE_Var_Est_6a_1}
\end{equation}%
Under Assumption \ref{A10}(i) and (\ref{P_TWFE_Xe_1}), we have $\mathbb{E}%
[\varepsilon _{2,i,t}]=0$. This implies that $\sigma _{2,i}^{2}=s_{2,i}^{2}$%
. Similarly, under Assumption \ref{A10}(i) and (\ref{P_GTFE_2_1}), $\sigma
_{1,i}^{2}=s_{1,i}^{2}$. Therefore, by Assumption \ref{A1}(i), Lemmas \ref%
{TWFE_Est4} and \ref{GTFE_Est3},%
\begin{equation}
	\widehat{\mathbb{E}[S_{n,T}]}-\mathbb{E}[S_{n,T}]=O_{p}\left(
	n^{-1}+\sum_{g\in \mathcal{G}_{1}}n_{g}^{-1}\right) .
	\label{P_TWE_Var_Est_6a_2}
\end{equation}%
The claim of the lemma follows from Assumptions \ref{A9}(i) and \ref{A10}%
(ii), and (\ref{P_TWE_Var_Est_6a_2}).\hfill $Q.E.D.$

\bigskip

\begin{lemma}
	\textit{\label{TWE_Var_Est_6}\ Suppose }Assumptions \textit{\ref{A1},\ }\ref%
	{A8}, \ref{A9} and \ref{A10} hold. Then under the null hypothesis,%
	\begin{equation*}
		(nT)^{-1}\sum_{i\leq n}\sum_{t\leq T}\frac{\hat{\varepsilon}_{2,i,t}^{2}-%
			\hat{\varepsilon}_{1,i,t}^{2}}{2}-(nT)^{-1/2}\widehat{\mathbb{E}[S_{n,T}]}%
		=O_{p}\left( n^{-1}\right) .
	\end{equation*}
\end{lemma}

\noindent \textsc{Proof of Lemma \ref{TWE_Var_Est_6}}. From Theorem \ref%
{TWFE_Vuong} and Lemma \ref{TWE_Var_Est_6a}, we have 
\begin{eqnarray}
	&&(nT)^{-1}\sum_{i\leq n}\sum_{t\leq T}\frac{\hat{\varepsilon}_{2,i,t}^{2}-%
		\hat{\varepsilon}_{1,i,t}^{2}}{2}-(nT)^{-1/2}\widehat{\mathbb{E}[S_{n,T}]} 
	\notag \\
	&=&\frac{\mathbb{E}[S_{n,T}]-\widehat{\mathbb{E}[S_{n,T}]}+\overline{QLR}%
		_{n,T}}{(nT)^{1/2}}+\frac{QLR_{n,T}-\mathbb{E}[S_{n,T}]-\overline{QLR}_{n,T}%
	}{(nT)^{1/2}}  \notag \\
	&=&\frac{\mathbb{E}[S_{n,T}]-\widehat{\mathbb{E}[S_{n,T}]}+\overline{QLR}%
		_{n,T}}{(nT)^{1/2}}+O_{p}(\omega _{n,T}(nT)^{-1/2})  \notag \\
	&=&(nT)^{-1/2}\overline{QLR}_{n,T}+O_{p}(\omega _{n,T}(nT)^{-1/2})
	\label{P_TWE_Var_Est_6_1}
\end{eqnarray}%
where%
\begin{equation*}
	\overline{QLR}_{n,T}=(nT)^{-1/2}\sum_{i\leq n}\sum_{t\leq T}\frac{%
		s_{2,i,t}^{2}-s_{1,i,t}^{2}}{2}.
\end{equation*}%
Since $\overline{QLR}_{n,T}=0$ under the null hypothesis, from (\ref%
{P_TWE_Var_Est_6_1}) it follows that 
\begin{equation}
	(nT)^{-1}\sum_{i\leq n}\sum_{t\leq T}\frac{\hat{\varepsilon}_{2,i,t}^{2}-%
		\hat{\varepsilon}_{1,i,t}^{2}}{2}-(nT)^{-1/2}\widehat{\mathbb{E}[S_{n,T}]}%
	=O_{p}(\omega _{n,T}(nT)^{-1/2}).  \label{P_TWE_Var_Est_6_2}
\end{equation}%
Moreover, from Lemma \ref{TWE_Bias&Var_Est_Alt}(iii), we have $\omega
_{n,T}=O(1)$. This along with Assumption \ref{A1}(i) and (\ref%
{P_TWE_Var_Est_6_2}) establishes the claim of the lemma.\hfill $Q.E.D.$

\bigskip

\begin{lemma}
	\textit{\label{TWE_Var_Est_7}\ Suppose }Assumptions \textit{\ref{A1},\ }\ref%
	{A8}, \ref{A9} and \ref{A10} hold. Then%
	\begin{equation*}
		\frac{\hat{\sigma}_{n,T}^{2}-(\sigma _{n,T}^{2}+2\sigma _{U,n,T}^{2})}{%
			\omega _{n,T}^{2}}=o_{p}(1),
	\end{equation*}%
	under the null hypothesis.
\end{lemma}

\noindent \textsc{Proof of Lemma \ref{TWE_Var_Est_7}}. Applying Assumptions %
\ref{A9}(i) and \ref{A10}(ii), Lemmas \ref{TWE_Var_Est_0}-\ref{TWE_Var_Est_2}%
, we obtain:%
\begin{eqnarray}
	(nT)^{-1}\sum_{i\leq n}\sum_{t\leq T}(\hat{\varepsilon}_{2,i,t}^{2}-\hat{%
		\varepsilon}_{1,i,t}^{2})^{2} &=&(nT)^{-1}\sum_{i\leq n}\sum_{t\leq
		T}(\varepsilon _{2,i,t}^{2}-\varepsilon _{1,i,t}^{2})^{2}  \notag \\
	&&+4n^{-1}\sum_{i\leq n}s_{2,i}^{2}\mathbb{E}[\bar{\varepsilon}_{2,i}^{\ast
		2}]+4n^{-3}\sum_{i\leq n}\sum_{i^{\prime }\leq n}\left( T^{-1}\sum_{t\leq
		T}s_{2,i,t}^{2}\sigma _{2,i^{\prime },t}^{2}\right)  \notag \\
	&&+4(nT)^{-1}\sum_{t\leq T}\sum_{g\in \mathcal{G}_{1}}\left(
	n_{g}^{-1}\sum_{i\in I_{g}}s_{1,i,t}^{2}\right) \left( n_{g}^{-1}\sum_{i\in
		I_{g}}\sigma _{1,i,t}^{2}\right)  \notag \\
	&&-8(n^{2}T)^{-1}\sum_{t\leq T}\sum_{g\in \mathcal{G}_{1}}n_{g}s_{1,2,g,t}%
	\sigma _{1,2,g,t}+o_{p}\left( \omega _{n,T}^{2}\right) .
	\label{P_TWE_Var_Est_7_1}
\end{eqnarray}%
Under Assumption \ref{A10}(i) and the null hypothesis, we have $%
s_{j,i,t}^{2}=\sigma _{j,i,t}^{2}$ for $j=1,2$, $s_{1,2,g,t}=\sigma
_{1,2,g,t}$ and $\mathbb{E}[\bar{\varepsilon}_{2,i}^{\ast 2}]=T^{-1}\sigma
_{2,i}^{2}$. Hence,%
\begin{eqnarray}
	(nT)^{-1}\sum_{i\leq n}\sum_{t\leq T}(\hat{\varepsilon}_{2,i,t}^{2}-\hat{%
		\varepsilon}_{1,i,t}^{2})^{2} &=&(nT)^{-1}\sum_{i\leq n}\sum_{t\leq
		T}(\varepsilon _{2,i,t}^{2}-\varepsilon _{1,i,t}^{2})^{2}  \notag \\
	&&+4(nT)^{-1}\sum_{i\leq n}\sigma _{2,i}^{4}+4n^{-3}\left( \sum_{i\leq
		n}\sigma _{2,i}^{2}\right) ^{2}+4n^{-1}\sum_{g\in \mathcal{G}_{1}}\left(
	n_{g}^{-1}\sum_{i\in I_{g}}\sigma _{1,i}^{2}\right) ^{2}  \notag \\
	&&-8n^{-2}\sum_{g\in \mathcal{G}_{1}}n_{g}^{-1}\left( \sum_{i\in
		I_{g}}\sigma _{1,2,i}\right) ^{2}+o_{p}\left( \omega _{n,T}^{2}\right) 
	\notag \\
	&=&(nT)^{-1}\sum_{i\leq n}\sum_{t\leq T}(\varepsilon
	_{2,i,t}^{2}-\varepsilon _{1,i,t}^{2})^{2}+8\sigma _{U,n,T}^{2}+o_{p}\left(
	\omega _{n,T}^{2}\right) .  \label{P_TWE_Var_Est_7_2}
\end{eqnarray}%
By Assumptions \ref{A1}, \ref{A9}(ii) and \ref{A10}(i), 
\begin{equation*}
	(nT)^{-1}\sum_{i\leq n}\sum_{t\leq T}(\varepsilon _{2,i,t}^{2}-\varepsilon
	_{1,i,t}^{2})^{2}=(nT)^{-1}\sum_{i\leq n}\sum_{t\leq T}\mathbb{E}%
	[(\varepsilon _{2,i,t}^{2}-\varepsilon _{1,i,t}^{2})^{2}]+O_{p}(\omega
	_{n,T}^{2}n^{-1}),
\end{equation*}%
which, together with (\ref{P_TWE_Var_Est_7_2}), yields 
\begin{equation}
	(4nT)^{-1}\sum_{i\leq n}\sum_{t\leq T}(\hat{\varepsilon}_{2,i,t}^{2}-\hat{%
		\varepsilon}_{1,i,t}^{2})^{2}=(4n)^{-1}\sum_{i\leq n}\mathbb{E}[(\varepsilon
	_{2,i,t}^{2}-\varepsilon _{1,i,t}^{2})^{2}]+2\sigma _{U,n,T}^{2}+o_{p}\left(
	\omega _{n,T}^{2}\right) .  \label{P_TWE_Var_Est_7_3}
\end{equation}%
Under the null hypothesis $\sigma _{n,T}^{2}=(4n)^{-1}\sum_{i\leq n}\mathbb{E%
}[(\varepsilon _{2,i,t}^{2}-\varepsilon _{1,i,t}^{2})^{2}]$, so (\ref%
{P_TWE_Var_Est_7_3}) implies 
\begin{equation}
	(4nT)^{-1}\sum_{i\leq n}\sum_{t\leq T}(\hat{\varepsilon}_{2,i,t}^{2}-\hat{%
		\varepsilon}_{1,i,t}^{2})^{2}=\sigma _{n,T}^{2}+2\sigma
	_{U,n,T}^{2}+o_{p}\left( \omega _{n,T}^{2}\right) .
	\label{P_TWE_Var_Est_7_4}
\end{equation}

Moreover, Lemma \ref{TWE_Var_Est_6} implies 
\begin{equation*}
	(nT)^{-1/2}MQLR_{n,T}=(nT)^{-1}\sum_{i\leq n}\sum_{t\leq T}\frac{\hat{%
			\varepsilon}_{2,i,t}^{2}-\hat{\varepsilon}_{1,i,t}^{2}}{2}-(nT)^{-1/2}%
	\widehat{\mathbb{E}[S_{n,T}]}=O_{p}\left( n^{-1}\right) ,
\end{equation*}%
which, together with (\ref{P_TWE_Var_Est_7_4}), yields: 
\begin{eqnarray}
	\hat{\sigma}_{n,T}^{2} &\equiv &(nT)^{-1}\sum_{i\leq n}\sum_{t\leq T}\frac{(%
		\hat{\varepsilon}_{2,i,t}^{2}-\hat{\varepsilon}_{1,i,t}^{2})^{2}}{4}%
	-(nT)^{-1}\left( MQLR_{n,T}\right) ^{2}  \notag \\
	&=&\sigma _{n,T}^{2}+2\sigma _{U,n,T}^{2}+O_{p}\left( n^{-2}\right)
	+o_{p}\left( \omega _{n,T}^{2}\right) =\sigma _{n,T}^{2}+2\sigma
	_{U,n,T}^{2}+o_{p}\left( \omega _{n,T}^{2}\right)  \label{P_TWE_Var_Est_7_5}
\end{eqnarray}%
where the second equality follows from Assumption \ref{A9}(i). The claim of
the lemma follows immediately from (\ref{P_TWE_Var_Est_7_5}).\hfill $Q.E.D.$

\bigskip

\begin{lemma}
	\textit{\label{TWE_Cov_Est_2}\ Suppose }Assumptions \textit{\ref{A1},\ }\ref%
	{A8} and \ref{A9} hold. Then%
	\begin{equation*}
		n^{-1}\sum_{g\in \mathcal{G}_{1}}n_{g}^{-1}\left( \sum_{i\in I_{g}}(\hat{%
			\sigma}_{1,2,i}-s_{1,2,i})\right) ^{2}=O_{p}\left( G_{1}n^{-1}\right) .
	\end{equation*}
\end{lemma}

\noindent \textsc{Proof of Lemma \ref{TWE_Cov_Est_2}}. By the definition of $%
\hat{\sigma}_{1,2,i}$, 
\begin{eqnarray}
	\hat{\sigma}_{1,2,i}-s_{1,2,i} &=&T^{-1}\sum_{t\leq T}\hat{\varepsilon}%
	_{1,i,t}\hat{\varepsilon}_{2,i,t}-T^{-1}\sum_{t\leq T}\mathbb{E}[\varepsilon
	_{1,i,t}\varepsilon _{2,i,t}]  \notag \\
	&=&T^{-1}\sum_{t\leq T}(\dot{\varepsilon}_{1,i,t}\ddot{\varepsilon}_{2,i,t}-%
	\mathbb{E}[\varepsilon _{1,i,t}\varepsilon _{2,i,t}])  \notag \\
	&&-T^{-1}\sum_{t\leq T}\ddot{\varepsilon}_{2,i,t}\dot{x}_{i,t}^{\top }(\hat{%
		\theta}_{1}-\theta _{1}^{\ast })-T^{-1}\sum_{t\leq T}\dot{\varepsilon}%
	_{1,i,t}\ddot{x}_{i,t}^{\top }(\hat{\theta}_{2}-\theta _{2}^{\ast })  \notag
	\\
	&&+(\hat{\theta}_{2}-\theta _{2}^{\ast })^{\top }\left( T^{-1}\sum_{t\leq T}%
	\ddot{x}_{i,t}\dot{x}_{i,t}^{\top }\right) (\hat{\theta}_{1}-\theta
	_{1}^{\ast })  \label{P_TWE_Cov_Est2_0}
\end{eqnarray}%
It follows that%
\begin{eqnarray}
	n^{-1}\sum_{g\in \mathcal{G}_{1}}n_{g}^{-1}\left( \sum_{i\in I_{g}}(\hat{%
		\sigma}_{1,2,i}-s_{1,2,i})\right) ^{2} &\leq &K(nT^{2})^{-1}\sum_{g\in 
		\mathcal{G}_{1}}n_{g}^{-1}\left( \sum_{i\in I_{g}}\sum_{t\leq T}(\dot{%
		\varepsilon}_{1,i,t}\ddot{\varepsilon}_{2,i,t}-\mathbb{E}[\varepsilon
	_{1,i,t}\varepsilon _{2,i,t}])\right) ^{2}  \notag \\
	&&+K(nT^{2})^{-1}\sum_{g\in \mathcal{G}_{1}}n_{g}^{-1}\left( \sum_{i\in
		I_{g}}\sum_{t\leq T}\ddot{\varepsilon}_{2,i,t}\dot{x}_{i,t}^{\top }(\hat{%
		\theta}_{1}-\theta _{1}^{\ast })\right) ^{2}  \notag \\
	&&+K(nT^{2})^{-1}\sum_{g\in \mathcal{G}_{1}}n_{g}^{-1}\left( \sum_{i\in
		I_{g}}\sum_{t\leq T}\dot{\varepsilon}_{1,i,t}\ddot{x}_{i,t}^{\top }(\hat{%
		\theta}_{2}-\theta _{2}^{\ast })\right) ^{2}  \notag \\
	&&+K(nT^{2})^{-1}\sum_{g\in \mathcal{G}_{1}}n_{g}^{-1}\left( (\hat{\theta}%
	_{2}-\theta _{2}^{\ast })^{\top }\sum_{i\in I_{g}}\sum_{t\leq T}\ddot{x}%
	_{i,t}\dot{x}_{i,t}^{\top }(\hat{\theta}_{1}-\theta _{1}^{\ast })\right)
	^{2}.  \notag \\
	&&  \label{P_TWE_Cov_Est_2_1}
\end{eqnarray}%
By Assumptions \ref{A1}(i) and \ref{A8}(ii), Lemma \ref{TWFE_Est1} and Lemma %
\ref{GTFE_Est1}, the\ Cauchy-Schwarz inequality and Markov's inequality, 
\begin{eqnarray}
	&&(nT^{2})^{-1}\sum_{g\in \mathcal{G}_{1}}n_{g}^{-1}\left( (\hat{\theta}%
	_{2}-\theta _{2}^{\ast })^{\top }\sum_{i\in I_{g}}\sum_{t\leq T}\ddot{x}%
	_{i,t}\dot{x}_{i,t}^{\top }(\hat{\theta}_{1}-\theta _{1}^{\ast })\right) ^{2}
	\notag \\
	&\leq &K||\hat{\theta}_{1}-\theta _{1}^{\ast }||^{2}||\hat{\theta}%
	_{2}-\theta _{2}^{\ast }||^{2}(nT)^{-1}\sum_{i\leq n}\sum_{t\leq T}||\ddot{x}%
	_{i,t}\dot{x}_{i,t}^{\top }||^{2}=O_{p}(n^{-4}).  \label{P_TWE_Cov_Est2_2}
\end{eqnarray}%
Similarly,%
\begin{equation}
	(nT^{2})^{-1}\sum_{g\in \mathcal{G}_{1}}n_{g}^{-1}\left( \sum_{i\in
		I_{g}}\sum_{t\leq T}\ddot{\varepsilon}_{2,i,t}\dot{x}_{i,t}^{\top }(\hat{%
		\theta}_{1}-\theta _{1}^{\ast })\right) ^{2}\leq \frac{||\hat{\theta}%
		_{1}-\theta _{1}^{\ast }||^{2}}{nT}\sum_{i\leq n}\sum_{t\leq T}||\ddot{%
		\varepsilon}_{2,i,t}\dot{x}_{i,t}^{\top }||^{2}=O_{p}(n^{-2})
	\label{P_TWE_Cov_Est2_3}
\end{equation}%
and%
\begin{equation}
	(nT^{2})^{-1}\sum_{g\in \mathcal{G}_{1}}n_{g}^{-1}\left( \sum_{i\in
		I_{g}}\sum_{t\leq T}\dot{\varepsilon}_{1,i,t}\ddot{x}_{i,t}^{\top }(\hat{%
		\theta}_{2}-\theta _{2}^{\ast })\right) ^{2}\leq \frac{||\hat{\theta}%
		_{1}-\theta _{1}^{\ast }||^{2}}{nT}\sum_{i\leq n}\sum_{t\leq T}\left\Vert 
	\dot{\varepsilon}_{1,i,t}\ddot{x}_{i,t}^{\top }\right\Vert
	^{2}=O_{p}(n^{-2}).  \label{P_TWE_Cov_Est_2_4}
\end{equation}%
Next, using the definitions of $\dot{\varepsilon}_{1,i,t}$ and $\ddot{%
	\varepsilon}_{2,i,t}$,%
\begin{eqnarray}
	\sum_{t\leq T}\dot{\varepsilon}_{1,i,t}\ddot{\varepsilon}_{2,i,t}
	&=&\sum_{t\leq T}(\varepsilon _{2,i,t}-\bar{\varepsilon}_{2,i}-\bar{%
		\varepsilon}_{2,t}+\bar{\varepsilon}_{2})(\varepsilon _{1,i,t}-\bar{%
		\varepsilon}_{1,g,t})  \notag \\
	&=&\sum_{t\leq T}\varepsilon _{1,i,t}\varepsilon _{2,i,t}-T^{-1}\sum_{t\leq
		T}\varepsilon _{1,i,t}(\bar{\varepsilon}_{2,i}+\bar{\varepsilon}_{2,t}-\bar{%
		\varepsilon}_{2})  \notag \\
	&&-\sum_{t\leq T}\varepsilon _{2,i,t}\bar{\varepsilon}_{1,g,t}+T^{-1}\sum_{t%
		\leq T}\bar{\varepsilon}_{1,g,t}(\bar{\varepsilon}_{2,i}+\bar{\varepsilon}%
	_{2,t}-\bar{\varepsilon}_{2}),  \label{P_TWE_Cov_Est_2_5}
\end{eqnarray}%
and hence%
\begin{eqnarray}
	&&(nT^{2})^{-1}\sum_{g\in \mathcal{G}_{1}}n_{g}^{-1}\left( \sum_{i\in
		I_{g}}\sum_{t\leq T}(\dot{\varepsilon}_{1,i,t}\ddot{\varepsilon}_{2,i,t}-%
	\mathbb{E}[\varepsilon _{1,i,t}\varepsilon _{2,i,t}])\right) ^{2}  \notag \\
	&\leq &K(nT^{2})^{-1}\sum_{g\in \mathcal{G}_{1}}n_{g}^{-1}\left( \sum_{i\in
		I_{g}}\sum_{t\leq T}(\varepsilon _{1,i,t}\varepsilon _{2,i,t}-\mathbb{E}%
	[\varepsilon _{1,i,t}\varepsilon _{2,i,t}])\right) ^{2}  \notag \\
	&&+K(nT^{2})^{-1}\sum_{g\in \mathcal{G}_{1}}n_{g}^{-1}\left( \sum_{i\in
		I_{g}}\sum_{t\leq T}\varepsilon _{1,i,t}(\bar{\varepsilon}_{2,i}+\bar{%
		\varepsilon}_{2,t}-\bar{\varepsilon}_{2})\right) ^{2}  \notag \\
	&&+K(nT^{2})^{-1}\sum_{g\in \mathcal{G}_{1}}n_{g}^{-1}\left( \sum_{i\in
		I_{g}}\sum_{t\leq T}\varepsilon _{2,i,t}\bar{\varepsilon}_{1,g,t}\right) ^{2}
	\notag \\
	&&+K(nT^{2})^{-1}\sum_{g\in \mathcal{G}_{1}}n_{g}^{-1}\left( \sum_{i\in
		I_{g}}\sum_{t\leq T}\bar{\varepsilon}_{1,g,t}(\bar{\varepsilon}_{2,i}+\bar{%
		\varepsilon}_{2,t}-\bar{\varepsilon}_{2})\right) ^{2}.
	\label{P_TWE_Cov_Est_2_6}
\end{eqnarray}%
By the Cauchy-Schwarz inequality,%
\begin{eqnarray}
	&&(nT^{2})^{-1}\sum_{g\in \mathcal{G}_{1}}n_{g}^{-1}\left( \sum_{i\in
		I_{g}}\sum_{t\leq T}\varepsilon _{1,i,t}(\bar{\varepsilon}_{2,i}+\bar{%
		\varepsilon}_{2,t}-\bar{\varepsilon}_{2})\right) ^{2}  \notag \\
	&=&(nT^{2})^{-1}\sum_{g\in \mathcal{G}_{1}}n_{g}^{-1}\left( \sum_{i\in
		I_{g}}\sum_{t\leq T}\varepsilon _{1,i,t}(\bar{\varepsilon}_{2,i}^{\ast }+%
	\bar{\varepsilon}_{2,t}^{\ast }-\bar{\varepsilon}_{2}^{\ast })\right) ^{2} 
	\notag \\
	&\leq &(nT)^{-1}\sum_{g\in \mathcal{G}_{1}}\sum_{i\in I_{g}}\sum_{t\leq
		T}\varepsilon _{1,i,t}^{2}(\bar{\varepsilon}_{2,i}^{\ast }+\bar{\varepsilon}%
	_{2,t}^{\ast }-\bar{\varepsilon}_{2}^{\ast })^{2}=O_{p}(n^{-1}),
	\label{P_TWE_Cov_Est_2_7}
\end{eqnarray}%
where the first equality follows from (\ref{P_TWFE_Xe_1}), and the last from
Lemma \ref{GTFE_1}(iii), (\ref{P_TWE_Var_Est_0_14}), H\"{o}lder's inequality
and Markov's inequality. Similarly, 
\begin{eqnarray}
	&&(nT^{2})^{-1}\sum_{g\in \mathcal{G}_{1}}n_{g}^{-1}\left( \sum_{i\in
		I_{g}}\sum_{t\leq T}\varepsilon _{2,i,t}\bar{\varepsilon}_{1,g,t}\right) ^{2}
	\notag \\
	&=&(nT^{2})^{-1}\sum_{g\in \mathcal{G}_{1}}n_{g}^{-1}\left( \sum_{i\in
		I_{g}}\sum_{t\leq T}\varepsilon _{2,i,t}\bar{\varepsilon}_{1,g,t}^{\ast
	}\right) ^{2}  \notag \\
	&\leq &(nT)^{-1}\sum_{g\in \mathcal{G}_{1}}\sum_{i\in I_{g}}\sum_{t\leq
		T}\varepsilon _{2,i,t}^{2}\bar{\varepsilon}_{1,g,t}^{\ast
		2}=O_{p}(G_{1}n^{-1})  \label{P_TWE_Cov_Est_2_8}
\end{eqnarray}%
and%
\begin{eqnarray}
	&&(nT^{2})^{-1}\sum_{g\in \mathcal{G}_{1}}n_{g}^{-1}\left( \sum_{i\in
		I_{g}}\sum_{t\leq T}\bar{\varepsilon}_{1,g,t}(\bar{\varepsilon}_{2,i}+\bar{%
		\varepsilon}_{2,t}-\bar{\varepsilon}_{2})\right) ^{2}  \notag \\
	&=&(nT^{2})^{-1}\sum_{g\in \mathcal{G}_{1}}n_{g}^{-1}\left( \sum_{i\in
		I_{g}}\sum_{t\leq T}\bar{\varepsilon}_{1,g,t}^{\ast }(\bar{\varepsilon}%
	_{2,i}^{\ast }+\bar{\varepsilon}_{2,t}^{\ast }-\bar{\varepsilon}_{2}^{\ast
	})\right) ^{2}  \notag \\
	&\leq &(nT)^{-1}\sum_{g\in \mathcal{G}_{1}}\sum_{i\in I_{g}}\sum_{t\leq T}%
	\bar{\varepsilon}_{1,g,t}^{\ast 2}(\bar{\varepsilon}_{2,i}^{\ast }+\bar{%
		\varepsilon}_{2,t}^{\ast }-\bar{\varepsilon}_{2}^{\ast })^{2}\leq
	O_{p}(G_{1}n^{-2}).  \label{P_TWE_Cov_Est_2_9}
\end{eqnarray}%
Finally, by Assumption \ref{A1}, Lemma \ref{TWFE_Para}(ii) and Lemma \ref%
{GTFE_1}(iii),%
\begin{equation}
	(nT^{2})^{-1}\sum_{g\in \mathcal{G}_{1}}n_{g}^{-1}\left( \sum_{i\in
		I_{g}}\sum_{t\leq T}(\varepsilon _{1,i,t}\varepsilon _{2,i,t}-\mathbb{E}%
	[\varepsilon _{1,i,t}\varepsilon _{2,i,t}])\right) ^{2}=O_{p}(G_{1}n^{-2}).
	\label{P_TWE_Cov_Est_2_10}
\end{equation}%
Combining (\ref{P_TWE_Cov_Est_2_6})-(\ref{P_TWE_Cov_Est_2_10}) yields 
\begin{equation}
	(nT^{2})^{-1}\sum_{g\in \mathcal{G}_{1}}n_{g}^{-1}\left( \sum_{i\in
		I_{g}}\sum_{t\leq T}(\dot{\varepsilon}_{1,i,t}\ddot{\varepsilon}_{2,i,t}-%
	\mathbb{E}[\varepsilon _{1,i,t}\varepsilon _{2,i,t}])\right)
	^{2}=O_{p}\left( G_{1}n^{-1}\right) .  \label{P_TWE_Cov_Est_2_11}
\end{equation}%
The claim of the lemma follows from (\ref{P_TWE_Cov_Est_2_1})-(\ref%
{P_TWE_Cov_Est_2_4}), and (\ref{P_TWE_Cov_Est_2_11}).\hfill $Q.E.D.$

\bigskip

\begin{lemma}
	\textit{\label{TWE_Var_Est_8}\ Suppose }Assumptions \textit{\ref{A1},\ }\ref%
	{A8} and \ref{A9} hold. Then $\hat{\sigma}%
	_{U,n,T}^{2}-s_{U,n,T}^{2}=O_{p}(n^{-1}(\sum_{g\in \mathcal{G}%
		_{1}}n_{g}^{-1})^{1/2})$, where%
	\begin{align*}
		s_{U,n,T}^{2}& \equiv (2nT)^{-1}\sum_{i\leq
			n}s_{2,i}^{4}+(2n)^{-1}\sum_{g\in \mathcal{G}_{1}}n_{g}^{-2}\left(
		\sum_{i\in I_{g}}s_{1,i}^{2}\right) ^{2} \\
		& +(2n^{3})^{-1}\left( \sum_{i\leq n}s_{2,i}^{2}\right)
		^{2}-n^{-2}\sum_{g\in \mathcal{G}_{1}}n_{g}^{-1}\left( \sum_{i\in
			I_{g}}s_{1,2,i}\right) ^{2}.
	\end{align*}
\end{lemma}

\noindent \textsc{Proof of Lemma \ref{TWE_Var_Est_8}}. From the definitions
of $\hat{\sigma}_{U,n,T}^{2}$ and $\sigma _{U,n,T}^{2}$, we can write%
\begin{eqnarray}
	\hat{\sigma}_{U,n,T}^{2}-s_{U,n,T}^{2} &=&(2nT)^{-1}\sum_{i\leq n}(\hat{%
		\sigma}_{2,i}^{4}-s_{2,i}^{4})+(2n)^{-1}\left( \left( n^{-1}\sum_{i\leq n}%
	\hat{\sigma}_{2,i}^{2}\right) ^{2}-\left( n^{-1}\sum_{i\leq
		n}s_{2,i}^{2}\right) ^{2}\right)  \notag \\
	&&+(2n)^{-1}\sum_{g\in \mathcal{G}_{1}}\left( \left( n_{g}^{-1}\sum_{i\in
		I_{g}}\hat{\sigma}_{1,i}^{2}\right) ^{2}-\left( n_{g}^{-1}\sum_{i\in
		I_{g}}s_{1,i}^{2}\right) ^{2}\right)  \notag \\
	&&+n^{-2}\sum_{g\in \mathcal{G}_{1}}n_{g}^{-1}\left( \left( \sum_{i\in I_{g}}%
	\hat{\sigma}_{1,2,i}\right) ^{2}-\left( \sum_{i\in I_{g}}s_{1,2,i}\right)
	^{2}\right) .  \label{P_TWE_Var_Est_8_1}
\end{eqnarray}%
From Lemma \ref{TWFE_Para}(ii), $s_{2,i}^{4}\leq K$. Hence, by Lemma \ref%
{TWFE_Est5} and the Cauchy-Schwarz inequality,%
\begin{eqnarray}
	\left\vert (nT)^{-1}\sum_{i\leq n}(\hat{\sigma}_{2,i}^{4}-s_{2,i}^{4})\right%
	\vert &\leq &2\left( (nT)^{-1}\sum_{i\leq n}(\hat{\sigma}%
	_{2,i}^{2}-s_{2,i}^{2})^{2}\right) ^{1/2}\left( (nT)^{-1}\sum_{i\leq
		n}s_{2,i}^{4}\right) ^{1/2}  \notag \\
	&&+(nT)^{-1}\sum_{i\leq n}(\hat{\sigma}_{2,i}^{2}-s_{2,i}^{2})^{2}\overset{}{%
		=}O_{p}((nT^{2})^{-1/2}).  \label{P_TWE_Var_Est_8_2}
\end{eqnarray}%
Similarly,%
\begin{eqnarray}
	&&\left\vert n^{-1}\left( \left( n^{-1}\sum_{i\leq n}\hat{\sigma}%
	_{2,i}^{2}\right) ^{2}-\left( n^{-1}\sum_{i\leq n}s_{2,i}^{2}\right)
	^{2}\right) \right\vert  \notag \\
	&=&n^{-3}\left\vert \sum_{i\leq n}(\hat{\sigma}_{2,i}^{2}-s_{2,i}^{2})\times
	\sum_{i\leq n}(\hat{\sigma}_{2,i}^{2}+s_{2,i}^{2})\right\vert  \notag \\
	&\leq &n^{-1}\left( n^{-1}\sum_{i\leq n}(\hat{\sigma}_{2,i}^{2}-s_{2,i}^{2})%
	\right) ^{2}  \notag \\
	&&+2n^{-3}\left\vert \sum_{i\leq n}(\hat{\sigma}_{2,i}^{2}-s_{2,i}^{2})%
	\times \sum_{i\leq n}s_{2,i}^{2}\right\vert \overset{}{=}O_{p}(n^{-2}).
	\label{P_TWE_Var_Est_8_3}
\end{eqnarray}%
Next, by the triangle inequality, the Cauchy-Schwarz inequality, and Lemma %
\ref{TWFE_Est5}, 
\begin{eqnarray}
	&&\left\vert n^{-1}\sum_{g\in \mathcal{G}_{1}}\left( \left(
	n_{g}^{-1}\sum_{i\in I_{g}}\hat{\sigma}_{1,i}^{2}\right) ^{2}-\left(
	n_{g}^{-1}\sum_{i\in I_{g}}s_{1,i}^{2}\right) ^{2}\right) \right\vert  \notag
	\\
	&\leq &2n^{-1}\left( \sum_{g\in \mathcal{G}_{1}}\left( n_{g}^{-1}\sum_{i\in
		I_{g}}(\hat{\sigma}_{1,i}^{2}-s_{1,i}^{2})\right) ^{2}\right) ^{1/2}\left(
	\sum_{g\in \mathcal{G}_{1}}\left( n_{g}^{-1}\sum_{i\in
		I_{g}}s_{1,i}^{2}\right) ^{2}\right) ^{1/2}  \notag \\
	&&+n^{-1}\sum_{g\in \mathcal{G}_{1}}\left( n_{g}^{-1}\sum_{i\in I_{g}}(\hat{%
		\sigma}_{1,i}^{2}-s_{1,i}^{2})\right) ^{2}\overset{}{=}O_{p}\left(
	n^{-1}\left( \sum_{g\in \mathcal{G}_{1}}n_{g}^{-1}\right) ^{1/2}\right)
	\label{P_TWE_Var_Est_8_4}
\end{eqnarray}%
Finally, by Lemma \ref{TWE_Cov_Est_2},%
\begin{eqnarray}
	&&\left\vert n^{-2}\sum_{g\in \mathcal{G}_{1}}n_{g}^{-1}\left( \left(
	\sum_{i\in I_{g}}\hat{\sigma}_{1,2,i}\right) ^{2}-\left( \sum_{i\in
		I_{g}}s_{1,2,i}\right) ^{2}\right) \right\vert  \notag \\
	&\leq &2n^{-1}\left( n^{-1}\sum_{g\in \mathcal{G}_{1}}n_{g}^{-1}\left(
	\sum_{i\in I_{g}}(\hat{\sigma}_{1,2,i}-s_{1,2,i})\right) ^{2}\right)
	^{1/2}\left( n^{-1}\sum_{g\in \mathcal{G}_{1}}n_{g}^{-1}\left( \sum_{i\in
		I_{g}}s_{1,2,i}\right) ^{2}\right) ^{1/2}  \notag \\
	&&+n^{-2}\sum_{g\in \mathcal{G}_{1}}n_{g}^{-1}\left( \sum_{i\in I_{g}}(\hat{%
		\sigma}_{1,2,i}-s_{1,2,i})\right) ^{2}\overset{}{=}%
	O_{p}(G_{1}^{1/2}n^{-3/2}).  \label{P_TWE_Var_Est_8_5}
\end{eqnarray}%
Since $\sum_{g\in \mathcal{G}_{1}}n_{g}^{-1}\geq G_{1}n^{-1}$, the claim of
the lemma follows from (\ref{P_TWE_Var_Est_8_1})-(\ref{P_TWE_Var_Est_8_5}%
).\hfill $Q.E.D.$

\subsection{Properties of Estimators in the Two Way Linear Model\label%
	{subsec:TWFX}}

Throughout this section, we let $x_{j,i,t}$, $\bar{x}_{j,i}$, $\bar{x}_{j,t}$
and $\bar{x}_{j}$ denote the $j$th entries of $x_{i,t}$, $\bar{x}_{i}$, $%
\bar{x}_{t}$ and $\bar{x}$ for $j=1,\ldots,d_{x}$\ respectively.

\begin{lemma}
	\textit{\label{TWFE_X} Under Assumptions }\ref{A1} and \ref{A8}(ii), we have:
	
	(i) $(nT)^{-1}\sum_{i\leq n}\sum_{t\leq T}\left( x_{i,t}x_{i,t}^{\top }-%
	\mathbb{E}[x_{i,t}x_{i,t}^{\top }]\right) =O_{p}((nT)^{-1/2})$;
	
	(ii) $n^{-1}\sum_{i\leq n}(\bar{x}_{i}\bar{x}_{i}^{\top }-\mathbb{E}\left[ 
	\bar{x}_{i}\right] \mathbb{E}[\bar{x}_{i}^{\top }])=O_{p}((nT)^{-1/2})$;
	
	(iii) $T^{-1}\sum_{t\leq T}(\bar{x}_{t}\bar{x}_{t}^{\top }-\mathbb{E}\left[ 
	\bar{x}_{t}\right] \mathbb{E}[\bar{x}_{t}^{\top }])=O_{p}((nT)^{-1/2})$;
	
	(iv) $\bar{x}\bar{x}^{\top}=\mathbb{E}\left[ \bar{x}\right] \mathbb{E}[\bar{x%
	}_{t}^{\top}]+O_{p}((nT)^{-1/2})$.
\end{lemma}

\noindent \textsc{Proof of Lemma \ref{TWFE_X}}. (i) Given Assumptions \ref%
{A1}(ii, iii, iv), \ref{A8}(ii) and Corollary A.2. in \cite{HallHeyde1980},
we obtain 
\begin{align}
	& \mathbb{E}\left[ \left\vert (nT)^{-1/2}\sum_{i\leq n}\sum_{t\leq T}\left(
	x_{j,i,t}x_{j^{\prime },i,t}-\mathbb{E}[x_{j,i,t}x_{j^{\prime },i,t}]\right)
	\right\vert ^{2}\right]  \notag \\
	& =n^{-1}\sum_{i\leq n}\mathbb{E}\left[ \left\vert T^{-1/2}\sum_{t\leq
		T}\left( x_{j,i,t}x_{j^{\prime },i,t}-\mathbb{E}[x_{j,i,t}x_{j^{\prime
		},i,t}]\right) \right\vert ^{2}\right]  \notag \\
	& \leq \max_{i\leq n}T^{-1}\sum_{t_{1},t_{2}\leq T}\mathbb{E}\left[
	(x_{j,i,t_{1}}x_{j^{\prime },i,t_{1}}-\mathbb{E}[x_{j,i,t_{1}}x_{j^{\prime
		},i,t_{1}}])(x_{j,i,t_{2}}x_{j^{\prime },i,t_{2}}-\mathbb{E}%
	[x_{j,i,t_{2}}x_{j^{\prime },i,t_{2}}])\right]  \notag \\
	& \leq K\max_{i\leq n,t\leq T}||x_{j,i,t}x_{j^{\prime },i,t}||_{2+\delta
		/2}^{2}\sum_{h=0}^{\infty }\left\vert \alpha _{i}(h)\right\vert ^{\delta
		/(4+\delta )}\leq K\max_{i\leq n,t\leq T}||x_{j,i,t}||_{4+\delta
	}^{2}||x_{j^{\prime },i,t}||_{4+\delta }^{2}\leq K,  \label{P_TWFE_X_0}
\end{align}%
for any $j,j^{\prime }=1,\ldots ,d_{x}$, which along with Markov's
inequality, establishes the given claim.

(ii) We begin by observing that%
\begin{align}
	n^{-1}\sum_{i\leq n}(\bar{x}_{i}\bar{x}_{i}^{\top }-\mathbb{E}\left[ \bar{x}%
	_{i}\right] \mathbb{E}[\bar{x}_{i}^{\top }])& =n^{-1}\sum_{i\leq n}(\bar{x}%
	_{i}-\mathbb{E}\left[ \bar{x}_{i}\right] )(\bar{x}_{i}^{\top }-\mathbb{E}[%
	\bar{x}_{i}^{\top }])  \notag \\
	& \text{ \ }+n^{-1}\sum_{i\leq n}\left( \mathbb{E}\left[ \bar{x}_{i}\right] (%
	\bar{x}_{i}^{\top }-\mathbb{E}[\bar{x}_{i}^{\top }])+(\bar{x}_{i}-\mathbb{E}%
	\left[ \bar{x}_{i}\right] )\mathbb{E}[\bar{x}_{i}^{\top }]\right) .
	\label{P_TWFE_X_1a}
\end{align}%
Relying on Assumptions \ref{A1}(ii, iv), \ref{A8}(ii) and Rosenthal's
inequality of strong mixing processes (e.g., (\ref{P_L0_2}) in the proof of
Lemma \ref{L0}), we can proceed:%
\begin{align}
	& \max_{i\leq n}\mathbb{E}[(\bar{x}_{j,i}-\mathbb{E}[\bar{x}_{j,i}])^{4}] 
	\notag \\
	& \leq K\max_{i\leq n}\max \left\{ \left( T^{-2}\sum_{t\leq T}\left\Vert
	x_{j,i,t}-\mathbb{E}[x_{j,i,t}]\right\Vert _{2+\delta }^{2}\right) ^{2},%
	\text{ }T^{-4}\sum_{t\leq T}\left\Vert x_{j,i,t}-\mathbb{E}%
	[x_{j,i,t}]\right\Vert _{4+\delta }^{4}\right\}  \notag \\
	& \leq K\left( T^{-2}\max_{i\leq n,t\leq T}\left\Vert x_{j,i,t}\right\Vert
	_{2+\delta }^{4}+T^{-3}\max_{i\leq n,t\leq T}\left\Vert x_{j,i,t}\right\Vert
	_{4+\delta }^{4}\right) \leq KT^{-2}.  \label{P_TWFE_X_1b}
\end{align}%
Hence by the triangle inequality and H\"{o}lder's inequality,%
\begin{align*}
	& \mathbb{E}\left[ \left\vert n^{-1}\sum_{i\leq n}(\bar{x}_{j,i}-\mathbb{E}%
	\left[ \bar{x}_{j,i}\right] )(\bar{x}_{j^{\prime }i}-\mathbb{E}[\bar{x}%
	_{j^{\prime }i}])\right\vert \right] \\
	& \leq n^{-1}\sum_{i\leq n}\mathbb{E}\left[ \left\vert (\bar{x}_{j,i}-%
	\mathbb{E}\left[ \bar{x}_{j,i}\right] )(\bar{x}_{j^{\prime }i}-\mathbb{E}[%
	\bar{x}_{j^{\prime }i}])\right\vert \right] \\
	& \leq \max_{i\leq n}(\mathbb{E}\left[ (\bar{x}_{j,i}-\mathbb{E}\left[ \bar{x%
	}_{j,i}\right] )^{2}\right] \mathbb{E}[(\bar{x}_{j^{\prime }i}-\mathbb{E}[%
	\bar{x}_{j^{\prime }i}])^{2}])^{1/2}\leq KT^{-1},
\end{align*}%
which together with Markov's inequality implies that the first term in the
RHS of (\ref{P_TWFE_X_1a}) satisfies 
\begin{equation}
	n^{-1}\sum_{i\leq n}(\bar{x}_{i}-\mathbb{E}\left[ \bar{x}_{i}\right] )(\bar{x%
	}_{i}^{\top }-\mathbb{E}[\bar{x}_{i}^{\top }])=O_{p}(T^{-1}).
	\label{P_TWFE_X_1c}
\end{equation}%
We next analyze the term in the second line of (\ref{P_TWFE_X_1a}). Since by
Assumption \ref{A8}(ii) and the Cauchy-Schwarz inequality, 
\begin{equation}
	\max_{i\leq n,t\leq T}\left( ||\mathbb{E}[\bar{x}_{i}]||^{2}+||\mathbb{E}[%
	\bar{x}_{t}]||^{2}+||\mathbb{E}[\bar{x}]||^{2}\right) \leq 3\max_{i\leq
		n,t\leq T}\mathbb{E}[||x_{i,t}||^{2}]\leq K,  \label{P_TWFE_X_1d}
\end{equation}%
we can employ Assumption \ref{A1}(iii) and (\ref{P_TWFE_X_1b}) to show that%
\begin{align*}
	\mathbb{E}\left[ \left\vert n^{-1}\sum_{i\leq n}\mathbb{E}[\bar{x}_{j,i}](%
	\bar{x}_{j^{\prime },i}-\mathbb{E}[\bar{x}_{j^{\prime },i}])\right\vert ^{2}%
	\right] & =n^{-2}\sum_{i\leq n}(\mathbb{E}[\bar{x}_{j,i}])^{2}\mathbb{E}%
	\left[ (\bar{x}_{j^{\prime },i}-\mathbb{E}[\bar{x}_{j^{\prime },i}])^{2}%
	\right] \\
	& \leq n^{-1}\max_{i\leq n}\left( (\mathbb{E}[\bar{x}_{j,i}])^{2}\mathbb{E}%
	\left[ (\bar{x}_{j^{\prime },i}-\mathbb{E}[\bar{x}_{j^{\prime },i}])^{2}%
	\right] \right) \\
	& \leq Kn^{-1}\max_{i\leq n}\mathbb{E}\left[ (\bar{x}_{j^{\prime },i}-%
	\mathbb{E}[\bar{x}_{j^{\prime },i}])^{2}\right] \leq K(nT)^{-1}.
\end{align*}%
Hence by Markov's inequality, 
\begin{equation}
	n^{-1}\sum_{i\leq n}\left( \mathbb{E}\left[ \bar{x}_{i}\right] (\bar{x}%
	_{i}^{\top }-\mathbb{E}[\bar{x}_{i}^{\top }])+(\bar{x}_{i}-\mathbb{E}\left[ 
	\bar{x}_{i}\right] )\mathbb{E}[\bar{x}_{i}^{\top }]\right)
	=O_{p}((nT)^{-1/2}).  \label{P_TWFE_X_1e}
\end{equation}%
Combining the results in (\ref{P_TWFE_X_1a}), (\ref{P_TWFE_X_1c}) and (\ref%
{P_TWFE_X_1e}),\ and applying Assumption \ref{A1}(i) establishes the lemma
claim in this part.

(iii) Let $x_{i,t}^{\ast }\equiv x_{i,t}-\mathbb{E}\left[ x_{i,t}\right] $
and $x_{j,i,t}^{\ast }$ denote the $j$th entry of $x_{i,t}^{\ast }$. By the
definition of $\bar{x}_{t}$, we can express%
\begin{equation}
	(\bar{x}_{j,t}-\mathbb{E}\left[ \bar{x}_{j,t}\right] )(\bar{x}_{j^{\prime
		},t}-\mathbb{E}[\bar{x}_{j^{\prime },t}])=n^{-2}\sum_{i\leq
		n}x_{j,i,t}^{\ast }x_{j^{\prime },i,t}^{\ast
	}+n^{-2}\sum_{i_{1}=2}^{n}\sum_{i_{2}=1}^{i_{1}-1}(x_{j,i_{1}t}^{\ast
	}x_{j^{\prime },i_{2}t}^{\ast }+x_{j^{\prime },i_{1}t}^{\ast
	}x_{j,i_{2}t}^{\ast }).  \label{P_TWFE_X_2a}
\end{equation}%
Therefore%
\begin{align}
	& T^{-1}\sum_{t\leq T}(\bar{x}_{j,t}-\mathbb{E}\left[ \bar{x}_{j,t}\right] )(%
	\bar{x}_{j^{\prime },t}-\mathbb{E}\left[ \bar{x}_{j^{\prime },t}\right] ) 
	\notag \\
	& =(n^{2}T)^{-1}\sum_{i\leq n}X_{j,i}^{\ast \top }X_{j^{\prime },i}^{\ast
	}+(n^{2}T)^{-1}\sum_{i_{1}=2}^{n}\sum_{i_{2}=1}^{i_{1}-1}(X_{j,i_{1}}^{\ast
		\top }X_{j^{\prime },i_{2}}^{\ast }+X_{j^{\prime },i_{1}}^{\ast \top
	}X_{j,i_{2}}^{\ast }),  \label{P_TWFE_X_2b}
\end{align}%
where\ $X_{j,i}^{\ast }\equiv (x_{j,i,t}^{\ast })_{t\leq T}$ for $j=1,\ldots
,d_{x}$. By Assumption \ref{A8}(ii), the triangle inequality and H\"{o}%
lder's inequality, 
\begin{align*}
	\mathbb{E}\left[ \left\vert (n^{2}T)^{-1}\sum_{i\leq n}X_{j,i}^{\ast \top
	}X_{j^{\prime },i}^{\ast }\right\vert \right] & \leq
	(n^{2}T)^{-1}\sum_{i\leq n}\mathbb{E}\left[ |X_{j,i}^{\ast \top
	}X_{j^{\prime },i}^{\ast }|\right] \\
	& \leq n^{-1}\max_{i\leq n,t\leq T}(\mathbb{E}\left[ x_{j,i,t}^{2}\right] 
	\mathbb{E}[x_{j^{\prime },i,t}^{2}])^{1/2}\leq Kn^{-1}.
\end{align*}%
Hence by Markov's inequality,%
\begin{equation}
	(n^{2}T)^{-1}\sum_{i\leq n}X_{j,i}^{\ast \top }X_{j^{\prime },i}^{\ast
	}=O_{p}(n^{-1}).  \label{P_TWFE_X_2c}
\end{equation}%
We next turn to the second term in the RHS of (\ref{P_TWFE_X_2b}). By
Assumption \ref{A1}(iii), 
\begin{align}
	& \mathbb{E}\left[ \left\vert
	\sum_{i_{1}=2}^{n}\sum_{i_{2}=1}^{i_{1}-1}X_{j,i_{1}}^{\ast \top
	}X_{j^{\prime },i_{2}}^{\ast }\right\vert ^{2}\right] =\sum_{i_{1}=2}^{n}%
	\sum_{i_{2}=1}^{i_{1}-1}\mathbb{E}[X_{j,i_{1}}^{\ast \top }\mathrm{Cov}%
	(X_{j^{\prime },i_{2}}^{\ast })X_{j,i_{1}}^{\ast }]  \notag \\
	& \leq Kn^{2}\left( \max_{i\leq n}\lambda _{\max }(\mathrm{Cov}(X_{j^{\prime
		},i}^{\ast }))\right) \left( \max_{i\leq n}\mathbb{E}\left[ \left\Vert
	X_{j,i}^{\ast }\right\Vert ^{2}\right] \right) ,  \label{P_TWFE_X_2e}
\end{align}%
where%
\begin{equation}
	\max_{j\leq d_{x}}\max_{i\leq n}\mathbb{E}\left[ \left\Vert X_{j,i}^{\ast
	}\right\Vert ^{2}\right] \leq \max_{j\leq d_{x}}\max_{i\leq n}\sum_{t\leq T}%
	\mathbb{E}[x_{j,i,t}^{2}]\leq KT  \label{P_TWFE_X_2f}
\end{equation}%
by Assumption \ref{A8}(ii). By Assumptions \ref{A1}(ii), (iv), and \ref{A8}%
(ii), and Corollary A.2. in \cite{HallHeyde1980},%
\begin{equation*}
	\max_{i\leq n}\max_{t\leq T}\sum_{t^{\prime }=1}^{T}\left\vert \mathbb{E}%
	\left[ x_{j,i,t}(x_{j,i,t^{\prime }}-\mathbb{E}[x_{j,i,t^{\prime }}])\right]
	\right\vert \leq \max_{i\leq n}\max_{t\leq T}||x_{j,i,t}||_{2+\delta
		/2}^{2}\sum_{h=0}^{\infty }\left\vert \alpha _{i}(h)\right\vert ^{\delta
		/(4+\delta )}\leq K.
\end{equation*}%
Hence by Ger\v{s}gorin's disc theorem%
\begin{equation}
	\max_{j\leq d_{x}}\max_{i\leq n}\lambda _{\max }(\mathrm{Cov}(X_{j,i}^{\ast
	}))\leq K,  \label{P_TWFE_X_2g}
\end{equation}%
which along with (\ref{P_TWFE_X_2e}) and (\ref{P_TWFE_X_2f}), and Markov's
inequality shows that for any $j,j^{\prime }=1,\ldots ,d_{x}$%
\begin{equation}
	(n^{2}T)^{-1}\sum_{i_{1}=2}^{n}\sum_{i_{2}=1}^{i_{1}-1}X_{j,i_{1}}^{\ast
		\top }X_{j^{\prime },i_{2}}^{\ast }=O_{p}((n^{2}T)^{-1/2}).
	\label{P_TWFE_X_2h}
\end{equation}%
Combining the results in (\ref{P_TWFE_X_2b}), (\ref{P_TWFE_X_2c}) and (\ref%
{P_TWFE_X_2h}) yields%
\begin{equation}
	T^{-1}\sum_{t\leq T}(\bar{x}_{j,t}-\mathbb{E}\left[ \bar{x}_{j,t}\right] )(%
	\bar{x}_{j^{\prime },t}-\mathbb{E}\left[ \bar{x}_{j^{\prime },t}\right]
	)=O_{p}(n^{-1}).  \label{P_TWFE_X_2i}
\end{equation}

Next note that%
\begin{equation}
	T^{-1}\sum_{t\leq T}\mathbb{E}\left[ \bar{x}_{j,t}\right] (\bar{x}%
	_{j^{\prime },t}-\mathbb{E}[\bar{x}_{j^{\prime },t}])=(nT)^{-1}\sum_{i\leq
		n}\sum_{t\leq T}\mathbb{E}\left[ \bar{x}_{j,t}\right] x_{j^{\prime
		},i,t}^{\ast }.  \label{P_TWFE_X_2j}
\end{equation}%
By Assumptions \ref{A1}(ii, iii, iv) and \ref{A8}(ii), (\ref{P_TWFE_X_1d})
and Corollary A.2. in \cite{HallHeyde1980}, 
\begin{align}
	\mathbb{E}\left[ \left\vert \sum_{i\leq n}\sum_{t\leq T}\mathbb{E}\left[ 
	\bar{x}_{j,t}\right] x_{j^{\prime },i,t}^{\ast }\right\vert ^{2}\right] &
	=\sum_{i\leq n}\mathbb{E}\left[ \left\vert \sum_{t\leq T}\mathbb{E}\left[ 
	\bar{x}_{j,t}\right] x_{j^{\prime },i,t}^{\ast }\right\vert ^{2}\right] 
	\notag \\
	& \leq n\max_{i\leq n}\sum_{t_{1},t_{2}=1}^{T}\mathbb{E}\left[ \bar{x}%
	_{j,t_{1}}\right] \mathbb{E}\left[ \bar{x}_{j,t_{2}}\right] \mathbb{E}%
	[x_{j^{\prime },i,t_{1}}^{\ast }x_{j^{\prime },i,t_{2}}^{\ast }]  \notag \\
	& \leq n\max_{j\leq d_{x}}\max_{t\leq T}(\mathbb{E}\left[ \bar{x}_{j,t}%
	\right] )^{2}\sum_{t_{1},t_{2}=1}^{T}\left\vert \mathbb{E}[x_{j^{\prime
		},i,t_{1}}^{\ast }x_{j^{\prime },i,t_{2}}^{\ast }]\right\vert  \notag \\
	& \leq KnT\max_{j\leq d_{x}}\max_{i\leq n,t\leq T}||x_{j,i,t}||_{2+\delta
		/2}^{2}\sum_{h=0}^{\infty }\left\vert \alpha _{i}(h)\right\vert ^{\delta
		/(4+\delta )}\leq KnT,  \label{P_TWFE_X_2l}
\end{align}%
which together with (\ref{P_TWFE_X_2j}) and Markov's inequality implies that%
\begin{equation}
	T^{-1}\sum_{t\leq T}\mathbb{E}\left[ \bar{x}_{j,t}\right] (\bar{x}%
	_{j^{\prime },t}-\mathbb{E}[\bar{x}_{j^{\prime },t}])=O_{p}((nT)^{-1/2}).
	\label{P_TWFE_X_2m}
\end{equation}%
Since 
\begin{align*}
	T^{-1}\sum_{t\leq T}(\bar{x}_{t}\bar{x}_{t}^{\top }-\mathbb{E}\left[ \bar{x}%
	_{t}\right] \mathbb{E}[\bar{x}_{t}^{\top }])& =T^{-1}\sum_{t\leq T}\mathbb{E}%
	\left[ \bar{x}_{t}\right] (\bar{x}_{t}^{\top }-\mathbb{E}[\bar{x}_{t}^{\top
	}])+T^{-1}\sum_{t\leq T}(\bar{x}_{t}-\mathbb{E}\left[ \bar{x}_{t}\right] )%
	\mathbb{E}[\bar{x}_{t}^{\top }] \\
	& +T^{-1}\sum_{t\leq T}(\bar{x}_{t}-\mathbb{E}\left[ \bar{x}_{t}\right] )(%
	\bar{x}_{t}^{\top }-\mathbb{E}[\bar{x}_{t}^{\top }]),
\end{align*}%
the claim of the lemma in this part follows from Assumption \ref{A1}(i), (%
\ref{P_TWFE_X_2i}) and (\ref{P_TWFE_X_2m}).

(iv) By the similar arguments as in the proofs of (\ref{P_TWFE_X_0})\ and\ (%
\ref{P_TWFE_X_1e}), we obtain%
\begin{equation}
	\mathrm{Var}(\bar{x}_{j})=O((nT)^{-1/2})\text{ \ \ and \ \ }\mathbb{E}[\bar {%
		x}_{j}]\left( \bar{x}_{j}-\mathbb{E}[\bar{x}_{j}]\right) =O_{p}((nT)^{-1/2})
	\label{P_TWFE_X_8}
\end{equation}
for any $j=1,\ldots,d_{x}$. Therefore, using (\ref{P_TWFE_X_8}), H\"{o}%
lder's inequality and Markov's inequality,%
\begin{align*}
	\bar{x}\bar{x}^{\top}-\mathbb{E}\left[ \bar{x}\right] \mathbb{E}[\bar {x}%
	^{\top}] & =(\bar{x}-\mathbb{E}\left[ \bar{x}\right] )(\bar {x}-\mathbb{E}[%
	\bar{x}])^{\top}+\mathbb{E}\left[ \bar{x}\right] (\bar {x}-\mathbb{E}[\bar{x}%
	])^{\top} \\
	& \text{ \ \ \ }+(\bar{x}-\mathbb{E}\left[ \bar{x}\right] )\mathbb{E}[\bar{x}%
	^{\top}]\overset{}{=}O_{p}((nT)^{-1/2}),
\end{align*}
which establishes the claim of the lemma in this part.\hfill$Q.E.D.$

\bigskip

\begin{lemma}
	\textit{\label{TWFE_Para}\ Under Assumptions }\ref{A1} and \ref{A8}(ii), we
	have:
	
	(i)\ $\left\Vert \theta_{2}^{\ast}\right\Vert +\max_{i\leq n}|\gamma
	_{2,i}^{\ast}|+\max_{t\leq T}|\gamma_{2,t}^{\ast}|\leq K$;
	
	(ii)\ $\max_{i\leq n,t\leq T}\mathbb{E}[\varepsilon _{2,i,t}^{8}]\leq K$.
\end{lemma}

\noindent \textsc{Proof of Lemma \ref{TWFE_Para}}. (i) By the definition of $%
\theta ^{\ast }$ and Assumption \ref{A8}, 
\begin{align}
	||\theta _{2}^{\ast }||^{2}& =\left( \sum_{i\leq n}\sum_{t\leq T}\mathbb{E}%
	[y_{i,t}\ddot{x}_{i,t}^{\ast \top }]\right) \left( \sum_{i\leq n}\sum_{t\leq
		T}\mathbb{E}[\ddot{x}_{i,t}^{\ast }\ddot{x}_{i,t}^{\ast \top }]\right)
	^{-2}\left( \sum_{i\leq n}\sum_{t\leq T}\mathbb{E}[\ddot{x}_{i,t}^{\ast
	}y_{i,t}]\right)  \notag \\
	& \leq \lambda _{\max }\left( \left( (nT)^{-1}\sum_{i\leq n}\sum_{t\leq T}%
	\mathbb{E}[\ddot{x}_{i,t}^{\ast }\ddot{x}_{i,t}^{\ast \top }]\right)
	^{-1}\right) (nT)^{-1}\sum_{i\leq n}\sum_{t\leq T}\mathbb{E}[y_{i,t}^{2}] 
	\notag \\
	& \leq K\max_{i\leq n,t\leq T}\mathbb{E}[y_{i,t}^{2}]\leq K.
	\label{P_TWFE_Para_4}
\end{align}%
By\ Assumption \ref{A8}(ii) and the triangle inequality, we have 
\begin{equation}
	\max_{i\leq n,t\leq T}\left( |\mathbb{E}[\bar{y}_{i}]|^{2}+|\mathbb{E}[\bar{y%
	}_{t}]|^{2}+|\mathbb{E}[\bar{y}]|^{2}\right) \leq 3\max_{i\leq n,t\leq T}%
	\mathbb{E}[y_{i,t}^{2}]\leq K.  \label{P_TWFE_Para_4c}
\end{equation}%
Combining the results in (\ref{P_TWFE_X_1d}),\ (\ref{P_TWFE_Para_4}), (\ref%
{P_TWFE_Para_4c}) yields 
\begin{equation}
	\max_{i\leq n}\left\vert \gamma _{2,i}^{\ast }\right\vert \leq \max_{i\leq
		n}\left\vert \mathbb{E}[\bar{y}_{i}]\right\vert +||\theta _{2}^{\ast
	}||\max_{i\leq n}\left\Vert \mathbb{E}[\bar{x}_{i}]\right\Vert \leq K
	\label{P_TWFE_Para_5}
\end{equation}%
and 
\begin{equation}
	\max_{t\leq T}\left\vert \gamma _{2,t}^{\ast }\right\vert \leq \max_{t\leq
		T}\left\vert \mathbb{E}[\bar{y}_{t}-\bar{y}]\right\vert +||\theta _{2}^{\ast
	}||\max_{i\leq n}\left\Vert \mathbb{E}[\bar{x}_{t}-\bar{x}]\right\Vert \leq
	K.  \label{P_TWFE_Para_6}
\end{equation}%
This completes the proof of part (i) of the lemma.

(ii) By Assumption \ref{A8}(ii), (\ref{P_TWFE_Para_4}), (\ref{P_TWFE_Para_5}%
), (\ref{P_TWFE_Para_6}), the triangle inequality, the $c_{r}$-inequality
and the Cauchy-Schwarz inequality%
\begin{equation}
	\max_{i\leq n,t\leq T}\mathbb{E}[\varepsilon _{2,i,t}^{8}]\leq K\max_{i\leq
		n,t\leq T}\mathbb{E}\left[ |y_{i,t}|^{8}+||x_{i,t}||^{8}||\theta ^{\ast
	}||^{8}+\left\vert \gamma _{1,i}^{\ast }\right\vert ^{8}+\left\vert \gamma
	_{1,t}^{\ast }\right\vert ^{8}\right] \leq K.  \label{P_TWFE_Para_7}
\end{equation}%
This completes the proof of part (ii) of the lemma.\hfill $Q.E.D.$

\bigskip

\begin{lemma}
	\textit{\label{TWFE_Xe}\ Under Assumptions }\ref{A1} and \ref{A8}, we have%
	\begin{equation}
		(nT)^{-1/2}\sum_{i\leq n}\sum_{t\leq T}\ddot{x}_{i,t}\varepsilon
		_{2,i,t}=(nT)^{-1/2}\sum_{i\leq n}\sum_{t\leq T}(\ddot{x}_{i,t}^{\ast
		}\varepsilon _{2,i,t}-\mathbb{E}[(\bar{x}_{i}+\bar{x}_{t})\varepsilon
		_{2,i,t}])+o_{p}(1)=O_{p}(1),  \label{TWFE_Xe_1}
	\end{equation}%
	where $\varepsilon _{2,i,t}\equiv y_{i,t}-x_{i,t}^{\top }\theta _{2}^{\ast
	}-\gamma _{2,i}^{\ast }-\gamma _{2,t}^{\ast }$.
\end{lemma}

\noindent \textsc{Proof of Lemma \ref{TWFE_Xe}.}\ We begin by providing some
axillary results which are useful for the proof. First by the first-order
conditions of $\theta ^{\ast }$, $\gamma _{1,i}^{\ast }$ and $\gamma
_{2,t}^{\ast }$, we obtain%
\begin{equation}
	\sum_{i\leq n}\sum_{t\leq T}\mathbb{E}[\ddot{x}_{i,t}^{\ast }\varepsilon
	_{2,i,t}]=0\text{ \ \ \ \ and \ \ \ \ }\mathbb{E}[\bar{\varepsilon}_{2,i}]=%
	\mathbb{E}[\bar{\varepsilon}_{2,t}]=0.  \label{P_TWFE_Xe_1}
\end{equation}%
Moreover in view of Lemma \ref{TWFE_Para}(i),\ Assumptions \ref{A1}(ii),
(iii), (iv) apply to $\left( x_{i,t}^{\top },\varepsilon _{2,i,t}\right) $.

We next prove (\ref{TWFE_Xe_1}).\ By the definition of $\ddot{x}_{i,t}$ and (%
\ref{P_TWFE_Xe_1}), we can express:%
\begin{equation}
	(nT)^{-1}\sum_{i\leq n}\sum_{t\leq T}\ddot{x}_{i,t}\varepsilon
	_{2,i,t}=(nT)^{-1}\sum_{i\leq n}\sum_{t\leq T}x_{i,t}\varepsilon
	_{2,i,t}-n^{-1}\sum_{i\leq n}\bar{x}_{i}\bar{\varepsilon}_{2,i}-T^{-1}%
	\sum_{t\leq T}\bar{x}_{t}\bar{\varepsilon}_{2,t}+\bar{x}\bar{\varepsilon}%
	_{2}.  \label{P_TWFE_Xe_2}
\end{equation}%
Starting with the first term in the RHS of the equality in (\ref{P_TWFE_Xe_2}%
),\ we can use similar arguments as in (\ref{P_TWFE_X_0}) to show that%
\begin{equation}
	\mathbb{E}\left[ \left\vert (nT)^{-1/2}\sum_{i\leq n}\sum_{t\leq
		T}(x_{j,i,t}\varepsilon _{2,i,t}-\mathbb{E}[x_{j,i,t}\varepsilon
	_{2,i,t}])\right\vert ^{2}\right] \leq K\max_{i\leq n,t\leq
		T}||x_{j,i,t}||_{^{4+\delta }}^{2}||\varepsilon _{2,i,t}||_{^{4+\delta
	}}^{2}\leq K,  \label{P_TWFE_Xe_3}
\end{equation}%
where the last inequality follows from Assumption \ref{A8}(ii) and Lemma \ref%
{TWFE_Para}(ii). Therefore, by (\ref{P_TWFE_Xe_1}) and Markov's inequality,%
\begin{equation}
	(nT)^{-1}\sum_{i\leq n}\sum_{t\leq T}x_{i,t}\varepsilon
	_{2,i,t}=(nT)^{-1}\sum_{i\leq n}\sum_{t\leq T}(x_{i,t}\varepsilon _{2,i,t}-%
	\mathbb{E}[x_{i,t}\varepsilon _{2,i,t}])=O_{p}((nT)^{-1/2}).
	\label{P_TWFE_Xe_4}
\end{equation}

To analyze the second term in the RHS of the equality in (\ref{P_TWFE_Xe_2}%
), we first express:%
\begin{equation}
	n^{-1}\sum_{i\leq n}\bar{x}_{i}\bar{\varepsilon}_{2,i}=n^{-1}\sum_{i\leq
		n}\left( (\bar{x}_{i}-\mathbb{E}[\bar{x}_{i}])\bar{\varepsilon}_{2,i}-%
	\mathbb{E}[\bar{x}_{i}\bar{\varepsilon}_{2,i}]\right) +n^{-1}\sum_{i\leq n}%
	\mathbb{E}[\bar{x}_{i}]\bar{\varepsilon}_{2,i}+n^{-1}\sum_{i\leq n}\mathbb{E}%
	[\bar{x}_{i}\bar{\varepsilon}_{2,i}].  \label{P_TWFE_Xe_4a}
\end{equation}%
We next study the three terms in the RHS of the equation in (\ref%
{P_TWFE_Xe_4a}). Since $\mathbb{E}[(\bar{x}_{j,i}-\mathbb{E}[\bar{x}_{j,i}])%
\bar{\varepsilon}_{2,i}]=\mathbb{E}[\bar{x}_{j,i}\bar{\varepsilon}_{2,i}]$\
in view of (\ref{P_TWFE_Xe_1}),\ by Assumption \ref{A1}(i) we obtain%
\begin{align}
	\mathbb{E}\left[ \left\vert n^{-1}\sum_{i\leq n}\left( (\bar{x}_{j,i}-%
	\mathbb{E}[\bar{x}_{j,i}])\bar{\varepsilon}_{2,i}-\mathbb{E}[\bar{x}_{j,i}%
	\bar{\varepsilon}_{2,i}]\right) \right\vert ^{2}\right] & \leq
	n^{-1}\max_{i\leq n}\mathbb{E}[((\bar{x}_{j,i}-\mathbb{E}[\bar{x}_{j,i}])%
	\bar{\varepsilon}_{2,i})^{2}]  \notag \\
	& \leq n^{-1}\max_{i\leq n}\left( \mathbb{E}[(\bar{x}_{j,i}-\mathbb{E}[\bar{x%
	}_{j,i}])^{4}]\mathbb{E}[\bar{\varepsilon}_{2,i}^{4}]\right) ^{1/2}.
	\label{P_TWFE_Xe_4b}
\end{align}%
Using similar arguments as in deriving (\ref{P_TWFE_X_1b}), we can show that 
\begin{equation}
	\max_{i\leq n}\mathbb{E}\left[ \bar{\varepsilon}_{2,i}^{4}\right]
	=\max_{i\leq n}\mathbb{E}\left[ (\bar{\varepsilon}_{2,i}-\mathbb{E}[\bar{%
		\varepsilon}_{2,i}])^{4}\right] \leq KT^{-2},  \label{P_TWFE_Xe_4c}
\end{equation}%
which together with (\ref{P_TWFE_X_1b}), (\ref{P_TWFE_Xe_4b}) and Markov's
inequality implies that%
\begin{equation}
	n^{-1}\sum_{i\leq n}\left( (\bar{x}_{j,i}-\mathbb{E}[\bar{x}_{j,i}])\bar{%
		\varepsilon}_{2,i}-\mathbb{E}[\bar{x}_{j,i}\bar{\varepsilon}_{2,i}]\right)
	=O_{p}((nT^{2})^{-1/2}).  \label{P_TWFE_Xe_4d}
\end{equation}%
For the second and the third terms in the RHS of equation in (\ref%
{P_TWFE_Xe_4a}), we can use Assumption \ref{A1}(iii), (\ref{P_TWFE_X_1d}), (%
\ref{P_TWFE_Xe_1}) and (\ref{P_TWFE_Xe_4c}) to show that%
\begin{equation}
	\mathbb{E}\left[ \left\vert n^{-1}\sum_{i\leq n}\mathbb{E}[\bar{x}_{j,i}]%
	\bar{\varepsilon}_{2,i}\right\vert ^{2}\right] =n^{-2}\sum_{i\leq n}(\mathbb{%
		E}[\bar{x}_{j,i}])^{2}\mathbb{E}[\bar{\varepsilon}_{2,i}^{2}]\leq K(nT)^{-1},
	\label{P_TWFE_Xe_4e}
\end{equation}%
and use H\"{o}lder's inequality, (\ref{P_TWFE_X_1b}), (\ref{P_TWFE_Xe_1})
and (\ref{P_TWFE_Xe_4c}) to obtain%
\begin{equation}
	\max_{i\leq n}\left\vert \mathbb{E}[\bar{x}_{j,i}\bar{\varepsilon}%
	_{2,i}]\right\vert =\max_{i\leq n}\left\vert \mathbb{E}[(\bar{x}_{j,i}-%
	\mathbb{E}[\bar{x}_{j,i}])\bar{\varepsilon}_{2,i}]\right\vert \leq
	\max_{i\leq n}\left( \mathbb{E}[(\bar{x}_{j,i}-\mathbb{E}[\bar{x}%
	_{j,i}])^{2}]\mathbb{E}[\bar{\varepsilon}_{2,i}^{2}]\right) ^{1/2}\leq
	K(nT)^{-1}.  \label{P_TWFE_Xe_4f}
\end{equation}%
Combining the results in (\ref{P_TWFE_Xe_4a}), (\ref{P_TWFE_Xe_4d})-(\ref%
{P_TWFE_Xe_4f}), and applying Markov's inequality yields%
\begin{equation}
	n^{-1}\sum_{i\leq n}\bar{x}_{i}\bar{\varepsilon}_{2,i}=n^{-1}\sum_{i\leq n}%
	\mathbb{E}[\bar{x}_{i}]\bar{\varepsilon}_{2,i}+n^{-1}\sum_{i\leq n}\mathbb{E}%
	[\bar{x}_{i}\bar{\varepsilon}%
	_{2,i}]+O_{p}((nT^{2})^{-1/2})=O_{p}((nT)^{-1/2}).  \label{P_TWFE_Xe_4g}
\end{equation}

We next turn to the third term in the RHS of the equality in (\ref%
{P_TWFE_Xe_2}). By\ (\ref{P_TWFE_Xe_1}), and the definitions of $\bar{x}_{t}$
and $\bar{\varepsilon}_{2,t}$, we can express:%
\begin{equation*}
	\bar{x}_{t}\bar{\varepsilon}_{2,t}=\bar{x}_{t}(\bar{\varepsilon}_{2,t}-%
	\mathbb{E}[\bar{\varepsilon}_{2,t}])=n^{-2}\sum_{i\leq n}x_{i,t}\varepsilon
	_{2,i,t}^{\ast
	}+n^{-2}\sum_{i_{1}=2}^{n}\sum_{i_{2}=1}^{i_{1}-1}(x_{i_{1}t}\varepsilon
	_{2,i_{2},t}^{\ast }+\varepsilon _{2,i_{1}t}^{\ast }x_{i_{2},t}),
\end{equation*}%
where\ $\varepsilon _{i,t}^{\ast }\equiv \varepsilon _{i,t}-\mathbb{E}%
[\varepsilon _{i,t}]$. Therefore,%
\begin{equation}
	n^{2}\sum_{t\leq T}(\bar{x}_{t}\bar{\varepsilon}_{2,t}-\mathbb{E}[\bar{x}_{t}%
	\bar{\varepsilon}_{2,t}])=\sum_{i\leq n}\sum_{t\leq T}(x_{i,t}\varepsilon
	_{2,i,t}^{\ast }-\mathbb{E}[x_{i,t}\varepsilon _{2,i,t}^{\ast
	}])+\sum_{i_{1}=2}^{n}\sum_{i_{2}=1}^{i_{1}-1}\sum_{t\leq
		T}(x_{i_{1},t}\varepsilon _{2,i_{2}t}^{\ast }+\varepsilon _{2,i_{1}t}^{\ast
	}x_{i_{2},t}).  \label{P_TWFE_Xe_5a}
\end{equation}%
By employing analogous arguments as demonstrated in the proof of (\ref%
{P_TWFE_X_0}), we can establish that%
\begin{equation}
	(n^{2}T)^{-1}\sum_{i\leq n}\sum_{t\leq T}(x_{j,i,t}\varepsilon
	_{2,i,t}^{\ast }-\mathbb{E}[x_{j,i,t}\varepsilon _{2,i,t}^{\ast
	}])=O_{p}((n^{3}T)^{-1/2}).  \label{P_TWFE_Xe_5b}
\end{equation}%
Let\ $e_{i}\equiv (\varepsilon _{2,i,t}^{\ast })_{t\leq T}$ and $%
X_{j,i}\equiv (x_{j,i,t})_{t\leq T}$. Then 
\begin{equation}
	\sum_{i_{1}=2}^{n}\sum_{i_{2}=1}^{i_{1}-1}\sum_{t\leq
		T}x_{j,i_{1}t}\varepsilon _{2,i_{2},t}^{\ast
	}=\sum_{i_{1}=2}^{n}\sum_{i_{2}=1}^{i_{1}-1}(X_{j,i_{1}}^{\top }-\mathbb{E}%
	[X_{j,i_{1}}^{\top }])e_{i_{2}}+\sum_{i_{1}=2}^{n}\sum_{i_{2}=1}^{i_{1}-1}%
	\mathbb{E}[X_{j,i_{1}}^{\top }]e_{i_{2}}.  \label{P_TWFE_Xe_5c}
\end{equation}%
By Assumption \ref{A1}(iii), (\ref{P_TWFE_X_2g}) and Lemma \ref{TWFE_Para}%
(ii), 
\begin{align}
	& \mathbb{E}\left[ \left\vert
	(n^{2}T)^{-1}\sum_{i_{1}=2}^{n}\sum_{i_{2}=1}^{i_{1}-1}(X_{j,i_{1}}^{\top }-%
	\mathbb{E}[X_{j,i_{1}}^{\top }])e_{i_{2}}\right\vert ^{2}\right]  \notag \\
	& =(n^{2}T)^{-2}\sum_{i_{1}=2}^{n}\sum_{i_{2}=1}^{i_{1}-1}\mathbb{E}%
	[e_{i_{2}}^{\top }\mathrm{Cov}(X_{j,i_{1}})e_{i_{2}}]  \notag \\
	& \leq (nT)^{-2}\max_{j\leq d_{x}}\max_{i\leq n}\lambda (\mathrm{Cov}%
	(X_{j,i}))\max_{i\leq n}\mathbb{E}[\left\Vert e_{i}]\right\Vert ^{2}]\leq
	K(n^{2}T)^{-1}.  \label{P_TWFE_Xe_5d}
\end{align}%
Consider the expression for the second term on the RHS of the equation in (%
\ref{P_TWFE_Xe_5c}): 
\begin{equation}
	\sum_{i_{1}=2}^{n}\sum_{i_{2}=1}^{i_{1}-1}\mathbb{E}[X_{j,i_{1}}^{\top
	}]e_{i_{2}}=\sum_{i=1}^{n-1}\left( \sum_{i^{\prime }=i+1}^{n}\mathbb{E}%
	[X_{j,i^{\prime }}^{\top }]\right) e_{i}.  \label{P_TWFE_Xe_5e}
\end{equation}%
Using arguments similar to those employed in deriving (\ref{P_TWFE_X_2g}) we
deduce 
\begin{equation}
	\max_{i\leq n}\lambda (\mathrm{Cov}(e_{i}))\leq K.  \label{P_TWFE_Xe_5f}
\end{equation}%
Therefore by Assumptions \ref{A1}(iii) and \ref{A8}(ii), (\ref{P_TWFE_Xe_5e}%
) and (\ref{P_TWFE_Xe_5f}),%
\begin{align}
	\mathbb{E}\left[ \left\vert \sum_{i_{1}=2}^{n}\sum_{i_{2}=1}^{i_{1}-1}%
	\mathbb{E}[X_{j,i_{1}}^{\top }]e_{i_{2}}\right\vert ^{2}\right] &
	=\sum_{i=1}^{n-1}\left( \sum_{i^{\prime }=i+1}^{n}\mathbb{E}[X_{j,i^{\prime
	}}^{\top }]\right) \mathrm{Cov}(e_{i})\left( \sum_{i^{\prime }=i+1}^{n}%
	\mathbb{E}[X_{j,i^{\prime }}]\right)  \notag \\
	& \leq K\sum_{i=1}^{n-1}\sum_{i_{1}^{\prime },i_{2}^{\prime }=i+1}^{n}%
	\mathbb{E}[X_{j,i_{1}^{\prime }}^{\top }]\mathbb{E}[X_{j,i_{2}^{\prime }}] 
	\notag \\
	& \leq K\sum_{i=1}^{n-1}\left( \sum_{i^{\prime }=i+1}^{n}||\mathbb{E}%
	[X_{j,i^{\prime }}]||\right) ^{2}\leq Kn^{3}T,  \label{P_TWFE_Xe_5g}
\end{align}%
where the last inequality is by $||\mathbb{E}[X_{j,i^{\prime }}]||^{2}\leq
\sum_{t\leq T}\mathbb{E}[x_{j,i^{\prime },t}^{2}]\leq KT$. Combining the
results from (\ref{P_TWFE_Xe_5c}), (\ref{P_TWFE_Xe_5d}) and (\ref%
{P_TWFE_Xe_5g}), and applying Markov's inequality\ yields: 
\begin{align}
	(n^{2}T)^{-1}\sum_{i_{1}=2}^{n}\sum_{i_{2}=1}^{i_{1}-1}\sum_{t\leq
		T}x_{j,i_{1},t}\varepsilon _{2,i_{2},t}^{\ast }&
	=(n^{2}T)^{-1}\sum_{i=1}^{n-1}\left( \sum_{i^{\prime }=i+1}^{n}\mathbb{E}%
	[X_{j,i^{\prime }}^{\top }]\right) e_{i}  \notag \\
	& \text{ \ \ \ \ }+O_{p}((n^{2}T)^{-1/2})\overset{}{=}O_{p}((nT)^{-1/2}).
	\label{P_TWFE_Xe_5h}
\end{align}%
Next consider the following expression: 
\begin{equation}
	\sum_{i_{1}=2}^{n}\sum_{i_{2}=1}^{i_{1}-1}\sum_{t\leq T}\varepsilon
	_{2,i_{1},t}^{\ast
	}x_{i_{2},t}=\sum_{i_{1}=2}^{n}\sum_{i_{2}=1}^{i_{1}-1}e_{i_{1}}^{\top
	}(X_{j,i_{2}}-\mathbb{E}[X_{j,i_{2}}])+\sum_{i_{1}=2}^{n}%
	\sum_{i_{2}=1}^{i_{1}-1}e_{i_{1}}^{\top }\mathbb{E}[X_{j,i_{2}}].
	\label{P_TWFE_Xe_5i}
\end{equation}%
Using the same arguments as in the proof of (\ref{P_TWFE_Xe_5d}), we can
establish: 
\begin{equation}
	\mathbb{E}\left[ \left\vert
	(n^{2}T)^{-1}\sum_{i_{1}=2}^{n}\sum_{i_{2}=1}^{i_{1}-1}e_{i}^{\top
	}(X_{j,i_{2}}-\mathbb{E}[X_{j,i_{2}}])\right\vert ^{2}\right] \leq
	K(n^{2}T)^{-1}.  \label{P_TWFE_Xe_5j}
\end{equation}%
By employing arguments similar to the proof of (\ref{P_TWFE_Xe_5g}), 
\begin{align}
	\mathbb{E}\left[ \left\vert
	(n^{2}T)^{-1}\sum_{i_{1}=2}^{n}\sum_{i_{2}=1}^{i_{1}-1}\mathbb{E}%
	[X_{j,i_{2}}^{\top }]e_{i_{1}}\right\vert ^{2}\right] &
	=(n^{2}T)^{-2}\sum_{i=2}^{n}\left( \sum_{i^{\prime }=1}^{i-1}\mathbb{E}%
	[X_{j,i^{\prime }}^{\top }]\right) \mathrm{Cov}(e_{i})\left( \sum_{i^{\prime
		}=1}^{i-1}\mathbb{E}[X_{j,i^{\prime }}]\right)  \notag \\
	& \leq K(n^{2}T)^{-2}\sum_{i=2}^{n}\sum_{i_{1}^{\prime },i_{2}^{\prime
		}=1}^{i-1}\mathbb{E}[X_{j,i_{1}^{\prime }}^{\top }]\mathbb{E}%
	[X_{j,i_{2}^{\prime }}]  \notag \\
	& \leq K\sum_{i=2}^{n}\left( \sum_{i^{\prime }=1}^{i-1}||\mathbb{E}%
	[X_{j,i^{\prime }}]||\right) ^{2}\leq K(nT)^{-1},  \label{P_TWFE_Xe_5k}
\end{align}%
which along with (\ref{P_TWFE_Xe_5i}), (\ref{P_TWFE_Xe_5j}) and (\ref%
{P_TWFE_Xe_5k}), and Markov's inequality implies that%
\begin{equation}
	(n^{2}T)^{-1}\sum_{i_{1}=2}^{n}\sum_{i_{2}=1}^{i_{1}-1}\sum_{t\leq
		T}\varepsilon _{2,i_{1},t}^{\ast
	}x_{i_{2},t}=(n^{2}T)^{-1}\sum_{i=2}^{n}\sum_{i^{\prime }=1}^{i-1}\mathbb{E}%
	[X_{j,i^{\prime }}^{\top }]e_{i}+O_{p}((n^{2}T)^{-1/2})=O_{p}((nT)^{-1/2}).
	\label{P_TWFE_Xe_5l}
\end{equation}%
By Assumptions \ref{A1}(iii) and \ref{A8}(ii), and (\ref{P_TWFE_Xe_5f}), 
\begin{align*}
	\mathbb{E}\left[ \left\vert (n^{2}T)^{-1}\sum_{i\leq n}\mathbb{E}%
	[X_{j,i}^{\top }]e_{i}\right\vert ^{2}\right] & =(n^{2}T)^{-2}\sum_{i\leq n}%
	\mathbb{E}[X_{j,i}^{\top }]\mathrm{Cov}(e_{i})\mathbb{E}[X_{j,i}] \\
	& \leq (n^{2}T)^{-2}\max_{i\leq n}\lambda (\mathrm{Cov}(e_{i}))\sum_{i\leq
		n}\left\Vert \mathbb{E}[X_{j,i}]\right\Vert ^{2}\leq K(n^{3}T)^{-1},
\end{align*}%
which together with Markov's inequality implies that%
\begin{equation}
	(n^{2}T)^{-1}\sum_{i\leq n}\mathbb{E}[X_{j,i}^{\top
	}]e_{i}=O_{p}((n^{3}T)^{-1/2}).  \label{P_TWFE_Xe_5m}
\end{equation}%
Combining (\ref{P_TWFE_Xe_5a}), (\ref{P_TWFE_Xe_5b}), (\ref{P_TWFE_Xe_5h}), (%
\ref{P_TWFE_Xe_5l}) and (\ref{P_TWFE_Xe_5m}) yields%
\begin{align}
	T^{-1}\sum_{t\leq T}(\bar{x}_{j,t}\bar{\varepsilon}_{2,t}-\mathbb{E}[\bar{x}%
	_{j,t}\bar{\varepsilon}_{2,t}])& =(n^{2}T)^{-1}\sum_{i=1}^{n-1}\left(
	\sum_{i^{\prime }=i+1}^{n}\mathbb{E}[X_{j,i^{\prime }}^{\top }]\right) e_{i}
	\notag \\
	& \text{ \ }+(n^{2}T)^{-1}\sum_{i=2}^{n}\sum_{i^{\prime }=1}^{i-1}\mathbb{E}%
	[X_{j,i^{\prime }}^{\top }]e_{i}+O_{p}((n^{2}T)^{-1/2})  \notag \\
	& =(n^{2}T)^{-1}\sum_{i\leq n}\sum_{i^{\prime }=1}^{n}\mathbb{E}%
	[X_{j,i^{\prime }}^{\top }]e_{i}+O_{p}((n^{2}T)^{-1/2})=O_{p}((nT)^{-1/2}).
	\label{P_TWFE_Xe_5n}
\end{align}%
By Assumptions \ref{A1}(i, iii)\ and \ref{A8}(ii), (\ref{P_TWFE_Xe_1})\ and H%
\"{o}lder's inequality,\ 
\begin{align}
	\max_{t\leq T}\left\vert \mathbb{E}[\bar{x}_{j,t}\bar{\varepsilon}%
	_{2,t}]\right\vert & =\max_{t\leq T}\left\vert \mathbb{E}[(\bar{x}_{j,t}-%
	\mathbb{E}[\bar{x}_{j,t}])\bar{\varepsilon}_{2,t}]\right\vert \leq
	\max_{i\leq n}\left( \mathbb{E}[(\bar{x}_{j,t}-\mathbb{E}[\bar{x}%
	_{j,t}])^{2}]\mathbb{E}[\bar{\varepsilon}_{2,t}^{2}]\right) ^{1/2}  \notag \\
	& \leq n^{-1}\max_{i\leq n,t\leq T}\left( \mathbb{E}[x_{j,i,t}^{2}]\right)
	^{1/2}\max_{i\leq n,t\leq T}\left( \mathbb{E}[\varepsilon _{i,t}^{2}]\right)
	^{1/2}\leq K(nT)^{-1/2}.  \label{P_TWFE_Xe_5o}
\end{align}%
Collecting the results in (\ref{P_TWFE_Xe_5m}), (\ref{P_TWFE_Xe_5n}) and (%
\ref{P_TWFE_Xe_5o}) yields 
\begin{align}
	T^{-1}\sum_{t\leq T}\bar{x}_{j,t}\bar{\varepsilon}_{2,t}& =T^{-1}\sum_{t\leq
		T}\mathbb{E}[\bar{x}_{j,t}\bar{\varepsilon}_{2,t}]+(n^{2}T)^{-1}\sum_{i\leq
		n}\sum_{i^{\prime }=1}^{n}\mathbb{E}[X_{j,i^{\prime }}^{\top
	}]e_{i}+O_{p}((n^{2}T)^{-1/2})  \notag \\
	& =T^{-1}\sum_{t\leq T}\mathbb{E}[\bar{x}_{j,t}\bar{\varepsilon}%
	_{2,t}]+T^{-1}\sum_{t\leq T}\mathbb{E}[\bar{x}_{j,t}]\bar{\varepsilon}%
	_{2,t}+O_{p}((n^{2}T)^{-1/2})=O_{p}((nT)^{-1/2}).  \label{P_TWFE_Xe_6}
\end{align}

We next turn to the last term in the RHS of the equality in (\ref%
{P_TWFE_Xe_2}). Since $\mathbb{E}[\bar{\varepsilon}_{2}]=0$ by (\ref%
{P_TWFE_Xe_1}), we can express%
\begin{align}
	(\bar{x}-\mathbb{E}[\bar{x}])\bar{\varepsilon}_{2}& =(\bar{x}-\mathbb{E}[%
	\bar{x}])(\bar{\varepsilon}_{2}-\mathbb{E}[\bar{\varepsilon}%
	_{2}])=n^{-2}\sum_{i_{1},i_{2}=1}^{n}(\bar{x}_{i_{1}}-\mathbb{E}[\bar{x}%
	_{i_{1}}])(\bar{\varepsilon}_{2,i_{2}}-\mathbb{E}[\bar{\varepsilon}%
	_{2,i_{2}}])  \notag \\
	& =n^{-2}\sum_{i\leq n}(\bar{x}_{i}-\mathbb{E}[\bar{x}_{i}])(\bar{\varepsilon%
	}_{2,i}-\mathbb{E}[\bar{\varepsilon}_{2,i}])+n^{-2}\sum_{i_{1}=2}^{n}%
	\sum_{i_{2}=1}^{i_{1}-1}(\bar{x}_{i_{1}}-\mathbb{E}[\bar{x}_{i_{1}}])(\bar{%
		\varepsilon}_{2,i_{2}}-\mathbb{E}[\bar{\varepsilon}_{2,i_{2}}])  \notag \\
	& \text{ \ \ \ }+n^{-2}\sum_{i_{1}=2}^{n}\sum_{i_{2}=1}^{i_{1}-1}(\bar{%
		\varepsilon}_{2,i_{1}}-\mathbb{E}[\bar{\varepsilon}_{2,i_{1}}])(\bar{x}%
	_{i_{2}}-\mathbb{E}[\bar{x}_{i_{2}}]).  \label{P_TWFE_Xe_14}
\end{align}%
By the triangle inequality and H\"{o}lder's inequality, we can show that%
\begin{align}
	\mathbb{E}\left[ \left\vert n^{-2}\sum_{i\leq n}(\bar{x}_{j,i}-\mathbb{E}[%
	\bar{x}_{j,i}])(\bar{\varepsilon}_{2,i}-\mathbb{E}[\bar{\varepsilon}%
	_{2,i}])\right\vert \right] & \leq n^{-2}\sum_{i\leq n}\mathbb{E}\left[
	\left\vert (\bar{x}_{j,i}-\mathbb{E}[\bar{x}_{j,i}])(\bar{\varepsilon}_{2,i}-%
	\mathbb{E}[\bar{\varepsilon}_{2,i}])\right\vert \right]  \notag \\
	& \leq n^{-1}\max_{i\leq n}(\mathbb{E}\left[ (\bar{x}_{j,i}-\mathbb{E}[\bar{x%
	}_{j,i}])^{2}\right] \mathbb{E}\left[ (\bar{\varepsilon}_{2,i}-\mathbb{E}[%
	\bar{\varepsilon}_{2,i}])^{2}\right] )^{1/2}  \notag \\
	& \leq K(nT)^{-1},  \label{P_TWFE_Xe_15}
\end{align}%
where the last inequality is by (\ref{P_TWFE_X_1b})\ and (\ref{P_TWFE_Xe_4c}%
). Similarly, we can show that%
\begin{align}
	& \mathbb{E}\left[ \left\vert
	n^{-2}\sum_{i_{1}=2}^{n}\sum_{i_{2}=1}^{i_{1}-1}(\bar{x}_{j,i_{1}}-\mathbb{E}%
	[\bar{x}_{j,i_{1}}])(\bar{\varepsilon}_{2,i_{2}}-\mathbb{E}[\bar{\varepsilon}%
	_{2,i_{2}}])\right\vert ^{2}\right]  \notag \\
	& =n^{-4}\sum_{i_{1}=2}^{n}\sum_{i_{2}=1}^{i_{1}-1}\mathbb{E}\left[
	\left\vert (\bar{x}_{j,i_{1}}-\mathbb{E}[\bar{x}_{j,i_{1}}])(\bar{\varepsilon%
	}_{2,i_{2}}-\mathbb{E}[\bar{\varepsilon}_{2,i_{2}}])\right\vert ^{2}\right] 
	\notag \\
	& \leq n^{-2}\left( \max_{i\leq n}(\mathbb{E}[\left\vert \bar{x}_{j,i}-%
	\mathbb{E}[\bar{x}_{j,i}]\right\vert ^{4})^{1/2}\max_{i\leq n}(\mathbb{E}%
	[\left\vert \bar{\varepsilon}_{2,i}-\mathbb{E}[\bar{\varepsilon}%
	_{2,i}]\right\vert ^{4}])^{1/2}\right) \leq K(nT)^{-2}  \label{P_TWFE_Xe_17}
\end{align}%
and%
\begin{equation}
	\mathbb{E}\left[ \left\vert n^{-2}\sum_{i_{1}=2}^{n}\sum_{i_{2}=1}^{i_{1}-1}(%
	\bar{\varepsilon}_{2,i_{1}}-\mathbb{E}[\bar{\varepsilon}_{2,i_{1}}])(\bar{x}%
	_{j,i_{2}}-\mathbb{E}[\bar{x}_{j,i_{2}}])\right\vert ^{2}\right] \leq
	K(nT)^{-2}.  \label{P_TWFE_Xe_18}
\end{equation}%
From (\ref{P_TWFE_Xe_14})-(\ref{P_TWFE_Xe_18}) and Markov's inequality, we
deduce that 
\begin{equation}
	(\bar{x}-\mathbb{E}[\bar{x}])\bar{\varepsilon}_{2}=O_{p}((nT)^{-1}).
	\label{P_TWFE_Xe_19}
\end{equation}%
Since $\bar{\varepsilon}_{2}=n^{-1}\sum_{i\leq n}(\bar{\varepsilon}_{2,i}-%
\mathbb{E}[\bar{\varepsilon}_{2,i}])$ by (\ref{P_TWFE_Xe_1}), using
Assumption \ref{A1}(iii) we get 
\begin{equation}
	\mathbb{E}[\left\vert \bar{\varepsilon}_{2}\right\vert
	^{2}]=n^{-2}\sum_{i\leq n}\mathbb{E}\left[ \left\vert \bar{\varepsilon}%
	_{2,i}-\mathbb{E}[\bar{\varepsilon}_{2,i}]\right\vert ^{2}\right] \leq
	n^{-1}\max_{i\leq n}\mathbb{E}\left[ \left\vert \bar{\varepsilon}_{2,i}-%
	\mathbb{E}[\bar{\varepsilon}_{2,i}]\right\vert ^{2}\right] \leq K(nT)^{-1},
	\label{P_TWFE_Xe_20}
\end{equation}%
where the second inequality is by (\ref{P_TWFE_Xe_4c}) and H\"{o}lder's
inequality. Therefore, by (\ref{P_TWFE_X_1d}), (\ref{P_TWFE_Xe_20}) and
Markov's inequality, we get%
\begin{equation*}
	\mathbb{E}[\bar{x}]\bar{\varepsilon}_{2}=O_{p}((nT)^{-1/2}).
\end{equation*}%
This, combined with (\ref{P_TWFE_Xe_19}) shows that 
\begin{equation}
	\bar{x}\bar{\varepsilon}_{2}=n^{-1}\sum_{i\leq n}\mathbb{E}[\bar{x}]\bar{%
		\varepsilon}_{2,i}+O_{p}((nT)^{-1})=O_{p}((nT)^{-1/2}).  \label{P_TWFE_Xe_21}
\end{equation}%
The claim of the lemma now follows from Assumption \ref{A1}(i), (\ref%
{P_TWFE_Xe_2}), (\ref{P_TWFE_Xe_4}), (\ref{P_TWFE_Xe_4g}), (\ref{P_TWFE_Xe_6}%
) and (\ref{P_TWFE_Xe_21}).\hfill $Q.E.D.$

\bigskip

\begin{lemma}
	\textit{\label{TWFE_Est1}\ Under Assumptions }\ref{A1} and \ref{A8}, we have:%
	\begin{equation}
		\hat{\theta}_{2}-\theta _{2}^{\ast }=\frac{\Sigma _{\ddot{x}}^{-1}}{nT}%
		\sum_{i\leq n}\sum_{t\leq T}\left( \ddot{x}_{i,t}^{\ast }\varepsilon
		_{2,i,t}-\mathbb{E}[(\bar{x}_{i}+\bar{x}_{t})\varepsilon _{2,i,t}]\right)
		+o_{p}((nT)^{-1/2})=O_{p}((nT)^{-1/2}),  \label{TWFE_Est1_1}
	\end{equation}%
	where $\varepsilon _{2,i,t}\equiv y_{i,t}-x_{i,t}^{\top }\theta _{2}^{\ast
	}-\gamma _{2,i}^{\ast }-\gamma _{2,t}^{\ast }$.
\end{lemma}

\noindent \textsc{Proof of Lemma \ref{TWFE_Est1}.} By the definitions of $%
\hat{\theta}_{2}$, $\theta _{2}^{\ast }$ and $\varepsilon _{2,i,t}$, we can
express:%
\begin{align}
	\hat{\theta}_{2}-\theta _{2}^{\ast }& =\hat{\Sigma}_{\ddot{x}%
	}^{-1}(nT)^{-1}\sum_{i\leq n}\sum_{t\leq T}\ddot{x}_{i,t}y_{i,t}-\theta
	^{\ast }=\hat{\Sigma}_{\ddot{x}}^{-1}(nT)^{-1}\sum_{i\leq n}\sum_{t\leq T}%
	\ddot{x}_{i,t}\varepsilon _{2,i,t}  \notag \\
	& =\Sigma _{\ddot{x}}^{-1}(nT)^{-1}\sum_{i\leq n}\sum_{t\leq T}\ddot{x}%
	_{i,t}\varepsilon _{2,i,t}+(\hat{\Sigma}_{\ddot{x}}^{-1}-\Sigma _{\ddot{x}%
	}^{-1})(nT)^{-1}\sum_{i\leq n}\sum_{t\leq T}\ddot{x}_{i,t}\varepsilon
	_{2,i,t}.  \label{P_TWFE_Est1_0}
\end{align}%
By the definition of $\hat{\Sigma}_{\ddot{x}}$ and Lemma \ref{TWFE_X},%
\begin{align}
	\hat{\Sigma}_{\ddot{x}}-\Sigma _{\ddot{x}}& =(nT)^{-1}\sum_{i\leq
		n}\sum_{t\leq T}(x_{i,t}x_{i,t}^{\top }-\mathbb{E}[x_{i,t}x_{i,t}^{\top
	}])-n^{-1}\sum_{i\leq n}(\bar{x}_{i}\bar{x}_{i}^{\top }-\mathbb{E}\left[ 
	\bar{x}_{i}\right] \mathbb{E}[\bar{x}_{i}^{\top }])  \notag \\
	& \text{ \ \ }-T^{-1}\sum_{t\leq T}(\bar{x}_{t}\bar{x}_{t}^{\top }-\mathbb{E}%
	\left[ \bar{x}_{t}\right] \mathbb{E}[\bar{x}_{t}^{\top }])+(\bar{x}\bar{x}%
	^{\top }-\mathbb{E}\left[ \bar{x}\right] \mathbb{E}[\bar{x}^{\top }])\overset%
	{}{=}O_{p}((nT)^{-1/2}).  \label{P_TWFE_Est1_1}
\end{align}%
This, combined with Assumption \ref{A8}(i), further implies that%
\begin{equation}
	\lambda _{\min }(\hat{\Sigma}_{\ddot{x}})\geq K^{-1},\text{ \ wpa1.}
	\label{P_TWFE_Est1_2}
\end{equation}%
Summing up the results in Lemma \ref{TWFE_Xe}, (\ref{P_TWFE_Est1_1}) and (%
\ref{P_TWFE_Est1_2}) yields%
\begin{equation*}
	\left\Vert \frac{\hat{\Sigma}_{\ddot{x}}^{-1}-\Sigma _{\ddot{x}}^{-1}}{nT}%
	\sum_{i\leq n}\sum_{t\leq T}\ddot{x}_{i,t}\varepsilon _{2,i,t}\right\Vert
	^{2}\leq \frac{||\hat{\Sigma}_{\ddot{x}}-\Sigma _{\ddot{x}}||^{2}}{\lambda
		_{\min }(\hat{\Sigma}_{\ddot{x}})\lambda _{\min }(\Sigma _{\ddot{x}})}%
	\left\Vert (nT)^{-1}\sum_{i\leq n}\sum_{t\leq T}\ddot{x}_{i,t}\varepsilon
	_{2,i,t}\right\Vert ^{2}=O_{p}((nT)^{-2}).
\end{equation*}%
This together with (\ref{P_TWFE_Est1_0}) shows that 
\begin{equation}
	\hat{\theta}-\theta ^{\ast }=\Sigma _{\ddot{x}}^{-1}(nT)^{-1}\sum_{i\leq
		n}\sum_{t\leq T}\ddot{x}_{i,t}\varepsilon _{2,i,t}+O_{p}((nT)^{-1}).
	\label{P_TWFE_Est1_3}
\end{equation}%
The claim of the lemma now follows from Assumption \ref{A8}(i), Lemma \ref%
{TWFE_Xe} and (\ref{P_TWFE_Est1_3}).\hfill $Q.E.D.$

\bigskip

\begin{lemma}
	\textit{\label{TWFE_Est2}\ Under Assumptions }\ref{A1} and \ref{A8}, we have:%
	\begin{equation*}
		(nT)^{-1}\sum_{i\leq n}\sum_{t\leq T}\frac{\varepsilon
			_{2,i,t}^{2}-(y_{i,t}-x_{i,t}^{\top }\hat{\theta}_{2}-\hat{\gamma}%
			_{2,i,t})^{2}}{2}=\frac{n^{-1}\sum_{i\leq n}(\bar{\varepsilon}_{2,i}^{\ast
			})^{2}+T^{-1}\sum_{t\leq T}(\bar{\varepsilon}_{2,t}^{\ast })^{2}}{2}%
		+O_{p}((nT)^{-1})
	\end{equation*}%
	where $\varepsilon _{2,i,t}\equiv y_{i,t}-x_{i,t}^{\top }\theta _{2}^{\ast
	}-\gamma _{2,i}^{\ast }-\gamma _{2,t}^{\ast }$, $\bar{\varepsilon}%
	_{2,i}^{\ast }\equiv T^{-1}\sum_{t\leq T}\varepsilon _{2,i,t}^{\ast }$, $%
	\bar{\varepsilon}_{2,t}^{\ast }\equiv n^{-1}\sum_{i\leq n}\varepsilon
	_{2,i,t}^{\ast }$ and $\varepsilon _{2,i,t}^{\ast }\equiv \varepsilon
	_{2,i,t}-\mathbb{E}[\varepsilon _{2,i,t}]$.
\end{lemma}

\noindent \textsc{Proof of Lemma \ref{TWFE_Est2}}. First note that $%
y_{i,t}-x_{i,t}^{\top }\hat{\theta}_{2}-\hat{\gamma}_{2,i,t}=\ddot{y}_{i,t}-%
\ddot{x}_{i,t}^{\top }\hat{\theta}_{2}=\ddot{\varepsilon}_{2,i,t}-\ddot{x}%
_{i,t}^{\top }(\hat{\theta}_{2}-\theta _{2}^{\ast })$, where $\ddot{%
	\varepsilon}_{2,i,t}\equiv \varepsilon _{2,i,t}-\bar{\varepsilon}_{2,i}-\bar{%
	\varepsilon}_{2,t}+\bar{\varepsilon}_{2}$. Therefore%
\begin{align}
	& (nT)^{-1}\sum_{i\leq n}\sum_{t\leq T}(-2)^{-1}\left[ (y_{i,t}-x_{i,t}^{%
		\top }\hat{\theta}_{2}-\hat{\gamma}_{2,i,t})^{2}-\varepsilon _{2,i,t}^{2}%
	\right]  \notag \\
	& =-(2nT)^{-1}\sum_{i\leq n}\sum_{t\leq T}(\ddot{\varepsilon}%
	_{2,i,t}^{2}-\varepsilon _{2,i,t}^{2})+(nT)^{-1}\sum_{i\leq n}\sum_{t\leq T}%
	\ddot{\varepsilon}_{2,i,t}\ddot{x}_{i,t}^{\top }(\hat{\theta}_{2}-\theta
	_{2}^{\ast })  \notag \\
	& \text{ \ \ }-(2nT)^{-1}(\hat{\theta}_{2}-\theta _{2}^{\ast })^{\top
	}\sum_{i\leq n}\sum_{t\leq T}\ddot{x}_{i,t}\ddot{x}_{i,t}^{\top }(\hat{\theta%
	}_{2}-\theta _{2}^{\ast }).  \label{P_TWFE_Est2_1}
\end{align}%
By Lemma \ref{TWFE_Xe} and Lemma \ref{TWFE_Est1},%
\begin{equation}
	(nT)^{-1}\sum_{i\leq n}\sum_{t\leq T}\ddot{\varepsilon}_{2,i,t}\ddot{x}%
	_{i,t}^{\top }(\hat{\theta}_{2}-\theta _{2}^{\ast })=(nT)^{-1}\sum_{i\leq
		n}\sum_{t\leq T}\varepsilon _{2,i,t}\ddot{x}_{i,t}^{\top }(\hat{\theta}%
	_{2}-\theta _{2}^{\ast })=O_{p}((nT)^{-1}).  \label{P_TWFE_Est2_2}
\end{equation}%
By Assumption \ref{A8}(ii) and (\ref{P_TWFE_X_1d}), $\left\Vert \Sigma _{%
	\ddot{x}}\right\Vert \leq K$ which together with (\ref{P_TWFE_Est1_1})
implies that 
\begin{equation}
	\hat{\Sigma}_{\ddot{x}}=O_{p}(1).  \label{P_TWFE_Est2_3}
\end{equation}%
This along with Lemma \ref{TWFE_Est1} implies that 
\begin{equation}
	(2nT)^{-1}(\hat{\theta}_{2}-\theta _{2}^{\ast })^{\top }\sum_{i\leq
		n}\sum_{t\leq T}\ddot{x}_{i,t}\ddot{x}_{i,t}^{\top }(\hat{\theta}_{2}-\theta
	_{2}^{\ast })=O_{p}((nT)^{-1}).  \label{P_TWFE_Est2_4}
\end{equation}%
By the definition of $\ddot{\varepsilon}_{i,t}$, we can express%
\begin{align}
	(nT)^{-1}\sum_{i\leq n}\sum_{t\leq T}\ddot{\varepsilon}_{2,i,t}^{2}&
	=(nT)^{-1}\sum_{i\leq n}\sum_{t\leq T}\ddot{\varepsilon}_{2,i,t}\varepsilon
	_{2,i,t}  \notag \\
	& =(nT)^{-1}\sum_{i\leq n}\sum_{t\leq T}\varepsilon
	_{2,i,t}^{2}-n^{-1}\sum_{i\leq n}\bar{\varepsilon}_{2,i}^{2}-T^{-1}\sum_{t%
		\leq T}\bar{\varepsilon}_{2,t}^{2}+\bar{\varepsilon}_{2}^{2}.
	\label{P_TWFE_Est2_5}
\end{align}%
Therefore by (\ref{P_TWFE_Xe_20}) and Markov's inequality%
\begin{equation}
	-(2nT)^{-1}\sum_{i\leq n}\sum_{t\leq T}(\ddot{\varepsilon}%
	_{2,i,t}^{2}-\varepsilon _{2,i,t}^{2})=(2n)^{-1}\sum_{i\leq n}\bar{%
		\varepsilon}_{2,i}^{2}+(2T)^{-1}\sum_{t\leq T}\bar{\varepsilon}%
	_{2,t}^{2}+O_{p}(nT)^{-1}.  \label{P_TWFE_Est2_6}
\end{equation}%
The claim and the lemma now follows from (\ref{P_TWFE_Xe_1}), (\ref%
{P_TWFE_Est2_1}), (\ref{P_TWFE_Est2_2}), (\ref{P_TWFE_Est2_4}) and (\ref%
{P_TWFE_Est2_6}).\hfill $Q.E.D.$

\bigskip

\begin{lemma}
	\textit{\label{TWFE_Est4}}\ Under Assumptions \ref{A1} and \ref{A8}, $%
	n^{-1}\sum_{i\leq n}(\hat{\sigma}_{2,i}^{2}-s_{2,i}^{2})=O_{p}(n^{-1})$,\
	where $s_{2,i}^{2}\equiv T^{-1}\sum_{t\leq T}\mathbb{E}[\varepsilon
	_{2,i,t}^{2}]$.
\end{lemma}

\noindent \textsc{Proof of Lemma \ref{TWFE_Est4}}. Using Lemma \ref%
{TWFE_Est2} and the definition of $\hat{\sigma}_{\varepsilon _{2},i}^{2}$,
we have%
\begin{eqnarray}
	(nT)^{-1}\sum_{i\leq n}\sum_{t\leq T}\varepsilon
	_{2,i,t}^{2}-n^{-1}\sum_{i\leq n}\hat{\sigma}_{2,i}^{2}
	&=&(nT)^{-1}\sum_{i\leq n}\sum_{t\leq T}(\varepsilon _{2,i,t}^{2}-\hat{%
		\varepsilon}_{2,i,t}^{2})  \notag \\
	&=&n^{-1}\sum_{i\leq n}(\bar{\varepsilon}_{2,i}^{\ast
	})^{2}+T^{-1}\sum_{t\leq T}(\bar{\varepsilon}_{2,t}^{\ast
	})^{2}+O_{p}((nT)^{-1}).  \label{P_TWFE_Est4_1}
\end{eqnarray}%
By Assumption \ref{A1} and Lemma \ref{TWFE_Para}(ii)%
\begin{equation}
	\mathbb{E}[\bar{\varepsilon}_{2,i}^{\ast 4}+\bar{\varepsilon}_{2,t}^{\ast
		4}]\leq Kn^{-2}\text{ \ \ and \ \ }\mathbb{E}[\bar{\varepsilon}_{2}^{\ast
		4}]\leq Kn^{-4}.  \label{P_TWFE_Est4_2}
\end{equation}%
Hence by Assumption \ref{A1}(i) and Markov's inequality, 
\begin{equation}
	n^{-1}\sum_{i\leq n}(\bar{\varepsilon}_{2,i}^{\ast })^{2}+T^{-1}\sum_{t\leq
		T}(\bar{\varepsilon}_{2,t}^{\ast })^{2}=O_{p}(n^{-1}).  \label{P_TWFE_Est4_3}
\end{equation}%
From (\ref{P_TWFE_Est4_1}) and (\ref{P_TWFE_Est4_3}), it follows that%
\begin{equation}
	n^{-1}\sum_{i\leq n}\hat{\sigma}_{2,i}^{2}=(nT)^{-1}\sum_{i\leq
		n}\sum_{t\leq T}\varepsilon _{2,i,t}^{2}+O_{p}(n^{-1}).
	\label{P_TWFE_Est4_4}
\end{equation}
Similarly, Assumption \ref{A1}, Lemma \ref{TWFE_Para}(ii), and Markov's
inequality imply%
\begin{equation}
	(nT)^{-1}\sum_{i\leq n}\sum_{t\leq T}(\varepsilon _{2,i,t}^{2}-\mathbb{E}%
	[\varepsilon _{2,i,t}^{2}])=O_{p}(n^{-1}).  \label{P_TWFE_Est4_5}
\end{equation}%
Combining (\ref{P_TWFE_Est4_4}), and (\ref{P_TWFE_Est4_2}) with the
definition of $s_{\varepsilon _{2},i}^{2}$ establishes the claim of the
lemma.\hfill $Q.E.D.$

\bigskip

\begin{lemma}
	\textit{\label{TWFE_Est5}}\ Under Assumptions \ref{A1} and \ref{A8},\ $%
	n^{-1}\sum_{i\leq n}(\hat{\sigma}_{2,i}^{2}-s_{2,i}^{2})^{2}=O_{p}(n^{-1})$.
\end{lemma}

\noindent \textsc{Proof of Lemma \ref{TWFE_Est5}}. By the definition of $%
\hat{\sigma}_{2,i}^{2}$, 
\begin{eqnarray*}
	\hat{\sigma}_{2,i}^{2} &=&T^{-1}\sum_{t\leq T}\hat{\varepsilon}%
	_{2,i,t}^{2}=T^{-1}\sum_{t\leq T}(\ddot{\varepsilon}_{2,i,t}-\ddot{x}%
	_{i,t}^{\top }(\hat{\theta}_{2}-\theta _{2}^{\ast }))^{2} \\
	&=&T^{-1}\sum_{t\leq T}\ddot{\varepsilon}_{2,i,t}^{2}-2T^{-1}\sum_{t\leq T}%
	\ddot{\varepsilon}_{2,i,t}\ddot{x}_{i,t}^{\top }(\hat{\theta}_{2}-\theta
	_{2}^{\ast })+(\hat{\theta}_{2}-\theta _{2}^{\ast })^{\top }\left(
	T^{-1}\sum_{t\leq T}\ddot{x}_{i,t}\ddot{x}_{i,t}^{\top }\right) (\hat{\theta}%
	_{2}-\theta _{2}^{\ast }).
\end{eqnarray*}%
Therefore, by the Cauchy-Schwarz inequality,%
\begin{eqnarray}
	n^{-1}\sum_{i\leq n}(\hat{\sigma}_{2,i}^{2}-s_{2,i}^{2})^{2}
	&=&n^{-1}\sum_{i\leq n}\left( T^{-1}\sum_{t\leq T}\hat{\varepsilon}%
	_{2,i,t}^{2}-s_{\varepsilon _{2},i}^{2}\right) ^{2}  \notag \\
	&\leq &Kn^{-1}\sum_{i\leq n}\left( T^{-1}\sum_{t\leq T}\ddot{\varepsilon}%
	_{2,i,t}^{2}-s_{\varepsilon _{2},i}^{2}\right) ^{2}+K(nT)^{-1}\sum_{i\leq
		n}\sum_{t\leq T}\left\Vert \ddot{\varepsilon}_{2,i,t}\ddot{x}_{i,t}^{\top
	}\right\Vert ^{2}||\hat{\theta}_{2}-\theta _{2}^{\ast }||^{2}  \notag \\
	&&+K(nT)^{-1}\sum_{i\leq n}\sum_{t\leq T}\left\Vert \ddot{x}_{i,t}\ddot{x}%
	_{i,t}^{\top }\right\Vert ^{2}||\hat{\theta}_{2}-\theta _{2}^{\ast }||^{4}.
	\label{P_TWFE_Est5_1a}
\end{eqnarray}%
From Assumptions \ref{A1} and \ref{A8}(ii), and Lemma \ref{TWFE_Para}(ii),
it follows that%
\begin{equation}
	\mathbb{E}[\left\Vert \ddot{x}_{i,t}\right\Vert ^{4}]+\mathbb{E}[\left\Vert 
	\ddot{\varepsilon}_{2,i,t}\right\Vert ^{4}]\leq K.  \label{P_TWFE_Est5_1b}
\end{equation}%
This together with Markov's inequality implies%
\begin{equation}
	(nT)^{-1}\sum_{i\leq n}\sum_{t\leq T}\left\Vert \ddot{x}_{i,t}\ddot{x}%
	_{i,t}^{\top }\right\Vert ^{2}+(nT)^{-1}\sum_{i\leq n}\sum_{t\leq
		T}\left\Vert \ddot{\varepsilon}_{2,i,t}\ddot{x}_{i,t}^{\top }\right\Vert
	^{2}=O_{p}(1).  \label{P_TWFE_Est5_1c}
\end{equation}%
Combining this with Assumption \ref{A1}(i), Lemma \ref{TWFE_Est1}, and (\ref%
{P_TWFE_Est5_1a}) yields 
\begin{equation}
	n^{-1}\sum_{i\leq n}(\hat{\sigma}_{2,i}^{2}-s_{2,i}^{2})^{2}\leq
	Kn^{-1}\sum_{i\leq n}\left( T^{-1}\sum_{t\leq T}\ddot{\varepsilon}%
	_{2,i,t}^{2}-s_{\varepsilon _{2},i}^{2}\right) ^{2}+O_{p}(n^{-2}).
	\label{P_TWFE_Est5_1}
\end{equation}%
Using (\ref{P_TWFE_Xe_1}) and the definition of $\ddot{\varepsilon}_{2,i,t}$%
, we may write%
\begin{equation*}
	\ddot{\varepsilon}_{2,i,t}=\varepsilon _{2,i,t}-\bar{\varepsilon}%
	_{2,i}^{\ast }-\bar{\varepsilon}_{2,t}^{\ast }+\bar{\varepsilon}_{2}^{\ast }.
\end{equation*}%
Hence 
\begin{eqnarray*}
	T^{-1}\sum_{t\leq T}\ddot{\varepsilon}_{2,i,t}^{2}-s_{\varepsilon
		_{2},i}^{2} &=&T^{-1}\sum_{t\leq T}(\varepsilon _{2,i,t}^{2}-\mathbb{E}%
	[\varepsilon _{2,i,t}^{2}])+T^{-1}\sum_{t\leq T}(\bar{\varepsilon}%
	_{2,i}^{\ast }+\bar{\varepsilon}_{2,t}^{\ast }-\bar{\varepsilon}_{2}^{\ast
	})^{2} \\
	&&-2T^{-1}\sum_{t\leq T}\varepsilon _{2,i,t}(\bar{\varepsilon}_{2,i}^{\ast }+%
	\bar{\varepsilon}_{2,t}^{\ast }-\bar{\varepsilon}_{2}^{\ast }).
\end{eqnarray*}%
Therefore,%
\begin{eqnarray}
	n^{-1}\sum_{i\leq n}\left( T^{-1}\sum_{t\leq T}\ddot{\varepsilon}%
	_{2,i,t}^{2}-s_{\varepsilon _{2},i}^{2}\right) ^{2} &\leq
	&Kn^{-1}\sum_{i\leq n}\left( T^{-1}\sum_{t\leq T}(\varepsilon _{2,i,t}^{2}-%
	\mathbb{E}[\varepsilon _{2,i,t}^{2}])\right) ^{2}  \notag \\
	&&+Kn^{-1}\sum_{i\leq n}\left( T^{-1}\sum_{t\leq T}(\bar{\varepsilon}%
	_{2,i}^{\ast }+\bar{\varepsilon}_{2,t}^{\ast }-\bar{\varepsilon}_{2}^{\ast
	})^{2}\right) ^{2}  \notag \\
	&&+Kn^{-1}\sum_{i\leq n}\left( T^{-1}\sum_{t\leq T}\varepsilon _{2,i,t}(\bar{%
		\varepsilon}_{2,i}^{\ast }+\bar{\varepsilon}_{2,t}^{\ast }-\bar{\varepsilon}%
	_{2}^{\ast })\right) ^{2}.  \label{P_TWFE_Est5_2}
\end{eqnarray}%
From (\ref{P_TWFE_Est4_2}) and Markov's inequality, 
\begin{equation}
	(nT)^{-1}\sum_{i\leq n}\sum_{t\leq T}(\bar{\varepsilon}_{2,i}^{\ast 4}+\bar{%
		\varepsilon}_{2,t}^{\ast 4})=O_{p}(n^{-2})\text{ \ \ and \ \ }\bar{%
		\varepsilon}_{2}^{\ast 4}=O_{p}(n^{-4}).  \label{P_TWFE_Est5_4}
\end{equation}%
Together with Assumption \ref{A1}(i) and the Cauchy--Schwarz inequality,
this implies 
\begin{equation}
	n^{-1}\sum_{i\leq n}\left( T^{-1}\sum_{t\leq T}(\bar{\varepsilon}%
	_{2,i}^{\ast }+\bar{\varepsilon}_{2,t}^{\ast }-\bar{\varepsilon}_{2}^{\ast
	})^{2}\right) ^{2}\leq K(nT)^{-1}\sum_{i\leq n}\sum_{t\leq T}(\bar{%
		\varepsilon}_{2,i}^{\ast 4}+\bar{\varepsilon}_{2,t}^{\ast 4}+\bar{\varepsilon%
	}_{2}^{\ast 4})=O_{p}(n^{-2}).  \label{P_TWFE_Est5_5}
\end{equation}%
Similarly, 
\begin{eqnarray}
	&&n^{-1}\sum_{i\leq n}\left( T^{-1}\sum_{t\leq T}\varepsilon _{2,i,t}(\bar{%
		\varepsilon}_{2,i}^{\ast }+\bar{\varepsilon}_{2,t}^{\ast }-\bar{\varepsilon}%
	_{2}^{\ast })\right) ^{2}  \notag \\
	&\leq &(nT)^{-1}\sum_{i\leq n}\sum_{t\leq T}\varepsilon _{2,i,t}^{2}(\bar{%
		\varepsilon}_{2,i}^{\ast }+\bar{\varepsilon}_{2,t}^{\ast }-\bar{\varepsilon}%
	_{2}^{\ast })^{2}  \notag \\
	&\leq &\sqrt{(nT)^{-1}\sum_{i\leq n}\sum_{t\leq T}\varepsilon _{2,i,t}^{4}}%
	\sqrt{(nT)^{-1}\sum_{i\leq n}\sum_{t\leq T}(\bar{\varepsilon}_{2,i}^{\ast }+%
		\bar{\varepsilon}_{2,t}^{\ast }-\bar{\varepsilon}_{2}^{\ast })^{4}}%
	=O_{p}(n^{-1}).  \label{P_TWFE_Est5_6}
\end{eqnarray}%
Finally, by Assumption \ref{A1} and Lemma \ref{TWFE_Para}(ii), 
\begin{equation*}
	\mathbb{E}\left[ n^{-1}\sum_{i\leq n}\left( T^{-1}\sum_{t\leq T}(\varepsilon
	_{2,i,t}^{2}-\mathbb{E}[\varepsilon _{2,i,t}^{2}])\right) ^{2}\right] \leq
	K(nT)^{-1}\sum_{i\leq n}\max_{t\leq T}\left\Vert \varepsilon
	_{2,i,t}^{2}\right\Vert _{2+\delta }^{2}\leq KT^{-1}.
\end{equation*}%
Thus, by Assumption \ref{A1}(i) and Markov's inequality 
\begin{equation}
	n^{-1}\sum_{i\leq n}\left( T^{-1}\sum_{t\leq T}(\varepsilon _{2,i,t}^{2}-%
	\mathbb{E}[\varepsilon _{2,i,t}^{2}])\right) ^{2}=O_{p}(n^{-1}).
	\label{P_TWFE_Est5_7}
\end{equation}%
The claim of the lemma follows from Assumption \ref{A1}(i),\ (\ref%
{P_TWFE_Est5_1}), (\ref{P_TWFE_Est5_2}), (\ref{P_TWFE_Est5_5})-(\ref%
{P_TWFE_Est5_7}).\hfill $Q.E.D.$

\subsection{Properties of Estimators from the Heterogeneous Time Effects
	Model\label{subsec:HTEM}}

\begin{lemma}
	\textit{\label{GTFE_1} Suppose that Assumptions }\ref{A1}, \ref{A8}(ii) and %
	\ref{A9}(i) hold. Then we have
	
	(i) $(nT)^{-1}\sum_{t\leq T}\sum_{g\in \mathcal{G}_{1}}n_{g}(\bar{x}%
	_{g,t}x_{g,t}^{\top }-\mathbb{E}[\bar{x}_{g,t}]\mathbb{E}[\bar{x}%
	_{g,t}^{\top }])=O_{p}((nT)^{-1/2});$
	
	(ii) $\left\Vert \theta_{1}^{\ast}\right\Vert +\max_{g\in\mathcal{G}%
		_{1},t\leq T}|\gamma_{1,g,t}^{\ast}|\leq K$;
	
	(iii) $\max_{i\leq n,t\leq T}\mathbb{E}[\varepsilon _{1,i,t}^{8}]\leq K$.
\end{lemma}

\noindent \textsc{Proof of Lemma \ref{GTFE_1}}. (i) By Assumptions \ref{A1}%
(iii) and \ref{A8}(ii), we can establish 
\begin{equation}
	\max_{g\in \mathcal{G}_{1},t\leq T}\left( ||\mathbb{E}[\bar{x}_{g,t}]||+|%
	\mathbb{E}[\bar{y}_{g,t}]|\right) \leq K,  \label{P_GTFE_1_1}
\end{equation}%
and 
\begin{equation}
	\max_{g\in \mathcal{G}_{1},t\leq T}\mathbb{E}[n_{g}(\bar{x}_{j,g,t}-\mathbb{E%
	}[\bar{x}_{j,g,t}])^{2}]=\max_{g\in \mathcal{G}_{1},t\leq
		T}n_{g}^{-1}\sum_{i\in I_{g}}\mathbb{E}[x_{j,i,t}^{2}]\leq K.
	\label{P_GTFE_1_2}
\end{equation}%
Consider the expression: 
\begin{align}
	\bar{x}_{g,t}x_{g,t}^{\top }-\mathbb{E}[\bar{x}_{g,t}]\mathbb{E}[\bar{x}%
	_{g,t}^{\top }]& =(\bar{x}_{g,t}-\mathbb{E}[\bar{x}_{g,t}])(\bar{x}%
	_{g,t}^{\top }-\mathbb{E}[\bar{x}_{g,t}^{\top }])  \notag \\
	& +\mathbb{E}[\bar{x}_{g,t}](\bar{x}_{g,t}^{\top }-\mathbb{E}[\bar{x}%
	_{g,t}^{\top }])+(\bar{x}_{g,t}-\mathbb{E}[\bar{x}_{g,t}])\mathbb{E}[\bar{x}%
	_{g,t}^{\top }].  \label{P_GTFE_1_3}
\end{align}%
By the triangle inequality, Holder's inequality and (\ref{P_GTFE_1_2}),%
\begin{align}
	& \mathbb{E}\left[ \left\vert (nT)^{-1}\sum_{t\leq T}\sum_{g\in \mathcal{G}%
		_{1}}n_{g}(\bar{x}_{j,g,t}-\mathbb{E}[\bar{x}_{j,g,t}])(\bar{x}_{j^{\prime
		},g,t}-\mathbb{E}[\bar{x}_{j^{\prime },g,t}])\right\vert \right]  \notag \\
	& \leq (nT)^{-1}\sum_{t\leq T}\sum_{g\in \mathcal{G}_{1}}n_{g}\mathbb{E}%
	\left[ \left\vert (\bar{x}_{j,g,t}-\mathbb{E}[\bar{x}_{j,g,t}])(\bar{x}%
	_{j^{\prime },g,t}-\mathbb{E}[\bar{x}_{j^{\prime },g,t}])\right\vert \right]
	\notag \\
	& \leq (nT)^{-1}\sum_{t\leq T}\sum_{g\in \mathcal{G}_{1}}\max_{j\leq d_{x}}%
	\mathbb{E}[n_{g}(\bar{x}_{j,g,t}-\mathbb{E}[\bar{x}_{j,g,t}])^{2}]  \notag \\
	& \leq G_{1}n^{-1}\max_{j\leq d_{x}}\max_{g\in \mathcal{G}_{1},t\leq T}%
	\mathbb{E}[n_{g}(\bar{x}_{j,g,t}-\mathbb{E}[\bar{x}_{j,g,t}])^{2}]\leq
	Kn^{-1}.  \label{P_GTFE_1_4}
\end{align}%
By the definition of $\bar{x}_{g,t}$, we can write%
\begin{equation*}
	(nT)^{-1}\sum_{t\leq T}\sum_{g\in \mathcal{G}_{1}}n_{g}\mathbb{E}[\bar{x}%
	_{g,t}](\bar{x}_{g,t}^{\top }-\mathbb{E}[\bar{x}_{g,t}^{\top
	}])=(nT)^{-1}\sum_{g\in \mathcal{G}_{1}}\sum_{i\in I_{g}}\sum_{t\leq T}%
	\mathbb{E}[\bar{x}_{g,t}](x_{i,t}^{\top }-\mathbb{E}[x_{i,t}^{\top }]).
\end{equation*}%
Hence by Assumption \ref{A1}(iii) and the Cauchy-Schwarz inequality,%
\begin{align}
	& \mathbb{E}\left[ \left\vert (nT)^{-1}\sum_{t\leq T}\sum_{g\in \mathcal{G}%
		_{1}}n_{g}\mathbb{E}[\bar{x}_{j,g,t}](\bar{x}_{j^{\prime },g,t}-\mathbb{E}[%
	\bar{x}_{j^{\prime },g,t}])\right\vert ^{2}\right]  \notag \\
	& =(nT)^{-2}\sum_{g\in \mathcal{G}_{1}}\sum_{i\in I_{g}}\mathbb{E}\left[
	\left\vert \sum_{t\leq T}\mathbb{E}[\bar{x}_{j,g,t}](x_{j^{\prime },i,t}-%
	\mathbb{E}[x_{j^{\prime },i,t}])\right\vert ^{2}\right]  \notag \\
	& \leq (n^{2}T)^{-1}\sum_{g\in \mathcal{G}_{1}}\sum_{i\in
		I_{g}}n_{g}^{-1}\sum_{i^{\prime }\in I_{g}}\mathbb{E}\left[ \left\vert
	T^{-1/2}\sum_{t\leq T}\mathbb{E}[x_{j,i^{\prime }t}](x_{j^{\prime },i,t}-%
	\mathbb{E}[x_{j^{\prime },i,t}])\right\vert ^{2}\right]  \notag \\
	& \leq (nT)^{-1}\max_{i,i^{\prime }\leq n}\mathbb{E}\left[ \left\vert
	T^{-1/2}\sum_{t\leq T}\mathbb{E}[x_{j,i^{\prime }t}](x_{j^{\prime },i,t}-%
	\mathbb{E}[x_{j^{\prime },i,t}])\right\vert ^{2}\right] .  \label{P_GTFE_1_5}
\end{align}%
By Assumptions \ref{A1}(ii, iv) and \ref{A8}(ii), and Corollary A.2. in \cite%
{HallHeyde1980}, we obtain%
\begin{equation*}
	\max_{i,i^{\prime }\leq n}\mathbb{E}\left[ \left\vert T^{-1/2}\sum_{t\leq T}%
	\mathbb{E}[x_{j,i^{\prime }t}](x_{j^{\prime },i,t}-\mathbb{E}[x_{j^{\prime
		},i,t}])\right\vert ^{2}\right] \leq K
\end{equation*}%
which along with (\ref{P_GTFE_1_5}) shows that%
\begin{equation}
	\mathbb{E}\left[ \left\vert (nT)^{-1}\sum_{t\leq T}\sum_{g\in \mathcal{G}%
		_{1}}n_{g}\mathbb{E}[\bar{x}_{j,g,t}](\bar{x}_{j^{\prime },g,t}-\mathbb{E}[%
	\bar{x}_{j^{\prime },g,t}])\right\vert ^{2}\right] \leq K(nT)^{-1}
	\label{P_GTFE_1_6}
\end{equation}%
for any $j,j^{\prime }=1,\ldots ,d_{x}$. The claim of the lemma in this part
now follows from (\ref{P_GTFE_1_3}), (\ref{P_GTFE_1_4}), (\ref{P_GTFE_1_6}),
Assumption \ref{A1}(i) and Markov's inequality.

(ii)\ By Assumptions \ref{A8}(ii) and \ref{A9}(i), we can use similar
arguments in (\ref{P_TWFE_Para_4}) to show that $\left\Vert \theta_{1}^{\ast
}\right\Vert \leq K$. This together with (\ref{P_GTFE_1_1}) further implies
that $\max_{g\in\mathcal{G}_{1},t\leq T}|\gamma_{1,g,t}^{\ast}|\leq K$.

(iii) This part can be proved using similar arguments in the proof of Lemma %
\ref{TWFE_Para}, and its proof is hence omitted.\hfill$Q.E.D.$

\bigskip

\begin{lemma}
	\textit{\label{GTFE_2} Suppose that Assumptions }\ref{A1}, \ref{A8}(ii) and %
	\ref{A9}(i) hold. Then we have%
	\begin{equation}
		(nT)^{-1/2}\sum_{i\leq n}\sum_{t\leq T}\dot{x}_{i,t}\varepsilon
		_{1,i,t}=(nT)^{-1/2}\sum_{t\leq T}\sum_{g\in \mathcal{G}_{1}}\sum_{i\in
			I_{g}}\left( \dot{x}_{i,t}^{\ast }\varepsilon _{1,i,t}-\mathbb{E}[\bar{x}%
		_{g,t}\varepsilon _{1,i,t}]\right) +O_{p}(n^{-1/2}),  \label{GTFE_2_1}
	\end{equation}%
	where%
	\begin{equation}
		(nT)^{-1/2}\sum_{t\leq T}\sum_{g\in \mathcal{G}_{1}}\sum_{i\in I_{g}}\left( 
		\dot{x}_{i,t}^{\ast }\varepsilon _{1,i,t}-\mathbb{E}[\bar{x}%
		_{g,t}\varepsilon _{1,i,t}]\right) =O_{p}(1).  \label{GTFE_2_2}
	\end{equation}
\end{lemma}

\noindent \textsc{Proof of Lemma \ref{GTFE_2}}. By the first-order
conditions of $\theta _{1}^{\ast }\ $and $\gamma _{1,g,t}^{\ast }$, we have 
\begin{equation}
	\sum_{i\leq n}\sum_{t\leq T}\mathbb{E}[\dot{x}_{i,t}^{\ast }\varepsilon
	_{1,i,t}]=0\text{ \ \ \ \ and \ \ \ \ }\mathbb{E}[\bar{\varepsilon}%
	_{1,g,t}]=0  \label{P_GTFE_2_1}
\end{equation}%
where $\bar{\varepsilon}_{1,g,t}\equiv n_{g}^{-1}\sum_{i\in
	I_{g}}\varepsilon _{1,i,t}$. Moreover in view of Lemma \ref{GTFE_1}(ii),\
Assumptions \ref{A1}(ii, iii, iv) apply to $\left( x_{i,t}^{\top
},\varepsilon _{1,i,t}\right) $.

Let $x_{i,t}^{\ast }\equiv x_{i,t}-\mathbb{E}[x_{i,t}]$, $X_{i}^{\ast
}\equiv (x_{i,t}^{\ast })_{t\leq T}$ and $e_{1,i}\equiv (\varepsilon
_{1,i,t})_{t\leq T}$. Then%
\begin{equation*}
	(\bar{x}_{g,t}-\mathbb{E}[\bar{x}_{g,t}])\bar{\varepsilon}%
	_{1,g,t}=n_{g}^{-2}\sum_{i\in I_{g}}x_{i,t}^{\ast }\varepsilon
	_{1,i,t}+n_{g}^{-2}\sum_{i=N_{g}+2}^{N_{g}+n_{g}}\sum_{i^{\prime
		}=N_{g}+1}^{i}(x_{i,t}^{\ast }\varepsilon _{1,i^{\prime }t}+x_{i^{\prime
		}t}^{\ast }\varepsilon _{1,i,t}).
\end{equation*}%
Since $\mathbb{E}[(\bar{x}_{g,t}-\mathbb{E}[\bar{x}_{g,t}])\bar{\varepsilon}%
_{1,g,t}]=\mathbb{E}[\bar{x}_{g,t}\bar{\varepsilon}_{1,g,t}]$\ by (\ref%
{P_GTFE_2_1}), and\ $\mathbb{E}[x_{i,t}^{\ast }\varepsilon _{1,i^{\prime
	}t}]=0$ and $\mathbb{E}[x_{i^{\prime }t}^{\ast }\varepsilon _{1,i,t}]=0$ for
any $i\neq i^{\prime }$, we can use the above expression to obtain 
\begin{align}
	& \sum_{t\leq T}\sum_{g\in \mathcal{G}_{1}}n_{g}\left[ (\bar{x}_{j,g,t}-%
	\mathbb{E}[\bar{x}_{j,g,t}])\bar{\varepsilon}_{1,g,t}-\mathbb{E}[\bar{x}%
	_{g,t}\bar{\varepsilon}_{1,g,t}]\right]  \notag \\
	& =\sum_{g\in \mathcal{G}_{1}}\sum_{i\in I_{g}}\frac{X_{j,i}^{\ast \top
		}e_{1,i}-\mathbb{E}[X_{j,i}^{\ast \top }e_{1,i}]}{n_{g}}+\sum_{g\in \mathcal{%
			G}_{1}}\sum_{i=N_{g}+2}^{N_{g}+n_{g}}\sum_{i^{\prime }=N_{g}+1}^{i}\frac{%
		X_{j,i}^{\ast \top }e_{1,i^{\prime }}+X_{j,i^{\prime }}^{\ast \top }e_{1,i}}{%
		n_{g}}.  \label{P_GTFE_2_2}
\end{align}%
By Assumptions \ref{A1}(ii, iii, iv) and \ref{A8}(ii), and Lemma \ref{GTFE_1}%
(iii), we can establish%
\begin{align}
	& \mathbb{E}\left[ \left\vert (nT)^{-1}\sum_{g\in \mathcal{G}_{1}}\sum_{i\in
		I_{g}}\frac{X_{j,i}^{\ast \top }e_{1,i}-\mathbb{E}[X_{j,i}^{\ast \top
		}e_{1,i}]}{n_{g}}\right\vert ^{2}\right]  \notag \\
	& =(nT)^{-2}\sum_{g\in \mathcal{G}_{1}}\sum_{i\in I_{g}}\frac{\mathbb{E}%
		\left[ \left\vert X_{j,i}^{\ast \top }e_{1,i}-\mathbb{E}[X_{j,i}^{\ast \top
		}e_{1,i}]\right\vert ^{2}\right] }{n_{g}^{2}}  \notag \\
	& \leq G_{1}(nT)^{-2}\max_{g\in \mathcal{G}_{1}}n_{g}^{-1}\max_{i\leq n}%
	\mathbb{E}\left[ \left\vert X_{j,i}^{\ast \top }e_{1,i}-\mathbb{E}%
	[X_{j,i}^{\ast \top }e_{1,i}]\right\vert ^{2}\right] \leq
	K(n^{2}T)^{-1}\max_{g\in \mathcal{G}_{1}}n_{g}^{-1}  \label{P_GTFE_2_2b}
\end{align}%
where the last inequality is by 
\begin{equation}
	\max_{i\leq n}\mathbb{E}\left[ T^{-1}\left\vert X_{j,i}^{\ast \top }e_{1,i}-%
	\mathbb{E}[X_{j,i}^{\ast \top }e_{1,i}]\right\vert ^{2}\right] \leq K
	\label{P_GTFE_2_2c}
\end{equation}%
which can be verified using Assumptions \ref{A1}(ii, iv) and \ref{A8}(ii),
and Lemma \ref{GTFE_1}(iii) and similar arguments in (\ref{P_TWFE_X_0}).
Therefore by (\ref{P_GTFE_2_2b}) and Markov's inequality, the first term in
the second line of (\ref{P_GTFE_2_2}) satisfies 
\begin{equation}
	(nT)^{-1}\sum_{g\in \mathcal{G}_{1}}\sum_{i\in I_{g}}\frac{X_{j,i}^{\ast
			\top }e_{1,i}-\mathbb{E}[X_{j,i}^{\ast \top }e_{1,i}]}{n_{g}}%
	=O_{p}((n^{2}T)^{-1/2}\max_{g\in \mathcal{G}_{1}}n_{g}^{-1/2}).
	\label{P_GTFE_2_3}
\end{equation}%
We next study the second term in the second line of (\ref{P_GTFE_2_2}). By
Assumption \ref{A1}(iii), Lemma \ref{GTFE_1}(iii) and (\ref{P_TWFE_X_2g}), 
\begin{align}
	& \mathbb{E}\left[ \left\vert (nT)^{-1}\sum_{g\in \mathcal{G}%
		_{1}}\sum_{i=N_{g}+2}^{N_{g}+n_{g}}\sum_{i^{\prime }=N_{g}+1}^{i}\frac{%
		X_{j,i}^{\ast \top }e_{1,i^{\prime }}}{n_{g}}\right\vert ^{2}\right]  \notag
	\\
	& =(nT)^{-2}\sum_{g\in \mathcal{G}_{1}}\sum_{i=N_{g}+2}^{N_{g}+n_{g}}%
	\sum_{i^{\prime }=N_{g}+1}^{i}\frac{\mathbb{E}\left[ e_{1,i^{\prime }}^{\top
		}X_{j,i}^{\ast }X_{j,i}^{\ast \top }e_{1,i^{\prime }}\right] }{n_{g}^{2}} 
	\notag \\
	& \leq \max_{j\leq d_{x}}\max_{i\leq n}\lambda _{\max }(\mathrm{Cov}%
	(X_{j,i}^{\ast })(nT)^{-2}\sum_{g\in \mathcal{G}_{1}}%
	\sum_{i=N_{g}+2}^{N_{g}+n_{g}}\sum_{i^{\prime }=N_{g}+1}^{i}\frac{\mathbb{E}%
		\left[ ||e_{1,i^{\prime }}||^{2}\right] }{n_{g}^{2}}  \notag \\
	& \leq KG_{1}(nT)^{-2}\max_{i\leq n}\mathbb{E}\left[ ||e_{1,i}||^{2}\right]
	\leq K(n^{2}T)^{-1}.  \label{P_GTFE_2_4}
\end{align}%
Therefore by Markov inequality, we have%
\begin{equation}
	(nT)^{-1}\sum_{g\in \mathcal{G}_{1}}\sum_{i=N_{g}+2}^{N_{g}+n_{g}}\sum_{i^{%
			\prime }=N_{g}+1}^{i}\frac{X_{j,i}^{\ast \top }e_{1,i^{\prime }}}{n_{g}}%
	=O_{p}((n^{2}T)^{-1/2}).  \label{P_GTFE_2_5}
\end{equation}%
Combining (\ref{P_GTFE_2_2}), (\ref{P_GTFE_2_3}) and (\ref{P_GTFE_2_5})
yields%
\begin{equation}
	(nT)^{-1/2}\sum_{t\leq T}\sum_{g\in \mathcal{G}_{1}}n_{g}\left[ (\bar{x}%
	_{j,g,t}-\mathbb{E}[\bar{x}_{j,g,t}])\bar{\varepsilon}_{1,g,t}-\mathbb{E}[%
	\bar{x}_{g,t}\bar{\varepsilon}_{1,g,t}]\right] =O_{p}(n^{-1/2}).
	\label{P_GTFE_2_6}
\end{equation}%
Since%
\begin{equation*}
	\sum_{i\leq n}\sum_{t\leq T}\dot{x}_{i,t}\varepsilon _{1,i,t}=\sum_{i\leq
		n}\sum_{t\leq T}\dot{x}_{i,t}^{\ast }\varepsilon _{1,i,t}-\sum_{t\leq
		T}\sum_{g\in \mathcal{G}_{1}}n_{g}(\bar{x}_{g,t}-\mathbb{E}[\bar{x}_{g,t}])%
	\bar{\varepsilon}_{1,g,t},
\end{equation*}%
the claim in (\ref{GTFE_2_1}) follows from (\ref{P_GTFE_2_6}).

To show (\ref{GTFE_2_2}), we note that the combination of (\ref{P_GTFE_2_2c}%
) and Markov's inequality leads to 
\begin{equation}
	(nT)^{-1/2}\sum_{i\leq n}\sum_{t\leq T}\dot{x}_{i,t}^{\ast }\varepsilon
	_{1,i,t}=O_{p}(1).  \label{P_GTFE_2_7}
\end{equation}%
Next, by Assumptions \ref{A1}(iii), \ref{A8}(ii) and Lemma \ref{GTFE_1}%
(iii), 
\begin{align*}
	\max_{t\leq T}\max_{g\in \mathcal{G}_{1}}\left\vert n_{g}\mathbb{E}[\bar{x}%
	_{j,g,t}\bar{\varepsilon}_{1,g,t}]\right\vert & =\max_{t\leq T}\max_{g\in 
		\mathcal{G}_{1}}\left\vert n_{g}\mathbb{E}[(\bar{x}_{j,g,t}-\mathbb{E}[\bar{x%
	}_{j,g,t}])\bar{\varepsilon}_{1,g,t}]\right\vert \\
	& \leq \max_{t\leq T}\max_{g\in \mathcal{G}_{1}}\left( n_{g}^{2}\mathbb{E}[(%
	\bar{x}_{j,g,t}-\mathbb{E}[\bar{x}_{j,g,t}])^{2}]\mathbb{E}[\bar{\varepsilon}%
	_{1,g,t}^{2}]\right) ^{1/2} \\
	& \leq \max_{t\leq T}\max_{g\in \mathcal{G}_{1}}\left( \mathbb{E}%
	[x_{j,i,t}^{2}]\mathbb{E}[\varepsilon _{1,i,t}^{2}]\right) ^{1/2}\leq K,
\end{align*}%
which together with Assumption \ref{A1}(i) implies that 
\begin{equation}
	\left\vert (nT)^{-1/2}\sum_{t\leq T}\sum_{g\in \mathcal{G}_{1}}n_{g}\mathbb{E%
	}[\bar{x}_{g,t}\bar{\varepsilon}_{1,g,t}]\right\vert \leq KG_{1}.
	\label{P_GTFE_2_8}
\end{equation}%
Since $G_{1}$ is fixed, the claim in (\ref{GTFE_2_2}) follows from (\ref%
{P_GTFE_2_7}) and (\ref{P_GTFE_2_8}).\hfill $Q.E.D.$

\bigskip

\begin{lemma}
	\textit{\label{GTFE_Est1}\ Suppose that Assumptions }\ref{A1}, \ref{A8}(ii)
	and \ref{A9}(i) hold. Then we have 
	\begin{equation*}
		(\hat{\theta}_{1}-\theta _{1}^{\ast })=\frac{\Sigma _{\dot{x}}^{-1}}{nT}%
		\sum_{t\leq T}\sum_{g\in \mathcal{G}_{1}}\sum_{i\in I_{g}}\left( \dot{x}%
		_{i,t}^{\ast }\varepsilon _{1,i,t}-\mathbb{E}[\bar{x}_{g,t}\varepsilon
		_{1,i,t}]\right) +o_{p}((nT)^{-1/2})=O_{p}((nT)^{-1/2}).
	\end{equation*}
\end{lemma}

\noindent \textsc{Proof of Lemma \ref{GTFE_Est1}}. By the definition of $%
\dot{x}_{i,t}$, we obtain 
\begin{equation*}
	(nT)^{-1}\sum_{i\leq n}\sum_{t\leq T}\dot{x}_{i,t}\dot{x}_{i,t}^{\top
	}=(nT)^{-1}\sum_{i\leq n}\sum_{t\leq T}x_{i,t}x_{i,t}^{\top
	}-(nT)^{-1}\sum_{t\leq T}\sum_{g\in \mathcal{G}_{1}}n_{g}\bar{x}_{g,t}\bar{x}%
	_{i,t}^{\top }.
\end{equation*}%
Hence by Assumption \ref{A9}(i), Lemma \ref{TWFE_X}(i) and Lemma \ref{GTFE_1}%
(i),%
\begin{equation}
	\hat{\Sigma}_{\dot{x}}-\Sigma _{\dot{x}}=O_{p}((nT)^{-1/2})=o_{p}(1)
	\label{P_GTFE_Est1_1}
\end{equation}%
which together with Assumption \ref{A9}(i) implies that 
\begin{equation}
	\lambda _{\min }(\hat{\Sigma}_{\dot{x}})\geq K^{-1},\text{ \ wpa1.}
	\label{P_GTFE_Est1_2}
\end{equation}%
The rest of the proof follows similar arguments in the proof of Lemma \ref%
{TWFE_Est1} and hence is omitted.\hfill $Q.E.D.$

\bigskip

\begin{lemma}
	\textit{\label{GTFE_Est2}\ Suppose that Assumptions }\ref{A1}, \ref{A8}(ii)
	and \ref{A9}(i) hold. Then we have%
	\begin{equation*}
		(nT)^{-1}\sum_{i\leq n}\sum_{t\leq T}\frac{\varepsilon
			_{1,i,t}^{2}-(y_{i,t}-x_{i,t}^{\top }\hat{\theta}_{1}-\hat{\gamma}%
			_{1,i,t})^{2}}{2}=(nT)^{-1}\sum_{t\leq T}\sum_{g\in \mathcal{G}_{1}}\frac{%
			n_{g}\bar{\varepsilon}_{1,g,t}^{\ast 2}}{2}+O_{p}((nT)^{-1})
	\end{equation*}%
	where $\bar{\varepsilon}_{1,g,t}^{\ast }\equiv n_{g}^{-1}\sum_{i\in
		I_{g}}\varepsilon _{1,i,t}^{\ast }$.
\end{lemma}

\noindent \textsc{Proof of Lemma \ref{GTFE_Est2}}. First, observe that $%
y_{i,t}-x_{i,t}^{\top }\hat{\theta}_{1}-\hat{\gamma}_{1,i,t}=\dot{\varepsilon%
}_{1,i,t}-\dot{x}_{i,t}^{\top }(\hat{\theta}_{1}-\theta _{1}^{\ast })$,
where $\dot{\varepsilon}_{1,i,t}\equiv \varepsilon _{1,i,t}-\bar{\varepsilon}%
_{1,g,t}$. Therefore, 
\begin{align}
	& (nT)^{-1}\sum_{i\leq n}\sum_{t\leq T}(-2)^{-1}\left[ (y_{i,t}-x_{i,t}^{%
		\top }\hat{\theta}_{1}-\hat{\gamma}_{1,i,t})^{2}-\varepsilon _{1,i,t}^{2}%
	\right]  \notag \\
	& =-(2nT)^{-1}\sum_{i\leq n}\sum_{t\leq T}(\dot{\varepsilon}%
	_{1,i,t}^{2}-\varepsilon _{1,i,t}^{2})+(nT)^{-1}\sum_{i\leq n}\sum_{t\leq T}%
	\dot{\varepsilon}_{1,i,t}\dot{x}_{i,t}^{\top }(\hat{\theta}_{1}-\theta
	_{1}^{\ast })  \notag \\
	& \text{ \ \ }-(2nT)^{-1}(\hat{\theta}_{1}-\theta _{1}^{\ast })^{\top
	}\sum_{i\leq n}\sum_{t\leq T}\dot{x}_{i,t}\dot{x}_{i,t}^{\top }(\hat{\theta}%
	_{1}-\theta _{1}^{\ast }).  \label{P_GTFE_Est2_1}
\end{align}%
Using Lemma \ref{GTFE_2} and Lemma \ref{GTFE_Est1}, and similar arguments as
those used to derive (\ref{P_TWFE_Est2_2}) and (\ref{P_TWFE_Est2_4}), we can
show that 
\begin{equation}
	(nT)^{-1}\sum_{i\leq n}\sum_{t\leq T}\dot{\varepsilon}_{1,i,t}\dot{x}%
	_{i,t}^{\top }(\hat{\theta}_{1}-\theta _{1}^{\ast })=(nT)^{-1}\sum_{i\leq
		n}\sum_{t\leq T}\varepsilon _{1,i,t}\dot{x}_{i,t}^{\top }(\hat{\theta}%
	_{1}-\theta _{1}^{\ast })=O_{p}((nT)^{-1})  \label{P_GTFE_Est2_2}
\end{equation}%
and 
\begin{equation}
	(2nT)^{-1}(\hat{\theta}_{1}-\theta _{1}^{\ast })^{\top }\sum_{i\leq
		n}\sum_{t\leq T}\dot{x}_{i,t}\dot{x}_{i,t}^{\top }(\hat{\theta}_{1}-\theta
	_{1}^{\ast })=O_{p}((nT)^{-1}).  \label{P_GTFE_Est2_3}
\end{equation}%
Combining (\ref{P_GTFE_Est2_1})-(\ref{P_GTFE_Est2_3}) yields 
\begin{align*}
	& (nT)^{-1}\sum_{i\leq n}\sum_{t\leq T}(-2)^{-1}\left[ (y_{i,t}-x_{i,t}^{%
		\top }\hat{\theta}_{1}-\hat{\gamma}_{1,i,t})^{2}-\varepsilon _{1,i,t}^{2}%
	\right] \\
	& =(nT)^{-1}\sum_{i\leq n}\sum_{t\leq T}(-2)^{-1}(\dot{\varepsilon}%
	_{1,i,t}\varepsilon _{1,i,t}-\varepsilon _{1,i,t}^{2})+O_{p}((nT)^{-1}) \\
	& =(nT)^{-1}\sum_{t\leq T}\sum_{g\in \mathcal{G}_{1}}\frac{n_{g}\bar{%
			\varepsilon}_{1,g,t}^{2}}{2}+O_{p}((nT)^{-1}),
\end{align*}%
which together with (\ref{P_GTFE_2_1}) shows the claim of the lemma.\hfill $%
Q.E.D.$

\bigskip

\begin{lemma}
	\textit{\label{GTFE_Est3}}\ Under Assumptions \ref{A1}, \ref{A8}(ii) and \ref%
	{A9}(i),\ $\sum_{g\in \mathcal{G}_{1}}n_{g}^{-1}\sum_{i\in I_{g}}(\hat{\sigma%
	}_{1,i}^{2}-s_{1,i}^{2})=O_{p}(\sum_{g\in \mathcal{G}_{1}}n_{g}^{-1})$,
	where $s_{1,i}^{2}\equiv T^{-1}\sum_{t\leq T}\mathbb{E}[\varepsilon
	_{1,i,t}^{2}]$.
\end{lemma}

\noindent \textsc{Proof of Lemma \ref{GTFE_Est3}}. Using the definition of $%
\hat{\sigma}_{\varepsilon _{1},i}^{2}$, we have%
\begin{eqnarray}
	\hat{\sigma}_{1,i}^{2}-T^{-1}\sum_{t\leq T}\varepsilon _{1,i,t}^{2}
	&=&T^{-1}\sum_{t\leq T}(\hat{\varepsilon}_{1,i,t}^{2}-\varepsilon
	_{1,i,t}^{2})  \notag \\
	&=&T^{-1}\sum_{t\leq T}(\dot{\varepsilon}_{1,i,t}^{2}-\varepsilon
	_{1,i,t}^{2})-2T^{-1}\sum_{t\leq T}\dot{\varepsilon}_{1,i,t}\dot{x}%
	_{i,t}^{\top }(\hat{\theta}_{1}-\theta _{1}^{\ast })  \notag \\
	&&+(\hat{\theta}_{1}-\theta _{1}^{\ast })^{\top }\left( T^{-1}\sum_{t\leq T}%
	\dot{x}_{i,t}\dot{x}_{i,t}^{\top }\right) (\hat{\theta}_{1}-\theta
	_{1}^{\ast }).  \label{P_GTFE_Est3_0}
\end{eqnarray}%
From this expression, we may write 
\begin{eqnarray}
	\sum_{g\in \mathcal{G}_{1}}n_{g}^{-1}\sum_{i\in I_{g}}\left( \hat{\sigma}%
	_{1,i}^{2}-T^{-1}\sum_{t\leq T}\varepsilon _{1,i,t}^{2}\right)
	&=&T^{-1}\sum_{g\in \mathcal{G}_{1}}n_{g}^{-1}\sum_{i\in I_{g}}\sum_{t\leq
		T}(\hat{\varepsilon}_{1,i,t}^{2}-\varepsilon _{1,i,t}^{2})  \notag \\
	&=&T^{-1}\sum_{g\in \mathcal{G}_{1}}n_{g}^{-1}\sum_{i\in I_{g}}\sum_{t\leq
		T}(\dot{\varepsilon}_{1,i,t}^{2}-\varepsilon _{1,i,t}^{2})  \notag \\
	&&-2T^{-1}\sum_{g\in \mathcal{G}_{1}}n_{g}^{-1}\sum_{i\in I_{g}}\sum_{t\leq
		T}\dot{\varepsilon}_{1,i,t}\dot{x}_{i,t}^{\top }(\hat{\theta}_{1}-\theta
	_{1}^{\ast })  \notag \\
	&&+T^{-1}(\hat{\theta}_{1}-\theta _{1}^{\ast })^{\top }\sum_{g\in \mathcal{G}%
		_{1}}n_{g}^{-1}\sum_{i\in I_{g}}\sum_{t\leq T}\dot{x}_{i,t}\dot{x}%
	_{i,t}^{\top }(\hat{\theta}_{1}-\theta _{1}^{\ast }).  \label{P_GTFE_Est3_1}
\end{eqnarray}%
By Assumptions \ref{A1} and \ref{A8}(ii) and Lemma \ref{GTFE_1}(iii)%
\begin{equation}
	\mathbb{E}[||\dot{x}_{i,t}||^{4}]\leq K\text{ \ \ \ \ \ \ \ and \ \ \ \ \ \
		\ }\mathbb{E}[||\dot{\varepsilon}_{1,i,t}||^{4}]\leq K.
	\label{P_GTFE_Est3_2}
\end{equation}%
Together with Lemma \ref{GTFE_Est1}, the triangle inequality, the
Cauchy-Schwarz inequality and Markov's inequality, this implies that the
quadratic term in (\ref{P_GTFE_Est3_1}) satisfies%
\begin{eqnarray}
	&&\left\vert (\hat{\theta}_{1}-\theta _{1}^{\ast })^{\top }\left(
	T^{-1}\sum_{g\in \mathcal{G}_{1}}n_{g}^{-1}\sum_{i\in I_{g}}\sum_{t\leq T}%
	\dot{x}_{i,t}\dot{x}_{i,t}^{\top }\right) (\hat{\theta}_{1}-\theta
	_{1}^{\ast })\right\vert  \notag \\
	&\leq &||\hat{\theta}_{1}-\theta _{1}^{\ast }||^{2}\times T^{-1}\sum_{g\in 
		\mathcal{G}_{1}}n_{g}^{-1}\sum_{i\in I_{g}}\sum_{t\leq T}||\dot{x}%
	_{i,t}||^{2}=O_{p}(G_{1}(nT)^{-1}).  \label{P_GTFE_Est3_3}
\end{eqnarray}%
Similarly,%
\begin{equation}
	\left\vert T^{-1}\sum_{g\in \mathcal{G}_{1}}n_{g}^{-1}\sum_{i\in
		I_{g}}\sum_{t\leq T}\dot{\varepsilon}_{1,i,t}\dot{x}_{i,t}^{\top }(\hat{%
		\theta}_{1}-\theta _{1}^{\ast })\right\vert \leq ||\hat{\theta}_{1}-\theta
	_{1}^{\ast }||\times T^{-1}\sum_{g\in \mathcal{G}_{1}}n_{g}^{-1}\sum_{i\in
		I_{g}}\sum_{t\leq T}||\dot{\varepsilon}_{1,i,t}\dot{x}_{i,t}^{\top
	}||=O_{p}(G_{1}n^{-1}).  \label{P_GTFE_Est3_4}
\end{equation}%
Next, since $\dot{\varepsilon}_{1,i,t}=\varepsilon _{1,i,t}-\bar{\varepsilon}%
_{1,g,t}$, we can decompose%
\begin{equation}
	T^{-1}\sum_{g\in \mathcal{G}_{1}}n_{g}^{-1}\sum_{i\in I_{g}}\sum_{t\leq T}%
	\dot{\varepsilon}_{1,i,t}^{2}=T^{-1}\sum_{g\in \mathcal{G}%
		_{1}}n_{g}^{-1}\sum_{i\in I_{g}}\sum_{t\leq T}\varepsilon
	_{1,i,t}^{2}-T^{-1}\sum_{g\in \mathcal{G}_{1}}\sum_{t\leq T}\bar{\varepsilon}%
	_{1,g,t}^{2}.  \label{P_GTFE_Est3_5}
\end{equation}%
By Assumptions \ref{A1}, Lemma \ref{GTFE_1}(iii), and (\ref{P_GTFE_2_1}),%
\begin{equation}
	\mathbb{E}[\bar{\varepsilon}_{1,g,t}^{4}]=\mathbb{E}[\bar{\varepsilon}%
	_{1,g,t}^{\ast 4}]\leq Kn_{g}^{-2}.  \label{P_GTFE_Est3_6}
\end{equation}%
Hence by (\ref{P_GTFE_Est3_5}) and Markov's inequality,%
\begin{equation}
	T^{-1}\sum_{g\in \mathcal{G}_{1}}n_{g}^{-1}\sum_{i\in I_{g}}\sum_{t\leq T}(%
	\dot{\varepsilon}_{1,i,t}^{2}-\varepsilon _{1,i,t}^{2})=T^{-1}\sum_{g\in 
		\mathcal{G}_{1}}\sum_{t\leq T}\bar{\varepsilon}_{1,g,t}^{2}=O_{p}\left(
	\sum_{g\in \mathcal{G}_{1}}n_{g}^{-1}\right) .  \label{P_GTFE_Est3_7}
\end{equation}%
Combining (\ref{P_GTFE_Est3_1}), (\ref{P_GTFE_Est3_3}), (\ref{P_GTFE_Est3_4}%
) and (\ref{P_GTFE_Est3_7}), we obtain 
\begin{equation}
	\sum_{g\in \mathcal{G}_{1}}n_{g}^{-1}\sum_{i\in I_{g}}\left( \hat{\sigma}%
	_{1,i}^{2}-T^{-1}\sum_{t\leq T}\varepsilon _{1,i,t}^{2}\right) =O_{p}\left(
	\sum_{g\in \mathcal{G}_{1}}n_{g}^{-1}+G_{1}n^{-1}\right) .
	\label{P_GTFE_Est3_8}
\end{equation}%
Finally, by Assumption \ref{A1}, Lemma \ref{GTFE_1}(iii), and Markov's
inequality,%
\begin{equation*}
	T^{-1}\sum_{g\in \mathcal{G}_{1}}n_{g}^{-1}\sum_{i\in I_{g}}\sum_{t\leq
		T}(\varepsilon _{1,i,t}^{2}-\mathbb{E}[\varepsilon
	_{1,i,t}^{2}])=O_{p}\left( n^{-1/2}\left( \sum_{g\in \mathcal{G}%
		_{1}}n_{g}^{-1}\right) ^{1/2}\right) .
\end{equation*}%
Together with (\ref{P_GTFE_Est3_8}), this implies%
\begin{equation}
	\sum_{g\in \mathcal{G}_{1}}n_{g}^{-1}\sum_{i\in I_{g}}\left( \hat{\sigma}%
	_{1,i}^{2}-T^{-1}\sum_{t\leq T}\mathbb{E}[\varepsilon _{1,i,t}^{2}]\right)
	=O_{p}\left( \sum_{g\in \mathcal{G}_{1}}n_{g}^{-1}+G_{1}n^{-1}+n^{-1/2}%
	\left( \sum_{g\in \mathcal{G}_{1}}n_{g}^{-1}\right) ^{1/2}\right) .
	\label{P_GTFE_Est3_9}
\end{equation}%
Since $\sum_{g\in \mathcal{G}_{1}}n_{g}^{-1}\geq G_{1}n^{-1}\geq n^{-1}$,
the claim of the lemma follows from (\ref{P_GTFE_Est3_9}) and the definition
of $s_{\varepsilon _{1},i}^{2}$.\hfill $Q.E.D.$

\bigskip

\begin{lemma}
	\textit{\label{GTFE_Est4}}\ Under Assumptions \ref{A1}, \ref{A8}(ii) and \ref%
	{A9}(i), $\sum_{g\in \mathcal{G}_{1}}(n_{g}^{-1}\sum_{i\in I_{g}}(\hat{\sigma%
	}_{1,i}^{2}-s_{1,i}^{2}))^{2}=O_{p}(\sum_{g\in \mathcal{G}_{1}}n_{g}^{-1})$.
\end{lemma}

\noindent \textsc{Proof of Lemma \ref{GTFE_Est4}}. Using \ref{A1} and the
Cauchy-Schwarz inequality, we have 
\begin{eqnarray}
	\sum_{g\in \mathcal{G}_{1}}\left( n_{g}^{-1}\sum_{i\in I_{g}}(\hat{\sigma}%
	_{1,i}^{2}-s_{1,i}^{2})\right) ^{2} &\leq &K\sum_{g\in \mathcal{G}%
		_{1}}(n_{g}T)^{-1}\sum_{i\in I_{g}}\sum_{t\leq T}(\dot{\varepsilon}%
	_{1,i,t}^{2}-\varepsilon _{1,i,t}^{2})^{2}  \notag \\
	&&+K||\hat{\theta}_{1}-\theta _{1}^{\ast }||^{2}\sum_{g\in \mathcal{G}%
		_{1}}(n_{g}T)^{-1}\sum_{i\in I_{g}}\sum_{t\leq T}\left\Vert \dot{\varepsilon}%
	_{1,i,t}\dot{x}_{i,t}^{\top }\right\Vert ^{2}  \notag \\
	&&+K||\hat{\theta}_{1}-\theta _{1}^{\ast }||^{4}\sum_{g\in \mathcal{G}%
		_{1}}(n_{g}T)^{-1}\sum_{i\in I_{g}}\sum_{t\leq T}||\dot{x}_{i,t}\dot{x}%
	_{i,t}^{\top }||^{2}.  \label{P_GTFE_Est4_1}
\end{eqnarray}%
By the similar arguments for showing (\ref{P_GTFE_Est3_3}) and (\ref%
{P_GTFE_Est3_4}), we have%
\begin{equation}
	||\hat{\theta}_{1}-\theta _{1}^{\ast }||^{4}\sum_{g\in \mathcal{G}%
		_{1}}(n_{g}T)^{-1}\sum_{i\in I_{g}}\sum_{t\leq T}||\dot{x}_{i,t}\dot{x}%
	_{i,t}^{\top }||^{2}=O_{p}(G_{1}n^{-4}),  \label{P_GTFE_Est4_2}
\end{equation}%
and 
\begin{equation}
	||\hat{\theta}_{1}-\theta _{1}^{\ast }||^{2}\sum_{g\in \mathcal{G}%
		_{1}}(n_{g}T)^{-1}\sum_{i\in I_{g}}\sum_{t\leq T}\left\Vert \dot{\varepsilon}%
	_{1,i,t}\dot{x}_{i,t}^{\top }\right\Vert ^{2}=O_{p}(G_{1}n^{-2}).
	\label{P_GTFE_Est4_3}
\end{equation}%
Next, since $\dot{\varepsilon}_{1,i,t}=\varepsilon _{1,i,t}-\bar{\varepsilon}%
_{1,g,t}$, we obtain%
\begin{eqnarray}
	\sum_{g\in \mathcal{G}_{1}}(n_{g}T)^{-1}\sum_{i\in I_{g}}\sum_{t\leq T}(\dot{%
		\varepsilon}_{1,i,t}^{2}-\varepsilon _{1,i,t}^{2})^{2} &\leq &2\sum_{g\in 
		\mathcal{G}_{1}}(n_{g}T)^{-1}\sum_{i\in I_{g}}\sum_{t\leq T}\varepsilon
	_{1,i,t}^{2}\bar{\varepsilon}_{1,g,t}^{2}  \notag \\
	&&+2\sum_{g\in \mathcal{G}_{1}}(n_{g}T)^{-1}\sum_{i\in I_{g}}\sum_{t\leq T}%
	\bar{\varepsilon}_{1,g,t}^{4}\overset{}{=}O_{p}\left( \sum_{g\in \mathcal{G}%
		_{1}}n_{g}^{-1}\right) ,  \label{P_GTFE_Est4_4}
\end{eqnarray}%
where the last equality holds by Lemma \ref{GTFE_1}(iii), (\ref%
{P_GTFE_Est3_6}) and Markov's inequality. The claim of the lemma now follows
from (\ref{P_GTFE_Est4_1})-(\ref{P_GTFE_Est4_4}).\hfill $Q.E.D.$

\section{Proofs for Appendix \protect\ref{Sec:AP2}\label%
	{Proof_App_General_Model}}

\noindent \textsc{Proof of Theorem \ref{Rep_Theta}}. By (\ref{L3A-2}) in
Lemma \ref{L3A}, Assumptions \ref{A3} and \ref{A5}(i), and Lemma \ref%
{Consistency}, 
\begin{equation}
	\max_{g\in \mathcal{G},m\in \mathcal{M}}\left\vert
	(n_{g}T_{m})^{-1}\sum_{i\in I_{g}}\sum_{t\in I_{m}}\left. \partial
	^{\left\vert \nu \right\vert }\psi (z_{i,t},\tilde{\phi}_{g,m})\right/
	(\partial \phi _{1}^{\nu _{1}}\cdots \partial \phi _{k}^{\nu
		_{k}})\right\vert =O_{p}(1)  \label{P_Rep_Theta_1}
\end{equation}%
for $\left\vert \nu \right\vert \leq 3$. Applying the\ Taylor expansion to
the second equation\ (\ref{Foc_1}) obtains%
\begin{align}
	0_{d_{\theta }\times 1}& =\sum_{g\in \mathcal{G}}\sum_{m\in \mathcal{M}%
	}(n_{g}T_{m})\left[ \hat{\Psi}_{\theta ,g,m}+\hat{\Psi}_{\theta \theta ,g,m}(%
	\hat{\theta}-\theta ^{\ast })+\hat{\Psi}_{\theta \gamma ,g,m}(\hat{\gamma}%
	_{g,m}-\gamma _{g,m}^{\ast })\right]  \notag \\
	& \text{ \ }+2^{-1}\left( \sum_{g\in \mathcal{G}}\sum_{m\in \mathcal{M}%
	}(n_{g}T_{m})(\hat{\phi}_{g,m}-\phi _{g,m}^{\ast })^{\top }\hat{\Psi}%
	_{\theta _{j}\phi \phi ,g,m}(\hat{\phi}_{g,m}-\phi _{g,m}^{\ast })\right)
	_{j=1,\ldots ,d_{\theta }}  \notag \\
	& \text{ \ }+O_{p}\left( \sum_{g\in \mathcal{G}}\sum_{m\in \mathcal{M}%
	}(n_{g}T_{m})||\hat{\phi}_{g,m}-\phi _{g,m}^{\ast }||^{3}\right) ,
	\label{P_Rep_Theta_2}
\end{align}%
where the last term is by (\ref{P_Rep_Theta_1}) and the Cauchy-Schwarz
inequality. By Lemma \ref{Rate_theta} and (\ref{P_Rate_gamma_6}) in Lemma %
\ref{Rate_gamma}%
\begin{align}
	\sum_{g\in \mathcal{G}}\sum_{m\in \mathcal{M}}(n_{g}T_{m})||\hat{\gamma}%
	_{g,m}-\gamma _{g,m}^{\ast }||^{3}& \leq \max_{g\in \mathcal{G},m\in 
		\mathcal{M}}||\hat{\gamma}_{g,m}-\gamma _{g,m}^{\ast }||\sum_{g\in \mathcal{G%
	}}\sum_{m\in \mathcal{M}}||\hat{\gamma}_{g,m}-\gamma _{g,m}^{\ast }||^{2} 
	\notag \\
	& =O_{p}\left( (GM)^{1/p}\max_{g\in \mathcal{G},m\in \mathcal{M}%
	}(n_{g}T_{m})^{-1/2}\right) O_{p}\left( GM+(nT)||\hat{\theta}-\theta ^{\ast
	}||^{2}\right)  \notag \\
	& =O_{p}\left( (GM)^{1+1/p}\max_{g\in \mathcal{G},m\in \mathcal{M}%
	}(n_{g}T_{m})^{-1/2}\right) ,  \label{P_Rep_Theta_3}
\end{align}%
which together with Lemma \ref{Rate_theta} implies that%
\begin{align}
	\sum_{g\in \mathcal{G}}\sum_{m\in \mathcal{M}}(n_{g}T_{m})||\hat{\phi}%
	_{g,m}-\phi _{g,m}^{\ast }||^{3}& \leq K\sum_{g\in \mathcal{G}}\sum_{m\in 
		\mathcal{M}}(n_{g}T_{m})||\hat{\gamma}_{g,m}-\gamma _{g,m}^{\ast
	}||^{3}+K(nT)||\hat{\theta}-\theta ^{\ast }||^{3}  \notag \\
	& =O_{p}\left( (GM)^{1+1/p}\max_{g\in \mathcal{G},m\in \mathcal{M}%
	}(n_{g}T_{m})^{-1/2}\right) .  \label{P_Rep_Theta_4}
\end{align}%
By (\ref{P_L3AB_2}) in Lemma \ref{L3AB} and Lemma \ref{Rate_theta},%
\begin{equation}
	(nT)^{-1}\sum_{g\in \mathcal{G}}\sum_{m\in \mathcal{M}}\hat{\Psi}_{\theta
		\theta ,g,m}(\hat{\theta}-\theta ^{\ast })=(nT)^{-1}\sum_{g\in \mathcal{G}%
	}\sum_{m\in \mathcal{M}}\Psi _{\theta \theta ,g,m}(\hat{\theta}-\theta
	^{\ast })+o_{p}\left( (nT)^{-1/2}\right) .  \label{P_Rep_Theta_5}
\end{equation}%
Combining the results in Lemma \ref{Rep_Theta_1}, Lemma \ref{Rep_Theta_2}, (%
\ref{P_Rep_Theta_2}), (\ref{P_Rep_Theta_4}) and (\ref{P_Rep_Theta_5}), and
applying Assumption \ref{A5}(ii) we get%
\begin{align*}
	0_{d_{\theta }\times 1}& =(nT)^{-1}\sum_{g\in \mathcal{G}}\sum_{m\in 
		\mathcal{M}}(n_{g}T_{m})(\hat{\Psi}_{\theta ,g,m}-\Psi _{\theta \gamma
		,g,m}\Psi _{\gamma \gamma ,g,m}^{-1}\hat{\Psi}_{\gamma ,g,m}) \\
	& \text{ \ \ }+(nT)^{-1}\sum_{g\in \mathcal{G}}\sum_{m\in \mathcal{M}}(\Psi
	_{\theta \theta ,g,m}-\Psi _{\theta \gamma ,g,m}\Psi _{\gamma \gamma
		,g,m}^{-1}\Psi _{\gamma \theta ,g,m})(\hat{\theta}-\theta ^{\ast }) \\
	& \text{ \ \ }-(nT)^{-1}\sum_{g\in \mathcal{G}}\sum_{i\in I_{g}}(n_{g}T_{m})%
	\mathbb{E}\left[ \hat{U}_{\gamma ,g,m}\Psi _{\gamma \gamma ,g,m}^{-1}\hat{%
		\Psi}_{\gamma ,g,m}\right] \\
	& \text{ \ \ }-\frac{(nT)^{-1}}{2}\sum_{g\in \mathcal{G}}\sum_{m\in \mathcal{%
			M}}(n_{g}T_{m})\Psi _{\theta \gamma ,g,m}\Psi _{\gamma \gamma
		,g,m}^{-1}\left( \mathbb{E}\left[ \hat{\Psi}_{\gamma ,g,m}^{\top }\Psi
	_{\gamma \gamma ,g,m}^{-1}\Psi _{\gamma _{j}\gamma \gamma ,g,m}\Psi _{\gamma
		\gamma ,g,m}^{-1}\hat{\Psi}_{\gamma ,g,m}\right] \right) _{j\leq d_{\gamma }}
	\\
	& \text{ \ \ }+\frac{(nT)^{-1}}{2}\sum_{g\in \mathcal{G}}\sum_{m\in \mathcal{%
			M}}(n_{g}T_{m})\left( \mathbb{E}\left[ \hat{\Psi}_{\gamma ,g,m}^{\top }\Psi
	_{\gamma \gamma ,g,m}^{-1}\Psi _{\theta _{j}\gamma \gamma ,g,m}\Psi _{\gamma
		\gamma ,g,m}^{-1}\hat{\Psi}_{\gamma ,g,m}\right] \right) _{j\leq d_{\theta
	}}+o_{p}\left( (nT)^{-1/2}\right) .
\end{align*}%
The assertion of the lemma follows from Assumption \ref{A4} and the above
equation.\hfill $Q.E.D.$

\bigskip

\noindent \textsc{Proof of Theorem \ref{Rep_L}}. First note that by Lemma %
\ref{Rep_L_C} and Lemma \ref{Rep_L_D}%
\begin{align*}
	& (nT)^{-1/2}\sum_{m\in \mathcal{M}}(n_{g}T_{m})(\hat{\phi}_{g,m}-\phi
	_{g,m}^{\ast })^{\top }\hat{\Psi}_{\phi \phi ,g,m}(\hat{\phi}_{g,m}-\phi
	_{g,m}^{\ast }) \\
	& =(nT)^{-1/2}\sum_{g\in \mathcal{G}}\sum_{m\in \mathcal{M}}(n_{g}T_{m})\hat{%
		\Psi}_{\gamma ,g,m}^{\top }\Psi _{\gamma \gamma ,g,m}^{-1}\hat{\Psi}_{\gamma
		,g,m}+o_{p}((GM)^{1/2}(nT)^{-1/2}).
\end{align*}%
which together with Lemma \ref{Rep_L_B} implies that 
\begin{align}
	& (nT)^{-1/2}\sum_{g\in \mathcal{G}}\sum_{m\in \mathcal{M}}(n_{g}T_{m})\left[
	\hat{\Psi}_{g,m}(\hat{\phi}_{g,m})-\hat{\Psi}_{g,m}(\phi _{g,m}^{\ast })%
	\right]  \notag \\
	& =(nT)^{-1/2}\sum_{g\in \mathcal{G}}\sum_{m\in \mathcal{M}}(n_{g}T_{m})\hat{%
		\Psi}_{\phi ,g,m}(\hat{\phi}_{g,m}-\phi _{g,m}^{\ast })  \notag \\
	& +\frac{(nT)^{-1/2}}{2}\sum_{g\in \mathcal{G}}\sum_{m\in \mathcal{M}%
	}(n_{g}T_{m})\hat{\Psi}_{\gamma ,g,m}^{\top }\Psi _{\gamma \gamma ,g,m}^{-1}%
	\hat{\Psi}_{\gamma ,g,m}+o_{p}((GM)^{1/2}(nT)^{-1/2}).  \label{P_Rep_L_1}
\end{align}%
By Lemma \ref{Rate_theta} and (\ref{L0-1}) in Lemma \ref{L0}, 
\begin{equation*}
	\left\vert (nT)^{-1/2}\sum_{g\in \mathcal{G}}\sum_{m\in \mathcal{M}%
	}(n_{g}T_{m})\hat{\Psi}_{\theta ,g,m}(\hat{\theta}-\theta ^{\ast
	})\right\vert \leq \left\Vert \hat{\theta}-\theta ^{\ast }\right\Vert
	\left\Vert (nT)^{-1/2}\sum_{g\in \mathcal{G}}\sum_{m\in \mathcal{M}%
	}(n_{g}T_{m})\hat{\Psi}_{\theta ,g,m}\right\Vert =O_{p}((nT)^{-1/2})
\end{equation*}%
which together with (\ref{P_Rep_L_1}) implies that 
\begin{align}
	& (nT)^{-1/2}\sum_{g\in \mathcal{G}}\sum_{m\in \mathcal{M}}(n_{g}T_{m})\left[
	\hat{\Psi}_{g,m}(\hat{\phi}_{g,m})-\hat{\Psi}_{g,m}(\phi _{g,m}^{\ast })%
	\right]  \notag \\
	& =(nT)^{-1/2}\sum_{g\in \mathcal{G}}\sum_{m\in \mathcal{M}}(n_{g}T_{m})\hat{%
		\Psi}_{\gamma ,g,m}(\hat{\gamma}_{g,m}-\gamma _{g,m}^{\ast })  \notag \\
	& +\frac{(nT)^{-1/2}}{2}\sum_{g\in \mathcal{G}}\sum_{m\in \mathcal{M}%
	}(n_{g}T_{m})\hat{\Psi}_{\gamma ,g,m}^{\top }\Psi _{\gamma \gamma ,g,m}^{-1}%
	\hat{\Psi}_{\gamma ,g,m}+o_{p}((GM)^{1/2}(nT)^{-1/2}).  \label{P_Rep_L_2}
\end{align}%
The assertion of the lemma follows from (\ref{P_Rep_L_C_20}) in the proof of
Lemma \ref{Rep_L_C} and (\ref{P_Rep_L_2}).\hfill $Q.E.D.$

\bigskip

\noindent \textsc{Proof of Theorem \ref{MGCLT_V}}. By the martingale CLT\
(e.g., Corollary 3.1 of \cite{HallHeyde1980}), to verify Assumption \ref{A6}%
(ii) it is sufficient to show that (\ref{C1_MGCLT}) and (\ref{C2_MGCLT})
hold.

To show (\ref{C1_MGCLT}), we first notice that%
\begin{equation}
	\mathbb{E}[\xi _{nT,i}^{2}|\mathcal{F}_{nT,i-1}]=\frac{\mathbb{E}[(\tilde{%
			\Psi}_{i}^{\ast }+\tilde{V}_{i}^{\ast })^{2}]}{n\omega _{n,T}^{2}}+\frac{%
		\mathbb{E}[\tilde{U}_{i}^{2}|\mathcal{F}_{nT,i-1}]}{n\omega _{n,T}^{2}}+%
	\frac{2\mathbb{E}[(\tilde{\Psi}_{i}^{\ast }+\tilde{V}_{i}^{\ast })\tilde{U}%
		_{i}|\mathcal{F}_{nT,i-1}]}{n\omega _{n,T}^{2}}.  \label{P_MGCLT_V_1}
\end{equation}%
By Assumptions \ref{A6}(i) and \ref{A7}(iii), and Lemma \ref{C1_MGCLT_L2}, 
\begin{equation}
	\frac{\sum_{i\leq n}\mathbb{E}[\tilde{U}_{i}^{2}|\mathcal{F}_{nT,i-1}]}{%
		n\omega _{n,T}^{2}}=\frac{\sum_{i\leq n}\mathbb{E}[\tilde{U}_{i}^{2}]}{%
		n\omega _{n,T}^{2}}+o_{p}\left( 1\right) .  \label{P_MGCLT_V_2}
\end{equation}%
By the independence of $\left\{ z_{i,t}\right\} _{t\leq T}$ across $i$, $%
\tilde{\Psi}_{i}^{\ast }+\tilde{V}_{i}^{\ast }$ and $\tilde{U}_{i}$ are
uncorrelated which implies that 
\begin{equation}
	n\omega _{n,T}^{2}=\sum_{i\leq n}\mathrm{Var}(\tilde{\Psi}_{i}+\tilde{V}_{i}+%
	\tilde{U}_{i})=\sum_{i\leq n}(\mathrm{Var}(\tilde{\Psi}_{i}+\tilde{V}_{i})+%
	\mathrm{Var}(\tilde{U}_{i}))=\sum_{i\leq n}\mathbb{E}[(\tilde{\Psi}%
	_{i}^{\ast }+\tilde{V}_{i}^{\ast })^{2}+\tilde{U}_{i}^{2}]
	\label{P_MGCLT_V_2b}
\end{equation}%
where the third equality is by $\mathbb{E}[\tilde{U}_{i}]=0$ for all $i$. By
(\ref{P_MGCLT_V_1}), (\ref{P_MGCLT_V_2}) and (\ref{P_MGCLT_V_2b}) we deduce
that%
\begin{align}
	\sum_{i\leq n}\mathbb{E}[\xi _{nT,i}^{2}|\mathcal{F}_{nT,i-1}]& =\frac{%
		\sum_{i\leq n}\mathbb{E}[(\tilde{\Psi}_{i}^{\ast }+\tilde{V}_{i}^{\ast
		})^{2}+\tilde{U}_{i}^{2}]}{n\omega _{n,T}^{2}}  \notag \\
	& \text{ \ \ \ \ }+\frac{2\sum_{i\leq n}\mathbb{E}[(\tilde{\Psi}_{i}^{\ast }+%
		\tilde{V}_{i}^{\ast })\tilde{U}_{i}|\mathcal{F}_{nT,i-1}]}{n\omega _{n,T}^{2}%
	}+o_{p}\left( 1\right)  \notag \\
	& =\frac{\sum_{i\leq n}\mathbb{E}[(\tilde{\Psi}_{i}^{\ast }+\tilde{V}%
		_{i}^{\ast })^{2}+\tilde{U}_{i}^{2}]}{n\omega _{n,T}^{2}}+o_{p}\left(
	1\right) =1+o_{p}(1),  \label{P_MGCLT_V_3}
\end{align}%
where the second equality follows by Lemma \ref{C1_MGCLT_L1} and Assumption %
\ref{A7}(iii). The first sufficient condition of the martingale CLT, i.e., (%
\ref{C1_MGCLT}) now follows from (\ref{P_MGCLT_V_3}).

To verify (\ref{C2_MGCLT}), we notice that for any $\varepsilon >0$,%
\begin{align}
	\sum_{i\leq n}\mathbb{E}[\xi _{nT,i}^{2}I\{|\xi _{nT,i}|\overset{}{>}%
	\varepsilon \}]& \leq \varepsilon ^{-2}\sum_{i\leq n}\mathbb{E}[|\xi
	_{nT,i}|^{4}]  \notag \\
	& \leq K\varepsilon ^{-2}\frac{\sum_{i\leq n}\mathbb{E}[(\tilde{\Psi}%
		_{i}^{\ast })^{4}]+\sum_{i\leq n}\mathbb{E}[(\tilde{V}_{i}^{\ast
		})^{4}]+\sum_{i\leq n}\mathbb{E}[\tilde{U}_{i}^{4}]}{(n\omega _{n,T}^{2})^{2}%
	}.  \label{P_MGCLT_V_4}
\end{align}%
By (\ref{P_C1_MGCLT_L1_8}) in the proof of Lemma \ref{C1_MGCLT_L1},%
\begin{equation}
	\frac{\sum_{i\leq n}\mathbb{E}[(\tilde{\Psi}_{i}^{\ast })^{4}]}{(n\omega
		_{n,T}^{2})^{2}}=\frac{\sum_{i\leq n}\mathbb{E}[|\tilde{\Psi}_{1,i}^{\ast }-%
		\tilde{\Psi}_{2,i}^{\ast }|^{4}]}{(n\omega _{n,T}^{2})^{2}}\leq
	n^{-1}\max_{i\leq n}\mathbb{E}\left[ \left( \frac{\tilde{\Psi}_{1,i}^{\ast }-%
		\tilde{\Psi}_{2,i}^{\ast }}{\omega _{n,T}}\right) ^{4}\right] \leq Kn^{-1}.
	\label{P_MGCLT_V_5}
\end{equation}%
By Assumption \ref{A6}(i) and (\ref{P_C1_MGCLT_L1_9}) in the proof of Lemma %
\ref{C1_MGCLT_L1}%
\begin{align}
	\frac{\sum_{i\leq n}\mathbb{E}[(\tilde{V}_{j,i}^{\ast })^{4}]}{(n\omega
		_{n,T}^{2})^{2}}& =\frac{\sum_{g\in \mathcal{G}_{j}}\sum_{i\in I_{g}}\mathbb{%
			E}[(\tilde{V}_{j,i}^{\ast })^{4}]}{(n\omega _{n,T}^{2})^{2}}\leq K\sum_{g\in 
		\mathcal{G}_{j}}\sum_{i\in I_{g}}(M_{j}^{2}n_{g}^{-4}T^{-2})  \notag \\
	& \leq KM_{j}^{2}T^{-2}\sum_{g\in \mathcal{G}_{j}}n_{g}^{-3}\leq
	KM_{j}^{2}G_{j}T^{-2}\max_{g\in \mathcal{G}_{j}}n_{g}^{-3}\leq
	KG_{j}^{-1}\max_{g\in \mathcal{G}_{j}}n_{g}^{-3}  \label{P_MGCLT_V_6}
\end{align}%
where the last inequality is by 
\begin{equation}
	M_{j}^{2}G_{j}T^{-2}=(M_{j}G_{j}(nT)^{-1/2})^{2}(nT^{-1})G_{j}^{-1}\leq
	KG_{j}^{-1},  \label{P_MGCLT_V_6b}
\end{equation}%
where the inequality follows since $nT^{-1}\leq K$ by Assumption \ref{A1}(i)
and $M_{j}G_{j}(nT)^{-1/2}\leq K$ by Assumption \ref{A5}(ii). By Assumption %
\ref{A6}(i) and Lemma \ref{C2_MGCLT_L1},%
\begin{align}
	\frac{\sum_{i\leq n}\mathbb{E}[|\tilde{U}_{j,i}|^{4}]}{(n\omega
		_{n,T}^{2})^{2}}& =\frac{\sum_{g\in \mathcal{G}_{j}}\sum_{i\in I_{g}}\mathbb{%
			E}[|\tilde{U}_{j,i}|^{4}]}{(n\omega _{n,T}^{2})^{2}}\leq K\sum_{g\in 
		\mathcal{G}_{j}}\sum_{i\in I_{g}}M_{j}^{2}T^{-2}n_{g}^{-2}  \notag \\
	& \leq KM_{j}^{2}G_{j}T^{-2}\max_{g\in \mathcal{G}_{j}}n_{g}^{-1}\leq
	KG_{j}^{-1}\max_{g\in \mathcal{G}_{j}}n_{g}^{-1},  \label{P_MGCLT_V_7}
\end{align}%
where the last equality is by (\ref{P_MGCLT_V_6b}). Collecting the results
in (\ref{P_MGCLT_V_4})-(\ref{P_MGCLT_V_7}), we deduce that 
\begin{equation*}
	\sum_{i\leq n}\mathbb{E}[\xi _{nT,i}^{2}I\{|\xi _{nT,i}|>\varepsilon \}]\leq
	K\varepsilon ^{-2}\left( n^{-1}+G_{j}^{-1}\max_{g\in \mathcal{G}%
		_{j}}n_{g}^{-1}\right) \rightarrow 0\text{, as }n,T\rightarrow \infty
\end{equation*}%
which shows (\ref{C2_MGCLT}).\hfill $Q.E.D.$

\bigskip

\begin{theorem}
	\textit{\label{Rate_gamma} }Under Assumptions \ref{A1}, \ref{A2}, \ref{A3}, %
	\ref{A4} and \ref{A5}, we have%
	\begin{equation}
		\max_{g\in \mathcal{G},m\in \mathcal{M}}\frac{\left\Vert \hat{\gamma}%
			_{g,m}-\gamma _{g,m}^{\ast }\right\Vert }{(GM)^{1/p}(n_{g}T_{m})^{-1/2}}%
		=O_{p}(1),  \label{P_Rate_gamma_6}
	\end{equation}%
	and%
	\begin{equation*}
		\max_{g\in \mathcal{G},m\in \mathcal{M}}\frac{\left\Vert \hat{\gamma}%
			_{g,m}-\gamma _{g,m}^{\ast }+\Psi _{\gamma \gamma ,g,m}^{-1}\hat{\Psi}%
			_{\gamma ,g,m}\right\Vert }{(GM)^{2/p}(n_{g}T_{m})^{-1}}=O_{p}(1),
	\end{equation*}%
	where $\hat{\Psi}_{\gamma ,g,m}\equiv (n_{g}T_{m})^{-1}\sum_{i\in
		I_{g}}\sum_{t\in I_{m}}\psi _{\gamma }(z_{i,t})$.
\end{theorem}

\noindent \textsc{Proof of Theorem \ref{Rate_gamma}}.\ By the triangle
inequality,%
\begin{align}
	& \max_{g\in \mathcal{G},m\in \mathcal{M}}\frac{\left\Vert \hat{\gamma}%
		_{g,m}-\gamma _{g,m}^{\ast }+\Psi _{\gamma \gamma ,g,m}^{-1}\hat{\Psi}%
		_{\gamma ,g,m}\right\Vert }{(GM)^{2/p}(n_{g}T_{m})^{-1}+||\hat{\phi}%
		_{g,m}-\phi _{g,m}^{\ast }||^{2}}  \notag \\
	& \leq \max_{g\in \mathcal{G},m\in \mathcal{M}}(\lambda _{\max }((\hat{\Psi}%
	_{\gamma \gamma ,g,m}^{-1})^{2}))^{1/2}\frac{\left\Vert \hat{\Psi}_{\gamma
			\gamma ,g,m}(\hat{\gamma}_{g,m}-\gamma _{g,m}^{\ast })+\hat{\Psi}_{\gamma
			\gamma ,g,m}\Psi _{\gamma \gamma ,g,m}^{-1}\hat{\Psi}_{\gamma
			,g,m}\right\Vert }{(GM)^{2/p}(n_{g}T_{m})^{-1}+||\hat{\phi}_{g,m}-\phi
		_{g,m}^{\ast }||^{2}}  \notag \\
	& \leq \max_{g\in \mathcal{G},m\in \mathcal{M}}(\lambda _{\max }((\hat{\Psi}%
	_{\gamma \gamma ,g,m}^{-1})^{2}))^{1/2}\frac{\left\Vert \hat{\Psi}_{\gamma
			\gamma ,g,m}(\hat{\gamma}_{g,m}-\gamma _{g,m}^{\ast })+\hat{\Psi}_{\gamma
			\theta ,g,m}(\hat{\theta}-\theta ^{\ast })+\hat{\Psi}_{\gamma
			,g,m}\right\Vert }{(GM)^{2/p}(n_{g}T_{m})^{-1}+||\hat{\phi}_{g,m}-\phi
		_{g,m}^{\ast }||^{2}}  \notag \\
	& \text{ \ \ }+\max_{g\in \mathcal{G},m\in \mathcal{M}}(\lambda _{\max }((%
	\hat{\Psi}_{\gamma \gamma ,g,m}^{-1})^{2}))^{1/2}\frac{\left\Vert (\hat{\Psi}%
		_{\gamma \gamma ,g,m}-\Psi _{\gamma \gamma ,g,m})\Psi _{\gamma \gamma
			,g,m}^{-1}\hat{\Psi}_{\gamma ,g,m}\right\Vert }{(GM)^{2/p}(n_{g}T_{m})^{-1}}
	\notag \\
	& \text{ \ \ }+\max_{g\in \mathcal{G},m\in \mathcal{M}}(\lambda _{\max }((%
	\hat{\Psi}_{\gamma \gamma ,g,m}^{-1})^{2}))^{1/2}(nT)^{1/2}\left\Vert \hat{%
		\Psi}_{\gamma \theta ,g,m}(\hat{\theta}-\theta ^{\ast })\right\Vert .
	\label{P_Rate_gamma_1}
\end{align}%
The first term on the far RHS of (\ref{P_Rate_gamma_1}) can be analyzed
using (\ref{P_Rate_theta_3}) of Lemma \ref{Rate_theta} and Lemma \ref{L3AB}%
(ii) to yield 
\begin{equation}
	\max_{g\in \mathcal{G},m\in \mathcal{M}}\frac{(\lambda _{\max }((\hat{\Psi}%
		_{\gamma \gamma ,g,m}^{-1})^{2}))^{1/2}\left\Vert \hat{\Psi}_{\gamma \gamma
			,g,m}(\hat{\gamma}_{g,m}-\gamma _{g,m}^{\ast })+\hat{\Psi}_{\gamma \theta
			,g,m}(\hat{\theta}-\theta ^{\ast })+\hat{\Psi}_{\gamma ,g,m}\right\Vert }{%
		(GM)^{2/p}(n_{g}T_{m})^{-1}+||\hat{\phi}_{g,m}-\phi _{g,m}^{\ast }||^{2}}%
	=O_{p}(1).  \label{P_Rate_gamma_2}
\end{equation}

In order to analyze the second term on the far RHS of (\ref{P_Rate_gamma_1}%
), we note%
\begin{align}
	& \max_{g\in \mathcal{G},m\in \mathcal{M}}(\lambda _{\max }((\hat{\Psi}%
	_{\gamma \gamma ,g,m}^{-1})^{2}))^{1/2}\frac{\left\Vert (\hat{\Psi}_{\gamma
			\gamma ,g,m}-\Psi _{\gamma \gamma ,g,m})\Psi _{\gamma \gamma ,g,m}^{-1}\hat{%
			\Psi}_{\gamma ,g,m}\right\Vert }{(GM)^{2/p}(n_{g}T_{m})^{-1}}  \notag \\
	& \leq \max_{g\in \mathcal{G},m\in \mathcal{M}}(\lambda _{\max }((\hat{\Psi}%
	_{\gamma \gamma ,g,m}^{-1})^{2})\lambda _{\max }((\Psi _{\gamma \gamma
		,g,m}^{-1})^{2}))^{1/2}\max_{g\in \mathcal{G},m\in \mathcal{M}}\frac{%
		\left\Vert \hat{\Psi}_{\gamma \gamma ,g,m}-\Psi _{\gamma \gamma
			,g,m}\right\Vert \left\Vert \hat{\Psi}_{\gamma ,g,m}\right\Vert }{%
		(GM)^{2/p}(n_{g}T_{m})^{-1}},  \label{P_Rate_gamma_3a}
\end{align}%
where $\max_{g\in \mathcal{G},m\in \mathcal{M}}(\lambda _{\max }((\hat{\Psi}%
_{\gamma \gamma ,g,m}^{-1})^{2})\lambda _{\max }((\Psi _{\gamma \gamma
	,g,m}^{-1})^{2}))^{1/2}=O_{p}(1)$ by Lemma \ref{L3AB}(ii) and Assumption \ref%
{A4}; 
\begin{equation}
	\max_{g\in \mathcal{G},m\in \mathcal{M}}\frac{\left\Vert \hat{\Psi}_{\gamma
			\gamma ,g,m}-\Psi _{\gamma \gamma ,g,m}\right\Vert }{%
		(GM)^{1/p}(n_{g}T_{m})^{-1/2}}=O_{p}(1)  \label{P_Rate_gamma_3b}
\end{equation}%
by Lemma \ref{P_L3AB_1}; and%
\begin{equation}
	\max_{g\in \mathcal{G},m\in \mathcal{M}}\frac{\left\Vert \hat{\Psi}_{\gamma
			,g,m}\right\Vert }{(GM)^{1/p}(n_{g}T_{m})^{-1/2}}=O_{p}(1)
	\label{P_Rate_gamma_3c}
\end{equation}%
by (\ref{L1-1}) in Lemma \ref{L1}. By (\ref{P_Rate_gamma_3a}), (\ref%
{P_Rate_gamma_3b}) and (\ref{P_Rate_gamma_3c}), we have%
\begin{equation}
	\max_{g\in \mathcal{G},m\in \mathcal{M}}(\lambda _{\max }((\hat{\Psi}%
	_{\gamma \gamma ,g,m}^{-1})^{2}))^{1/2}\frac{\left\Vert (\hat{\Psi}_{\gamma
			\gamma ,g,m}-\Psi _{\gamma \gamma ,g,m})\Psi _{\gamma \gamma ,g,m}^{-1}\hat{%
			\Psi}_{\gamma ,g,m}\right\Vert }{(GM)^{2/p}(n_{g}T_{m})^{-1}}=O_{p}(1).
	\label{P_Rate_gamma_3}
\end{equation}

In order to analyze the third term on the far RHS of (\ref{P_Rate_gamma_1}),
we use the Cauchy-Schwarz inequality, Lemma \ref{Rate_theta}, Lemma \ref%
{L3AB}(ii), and (\ref{P_L3AB_3}) of Lemma \ref{L3AB}, and conclude 
\begin{equation}
	\max_{g\in \mathcal{G},m\in \mathcal{M}}(\lambda _{\max }((\hat{\Psi}%
	_{\gamma \gamma ,g,m}^{-1})^{2}))^{1/2}(nT)^{1/2}\left\Vert \Psi _{\gamma
		\theta ,g,m}(\hat{\theta}-\theta ^{\ast })\right\Vert =O_{p}(1).
	\label{P_Rate_gamma_4}
\end{equation}

From (\ref{P_Rate_gamma_1}), (\ref{P_Rate_gamma_2}), (\ref{P_Rate_gamma_3})
and (\ref{P_Rate_gamma_4}), we have 
\begin{equation}
	\max_{g\in \mathcal{G},m\in \mathcal{M}}\frac{\left\Vert \hat{\gamma}%
		_{g,m}-\gamma _{g,m}^{\ast }+\Psi _{\gamma \gamma ,g,m}^{-1}\hat{\Psi}%
		_{\gamma ,g,m}\right\Vert }{(GM)^{2/p}(n_{g}T_{m})^{-1}+||\hat{\phi}%
		_{g,m}-\phi _{g,m}^{\ast }||^{2}}=O_{p}(1).  \label{P_Rate_gamma_5}
\end{equation}

By the triangle inequality,%
\begin{align}
	\max_{g\in \mathcal{G},m\in \mathcal{M}}\frac{\left\Vert \hat{\gamma}%
		_{g,m}-\gamma _{g,m}^{\ast }\right\Vert }{(GM)^{1/p}(n_{g}T_{m})^{-1/2}}&
	\leq \max_{g\in \mathcal{G},m\in \mathcal{M}}\frac{\left\Vert \hat{\gamma}%
		_{g,m}-\gamma _{g,m}^{\ast }+\Psi _{\gamma \gamma ,g,m}^{-1}\hat{\Psi}%
		_{\gamma ,g,m}\right\Vert }{(GM)^{1/p}(n_{g}T_{m})^{-1/2}}  \notag \\
	& \text{ \ }+\max_{g\in \mathcal{G},m\in \mathcal{M}}\frac{\left\Vert \Psi
		_{\gamma \gamma ,g,m}^{-1}\hat{\Psi}_{\gamma ,g,m}\right\Vert }{%
		(GM)^{1/p}(n_{g}T_{m})^{-1/2}},  \label{P_Rate_gamma_6a}
\end{align}%
where%
\begin{equation}
	\max_{g\in \mathcal{G},m\in \mathcal{M}}\frac{\left\Vert \Psi _{\gamma
			\gamma ,g,m}^{-1}\hat{\Psi}_{\gamma ,g,m}\right\Vert }{%
		(GM)^{1/p}(n_{g}T_{m})^{-1/2}}=O_{p}(1)  \label{P_Rate_gamma_6b}
\end{equation}%
by Assumption \ref{A4} and (\ref{P_Rate_gamma_3c}). Since $||\hat{\phi}%
_{g,m}-\phi _{g,m}^{\ast }||^{2}=||\hat{\theta}-\theta ^{\ast }||^{2}+||\hat{%
	\gamma}_{g,m}-\gamma _{g,m}^{\ast }||^{2}$ and $n_{g}T_{m}\leq nT$,%
\begin{align}
	& \max_{g\in \mathcal{G},m\in \mathcal{M}}\frac{%
		(GM)^{2/p}(n_{g}T_{m})^{-1}+||\hat{\phi}_{g,m}-\phi _{g,m}^{\ast }||^{2}}{%
		(GM)^{1/p}(n_{g}T_{m})^{-1/2}}  \notag \\
	& \leq (GM)^{1/p}\max_{g\in \mathcal{G},m\in \mathcal{M}}(n_{g}T_{m})^{-1/2}+%
	\frac{(nT)^{1/2}||\hat{\theta}-\theta ^{\ast }||^{2}}{(GM)^{1/p}}+\max_{g\in 
		\mathcal{G},m\in \mathcal{M}}\frac{||\hat{\gamma}_{g,m}-\gamma _{g,m}^{\ast
		}||^{2}}{(GM)^{1/p}(n_{g}T_{m})^{-1/2}}  \notag \\
	& =\max_{g\in \mathcal{G},m\in \mathcal{M}}||\hat{\gamma}_{g,m}-\gamma
	_{g,m}^{\ast }||\times \max_{g\in \mathcal{G},m\in \mathcal{M}}\frac{||\hat{%
			\gamma}_{g,m}-\gamma _{g,m}^{\ast }||}{(GM)^{1/p}(n_{g}T_{m})^{-1/2}}%
	+o_{p}(1)  \notag \\
	& =o_{p}(1)\times \max_{g\in \mathcal{G},m\in \mathcal{M}}\frac{||\hat{\gamma%
		}_{g,m}-\gamma _{g,m}^{\ast }||}{(GM)^{1/p}(n_{g}T_{m})^{-1/2}}+o_{p}(1)
	\label{P_Rate_gamma_6c}
\end{align}%
where the first equality is by Assumption \ref{A5}(i) and Lemma \ref%
{Rate_theta}, the second equality is by Lemma \ref{Consistency}. By (\ref%
{P_Rate_gamma_5}) and (\ref{P_Rate_gamma_6c})%
\begin{align}
	& \max_{g\in \mathcal{G},m\in \mathcal{M}}\frac{\left\Vert \hat{\gamma}%
		_{g,m}-\gamma _{g,m}^{\ast }+\Psi _{\gamma \gamma ,g,m}^{-1}\hat{\Psi}%
		_{\gamma ,g,m}\right\Vert }{(GM)^{1/p}(n_{g}T_{m})^{-1/2}}  \notag \\
	& =\max_{g\in \mathcal{G},m\in \mathcal{M}}\frac{\left\Vert \hat{\gamma}%
		_{g,m}-\gamma _{g,m}^{\ast }+\Psi _{\gamma \gamma ,g,m}^{-1}\hat{\Psi}%
		_{\gamma ,g,m}\right\Vert }{(GM)^{2/p}(n_{g}T_{m})^{-1}+||\hat{\phi}%
		_{g,m}-\phi _{g,m}^{\ast }||^{2}}\frac{(GM)^{2/p}(n_{g}T_{m})^{-1}+||\hat{%
			\phi}_{g,m}-\phi _{g,m}^{\ast }||^{2}}{(GM)^{1/p}(n_{g}T_{m})^{-1/2}}  \notag
	\\
	& =o_{p}(1)\times \max_{g\in \mathcal{G},m\in \mathcal{M}}\frac{||\hat{\gamma%
		}_{g,m}-\gamma _{g,m}^{\ast }||}{(GM)^{1/p}(n_{g}T_{m})^{-1/2}}+o_{p}(1).
	\label{P_Rate_gamma_6d}
\end{align}%
Collecting the results in (\ref{P_Rate_gamma_6a}), (\ref{P_Rate_gamma_6b})
and (\ref{P_Rate_gamma_6d}), we get%
\begin{equation*}
	\max_{g\in \mathcal{G},m\in \mathcal{M}}\frac{\left\Vert \hat{\gamma}%
		_{g,m}-\gamma _{g,m}^{\ast }\right\Vert }{(GM)^{1/p}(n_{g}T_{m})^{-1/2}}\leq
	o_{p}(1)\times \max_{g\in \mathcal{G},m\in \mathcal{M}}\frac{||\hat{\gamma}%
		_{g,m}-\gamma _{g,m}^{\ast }||}{(GM)^{1/p}(n_{g}T_{m})^{-1/2}}+O_{p}(1),
\end{equation*}%
which implies that (\ref{P_Rate_gamma_6}) holds. Therefore, by Lemma \ref%
{Rate_theta} and (\ref{P_Rate_gamma_6}),%
\begin{align}
	& \max_{g\in \mathcal{G},m\in \mathcal{M}}\frac{%
		(GM)^{2/p}(n_{g}T_{m})^{-1}+||\hat{\phi}_{g,m}-\phi _{g,m}^{\ast }||^{2}}{%
		(GM)^{2/p}(n_{g}T_{m})^{-1}}  \notag \\
	& =\max_{g\in \mathcal{G},m\in \mathcal{M}}\left( 1+\frac{(nT)^{-1}}{%
		(GM)^{2/p}(n_{g}T_{m})^{-1}}(nT)||\hat{\theta}-\theta ^{\ast }||^{2}+\frac{%
		\left\Vert \hat{\gamma}_{g,m}-\gamma _{g,m}^{\ast }\right\Vert ^{2}}{%
		(GM)^{2/p}(n_{g}T_{m})^{-1}}\right) =O_{p}(1).  \label{P_Rate_gamma_7}
\end{align}%
The assertion of the lemma now follows from (\ref{P_Rate_gamma_5}) and (\ref%
{P_Rate_gamma_7}).\hfill $Q.E.D.$

\subsection{Auxiliary Lemmas}

\begin{lemma}
	\textit{\label{L00} }Suppose $X_{1}$ and $X_{2}$ are zero mean random
	vectors with $\max_{j=1,2}\lambda_{\max}(\mathbb{E}[X_{j}X_{j}^{\top}])<%
	\infty$. Then%
	\begin{equation}
		\lambda_{\max}(\mathbb{E}[X_{1}X_{2}^{\top}]\mathbb{E}[X_{2}X_{1}^{\top}])%
		\leq\lambda_{\max}(\mathbb{E}[X_{1}X_{1}^{\top}])\lambda_{\max}(\mathbb{E}%
		[X_{2}X_{2}^{\top}]).  \label{L00_1}
	\end{equation}
\end{lemma}

\noindent\textsc{Proof of Lemma \ref{L00}}. Let $\Sigma_{X_{j}}\equiv 
\mathbb{E}[X_{j}X_{j}^{\top}]$ and $k_{j}$ denote the dimension of $X_{j}$ ($%
j=1,2$). For any $k_{1}\times1$ real vectors $a_{1}$, we shall show 
\begin{equation}
	a_{1}^{\top}(\mathbb{E}[X_{1}X_{2}^{\top}]\mathbb{E}[X_{2}X_{1}^{%
		\top}])a_{1}\leq\lambda_{\max}(\Sigma_{X_{1}})\lambda_{\max}(%
	\Sigma_{X_{2}})a_{1}^{\top}a_{1},  \label{P_L00_1}
\end{equation}
which directly shows (\ref{L00_1}). Since (\ref{P_L00_1}) holds trivially if 
$a_{1}^{\top}\Sigma_{X_{1}}a_{1}=0$, we only need to consider the case that $%
a_{1}^{\top}\Sigma_{X_{1}}a_{1}>0$.\ Let $(\Sigma_{X_{2}}^{1/2})^{+}$ denote
Moore-Penrose inverse of $\Sigma_{X_{2}}^{1/2}$ and $\Sigma_{X_{2}}^{1/2}$
is the unique symmetric matrix square root of $\Sigma_{X_{2}}$.\ Then by
Lemma E.1.(b, c) in \cite{Liao&Shi2020},%
\begin{align*}
	a_{1}^{\top}(\mathbb{E}[X_{1}X_{2}^{\top}]\mathbb{E}[X_{2}X_{1}^{%
		\top}])a_{1} & =(\mathrm{Cov}(a_{1}^{\top}X_{1},X_{2}))^{\top}(\mathrm{Cov}%
	(a_{1}^{\top }X_{1},X_{2})) \\
	& =\left( (\Sigma_{X_{2}}^{1/2})^{+}\mathrm{Cov}(a_{1}^{\top}X_{1},X_{2})%
	\right) ^{\top}\Sigma_{X_{2}}\left( (\Sigma_{X_{2}}^{1/2})^{+}\mathrm{Cov}%
	(a_{1}^{\top}X_{1},X_{2})\right) \\
	& \leq\lambda_{\max}(\Sigma_{X_{2}})\mathrm{Var}(a_{1}^{\top}X_{1})\leq%
	\lambda_{\max}(\Sigma_{X_{1}})\lambda_{\max}(\Sigma_{X_{2}})a_{1}^{\top
	}a_{1},
\end{align*}
which proves (\ref{P_L00_1}).\hfill$Q.E.D.$

\bigskip

\begin{lemma}
	\textit{\label{L0}}Suppose that: (i) $\varphi(z_{i,t};\phi)$ is a function
	indexed by the parameter $\phi\in\Phi$ where $\Phi$ is a compact, convex
	subset of $\mathbb{R}^{k}$ and $k\equiv\dim\left( \phi\right) $; (ii) there
	exists a function\ $D\left( z_{i,t}\right) $ such that $\left\vert
	\varphi\left( z_{i,t},\phi_{1}\right) -\varphi\left( z_{i,t},\phi
	_{2}\right) \right\vert \leq D\left( z_{i,t}\right) \left\Vert \phi
	_{1}-\phi_{2}\right\Vert $ for all $\phi_{1},\phi_{2}\in\Phi$; (iii)\ the
	function $D(z_{i,t})$ satisfies $\sup_{i,t}\mathbb{E}\left[ \left\vert
	D(z_{i,t})\right\vert ^{p+\delta}\right] <K$ for some integer $p\geq
	\max\{2,k+\delta\}$ and for some $\delta>0$; (iv) $\sup_{i,t}\sup_{\phi\in
		\Phi}\mathbb{E}\left[ \left\vert \varphi\left( z_{i,t},\phi\right)
	\right\vert ^{p+\delta}\right] <K$. Then under Assumption \ref{A1}, we have 
	\begin{equation}
		\max_{g\in\mathcal{G},m\in\mathcal{M}}\mathbb{E}\left[ \left\vert
		(n_{g}T_{m})^{-1/2}\sum_{i\in I_{g}}\sum_{t\in I_{m}}\varphi^{\ast}\left(
		z_{i,t};\phi\right) \right\vert ^{p}\right] \leq K  \label{L0-1}
	\end{equation}
	for any $\phi\in\Phi$, where $\varphi^{\ast}\left( z_{i,t};\phi\right)
	\equiv\varphi\left( z_{i,t};\phi\right) -\mathbb{E}\left[ \varphi\left(
	z_{i,t};\phi\right) \right] $. Moreover for any $\phi_{1},\phi_{2}\in\Phi$%
	\begin{equation}
		\max_{g\in\mathcal{G},m\in\mathcal{M}}\mathbb{E}\left[ \left\vert
		(n_{g}T_{m})^{-1/2}\sum_{i\in I_{g}}\sum_{t\in I_{m}}\left(
		\varphi^{\ast}\left( z_{i,t};\phi_{1}\right) -\varphi^{\ast}\left(
		z_{i,t};\phi_{2}\right) \right) \right\vert ^{p}\right] \leq K\left\Vert
		\phi_{1}-\phi _{2}\right\Vert ^{p}.  \label{L0-2}
	\end{equation}
\end{lemma}

\noindent\textsc{Proof of Lemma \ref{L0}}. Since $\left\{
z_{i,t},t=1,2,\ldots\right\} $\ are independent across $i$, we can use
Rosenthal's inequality of independent random variables (see, e.g., Theorem
2.12 in \cite{HallHeyde1980}) to\ obtain 
\begin{equation}
	\mathbb{E}\left[ \left\vert \sum_{i\in I_{g}}\sum_{t\in I_{m}}\varphi^{\ast
	}\left( z_{i,t};\phi\right) \right\vert ^{p}\right] \leq K\left( \left(
	\sum_{i\in I_{g}}\left\Vert \sum_{t\in I_{m}}\varphi^{\ast}\left(
	z_{i,t};\phi\right) \right\Vert _{2}^{2}\right) ^{p/2}+\sum_{i\in
		I_{g}}\left\Vert \sum_{t\in I_{m}}\varphi^{\ast}\left( z_{i,t};\phi\right)
	\right\Vert _{p}^{p}\right) .  \label{P_L0_1}
\end{equation}
For any $\tilde{p}\geq2$, applying Rosenthal's inequality of strong mixing
processes (see, e.g., Theorem 2 in Section 1.4 of \cite{doukhan1994mixing}),
we get a moment bound on $\sum_{t\in I_{m}}\varphi^{\ast}\left(
z_{i,t};\phi\right) $:%
\begin{equation}
	\mathbb{E}\left[ \left\vert \sum_{t\in I_{m}}\varphi^{\ast}\left(
	z_{i,t};\phi\right) \right\vert ^{\tilde{p}}\right] \leq K\max\left\{ \left(
	\sum_{t\in I_{m}}\left\Vert \varphi^{\ast}\left( z_{i,t};\phi\right)
	\right\Vert _{2+\delta}^{2}\right) ^{\tilde{p}/2},\text{ \ }\sum_{t\in
		I_{m}}\left\Vert \varphi^{\ast}\left( z_{i,t};\phi\right) \right\Vert _{%
		\tilde {p}+\delta}^{\tilde{p}}\right\}  \label{P_L0_2}
\end{equation}
as long as $\mathbb{E}[\left\vert \varphi\left( z_{i,t},\phi\right)
\right\vert ^{\tilde{p}+\delta}]<\infty$. In the rest of the proof, we shall
consider $\tilde{p}=2\ $or $p$. Since $\left\Vert \varphi^{\ast}\left(
z_{i,t};\phi\right) \right\Vert _{\tilde{p}+\delta}\leq K$ by condition (iv)
of the lemma, (\ref{P_L0_2}) further implies that 
\begin{equation}
	\sum_{i\in I_{g}}\mathbb{E}\left[ \left\vert \sum_{t\in I_{m}}\varphi^{\ast
	}\left( z_{i,t};\phi\right) \right\vert ^{\tilde{p}}\right] \leq
	K(n_{g}T_{m}^{\tilde{p}/2}).  \label{P_L0_3}
\end{equation}
Combining the results in (\ref{P_L0_1}) and (\ref{P_L0_3}), we obtain%
\begin{equation}
	\mathbb{E}\left[ \left\vert (n_{g}T_{m})^{-1/2}\sum_{i\in I_{g}}\sum_{t\in
		I_{m}}\varphi^{\ast}\left( z_{i,t};\phi\right) \right\vert ^{p}\right] \leq
	K\left( 1+n_{g}^{1-p/2}\right) \leq K  \label{P_L0_4}
\end{equation}
where the second inequality is due to $p\geq2$. The assertion in (\ref{L0-1}%
) now follows from (\ref{P_L0_4}).

For ease of notation, let $u\left( z_{i,t};\phi_{1},\phi_{2}\right)
\equiv\varphi^{\ast}\left( z_{i,t};\phi_{1}\right) -\varphi^{\ast}\left(
z_{i,t};\phi_{2}\right) $. By the triangle inequality and condition (ii) of
the lemma,%
\begin{equation*}
	\left\vert u\left( z_{i,t};\phi_{1},\phi_{2}\right) \right\vert \leq(D\left(
	z_{i,t}\right) +\mathbb{E}\left[ D\left( z_{i,t}\right) \right] )\left\Vert
	\phi_{1}-\phi_{2}\right\Vert ,
\end{equation*}
which together with condition (iii) of the lemma implies that for $\tilde {p}%
=2\ $or $p$, 
\begin{equation}
	\sup_{i,t}\left\Vert u\left( z_{i,t};\phi_{1},\phi_{2}\right) \right\Vert _{%
		\tilde{p}+\delta}\leq\left\Vert \phi_{1}-\phi_{2}\right\Vert .
	\label{P_L0_5}
\end{equation}
By the same arguments for deriving (\ref{P_L0_1}) and (\ref{P_L0_2}), we can
show that%
\begin{equation*}
	\mathbb{E}\left[ \left\vert \sum_{i\in I_{g}}\sum_{t\in I_{m}}u\left(
	z_{i,t};\phi_{1},\phi_{2}\right) \right\vert ^{p}\right] \leq K\left( \left(
	\sum_{i\in I_{g}}\left\Vert \sum_{t\in I_{m}}u\left( z_{i,t};\phi
	_{1},\phi_{2}\right) \right\Vert _{2}^{2}\right) ^{p/2}+\sum_{i\in
		I_{g}}\left\Vert \sum_{t\in I_{m}}u\left( z_{i,t};\phi_{1},\phi_{2}\right)
	\right\Vert _{p}^{p}\right)
\end{equation*}
and%
\begin{equation*}
	\mathbb{E}\left[ \left\vert \sum_{t\in I_{m}}u\left(
	z_{i,t};\phi_{1},\phi_{2}\right) \right\vert ^{\tilde{p}}\right] \leq
	K\max\left\{ \left( \sum_{t\in I_{m}}\left\Vert u\left(
	z_{i,t};\phi_{1},\phi_{2}\right) \right\Vert _{2+\delta}^{2}\right) ^{\tilde{%
			p}/2},\sum_{t\in I_{m}}\left\Vert u\left( z_{i,t};\phi_{1},\phi_{2}\right)
	\right\Vert _{\tilde{p}+\delta }^{\tilde{p}}\right\} ,
\end{equation*}
respectively, which combined with (\ref{P_L0_5}) yields%
\begin{equation}
	\mathbb{E}\left[ \left\vert \sum_{i\in I_{g}}\sum_{t\in I_{m}}u\left(
	z_{i,t};\phi_{1},\phi_{2}\right) \right\vert ^{p}\right] \leq K\left(
	1+n_{g}^{1-p/2}\right) (n_{g}T_{m})^{p/2}\left\Vert \phi_{1}-\phi
	_{2}\right\Vert ^{p}.  \label{P_L0_6}
\end{equation}
The claim in (\ref{L0-2}) follows directly from (\ref{P_L0_6}).\hfill$Q.E.D.$

\bigskip

\begin{lemma}
	\textit{\label{P_L3C_3&4}}Under Assumptions \ref{A1}, \ref{A3}, \ref{A4} and %
	\ref{A5}, we have%
	\begin{equation}
		\max_{g\in \mathcal{G},m\in \mathcal{M}}\mathbb{E}\left[ \left\Vert
		(n_{g}T_{m})^{1/2}(\hat{\Psi}_{\gamma \gamma ,g,m}-\Psi _{\gamma \gamma
			,g,m})\right\Vert ^{4}\right] \leq K  \label{P_L3C_3}
	\end{equation}%
	and 
	\begin{equation}
		\max_{g\in \mathcal{G},m\in \mathcal{M}}\mathbb{E}\left[ \left\Vert
		(n_{g}T_{m})^{1/2}\hat{\Psi}_{\gamma ,g,m}\right\Vert ^{2}\right] \leq K,
		\label{P_L3C_4}
	\end{equation}
\end{lemma}

\noindent\textsc{Proof of Lemma \ref{P_L3C_3&4} }. Follows from Assumption %
\ref{A3}(ii) and (\ref{L0-1}) in Lemma \ref{L0}.\hfill$Q.E.D.$

\bigskip

\begin{lemma}
	\textit{\label{L1}}Under the conditions of Lemma \ref{L0}, we have 
	\begin{equation}
		\max_{g\in\mathcal{G},m\in\mathcal{M}}\sup_{\phi\in\Phi}\left\vert
		(n_{g}T_{m})^{-1/2}\sum_{i\in I_{g}}\sum_{t\in I_{m}}\varphi^{\ast}\left(
		z_{i,t};\phi\right) \right\vert =O_{p}((GM)^{1/p}),  \label{L1-1}
	\end{equation}
	and for any $\phi^{\prime}\in\limfunc{int}\Phi$ 
	\begin{equation}
		\max_{g\in\mathcal{G},m\in\mathcal{M}}\sup_{\phi\in\Phi}\frac{\left\vert
			(n_{g}T_{m})^{-1/2}\sum_{i\in I_{g}}\sum_{t\in I_{m}}(\varphi\left(
			z_{i,t};\phi\right) -\mathbb{E}\left[ \varphi\left( z_{i,t};\phi^{\ast
			}\right) \right] )\right\vert }{(GM)^{1/p}+\left\Vert \phi-\phi^{\prime
			}\right\Vert }=O_{p}(1).  \label{L1-2}
	\end{equation}
\end{lemma}

\noindent\textsc{Proof of Lemma \ref{L1}}. To prove the claim in (\ref{L1-1}%
), we first show that 
\begin{equation}
	\max_{g\in\mathcal{G},m\in\mathcal{M}}\left\Vert
	\sup_{\phi\in\Phi}(n_{g}T_{m})^{-1/2}\sum_{i\in I_{g}}\sum_{t\in
		I_{m}}\varphi^{\ast}\left( z_{i,t};\phi\right) \right\Vert _{p}\leq K.
	\label{P_L1_2}
\end{equation}
Note that by a maximal inequality under the $L_{p}$-norm (see, e.g., Lemma
2.2.2 of \cite{vanderVaartWellner1996}) and (\ref{P_L1_2}),%
\begin{align}
	& \left\Vert \max_{g\in\mathcal{G},m\in\mathcal{M}}\sup_{\phi\in%
		\Phi}(n_{g}T_{m})^{-1/2}\sum_{i\in I_{g}}\sum_{t\in
		I_{m}}\varphi^{\ast}\left( z_{i,t};\phi\right) \right\Vert _{p}  \notag \\
	& \leq K(GM)^{1/p}\max_{g\in\mathcal{G},m\in\mathcal{M}}\left\Vert \sup
	_{\phi\in\Phi}(n_{g}T_{m})^{-1/2}\sum_{i\in I_{g}}\sum_{t\in
		I_{m}}\varphi^{\ast}\left( z_{i,t};\phi\right) \right\Vert _{p}\leq
	K(GM)^{1/p}  \label{P_L1_1}
\end{align}
which together with Markov's inequality implies that 
\begin{equation}
	\max_{g\in\mathcal{G},m\in\mathcal{M}}\sup_{\phi\in\Phi}(n_{g}T_{m})^{-1/2}%
	\sum_{i\in I_{g}}\sum_{t\in I_{m}}\varphi^{\ast}\left( z_{i,t};\phi\right)
	=O_{p}((GM)^{1/p}).  \label{P_L1_3}
\end{equation}
The result in (\ref{P_L1_3}) is then used to show (\ref{L1-1}).

For ease of notation, we set for each $g\in \mathcal{G}\ $and $m\in \mathcal{%
	M}$, 
\begin{equation*}
	\pi _{g,m}(\phi )\equiv (n_{g}T_{m})^{-1/2}\sum_{i\in I_{g}}\sum_{t\in
		I_{m}}\varphi ^{\ast }\left( z_{i,t};\phi \right) ,\quad \phi \in \Phi .
\end{equation*}%
Construct nested sets $\Phi _{0}\subset \Phi _{1}\cdots \subset \Phi $ such
that $\Phi _{0}=\{\phi _{0}\}$ for any given $\phi _{0}$ $\in \Phi $, and
for each $j\geq 1$, $\Phi _{j}$ is a maximal set of points such that each
pair of distinct elements in $\Phi $ has distance greater than $2^{-j}$.
Since $\Phi $ is compact, the number of points in $\Phi _{j}$ is less than $%
K(2^{j})^{k}$. Link every point $\phi _{j+1}\in \Phi _{j+1}$ to a unique $%
\phi _{j}\in \Phi _{j}$ such that $\left\Vert \phi _{j+1}-\phi
_{j}\right\Vert \leq 2^{-j}$. Then for any $J\geq 0$ and $\phi _{J+1}\in
\phi _{J+1}$, we can construct a chain $\phi _{J+1},\ldots ,\phi _{0}$, and
hence, by the triangle inequality%
\begin{equation}
	\left\vert \pi _{g,m}(\phi _{J+1})-\pi _{g,m}(\phi _{0})\right\vert
	=\left\vert \sum_{j=0}^{J}\left[ \pi _{g,m}(\phi _{j+1})-\pi _{g,m}(\phi
	_{j})\right] \right\vert \leq \sum_{j=0}^{J}\max \left\vert \pi _{g,m}(\phi
	_{j+1})-\pi _{g,m}(\phi _{j})\right\vert  \label{P_L1_4}
\end{equation}%
where, for each $j$, the maximum is taken over all links $(\phi _{j+1},\phi
_{j})$ from $\Phi _{j+1}$ to $\Phi _{j}$ (with the total number less than $%
K(2^{j+1})^{k}$). We then observe%
\begin{align*}
	\left\Vert \max_{\phi \in \Phi _{J+1}}\left\vert \pi _{g,m}(\phi
	)\right\vert \right\Vert _{p}& \leq \sum_{j=0}^{J}\left\Vert \max \left\vert
	\pi _{g,m}(\phi _{j+1})-\pi _{g,m}(\phi _{j})\right\vert \right\Vert
	_{p}+\left\Vert \pi _{g,m}(\phi _{0})\right\Vert _{p} \\
	& \leq K\sum_{j=0}^{J}(2^{j+1})^{k/p}\max \left\Vert \pi _{g,m}(\phi
	_{j+1})-\pi _{g,m}(\phi _{j})\right\Vert _{p}+K \\
	& \leq K\sum_{j=0}^{J}(2^{j})^{k/p}2^{-j}+\left\Vert \pi _{g,m}(\phi
	_{0})\right\Vert _{p}\leq K,
\end{align*}%
where the first inequality is by (\ref{P_L1_4}); the second inequality is by
the maximal inequality under the $L_{p}$-norm and (\ref{L0-1}); the third
inequality follows from (\ref{L0-2}) in Lemma \ref{L0}; and the last
inequality holds because $\sum_{j}(2^{j})^{k/p}2^{-j}<\infty $ as implied by 
$p>k$. Since the stochastic process $\pi _{g,m}(${$\phi $}$)$ indexed by\ {$%
	\phi $} is separable, by letting $J\rightarrow \infty $, we further have 
\begin{equation*}
	\left\Vert \sup_{{\phi }\in \Phi }\left\vert \pi _{g,m}({\phi })\right\vert
	\right\Vert _{p}\leq K
\end{equation*}%
for any $g\in \mathcal{G}\ $and $m\in \mathcal{M}$, which shows (\ref{P_L1_2}%
). Using the same arguments, we can show that%
\begin{equation}
	\max_{g\in \mathcal{G},m\in \mathcal{M}}\sup_{\phi \in \Phi
	}(n_{g}T_{m})^{-1/2}\sum_{i\in I_{g}}\sum_{t\in I_{m}}-\varphi ^{\ast
	}\left( z_{i,t};\phi \right) =O_{p}((GM)^{1/p}).  \label{P_L1_5}
\end{equation}%
Since%
\begin{equation}
	\left\vert \sum_{i\in I_{g}}\sum_{t\in I_{m}}\varphi ^{\ast }\left(
	z_{i,t};\phi \right) \right\vert \leq \max \left\{ \sum_{i\in
		I_{g}}\sum_{t\in I_{m}}\varphi ^{\ast }\left( z_{i,t};\phi \right) ,\text{ }%
	\sum_{i\in I_{g}}\sum_{t\in I_{m}}-\varphi ^{\ast }\left( z_{i,t};\phi
	\right) \right\} ,  \label{P_L1_6}
\end{equation}%
the claim in (\ref{L1-1}) follows from (\ref{P_L1_3}), (\ref{P_L1_5}) and (%
\ref{P_L1_6}).

To show (\ref{L1-2}), we note that%
\begin{align*}
	& \max_{g\in\mathcal{G},m\in\mathcal{M}}\sup_{\phi\in\Phi}\frac{\left\vert
		(n_{g}T_{m})^{-1}\sum_{i\in I_{g}}\sum_{t\in I_{m}}(\varphi\left(
		z_{i,t};\phi\right) -\mathbb{E}\left[ \varphi\left( z_{i,t};\phi^{\prime
		}\right) \right] )\right\vert }{(n_{g}T_{m})^{-1/2}(GM)^{1/p}+\left\Vert
		\phi-\phi^{\prime}\right\Vert } \\
	& \leq\max_{g\in\mathcal{G},m\in\mathcal{M}}\sup_{\phi\in\Phi}\frac {%
		\left\vert (n_{g}T_{m})^{-1/2}\sum_{i\in I_{g}}\sum_{t\in I_{m}}\varphi
		^{\ast}\left( z_{i,t};\phi\right) \right\vert }{(GM)^{1/p}} \\
	& \text{ \ \ }+\max_{g\in\mathcal{G},m\in\mathcal{M}}\sup_{\phi\in\Phi}\frac{%
		\left\vert (n_{g}T_{m})^{-1}\sum_{i\in I_{g}}\sum_{t\in I_{m}}\mathbb{E}%
		\left[ \varphi\left( z_{i,t};\phi\right) -\varphi\left(
		z_{i,t};\phi^{\prime}\right) \right] \right\vert }{%
		(n_{g}T_{m})^{-1/2}(GM)^{1/p}+\left\Vert \phi-\phi^{\prime}\right\Vert } \\
	& \leq\max_{g\in\mathcal{G},m\in\mathcal{M}}\sup_{\phi\in\Phi}\frac {%
		\sup_{i,t}\left\vert \mathbb{E}\left[ \left\vert D(z_{i,t}\right\vert )%
		\right] \right\vert \left\Vert \phi-\phi^{\prime}\right\Vert }{%
		(n_{g}T_{m})^{-1/2}(GM)^{1/p}+\left\Vert \phi-\phi^{\prime}\right\Vert }%
	+O_{p}(1)\overset{}{=}O_{p}(1),
\end{align*}
where the first inequality is by the triangle inequality, the second
inequality is by condition\ (ii) of Lemma \ref{L0} and (\ref{L1-1}), the
equality is by condition\ (iii) of Lemma \ref{L0}.\hfill$Q.E.D.$

\bigskip

For the sake of simplicity in notation, we assume that the observations
across $i$ are arranged in a manner where the group assignment function $%
g(\cdot)$ is non-decreasing. This arrangement is inconsequential because of
the independence assumption stated in Assumption \ref{A1}.(iii).

\begin{lemma}
	\textit{\label{L2}Let }$\varphi_{j}(\cdot)$: $\mathcal{Z}\longrightarrow 
	\mathbb{R}$ be functions on the support $\mathcal{Z}$ of $z_{i,t}$ for $%
	j=1,2 $. Suppose that for $j=1,2$: (i) $\mathbb{E}\left[ \varphi_{j}\left(
	z_{i,t}\right) \right] =0$; (ii) $\sup_{i,t}\mathbb{E}[\left\vert
	\varphi_{j}\left( z_{i,t}\right) \right\vert ^{4+\delta}]<K$ for some $%
	\delta>0$. Then under Assumption \ref{A1}, we have%
	\begin{equation}
		\mathrm{Var}\left( M^{-1/2}\sum_{m\in\mathcal{M}}T_{m}^{-1}\sum_{t_{1}\in
			I_{m}}\sum_{t_{2}\in
			I_{m}}\varphi_{1}(z_{i,t_{1}})\varphi_{2}(z_{i,t_{2}})\right) \leq K,
		\label{L2-1}
	\end{equation}
	and%
	\begin{equation}
		\mathbb{E}\left[ \left\Vert \sum_{g\in\mathcal{G}}n_{g}^{-1}%
		\sum_{i_{1}=i_{g}^{\min}+1}^{i_{g}^{\max}}\sum_{i_{2}=i_{g}^{%
				\min}}^{i_{1}-1}\left( \sum_{m\in\mathcal{M}}T_{m}^{-1}\left( \sum_{t\in
			I_{m}}\varphi_{1}(z_{i_{1},t})\right) \left( \sum_{t\in
			I_{m}}\varphi_{2}(z_{i_{2},t})\right) \right) \right\Vert ^{2}\right] \leq
		K(GM),  \label{L2-2}
	\end{equation}
	where $i_{g}^{\max}\equiv\max(I_{g})$ and $i_{g}^{\min}\equiv\min(I_{g})$
	for any $g\in\mathcal{G}$.
\end{lemma}

\noindent\textsc{Proof of Lemma \ref{L2}}. For ease of notation, let\ $%
Y_{m}\equiv T_{m}^{-1}\sum_{t_{1}\in I_{m}}\sum_{t_{2}\in
	I_{m}}\varphi_{1}(z_{i,t_{1}})\varphi_{2}(z_{i,t_{2}})$ for any $m\in%
\mathcal{M}$ and $\delta^{\prime}\equiv\delta/3$.\ For any $i$, let $%
\mathcal{F}_{-\infty }^{i,t}$ denote the sigma field generated by $\left\{
z_{is}\right\} _{s=-\infty}^{t}$ and $\mathcal{F}_{t}^{i,+\infty}$ denote
the sigma field generated by $\left\{ z_{is}\right\} _{s=t}^{+\infty}$.
Consider any $m_{1},m_{2}\in\mathcal{M}$ with $m_{2}>m_{1}$. It is clear
that $Y_{m_{1}}$ is $\mathcal{F}_{-\infty}^{i,t_{m_{1}}^{\max}}$ measurable
and $Y_{m_{2}}$ is $\mathcal{F}_{t_{m_{2}}^{\min}}^{i,+\infty}$ measurable
where $t_{m}^{\max }\equiv\max(I_{m})$ and $t_{m}^{\min}\equiv\min(I_{m})$
for any $m\in \mathcal{M}$. By Corollary A.2. in \cite{HallHeyde1980},%
\begin{align*}
	\left\vert \mathrm{Cov}\left( Y_{m_{1}},Y_{m_{2}}\right) \right\vert & \leq
	K\left( \alpha_{i}(t_{m_{2}}^{\min}-t_{m_{1}}^{\max})\right)
	^{\delta^{\prime}/(2+\delta^{\prime})}\left\Vert Y_{m_{1}}\right\Vert
	_{2+\delta^{\prime}}\left\Vert Y_{m_{2}}\right\Vert _{2+\delta^{\prime}} \\
	& \leq Ka^{(t_{m_{2}}^{\min}-t_{m_{1}}^{\max})\delta^{\prime}/(2+\delta
		^{\prime})}\max_{m\in\mathcal{M}}\left\Vert Y_{m}\right\Vert _{2+\delta
		^{\prime}}^{2} \\
	& \leq Ka^{(m_{2}-m_{1})\delta^{\prime}/(2+\delta^{\prime})}\max _{m\in%
		\mathcal{M}}\left\Vert Y_{m}\right\Vert _{2+\delta^{\prime}}^{2}
\end{align*}
where the second inequality is by Assumption \ref{A1}(iv), and the last
inequality is by $t_{m_{2}}^{\min}-t_{m_{1}}^{\max}\geq m_{2}-m_{1}$ and $%
a\in(0,1)$. Therefore%
\begin{align}
	\mathrm{Var}\left( M^{-1/2}\sum_{m\in\mathcal{M}}Y_{m}\right) &
	=M^{-1}\sum_{m\in\mathcal{M}}\mathrm{Var}\left( Y_{m}\right)
	+2M^{-1}\sum_{l=1}^{M-1}\sum_{j=1}^{M-l}\mathrm{Cov}\left(
	Y_{j},Y_{j+l}\right)  \notag \\
	& \leq K\left(
	1+2\sum_{l=1}^{M-1}(1-lM^{-1})a^{l\delta^{\prime}/(2+\delta^{\prime})}%
	\right) \max_{m\in\mathcal{M}}\left\Vert Y_{m}\right\Vert
	_{2+\delta^{\prime}}^{2}  \notag \\
	& \leq K\left(
	1+2\sum_{l=1}^{M-1}a^{l\delta^{\prime}/(2+\delta^{\prime})}\right) \max_{m\in%
		\mathcal{M}}\left\Vert Y_{m}\right\Vert _{2+\delta ^{\prime}}^{2}\leq
	K\max_{m\in\mathcal{M}}\left\Vert Y_{m}\right\Vert _{2+\delta^{\prime}}^{2}.
	\label{P_L2_1}
\end{align}
By H\"{o}lder's inequality,%
\begin{align}
	\left\Vert Y_{m}\right\Vert _{2+\delta^{\prime}}^{2} & =\left( \mathbb{E}%
	\left[ \left\vert \left( T_{m}^{-1/2}\sum_{t\in I_{m}}\varphi
	_{1}(z_{i,t})\right) \left( T_{m}^{-1/2}\sum_{t\in
		I_{m}}\varphi_{2}(z_{i,t})\right) \right\vert ^{2+\delta^{\prime}}\right]
	\right) ^{\frac {2}{2+\delta^{\prime}}}  \notag \\
	& \leq\left( \mathbb{E}\left[ \left\vert T_{m}^{-1/2}\sum_{t\in
		I_{m}}\varphi_{1}(z_{i,t})\right\vert ^{4+2\delta^{\prime}}\right] \mathbb{E}%
	\left[ \left\vert T_{m}^{-1/2}\sum_{t\in
		I_{m}}\varphi_{2}(z_{i,t})\right\vert ^{4+2\delta^{\prime}}\right] \right) ^{%
		\frac{1}{2+\delta ^{\prime}}}.  \label{P_L2_2}
\end{align}
By the same arguments for deriving (\ref{P_L0_2}), we can show that%
\begin{equation}
	\mathbb{E}\left[ \left\vert \sum_{t\in I_{m}}\varphi_{j}(z_{i,t})\right\vert
	^{4+2\delta^{\prime}}\right] \leq\max\left\{ \sum_{t\in I_{m}}\left\Vert
	\varphi_{j}(z_{i,t})\right\Vert
	_{4+3\delta^{\prime}}^{4+2\delta^{\prime}},\left( \sum_{t\in
		I_{m}}\left\Vert \varphi_{j}(z_{i,t})\right\Vert
	_{2+\delta^{\prime}}^{2}\right) ^{2+\delta^{\prime}}\right\} \leq
	KT_{m}^{2+\delta^{\prime}}  \label{P_L2_3}
\end{equation}
where the second inequality is by condition (ii) of the lemma. Combining the
results in (\ref{P_L2_2}) and (\ref{P_L2_3}), we have $\max_{m\in\mathcal{M}%
}\left\Vert Y_{m}\right\Vert _{2+\delta^{\prime}}^{2}\leq K$, which together
with (\ref{P_L2_1}) shows (\ref{L2-1}).

To show (\ref{L2-2}), we first use Assumption \ref{A1}(iii) and conditions
(i, ii) of the lemma to observe that 
\begin{align}
	& \mathbb{E}\left[ \left\vert \sum_{g\in\mathcal{G}}n_{g}^{-1}\sum
	_{i_{1}=i_{g}^{\min}+1}^{i_{g}^{\max}}\sum_{i_{2}=i_{g}^{\min}}^{i_{1}-1}%
	\left( \sum_{m\in\mathcal{M}}T_{m}^{-1}\left( \sum_{t\in
		I_{m}}\varphi_{1}(z_{i_{1},t})\right) \left( \sum_{t\in
		I_{m}}\varphi_{2}(z_{i_{2},t})\right) \right) \right\vert ^{2}\right]  \notag
	\\
	& =\sum_{g\in\mathcal{G}}n_{g}^{-2}\mathbb{E}\left[ \left\vert \sum
	_{i_{1}=i_{g}^{\min}+1}^{i_{g}^{\max}}\sum_{i_{2}=i_{g}^{\min}}^{i_{1}-1}%
	\left( \sum_{m\in\mathcal{M}}T_{m}^{-1}\left( \sum_{t\in
		I_{m}}\varphi_{1}(z_{i_{1},t})\right) \left( \sum_{t\in
		I_{m}}\varphi_{2}(z_{i_{2},t})\right) \right) \right\vert ^{2}\right]  \notag
	\\
	& =\sum_{g\in\mathcal{G}}n_{g}^{-2}\sum_{i_{1}=i_{g}^{\min}+1}^{i_{g}^{%
			\max}}\sum_{i_{2}=i_{g}^{\min}}^{i_{1}-1}\mathbb{E}\left[ \left\vert \sum
	_{m\in\mathcal{M}}T_{m}^{-1}\left( \sum_{t\in
		I_{m}}\varphi_{1}(z_{i_{1},t})\right) \left( \sum_{t\in
		I_{m}}\varphi_{2}(z_{i_{2},t})\right) \right\vert ^{2}\right] .
	\label{P_L2_4}
\end{align}
For any $m_{1},m_{2}\in\mathcal{M}$,%
\begin{align}
	& \left\vert \mathbb{E}\left[ (T_{m_{1}}T_{m_{2}})^{-1}\left( \sum_{t\in
		I_{m_{1}}}\varphi_{1}(z_{i_{1},t})\right) \left( \sum_{t\in
		I_{m_{1}}}\varphi_{2}(z_{i_{2},t})\right) \left( \sum_{t\in
		I_{m_{2}}}\varphi _{1}(z_{i_{1},t})\right) \left( \sum_{t\in
		I_{m_{2}}}\varphi_{2}(z_{i_{2},t})\right) \right] \right\vert  \notag \\
	& =\left\vert T_{m_{1}}^{-1}\mathrm{Cov}\left( \sum_{t\in
		I_{m_{1}}}\varphi_{1}(z_{i_{1},t}),\sum_{t\in
		I_{m_{2}}}\varphi_{1}(z_{i_{1},t})\right) \right\vert \left\vert
	T_{m_{2}}^{-1}\mathrm{Cov}\left( \sum_{t\in
		I_{m_{1}}}\varphi_{2}(z_{i_{2},t}),\sum_{t\in
		I_{m_{2}}}\varphi_{2}(z_{i_{2},t})\right) \right\vert  \notag \\
	& \leq Ka^{2|m_{2}-m_{1}|\delta^{\prime}/(2+\delta^{\prime})}\max _{m\in%
		\mathcal{M}}\left\Vert T_{m}^{-1/2}\sum_{t\in
		I_{m}}\varphi_{1}(z_{i_{1},t})\right\Vert _{2+\delta^{\prime}}^{2}\max_{m\in%
		\mathcal{M}}\left\Vert T_{m}^{-1/2}\sum_{t\in
		I_{m}}\varphi_{2}(z_{i_{2},t})\right\Vert _{2+\delta^{\prime}}^{2}  \notag \\
	& \leq Ka^{2|m_{2}-m_{1}|\delta^{\prime}/(2+\delta^{\prime})}  \label{P_L2_5}
\end{align}
where the first inequality is by Corollary A.2. in \cite{HallHeyde1980} and
Assumption \ref{A1}(iv), the second inequality is by (\ref{P_L2_3}) and H%
\"{o}lder's inequality. Using (\ref{P_L2_5}), we can show that%
\begin{equation}
	\mathbb{E}\left[ \left\vert \sum_{m\in\mathcal{M}}T_{m}^{-1}\left(
	\sum_{t\in I_{m}}\varphi_{1}(z_{i_{1},t})\right) \left( \sum_{t\in
		I_{m}}\varphi_{2}(z_{i_{2},t})\right) \right\vert ^{2}\right] \leq KM.
	\label{P_L2_6}
\end{equation}
The assertion in (\ref{L2-2}) follows from (\ref{P_L2_4}) and (\ref{P_L2_6}%
).\hfill$Q.E.D.$

\bigskip

\begin{lemma}
	\label{L2A}\textit{Let }$\hat{\varphi}_{j,g,m}\equiv
	(n_{g}T_{m})^{-1}\sum_{i\in I_{g}}\sum_{t\in I_{m}}\varphi _{j}\left(
	z_{i,t}\right) $,\textit{\ where }$\varphi _{j}(\cdot )$: $\mathcal{Z}%
	\longrightarrow \mathbb{R}$ are functions on the support $\mathcal{Z}$ of $%
	z_{i,t}$ for $j=1,2$. Then under the conditions of Lemma \ref{L2}, we have%
	\begin{equation}
		\sum_{g\in \mathcal{G}}\sum_{m\in \mathcal{M}}(n_{g}T_{m})\hat{\varphi}%
		_{1,g,m}\hat{\varphi}_{2,g,m}=\sum_{g\in \mathcal{G}}\sum_{m\in \mathcal{M}%
		}(n_{g}T_{m})\mathbb{E}\left[ \hat{\varphi}_{1,g,m}\hat{\varphi}_{2,g,m}%
		\right] +O_{p}((GM)^{1/2})=O_{p}(GM).  \label{L2A-1}
	\end{equation}
\end{lemma}

\noindent \textsc{Proof of Lemma \ref{L2A}}. First note that by Assumption %
\ref{A1}(iii), 
\begin{equation*}
	\sum_{g\in \mathcal{G}}\sum_{m\in \mathcal{M}}(n_{g}T_{m})\mathbb{E}\left[ 
	\hat{\varphi}_{1,g,m}\hat{\varphi}_{2,g,m}\right] =n_{g}^{-1}\mathbb{E}\left[
	T_{m}^{-1}\sum_{t_{1},t_{2}\in I_{m}}\varphi _{1}(z_{i,t_{1}})\varphi
	_{2}(z_{i,t_{2}})\right]
\end{equation*}%
where $\sum_{t_{1},t_{2}\in I_{m}}$ denotes the double summation over $%
t_{1},t_{2}\in I_{m}$. Therefore, for any $g\in \mathcal{G}$ and $m\in 
\mathcal{M}$, we can express the equations%
\begin{align}
	& \sum_{g\in \mathcal{G}}\sum_{m\in \mathcal{M}}(n_{g}T_{m})\hat{\varphi}%
	_{1,g,m}\hat{\varphi}_{2,g,m}-\sum_{g\in \mathcal{G}}\sum_{m\in \mathcal{M}%
	}(n_{g}T_{m})\mathbb{E}\left[ \hat{\varphi}_{1,g,m}\hat{\varphi}_{2,g,m}%
	\right]  \notag \\
	& =\sum_{g\in \mathcal{G}}n_{g}^{-1}\sum_{i\in I_{g}}\sum_{m\in \mathcal{M}%
	}T_{m}^{-1}\sum_{t_{1},t_{2}\in I_{m}}\left( \varphi
	_{1}(z_{i,t_{1}})\varphi _{2}(z_{i,t_{2}})\right) ^{\ast }  \notag \\
	& \text{ \ \ }+\sum_{g\in \mathcal{G}}n_{g}^{-1}\sum_{i_{1}=i_{g}^{\min
		}+1}^{i_{g}^{\max }}\sum_{i_{2}=i_{g}^{\min }}^{i_{1}-1}\left( \sum_{m\in 
		\mathcal{M}}T_{m}^{-1}\left( \sum_{t\in I_{m}}\varphi
	_{1}(z_{i_{1},t})\right) \left( \sum_{t\in I_{m}}\varphi
	_{2}(z_{i_{2},t})\right) \right)  \notag \\
	& \text{ \ \ }+\sum_{g\in \mathcal{G}}n_{g}^{-1}\sum_{i_{1}=i_{g}^{\min
		}+1}^{i_{g}^{\max }}\sum_{i_{2}=i_{g}^{\min }}^{i_{1}-1}\left( \sum_{m\in 
		\mathcal{M}}T_{m}^{-1}\left( \sum_{t\in I_{m}}\varphi
	_{2}(z_{i_{1},t})\right) \left( \sum_{t\in I_{m}}\varphi
	_{1}(z_{i_{2},t})\right) \right)  \label{P_L2A_1}
\end{align}%
where $\left( \varphi _{1}(z_{i,t_{1}})\varphi _{2}(z_{i,t_{2}})\right)
^{\ast }\equiv \varphi _{1}(z_{i,t_{1}})\varphi _{2}(z_{i,t_{2}})-\mathbb{E}%
\left[ \varphi _{1}(z_{i,t_{1}})\varphi _{2}(z_{i,t_{2}})\right] $. For any $%
m\in \mathcal{M}$, we can use the triangle inequality, H\"{o}lder's
inequality, Corollary A.2. in \cite{HallHeyde1980} and Assumption \ref{A1}%
(iv)\ to deduce that%
\begin{align}
	& \left\vert \mathbb{E}\left[ T_{m}^{-1}\sum_{t_{1},t_{2}\in I_{m}}\varphi
	_{1}(z_{i,t_{1}})\varphi _{2}(z_{i,t_{2}})\right] \right\vert  \notag \\
	& \leq T_{m}^{-1}\sum_{t\in I_{m}}\left\vert \mathbb{E}\left[ \varphi
	_{1}(z_{i,t})\varphi _{2}(z_{i,t})\right] \right\vert
	+T_{m}^{-1}\sum_{l=1}^{T_{m}-1}\sum_{t=t_{m}^{\min }}^{t_{m}^{\max
		}-l}\left\vert \mathbb{E}\left[ \varphi _{1}(z_{i,t})\varphi _{2}(z_{i,t+l})%
	\right] \right\vert  \notag \\
	& \text{ \ \ }+T_{m}^{-1}\sum_{l=1}^{T_{m}-1}\sum_{t=t_{m}^{\min
	}}^{t_{m}^{\max }-l}\left\vert \mathbb{E}\left[ \varphi _{2}(z_{i,t})\varphi
	_{1}(z_{i,t+l})\right] \right\vert  \notag \\
	& \leq K\left( 1+\sum_{l=1}^{T_{m}-1}(1-lT_{m}^{-1})a^{l\delta /(2+\delta
		)}\right) \max_{j=1,2}\sup_{i,t}\left\Vert \varphi _{j}(z_{i,t})\right\Vert
	_{2+\delta }^{2}\leq K,  \label{P_L2A_2}
\end{align}%
where $t_{m}^{\max }\equiv \max (I_{m})$ and $t_{m}^{\min }\equiv \min
(I_{m})$, and the last inequality is by condition (ii) of Lemma \ref{L2}.
This implies that 
\begin{equation}
	\left\vert \sum_{g\in \mathcal{G}}\sum_{m\in \mathcal{M}}(n_{g}T_{m})\mathbb{%
		E}\left[ \hat{\varphi}_{1,g,m}\hat{\varphi}_{2,g,m}\right] \right\vert \leq
	K(GM).  \label{P_L2A_2B}
\end{equation}%
By Assumption \ref{A1}(iii) and (\ref{L2-1}) in Lemma \ref{L2},%
\begin{align*}
	& \mathrm{Var}\left( \sum_{g\in \mathcal{G}}n_{g}^{-1}\sum_{i\in
		I_{g}}\sum_{m\in \mathcal{M}}T_{m}^{-1}\sum_{t_{1},t_{2}\in I_{m}}\varphi
	_{1}(z_{i,t_{1}})\varphi _{2}(z_{i,t_{2}})\right) \\
	& =\sum_{g\in \mathcal{G}}n_{g}^{-2}\sum_{i\in I_{g}}\mathrm{Var}\left(
	\sum_{m\in \mathcal{M}}T_{m}^{-1}\sum_{t_{1},t_{2}\in I_{m}}\varphi
	_{1}(z_{i,t_{1}})\varphi _{2}(z_{i,t_{2}})\right) \leq K(GM)\max_{g\in 
		\mathcal{G}}n_{g}^{-1}
\end{align*}%
which together with Markov's inequality and (\ref{P_L2A_2}) implies that%
\begin{equation}
	\sum_{g\in \mathcal{G}}n_{g}^{-1}\sum_{i\in I_{g}}\sum_{m\in \mathcal{M}%
	}T_{m}^{-1}\sum_{t_{1},t_{2}\in I_{m}}\left( \varphi
	_{1}(z_{i,t_{1}})\varphi _{2}(z_{i,t_{2}})\right) ^{\ast
	}=O_{p}((GM)^{1/2}\max_{g\in \mathcal{G}}n_{g}^{-1/2}).  \label{P_L2A_3}
\end{equation}%
By Markov's inequality and (\ref{L2-2}) in Lemma \ref{L2}, we have 
\begin{equation}
	(GM)^{-1/2}\sum_{g\in \mathcal{G}}n_{g}^{-1}\sum_{i_{1}=i_{g}^{\min
		}+1}^{i_{g}^{\max }}\sum_{i_{2}=i_{g}^{\min }}^{i_{1}-1}\left( \sum_{m\in 
		\mathcal{M}}T_{m}^{-1}\left( \sum_{t\in I_{m}}\varphi
	_{1}(z_{i_{1},t})\right) \left( \sum_{t\in I_{m}}\varphi
	_{2}(z_{i_{2},t})\right) \right) =O_{p}(1)  \label{P_L2A_4}
\end{equation}%
and 
\begin{equation}
	(GM)^{-1/2}\sum_{g\in \mathcal{G}}n_{g}^{-1}\sum_{i_{1}=i_{g}^{\min
		}+1}^{i_{g}^{\max }}\sum_{i_{2}=i_{g}^{\min }}^{i_{1}-1}\left( \sum_{m\in 
		\mathcal{M}}T_{m}^{-1}\left( \sum_{t\in I_{m}}\varphi
	_{2}(z_{i_{1},t})\right) \left( \sum_{t\in I_{m}}\varphi
	_{1}(z_{i_{2},t})\right) \right) =O_{p}(1),  \label{P_L2A_5}
\end{equation}%
which together with (\ref{P_L2A_1}) and (\ref{P_L2A_3}) shows the first
equality in (\ref{L2A-1}). The second equality in (\ref{L2A-1}) holds by (%
\ref{P_L2A_2B}).\hfill $Q.E.D.$

\bigskip

\begin{lemma}
	\textit{\label{L2B}Let }$\varphi _{j}(\cdot )$: $\mathcal{Z}\longrightarrow 
	\mathbb{R}$ be functions on the support $\mathcal{Z}$ of $z_{i,t}$ for $%
	j=1,2,3$. Suppose that for $j=1,2,3$: (i) $\mathbb{E}\left[ \varphi
	_{j}\left( z_{i,t}\right) \right] =0$; (ii) $\sup_{i,t}\mathbb{E}[\left\vert
	\varphi _{j}\left( z_{i,t}\right) \right\vert ^{6+\delta }]<K$ for some $%
	\delta >0$. Then under Assumption \ref{A1}, we have%
	\begin{equation*}
		(GM)^{-1/2}\sum_{g\in \mathcal{G}}\sum_{m\in \mathcal{M}}(n_{g}T_{m})\hat{%
			\varphi}_{1,g,m}\hat{\varphi}_{2,g,m}\hat{\varphi}_{3,g,m}=O_{p}\left(
		\max_{g\in \mathcal{G},m\in \mathcal{M}}\left(
		(n_{g}T_{m})^{-1/2}+(GM)^{1/2}(n_{g}T_{m})^{-1}\right) \right)
	\end{equation*}%
	where $\hat{\varphi}_{j,g,m}\equiv (n_{g}T_{m})^{-1}\sum_{i\in
		I_{g}}\sum_{t\in I_{m}}\varphi _{j}\left( z_{i,t}\right) $ for $j=1,2,3$.
\end{lemma}

\noindent \textsc{Proof of Lemma \ref{L2B}}. To prove the lemma, it is
sufficient to show that 
\begin{equation}
	\max_{g\in \mathcal{G},m\in \mathcal{M}}\left\vert (n_{g}T_{m})\mathbb{E}%
	\left[ \hat{\varphi}_{1,g,m}\hat{\varphi}_{2,g,m}\hat{\varphi}_{3,g,m}\right]
	\right\vert \leq K\max_{g\in \mathcal{G},m\in \mathcal{M}}(n_{g}T_{m})^{-1}
	\label{L2B_1}
\end{equation}%
and%
\begin{equation}
	\mathrm{Var}\left( \sum_{g\in \mathcal{G}}\sum_{m\in \mathcal{M}}(n_{g}T_{m})%
	\hat{\varphi}_{1,g,m}\hat{\varphi}_{2,g,m}\hat{\varphi}_{3,g,m}\right) \leq
	K(GM)\max_{g\in \mathcal{G},m\in \mathcal{M}}(n_{g}T_{m})^{-1}.
	\label{L2B_2}
\end{equation}%
Because by the triangle inequality,%
\begin{align}
	& \left\vert (GM)^{-1/2}\sum_{g\in \mathcal{G}}\sum_{m\in \mathcal{M}%
	}(n_{g}T_{m})\hat{\varphi}_{1,g,m}\hat{\varphi}_{2,g,m}\hat{\varphi}%
	_{3,g,m}\right\vert  \notag \\
	& \leq \left\vert (GM)^{-1/2}\sum_{g\in \mathcal{G}}\sum_{m\in \mathcal{M}%
	}(n_{g}T_{m})\left( \hat{\varphi}_{1,g,m}\hat{\varphi}_{2,g,m}\hat{\varphi}%
	_{3,g,m}-\mathbb{E}\left[ \hat{\varphi}_{1,g,m}\hat{\varphi}_{2,g,m}\hat{%
		\varphi}_{3,g,m}\right] \right) \right\vert  \notag \\
	& \text{ \ \ }+(GM)^{-1/2}\sum_{g\in \mathcal{G}}\sum_{m\in \mathcal{M}%
	}\left\vert (n_{g}T_{m})\mathbb{E}\left[ \hat{\varphi}_{1,g,m}\hat{\varphi}%
	_{2,g,m}\hat{\varphi}_{3,g,m}\right] \right\vert  \notag \\
	& =O_{p}\left( \max_{g\in \mathcal{G},m\in \mathcal{M}}\left(
	(n_{g}T_{m})^{-1/2}+(GM)^{1/2}(n_{g}T_{m})^{-1}\right) \right) ,
	\label{L2B_3}
\end{align}%
where equality follows by (\ref{L2B_1}), (\ref{L2B_2}) and Markov's
inequality. The assertion of the lemma follows directly from (\ref{L2B_3}).

We first show (\ref{L2B_1}). By the independence assumption in Assumption %
\ref{A1} and condition (i) in the lemma, we have%
\begin{equation}
	\mathbb{E}\left[ (n_{g}T_{m})^{2}\hat{\varphi}_{1,g,m}\hat{\varphi}_{2,g,m}%
	\hat{\varphi}_{3,g,m}\right] =(n_{g}T_{m})^{-1}\sum_{i\in I_{g}}\mathbb{E}%
	\left[ \sum_{t_{1},t_{2},t_{3}\in I_{m}}\varphi _{1}\left(
	z_{i,t_{1}}\right) \varphi _{2}\left( z_{i,t_{2}}\right) \varphi _{2}\left(
	z_{i,t_{3}}\right) \right] ,  \label{P_L2B_1}
\end{equation}%
where $\sum_{t_{1},t_{2},t_{3}\in I_{m}}$ denotes the triple summation $%
\sum_{t_{1}\in I_{m}}\sum_{t_{2}\in I_{m}}\sum_{t_{3}\in I_{m}}$. Next, we
write%
\begin{align}
	\sum_{t_{1},t_{2},t_{3}\in I_{m}}\varphi _{1}\left( z_{i,t_{1}}\right)
	\varphi _{2}\left( z_{i,t_{2}}\right) \varphi _{2}\left( z_{i,t_{3}}\right)
	& =\sum_{t\in I_{m}}\varphi _{1}\left( z_{i,t}\right) \varphi _{2}\left(
	z_{i,t}\right) \varphi _{3}\left( z_{i,t}\right)  \notag \\
	& +\sum_{(a,b,c)\in S_{1,m}}\sum_{t_{1}<t_{2}<t_{3}}\varphi _{a}\left(
	z_{i,t_{1}}\right) \varphi _{b}\left( z_{i,t_{2}}\right) \varphi _{c}\left(
	z_{i,t_{3}}\right)  \notag \\
	& +2^{-1}\sum_{(a,b,c)\in S_{1,m}}\sum_{t_{1}\neq t_{2}}\varphi _{a}\left(
	z_{i,t_{1}}\right) \varphi _{b}\left( z_{i,t_{1}}\right) \varphi _{c}\left(
	z_{i,t_{2}}\right) ,  \label{P_L2B_2}
\end{align}%
where $S_{1,m}\equiv \left\{ (a,b,c):a,b,c\in \{1,2,3\}\text{ and }a\neq
b\neq c\right\} $, $\sum_{t_{1}<t_{2}<t_{3}}$ denotes the summation over $%
t_{1},t_{2},t_{3}\in I_{m}$ with $t_{1}<t_{2}<t_{3}$, and similar definition
applies to $\sum_{t_{1}\neq t_{2}}$. By condition (ii) of the lemma, the
triangle inequality and H\"{o}lder's inequality 
\begin{equation}
	\left\vert \mathbb{E}\left[ \sum_{t\in I_{m}}\varphi _{1}\left(
	z_{i,t}\right) \varphi _{2}\left( z_{i,t}\right) \varphi _{3}\left(
	z_{i,t}\right) \right] \right\vert \leq \sum_{t\in I_{m}}\left\vert \mathbb{E%
	}\left[ \varphi _{1}\left( z_{i,t}\right) \varphi _{2}\left( z_{i,t}\right)
	\varphi _{3}\left( z_{i,t}\right) \right] \right\vert \leq KT_{m}.
	\label{P_L2B_3}
\end{equation}%
Consider any $(a,b,c)\in S_{1,m}$, we next study the moment of $\varphi
_{a}\left( z_{i,t_{1}}\right) \varphi _{b}\left( z_{i,t_{2}}\right) \varphi
_{c}\left( z_{i,t_{3}}\right) $ for any $t_{1}<t_{2}<t_{3}$. For this
purpose, let $l\equiv \max \{t_{2}-t_{1},t_{3}-t_{2}\}$. By Corollary A.2.
in \cite{HallHeyde1980}\ and Assumption \ref{A1}(iv), we have 
\begin{equation*}
	\left\vert \mathbb{E}\left[ \varphi _{a}\left( z_{i,t_{1}}\right) \varphi
	_{b}\left( z_{i,t_{2}}\right) \varphi _{c}\left( z_{i,t_{3}}\right) \right]
	\right\vert \leq Ka^{l\delta /(4+\delta )}\left\Vert \varphi _{a}\left(
	z_{i,t_{1}}\right) \right\Vert _{2+\delta /2}\left\Vert \varphi _{b}\left(
	z_{i,t_{2}}\right) \varphi _{c}\left( z_{i,t_{3}}\right) \right\Vert
	_{2+\delta /2}\leq Ka^{l\delta /(4+\delta )}
\end{equation*}%
when $l=t_{2}-t_{1}$. The same upper bound $Ka^{l\delta /(4+\delta )}$
applies to $\left\vert \mathbb{E}\left[ \varphi _{a}\left(
z_{i,t_{1}}\right) \varphi _{b}\left( z_{i,t_{2}}\right) \varphi _{c}\left(
z_{i,t_{3}}\right) \right] \right\vert $ when $l=t_{3}-t_{2}$. Therefore for
any $t_{1}<t_{2}<t_{3}$ with $l\equiv \max \{t_{2}-t_{1},t_{3}-t_{2}\}$, we
have%
\begin{equation}
	\left\vert \mathbb{E}\left[ \varphi _{a}\left( z_{i,t_{1}}\right) \varphi
	_{b}\left( z_{i,t_{2}}\right) \varphi _{c}\left( z_{i,t_{3}}\right) \right]
	\right\vert \leq Ka^{l\delta /(4+\delta )}.  \label{P_L2B_4}
\end{equation}%
Since $t_{1}<t_{2}<t_{3}$, it is clear that $1\leq l\leq T_{m}-1$. Let 
\begin{equation*}
	S_{2,m,l}\equiv \left\{ (t_{1},t_{2},t_{3}):\text{ }t_{1},t_{2},t_{3}\in
	I_{m}\text{, }t_{1}<t_{2}<t_{3}\text{ and }l=\max
	\{t_{2}-t_{1},t_{3}-t_{2}\}\right\} .
\end{equation*}%
Then $\#S_{2,m,l}\leq 2l(T_{m}-l)$. Therefore, 
\begin{align}
	\left\vert \mathbb{E}\left[ \sum_{t_{1}<t_{2}<t_{3}}\varphi _{a}\left(
	z_{i,t_{1}}\right) \varphi _{b}\left( z_{i,t_{2}}\right) \varphi _{c}\left(
	z_{i,t_{3}}\right) \right] \right\vert & \leq \sum_{t_{1}<t_{2}<t_{3}}%
	\mathbb{E}\left[ \left\vert \varphi _{a}\left( z_{i,t_{1}}\right) \varphi
	_{b}\left( z_{i,t_{2}}\right) \varphi _{c}\left( z_{i,t_{3}}\right)
	\right\vert \right]  \notag \\
	& =\sum_{l=1}^{T_{m}-1}\sum_{(t_{1},t_{2},t_{3})\in S_{2,m,l}}\mathbb{E}%
	\left[ \left\vert \varphi _{a}\left( z_{i,t_{1}}\right) \varphi _{b}\left(
	z_{i,t_{2}}\right) \varphi _{c}\left( z_{i,t_{3}}\right) \right\vert \right]
	\notag \\
	& \leq K\sum_{l=1}^{T_{m}-1}l(T_{m}-l)a^{l\delta /(4+\delta )}\leq KT_{m}
	\label{P_L2B_5}
\end{align}%
where the second inequality is by (\ref{P_L2B_4}), and the last inequality
is by $a\in (0,1)$. By the same arguments for deriving (\ref{P_L2B_4}),%
\begin{equation*}
	\left\vert \mathbb{E}\left[ \varphi _{a}\left( z_{i,t_{1}}\right) \varphi
	_{b}\left( z_{i,t_{1}}\right) \varphi _{c}\left( z_{i,t_{2}}\right) \right]
	\right\vert \leq Ka^{|t_{1}-t_{2}|\delta /(4+\delta )}
\end{equation*}%
which implies that%
\begin{align}
	\left\vert \mathbb{E}\left[ \sum_{t_{1}\neq t_{2}}\varphi _{a}\left(
	z_{i,t_{1}}\right) \varphi _{b}\left( z_{i,t_{1}}\right) \varphi _{c}\left(
	z_{i,t_{2}}\right) \right] \right\vert & \leq \sum_{t_{1}\neq t_{2}}\mathbb{E%
	}\left[ \left\vert \varphi _{a}\left( z_{i,t_{1}}\right) \varphi _{b}\left(
	z_{i,t_{1}}\right) \varphi _{c}\left( z_{i,t_{2}}\right) \right\vert \right]
	\notag \\
	& \leq K\sum_{t_{1}\neq t_{2}}a^{|t_{1}-t_{2}|\delta /(4+\delta )}  \notag \\
	& \leq K\sum_{l=1}^{T_{m}-1}(T_{m}-l)a^{l\delta /(4+\delta )}\leq KT_{m}.
	\label{P_L2B_6}
\end{align}%
Since the size of $\#S_{1,m}=6$, combining the results in (\ref{P_L2B_2}), (%
\ref{P_L2B_3}), (\ref{P_L2B_5}) and (\ref{P_L2B_6}) we get 
\begin{equation*}
	\left\vert \mathbb{E}\left[ T_{m}^{-1}\left( \sum_{t\in I_{m}}\varphi
	_{1}\left( z_{i,t}\right) \right) \left( \sum_{t\in I_{m}}\varphi _{2}\left(
	z_{i,t}\right) \right) \left( \sum_{t\in I_{m}}\varphi _{3}\left(
	z_{i,t}\right) \right) \right] \right\vert \leq K,
\end{equation*}%
which together with (\ref{P_L2B_1}) and the triangle inequality shows that%
\begin{equation}
	\max_{g\in \mathcal{G},m\in \mathcal{M}}\left\vert (n_{g}T_{m})^{2}\mathbb{E}%
	\left[ \hat{\varphi}_{1,g,m}\hat{\varphi}_{2,g,m}\hat{\varphi}_{3,g,m}\right]
	\right\vert \leq K.  \label{P_L2B_6B}
\end{equation}%
The desired result in (\ref{L2B_1}) follows by (\ref{P_L2B_6B}).

We next prove (\ref{L2B_2}). By the same arguments for showing (\ref{P_L2_1}%
), 
\begin{align}
	& \mathrm{Var}\left( M^{-1/2}\sum_{m\in \mathcal{M}}(n_{g}T_{m})\hat{\varphi}%
	_{1,g,m}\hat{\varphi}_{2,g,m}\hat{\varphi}_{3,g,m}\right)  \notag \\
	& \leq K\max_{g\in \mathcal{G},m\in \mathcal{M}}\left\Vert (n_{g}T_{m})\hat{%
		\varphi}_{1,g,m}\hat{\varphi}_{2,g,m}\hat{\varphi}_{3,g,m}\right\Vert
	_{2+\delta ^{\prime }}^{2}  \notag \\
	& \leq K\max_{g\in \mathcal{G},m\in \mathcal{M}}(n_{g}T_{m})^{-1}\max_{g\in 
		\mathcal{G},m\in \mathcal{M}}\left\Vert (n_{g}T_{m})^{3/2}\hat{\varphi}%
	_{1,g,m}\hat{\varphi}_{2,g,m}\hat{\varphi}_{3,g,m}\right\Vert _{2+\delta
		^{\prime }}^{2}  \label{P_L2B_7}
\end{align}%
where $\delta ^{\prime }=\delta /3$. Using similar arguments for deriving (%
\ref{P_L0_4}), we can show that%
\begin{equation*}
	\max_{g\in \mathcal{G},m\in \mathcal{M}}\mathbb{E}\left[ \left\vert
	(n_{g}T_{m})^{1/2}\hat{\varphi}_{j,g,m}\right\vert ^{6+\delta }\right] \leq K
\end{equation*}%
which together with (\ref{P_L2B_7}) and H\"{o}lder's inequality implies that 
\begin{equation}
	\max_{g\in \mathcal{G}}\mathrm{Var}\left( M^{-1/2}\sum_{m\in \mathcal{M}%
	}(n_{g}T_{m})\hat{\varphi}_{1,g,m}\hat{\varphi}_{2,g,m}\hat{\varphi}%
	_{3,g,m}\right) \leq K\max_{g\in \mathcal{G},m\in \mathcal{M}%
	}(n_{g}T_{m})^{-1}.  \label{P_L2B_8}
\end{equation}%
By Assumption \ref{A1}(iii) and (\ref{P_L2B_8}) 
\begin{align*}
	\mathrm{Var}\left( \sum_{g\in \mathcal{G}}\sum_{m\in \mathcal{M}}(n_{g}T_{m})%
	\hat{\varphi}_{1,g,m}\hat{\varphi}_{2,g,m}\hat{\varphi}_{3,g,m}\right) &
	=\sum_{g\in \mathcal{G}}\mathrm{Var}\left( \sum_{m\in \mathcal{M}%
	}(n_{g}T_{m})\hat{\varphi}_{1,g,m}\hat{\varphi}_{2,g,m}\hat{\varphi}%
	_{3,g,m}\right) \\
	& \leq K(GM)\max_{g\in \mathcal{G},m\in \mathcal{M}}(n_{g}T_{m})^{-1},
\end{align*}%
which shows (\ref{L2B_2}).\hfill $Q.E.D.$

\bigskip

\begin{lemma}
	\textit{\label{L3B}}Under Assumptions \ref{A1}, \ref{A3}, \ref{A4} and \ref%
	{A5}, we have%
	\begin{align}
		& (nT)^{-1}\sum_{g\in \mathcal{G}}\sum_{m\in \mathcal{M}}(n_{g}T_{m})(\hat{%
			\Psi}_{\theta \gamma ,g,m}-\Psi _{\theta \gamma ,g,m})\Psi _{\gamma \gamma
			,g,m}^{-1}\hat{\Psi}_{\gamma ,g,m}  \notag \\
		& \text{ \ \ \ \ \ }-(nT)^{-1}\sum_{g\in \mathcal{G}}\sum_{m\in \mathcal{M}%
		}(n_{g}T_{m})\mathbb{E}\left[ (\hat{\Psi}_{\theta \gamma ,g,m}-\Psi _{\theta
			\gamma ,g,m})\Psi _{\gamma \gamma ,g,m}^{-1}\hat{\Psi}_{\gamma ,g,m}\right]
		=O_{p}((GM)^{1/2}(nT)^{-1}),  \label{L3B_1}
	\end{align}%
	where $\hat{\Psi}_{\gamma ,g,m}\equiv (n_{g}T_{m})^{-1}\sum_{i\in
		I_{g}}\sum_{t\in I_{m}}\psi _{\gamma }(z_{i,t})$ for any $g\in \mathcal{G}$
	and $m\in \mathcal{M}$.
\end{lemma}

The proof of (\ref{L3B_1}) of Lemma \ref{L3B} is omitted since it directly
follows from (\ref{L2A-1}) of Lemma \ref{L2A}.

\bigskip

\begin{lemma}
	\textit{\label{L3C}}Under Assumptions \ref{A1}, \ref{A3}, \ref{A4} and \ref%
	{A5}, we have%
	\begin{align}
		& (nT)^{-1}\sum_{g\in \mathcal{G}}\sum_{m\in \mathcal{M}}(n_{g}T_{m})\Psi
		_{\theta \gamma ,g,m}(\hat{\Psi}_{\gamma \gamma ,g,m}^{-1}-\Psi _{\gamma
			\gamma ,g,m}^{-1})\hat{\Psi}_{\gamma ,g,m}  \notag \\
		& \text{ \ \ \ }+(nT)^{-1}\sum_{g\in \mathcal{G}}\sum_{m\in \mathcal{M}%
		}(n_{g}T_{m})\Psi _{\theta \gamma ,g,m}\Psi _{\gamma \gamma ,g,m}^{-1}%
		\mathbb{E}\left[ (\hat{\Psi}_{\gamma \gamma ,g,m}-\Psi _{\gamma \gamma
			,g,m})\Psi _{\gamma \gamma ,g,m}^{-1}\hat{\Psi}_{\gamma ,g,m}\right]  \notag
		\\
		& =O_{p}\left( (1+(GM)^{1/2}\max_{g\in \mathcal{G},m\in \mathcal{M}%
		}(n_{g}T_{m})^{-1/2})(GM)^{1/2}(nT)^{-1}\right) .  \label{L3C_1}
	\end{align}
\end{lemma}

\noindent \textsc{Proof of Lemma \ref{L3C}}.\ Since 
\begin{equation*}
	\hat{\Psi}_{\gamma \gamma ,g,m}^{-1}-\Psi _{\gamma \gamma ,g,m}^{-1}=-\Psi
	_{\gamma \gamma ,g,m}^{-1}(\hat{\Psi}_{\gamma \gamma ,g,m}-\Psi _{\gamma
		\gamma ,g,m})\Psi _{\gamma \gamma ,g,m}^{-1}-(\hat{\Psi}_{\gamma \gamma
		,g,m}^{-1}-\Psi _{\gamma \gamma ,g,m}^{-1})(\hat{\Psi}_{\gamma \gamma
		,g,m}-\Psi _{\gamma \gamma ,g,m})\Psi _{\gamma \gamma ,g,m}^{-1},
\end{equation*}%
we can derive the following equations: 
\begin{align}
	& \sum_{g\in \mathcal{G}}\sum_{m\in \mathcal{M}}(n_{g}T_{m})\Psi _{\theta
		\gamma ,g,m}(\hat{\Psi}_{\gamma \gamma ,g,m}^{-1}-\Psi _{\gamma \gamma
		,g,m}^{-1})\hat{\Psi}_{\gamma ,g,m}  \notag \\
	& =-\sum_{g\in \mathcal{G}}\sum_{m\in \mathcal{M}}(n_{g}T_{m})\Psi _{\theta
		\gamma ,g,m}\Psi _{\gamma \gamma ,g,m}^{-1}(\hat{\Psi}_{\gamma \gamma
		,g,m}-\Psi _{\gamma \gamma ,g,m})\Psi _{\gamma \gamma ,g,m}^{-1}\hat{\Psi}%
	_{\gamma ,g,m}  \notag \\
	& \text{ \ \ \ }-\sum_{g\in \mathcal{G}}\sum_{m\in \mathcal{M}%
	}(n_{g}T_{m})\Psi _{\theta \gamma ,g,m}(\hat{\Psi}_{\gamma \gamma
		,g,m}^{-1}-\Psi _{\gamma \gamma ,g,m}^{-1})(\hat{\Psi}_{\gamma \gamma
		,g,m}-\Psi _{\gamma \gamma ,g,m})\Psi _{\gamma \gamma ,g,m}^{-1}\hat{\Psi}%
	_{\gamma ,g,m}  \label{P_L3C_1}
\end{align}%
where%
\begin{align}
	& \sum_{g\in \mathcal{G}}\sum_{m\in \mathcal{M}}(n_{g}T_{m})\Psi _{\theta
		\gamma ,g,m}\Psi _{\gamma \gamma ,g,m}^{-1}(\hat{\Psi}_{\gamma \gamma
		,g,m}-\Psi _{\gamma \gamma ,g,m})\Psi _{\gamma \gamma ,g,m}^{-1}\hat{\Psi}%
	_{\gamma ,g,m}  \notag \\
	& \text{ \ \ }-\sum_{g\in \mathcal{G}}\sum_{m\in \mathcal{M}%
	}(n_{g}T_{m})\Psi _{\theta \gamma ,g,m}\Psi _{\gamma \gamma ,g,m}^{-1}%
	\mathbb{E}\left[ (\hat{\Psi}_{\gamma \gamma ,g,m}-\Psi _{\gamma \gamma
		,g,m})\Psi _{\gamma \gamma ,g,m}^{-1}\hat{\Psi}_{\gamma ,g,m}\right] \overset%
	{}{=}O_{p}((GM)^{1/2})  \label{P_L3C_1B}
\end{align}%
by Lemma \ref{L2A}. Next, we will examine the second term on the right-hand
side (RHS) of equation (\ref{P_L3C_1}). By the triangle inequality and the
Cauchy-Schwarz inequality,%
\begin{align}
	& \left\Vert \sum_{g\in \mathcal{G}}\sum_{m\in \mathcal{M}}(n_{g}T_{m})\Psi
	_{\theta \gamma ,g,m}(\hat{\Psi}_{\gamma \gamma ,g,m}^{-1}-\Psi _{\gamma
		\gamma ,g,m}^{-1})(\hat{\Psi}_{\gamma \gamma ,g,m}-\Psi _{\gamma \gamma
		,g,m})\Psi _{\gamma \gamma ,g,m}^{-1}\hat{\Psi}_{\gamma ,g,m}\right\Vert 
	\notag \\
	& \leq \sum_{g\in \mathcal{G}}\sum_{m\in \mathcal{M}}\left\Vert \Psi
	_{\theta \gamma ,g,m}(\hat{\Psi}_{\gamma \gamma ,g,m}^{-1}-\Psi _{\gamma
		\gamma ,g,m}^{-1})(\hat{\Psi}_{\gamma \gamma ,g,m}-\Psi _{\gamma \gamma
		,g,m})\Psi _{\gamma \gamma ,g,m}^{-1}\sum_{i\in I_{g}}\sum_{t\in I_{m}}\psi
	_{\gamma }(z_{i,t})\right\Vert  \notag \\
	& =\sum_{g\in \mathcal{G}}\sum_{m\in \mathcal{M}}\left\Vert \Psi _{\theta
		\gamma ,g,m}\hat{\Psi}_{\gamma \gamma ,g,m}^{-1}(\hat{\Psi}_{\gamma \gamma
		,g,m}-\Psi _{\gamma \gamma ,g,m})\Psi _{\gamma \gamma ,g,m}^{-1}(\hat{\Psi}%
	_{\gamma \gamma ,g,m}-\Psi _{\gamma \gamma ,g,m})\Psi _{\gamma \gamma
		,g,m}^{-1}\sum_{i\in I_{g}}\sum_{t\in I_{m}}\psi _{\gamma
	}(z_{i,t})\right\Vert  \notag \\
	& \leq \max_{g\in \mathcal{G},m\in \mathcal{M}}(\lambda _{\max }(\Psi
	_{\theta \gamma ,g,m}\Psi _{\gamma \theta ,g,m})\lambda _{\max }((\hat{\Psi}%
	_{\gamma \gamma ,g,m}^{-1})^{2}))^{1/2}\lambda _{\max }((\Psi _{\gamma
		\gamma ,g,m}^{-1})^{2})  \notag \\
	& \text{ \ \ \ }\times \sum_{g\in \mathcal{G}}\sum_{m\in \mathcal{M}%
	}(n_{g}T_{m})\left\Vert \hat{\Psi}_{\gamma \gamma ,g,m}-\Psi _{\gamma \gamma
		,g,m}\right\Vert ^{2}\left\Vert \hat{\Psi}_{\gamma ,g,m}\right\Vert ,
	\label{P_L3C_2}
\end{align}%
where $\max_{g\in \mathcal{G},m\in \mathcal{M}}(\lambda _{\max }(\Psi
_{\theta \gamma ,g,m}\Psi _{\gamma \theta ,g,m}))^{1/2}\lambda _{\max
}((\Psi _{\gamma \gamma ,g,m}^{-1})^{2})\leq K$ because Assumption \ref{A4},
(\ref{P_L3AB_3}) of Lemma \ref{L3AB}, and $\lambda _{\max }((\hat{\Psi}%
_{\gamma \gamma ,g,m}^{-1})^{2})=O_{p}(1)$ implied by Lemma \ref{L3AB}(ii).

Equations (\ref{P_L3C_3}) and (\ref{P_L3C_4}) in Lemma \ref{P_L3C_3&4}
together with Markov's inequality and H\"{o}lder's inequality imply that 
\begin{equation}
	\sum_{g\in \mathcal{G}}\sum_{m\in \mathcal{M}}(n_{g}T_{m})\left\Vert \hat{%
		\Psi}_{\gamma \gamma ,g,m}-\Psi _{\gamma \gamma ,g,m}\right\Vert
	^{2}\left\Vert \hat{\Psi}_{\gamma ,g,m}\right\Vert =O_{p}((GM)\max_{g\in 
		\mathcal{G},m\in \mathcal{M}}(n_{g}T_{m})^{-1/2}).  \label{P_L3C_5}
\end{equation}%
Collecting the results in (\ref{P_L3C_2}) and (\ref{P_L3C_5}) obtains%
\begin{align}
	& \sum_{g\in \mathcal{G}}\sum_{m\in \mathcal{M}}(n_{g}T_{m})\Psi _{\theta
		\gamma ,g,m}(\hat{\Psi}_{\gamma \gamma ,g,m}^{-1}-\Psi _{\gamma \gamma
		,g,m}^{-1})(\hat{\Psi}_{\gamma \gamma ,g,m}-\Psi _{\gamma \gamma ,g,m})\Psi
	_{\gamma \gamma ,g,m}^{-1}\hat{\Psi}_{\gamma ,g,m}  \notag \\
	& =O_{p}((GM)\max_{g\in \mathcal{G},m\in \mathcal{M}}(n_{g}T_{m})^{-1/2}).
	\label{P_L3C_6}
\end{align}%
The assertion of the lemma now follows from (\ref{P_L3C_1}), (\ref{P_L3C_1B}%
) and (\ref{P_L3C_6}).\hfill $Q.E.D.$

\bigskip

\begin{lemma}
	\textit{\label{L3D}}Under Assumptions \ref{A1}, \ref{A3}, \ref{A4} and \ref%
	{A5}, we have%
	\begin{equation}
		\sum_{g\in \mathcal{G}}\sum_{m\in \mathcal{M}}(n_{g}T_{m})\mathbb{E}\left[ (%
		\hat{\Psi}_{\theta \gamma ,g,m}-\Psi _{\theta \gamma ,g,m})\Psi _{\gamma
			\gamma ,g,m}^{-1}\hat{\Psi}_{\gamma ,g,m}\right] =O(GM)  \label{L3D-1}
	\end{equation}%
	and 
	\begin{equation}
		\sum_{g\in \mathcal{G}}\sum_{m\in \mathcal{M}}(n_{g}T_{m})\Psi _{\theta
			\gamma ,g,m}\Psi _{\gamma \gamma ,g,m}^{-1}\mathbb{E}\left[ (\hat{\Psi}%
		_{\gamma \gamma ,g,m}-\Psi _{\gamma \gamma ,g,m})\Psi _{\gamma \gamma
			,g,m}^{-1}\hat{\Psi}_{\gamma ,g,m}\right] =O(GM).  \label{L3D-2}
	\end{equation}
\end{lemma}

\noindent The proofs of (\ref{L3D-1}) and (\ref{L3D-2}) of Lemma \ref{L3D}
are omitted since they directly follow from (\ref{P_L2A_2B}) in the proof of
Lemma \ref{L2A}.

\bigskip

\begin{lemma}
	\textit{\ \label{L3E}}Under Assumptions \ref{A1}, \ref{A3}, \ref{A4} and \ref%
	{A5}, we have%
	\begin{align*}
		& \sum_{g\in \mathcal{G}}\sum_{m\in \mathcal{M}}(n_{g}T_{m})(\hat{\Psi}%
		_{\theta \gamma ,g,m}\hat{\Psi}_{\gamma \gamma ,g,m}^{-1}-\Psi _{\theta
			\gamma ,g,m}\Psi _{\gamma \gamma ,g,m}^{-1})\hat{\Psi}_{\gamma ,g,m} \\
		& =\sum_{g\in \mathcal{G}}\sum_{i\in I_{g}}(n_{g}T_{m})\mathbb{E}\left[ (%
		\hat{\Psi}_{\theta \gamma ,g,m}-\Psi _{\theta \gamma ,g,m}-\Psi _{\theta
			\gamma ,g,m}\Psi _{\gamma \gamma ,g,m}^{-1}(\hat{\Psi}_{\gamma \gamma
			,g,m}-\Psi _{\gamma \gamma ,g,m}))\Psi _{\gamma \gamma ,g,m}^{-1}\hat{\Psi}%
		_{\gamma ,g,m}\right] \\
		& \text{ \ \ \ }+O_{p}\left( (1+(GM)^{1/2}\max_{g\in \mathcal{G},m\in 
			\mathcal{M}}(n_{g}T_{m})^{-1/2})(GM)^{1/2}\right) \overset{}{=}O_{p}\left(
		GM\right) .
	\end{align*}
\end{lemma}

\noindent \textsc{Proof of Lemma \ref{L3E}}.\ By the triangle inequality,
the Cauchy-Schwarz inequality, and (\ref{P_L3AB_3}) of Lemma \ref{L3AB}, we
have%
\begin{align}
	& \left\Vert \sum_{g\in \mathcal{G}}\sum_{m\in \mathcal{M}}(n_{g}T_{m})(\hat{%
		\Psi}_{\theta \gamma ,g,m}-\Psi _{\theta \gamma ,g,m})(\hat{\Psi}_{\gamma
		\gamma ,g,m}^{-1}-\Psi _{\gamma \gamma ,g,m}^{-1})\hat{\Psi}_{\gamma
		,g,m}\right\Vert  \notag \\
	& \leq \sum_{g\in \mathcal{G}}\sum_{m\in \mathcal{M}}\left\Vert (n_{g}T_{m})(%
	\hat{\Psi}_{\theta \gamma ,g,m}-\Psi _{\theta \gamma ,g,m})\hat{\Psi}%
	_{\gamma \gamma ,g,m}^{-1}(\hat{\Psi}_{\gamma \gamma ,g,m}-\Psi _{\gamma
		\gamma ,g,m})\Psi _{\gamma \gamma ,g,m}^{-1}\hat{\Psi}_{\gamma
		,g,m}\right\Vert  \notag \\
	& \leq K\max_{g\in \mathcal{G},m\in \mathcal{M}}(\lambda _{\max }((\hat{\Psi}%
	_{\gamma \gamma ,g,m}^{-1})^{2}))^{1/2}  \notag \\
	& \text{ \ \ \ }\times \sum_{g\in \mathcal{G}}\sum_{m\in \mathcal{M}%
	}\left\Vert \hat{\Psi}_{\theta \gamma ,g,m}-\Psi _{\theta \gamma
		,g,m}\right\Vert \left\Vert \hat{\Psi}_{\gamma \gamma ,g,m}-\Psi _{\gamma
		\gamma ,g,m}\right\Vert \left\Vert (n_{g}T_{m})\hat{\Psi}_{\gamma
		,g,m}\right\Vert .  \label{P_L3E_1}
\end{align}%
By Assumption \ref{A3}(ii), and (\ref{L0-1}) in Lemma \ref{L0}, we have%
\begin{equation}
	\max_{g\in \mathcal{G},m\in \mathcal{M}}\mathbb{E}\left[ \left\Vert
	(n_{g}T_{m})^{1/2}(\hat{\Psi}_{\theta \gamma ,g,m}-\Psi _{\theta \gamma
		,g,m})\right\Vert ^{4}\right] \leq K  \label{P_L3E_2}
\end{equation}%
which together with Lemma \ref{P_L3C_3&4}, Markov's inequality and H\"{o}%
lder's inequality implies that%
\begin{eqnarray}
	&&\sum_{g\in \mathcal{G}}\sum_{m\in \mathcal{M}}\left\Vert \hat{\Psi}%
	_{\theta \gamma ,g,m}-\Psi _{\theta \gamma ,g,m}\right\Vert \left\Vert \hat{%
		\Psi}_{\gamma \gamma ,g,m}-\Psi _{\gamma \gamma ,g,m}\right\Vert \left\Vert
	(n_{g}T_{m})\hat{\Psi}_{\gamma ,g,m}\right\Vert  \notag \\
	&=&O_{p}((GM)\max_{g\in \mathcal{G},m\in \mathcal{M}}(n_{g}T_{m})^{-1/2}).
	\label{P_L3E_3}
\end{eqnarray}%
Therefore by Lemma \ref{L3AB}(ii), (\ref{P_L3E_1}) and (\ref{P_L3E_3}) 
\begin{eqnarray}
	&&\sum_{g\in \mathcal{G}}\sum_{m\in \mathcal{M}}(n_{g}T_{m})(\hat{\Psi}%
	_{\theta \gamma ,g,m}-\Psi _{\theta \gamma ,g,m})(\hat{\Psi}_{\gamma \gamma
		,g,m}^{-1}-\Psi _{\gamma \gamma ,g,m}^{-1})\hat{\Psi}_{\gamma ,g,m}  \notag
	\\
	&=&O_{p}((GM)\max_{g\in \mathcal{G},m\in \mathcal{M}}(n_{g}T_{m})^{-1/2}).
	\label{P_L3E_4}
\end{eqnarray}%
The claim of the lemma follows from Lemma \ref{L3B}, Lemma \ref{L3C}, Lemma %
\ref{L3D} and (\ref{P_L3E_4}).\hfill $Q.E.D.$

\subsection{Intermediate Results}

\begin{lemma}
	\textit{\label{L3A}}Under Assumptions \ref{A1}, \ref{A3} and \ref{A5}, we
	have%
	\begin{equation}
		\max_{g\in\mathcal{G},m\in\mathcal{M}}\sup_{\phi\in\Phi}\left\vert
		(n_{g}T_{m})^{-1/2}\sum_{i\in I_{g}}\sum_{t\in I_{m}}\left(
		D^{\nu}\psi\left( z_{i,t},\phi\right) -\mathbb{E}\left[ D^{\nu}\psi\left(
		z_{i,t},\phi\right) \right] \right) \right\vert =O_{p}((GM)^{1/p}),
		\label{L3A-1}
	\end{equation}
	and%
	\begin{equation}
		\max_{g\in\mathcal{G},m\in\mathcal{M}}\sup_{\phi\in\Phi}\left\vert \frac{%
			(n_{g}T_{m})^{-1}\sum_{i\in I_{g}}\sum_{t\in I_{m}}\left( D^{\nu}\psi\left(
			z_{i,t},\phi\right) -\mathbb{E}\left[ D^{\nu}\psi\left(
			z_{i,t},\phi^{\ast}\right) \right] \right) }{(GM)^{1/p}(n_{g}T_{m})^{-1/2}+%
			\left\Vert \phi-\phi^{\ast}\right\Vert }\right\vert =O_{p}(1)  \label{L3A-2}
	\end{equation}
	for any $\nu\equiv\left( \nu_{1},...,\nu_{k}\right) ^{\top}$ with $%
	\left\vert \nu\right\vert =\sum_{j=1}^{k}\nu_{j}\leq3$.
\end{lemma}

\noindent The proofs of (\ref{L3A-1}) and (\ref{L3A-2}) of Lemma \ref{L3A}
are omitted since they directly follow from Lemma \ref{L1}.

\bigskip

\begin{lemma}
	\label{Consistency} Under\ Assumptions \ref{A1}, \ref{A2}, \ref{A3} and \ref%
	{A5}(i), $||\hat{\theta}-\theta ^{\ast }||+\max_{g\in \mathcal{G},m\in 
		\mathcal{M}}\left\Vert \hat{\gamma}_{g,m}-\gamma _{g,m}^{\ast }\right\Vert
	=o_{p}(1)$.
\end{lemma}

\noindent \textsc{Proof of Lemma \ref{Consistency}}. Let $\psi ^{\ast
}\left( z_{i,t};\theta ,\gamma \right) \equiv \psi \left( z_{i,t};\theta
,\gamma \right) -\mathbb{E}\left[ \psi \left( z_{i,t};\theta ,\gamma \right) %
\right] $. Without loss of generality, let $\Phi =\Phi _{\theta }\times \Phi
_{\gamma }$ where $\Phi _{\theta }$ and $\Phi _{\gamma }$ include $\theta
^{\ast }$ and $\gamma _{g,m}^{\ast }$, respectively. By the triangle
inequality,%
\begin{align}
	& \sup_{\theta ,\{\gamma _{g,m}\}_{g\in \mathcal{G},m\in \mathcal{M}%
	}}\left\vert (nT)^{-1}\sum_{g\in \mathcal{G}}\sum_{m\in \mathcal{M}%
	}\sum_{i\in I_{g}}\sum_{t\in I_{m}}\psi ^{\ast }\left( z_{i,t};\theta
	,\gamma _{g,m}\right) \right\vert  \notag \\
	& \leq (nT)^{-1}\sum_{g\in \mathcal{G}}\sum_{m\in \mathcal{M}%
	}(n_{g}T_{m})^{1/2}\sup_{\phi \in \Phi }\left\vert
	(n_{g}T_{m})^{-1/2}\sum_{i\in I_{g}}\sum_{t\in I_{m}}\psi ^{\ast }\left(
	z_{i,t};\phi \right) \right\vert  \notag \\
	& \leq \max_{g\in \mathcal{G},m\in \mathcal{M}}\sup_{\phi \in \Phi
	}\left\vert (n_{g}T_{m})^{-1/2}\sum_{i\in I_{g}}\sum_{t\in I_{m}}\psi ^{\ast
	}\left( z_{i,t};\phi \right) \right\vert (nT)^{-1}\sum_{g\in \mathcal{G}%
	}\sum_{m\in \mathcal{M}}(n_{g}T_{m})^{1/2},  \label{P_Cons_1a}
\end{align}%
where $\sup_{\theta ,\{\gamma _{g,m}\}_{g\in \mathcal{G},m\in \mathcal{M}}}$
is taken over $\theta \in \Phi _{\theta }$ and $\gamma _{g,m}\in \Phi
_{\gamma }$ for $g\in \mathcal{G}$ and $m\in \mathcal{M}$. Under\
Assumptions \ref{A1} and \ref{A3}, we can apply (\ref{L3A-1}) of Lemma \ref%
{L3A} to obtain 
\begin{equation}
	\max_{g\in \mathcal{G},m\in \mathcal{M}}\sup_{\phi \in \Phi }\left\vert
	(n_{g}T_{m})^{-1/2}\sum_{i\in I_{g}}\sum_{t\in I_{m}}\psi ^{\ast }\left(
	z_{i,t};\phi \right) \right\vert =O_{p}((GM)^{1/p}).  \label{P_Cons_1b}
\end{equation}%
Since $\sum_{g\in \mathcal{G}}\sum_{m\in \mathcal{M}}n_{g}T_{m}=nT$, we have 
\begin{equation*}
	(nT)^{-1}\sum_{g\in \mathcal{G}}\sum_{m\in \mathcal{M}}(n_{g}T_{m})^{1/2}%
	\leq \max_{g\in \mathcal{G},m\in \mathcal{M}}(n_{g}T_{m})^{-1/2},
\end{equation*}%
which together with (\ref{P_Cons_1a}) and (\ref{P_Cons_1b}), and Assumption %
\ref{A5}(i) implies that%
\begin{equation}
	\sup_{\theta ,\{\gamma _{g,m}\}_{g\in \mathcal{G},m\in \mathcal{M}%
	}}\left\vert (nT)^{-1}\sum_{g\in \mathcal{G}}\sum_{m\in \mathcal{M}%
	}\sum_{i\in I_{g}}\sum_{t\in I_{m}}\psi ^{\ast }\left( z_{i,t};\theta
	,\gamma _{g,m}\right) \right\vert =o_{p}(1).  \label{P_Cons_1}
\end{equation}%
Recall that $\Psi _{g,m}(\theta ,\gamma )\equiv (n_{g}T_{m})^{-1}\sum_{i\in
	I_{g}}\sum_{t\in I_{m}}\mathbb{E}\left[ \psi \left( z_{i,t};\theta ,\gamma
\right) \right] $. For any $\eta >0$,%
\begin{align*}
	& \mathbb{P}\left( ||\hat{\theta}-\theta ^{\ast }||\geq \eta \right) \\
	& \leq \mathbb{P}\left( \sum_{g\in \mathcal{G}}\sum_{m\in \mathcal{M}%
	}\sum_{i\in I_{g}}\sum_{t\in I_{m}}\psi \left( z_{i,t};\theta ^{\ast
	},\gamma _{g,m}^{\ast }\right) \leq \sup_{\left\Vert \theta -\theta ^{\ast
		}\right\Vert \geq \eta ,\{\gamma _{g,m}\}_{g,m}}\sum_{g\in \mathcal{G}%
	}\sum_{m\in \mathcal{M}}\sum_{i\in I_{g}}\sum_{t\in I_{m}}\psi \left(
	z_{i,t};\theta ,\gamma _{g,m}\right) \right) \\
	& \leq \mathbb{P}\left( o_{p}(1)\leq \sup_{\left\Vert \theta -\theta ^{\ast
		}\right\Vert \geq \eta ,\{\gamma _{g,m}\}_{g,m}}(nT)^{-1}\sum_{g\in \mathcal{%
			G}}\sum_{m\in \mathcal{M}}(n_{g}T_{m})\left[ \Psi _{g,m}(\theta ,\gamma
	_{g,m})-\Psi _{g,m}(\theta ^{\ast },\gamma _{g,m}^{\ast })\right] \right) \\
	& \leq \mathbb{P}\left( o_{p}(1)\leq (nT)^{-1}\sum_{g\in \mathcal{G}%
	}\sum_{m\in \mathcal{M}}(n_{g}T_{m})\left[ \sup_{\{\phi \in \Phi :\text{ }%
		||\phi -\phi ^{\ast }||>\eta \}}\Psi _{g,m}(\phi )-\Psi _{g,m}(\phi ^{\ast })%
	\right] \right) \\
	& \leq \mathbb{P}\left( o_{p}(1)\leq \max_{g\in \mathcal{G},m\in \mathcal{M}}%
	\left[ \sup_{\{\phi \in \Phi :\text{ }||\phi -\phi ^{\ast }||>\eta \}}\Psi
	_{g,m}(\phi )-\Psi _{g,m}(\phi ^{\ast })\right] \right) \rightarrow 0,\ \ 
	\text{as }n,T\rightarrow \infty ,
\end{align*}%
where $\sup_{\left\Vert \theta -\theta ^{\ast }\right\Vert \geq \eta
	,\{\gamma _{g,m}\}_{g,m}}$ is taken over $\theta \in \Phi _{\theta }$ with $%
\left\Vert \theta -\theta ^{\ast }\right\Vert \geq \eta $ and over $\gamma
_{g,m}\in \Phi _{\gamma }$ for any $g\in \mathcal{G}\ $and any $m\in 
\mathcal{M}$, the second inequality is by\ (\ref{P_Cons_1}), and the
convergence is by Assumption\ \ref{A2}. This shows the consistency of $\hat{%
	\theta}$.

To show the consistency of $\hat{\gamma}_{g,m}$, we first note that by the
definition of $\hat{\gamma}_{g,m}$ 
\begin{equation*}
	\hat{\gamma}_{g,m}=\arg \max_{\gamma _{g,m}\in \Phi _{\gamma }}\sum_{i\in
		I_{g}}\sum_{t\in I_{m}}\psi (z_{i,t};\hat{\theta},\gamma _{g,m})
\end{equation*}%
for any $g\in \mathcal{G}\ $and any $m\in \mathcal{M}$. Therefore for any $%
\eta >0$, we have 
\begin{align}
	& \mathbb{P}\left( \max_{g\in \mathcal{G},m\in \mathcal{M}}\left\Vert \hat{%
		\gamma}_{g,m}-\gamma _{g,m}^{\ast }\right\Vert \geq \eta \right)  \notag \\
	& \leq \mathbb{P}\left( \max_{g\in \mathcal{G},m\in \mathcal{M}%
	}\sup_{\{\gamma _{g,m}\in \Phi _{\gamma }:\text{ }\left\Vert \gamma
		_{g,m}-\gamma _{g,m}^{\ast }\right\Vert \geq \eta \}}\sum_{i\in
		I_{g}}\sum_{t\in I_{m}}\left[ \psi (z_{i,t};\hat{\theta},\gamma _{g,m})-\psi
	(z_{i,t};\hat{\theta},\gamma _{g,m}^{\ast })\right] \geq 0\right)
	\label{P_Cons_2}
\end{align}%
By Assumption \ref{A3}(i) and the triangle inequality,%
\begin{align}
	& \max_{g\in \mathcal{G},m\in \mathcal{M}}\sup_{\gamma _{g,m}\in \Phi
		_{\gamma }}\left\vert (n_{g}T_{m})^{-1}\sum_{i\in I_{g}}\sum_{t\in I_{m}}%
	\left[ \psi (z_{i,t};\hat{\theta},\gamma _{g,m})-\psi (z_{i,t};\theta ^{\ast
	},\gamma _{g,m})\right] \right\vert  \notag \\
	& \leq ||\hat{\theta}-\theta ^{\ast }||\max_{g\in \mathcal{G},m\in \mathcal{M%
	}}(n_{g}T_{m})^{-1}\sum_{i\in I_{g}}\sum_{t\in I_{m}}M\left( z_{i,t}\right) 
	\notag \\
	& \leq ||\hat{\theta}-\theta ^{\ast }||\left( \sup_{i,t}\mathbb{E}\left[
	M(z_{i,t})\right] +\max_{g\in \mathcal{G},m\in \mathcal{M}}\left\vert
	(n_{g}T_{m})^{-1}\sum_{i\in I_{g}}\sum_{t\in I_{m}}M^{\ast }\left(
	z_{i,t}\right) \right\vert \right) .  \label{P_Cons_3}
\end{align}%
where $M^{\ast }\left( z_{i,t}\right) \equiv M\left( z_{i,t}\right) -\mathbb{%
	E}\left[ M(z_{i,t})\right] $. Applying (\ref{L3A-1}) of Lemma \ref{L3A}
yields 
\begin{equation}
	\max_{g\in \mathcal{G},m\in \mathcal{M}}\left\vert
	(n_{g}T_{m})^{-1}\sum_{i\in I_{g}}\sum_{t\in I_{m}}M^{\ast }\left(
	z_{i,t}\right) \right\vert =O_{p}\left( (GM)^{1/p}\max_{g\in \mathcal{G}%
		,m\in \mathcal{M}}(n_{g}T_{m})^{-1/2}\right) =o_{p}(1)  \label{P_Cons_4}
\end{equation}%
where the second equality is by Assumption \ref{A5}(i). Since $\sup_{i,t}%
\mathbb{E}\left[ M(z_{i,t})\right] \leq K$ by Assumption \ref{A3}(ii), we
can deduce from (\ref{P_Cons_3}), (\ref{P_Cons_4}) and the consistency of $%
\hat{\theta}$ that%
\begin{equation}
	\max_{g\in \mathcal{G},m\in \mathcal{M}}\sup_{\gamma _{g,m}\in \Phi _{\gamma
	}}\left\vert (n_{g}T_{m})^{-1}\sum_{i\in I_{g}}\sum_{t\in I_{m}}\left[ \psi
	(z_{i,t};\hat{\theta},\gamma _{g,m})-\psi (z_{i,t};\theta ^{\ast },\gamma
	_{g,m})\right] \right\vert =o_{p}(1).  \label{P_Cons_5}
\end{equation}%
Invoking (\ref{L3A-1}) of Lemma \ref{L3A} and Assumption \ref{A5}(i) obtains%
\begin{equation*}
	\max_{g\in \mathcal{G},m\in \mathcal{M}}\sup_{\gamma _{g,m}\in \Phi _{\gamma
	}}\left\vert (n_{g}T_{m})^{-1/2}\sum_{i\in I_{g}}\sum_{t\in I_{m}}\psi
	^{\ast }\left( z_{i,t};\theta ^{\ast },\gamma _{g,m}\right) \right\vert
	=o_{p}(1)
\end{equation*}%
which together with (\ref{P_Cons_2}), (\ref{P_Cons_5}) and Assumption\ \ref%
{A2} implies that%
\begin{align*}
	& \mathbb{P}\left( \max_{g\in \mathcal{G},m\in \mathcal{M}}\left\Vert \hat{%
		\gamma}_{g,m}-\gamma _{g,m}^{\ast }\right\Vert \geq \eta \right) \\
	& \leq \mathbb{P}\left( \max_{g\in \mathcal{G},m\in \mathcal{M}%
	}\sup_{\{\gamma _{g,m}\in \Phi _{\gamma }:\text{ }\left\Vert \gamma
		_{g,m}-\gamma _{g,m}^{\ast }\right\Vert \geq \eta
		\}}(n_{g}T_{m})^{-1}\sum_{i\in I_{g}}\sum_{t\in I_{m}}\left[ \psi
	(z_{i,t};\theta ^{\ast },\gamma _{g,m})-\psi (z_{i,t};\theta ^{\ast },\gamma
	_{g,m}^{\ast })\right] \geq o_{p}(1)\right) \\
	& \leq \mathbb{P}\left( \max_{g\in \mathcal{G},m\in \mathcal{M}%
	}\sup_{\{\gamma _{g,m}\in \Phi _{\gamma }:\text{ }\left\Vert \gamma
		_{g,m}-\gamma _{g,m}^{\ast }\right\Vert \geq \eta \}}\left[ \Psi
	_{g,m}(\theta ^{\ast },\gamma )-\Psi _{g,m}(\theta ^{\ast },\gamma
	_{g,m}^{\ast })\right] \geq o_{p}(1)\right) \\
	& \leq \mathbb{P}\left( \min_{g\in \mathcal{G},m\in \mathcal{M}}\left[ \Psi
	_{g,m}(\phi ^{\ast })-\sup_{\{\phi \in \Phi :\text{ }||\phi -\phi ^{\ast
		}||>\eta \}}\Psi _{g,m}(\phi )\right] \leq o_{p}(1)\right) \rightarrow 0%
	\text{, as }n,T\rightarrow \infty ,
\end{align*}%
which shows $\max_{g\in \mathcal{G},m\in \mathcal{M}}\left\Vert \hat{\gamma}%
_{g,m}-\gamma _{g,m}^{\ast }\right\Vert =o_{p}(1)$.\hfill $Q.E.D.$

\bigskip

\begin{lemma}
	\label{P_L3AB_1} Under Assumptions \ref{A1}, \ref{A3}, \ref{A4} and \ref{A5}%
	, we have 
	\begin{equation*}
		\max_{g\in \mathcal{G},m\in \mathcal{M}}\frac{\left\Vert \hat{\Psi}_{\theta
				\theta ,g,m}-\Psi _{\theta \theta ,g,m}\right\Vert +\left\Vert \hat{\Psi}%
			_{\theta \gamma ,g,m}-\Psi _{\theta \gamma ,g,m}\right\Vert +\left\Vert \hat{%
				\Psi}_{\gamma \gamma ,g,m}-\Psi _{\gamma \gamma ,g,m}\right\Vert }{%
			(GM)^{1/p}(n_{g}T_{m})^{-1/2}}=O_{p}\left( 1\right) .
	\end{equation*}
\end{lemma}

\noindent\textsc{Proof of Lemma \ref{P_L3AB_1}}. Follows from (\ref{L3A-1})
in Lemma \ref{L3A}. \hfill$Q.E.D.$

\bigskip

\begin{lemma}
	\textit{\label{L3AB}}Under Assumptions \ref{A1}, \ref{A3}, \ref{A4} and \ref%
	{A5}, we have%
	\begin{equation}
		\max_{g\in \mathcal{G},m\in \mathcal{M}}\left( \left\Vert \Psi _{\theta
			\theta ,g,m}\right\Vert +\left\Vert \Psi _{\theta \gamma ,g,m}\right\Vert
		+\left\Vert \Psi _{\gamma \gamma ,g,m}\right\Vert \right) \leq K.
		\label{P_L3AB_3}
	\end{equation}%
	We also have with probability approaching 1 (wpa1),
	
	(i) $\max_{g\in \mathcal{G},m\in \mathcal{M}}\left( ||\hat{\Psi}_{\theta
		\theta ,g,m}||+||\hat{\Psi}_{\theta \gamma ,g,m}||+||\hat{\Psi}_{\gamma
		\gamma ,g,m}||\right) \leq K$;
	
	(ii) $\min_{g\in \mathcal{G},m\in \mathcal{M}}\lambda _{\min }(-\hat{\Psi}%
	_{\gamma \gamma ,g,m})>K^{-1}$;
	
	(iii)\ $\min_{g\in \mathcal{G},m\in \mathcal{M}}\lambda _{\min }(\hat{\Psi}%
	_{\theta \gamma ,g,m}\hat{\Psi}_{\gamma \gamma ,g,m}^{-1}\hat{\Psi}_{\gamma
		\theta ,g,m}-\hat{\Psi}_{\theta \theta ,g,m})>K^{-1}$.
\end{lemma}

\noindent \textsc{Proof of Lemma \ref{L3AB}}.\ (i) Equation (\ref{P_L3AB_3})
follows from Assumption \ref{A3}(iii). Lemma \ref{P_L3AB_1} and Assumption %
\ref{A5}(i) imply that 
\begin{equation}
	\max_{g\in \mathcal{G},m\in \mathcal{M}}\left\Vert \hat{\Psi}_{\theta \theta
		,g,m}-\Psi _{\theta \theta ,g,m}\right\Vert +\left\Vert \hat{\Psi}_{\theta
		\gamma ,g,m}-\Psi _{\theta \gamma ,g,m}\right\Vert +\left\Vert \hat{\Psi}%
	_{\gamma \gamma ,g,m}-\Psi _{\gamma \gamma ,g,m}\right\Vert =o_{p}(1).
	\label{P_L3AB_2}
\end{equation}%
The assertion in part (i) of the lemma now follows from (\ref{P_L3AB_2}) and
(\ref{P_L3AB_3}).

(ii) Since $\min_{g\in \mathcal{G},m\in \mathcal{M}}\lambda _{\min }(-\Psi
_{\gamma \gamma ,g,m})>K^{-1}$ by Assumption \ref{A4}, the assertion in this
part of the lemma follows from (\ref{P_L3AB_2}).

(iii)\ By the assertions in parts (i) and (ii) of the lemma, (\ref{P_L3AB_2}%
) and (\ref{P_L3AB_3}), we can conclude that%
\begin{equation*}
	\max_{g\in \mathcal{G},m\in \mathcal{M}}\left\Vert (\hat{\Psi}_{\theta
		\gamma ,g,m}\hat{\Psi}_{\gamma \gamma ,g,m}^{-1}\hat{\Psi}_{\gamma \theta
		,g,m}-\hat{\Psi}_{\theta \theta ,g,m})-(\Psi _{\theta \gamma ,g,m}\Psi
	_{\gamma \gamma ,g,m}^{-1}\Psi _{\gamma \theta ,g,m}-\Psi _{\theta \theta
		,g,m})\right\Vert =o_{p}(1).
\end{equation*}%
This, combined with Assumption \ref{A4}, shows the assertion in part (iii)
of the lemma.\hfill $Q.E.D.$

\bigskip

\begin{lemma}
	\textit{\label{FOCs}}Let $\psi _{\theta }(z_{i,t};\phi )\equiv \frac{%
		\partial \psi (z_{i,t};\phi )}{\partial \theta }$ and $\psi _{\gamma
	}(z_{i,t};\phi )\equiv \frac{\partial \psi (z_{i,t};\phi )}{\partial \gamma }
	$.\ The\ M-estimator $(\theta ^{\top },\gamma _{\mathcal{G},1}^{\top
	},\ldots ,\gamma _{\mathcal{G},M}^{\top })^{\top }$ satisfies the
	first-order condition%
	\begin{equation}
		\sum_{g\in \mathcal{G}}\sum_{m\in \mathcal{M}}\sum_{i\in I_{g}}\sum_{t\in
			I_{m}}\psi _{\theta }(z_{i,t};\hat{\phi}_{g,m})=0_{d_{\theta }\times 1},
		\label{Foc_1}
	\end{equation}%
	and hence,%
	\begin{equation}
		0_{d_{\gamma }\times 1}=\hat{\Psi}_{\gamma ,g,m}+\hat{\Psi}_{\gamma \phi
			,g,m}(\hat{\phi}_{g,m}-\phi _{g,m}^{\ast })+2^{-1}\left( (\hat{\phi}%
		_{g,m}-\phi _{g,m}^{\ast })^{\top }\tilde{\Psi}_{\gamma _{j}\phi \phi ,g,m}(%
		\hat{\phi}_{g,m}-\phi _{g,m}^{\ast })\right) _{j\leq d_{\gamma }},
		\label{P_Rate_theta_1}
	\end{equation}%
	where $\tilde{\Psi}_{\gamma _{j}\phi \phi ,g,m}\equiv
	(n_{g}T_{m})^{-1}\sum_{i\in I_{g}}\sum_{t\in I_{m}}\frac{\partial ^{2}}{%
		\partial \phi \partial \phi ^{\top }}\left( \frac{\partial \psi (z_{i,t};%
		\tilde{\phi}_{g,m})}{\partial \gamma _{j}}\right) $ for $j=1,\ldots
	,d_{\gamma }$ and $\tilde{\phi}_{g,m}$ lies between $\hat{\phi}_{g,m}$ and $%
	\phi _{g,m}^{\ast }$. Under Assumptions \ref{A1}, \ref{A2}, \ref{A3}, \ref%
	{A4} and \ref{A5}, we also have 
	\begin{equation}
		\max_{j\leq d_{\gamma }}\max_{g\in \mathcal{G},m\in \mathcal{M}}\left\Vert 
		\frac{\tilde{\Psi}_{\gamma _{j}\phi \phi ,g,m}-\Psi _{\gamma _{j}\phi \phi
				,g,m}}{(GM)^{1/p}(n_{g}T_{m})^{-1/2}+||\hat{\phi}_{g,m}-\phi _{g,m}^{\ast }||%
		}\right\Vert =O_{p}(1).  \label{P_Rate_theta_2a}
	\end{equation}
\end{lemma}

\noindent \textsc{Proof of Lemma \ref{Rate_theta}}.\ The M-estimator also
satisfies 
\begin{equation}
	\sum_{i\in I_{g}}\sum_{t\in I_{m}}\psi _{\gamma }(z_{i,t};\hat{\phi}%
	_{g,m})=0_{d_{\gamma }\times 1}  \label{Foc_2}
\end{equation}%
for any $g\in \mathcal{G}$ and$\ m\in \mathcal{M}$. Applying the\ Taylor
expansion to (\ref{Foc_2}), we obtain (\ref{P_Rate_theta_1}). By (\ref{L3A-2}%
) in Lemma \ref{L3A}, we obtain (\ref{P_Rate_theta_2a}).\hfill $Q.E.D.$

\bigskip

\begin{lemma}
	\textit{\label{Rate_theta}}Under Assumptions \ref{A1}, \ref{A2}, \ref{A3}, %
	\ref{A4} and \ref{A5}, we have%
	\begin{align}
		& \max_{g\in \mathcal{G},m\in \mathcal{M}}\frac{\left\Vert \hat{\Psi}%
			_{\gamma ,g,m}+\hat{\Psi}_{\gamma \theta ,g,m}(\hat{\theta}-\theta ^{\ast })+%
			\hat{\Psi}_{\gamma \gamma ,g,m}(\hat{\gamma}_{g,m}-\gamma _{g,m}^{\ast
			})\right\Vert }{(nT)^{-1}+||\hat{\phi}_{g,m}-\phi _{g,m}^{\ast }||^{2}}%
		=O_{p}(1),  \label{P_Rate_theta_3} \\
		& (nT)^{-1}\sum_{g\in \mathcal{G}}\sum_{m\in \mathcal{M}}(n_{g}T_{m})||\hat{%
			\gamma}_{g,m}-\gamma _{g,m}^{\ast }||^{2}=O_{p}\left( (GM)(nT)^{-1}+||\hat{%
			\theta}-\theta ^{\ast }||^{2}\right) ,  \label{P_Rate_theta_11}
	\end{align}%
	and 
	\begin{equation*}
		\hat{\theta}-\theta ^{\ast }=O_{p}((nT)^{-1/2}).
	\end{equation*}
\end{lemma}

\noindent \textsc{Proof of Lemma \ref{Rate_theta}}.\ By (\ref%
{P_Rate_theta_2a}) of Lemma \ref{FOCs}, Assumptions \ref{A3}(iii) and \ref%
{A5}(i), and Lemma \ref{Consistency}, 
\begin{equation}
	\max_{j\leq d_{\gamma }}\max_{g\in \mathcal{G},m\in \mathcal{M}}\left\Vert 
	\tilde{\Psi}_{\gamma _{j}\phi \phi ,g,m}\right\Vert =O_{p}(1)
	\label{P_Rate_theta_2}
\end{equation}%
which together with (\ref{P_Rate_theta_1}) of Lemma \ref{FOCs} and the\
Cauchy-Schwarz inequality implies that (\ref{P_Rate_theta_3}) holds.

Applying the\ Taylor expansion to (\ref{Foc_1}) obtains%
\begin{align}
	0_{d_{\theta }\times 1}& =\sum_{g\in \mathcal{G}}\sum_{m\in \mathcal{M}%
	}(n_{g}T_{m})\left[ \hat{\Psi}_{\theta ,g,m}+\hat{\Psi}_{\theta \theta ,g,m}(%
	\hat{\theta}-\theta ^{\ast })+\hat{\Psi}_{\theta \gamma ,g,m}(\hat{\gamma}%
	_{g,m}-\gamma _{g,m}^{\ast })\right]  \notag \\
	& \text{ \ \ }+2^{-1}\left( \sum_{g\in \mathcal{G}}\sum_{m\in \mathcal{M}%
	}(n_{g}T_{m})(\hat{\phi}_{g,m}-\phi _{g,m}^{\ast })^{\top }\tilde{\Psi}%
	_{\theta _{j}\phi \phi ,g,m}(\hat{\phi}_{g,m}-\phi _{g,m}^{\ast })\right)
	_{j=1,\ldots ,d_{\theta }}  \notag \\
	& =\sum_{g\in \mathcal{G}}\sum_{m\in \mathcal{M}}(n_{g}T_{m})\left[ \hat{\Psi%
	}_{\theta ,g,m}+\hat{\Psi}_{\theta \theta ,g,m}(\hat{\theta}-\theta ^{\ast
	})+\hat{\Psi}_{\theta \gamma ,g,m}(\hat{\gamma}_{g,m}-\gamma _{g,m}^{\ast })%
	\right]  \notag \\
	& \text{ \ \ }+O_{p}\left( \sum_{g\in \mathcal{G}}\sum_{m\in \mathcal{M}%
	}(n_{g}T_{m})||\hat{\phi}_{g,m}-\phi _{g,m}^{\ast }||^{2}\right) ,
	\label{P_Rate_theta_4}
\end{align}%
where $\tilde{\psi}_{\theta _{j}\phi \phi }(z_{i,t})\equiv \psi _{\theta
	_{j}\phi \phi }(z_{i,t};\tilde{\phi}_{g,m})$, $\tilde{\psi}_{\theta _{j}\phi
	\phi }(z_{i,t};\phi )\equiv \frac{\partial ^{2}}{\partial \phi \partial \phi
	^{\top }}\left( \frac{\partial \psi (z_{i,t};\phi )}{\partial \theta _{j}}%
\right) $ for $j\leq d_{\theta }$ and $\tilde{\phi}_{g,m}$ lies between $%
\hat{\phi}_{g,m}$ and $\phi _{g,m}^{\ast }$, the second equality is by the
Cauchy-Schwarz inequality and%
\begin{equation}
	\max_{j\leq d_{\theta }}\max_{g\in \mathcal{G},m\in \mathcal{M}}\left\Vert 
	\tilde{\Psi}_{\theta _{j}\phi \phi ,g,m}\right\Vert =O_{p}(1),
	\label{P_Rate_theta_5}
\end{equation}%
which can be established using the same arguments for showing (\ref%
{P_Rate_theta_2}).

By (\ref{P_Rate_theta_3}) of Lemma \ref{Rate_theta} and Lemma \ref{L3AB}(i,
ii),%
\begin{align}
	\sum_{g\in \mathcal{G}}\sum_{m\in \mathcal{M}}(n_{g}T_{m})\hat{\Psi}_{\theta
		\gamma ,g,m}(\hat{\gamma}_{g,m}-\gamma _{g,m}^{\ast })& =-\sum_{g\in 
		\mathcal{G}}\sum_{m\in \mathcal{M}}(n_{g}T_{m})\hat{\Psi}_{\theta \gamma
		,g,m}\hat{\Psi}_{\gamma \gamma ,g,m}^{-1}(\hat{\Psi}_{\gamma ,g,m}+\hat{\Psi}%
	_{\gamma \theta ,g,m}(\hat{\theta}-\theta ^{\ast }))  \notag \\
	& \text{ \ }+O_{p}\left( 1+\sum_{g\in \mathcal{G}}\sum_{m\in \mathcal{M}%
	}(n_{g}T_{m})||\hat{\phi}_{g,m}-\phi _{g,m}^{\ast }||^{2}\right) .
	\label{P_Rate_theta_6}
\end{align}%
Applying the above representation in (\ref{P_Rate_theta_4}), we obtain%
\begin{align}
	0_{d_{\theta }\times 1}& =(nT)^{-1}\sum_{g\in \mathcal{G}}\sum_{m\in 
		\mathcal{M}}(n_{g}T_{m})(\hat{\Psi}_{\theta ,g,m}-\hat{\Psi}_{\theta \gamma
		,g,m}\hat{\Psi}_{\gamma \gamma ,g,m}^{-1}\hat{\Psi}_{\gamma ,g,m})  \notag \\
	& \text{ \ \ \ \ }+(nT)^{-1}\sum_{g\in \mathcal{G}}\sum_{m\in \mathcal{M}%
	}(n_{g}T_{m})(\hat{\Psi}_{\theta \theta ,g,m}-\hat{\Psi}_{\theta \gamma ,g,m}%
	\hat{\Psi}_{\gamma \gamma ,g,m}^{-1}\Psi _{\gamma \theta ,g,m})(\hat{\theta}%
	-\theta ^{\ast })  \notag \\
	& \text{ \ \ \ \ }+O_{p}\left( (nT)^{-1}+||\hat{\theta}-\theta ^{\ast
	}||^{2}+(nT)^{-1}\sum_{g\in \mathcal{G}}\sum_{m\in \mathcal{M}}(n_{g}T_{m})||%
	\hat{\gamma}_{g,m}-\gamma _{g,m}^{\ast }||^{2}\right) ,
	\label{P_Rate_theta_7}
\end{align}%
where by Lemma \ref{L3E}, Assumptions \ref{A1} and \ref{A3}(iii), 
\begin{equation}
	(nT)^{-1}\sum_{g\in \mathcal{G}}\sum_{m\in \mathcal{M}}(n_{g}T_{m})(\hat{\Psi%
	}_{\theta ,g,m}-\hat{\Psi}_{\theta \gamma ,g,m}\hat{\Psi}_{\gamma \gamma
		,g,m}^{-1}\hat{\Psi}_{\gamma ,g,m})=O_{p}\left(
	(nT)^{-1/2}+(GM)(nT)^{-1}\right) .  \label{P_Rate_theta_8}
\end{equation}%
By (\ref{P_Rate_theta_3}) of Lemma \ref{Rate_theta}, Lemma \ref{L3AB}(ii)
and the Cauchy-Schwarz inequality%
\begin{align}
	\sum_{g\in \mathcal{G}}\sum_{m\in \mathcal{M}}(n_{g}T_{m})||\hat{\gamma}%
	_{g,m}-\gamma _{g,m}^{\ast }||^{2}& \leq K\sum_{g\in \mathcal{G}}\sum_{m\in 
		\mathcal{M}}\left\Vert \hat{\Psi}_{\gamma \gamma ,g,m}^{-1}(n_{g}T_{m})^{1/2}%
	\hat{\Psi}_{\gamma ,g,m}\right\Vert ^{2}  \notag \\
	& \text{ \ \ }+K||\hat{\theta}-\theta ^{\ast }||^{2}\sum_{g\in \mathcal{G}%
	}\sum_{m\in \mathcal{M}}(n_{g}T_{m})\left\Vert \hat{\Psi}_{\gamma \gamma
		,g,m}^{-1}\hat{\Psi}_{\gamma \theta ,g,m}\right\Vert ^{2}  \notag \\
	& \text{ \ \ }+O_{p}(1)\sum_{g\in \mathcal{G}}\sum_{m\in \mathcal{M}%
	}(n_{g}T_{m})\left( (nT)^{-2}+||\hat{\phi}_{g,m}-\phi _{g,m}^{\ast
	}||^{4}\right) .  \label{P_Rate_theta_9}
\end{align}%
Note that%
\begin{equation}
	\sum_{g\in \mathcal{G}}\sum_{m\in \mathcal{M}}(n_{g}T_{m})\left\Vert \hat{%
		\Psi}_{\gamma \gamma ,g,m}^{-1}\hat{\Psi}_{\gamma ,g,m}\right\Vert ^{2}\leq
	\max_{g\in \mathcal{G},m\in \mathcal{M}}\lambda _{\max }((\hat{\Psi}_{\gamma
		\gamma ,g,m}^{-1})^{2})\sum_{g\in \mathcal{G}}\sum_{m\in \mathcal{M}%
	}(n_{g}T_{m})\left\Vert \hat{\Psi}_{\gamma ,g,m}\right\Vert ^{2}.
	\label{P_Rate_theta_10a}
\end{equation}%
By (\ref{P_L3C_4}) in Lemma \ref{P_L3C_3&4}, 
\begin{equation*}
	\mathbb{E}\left[ \sum_{g\in \mathcal{G}}\sum_{m\in \mathcal{M}}\left\Vert
	(n_{g}T_{m})^{1/2}\hat{\Psi}_{\gamma ,g,m}\right\Vert ^{2}\right] \leq K(GM)
\end{equation*}%
which together with Markov's inequality implies that%
\begin{equation}
	\sum_{g\in \mathcal{G}}\sum_{m\in \mathcal{M}}\left\Vert (n_{g}T_{m})^{1/2}%
	\hat{\Psi}_{\gamma ,g,m}\right\Vert ^{2}=O_{p}(GM).  \label{P_Rate_theta_10b}
\end{equation}%
Hence by Lemma \ref{L3AB}(ii), (\ref{P_Rate_theta_10a}) and (\ref%
{P_Rate_theta_10b}), 
\begin{equation}
	\sum_{g\in \mathcal{G}}\sum_{m\in \mathcal{M}}\left\Vert \hat{\Psi}_{\gamma
		\gamma ,g,m}^{-1}(n_{g}T_{m})^{1/2}\hat{\Psi}_{\gamma ,g,m}\right\Vert
	^{2}=O_{p}(GM).  \label{P_Rate_theta_10c}
\end{equation}%
By Lemma \ref{L3AB}(i, ii),%
\begin{equation}
	\sum_{g\in \mathcal{G}}\sum_{m\in \mathcal{M}}(n_{g}T_{m})\left\Vert \hat{%
		\Psi}_{\gamma \gamma ,g,m}^{-1}\hat{\Psi}_{\gamma \theta ,g,m}\right\Vert
	^{2}=O_{p}(nT).  \label{P_Rate_theta_10d}
\end{equation}%
Combining the results in Lemma \ref{Consistency}, (\ref{P_Rate_theta_9}), (%
\ref{P_Rate_theta_10c}) and (\ref{P_Rate_theta_10d}), we get (\ref%
{P_Rate_theta_11}).

Lemma \ref{L3AB}(iii) implies that wpa1, 
\begin{equation}
	\lambda _{\min }\left( -(nT)^{-1}\sum_{g\in \mathcal{G}}\sum_{m\in \mathcal{M%
	}}(n_{g}T_{m})(\hat{\Psi}_{\theta \theta ,g,m}-\hat{\Psi}_{\theta \gamma
		,g,m}\hat{\Psi}_{\gamma \gamma ,g,m}^{-1}\Psi _{\gamma \theta ,g,m})\right)
	>K^{-1},  \label{P_Rate_theta_12}
\end{equation}%
which together with (\ref{P_Rate_theta_7}), (\ref{P_Rate_theta_8}) and (\ref%
{P_Rate_theta_11}) shows that%
\begin{equation}
	||\hat{\theta}-\theta ^{\ast }||=O_{p}((nT)^{-1/2}+(GM)(nT)^{-1}).
	\label{P_Rate_theta_13}
\end{equation}%
The claim of the lemma follows from (\ref{P_Rate_theta_13}) and Assumption %
\ref{A5}(ii).\hfill $Q.E.D.$

\bigskip

\begin{lemma}
	\textit{\label{Rep_Theta_1}}Under Assumptions \ref{A1}, \ref{A2}, \ref{A3}, %
	\ref{A4} and \ref{A5}, we have%
	\begin{align*}
		& (nT)^{-1/2}\sum_{g\in \mathcal{G}}\sum_{m\in \mathcal{M}}(n_{g}T_{m})(\hat{%
			\phi}_{g,m}-\phi _{g,m}^{\ast })^{\top }\hat{\Psi}_{\theta _{j}\phi \phi
			,g,m}(\hat{\phi}_{g,m}-\phi _{g,m}^{\ast }) \\
		& =(nT)^{-1/2}\sum_{g\in \mathcal{G}}\sum_{m\in \mathcal{M}}(n_{g}T_{m})%
		\mathbb{E}\left[ \hat{\Psi}_{\gamma ,g,m}^{\top }\Psi _{\gamma \gamma
			,g,m}^{-1}\Psi _{\theta _{j}\gamma \gamma ,g,m}\Psi _{\gamma \gamma
			,g,m}^{-1}\hat{\Psi}_{\gamma ,g,m}\right] +o_{p}\left( 1\right) ,
	\end{align*}%
	where $\hat{\Psi}_{\theta _{j}\phi \phi ,g,m}\equiv
	(n_{g}T_{m})^{-1}\sum_{i\in I_{g}}\sum_{t\in I_{m}}\frac{\partial ^{2}}{%
		\partial \phi \partial \phi ^{\top }}\left( \frac{\partial \psi
		(z_{i,t};\phi _{g,m}^{\ast })}{\partial \theta _{j}}\right) $ for any $j\leq
	d_{\theta }$.
\end{lemma}

\noindent \textsc{Proof of Lemma \ref{Rep_Theta_1}}. By (\ref{L3A-1}) in
Lemma \ref{L3A}, 
\begin{equation}
	\max_{j=1,\ldots ,d_{\theta }}\max_{g\in \mathcal{G},m\in \mathcal{M}%
	}\left\Vert (n_{g}T_{m})^{1/2}(\hat{\Psi}_{\theta _{j}\phi \phi ,g,m}-\Psi
	_{\theta _{j}\phi \phi ,g,m})\right\Vert =O_{p}((GM)^{1/p}).
	\label{P_Rep_Theta_1_1}
\end{equation}%
Since%
\begin{equation}
	\max_{j\leq d_{\theta }}\left\Vert \mathbb{E}\left[ \psi _{\theta _{j}\phi
		\phi }(z_{i,t})\right] \right\Vert \leq K  \label{P_Rep_Theta_1_2}
\end{equation}%
by Assumption \ref{A3}(iii), from (\ref{P_Rep_Theta_1_1}) and Assumption \ref%
{A5}(i), we have%
\begin{equation}
	\max_{j\leq d_{\theta }}\max_{g\in \mathcal{G},m\in \mathcal{M}}\left\Vert 
	\hat{\Psi}_{\theta _{j}\phi \phi ,g,m}\right\Vert \leq K
	\label{P_Rep_Theta_1_3}
\end{equation}%
wpa1. By the definition of $\hat{\Psi}_{\theta _{j}\phi \phi ,g,m}$, we
obtain the following expression:%
\begin{align}
	(\hat{\phi}_{g,m}-\phi _{g,m}^{\ast })^{\top }\hat{\Psi}_{\theta _{j}\phi
		\phi ,g,m}(\hat{\phi}_{g,m}-\phi _{g,m}^{\ast })& =(\hat{\theta}-\theta
	^{\ast })^{\top }\hat{\Psi}_{\theta _{j}\theta \theta ,g,m}(\hat{\theta}%
	-\theta ^{\ast })  \notag \\
	& +(\hat{\gamma}_{g,m}-\gamma _{g,m}^{\ast })^{\top }\hat{\Psi}_{\theta
		_{j}\gamma \gamma ,g,m}(\hat{\gamma}_{g,m}-\gamma _{g,m}^{\ast })  \notag \\
	& +2(\hat{\theta}-\theta ^{\ast })^{\top }\hat{\Psi}_{\theta _{j}\theta
		\gamma ,g,m}(\hat{\gamma}_{g,m}-\gamma _{g,m}^{\ast }).
	\label{P_Rep_Theta_1_4}
\end{align}%
Applying Lemma \ref{Rate_theta} and (\ref{P_Rep_Theta_1_3}) yields%
\begin{equation}
	\sum_{g\in \mathcal{G}}\sum_{m\in \mathcal{M}}(n_{g}T_{m})(\hat{\theta}%
	-\theta ^{\ast })^{\top }\hat{\Psi}_{\theta _{j}\theta \theta ,g,m}(\hat{%
		\theta}-\theta ^{\ast })=O_{p}(1).  \label{P_Rep_Theta_1_5}
\end{equation}%
By the similar arguments for deriving (\ref{P_Rate_theta_6}), we can show
that%
\begin{align}
	& \sum_{g\in \mathcal{G}}\sum_{m\in \mathcal{M}}(n_{g}T_{m})\hat{\Psi}%
	_{\theta _{j}\theta \gamma ,g,m}(\hat{\gamma}_{g,m}-\gamma _{g,m}^{\ast }) 
	\notag \\
	& =-\sum_{g\in \mathcal{G}}\sum_{m\in \mathcal{M}}(n_{g}T_{m})\hat{\Psi}%
	_{\theta _{j}\theta \gamma ,g,m}\hat{\Psi}_{\gamma \gamma ,g,m}^{-1}\left( 
	\hat{\Psi}_{\gamma ,g,m}+\hat{\Psi}_{\gamma \theta ,g,m}(\hat{\theta}-\theta
	^{\ast })\right)  \notag \\
	& \text{ \ \ \ }+O_{p}\left( 1+\sum_{g\in \mathcal{G}}\sum_{m\in \mathcal{M}%
	}(n_{g}T_{m})||\hat{\phi}_{g,m}-\phi _{g,m}^{\ast }||^{2}\right)  \notag \\
	& =-\sum_{g\in \mathcal{G}}\sum_{m\in \mathcal{M}}(n_{g}T_{m})\hat{\Psi}%
	_{\theta _{j}\theta \gamma ,g,m}\hat{\Psi}_{\gamma \gamma ,g,m}^{-1}\hat{\Psi%
	}_{\gamma ,g,m}+O_{p}((nT)^{1/2}),  \label{P_Rep_Theta_1_6}
\end{align}%
where the second equality is by Assumption \ref{A5}(ii), Lemma \ref%
{Rate_theta}, Lemma \ref{L3AB}(i), (\ref{P_Rate_theta_11}) of Lemma \ref%
{Rate_theta} and (\ref{P_Rep_Theta_1_3}). By the similar arguments for
showing Lemma \ref{L3E}, we have%
\begin{equation*}
	\sum_{g\in \mathcal{G}}\sum_{m\in \mathcal{M}}(n_{g}T_{m})(\hat{\Psi}%
	_{\theta _{j}\theta \gamma ,g,m}\hat{\Psi}_{\gamma \gamma ,g,m}^{-1}-\Psi
	_{\theta _{j}\theta \gamma ,g,m}\Psi _{\gamma \gamma ,g,m}^{-1})\hat{\Psi}%
	_{\gamma ,g,m}=O_{p}\left( GM\right) ,
\end{equation*}%
which together with $\sum_{g\in \mathcal{G}}\sum_{m\in \mathcal{M}}\Psi
_{\theta _{j}\theta \gamma ,g,m}\Psi _{\gamma \gamma ,g,m}^{-1}\hat{\Psi}%
_{\gamma ,g,m}=O_{p}\left( (nT)^{1/2}\right) $ and Assumption \ref{A5}(ii)
implies that 
\begin{equation}
	\sum_{g\in \mathcal{G}}\sum_{m\in \mathcal{M}}(n_{g}T_{m})\hat{\Psi}_{\theta
		_{j}\theta \gamma ,g,m}\hat{\Psi}_{\gamma \gamma ,g,m}\hat{\Psi}_{\gamma
		,g,m}=O_{p}((nT)^{1/2}).  \label{P_Rep_Theta_1_7}
\end{equation}%
Therefore by Lemma \ref{Rate_theta}, (\ref{P_Rep_Theta_1_6}) and (\ref%
{P_Rep_Theta_1_7}),%
\begin{equation}
	\sum_{g\in \mathcal{G}}\sum_{m\in \mathcal{M}}(n_{g}T_{m})(\hat{\theta}%
	-\theta ^{\ast })^{\top }\hat{\Psi}_{\theta _{j}\theta \gamma ,g,m}(\hat{%
		\gamma}_{g,m}-\gamma _{g,m}^{\ast })=O_{p}(1).  \label{P_Rep_Theta_1_8}
\end{equation}%
Next note that 
\begin{align}
	& \sum_{g\in \mathcal{G}}\sum_{m\in \mathcal{M}}(n_{g}T_{m})(\hat{\gamma}%
	_{g,m}-\gamma _{g,m}^{\ast })^{\top }\hat{\Psi}_{\theta _{j}\gamma \gamma
		,g,m}(\hat{\gamma}_{g,m}-\gamma _{g,m}^{\ast })  \notag \\
	& \leq \sum_{g\in \mathcal{G}}\sum_{m\in \mathcal{M}}(n_{g}T_{m})(\hat{\gamma%
	}_{g,m}-\gamma _{g,m}^{\ast })^{\top }\Psi _{\theta _{j}\gamma \gamma ,g,m}(%
	\hat{\gamma}_{g,m}-\gamma _{g,m}^{\ast })  \notag \\
	& \text{ \ \ \ }+\max_{j=1,\ldots ,d_{\theta }}\max_{g\in \mathcal{G},m\in 
		\mathcal{M}}\left\Vert \hat{\Psi}_{\theta _{j}\gamma \gamma ,g,m}-\Psi
	_{\theta _{j}\gamma \gamma ,g,m}\right\Vert \sum_{g\in \mathcal{G}%
	}\sum_{m\in \mathcal{M}}(n_{g}T_{m})||\hat{\gamma}_{g,m}-\gamma _{g,m}^{\ast
	}||^{2}  \notag \\
	& =\sum_{g\in \mathcal{G}}\sum_{m\in \mathcal{M}}(n_{g}T_{m})(\hat{\gamma}%
	_{g,m}-\gamma _{g,m}^{\ast })^{\top }\Psi _{\theta _{j}\gamma \gamma ,g,m}(%
	\hat{\gamma}_{g,m}-\gamma _{g,m}^{\ast })  \notag \\
	& \text{ \ \ \ }+O_{p}\left( (GM)^{1+1/p}\max_{g\in \mathcal{G},m\in 
		\mathcal{M}}(n_{g}T_{m})^{-1/2}\right)  \label{P_Rep_Theta_1_9}
\end{align}%
where the equality is by Lemma \ref{Rate_theta}, (\ref{P_Rate_theta_11}) of
Lemma \ref{Rate_theta} and 
\begin{equation}
	\max_{j=1,\ldots ,d_{\theta }}\max_{g\in \mathcal{G},m\in \mathcal{M}%
	}\left\Vert \hat{\Psi}_{\theta _{j}\gamma \gamma ,g,m}-\Psi _{\theta
		_{j}\gamma \gamma ,g,m}\right\Vert =O_{p}\left( (GM)^{1/p}\max_{g\in 
		\mathcal{G},m\in \mathcal{M}}(n_{g}T_{m})^{-1/2}\right)
	\label{P_Rep_Theta_1_10}
\end{equation}%
which holds by the similar arguments for showing Lemma \ref{P_L3AB_1}. By
the Cauchy-Schwarz inequality, (\ref{P_Rep_Theta_1_2}) and Lemma \ref%
{Rate_gamma}%
\begin{align}
	& \sum_{g\in \mathcal{G}}\sum_{m\in \mathcal{M}}(n_{g}T_{m})(\hat{\gamma}%
	_{g,m}-\gamma _{g,m}^{\ast }+\Psi _{\gamma \gamma ,g,m}^{-1}\hat{\Psi}%
	_{\gamma ,g,m})^{\top }\Psi _{\theta _{j}\gamma \gamma ,g,m}(\hat{\gamma}%
	_{g,m}-\gamma _{g,m}^{\ast }+\Psi _{\gamma \gamma ,g,m}^{-1}\hat{\Psi}%
	_{\gamma ,g,m})  \notag \\
	& \leq K\sum_{g\in \mathcal{G}}\sum_{m\in \mathcal{M}}(n_{g}T_{m})\left\Vert 
	\hat{\gamma}_{g,m}-\gamma _{g,m}^{\ast }+\Psi _{\gamma \gamma ,g,m}^{-1}\hat{%
		\Psi}_{\gamma ,g,m}\right\Vert ^{2}=O_{p}\left( (GM)^{1+4/p}\max_{g\in 
		\mathcal{G},m\in \mathcal{M}}(n_{g}T_{m})^{-1}\right) .
	\label{P_Rep_Theta_1_11}
\end{align}%
Similarly by the triangle inequality, Markov's inequality, Assumption \ref%
{A4}, Lemma \ref{Rate_gamma}, Lemma \ref{P_L3C_3&4} and (\ref%
{P_Rep_Theta_1_2}) 
\begin{align}
	& \left\Vert \sum_{g\in \mathcal{G}}\sum_{m\in \mathcal{M}}(n_{g}T_{m})(\hat{%
		\gamma}_{g,m}-\gamma _{g,m}^{\ast }+\Psi _{\gamma \gamma ,g,m}^{-1}\hat{\Psi}%
	_{\gamma ,g,m})^{\top }\Psi _{\theta _{j}\gamma \gamma ,g,m}\Psi _{\gamma
		\gamma ,g,m}^{-1}\hat{\Psi}_{\gamma ,g,m}\right\Vert  \notag \\
	& \leq K\sum_{g\in \mathcal{G}}\sum_{m\in \mathcal{M}}(n_{g}T_{m})\left\Vert 
	\hat{\gamma}_{g,m}-\gamma _{g,m}^{\ast }+\Psi _{\gamma \gamma ,g,m}^{-1}\hat{%
		\Psi}_{\gamma ,g,m}\right\Vert \left\Vert \hat{\Psi}_{\gamma ,g,m}\right\Vert
	\notag \\
	& \leq O_{p}((GM)^{2/p})\sum_{g\in \mathcal{G}}\sum_{m\in \mathcal{M}%
	}\left\Vert \hat{\Psi}_{\gamma ,g,m}\right\Vert =O_{p}\left(
	(GM)^{1+2/p}\max_{g\in \mathcal{G},m\in \mathcal{M}}(n_{g}T_{m})^{-1/2}%
	\right)  \label{P_Rep_Theta_1_12}
\end{align}%
By the similar arguments for showing Lemma \ref{L3B}, we have 
\begin{align}
	& \sum_{g\in \mathcal{G}}\sum_{m\in \mathcal{M}}(n_{g}T_{m})\hat{\Psi}%
	_{\gamma ,g,m}^{\top }\Psi _{\gamma \gamma ,g,m}^{-1}\Psi _{\theta
		_{j}\gamma \gamma ,g,m}\Psi _{\gamma \gamma ,g,m}^{-1}\hat{\Psi}_{\gamma
		,g,m}  \notag \\
	& =\sum_{g\in \mathcal{G}}\sum_{m\in \mathcal{M}}(n_{g}T_{m})\mathbb{E}\left[
	\hat{\Psi}_{\gamma ,g,m}^{\top }\Psi _{\gamma \gamma ,g,m}^{-1}\Psi _{\theta
		_{j}\gamma \gamma ,g,m}\Psi _{\gamma \gamma ,g,m}^{-1}\hat{\Psi}_{\gamma
		,g,m}\right] +O_{p}((GM)^{1/2}).  \label{P_Rep_Theta_1_13}
\end{align}%
Combining the results in (\ref{P_Rep_Theta_1_11}), (\ref{P_Rep_Theta_1_12})
and (\ref{P_Rep_Theta_1_13}), and applying Assumption \ref{A5}, we deduce
that%
\begin{align}
	& (nT)^{-1/2}\sum_{g\in \mathcal{G}}\sum_{m\in \mathcal{M}}(n_{g}T_{m})(\hat{%
		\gamma}_{g,m}-\gamma _{g,m}^{\ast })^{\top }\Psi _{\theta _{j}\gamma \gamma
		,g,m}(\hat{\gamma}_{g,m}-\gamma _{g,m}^{\ast })  \notag \\
	& =(nT)^{-1/2}\sum_{g\in \mathcal{G}}\sum_{m\in \mathcal{M}}(n_{g}T_{m})%
	\mathbb{E}\left[ \hat{\Psi}_{\gamma ,g,m}^{\top }\Psi _{\gamma \gamma
		,g,m}^{-1}\Psi _{\theta _{j}\gamma \gamma ,g,m}\Psi _{\gamma \gamma
		,g,m}^{-1}\hat{\Psi}_{\gamma ,g,m}\right] +O_{p}\left(
	(GM)^{1/2}(nT)^{-1/2}\right)  \notag \\
	& \text{ \ \ \ }+O_{p}\left( ((GM)^{4/p}\max_{g\in \mathcal{G},m\in \mathcal{%
			M}}(n_{g}T_{m})^{-1}+(GM)^{2/p}\max_{g\in \mathcal{G},m\in \mathcal{M}%
	}(n_{g}T_{m})^{-1/2})(GM)(nT)^{-1/2}\right)  \notag \\
	& =(nT)^{-1/2}\sum_{g\in \mathcal{G}}\sum_{m\in \mathcal{M}}(n_{g}T_{m})%
	\mathbb{E}\left[ \hat{\Psi}_{\gamma ,g,m}^{\top }\Psi _{\gamma \gamma
		,g,m}^{-1}\Psi _{\theta _{j}\gamma \gamma ,g,m}\Psi _{\gamma \gamma
		,g,m}^{-1}\hat{\Psi}_{\gamma ,g,m}\right] +o_{p}(1)  \label{P_Rep_Theta_1_14}
\end{align}%
where the second equality is by Assumption \ref{A5}. The assertion of the
lemma now follows from (\ref{P_Rep_Theta_1_4}), (\ref{P_Rep_Theta_1_5}), (%
\ref{P_Rep_Theta_1_8}) and (\ref{P_Rep_Theta_1_14}). \hfill $Q.E.D.$

\bigskip

\begin{lemma}
	\textit{\label{Rep_Theta_2}}Under Assumptions \ref{A1}, \ref{A2}, \ref{A3}, %
	\ref{A4} and \ref{A5}, we have%
	\begin{align*}
		& (nT)^{-1/2}\sum_{g\in \mathcal{G}}\sum_{m\in \mathcal{M}}(n_{g}T_{m})\hat{%
			\Psi}_{\theta \gamma ,g,m}(\hat{\gamma}_{g,m}-\gamma _{g,m}^{\ast }) \\
		& =-(nT)^{-1/2}\sum_{g\in \mathcal{G}}\sum_{m\in \mathcal{M}}(n_{g}T_{m})%
		\left[ \Psi _{\theta \gamma ,g,m}\Psi _{\gamma \gamma ,g,m}^{-1}\hat{\Psi}%
		_{\gamma ,g,m}+\Psi _{\theta \gamma ,g,m}\Psi _{\gamma \gamma ,g,m}^{-1}\Psi
		_{\gamma \theta ,g,m}(\hat{\theta}-\theta ^{\ast })\right] \\
		& \text{ \ \ }-(nT)^{-1/2}\sum_{g\in \mathcal{G}}\sum_{i\in
			I_{g}}(n_{g}T_{m})\mathbb{E}\left[ \hat{U}_{\gamma ,g,m}\Psi _{\gamma \gamma
			,g,m}^{-1}\hat{\Psi}_{\gamma ,g,m}\right] \\
		& \text{ \ \ }-\frac{(nT)^{-1/2}}{2}\sum_{g\in \mathcal{G}}\sum_{m\in 
			\mathcal{M}}(n_{g}T_{m})\Psi _{\theta \gamma ,g,m}\Psi _{\gamma \gamma
			,g,m}^{-1}\left( \mathbb{E}\left[ \hat{\Psi}_{\gamma ,g,m}^{\top }\Psi
		_{\gamma \gamma ,g,m}^{-1}\Psi _{\gamma _{j}\gamma \gamma ,g,m}\Psi _{\gamma
			\gamma ,g,m}^{-1}\hat{\Psi}_{\gamma ,g,m}\right] \right) _{j\leq d_{\gamma
		}}+o_{p}(1)
	\end{align*}%
	where $\hat{U}_{\gamma ,g,m}\equiv \hat{\Psi}_{\theta \gamma ,g,m}-\Psi
	_{\theta \gamma ,g,m}\Psi _{\gamma \gamma ,g,m}^{-1}\hat{\Psi}_{\gamma
		\gamma ,g,m}$.
\end{lemma}

\noindent \textsc{Proof of Lemma \ref{Rep_Theta_2}}.\ By the definition of $%
\Psi _{\gamma _{j}\phi \phi ,g,m}$ and the triangle inequality, we obtain:%
\begin{align}
	& \max_{g\in \mathcal{G},m\in \mathcal{M}}\left\vert \frac{(\hat{\phi}%
		_{g,m}-\phi _{g,m}^{\ast })^{\top }\Psi _{\gamma _{j}\phi \phi ,g,m}(\hat{%
			\phi}_{g,m}-\phi _{g,m}^{\ast })-(\hat{\gamma}_{g,m}-\gamma _{g,m}^{\ast
		})^{\top }\Psi _{\gamma _{j}\gamma \gamma ,g,m}(\hat{\gamma}_{g,m}-\gamma
		_{g,m}^{\ast })}{(nT)^{-1}+(nT)^{-1/2}||\hat{\gamma}_{g,m}-\gamma
		_{g,m}^{\ast }||}\right\vert  \notag \\
	& \leq \max_{g\in \mathcal{G},m\in \mathcal{M}}\frac{\left\vert (\hat{\theta}%
		-\theta ^{\ast })^{\top }\Psi _{\gamma _{j}\theta \theta ,g,m}(\hat{\theta}%
		-\theta ^{\ast })\right\vert +2\left\vert (\hat{\theta}-\theta ^{\ast
		})^{\top }\Psi _{\gamma _{j}\theta \gamma ,g,m}(\hat{\gamma}_{g,m}-\gamma
		_{g,m}^{\ast })\right\vert }{(nT)^{-1}+(nT)^{-1/2}||\hat{\gamma}%
		_{g,m}-\gamma _{g,m}^{\ast }||}=O_{p}(1)  \label{P_Rep_Theta_2_1A}
\end{align}%
where the last equality is by Assumption \ref{A3}(iii), the Cauchy-Schwarz
inequality and Lemma \ref{Rate_theta}. By (\ref{P_Rate_theta_1}) and (\ref%
{P_Rate_theta_2a}) of Lemma \ref{FOCs}, we have 
\begin{equation}
	\max_{g\in \mathcal{G},m\in \mathcal{M}}\frac{\left\Vert \hat{\Psi}_{\gamma
			,g,m}+\hat{\Psi}_{\gamma \phi ,g,m}(\hat{\phi}_{g,m}-\phi _{g,m}^{\ast })+%
		\frac{\left( (\hat{\phi}_{g,m}-\phi _{g,m}^{\ast })^{\top }\Psi _{\gamma
				_{j}\phi \phi ,g,m}(\hat{\phi}_{g,m}-\phi _{g,m}^{\ast })\right) _{j\leq
				d_{\gamma }}}{2}\right\Vert }{(nT)^{-1}+(GM)^{1/p}(n_{g}T_{m})^{-1/2}||\hat{%
			\phi}_{g,m}-\phi _{g,m}^{\ast }||^{2}+||\hat{\phi}_{g,m}-\phi _{g,m}^{\ast
		}||^{3}}=O_{p}(1).  \label{P_Rep_Theta_2_1B}
\end{equation}%
Applying the triangle inequality, Assumption \ref{A5}(i), Lemma \ref%
{Rate_gamma} obtains%
\begin{align}
	\max_{g\in \mathcal{G},m\in \mathcal{M}}||(n_{g}T_{m})^{1/2}(\hat{\gamma}%
	_{g,m}-\gamma _{g,m}^{\ast })||& \leq \max_{g\in \mathcal{G},m\in \mathcal{M}%
	}\left\Vert (n_{g}T_{m})^{1/2}\Psi _{\gamma \gamma ,g,m}^{-1}\hat{\Psi}%
	_{\gamma ,g,m}\right\Vert  \notag \\
	& +\max_{g\in \mathcal{G},m\in \mathcal{M}}\left\Vert (n_{g}T_{m})^{1/2}(%
	\hat{\gamma}_{g,m}-\gamma _{g,m}^{\ast }+\Psi _{\gamma \gamma ,g,m}^{-1}\hat{%
		\Psi}_{\gamma ,g,m})\right\Vert  \notag \\
	& =\max_{g\in \mathcal{G},m\in \mathcal{M}}\left\Vert (n_{g}T_{m})^{1/2}\Psi
	_{\gamma \gamma ,g,m}^{-1}\hat{\Psi}_{\gamma ,g,m}\right\Vert +o_{p}(1).
	\label{P_Rep_Theta_2_1C}
\end{align}%
By the maximum inequality under $L_{p}$-norm, Markov's inequality,
Assumption \ref{A3}(iii) and (\ref{L0-1}) in Lemma \ref{L0}, 
\begin{equation}
	\max_{g\in \mathcal{G},m\in \mathcal{M}}\left\Vert (n_{g}T_{m})^{1/2}\Psi
	_{\gamma \gamma ,g,m}^{-1}\hat{\Psi}_{\gamma ,g,m}\right\Vert
	=O_{p}((GM)^{1/p})  \label{P_Rep_Theta_2_1D}
\end{equation}%
which together with (\ref{P_Rep_Theta_2_1C}) implies that 
\begin{equation}
	\max_{g\in \mathcal{G},m\in \mathcal{M}}||(n_{g}T_{m})^{1/2}(\hat{\gamma}%
	_{g,m}-\gamma _{g,m}^{\ast })||=O_{p}((GM)^{1/p}).  \label{P_Rep_Theta_2_1E}
\end{equation}%
By Lemma \ref{Rate_theta} and (\ref{P_Rep_Theta_2_1E}),%
\begin{align*}
	\max_{g\in \mathcal{G},m\in \mathcal{M}}||(n_{g}T_{m})^{1/2}(\hat{\phi}%
	_{g,m}-\phi _{g,m}^{\ast })||& \leq \max_{g\in \mathcal{G},m\in \mathcal{M}%
	}||(n_{g}T_{m})^{1/2}(\hat{\theta}-\theta ^{\ast })|| \\
	& +\max_{g\in \mathcal{G},m\in \mathcal{M}}||(n_{g}T_{m})^{1/2}(\hat{\gamma}%
	_{g,m}-\gamma _{g,m}^{\ast })|| \\
	& =O_{p}((GM)^{1/p})
\end{align*}%
which implies that 
\begin{equation}
	\max_{g\in \mathcal{G},m\in \mathcal{M}}\frac{%
		(nT)^{-1}+(GM)^{1/p}(n_{g}T_{m})^{-1/2}||\hat{\phi}_{g,m}-\phi _{g,m}^{\ast
		}||^{2}+||\hat{\phi}_{g,m}-\phi _{g,m}^{\ast }||^{3}}{%
		(nT)^{-1}+(GM)^{1/p}(n_{g}T_{m})^{-1/2}||\hat{\gamma}_{g,m}-\gamma
		_{g,m}^{\ast }||^{2}+(nT)^{-1/2}||\hat{\gamma}_{g,m}-\gamma _{g,m}^{\ast }||}%
	=O_{p}(1).  \label{P_Rep_Theta_2_1F}
\end{equation}%
Collecting the results in (\ref{P_Rep_Theta_2_1A}), (\ref{P_Rep_Theta_2_1B})
and (\ref{P_Rep_Theta_2_1F}), we deduce that%
\begin{equation}
	\max_{g\in \mathcal{G},m\in \mathcal{M}}\frac{\left\Vert \hat{\Psi}_{\gamma
			,g,m}+\hat{\Psi}_{\gamma \phi ,g,m}(\hat{\phi}_{g,m}-\phi _{g,m}^{\ast })+%
		\frac{\left( (\hat{\gamma}_{g,m}-\gamma _{g,m}^{\ast })^{\top }\Psi _{\gamma
				_{j}\gamma \gamma ,g,m}(\hat{\gamma}_{g,m}-\gamma _{g,m}^{\ast })\right)
			_{j\leq d_{\gamma }}}{2}\right\Vert }{%
		(nT)^{-1}+(GM)^{1/p}(n_{g}T_{m})^{-1/2}||\hat{\gamma}_{g,m}-\gamma
		_{g,m}^{\ast }||^{2}+(nT)^{-1/2}||\hat{\gamma}_{g,m}-\gamma _{g,m}^{\ast }||}%
	=O_{p}(1).  \label{P_Rep_Theta_2_1}
\end{equation}

Next note that by Lemma \ref{Rate_theta} 
\begin{equation*}
	(nT)^{-1/2}\sum_{g\in \mathcal{G}}\sum_{m\in \mathcal{M}}(n_{g}T_{m})||\hat{%
		\gamma}_{g,m}-\gamma _{g,m}^{\ast }||^{2}=O_{p}\left( (GM)(nT)^{-1/2}\right)
\end{equation*}%
which together with Assumption \ref{A5} and (\ref{P_Rep_Theta_2_1E}) implies
that%
\begin{align}
	& (nT)^{-1/2}\sum_{g\in \mathcal{G}}\sum_{m\in \mathcal{M}%
	}(n_{g}T_{m})\left( (nT)^{-1}+(GM)^{1/p}(n_{g}T_{m})^{-1/2}||\hat{\gamma}%
	_{g,m}-\gamma _{g,m}^{\ast }||^{2}+(nT)^{-1/2}||\hat{\gamma}_{g,m}-\gamma
	_{g,m}^{\ast }||\right)  \notag \\
	& \leq (nT)^{-1/2}+\max_{g\in \mathcal{G},m\in \mathcal{M}}||\hat{\gamma}%
	_{g,m}-\gamma _{g,m}^{\ast }||  \notag \\
	& +(GM)^{1/p}\max_{g\in \mathcal{G},m\in \mathcal{M}%
	}(n_{g}T_{m})^{-1/2}(nT)^{-1/2}\sum_{g\in \mathcal{G}}\sum_{m\in \mathcal{M}%
	}(n_{g}T_{m})||\hat{\gamma}_{g,m}-\gamma _{g,m}^{\ast }||^{2}  \notag \\
	& =O_{p}\left( (nT)^{-1/2}+(1+(GM)(nT)^{-1/2})(GM)^{1/p}\max_{g\in \mathcal{G%
		},m\in \mathcal{M}}(n_{g}T_{m})^{-1/2}\right) =o_{p}(1).
	\label{P_Rep_Theta_2_2}
\end{align}%
Therefore, by (\ref{P_Rep_Theta_2_1}) and (\ref{P_Rep_Theta_2_2}), we have%
\begin{align}
	& (nT)^{-1/2}\sum_{g\in \mathcal{G}}\sum_{m\in \mathcal{M}}(n_{g}T_{m})\hat{%
		\Psi}_{\theta \gamma ,g,m}\hat{\Psi}_{\gamma \gamma ,g,m}^{-1}\hat{\Psi}%
	_{\gamma \gamma ,g,m}(\hat{\gamma}_{g,m}-\gamma _{g,m}^{\ast })  \notag \\
	& =-(nT)^{-1/2}\sum_{g\in \mathcal{G}}\sum_{m\in \mathcal{M}}(n_{g}T_{m})%
	\hat{\Psi}_{\theta \gamma ,g,m}\hat{\Psi}_{\gamma \gamma ,g,m}^{-1}\hat{\Psi}%
	_{\gamma ,g,m}  \notag \\
	& \text{ \ \ }-(nT)^{-1/2}\sum_{g\in \mathcal{G}}\sum_{m\in \mathcal{M}%
	}(n_{g}T_{m})\hat{\Psi}_{\theta \gamma ,g,m}\hat{\Psi}_{\gamma \gamma
		,g,m}^{-1}\Psi _{\gamma \theta ,g,m}(\hat{\theta}-\theta ^{\ast })  \notag \\
	& \text{ \ \ }-\frac{(nT)^{-1/2}}{2}\sum_{g\in \mathcal{G}}\sum_{m\in 
		\mathcal{M}}(n_{g}T_{m})\hat{\Psi}_{\theta \gamma ,g,m}\hat{\Psi}_{\gamma
		\gamma ,g,m}^{-1}((\hat{\gamma}_{g,m}-\gamma _{g,m}^{\ast })^{\top }\hat{\Psi%
	}_{\gamma _{j}\gamma \gamma ,g,m}(\hat{\gamma}_{g,m}-\gamma _{g,m}^{\ast
	}))_{j}+o_{p}(1).  \label{P_Rep_Theta_2_3}
\end{align}%
Since%
\begin{align*}
	\hat{\Psi}_{\theta \gamma ,g,m}\hat{\Psi}_{\gamma \gamma ,g,m}^{-1}-\Psi
	_{\theta \gamma ,g,m}\Psi _{\gamma \gamma ,g,m}^{-1}& =(\hat{\Psi}_{\theta
		\gamma ,g,m}-\Psi _{\theta \gamma ,g,m})\Psi _{\gamma \gamma ,g,m}^{-1}+\Psi
	_{\theta \gamma ,g,m}(\hat{\Psi}_{\gamma \gamma ,g,m}^{-1}-\Psi _{\gamma
		\gamma ,g,m}^{-1}) \\
	& \text{ \ \ }+(\hat{\Psi}_{\theta \gamma ,g,m}-\Psi _{\theta \gamma ,g,m})(%
	\hat{\Psi}_{\gamma \gamma ,g,m}^{-1}-\Psi _{\gamma \gamma ,g,m}^{-1}),
\end{align*}%
and%
\begin{equation*}
	\hat{\Psi}_{\gamma \gamma ,g,m}^{-1}-\Psi _{\gamma \gamma ,g,m}^{-1}=-\hat{%
		\Psi}_{\gamma \gamma ,g,m}^{-1}\left( \hat{\Psi}_{\gamma \gamma ,g,m}-\Psi
	_{\gamma \gamma ,g,m}\right) \Psi _{\gamma \gamma ,g,m}^{-1},
\end{equation*}%
by the triangle inequality and the Cauchy-Schwarz inequality,%
\begin{align}
	& \left\Vert (nT)^{-1}\sum_{g\in \mathcal{G}}\sum_{m\in \mathcal{M}%
	}(n_{g}T_{m})(\hat{\Psi}_{\theta \gamma ,g,m}\hat{\Psi}_{\gamma \gamma
		,g,m}^{-1}-\Psi _{\theta \gamma ,g,m}\Psi _{\gamma \gamma
		,g,m}^{-1})\right\Vert  \notag \\
	& \leq K\max_{g\in \mathcal{G},m\in \mathcal{M}}(\lambda _{\max }((\hat{\Psi}%
	_{\gamma \gamma ,g,m}^{-1})^{2}))^{1/2}(nT)^{-1}\sum_{g\in \mathcal{G}%
	}\sum_{m\in \mathcal{M}}(n_{g}T_{m})\left\Vert \hat{\Psi}_{\theta \gamma
		,g,m}-\Psi _{\theta \gamma ,g,m}\right\Vert \left\Vert \hat{\Psi}_{\gamma
		\gamma ,g,m}-\Psi _{\gamma \gamma ,g,m}\right\Vert  \notag \\
	& \text{ \ \ }+K(nT)^{-1}\sum_{g\in \mathcal{G}}\sum_{m\in \mathcal{M}%
	}(n_{g}T_{m})\left\Vert \hat{\Psi}_{\theta \gamma ,g,m}-\Psi _{\theta \gamma
		,g,m}\right\Vert  \notag \\
	& \text{ \ \ }+K\max_{g\in \mathcal{G},m\in \mathcal{M}}(\lambda _{\max }((%
	\hat{\Psi}_{\gamma \gamma ,g,m}^{-1})^{2})^{1/2}(nT)^{-1}\sum_{g\in \mathcal{%
			G}}\sum_{m\in \mathcal{M}}(n_{g}T_{m})\left\Vert \hat{\Psi}_{\gamma \gamma
		,g,m}-\Psi _{\gamma \gamma ,g,m}\right\Vert .  \label{P_Rep_Theta_2_4}
\end{align}%
By (\ref{L0-1}) in Lemma \ref{L0}, 
\begin{equation}
	\mathbb{E}\left[ \left\Vert (n_{g}T_{m})^{1/2}(\hat{\Psi}_{\theta \gamma
		,g,m}-\Psi _{\theta \gamma ,g,m})\right\Vert ^{2}\right] \leq K\text{ and }%
	\mathbb{E}\left[ \left\Vert (n_{g}T_{m})^{1/2}(\hat{\Psi}_{\gamma \gamma
		,g,m}-\Psi _{\gamma \gamma ,g,m})\right\Vert ^{2}\right] \leq K,
	\label{P_Rep_Theta_2_5}
\end{equation}%
which together with Markov's inequality, Assumption \ref{A5}(i), Lemma \ref%
{L3AB}(ii) and (\ref{P_Rep_Theta_2_4}) implies that%
\begin{equation}
	(nT)^{-1}\sum_{g\in \mathcal{G}}\sum_{m\in \mathcal{M}}(n_{g}T_{m})(\hat{\Psi%
	}_{\theta \gamma ,g,m}\hat{\Psi}_{\gamma \gamma ,g,m}^{-1}-\Psi _{\theta
		\gamma ,g,m}\Psi _{\gamma \gamma ,g,m}^{-1})=o_{p}\left( 1\right) .
	\label{P_Rep_Theta_2_6}
\end{equation}%
Therefore by Lemma \ref{Rate_theta} and (\ref{P_Rep_Theta_2_6}), we have 
\begin{align}
	& (nT)^{-1/2}\sum_{g\in \mathcal{G}}\sum_{m\in \mathcal{M}}(n_{g}T_{m})\hat{%
		\Psi}_{\theta \gamma ,g,m}\hat{\Psi}_{\gamma \gamma ,g,m}^{-1}\Psi _{\gamma
		\theta ,g,m}(\hat{\theta}-\theta ^{\ast })  \notag \\
	& =(nT)^{-1/2}\sum_{g\in \mathcal{G}}\sum_{m\in \mathcal{M}}(n_{g}T_{m})\Psi
	_{\theta \gamma ,g,m}\Psi _{\gamma \gamma ,g,m}^{-1}\Psi _{\gamma \theta
		,g,m}(\hat{\theta}-\theta ^{\ast })+o_{p}(1).  \label{P_Rep_Theta_2_7}
\end{align}%
By Lemma \ref{L3E},%
\begin{align}
	& (nT)^{-1/2}\sum_{g\in \mathcal{G}}\sum_{m\in \mathcal{M}}(n_{g}T_{m})\hat{%
		\Psi}_{\theta \gamma ,g,m}\hat{\Psi}_{\gamma \gamma ,g,m}^{-1}\hat{\Psi}%
	_{\gamma ,g,m}  \notag \\
	& =(nT)^{-1/2}\sum_{g\in \mathcal{G}}\sum_{m\in \mathcal{M}}(n_{g}T_{m})\Psi
	_{\theta \gamma ,g,m}\Psi _{\gamma \gamma ,g,m}^{-1}\hat{\Psi}_{\gamma ,g,m}
	\notag \\
	& \text{ \ \ }+(nT)^{-1/2}\sum_{g\in \mathcal{G}}\sum_{i\in
		I_{g}}(n_{g}T_{m})\mathbb{E}\left[ \hat{U}_{\gamma ,g,m}\Psi _{\gamma \gamma
		,g,m}^{-1}\hat{\Psi}_{\gamma ,g,m}\right] +o_{p}\left( 1\right) .
	\label{P_Rep_Theta_2_8}
\end{align}%
By Lemma \ref{L3AB}(i, ii), Lemma \ref{Rate_theta}, Lemma \ref{P_L3AB_1} and
(\ref{P_Rate_theta_11}) of Lemma \ref{Rate_theta}, 
\begin{align}
	& \left\Vert \sum_{g\in \mathcal{G}}\sum_{m\in \mathcal{M}}(n_{g}T_{m})(\hat{%
		\Psi}_{\theta \gamma ,g,m}\hat{\Psi}_{\gamma \gamma ,g,m}-\Psi _{\theta
		\gamma ,g,m}\Psi _{\gamma \gamma ,g,m}^{-1})((\hat{\gamma}_{g,m}-\gamma
	_{g,m}^{\ast })^{\top }\hat{\Psi}_{\gamma _{j}\gamma \gamma ,g,m}(\hat{\gamma%
	}_{g,m}-\gamma _{g,m}^{\ast }))_{j\leq d_{\gamma }}\right\Vert  \notag \\
	& \leq K\max_{g\in \mathcal{G},m\in \mathcal{M}}\left( \left\Vert \hat{\Psi}%
	_{\theta \gamma ,g,m}\hat{\Psi}_{\gamma \gamma ,g,m}-\Psi _{\theta \gamma
		,g,m}\Psi _{\gamma \gamma ,g,m}^{-1}\right\Vert \max_{j\leq d_{\gamma
	}}\left\Vert \hat{\Psi}_{\gamma _{j}\gamma \gamma ,g,m}\right\Vert \right) 
	\notag \\
	& \text{ \ \ }\times \sum_{g\in \mathcal{G}}\sum_{m\in \mathcal{M}%
	}(n_{g}T_{m})||\hat{\gamma}_{g,m}-\gamma _{g,m}^{\ast }||^{2}  \notag \\
	& =O_{p}((GM)^{1+2/p}\max_{g\in \mathcal{G},m\in \mathcal{M}%
	}(n_{g}T_{m})^{-1})=o_{p}((nT)^{1/2}).  \label{P_Rep_Theta_2_9}
\end{align}%
Similarly, we can show that 
\begin{equation}
	(nT)^{-1/2}\sum_{g\in \mathcal{G}}\sum_{m\in \mathcal{M}}\Psi _{\theta
		\gamma ,g,m}\Psi _{\gamma \gamma ,g,m}^{-1}((\hat{\gamma}_{g,m}-\gamma
	_{g,m}^{\ast })^{\top }\hat{\Psi}_{\gamma _{j}\gamma \gamma ,g,m}^{\ast }(%
	\hat{\gamma}_{g,m}-\gamma _{g,m}^{\ast }))_{j\leq d_{\gamma }}=o_{p}(1)
	\label{P_Rep_Theta_2_10}
\end{equation}%
where $\hat{\Psi}_{\gamma _{j}\gamma \gamma ,g,m}^{\ast }\equiv \hat{\Psi}%
_{\gamma _{j}\gamma \gamma ,g,m}-\Psi _{\gamma _{j}\gamma \gamma ,g,m}$. By
the similar arguments for deriving (\ref{P_Rep_Theta_1_14}), we can show
that 
\begin{align}
	& (nT)^{-1/2}\sum_{g\in \mathcal{G}}\sum_{m\in \mathcal{M}}(n_{g}T_{m})\Psi
	_{\theta \gamma ,g,m}\Psi _{\gamma \gamma ,g,m}^{-1}((\hat{\gamma}%
	_{g,m}-\gamma _{g,m}^{\ast })^{\top }\Psi _{\gamma _{j}\gamma \gamma ,g,m}(%
	\hat{\gamma}_{g,m}-\gamma _{g,m}^{\ast }))_{j\leq d_{\gamma }}  \notag \\
	& =(nT)^{-1/2}\sum_{g\in \mathcal{G}}\sum_{m\in \mathcal{M}}(n_{g}T_{m})\Psi
	_{\theta \gamma ,g,m}\Psi _{\gamma \gamma ,g,m}^{-1}(\mathbb{E}[\hat{\Psi}%
	_{\gamma ,g,m}^{\top }\Psi _{\gamma \gamma ,g,m}^{-1}\Psi _{\gamma
		_{j}\gamma \gamma ,g,m}\Psi _{\gamma \gamma ,g,m}^{-1}\hat{\Psi}_{\gamma
		,g,m}])_{j\leq d_{\gamma }}+o_{p}\left( 1\right) .  \label{P_Rep_Theta_2_11}
\end{align}%
The assertion of the lemma follows from (\ref{P_Rep_Theta_2_3}), (\ref%
{P_Rep_Theta_2_7}) and (\ref{P_Rep_Theta_2_8})-(\ref{P_Rep_Theta_2_11}%
).\hfill $Q.E.D.$

\bigskip

\begin{lemma}
	\textit{\label{Rep_L_A}}Under Assumptions \ref{A1}, \ref{A2}, \ref{A3}, \ref%
	{A4} and \ref{A5}, we have 
	\begin{equation*}
		(nT)^{-1/2}\sum_{g\in \mathcal{G}}\sum_{m\in \mathcal{M}%
		}||(n_{g}T_{m})^{1/2}(\hat{\gamma}_{g,m}-\gamma _{g,m}^{\ast
		})||^{s}=O_{p}\left( (GM)(nT)^{-1/2}\right)
	\end{equation*}%
	for any $s\leq p$.
\end{lemma}

\noindent \textsc{Proof of Lemma \ref{Rep_L_A}}.\ By the triangle inequality
and the $c_{r}$-inequality,%
\begin{eqnarray}
	\sum_{g\in \mathcal{G}}\sum_{m\in \mathcal{M}}||(n_{g}T_{m})^{1/2}(\hat{%
		\gamma}_{g,m}-\gamma _{g,m}^{\ast })||^{s} &\leq &K\sum_{g\in \mathcal{G}%
	}\sum_{m\in \mathcal{M}}||(n_{g}T_{m})^{1/2}(\hat{\gamma}_{g,m}-\gamma
	_{g,m}^{\ast }+\Psi _{\gamma \gamma ,g,m}^{-1}\hat{\Psi}_{\gamma ,g,m})||^{s}
	\notag \\
	&&+K\sum_{g\in \mathcal{G}}\sum_{m\in \mathcal{M}}||(n_{g}T_{m})^{1/2}\Psi
	_{\gamma \gamma ,g,m}^{-1}\hat{\Psi}_{\gamma ,g,m}||^{s}.
	\label{P_Rep_L_A_1}
\end{eqnarray}%
By Lemma \ref{Rate_gamma},%
\begin{align}
	& (nT)^{-1/2}\sum_{g\in \mathcal{G}}\sum_{m\in \mathcal{M}%
	}||(n_{g}T_{m})^{1/2}(\hat{\gamma}_{g,m}-\gamma _{g,m}^{\ast }+\Psi _{\gamma
		\gamma ,g,m}^{-1}\hat{\Psi}_{\gamma ,g,m})||^{s}  \notag \\
	& \leq \max_{g\in \mathcal{G},m\in \mathcal{M}}\frac{\left\Vert \hat{\gamma}%
		_{g,m}-\gamma _{g,m}^{\ast }+\Psi _{\gamma \gamma ,g,m}^{-1}\hat{\Psi}%
		_{\gamma ,g,m}\right\Vert ^{s}}{(GM)^{2s/p}(n_{g}T_{m})^{-s}}%
	(nT)^{-1/2}\sum_{g\in \mathcal{G}}\sum_{m\in \mathcal{M}%
	}(GM)^{2s/p}(n_{g}T_{m})^{-s/2}  \notag \\
	& =O_{p}\left( (GM)^{1+2s/p}(nT)^{-1/2}\max_{g\in \mathcal{G},m\in \mathcal{M%
	}}(n_{g}T_{m})^{-s/2}\right) =o_{p}\left( (GM)(nT)^{-1/2}\right)
	\label{P_Rep_L_A_2}
\end{align}%
where the second equality is by Assumption \ref{A5}(ii). By Assumption \ref%
{A4} and (\ref{L0-1}) in Lemma \ref{L0}, 
\begin{equation}
	\max_{g\in \mathcal{G},m\in \mathcal{M}}\mathbb{E}\left[
	||(n_{g}T_{m})^{1/2}\Psi _{\gamma \gamma ,g,m}^{-1}\hat{\Psi}_{\gamma
		,g,m}||^{s}\right] \leq K\max_{g\in \mathcal{G},m\in \mathcal{M}}\mathbb{E}%
	\left[ ||(n_{g}T_{m})^{1/2}\hat{\Psi}_{\gamma ,g,m}||^{s}\right] \leq K
	\label{P_Rep_L_A_3}
\end{equation}%
which together with Markov's inequality implies that 
\begin{equation}
	(nT)^{-1/2}\sum_{g\in \mathcal{G}}\sum_{m\in \mathcal{M}%
	}||(n_{g}T_{m})^{1/2}\Psi _{\gamma \gamma ,g,m}^{-1}\hat{\Psi}_{\gamma
		,g,m}||^{s}=O_{p}((GM)(nT)^{-1/2}).  \label{P_Rep_L_A_4}
\end{equation}%
The assertion of the lemma now follows from (\ref{P_Rep_L_A_2}) and (\ref%
{P_Rep_L_A_4}).\hfill $Q.E.D.$

\bigskip

\begin{lemma}
	\textit{\label{Rep_L_B}}Under Assumptions \ref{A1}, \ref{A2}, \ref{A3}, \ref%
	{A4} and \ref{A5}, we have%
	\begin{align*}
		& (nT)^{-1/2}\sum_{g\in \mathcal{G}}\sum_{m\in \mathcal{M}}(n_{g}T_{m})\left[
		\hat{\Psi}_{g,m}(\hat{\phi}_{g,m})-\hat{\Psi}_{g,m}(\phi _{g,m}^{\ast })%
		\right] \\
		& =(nT)^{-1/2}\sum_{g\in \mathcal{G}}\sum_{m\in \mathcal{M}}(n_{g}T_{m})\hat{%
			\Psi}_{\phi ,g,m}(\hat{\phi}_{g,m}-\phi _{g,m}^{\ast }) \\
		& \text{ \ \ }+\frac{(nT)^{-1/2}}{2}\sum_{g\in \mathcal{G}}\sum_{m\in 
			\mathcal{M}}(n_{g}T_{m})(\hat{\phi}_{g,m}-\phi _{g,m}^{\ast })^{\top }\hat{%
			\Psi}_{\phi \phi ,g,m}(\hat{\phi}_{g,m}-\phi _{g,m}^{\ast }) \\
		& \text{ \ \ }+O_{p}\left( (GM)^{1/2}(nT)^{-1/2}\left(
		(GM)^{1/2+2/p}\max_{g\in \mathcal{G},m\in \mathcal{M}}(n_{g}T_{m})^{-1}+%
		\max_{g\in \mathcal{G},m\in \mathcal{M}}(n_{g}T_{m})^{-1/2}\right) \right) .
	\end{align*}
\end{lemma}

\noindent \textsc{Proof of Lemma \ref{Rep_L_B}}.\ Applying the Taylor
expansion obtains%
\begin{align}
	& (nT)^{-1/2}\sum_{g\in \mathcal{G}}\sum_{m\in \mathcal{M}}(n_{g}T_{m})\left[
	\hat{\Psi}_{g,m}(\hat{\phi}_{g,m})-\hat{\Psi}_{g,m}(\phi _{g,m}^{\ast })%
	\right]  \notag \\
	& =(nT)^{-1/2}\sum_{g\in \mathcal{G}}\sum_{m\in \mathcal{M}}(n_{g}T_{m})\hat{%
		\Psi}_{\phi ,g,m}(\hat{\phi}_{g,m}-\phi _{g,m}^{\ast })  \notag \\
	& \text{ \ \ }+\frac{(nT)^{-1/2}}{2}\sum_{g\in \mathcal{G}}\sum_{m\in 
		\mathcal{M}}(n_{g}T_{m})(\hat{\phi}_{g,m}-\phi _{g,m}^{\ast })^{\top }\hat{%
		\Psi}_{\phi \phi ,g,m}(\hat{\phi}_{g,m}-\phi _{g,m}^{\ast })  \notag \\
	& \text{ \ \ }+\frac{(nT)^{-1/2}}{6}\sum_{g\in \mathcal{G}}\sum_{m\in 
		\mathcal{M}}(n_{g}T_{m})((\hat{\phi}_{g,m}-\phi _{g,m}^{\ast })^{\top }%
	\tilde{\Psi}_{\phi \phi \phi _{j},g,m}(\hat{\phi}_{g,m}-\phi _{g,m}^{\ast
	}))_{j\leq d_{\phi }}^{\top }(\hat{\phi}_{g,m}-\phi _{g,m}^{\ast }),
	\label{P_Rep_L_B_1}
\end{align}%
where%
\begin{equation*}
	\tilde{\Psi}_{\phi \phi \phi _{j},g,m}\equiv (n_{g}T_{m})^{-1}\sum_{i\in
		I_{g}}\sum_{t\in I_{m}}\frac{\partial }{\partial \phi _{j}}\left( \frac{%
		\partial ^{2}\psi (z_{i,t};\tilde{\phi}_{g,m})}{\partial \phi \partial \phi
		^{\top }}\right)
\end{equation*}%
for any $j\leq k$, and $\tilde{\phi}_{g,m}$ is between $\hat{\phi}_{g,m}$
and $\phi _{g,m}^{\ast }$. By (\ref{L1-2}) in Lemma \ref{L1}, we can show
that%
\begin{equation}
	\max_{g\in \mathcal{G},m\in \mathcal{M}}\left\Vert \frac{\tilde{\Psi}_{\phi
			\phi \phi _{j},g,m}-\Psi _{\phi \phi \phi _{j},g,m}}{%
		(GM)^{1/p}(n_{g}T_{m})^{-1/2}+||\hat{\phi}_{g,m}-\phi _{g,m}^{\ast }||}%
	\right\Vert =O_{p}(1),  \label{P_Rep_L_B_2}
\end{equation}%
where $\Psi _{\phi \phi \phi _{j},g,m}\equiv \mathbb{E}\left[ \frac{\partial 
}{\partial \phi _{j}}\left( \frac{\partial ^{2}\psi (z_{i,t};\phi
	_{g,m}^{\ast })}{\partial \phi \partial \phi ^{\top }}\right) \right] $. By
Assumptions \ref{A3}(iii), \ref{A5}(i) and Lemma \ref{Consistency}%
\begin{align}
	\max_{j\leq k}\max_{g\in \mathcal{G},m\in \mathcal{M}}\left\Vert \tilde{\Psi}%
	_{\phi \phi \phi _{j},g,m}\right\Vert & \leq \max_{j\leq k}\max_{g\in 
		\mathcal{G},m\in \mathcal{M}}\left\Vert \Psi _{\phi \phi \phi
		_{j},g,m}\right\Vert  \notag \\
	& \text{ \ \ }+O_{p}\left( ((GM)^{1/p}+||\hat{\phi}_{g,m}-\phi ^{\ast
	}||)\max_{g\in \mathcal{G},m\in \mathcal{M}}(n_{g}T_{m})^{-1/2}\right) 
	\notag \\
	& =O_{p}(1).  \label{P_Rep_L_B_3}
\end{align}

By the triangle inequality and the Cauchy-Schwarz inequality, 
\begin{align}
	& \left\vert (nT)^{-1/2}\sum_{g\in \mathcal{G}}\sum_{m\in \mathcal{M}%
	}(n_{g}T_{m})((\hat{\phi}_{g,m}-\phi _{g,m}^{\ast })^{\top }\tilde{\Psi}%
	_{\phi \phi \theta _{j},g,m}(\hat{\phi}_{g,m}-\phi _{g,m}^{\ast }))_{j\leq
		d_{\theta }}^{\top }(\hat{\theta}-\theta ^{\ast })\right\vert  \notag \\
	& \leq \left( \max_{j\leq d_{\theta }}\max_{g\in \mathcal{G},m\in \mathcal{M}%
	}\left\Vert \tilde{\Psi}_{\phi \phi \theta _{j},g,m}\right\Vert \right) ||%
	\hat{\theta}-\theta ^{\ast }||(nT)^{-1/2}\sum_{g\in \mathcal{G}}\sum_{m\in 
		\mathcal{M}}(n_{g}T_{m})||\hat{\phi}_{g,m}-\phi _{g,m}^{\ast }||^{2}.
	\label{P_Rep_L_B_4}
\end{align}%
From Lemma \ref{Rate_theta} and Lemma \ref{Rep_L_B},%
\begin{align*}
	& (nT)^{-1/2}\sum_{g\in \mathcal{G}}\sum_{m\in \mathcal{M}}(n_{g}T_{m})||%
	\hat{\phi}_{g,m}-\phi _{g,m}^{\ast }||^{2} \\
	& =(nT)^{1/2}||\hat{\theta}-\theta ^{\ast }||^{2}+(nT)^{-1/2}\sum_{g\in 
		\mathcal{G}}\sum_{m\in \mathcal{M}}(n_{g}T_{m})||\hat{\gamma}_{g,m}-\gamma
	_{g,m}^{\ast }||^{2} \\
	& =O_{p}((nT)^{-1/2}+(GM)(nT)^{-1/2})=O_{p}((GM)(nT)^{-1/2}),
\end{align*}%
which together with (\ref{P_Rep_L_B_4}) and Lemma \ref{Rate_theta} implies
that 
\begin{equation}
	\sum_{g\in \mathcal{G}}\sum_{m\in \mathcal{M}}(n_{g}T_{m})((\hat{\phi}%
	_{g,m}-\phi _{g,m}^{\ast })^{\top }\tilde{\Psi}_{\phi \phi \theta _{j},g,m}(%
	\hat{\phi}_{g,m}-\phi _{g,m}^{\ast }))_{j\leq d_{\theta }}^{\top }(\hat{%
		\theta}-\theta ^{\ast })=O_{p}((GM)(nT)^{-1/2}).  \label{P_Rep_L_B_5}
\end{equation}%
Similarly, we can show that%
\begin{eqnarray}
	&&\sum_{g\in \mathcal{G}}\sum_{m\in \mathcal{M}}(n_{g}T_{m})((\hat{\theta}%
	-\theta ^{\ast })^{\top }\tilde{\Psi}_{\theta \theta \gamma _{j},g,m}(\hat{%
		\theta}-\theta ^{\ast }))_{j\leq d_{\gamma }}^{\top }(\hat{\gamma}%
	_{g,m}-\gamma _{g,m}^{\ast })  \notag \\
	&=&O_{p}\left( (GM)(nT)^{-1}\max_{g\in \mathcal{G},m\in \mathcal{M}%
	}(n_{g}T_{m})^{1/2}\right)  \label{P_Rep_L_B_6}
\end{eqnarray}%
and 
\begin{eqnarray}
	&&\sum_{g\in \mathcal{G}}\sum_{m\in \mathcal{M}}(n_{g}T_{m})((\hat{\theta}%
	-\theta ^{\ast })^{\top }\tilde{\Psi}_{\theta \gamma \gamma _{j},g,m}(\hat{%
		\gamma}_{g,m}-\gamma _{g,m}^{\ast }))_{j\leq d_{\gamma }}^{\top }(\hat{\gamma%
	}_{g,m}-\gamma _{g,m}^{\ast })  \notag \\
	&=&O_{p}((GM)(nT)^{-1/2}).  \label{P_Rep_L_B_7}
\end{eqnarray}%
By the triangle inequality, Cauchy-Schwarz inequality, Lemma \ref%
{Consistency} and (\ref{P_Rep_L_B_2}),%
\begin{align}
	& \left\vert (nT)^{-1/2}\sum_{g\in \mathcal{G}}\sum_{m\in \mathcal{M}%
	}(n_{g}T_{m})((\hat{\gamma}_{g,m}-\gamma _{g,m}^{\ast })^{\top }(\tilde{\Psi}%
	_{\gamma \gamma \gamma _{j},g,m}-\Psi _{\gamma \gamma \gamma _{j},g,m})(\hat{%
		\gamma}_{g,m}-\gamma _{g,m}^{\ast }))_{j\leq d_{\gamma }}^{\top }(\hat{\gamma%
	}_{g,m}-\gamma _{g,m}^{\ast })\right\vert  \notag \\
	& \leq (nT)^{-1/2}\sum_{g\in \mathcal{G}}\sum_{m\in \mathcal{M}%
	}(n_{g}T_{m})||\hat{\gamma}_{g,m}-\gamma _{g,m}^{\ast }||^{3}\left\Vert 
	\tilde{\Psi}_{\gamma \gamma \gamma _{j},g,m}-\Psi _{\gamma \gamma \gamma
		_{j},g,m}\right\Vert  \notag \\
	& \leq O_{p}(1)(nT)^{-1/2}\sum_{g\in \mathcal{G}}\sum_{m\in \mathcal{M}%
	}(n_{g}T_{m})^{1/2}||\hat{\gamma}_{g,m}-\gamma _{g,m}^{\ast
	}||^{3}((GM)^{1/p}+||\hat{\gamma}_{g,m}-\gamma _{g,m}^{\ast }||+||\hat{\theta%
	}-\theta ^{\ast }||)  \notag \\
	& =O_{p}((GM)^{1/p})(nT)^{-1/2}\sum_{g\in \mathcal{G}}\sum_{m\in \mathcal{M}%
	}(n_{g}T_{m})^{1/2}||\hat{\gamma}_{g,m}-\gamma _{g,m}^{\ast }||^{3}  \notag
	\\
	& =O_{p}\left( (GM)^{1+1/p}(nT)^{-1/2}\max_{g\in \mathcal{G},m\in \mathcal{M}%
	}(n_{g}T_{m})^{-1}\right) ,  \label{P_Rep_L_B_8}
\end{align}%
where the last equality is by Lemma \ref{Rep_L_A}. By Assumption \ref{A3}%
(iii), Lemma \ref{Rate_gamma} and Lemma \ref{P_L3C_3&4}%
\begin{align}
	& \left\vert 
	\begin{array}{c}
		\sum_{g\in \mathcal{G}}\sum_{m\in \mathcal{M}}(n_{g}T_{m})\left( (\hat{\gamma%
		}_{g,m}-\gamma _{g,m}^{\ast })^{\top }\Psi _{\gamma \gamma \gamma _{j},g,m}(%
		\hat{\gamma}_{g,m}-\gamma _{g,m}^{\ast })\right) _{j\leq d_{\gamma }}^{\top
		}(\hat{\gamma}_{g,m}-\gamma _{g,m}^{\ast }) \\ 
		-\sum_{g\in \mathcal{G}}\sum_{m\in \mathcal{M}}(n_{g}T_{m})(\hat{\Psi}%
		_{\gamma ,g,m}^{\top }\Psi _{\gamma \gamma ,g,m}^{-1}\Psi _{\gamma \gamma
			\gamma _{j},g,m}\Psi _{\gamma \gamma ,g,m}^{-1}\hat{\Psi}_{\gamma
			,g,m})_{j\leq d_{\gamma }}^{\top }\Psi _{\gamma \gamma ,g,m}^{-1}\hat{\Psi}%
		_{\gamma ,g,m}%
	\end{array}%
	\right\vert  \notag \\
	& \leq \max_{j\leq d_{\gamma }}\max_{g\in \mathcal{G},m\in \mathcal{M}%
	}\left\Vert \Psi _{\gamma \gamma \gamma _{j},g,m}\right\Vert \sum_{g\in 
		\mathcal{G}}\sum_{m\in \mathcal{M}}(n_{g}T_{m})\left\Vert \Psi _{\gamma
		\gamma ,g,m}^{-1}\hat{\Psi}_{\gamma ,g,m}\right\Vert \left\Vert \hat{\gamma}%
	_{g,m}-\gamma _{g,m}^{\ast }+\Psi _{\gamma \gamma ,g,m}^{-1}\hat{\Psi}%
	_{\gamma ,g,m}\right\Vert ^{2}  \notag \\
	& \text{ \ \ }+\max_{j\leq d_{\gamma }}\max_{g\in \mathcal{G},m\in \mathcal{M%
	}}\left\Vert \Psi _{\gamma \gamma \gamma _{j},g,m}\right\Vert \sum_{g\in 
		\mathcal{G}}\sum_{m\in \mathcal{M}}(n_{g}T_{m})\left\Vert \Psi _{\gamma
		\gamma ,g,m}^{-1}\hat{\Psi}_{\gamma ,g,m}\right\Vert ^{2}\left\Vert \hat{%
		\gamma}_{g,m}-\gamma _{g,m}^{\ast }+\Psi _{\gamma \gamma ,g,m}^{-1}\hat{\Psi}%
	_{\gamma ,g,m}\right\Vert  \notag \\
	& =O_{p}(1)(nT)^{-1/2}\sum_{g\in \mathcal{G}}\sum_{m\in \mathcal{M}}\left(
	(GM)^{4/p}(n_{g}T_{m})^{-1}\left\Vert \Psi _{\gamma \gamma ,g,m}^{-1}\hat{%
		\Psi}_{\gamma ,g,m}\right\Vert +(GM)^{2/p}\left\Vert \Psi _{\gamma \gamma
		,g,m}^{-1}\hat{\Psi}_{\gamma ,g,m}\right\Vert ^{2}\right)  \notag \\
	& =O_{p}\left( (GM)^{1+2/p}(nT)^{-1/2}\max_{g\in \mathcal{G},m\in \mathcal{M}%
	}(n_{g}T_{m})^{-1}\right) ,  \label{P_Rep_L_B_9}
\end{align}%
where the last equality is by Assumption \ref{A5}(i). Applying Lemma \ref%
{L2B}, we get%
\begin{align}
	& (nT)^{-1/2}\sum_{g\in \mathcal{G}}\sum_{m\in \mathcal{M}}(n_{g}T_{m})(\hat{%
		\Psi}_{\gamma ,g,m}^{\top }\Psi _{\gamma \gamma ,g,m}^{-1}\Psi _{\gamma
		\gamma \gamma _{j},g,m}\Psi _{\gamma \gamma ,g,m}^{-1}\hat{\Psi}_{\gamma
		,g,m})_{j\leq d_{\gamma }}^{\top }\Psi _{\gamma \gamma ,g,m}^{-1}\hat{\Psi}%
	_{\gamma ,g,m}  \notag \\
	& =O_{p}\left( (GM)^{1/2}(nT)^{-1/2}\left( \max_{g\in \mathcal{G},m\in 
		\mathcal{M}}(n_{g}T_{m})^{-1/2}+(GM)^{1/2}\max_{g\in \mathcal{G},m\in 
		\mathcal{M}}(n_{g}T_{m})^{-1}\right) \right) .  \label{P_Rep_L_B_10}
\end{align}%
Collecting the results in (\ref{P_Rep_L_B_5})-(\ref{P_Rep_L_B_10}), we
deduce that%
\begin{align}
	& (nT)^{-1/2}\sum_{g\in \mathcal{G}}\sum_{m\in \mathcal{M}%
	}(n_{g}T_{m})\left( (\hat{\phi}_{g,m}-\phi _{g,m}^{\ast })^{\top }\tilde{\Psi%
	}_{\phi \phi \phi _{j},g,m}(\hat{\phi}_{g,m}-\phi _{g,m}^{\ast })\right)
	_{j\leq k}^{\top }(\hat{\phi}_{g,m}-\phi _{g,m}^{\ast })  \notag \\
	& =O_{p}\left( (GM)^{1/2}(nT)^{-1/2}\left( (GM)^{1/2+2/p}\max_{g\in \mathcal{%
			G},m\in \mathcal{M}}(n_{g}T_{m})^{-1}+\max_{g\in \mathcal{G},m\in \mathcal{M}%
	}(n_{g}T_{m})^{-1/2}\right) \right)  \label{P_Rep_L_B_11}
\end{align}%
which together with (\ref{P_Rep_L_B_1}) shows the assertion of the
lemma.\hfill $Q.E.D.$

\bigskip\ 

\begin{lemma}
	\textit{\label{Rep_L_C}}Under Assumptions \ref{A1}, \ref{A2}, \ref{A3}, \ref%
	{A4} and \ref{A5}, we have%
	\begin{align*}
		& (nT)^{-1/2}\sum_{g\in \mathcal{G}}\sum_{m\in \mathcal{M}}(n_{g}T_{m})(\hat{%
			\gamma}_{g,m}-\gamma _{g,m}^{\ast })^{\top }\hat{\Psi}_{\gamma \gamma ,g,m}(%
		\hat{\gamma}_{g,m}-\gamma _{g,m}^{\ast }) \\
		& =(nT)^{-1/2}\sum_{g\in \mathcal{G}}\sum_{m\in \mathcal{M}}(n_{g}T_{m})\hat{%
			\Psi}_{\gamma ,g,m}^{\top }\Psi _{\gamma \gamma ,g,m}^{-1}\hat{\Psi}_{\gamma
			,g,m}+o_{p}\left( (GM)^{1/2}(nT)^{-1/2}\right) .
	\end{align*}
\end{lemma}

\noindent \textsc{Proof of Lemma \ref{Rep_L_C}}.\ By Lemma \ref{Rate_theta}
and (\ref{P_Rep_Theta_2_1E}),%
\begin{align*}
	& \max_{g\in \mathcal{G},m\in \mathcal{M}}\frac{%
		(nT)^{-1}+(GM)^{1/p}(n_{g}T_{m})^{-1/2}||\hat{\phi}_{g,m}-\phi _{g,m}^{\ast
		}||^{2}+||\hat{\phi}_{g,m}-\phi _{g,m}^{\ast }||^{3}}{%
		(nT)^{-1}+(GM)^{1/p}(n_{g}T_{m})^{-1/2}||\hat{\gamma}_{g,m}-\gamma
		_{g,m}^{\ast }||^{2}} \\
	& \leq \max_{g\in \mathcal{G},m\in \mathcal{M}}\frac{%
		(nT)^{-1}+(GM)^{1/p}(n_{g}T_{m})^{-1/2}||\hat{\gamma}_{g,m}-\gamma
		_{g,m}^{\ast }||^{2}+||\hat{\gamma}_{g,m}-\gamma _{g,m}^{\ast }||^{3}}{%
		(nT)^{-1}+(GM)^{1/p}(n_{g}T_{m})^{-1/2}||\hat{\gamma}_{g,m}-\gamma
		_{g,m}^{\ast }||^{2}} \\
	& +\max_{g\in \mathcal{G},m\in \mathcal{M}}\frac{%
		(GM)^{1/p}(n_{g}T_{m})^{-1/2}||\hat{\theta}-\theta ^{\ast }||^{2}+||\hat{%
			\theta}-\theta ^{\ast }||^{3}}{(nT)^{-1}+(GM)^{1/p}(n_{g}T_{m})^{-1/2}||\hat{%
			\gamma}_{g,m}-\gamma _{g,m}^{\ast }||^{2}} \\
	& \leq 1+\max_{g\in \mathcal{G},m\in \mathcal{M}}\frac{\left( \max_{g\in 
			\mathcal{G},m\in \mathcal{M}}||\hat{\gamma}_{g,m}-\gamma _{g,m}^{\ast
		}||\right) ||\hat{\gamma}_{g,m}-\gamma _{g,m}^{\ast }||^{2}}{%
		(nT)^{-1}+(GM)^{1/p}(n_{g}T_{m})^{-1/2}||\hat{\gamma}_{g,m}-\gamma
		_{g,m}^{\ast }||^{2}}+o_{p}(1)=O_{p}(1).
\end{align*}%
Therefore, we can use (\ref{P_Rep_Theta_2_1B}) to obtain%
\begin{equation}
	\max_{g\in \mathcal{G},m\in \mathcal{M}}\frac{\left\Vert \hat{\Psi}_{\gamma
			,g,m}+\hat{\Psi}_{\gamma \phi ,g,m}(\hat{\phi}_{g,m}-\phi _{g,m}^{\ast })+%
		\frac{\left( (\hat{\phi}_{g,m}-\phi _{g,m}^{\ast })^{\top }\Psi _{\gamma
				_{j}\phi \phi ,g,m}(\hat{\phi}_{g,m}-\phi _{g,m}^{\ast })\right) _{j\leq
				d_{\gamma }}}{2}\right\Vert }{(nT)^{-1}+(GM)^{1/p}(n_{g}T_{m})^{-1/2}||\hat{%
			\gamma}_{g,m}-\gamma _{g,m}^{\ast }||^{2}}=O_{p}(1).  \label{P_Rep_L_C_1}
\end{equation}%
To employ the above expansion, we write%
\begin{align}
	& (nT)^{-1/2}\sum_{g\in \mathcal{G}}\sum_{m\in \mathcal{M}}(n_{g}T_{m})(\hat{%
		\gamma}_{g,m}-\gamma _{g,m}^{\ast })^{\top }\hat{\Psi}_{\gamma \gamma ,g,m}(%
	\hat{\gamma}_{g,m}-\gamma _{g,m}^{\ast })  \notag \\
	& =-(nT)^{-1/2}\sum_{g\in \mathcal{G}}\sum_{m\in \mathcal{M}}(n_{g}T_{m})(%
	\hat{\gamma}_{g,m}-\gamma _{g,m}^{\ast })^{\top }(\hat{\Psi}_{\gamma ,g,m}+%
	\hat{\Psi}_{\gamma \theta ,g,m}(\hat{\theta}-\theta ^{\ast }))  \notag \\
	& \text{ \ \ }-2^{-1}(nT)^{-1/2}\sum_{g\in \mathcal{G}}\sum_{m\in \mathcal{M}%
	}(n_{g}T_{m})(\hat{\gamma}_{g,m}-\gamma _{g,m}^{\ast })^{\top }((\hat{\phi}%
	_{g,m}-\phi _{g,m}^{\ast })^{\top }\Psi _{\gamma _{j}\phi \phi ,g,m}(\hat{%
		\phi}_{g,m}-\phi _{g,m}^{\ast }))_{j\leq d_{\gamma }}+R_{1,n,T}
	\label{P_Rep_L_C_2}
\end{align}%
where%
\begin{equation*}
	R_{1,n,T}\equiv (nT)^{-1/2}\sum_{g\in \mathcal{G}}\sum_{m\in \mathcal{M}%
	}(n_{g}T_{m})(\hat{\gamma}_{g,m}-\gamma _{g,m}^{\ast })^{\top }\left( 
	\begin{array}{c}
		\hat{\Psi}_{\gamma ,g,m}+\hat{\Psi}_{\gamma \phi ,g,m}(\hat{\phi}_{g,m}-\phi
		_{g,m}^{\ast }) \\ 
		+2^{-1}\left( (\hat{\phi}_{g,m}-\phi _{g,m}^{\ast })^{\top }\Psi _{\gamma
			_{j}\phi \phi ,g,m}(\hat{\phi}_{g,m}-\phi _{g,m}^{\ast })\right) _{j\leq
			d_{\gamma }}%
	\end{array}%
	\right) .
\end{equation*}%
We can apply the Cauchy-Schwarz inequality, Lemma \ref{Rep_L_A}, (\ref%
{P_Rep_L_C_1}) and (\ref{P_Rep_Theta_2_1E}) to bound $R_{1,n,T}$%
\begin{align}
	\left\vert R_{1,n,T}\right\vert & \leq (nT)^{-1/2}\sum_{g\in \mathcal{G}%
	}\sum_{m\in \mathcal{M}}(n_{g}T_{m})||\hat{\gamma}_{g,m}-\gamma _{g,m}^{\ast
	}||((nT)^{-1}+(GM)^{1/p}(n_{g}T_{m})^{-1/2}||\hat{\gamma}_{g,m}-\gamma
	_{g,m}^{\ast }||^{2})  \notag \\
	& \leq (nT)^{-1/2}\max_{g\in \mathcal{G},m\in \mathcal{M}}||\hat{\gamma}%
	_{g,m}-\gamma _{g,m}^{\ast }||+(GM)^{1/p}(nT)^{-1/2}\sum_{g\in \mathcal{G}%
	}\sum_{m\in \mathcal{M}}(n_{g}T_{m})^{1/2}||\hat{\gamma}_{g,m}-\gamma
	_{g,m}^{\ast }||^{3}  \notag \\
	& =O_{p}\left( (GM)^{1/p}(nT)^{-1/2}\max_{g\in \mathcal{G},m\in \mathcal{M}%
	}(n_{g}T_{m})^{-1/2}+(GM)^{1+1/p}(nT)^{-1/2}\max_{g\in \mathcal{G},m\in 
		\mathcal{M}}(n_{g}T_{m})^{-1}\right) .  \label{P_Rep_L_C_3}
\end{align}%
Next, we bound the second term in the RHS of (\ref{P_Rep_L_C_2}). For this
purpose, we write 
\begin{align}
	& (nT)^{-1/2}\sum_{g\in \mathcal{G}}\sum_{m\in \mathcal{M}}(n_{g}T_{m})(\hat{%
		\gamma}_{g,m}-\gamma _{g,m}^{\ast })^{\top }((\hat{\phi}_{g,m}-\phi
	_{g,m}^{\ast })^{\top }\Psi _{\gamma _{j}\phi \phi ,g,m}(\hat{\phi}%
	_{g,m}-\phi _{g,m}^{\ast }))_{j\leq d_{\gamma }}  \notag \\
	& =(nT)^{-1/2}\sum_{g\in \mathcal{G}}\sum_{m\in \mathcal{M}}(n_{g}T_{m})(%
	\hat{\gamma}_{g,m}-\gamma _{g,m}^{\ast })^{\top }((\hat{\theta}-\theta
	^{\ast })^{\top }\Psi _{\gamma _{j}\theta \theta ,g,m}(\hat{\theta}-\theta
	^{\ast }))_{j\leq d_{\gamma }}  \notag \\
	& +2(nT)^{-1/2}\sum_{g\in \mathcal{G}}\sum_{m\in \mathcal{M}}(n_{g}T_{m})(%
	\hat{\gamma}_{g,m}-\gamma _{g,m}^{\ast })^{\top }((\hat{\theta}-\theta
	^{\ast })^{\top }\Psi _{\gamma _{j}\theta \gamma ,g,m}(\hat{\gamma}%
	_{g,m}-\gamma _{g,m}^{\ast }))_{j\leq d_{\gamma }}  \notag \\
	& +(nT)^{-1/2}\sum_{g\in \mathcal{G}}\sum_{m\in \mathcal{M}}(n_{g}T_{m})(%
	\hat{\gamma}_{g,m}-\gamma _{g,m}^{\ast })^{\top }((\hat{\gamma}_{g,m}-\gamma
	_{g,m}^{\ast })^{\top }\Psi _{\gamma _{j}\gamma \gamma ,g,m}(\hat{\gamma}%
	_{g,m}-\gamma _{g,m}^{\ast }))_{j\leq d_{\gamma }}.  \label{P_Rep_L_C_4}
\end{align}%
By the Cauchy-Schwarz inequality, Assumption \ref{A3}(iii), Lemma \ref%
{Rate_theta} and (\ref{P_Rep_Theta_2_1E}), 
\begin{align}
	& \left\vert (nT)^{-1/2}\sum_{g\in \mathcal{G}}\sum_{m\in \mathcal{M}%
	}(n_{g}T_{m})(\hat{\gamma}_{g,m}-\gamma _{g,m}^{\ast })^{\top }((\hat{\theta}%
	-\theta ^{\ast })^{\top }\Psi _{\gamma _{j}\theta \theta ,g,m}(\hat{\theta}%
	-\theta ^{\ast }))_{j\leq d_{\gamma }}\right\vert  \notag \\
	& \leq \left( \max_{j\leq d_{\gamma }}\max_{g\in \mathcal{G},m\in \mathcal{M}%
	}\left\Vert \Psi _{\gamma _{j}\theta \theta ,g,m}\right\Vert \right)
	\left\Vert \hat{\theta}-\theta ^{\ast }\right\Vert ^{2}(nT)^{-1/2}\sum_{g\in 
		\mathcal{G}}\sum_{m\in \mathcal{M}}(n_{g}T_{m})||\hat{\gamma}_{g,m}-\gamma
	_{g,m}^{\ast }||  \notag \\
	& =O_{p}\left( (nT)^{-1/2}\max_{g\in \mathcal{G},m\in \mathcal{M}}||\hat{%
		\gamma}_{g,m}-\gamma _{g,m}^{\ast }||\right)  \notag \\
	& =O_{p}\left( (GM)^{1/p}(nT)^{-1/2}\max_{g\in \mathcal{G},m\in \mathcal{M}%
	}(n_{g}T_{m})^{-1/2}\right) .  \label{P_Rep_L_C_5}
\end{align}%
Similarly,%
\begin{align}
	& \left\vert (nT)^{-1/2}\sum_{g\in \mathcal{G}}\sum_{m\in \mathcal{M}%
	}(n_{g}T_{m})\left( (\hat{\theta}-\theta ^{\ast })^{\top }\Psi _{\theta
		\gamma \gamma _{j},g,m}(\hat{\gamma}_{g,m}-\gamma _{g,m}^{\ast })\right)
	_{j\leq d_{\gamma }}^{\top }(\hat{\gamma}_{g,m}-\gamma _{g,m}^{\ast
	})\right\vert  \notag \\
	& =\left\vert (\hat{\theta}-\theta ^{\ast })^{\top }(nT)^{-1/2}\sum_{g\in 
		\mathcal{G}}\sum_{m\in \mathcal{M}}(n_{g}T_{m})\left( (\hat{\gamma}%
	_{g,m}-\gamma _{g,m}^{\ast })^{\top }\Psi _{\theta _{j}\gamma \gamma ,g,m}(%
	\hat{\gamma}_{g,m}-\gamma _{g,m}^{\ast })\right) _{j\leq d_{\gamma
	}}\right\vert  \notag \\
	& \leq K\left\Vert \hat{\theta}-\theta ^{\ast }\right\Vert
	(nT)^{-1/2}\sum_{g\in \mathcal{G}}\sum_{m\in \mathcal{M}}(n_{g}T_{m})||\hat{%
		\gamma}_{g,m}-\gamma _{g,m}^{\ast }||^{2}=O_{p}\left( (GM)(nT)^{-1}\right) ,
	\label{P_Rep_L_C_6}
\end{align}%
where the equality is by Lemma \ref{Rate_theta} and Lemma \ref{Rep_L_A}.
Applying (\ref{P_Rep_L_B_9}) and (\ref{P_Rep_L_B_10}), we obtain 
\begin{align}
	& (nT)^{-1/2}\sum_{g\in \mathcal{G}}\sum_{m\in \mathcal{M}}(n_{g}T_{m})(\hat{%
		\gamma}_{g,m}-\gamma _{g,m}^{\ast })^{\top }((\hat{\gamma}_{g,m}-\gamma
	_{g,m}^{\ast })^{\top }\Psi _{\gamma _{j}\gamma \gamma ,g,m}(\hat{\gamma}%
	_{g,m}-\gamma _{g,m}^{\ast }))_{j\leq d_{\gamma }}  \notag \\
	& =O_{p}\left( (GM)^{1/2}(nT)^{-1/2}\left( \max_{g\in \mathcal{G},m\in 
		\mathcal{M}}(n_{g}T_{m})^{-1/2}+(GM)^{1/2+2/p}\max_{g\in \mathcal{G},m\in 
		\mathcal{M}}(n_{g}T_{m})^{-1}\right) \right) .  \label{P_Rep_L_C_7}
\end{align}%
Collecting the results in (\ref{P_Rep_L_C_4}), (\ref{P_Rep_L_C_5}), (\ref%
{P_Rep_L_C_6}) and (\ref{P_Rep_L_C_7}) and applying Assumption \ref{A5} yield%
\begin{eqnarray}
	&&(nT)^{-1/2}\sum_{g\in \mathcal{G}}\sum_{m\in \mathcal{M}}(n_{g}T_{m})(\hat{%
		\gamma}_{g,m}-\gamma _{g,m}^{\ast })^{\top }((\hat{\phi}_{g,m}-\phi
	_{g,m}^{\ast })^{\top }\Psi _{\gamma _{j}\phi \phi ,g,m}(\hat{\phi}%
	_{g,m}-\phi _{g,m}^{\ast }))_{j\leq d_{\gamma }}  \notag \\
	&=&o_{p}((GM)^{1/2}(nT)^{-1/2}).  \label{P_Rep_L_C_8}
\end{eqnarray}%
Next, by Lemma \ref{Rate_gamma}, Lemma \ref{L2A} and Lemma \ref{Rate_theta}%
\begin{align*}
	& (nT)^{-1/2}\sum_{g\in \mathcal{G}}\sum_{m\in \mathcal{M}}(n_{g}T_{m})(\hat{%
		\gamma}_{g,m}-\gamma _{g,m}^{\ast })^{\top }\hat{\Psi}_{\gamma \theta ,g,m}(%
	\hat{\theta}-\theta ^{\ast }) \\
	& =(nT)^{-1/2}\sum_{g\in \mathcal{G}}\sum_{m\in \mathcal{M}}(n_{g}T_{m})\hat{%
		\Psi}_{\gamma ,g,m}^{\top }\Psi _{\gamma \gamma ,g,m}^{-1}\Psi _{\gamma
		\theta ,g,m}(\hat{\theta}-\theta ^{\ast }) \\
	& +(nT)^{-1/2}\sum_{g\in \mathcal{G}}\sum_{m\in \mathcal{M}}(n_{g}T_{m})\hat{%
		\Psi}_{\gamma ,g,m}^{\top }\Psi _{\gamma \gamma ,g,m}^{-1}(\hat{\Psi}%
	_{\gamma \theta ,g,m}-\Psi _{\gamma \theta ,g,m})(\hat{\theta}-\theta ^{\ast
	})+O_{p}\left( (GM)^{1+2/p}(nT)^{-1}\right) \\
	& =O_{p}\left( (nT)^{-1/2}+(GM)^{1+2/p}(nT)^{-1}\right) =o_{p}\left(
	(GM)^{1/2}(nT)^{-1/2}\right) ,
\end{align*}%
which together with (\ref{P_Rep_L_C_3}) and (\ref{P_Rep_L_C_8}) implies that%
\begin{align}
	& (nT)^{-1/2}\sum_{g\in \mathcal{G}}\sum_{m\in \mathcal{M}}(n_{g}T_{m})(\hat{%
		\gamma}_{g,m}-\gamma _{g,m}^{\ast })^{\top }\hat{\Psi}_{\gamma \gamma ,g,m}(%
	\hat{\gamma}_{g,m}-\gamma _{g,m}^{\ast })  \notag \\
	& =-(nT)^{-1/2}\sum_{g\in \mathcal{G}}\sum_{m\in \mathcal{M}}(n_{g}T_{m})(%
	\hat{\gamma}_{g,m}-\gamma _{g,m}^{\ast })^{\top }\hat{\Psi}_{\gamma
		,g,m}+o_{p}\left( (GM)^{1/2}(nT)^{-1/2}\right) .  \label{P_Rep_L_C_9}
\end{align}

We shall apply (\ref{P_Rep_L_C_1}) to the first term in the RHS of (\ref%
{P_Rep_L_C_9}) to show the claim of the lemma. For this purpose, we write 
\begin{align}
	& (nT)^{-1/2}\sum_{g\in \mathcal{G}}\sum_{m\in \mathcal{M}}(n_{g}T_{m})(\hat{%
		\gamma}_{g,m}-\gamma _{g,m}^{\ast })^{\top }\hat{\Psi}_{\gamma ,g,m}  \notag
	\\
	& =R_{2,n,T}-(nT)^{-1/2}\sum_{g\in \mathcal{G}}\sum_{m\in \mathcal{M}%
	}(n_{g}T_{m})(\hat{\Psi}_{\gamma ,g,m}+\hat{\Psi}_{\gamma \theta ,g,m}(\hat{%
		\theta}-\theta ^{\ast }))^{\top }\hat{\Psi}_{\gamma \gamma ,g,m}^{-1}\hat{%
		\Psi}_{\gamma ,g,m}  \notag \\
	& -(4nT)^{-1/2}\sum_{g\in \mathcal{G}}\sum_{m\in \mathcal{M}}(n_{g}T_{m})((%
	\hat{\phi}_{g,m}-\phi _{g,m}^{\ast })^{\top }\Psi _{\gamma _{j}\phi \phi
		,g,m}(\hat{\phi}_{g,m}-\phi _{g,m}^{\ast }))_{j\leq d_{\gamma }}^{\top }\hat{%
		\Psi}_{\gamma \gamma ,g,m}^{-1}\hat{\Psi}_{\gamma ,g,m},
	\label{P_Rep_L_C_10}
\end{align}%
where%
\begin{equation*}
	R_{2,n,T}\equiv (nT)^{-1/2}\sum_{g\in \mathcal{G}}\sum_{m\in \mathcal{M}%
	}(n_{g}T_{m})\hat{\Psi}_{\gamma ,g,m}^{\top }\hat{\Psi}_{\gamma \gamma
		,g,m}^{-1}\left( 
	\begin{array}{c}
		\hat{\Psi}_{\gamma ,g,m}+\hat{\Psi}_{\gamma \phi ,g,m}(\hat{\phi}_{g,m}-\phi
		_{g,m}^{\ast }) \\ 
		+2^{-1}\left( (\hat{\phi}_{g,m}-\phi _{g,m}^{\ast })^{\top }\Psi _{\gamma
			_{j}\phi \phi ,g,m}(\hat{\phi}_{g,m}-\phi _{g,m}^{\ast })\right) _{j\leq
			d_{\gamma }}%
	\end{array}%
	\right) .
\end{equation*}%
By the Cauchy-Schwarz inequality%
\begin{eqnarray}
	\left\vert R_{2,n,T}\right\vert &\leq &\max_{g\in \mathcal{G},m\in \mathcal{M%
	}}||\hat{\Psi}_{\gamma \gamma ,g,m}^{-1}||  \notag \\
	&&\times (nT)^{-1/2}\sum_{g\in \mathcal{G}}\sum_{m\in \mathcal{M}%
	}(n_{g}T_{m})||\hat{\Psi}_{\gamma
		,g,m}||((nT)^{-1}+(GM)^{1/p}(n_{g}T_{m})^{-1/2}||\hat{\gamma}_{g,m}-\gamma
	_{g,m}^{\ast }||^{2}).  \notag \\
	&&  \label{P_Rep_L_C_11}
\end{eqnarray}%
Since $\max_{g\in \mathcal{G},m\in \mathcal{M}}\mathbb{E}\left[
||(n_{g}T_{m})^{1/2}\hat{\Psi}_{\gamma ,g,m}||^{p}\right] \leq K$ by (\ref%
{L0-1}) in Lemma \ref{L0-1}, we can use Holder's inequality Markov's
inequality to obtain%
\begin{equation}
	\sum_{g\in \mathcal{G}}\sum_{m\in \mathcal{M}}||(n_{g}T_{m})^{1/2}\hat{\Psi}%
	_{\gamma ,g,m}||^{s}=O_{p}\left( GM\right)  \label{P_Rep_L_C_12}
\end{equation}%
for any positive $s$ less than or equal to$\ p$, which together with Lemma %
\ref{L3AB}(ii) implies that 
\begin{equation}
	\max_{g\in \mathcal{G},m\in \mathcal{M}}||\hat{\Psi}_{\gamma \gamma
		,g,m}^{-1}||(nT)^{-3/2}\sum_{g\in \mathcal{G}}\sum_{m\in \mathcal{M}%
	}(n_{g}T_{m})||\hat{\Psi}_{\gamma ,g,m}||=O_{p}\left( (nT)^{-1/2}\max_{g\in 
		\mathcal{G},m\in \mathcal{M}}(n_{g}T_{m})^{-1/2}\right) .
	\label{P_Rep_L_C_13}
\end{equation}%
By the Cauchy-Schwarz inequality%
\begin{align}
	& \sum_{g\in \mathcal{G}}\sum_{m\in \mathcal{M}}||(n_{g}T_{m})^{1/2}\hat{\Psi%
	}_{\gamma ,g,m}||\times ||\hat{\gamma}_{g,m}-\gamma _{g,m}^{\ast }||^{2} 
	\notag \\
	& \leq \left( \sum_{g\in \mathcal{G}}\sum_{m\in \mathcal{M}%
	}||(n_{g}T_{m})^{1/2}\hat{\Psi}_{\gamma ,g,m}||^{2}\right) ^{1/2}\left(
	\sum_{g\in \mathcal{G}}\sum_{m\in \mathcal{M}}||\hat{\gamma}_{g,m}-\gamma
	_{g,m}^{\ast }||^{4}\right) ^{1/2}  \notag \\
	& =O_{p}\left( (GM)\max_{g\in \mathcal{G},m\in \mathcal{M}%
	}(n_{g}T_{m})^{-1}\right) ,  \label{P_Rep_L_C_14}
\end{align}%
where the equality is by (\ref{P_Rep_L_C_12}) and Lemma \ref{Rep_L_A}. Using
Lemma \ref{L3AB}(ii), (\ref{P_Rep_L_C_11}), (\ref{P_Rep_L_C_13}) and (\ref%
{P_Rep_L_C_14}), we get%
\begin{eqnarray}
	R_{2,n,T} &=&O_{p}\left( \left( \max_{g\in \mathcal{G},m\in \mathcal{M}%
	}(n_{g}T_{m})^{-1/2}+(GM)^{1+1/p}\max_{g\in \mathcal{G},m\in \mathcal{M}%
	}(n_{g}T_{m})^{-1}\right) (nT)^{-1/2}\right)  \notag \\
	&=&o_{p}\left( (GM)^{1/2}(nT)^{-1/2}\right) ,  \label{P_Rep_L_C_15}
\end{eqnarray}%
where the second equality is by Assumption \ref{A5}. By the similar
arguments for showing (\ref{P_Rep_L_C_8}),%
\begin{eqnarray}
	&&(nT)^{-1/2}\sum_{g\in \mathcal{G}}\sum_{m\in \mathcal{M}}(n_{g}T_{m})((%
	\hat{\phi}_{g,m}-\phi _{g,m}^{\ast })^{\top }\Psi _{\gamma _{j}\phi \phi
		,g,m}(\hat{\phi}_{g,m}-\phi _{g,m}^{\ast }))_{j\leq d_{\gamma }}^{\top }\hat{%
		\Psi}_{\gamma \gamma ,g,m}^{-1}\hat{\Psi}_{\gamma ,g,m}  \notag \\
	&=&o_{p}((GM)^{1/2}(nT)^{-1/2}).  \label{P_Rep_L_C_16}
\end{eqnarray}%
By the triangle inequality,%
\begin{align}
	& \left\Vert (nT)^{-1/2}\sum_{g\in \mathcal{G}}\sum_{m\in \mathcal{M}%
	}(n_{g}T_{m})\hat{\Psi}_{\theta \gamma ,g,m}\hat{\Psi}_{\gamma \gamma
		,g,m}^{-1}\hat{\Psi}_{\gamma ,g,m}\right\Vert  \notag \\
	& \leq \left\Vert (nT)^{-1/2}\sum_{g\in \mathcal{G}}\sum_{m\in \mathcal{M}%
	}(n_{g}T_{m})\Psi _{\theta \gamma ,g,m}\Psi _{\gamma \gamma ,g,m}^{-1}\hat{%
		\Psi}_{\gamma ,g,m}\right\Vert  \notag \\
	& +\left\Vert (nT)^{-1/2}\sum_{g\in \mathcal{G}}\sum_{m\in \mathcal{M}%
	}(n_{g}T_{m})(\hat{\Psi}_{\theta \gamma ,g,m}-\Psi _{\theta \gamma ,g,m})%
	\hat{\Psi}_{\gamma \gamma ,g,m}^{-1}\hat{\Psi}_{\gamma ,g,m}\right\Vert 
	\notag \\
	& +\left\Vert (nT)^{-1/2}\sum_{g\in \mathcal{G}}\sum_{m\in \mathcal{M}%
	}(n_{g}T_{m})\Psi _{\theta \gamma ,g,m}(\hat{\Psi}_{\gamma \gamma
		,g,m}^{-1}-\Psi _{\gamma \gamma ,g,m}^{-1})\hat{\Psi}_{\gamma
		,g,m}\right\Vert  \notag \\
	& \leq O_{p}(1)+\left( \max_{g\in \mathcal{G},m\in \mathcal{M}}||\hat{\Psi}%
	_{\gamma \gamma ,g,m}^{-1}||\right) (nT)^{-1/2}\sum_{g\in \mathcal{G}%
	}\sum_{m\in \mathcal{M}}(n_{g}T_{m})||\hat{\Psi}_{\theta \gamma ,g,m}-\Psi
	_{\theta \gamma ,g,m}||\times ||\hat{\Psi}_{\gamma ,g,m}||  \notag \\
	& +\left( \max_{g\in \mathcal{G},m\in \mathcal{M}}||\hat{\Psi}_{\gamma
		\gamma ,g,m}^{-1}||||\Psi _{\gamma \gamma ,g,m}^{-1}||\right)
	(nT)^{-1/2}\sum_{g\in \mathcal{G}}\sum_{m\in \mathcal{M}}(n_{g}T_{m})||\hat{%
		\Psi}_{\gamma \gamma ,g,m}-\Psi _{\gamma \gamma ,g,m}||\times ||\hat{\Psi}%
	_{\gamma ,g,m}||  \notag \\
	& =O_{p}(1+(GM)(nT)^{-1/2})=O_{p}(1),  \label{P_Rep_L_C_17}
\end{align}%
where the second inequality is by the Cauchy-Schwarz inequality and 
\begin{equation*}
	(nT)^{-1/2}\sum_{g\in \mathcal{G}}\sum_{m\in \mathcal{M}}(n_{g}T_{m})\Psi
	_{\theta \gamma ,g,m}\Psi _{\gamma \gamma ,g,m}^{-1}\hat{\Psi}_{\gamma
		,g,m}=O_{p}(1)
\end{equation*}%
which holds by \ref{L0-1} in Lemma \ref{L0}; the first equality is by Lemma %
\ref{L3AB}(ii), (\ref{P_Rep_L_C_12}), (\ref{P_Rep_Theta_2_5}) and Markov's
inequality; and the second equality is by Assumption \ref{A5}. By the
Cauchy-Schwarz inequality, Lemma \ref{Rate_theta} and (\ref{P_Rep_L_C_17})%
\begin{equation}
	(nT)^{-1/2}\sum_{g\in \mathcal{G}}\sum_{m\in \mathcal{M}}(n_{g}T_{m})(\hat{%
		\theta}-\theta ^{\ast })^{\top }\hat{\Psi}_{\theta \gamma ,g,m}\hat{\Psi}%
	_{\gamma \gamma ,g,m}^{-1}\hat{\Psi}_{\gamma ,g,m}=O_{p}((nT)^{-1/2}).
	\label{P_Rep_L_C_18}
\end{equation}%
Next note that 
\begin{align}
	& \left\vert (nT)^{-1/2}\sum_{g\in \mathcal{G}}\sum_{m\in \mathcal{M}%
	}(n_{g}T_{m})\hat{\Psi}_{\gamma ,g,m}^{\top }(\hat{\Psi}_{\gamma \gamma
		,g,m}^{-1}-\Psi _{\gamma \gamma ,g,m}^{-1})\hat{\Psi}_{\gamma
		,g,m}\right\vert  \notag \\
	& \leq \left\vert (nT)^{-1/2}\sum_{g\in \mathcal{G}}\sum_{m\in \mathcal{M}%
	}(n_{g}T_{m})\hat{\Psi}_{\gamma ,g,m}^{\top }\Psi _{\gamma \gamma ,g,m}^{-1}(%
	\hat{\Psi}_{\gamma \gamma ,g,m}-\Psi _{\gamma \gamma ,g,m})\Psi _{\gamma
		\gamma ,g,m}^{-1}\hat{\Psi}_{\gamma ,g,m}\right\vert  \notag \\
	& +\left\vert (nT)^{-1/2}\sum_{g\in \mathcal{G}}\sum_{m\in \mathcal{M}%
	}(n_{g}T_{m})\hat{\Psi}_{\gamma ,g,m}^{\top }(\hat{\Psi}_{\gamma \gamma
		,g,m}^{-1}-\Psi _{\gamma \gamma ,g,m}^{-1})(\hat{\Psi}_{\gamma \gamma
		,g,m}-\Psi _{\gamma \gamma ,g,m})\Psi _{\gamma \gamma ,g,m}^{-1}\hat{\Psi}%
	_{\gamma ,g,m}\right\vert  \notag \\
	& \leq \left( \max_{g\in \mathcal{G},m\in \mathcal{M}}||\hat{\Psi}_{\gamma
		\gamma ,g,m}^{-1}||||\Psi _{\gamma \gamma ,g,m}^{-1}||||\hat{\Psi}_{\gamma
		\gamma ,g,m}-\Psi _{\gamma \gamma ,g,m}||^{2}\right) (nT)^{-1/2}\sum_{g\in 
		\mathcal{G}}\sum_{m\in \mathcal{M}}(n_{g}T_{m})||\hat{\Psi}_{\gamma
		,g,m}||^{2}  \notag \\
	& +O_{p}\left( (GM)^{1/2}(nT)^{-1/2}\max_{g\in \mathcal{G},m\in \mathcal{M}%
	}\left( (n_{g}T_{m})^{-1/2}+(GM)^{1/2}(n_{g}T_{m})^{-1}\right) \right) 
	\notag \\
	& =O_{p}\left( (GM)^{1/2}(nT)^{-1/2}\max_{g\in \mathcal{G},m\in \mathcal{M}%
	}\left( (n_{g}T_{m})^{-1/2}+(GM)^{1/2+2/p}(n_{g}T_{m})^{-1}\right) \right)
	=o_{p}((GM)^{1/2}(nT)^{-1/2})  \label{P_Rep_L_C_19}
\end{align}%
where the second inequality is by Lemma \ref{L2B} and the Cauchy-Schwarz
inequality, the first equality is by Lemma \ref{L3AB}(ii), Lemma \ref%
{P_L3AB_1} and (\ref{P_Rep_L_C_12}) and the second equality is by Assumption %
\ref{A5}.

Collecting the results in (\ref{P_Rep_L_C_10}), (\ref{P_Rep_L_C_11}), (\ref%
{P_Rep_L_C_16}), (\ref{P_Rep_L_C_18}) and (\ref{P_Rep_L_C_19}), we have%
\begin{align}
	& (nT)^{-1/2}\sum_{g\in \mathcal{G}}\sum_{m\in \mathcal{M}}(n_{g}T_{m})(\hat{%
		\gamma}_{g,m}-\gamma _{g,m}^{\ast })^{\top }\hat{\Psi}_{\gamma ,g,m}  \notag
	\\
	& =-(nT)^{-1/2}\sum_{g\in \mathcal{G}}\sum_{m\in \mathcal{M}}(n_{g}T_{m})%
	\hat{\Psi}_{\gamma ,g,m}^{\top }\hat{\Psi}_{\gamma \gamma ,g,m}^{-1}\hat{\Psi%
	}_{\gamma ,g,m}+o_{p}((GM)^{1/2}(nT)^{-1/2})  \notag \\
	& =-(nT)^{-1/2}\sum_{g\in \mathcal{G}}\sum_{m\in \mathcal{M}}(n_{g}T_{m})%
	\hat{\Psi}_{\gamma ,g,m}^{\top }\Psi _{\gamma \gamma ,g,m}^{-1}\hat{\Psi}%
	_{\gamma ,g,m}+o_{p}((GM)^{1/2}(nT)^{-1/2})  \label{P_Rep_L_C_20}
\end{align}%
which together with (\ref{P_Rep_L_C_9}) shows the claim of the lemma.\hfill $%
Q.E.D.$

\bigskip

\begin{lemma}
	\textit{\label{Rep_L_D}}Under Assumptions \ref{A1}, \ref{A2}, \ref{A3}, \ref%
	{A4} and \ref{A5}, we have%
	\begin{equation}
		(nT)^{-1/2}\sum_{g\in \mathcal{G}}\sum_{m\in \mathcal{M}}(n_{g}T_{m})(\hat{%
			\theta}-\theta ^{\ast })^{\top }\hat{\Psi}_{\theta \theta ,g,m}(\hat{\theta}%
		-\theta ^{\ast })=O_{p}\left( (nT)^{-1/2}\right)  \label{Rep_L_D-1}
	\end{equation}%
	and 
	\begin{equation}
		(nT)^{-1/2}\sum_{g\in \mathcal{G}}\sum_{m\in \mathcal{M}}(n_{g}T_{m})(\hat{%
			\theta}-\theta ^{\ast })^{\top }\hat{\Psi}_{\theta \gamma ,g,m}(\hat{\gamma}%
		_{g,m}-\gamma _{g,m}^{\ast })=O_{p}\left(
		(nT)^{-1/2}+(GM)^{1+2/p}(nT)^{-1}\right) .  \label{Rep_L_D-2}
	\end{equation}
\end{lemma}

\noindent \textsc{Proof of Lemma \ref{Rep_L_D}}.\ By the triangle inequality
and the Cauchy-Schwarz inequality%
\begin{align}
	& \left\vert (nT)^{-1/2}\sum_{g\in \mathcal{G}}\sum_{m\in \mathcal{M}%
	}(n_{g}T_{m})(\hat{\theta}-\theta ^{\ast })^{\top }\hat{\Psi}_{\theta \theta
		,g,m}(\hat{\theta}-\theta ^{\ast })\right\vert  \notag \\
	& \leq \left\Vert \hat{\theta}-\theta ^{\ast }\right\Vert
	^{2}(nT)^{-1/2}\sum_{g\in \mathcal{G}}\sum_{m\in \mathcal{M}}(n_{g}T_{m})||%
	\hat{\Psi}_{\theta \theta ,g,m}||  \notag \\
	& \leq \left\Vert \hat{\theta}-\theta ^{\ast }\right\Vert
	^{2}(nT)^{1/2}\max_{g\in \mathcal{G},m\in \mathcal{M}}||\hat{\Psi}_{\theta
		\theta ,g,m}||=O_{p}((nT)^{-1/2})  \label{P_Rep_L_D_1}
\end{align}%
where the equality is by Lemma \ref{Rate_theta} and Lemma \ref{L3AB}(i).
This shows the assertion in (\ref{Rep_L_D-1}). To show (\ref{Rep_L_D-2}), we
note that by the triangle inequality%
\begin{align}
	& \left\Vert (nT)^{-1/2}\sum_{g\in \mathcal{G}}\sum_{m\in \mathcal{M}%
	}(n_{g}T_{m}\hat{\Psi}_{\theta \gamma ,g,m}(\hat{\gamma}_{g,m}-\gamma
	_{g,m}^{\ast })\right\Vert  \notag \\
	& \leq \left\Vert (nT)^{-1/2}\sum_{g\in \mathcal{G}}\sum_{m\in \mathcal{M}%
	}(n_{g}T_{m})\Psi _{\theta \gamma ,g,m}\Psi _{\gamma \gamma ,g,m}^{-1}\hat{%
		\Psi}_{\gamma ,g,m}\right\Vert  \notag \\
	& +\left\Vert (nT)^{-1/2}\sum_{g\in \mathcal{G}}\sum_{m\in \mathcal{M}%
	}(n_{g}T_{m})(\hat{\Psi}_{\theta \gamma ,g,m}-\Psi _{\theta \gamma
		,g,m})\Psi _{\gamma \gamma ,g,m}^{-1}\hat{\Psi}_{\gamma ,g,m}\right\Vert 
	\notag \\
	& +\left\Vert (nT)^{-1/2}\sum_{g\in \mathcal{G}}\sum_{m\in \mathcal{M}%
	}(n_{g}T_{m})\hat{\Psi}_{\theta \gamma ,g,m}(\hat{\gamma}_{g,m}-\gamma
	_{g,m}^{\ast }+\Psi _{\gamma \gamma ,g,m}^{-1}\hat{\Psi}_{\gamma
		,g,m})\right\Vert  \notag \\
	& \leq (nT)^{-1/2}\sum_{g\in \mathcal{G}}\sum_{m\in \mathcal{M}%
	}(n_{g}T_{m})||\hat{\Psi}_{\theta \gamma ,g,m}||\left\Vert \hat{\gamma}%
	_{g,m}-\gamma _{g,m}^{\ast }+\Psi _{\gamma \gamma ,g,m}^{-1}\hat{\Psi}%
	_{\gamma ,g,m}\right\Vert  \notag \\
	& +O_{p}(1+(GM)(nT)^{-1/2})  \notag \\
	& =O_{p}(1+(GM)^{1+2/p}(nT)^{-1/2})  \label{P_Rep_L_D_2}
\end{align}%
where the second inequality is by (\ref{L0-1}) in Lemma \ref{L0} and Lemma %
\ref{L2A}, the equality is by Lemma \ref{L3AB}(i) and Lemma \ref{Rate_gamma}%
. The assertion in (\ref{Rep_L_D-2}) now follows from Lemma \ref{Rate_theta}%
, (\ref{P_Rep_L_D_2}) and the Cauchy-Schwarz inequality.\hfill $Q.E.D.$

\bigskip

\begin{lemma}
	\textit{\label{C1_MGCLT_L1}\ Under the conditions of Theorem }\ref{MGCLT_V},
	we have%
	\begin{equation}
		(n\omega _{n,T}^{2})^{-1}\sum_{i\leq n}\mathbb{E}\left[ (\tilde{\Psi}_{i}+%
		\tilde{V}_{i})\tilde{U}_{j,i}|\mathcal{F}_{nT,i-1}\right] =O_{p}\left(
		T^{-1/2}+G_{j}^{-1/2}\max_{g\in \mathcal{G}_{j}}n_{g}^{-1}\right) .
		\label{C1_MGCLT_L1_1}
	\end{equation}
\end{lemma}

\noindent \textsc{Proof of Lemma \ref{C1_MGCLT_L1}}.\ For any $i=1,\ldots ,n$%
, by the independence of $\{z_{i,t}\}_{t\leq T}$ across $i$ we have%
\begin{equation*}
	\frac{n_{g_{j}(i)}T^{1/2}\mathbb{E}\left[ (\tilde{\Psi}_{i}+\tilde{V}_{i})%
		\tilde{U}_{j,i}|\mathcal{F}_{nT,i-1}\right] }{\omega _{n,T}}=\mathbb{E}\left[
	\frac{(\tilde{\Psi}_{i}+\tilde{V}_{i})A_{j,i}^{\top }}{\omega _{n,T}}\right]
	\sum_{\{i^{\prime }\in I_{g_{j}(i)}:\text{ }i^{\prime }<i\}}A_{j,i^{\prime
	}}=\sum_{\{i^{\prime }\in I_{g_{j}(i)}:\text{ }i^{\prime }<i\}}c_{j,i}^{\top
	}A_{j,i^{\prime }},
\end{equation*}%
where\ $c_{j,i}\equiv \mathbb{E}[(\tilde{\Psi}_{i}+\tilde{V}%
_{i})A_{j,i}/\omega _{n,T}]$. Let $C_{j,i}\equiv \sum_{\{i^{\prime }\in
	I_{g_{j}(i)}:\text{ }i^{\prime }>i\}}c_{j,i^{\prime }}$. Then the above
expression implies that%
\begin{equation}
	\frac{\sum_{i\leq n}\mathbb{E}\left[ (\tilde{\Psi}_{i}+\tilde{V}_{i})\tilde{U%
		}_{j,i}|\mathcal{F}_{nT,i-1}\right] }{n\omega _{n,T}^{2}}=\sum_{i\leq n}%
	\frac{C_{j,i}^{\top }A_{j,i}}{n_{g_{j}(i)}T^{1/2}n\omega _{n,T}}.
	\label{P_C1_MGCLT_L1_2}
\end{equation}%
Since $\mathbb{E}\left[ A_{j,i}\right] =0$ and $A_{j,i}$ are independent
across $i$,%
\begin{align}
	\mathbb{E}\left[ \left\vert \sum_{i\leq n}\frac{C_{j,i}^{\top }A_{j,i}}{%
		n_{g_{j}(i)}n\omega _{n,T}T^{1/2}}\right\vert ^{2}\right] & =\sum_{i\leq n}%
	\frac{C_{j,i}^{\top }\mathbb{E}[A_{j,i}A_{j,i}^{\top }]C_{j,i}}{%
		n_{g_{j}(i)}^{2}n^{2}\omega _{n,T}^{2}T}  \notag \\
	& \leq \frac{K\max_{i\leq n}\lambda _{\max }(\mathbb{E}[A_{j,i}A_{j,i}^{\top
		}])}{n^{2}\omega _{n,T}^{2}T}\sum_{i\leq n}n_{g_{j}(i)}^{-2}C_{j,i}^{\top
	}C_{j,i}.  \label{P_C1_MGCLT_L1_3}
\end{align}%
Applying Assumptions \ref{A1}, \ref{A4} and \ref{A7}(ii), and Rosenthal's
inequality of strong mixing processes (e.g., (\ref{P_L0_2}) with $\tilde{p}%
=4+\delta $ in the proof of Lemma \ref{L0}), we obtain%
\begin{equation}
	\max_{i\leq n}\max_{m\in \mathcal{M}_{j}}\mathbb{E}[|\tilde{\Psi}_{j\gamma
		,im}|^{4+\delta }]=\max_{i\leq n}\max_{m\in \mathcal{M}_{j}}\mathbb{E}\left[
	\left( T_{m}^{-1/2}\sum_{t\in I_{m}}\tilde{\psi}_{j\gamma }(z_{i,t};\phi
	_{j,g,m}^{\ast })\right) ^{4+\delta }\right] \leq K.  \label{P_C1_MGCLT_L1_4}
\end{equation}%
By Corollary A.2. in \cite{HallHeyde1980},%
\begin{eqnarray}
	&&\max_{i\leq n}\max_{m^{\prime }\in \mathcal{M}_{j}}\sum_{m\in \mathcal{M}%
		_{j}}\left\vert \mathbb{E}[\tilde{\Psi}_{j\gamma ,im}\tilde{\Psi}_{j\gamma
		,im^{\prime }}]\right\vert  \notag \\
	&\leq &K\max_{i\leq n}\max_{m\in \mathcal{M}_{j}}(\mathbb{E}[|\tilde{\Psi}%
	_{j\gamma ,im}|^{2+\delta }])^{2}\sum_{h=0}^{\infty }\left\vert \alpha
	_{i}(h)\right\vert ^{\delta /(2+\delta )}\leq K,  \label{P_C1_MGCLT_L1_5}
\end{eqnarray}%
where the second inequality is by Assumption \ref{A1}(iv) and (\ref%
{P_C1_MGCLT_L1_4}). In view of the upper bound in (\ref{P_C1_MGCLT_L1_5}),
we can use Ger\v{s}gorin's disc theorem (see, e.g., Theorem 6.1.1 in \cite%
{horn2012matrix}) to show that 
\begin{equation}
	\max_{j=1,2}\max_{i\leq n}\lambda _{\max }(\mathbb{E}[A_{j,i}A_{j,i}^{\top
	}])\leq K.  \label{P_C1_MGCLT_L1_6}
\end{equation}%
Using Lemma \ref{L00} (with $X_{1}=(\tilde{\Psi}_{i}^{\ast }+\tilde{V}%
_{i}^{\ast })/\omega _{n,T}$ and $X_{2}=A_{j,i}$) and the inequality above
obtains%
\begin{align}
	\max_{i\leq n}\left\Vert c_{j,i}\right\Vert ^{2}& \leq \max_{i\leq
		n}(\lambda _{\max }(\mathbb{E}[A_{j,i}A_{j,i}^{\top }]))\max_{i\leq n}%
	\mathbb{E}\left[ \frac{(\tilde{\Psi}_{1,i}^{\ast }-\tilde{\Psi}_{2,i}^{\ast
		}+\tilde{V}_{1,i}^{\ast }-\tilde{V}_{2,i}^{\ast })^{2}}{\omega _{n,T}^{2}}%
	\right]  \notag \\
	& \leq K\max_{i\leq n}\mathbb{E}\left[ \frac{(\tilde{\Psi}_{1,i}^{\ast }-%
		\tilde{\Psi}_{2,i}^{\ast })^{2}}{\omega _{n,T}^{2}}\right] +K\frac{%
		\max_{j=1,2}\max_{i\leq n}\mathbb{E}[|\tilde{V}_{j,i}^{\ast }|^{2}]}{\omega
		_{n,T}^{2}}.  \label{P_C1_MGCLT_L1_7}
\end{align}%
Applying Assumptions \ref{A1} and \ref{A7}(i), and Rosenthal's inequality of
strong mixing processes, we can show that 
\begin{equation}
	\max_{i\leq n}\mathbb{E}\left[ \left( \frac{\tilde{\Psi}_{1,i}^{\ast }-%
		\tilde{\Psi}_{2,i}^{\ast }}{\omega _{n,T}}\right) ^{4}\right] =\max_{i\leq n}%
	\mathbb{E}\left[ \left( T^{-1/2}\sum_{t\leq T}\frac{\psi _{1}^{\ast
		}(z_{i,t};\phi _{1,i,t}^{\ast })-\psi _{2}^{\ast }(z_{i,t};\phi
		_{2,i,t}^{\ast })}{\omega _{n,T}}\right) ^{4}\right] \leq K,
	\label{P_C1_MGCLT_L1_8}
\end{equation}%
where $\psi _{j}^{\ast }(z_{i,t};\phi _{j,i,t}^{\ast })\equiv \psi
_{j}(z_{i,t};\phi _{j,i,t}^{\ast })-\mathbb{E}[\psi _{j}(z_{i,t};\phi
_{j,i,t}^{\ast })]$. Similarly for any $\delta ^{\prime }\in \lbrack
0,2+\delta /2]$,%
\begin{equation}
	\mathbb{E}[|\tilde{V}_{j,i}^{\ast }|^{2+\delta ^{\prime }}]=\frac{\mathbb{E}%
		\left[ \left( \sum_{m\in \mathcal{M}_{j}}(\tilde{\Psi}_{j\gamma ,im}^{2}-%
		\mathbb{E}[\tilde{\Psi}_{j\gamma ,im}^{2}])\right) ^{2+\delta ^{\prime }}%
		\right] }{(2n_{g_{j}(i)}T^{1/2})^{2+\delta ^{\prime }}}\leq \frac{%
		KM_{j}^{1+\delta ^{\prime }/2}}{(n_{g_{j}(i)}T^{1/2})^{2+\delta ^{\prime }}}.
	\label{P_C1_MGCLT_L1_9}
\end{equation}%
Therefore by (\ref{P_C1_MGCLT_L1_7}), (\ref{P_C1_MGCLT_L1_8}) and (\ref%
{P_C1_MGCLT_L1_9}) with $\delta ^{\prime }=0$, and Assumptions \ref{A1}(i)
and \ref{A6}(i), we can deduce that 
\begin{equation}
	\max_{i\leq n}\left\Vert c_{j,i}\right\Vert ^{2}\leq K\left(
	1+M_{j}\max_{g\in \mathcal{G}_{j}}n_{g}^{-2}\right) .
	\label{P_C1_MGCLT_L1_10}
\end{equation}%
Using the Cauchy-Schwarz inequality and (\ref{P_C1_MGCLT_L1_10}) yields%
\begin{align}
	\max_{i\leq n}\frac{C_{j,i}^{\top }C_{j,i}}{n_{g_{j}(i)}^{2}}& =\max_{i\leq
		n}\frac{\sum_{i_{1},i_{2}\in \{i^{\prime }\in I_{g_{j}(i)}:\text{ }i^{\prime
			}>i\}}|c_{j,i_{1}}^{\top }c_{j,i_{2}}|}{n_{g_{j}(i)}^{2}}  \notag \\
	& \leq \max_{i\leq n}\left( \frac{\sum_{\{i^{\prime }\in I_{g_{j}(i)}:\text{ 
			}i^{\prime }>i\}}\left\Vert c_{j,i^{\prime }}\right\Vert }{n_{g_{j}(i)}}%
	\right) ^{2}\leq K\left( 1+M_{j}\max_{g\in \mathcal{G}_{j}}n_{g}^{-2}\right)
	.  \label{P_C1_MGCLT_L1_11}
\end{align}%
Collecting the results in (\ref{P_C1_MGCLT_L1_3}), (\ref{P_C1_MGCLT_L1_6})
and (\ref{P_C1_MGCLT_L1_11}), and applying Assumption \ref{A6}(i) obtains%
\begin{equation}
	\mathbb{E}\left[ \left\vert \sum_{i\leq n}\frac{C_{j,i}^{\top }A_{j,i}}{%
		n_{g_{j}(i)}n\omega _{n,T}T^{1/2}}\right\vert ^{2}\right] \leq \frac{K}{%
		Tn^{2}\omega _{n,T}^{2}}\sum_{i\leq n}\frac{C_{j,i}^{\top }C_{j,i}}{%
		n_{g_{j}(i)}^{2}}\leq KT^{-1}\left( 1+M_{j}\max_{g\in \mathcal{G}%
		_{j}}n_{g}^{-2}\right) .  \label{P_C1_MGCLT_L1_12}
\end{equation}%
The desired result in (\ref{C1_MGCLT_L1_1}) now follows from (\ref%
{P_C1_MGCLT_L1_2}), (\ref{P_C1_MGCLT_L1_12}), Assumption \ref{A5}(ii) and
Markov's inequality.\hfill $Q.E.D.$

\bigskip

\begin{lemma}
	\textit{\label{C1_MGCLT_L2}\ Under the conditions of Theorem }\ref{MGCLT_V},
	we have%
	\begin{equation}
		(n\omega_{n,T}^{2})^{-1}\sum_{i\leq n}\left( \mathbb{E}[\tilde{U}_{i}^{2}|%
		\mathcal{F}_{nT,i-1}]-\mathbb{E}[\tilde{U}_{i}^{2}]\right) =O_{p}\left(
		T^{-1/2}+\max_{j=1,2}\max_{g\in\mathcal{G}_{j}}(G_{j}n_{g})^{-1/2}\right) .
		\label{C1_MGCLT_L1_2}
	\end{equation}
\end{lemma}

\noindent \textsc{Proof of Lemma \ref{C1_MGCLT_L2}}.\ Since 
\begin{equation*}
	\mathbb{E}\left[ \tilde{U}_{i}^{2}|\mathcal{F}_{nT,i-1}\right] =\mathbb{E}%
	\left[ \tilde{U}_{1,i}^{2}|\mathcal{F}_{nT,i-1}\right] +\mathbb{E}\left[ 
	\tilde{U}_{2,i}^{2}|\mathcal{F}_{nT,i-1}\right] -2\mathbb{E}\left[ \tilde{U}%
	_{1,i}\tilde{U}_{2,i}|\mathcal{F}_{nT,i-1}\right] ,
\end{equation*}%
the claim of the lemma follows if%
\begin{align}
	\sum_{i\leq n}\left( \mathbb{E}[\tilde{U}_{j,i}^{2}|\mathcal{F}_{nT,i-1}]-%
	\mathbb{E}[\tilde{U}_{j,i}^{2}]\right) & =O_{p}\left( T^{-1/2}+\max_{g\in 
		\mathcal{G}_{j}}(G_{j}n_{g})^{-1/2}\right) ,\text{ \ and}
	\label{P_C1_MGCLT_L2_1} \\
	\sum_{i\leq n}\left( \mathbb{E}[\tilde{U}_{1,i}\tilde{U}_{2,i}|\mathcal{F}%
	_{nT,i-1}]-\mathbb{E}[\tilde{U}_{1,i}\tilde{U}_{2,i}]\right) & =O_{p}\left(
	T^{-1/2}+\max_{j=1,2}\max_{g\in \mathcal{G}_{j}}(G_{j}n_{g})^{-1/2}\right) .
	\label{P_C1_MGCLT_L2_2}
\end{align}%
We next verify\ (\ref{P_C1_MGCLT_L2_1}) and (\ref{P_C1_MGCLT_L2_2}).

We first show (\ref{P_C1_MGCLT_L2_1}) for $j=1$. The proof for $j=2$ follows
the same arguments. For simplicity of notations, we suppose that for any $%
g\in \mathcal{G}_{1}$, $I_{g}=\{N_{g}+1,\ldots ,N_{g}+n_{g}\}$ where $%
N_{g}\equiv \sum_{g^{\prime }<g}n_{g^{\prime }}$. Following these notations,
for any $i\in I_{g}$ and any $g\in \mathcal{G}_{1}$ we can write%
\begin{equation*}
	\tilde{U}_{1,i}=(n_{g}T^{1/2})^{-1}\sum_{i^{\prime
		}=N_{g}+1}^{i-1}A_{1,i}^{\top }A_{1,i^{\prime }}.
\end{equation*}%
Since $\sum_{i^{\prime }=N_{g}+1}^{i-1}A_{1,i^{\prime }}$ is non-random
given $\mathcal{F}_{nT,i-1}$ and $A_{1,i}$ is independent of $%
\sum_{i^{\prime }=N_{g}+1}^{i-1}A_{1,i^{\prime }}$,\ for any $i\in I_{g}$
and any $g\in \mathcal{G}_{1}$ 
\begin{align*}
	n_{g}^{2}T\mathbb{E}[\tilde{U}_{1,i}^{2}|\mathcal{F}_{nT,i-1}]& =\mathbb{E}%
	\left[ \left. \left( \sum_{i^{\prime }=N_{g}+1}^{i-1}A_{1,i}^{\top
	}A_{1,i^{\prime }}\right) ^{2}\right\vert \mathcal{F}_{nT,i-1}\right] \\
	& =\sum_{i_{1},i_{2}=N_{g}+1}^{i-1}A_{1,i_{1}}^{\top }(\mathbb{E}%
	[A_{1,i}A_{1,i}^{\top }])A_{1,i_{2}} \\
	& =\sum_{i^{\prime }=N_{g}+1}^{i-1}A_{1,i^{\prime }}^{\top }(\mathbb{E}%
	[A_{1,i}A_{1,i}^{\top }])A_{1,i^{\prime
	}}+2\sum_{i_{1}=N_{g}+2}^{i-1}\sum_{i_{2}=N_{g}+1}^{i_{1}-1}A_{1,i_{1}}^{%
		\top }(\mathbb{E}[A_{1,i}A_{1,i}^{\top }])A_{1,i_{2}} \\
	& =\mathrm{Tr}\left( \mathbb{E}[A_{1,i}A_{1,i}^{\top }]\left(
	\sum_{i^{\prime }=N_{g}+1}^{i-1}A_{1,i^{\prime }}A_{1,i^{\prime }}^{\top
	}+2\sum_{i_{1}=N_{g}+2}^{i-1}%
	\sum_{i_{2}=N_{g}+1}^{i_{1}-1}A_{1,i_{2}}A_{1,i_{1}}^{\top }\right) \right) ,
\end{align*}%
which implies that%
\begin{equation}
	\sum_{i\leq n}\mathbb{E}[\tilde{U}_{1,i}^{2}|\mathcal{F}_{nT,i-1}]=\sum_{g%
		\in \mathcal{G}_{1}}\sum_{i=N_{g}+1}^{N_{g}+n_{g}-1}\frac{A_{1,i}^{\top
		}D_{1,i}A_{1,i}}{n_{g}^{2}T}+2\sum_{g\in \mathcal{G}_{1}}%
	\sum_{i=N_{g}+2}^{N_{g}+n_{g}-1}\sum_{i^{\prime }=N_{g}+1}^{i-1}\frac{%
		A_{1,i}^{\top }D_{1,i}A_{1,i^{\prime }}}{n_{g}^{2}T},
	\label{P_C1_MGCLT_L2_4}
\end{equation}%
where 
\begin{equation*}
	D_{1,i}\equiv \sum_{i^{\prime }=i+1}^{N_{g}+n_{g}}\mathbb{E}[A_{1,i^{\prime
	}}A_{1,i^{\prime }}^{\top }]
\end{equation*}%
for any $i\in I_{g}$. Since $\mathbb{E}[A_{1,i}^{\top }D_{1,i}A_{1,i^{\prime
}}]=0$ for any $i,i^{\prime }\in I_{g}$ with $i\neq i^{\prime }$, by (\ref%
{P_C1_MGCLT_L2_4}) and the law of iterated expectation, 
\begin{equation}
	\sum_{i\leq n}\mathbb{E}[\tilde{U}_{1,i}^{2}]=\sum_{g\in \mathcal{G}%
		_{1}}\sum_{i=N_{g}+1}^{N_{g}+n_{g}-1}\frac{\mathbb{E}[A_{1,i}^{\top
		}D_{1,i}A_{1,i}]}{n_{g}^{2}T}.  \label{P_C1_MGCLT_L2_4a}
\end{equation}%
By (\ref{P_C1_MGCLT_L2_4}) and (\ref{P_C1_MGCLT_L2_4a}), 
\begin{align}
	\sum_{i\leq n}\left( \mathbb{E}[\tilde{U}_{j,i}^{2}|\mathcal{F}_{nT,i-1}]-%
	\mathbb{E}[\tilde{U}_{j,i}^{2}]\right) & =\sum_{g\in \mathcal{G}%
		_{1}}\sum_{i=N_{g}+1}^{N_{g}+n_{g}-1}\frac{A_{1,i}^{\top }D_{1,i}A_{1,i}-%
		\mathbb{E}[A_{1,i}^{\top }D_{1,i}A_{1,i}]}{n_{g}^{2}T}  \notag \\
	& \text{ \ \ \ \ \ \ \ \ }+2\sum_{g\in \mathcal{G}_{1}}%
	\sum_{i=N_{g}+2}^{N_{g}+n_{g}-1}\sum_{i^{\prime }=N_{g}+1}^{i-1}\frac{%
		A_{1,i}^{\top }D_{1,i}A_{1,i^{\prime }}}{n_{g}^{2}T}.
	\label{P_C1_MGCLT_L2_4b}
\end{align}%
In view of the decomposition in (\ref{P_C1_MGCLT_L2_4b}), it is clear that (%
\ref{P_C1_MGCLT_L2_1}) holds for $j=1$ by Markov's inequality if 
\begin{equation}
	\mathrm{Var}\left( \sum_{g\in \mathcal{G}_{1}}%
	\sum_{i=N_{g}+1}^{N_{g}+n_{g}-1}\frac{A_{1,i}^{\top }D_{1,i}A_{1,i}}{%
		n_{g}^{2}T}\right) \leq KG_{1}^{-1}\max_{g\in \mathcal{G}_{1}}n_{g}^{-1}
	\label{P_C1_MGCLT_L2_1a}
\end{equation}%
and 
\begin{equation}
	\mathrm{Var}\left( \sum_{g\in \mathcal{G}_{1}}%
	\sum_{i=N_{g}+2}^{N_{g}+n_{g}-1}\sum_{i^{\prime }=N_{g}+1}^{i-1}\frac{%
		A_{1,i}^{\top }D_{1,i}A_{1,i^{\prime }}}{n_{g}^{2}T}\right) \leq KT^{-1}.
	\label{P_C1_MGCLT_L2_1b}
\end{equation}%
We next prove (\ref{P_C1_MGCLT_L2_1a}) and (\ref{P_C1_MGCLT_L2_1b}).

First note that the upper bound on the eigenvalues of $\mathbb{E}%
[A_{j,i}A_{j,i}^{\top}]$ established in (\ref{P_C1_MGCLT_L1_6}) implies that 
\begin{equation}
	\max_{i\leq n}\mathbb{E}[A_{j,i}^{\top}A_{j,i}]=\max_{i\leq n}\mathrm{Tr}(%
	\mathbb{E}[A_{j,i}A_{j,i}^{\top}])\leq M_{j}\max_{i\leq n}\lambda_{\max }(%
	\mathbb{E}[A_{j,i}A_{j,i}^{\top}])\leq KM_{j},  \label{P_C1_MGCLT_L2_5a}
\end{equation}
where the first inequality is because $\mathrm{Tr}(\mathbb{E}%
[A_{j,i}A_{j,i}^{\top}])$ is equal to the sum of the eigenvalues of $\mathbb{%
	E}[A_{j,i}A_{j,i}^{\top}]$, which is bounded by $M_{j}\lambda_{\max}(\mathbb{%
	E}[A_{j,i}A_{j,i}^{\top}])$. Also in (\ref{P_C1_MGCLT_L1_9}), we have shown
that%
\begin{equation}
	\max_{i\leq n}\mathbb{E}\left[ \left( M_{j}^{-1/2}\sum_{m\in\mathcal{M}_{j}}(%
	\tilde{\Psi}_{j\gamma,im}^{2}-\mathbb{E}[\tilde{\Psi}_{j\gamma,im}^{2}])%
	\right) ^{2}\right] \leq K.  \label{P_C1_MGCLT_L2_5b}
\end{equation}
By (\ref{P_C1_MGCLT_L2_5a}) and (\ref{P_C1_MGCLT_L2_5b}), and the
Cauchy-Schwarz inequality,%
\begin{align}
	\max_{i\leq n}\mathbb{E}[(A_{j,i}^{\top}A_{j,i})^{2}] & =\max_{i\leq n}%
	\mathbb{E}\left[ \left( \sum_{m\in\mathcal{M}_{j}}\tilde{\Psi}%
	_{j\gamma,im}^{2}\right) ^{2}\right]  \notag \\
	& \leq2\max_{i\leq n}\left[ \mathbb{E}\left[ \left( \sum_{m\in \mathcal{M}%
		_{j}}(\tilde{\Psi}_{j\gamma,im}^{2}-\mathbb{E}[\tilde{\Psi }%
	_{j\gamma,im}^{2}])\right) ^{2}\right] +\left( \sum_{m\in\mathcal{M}_{j}}%
	\mathbb{E}[\tilde{\Psi}_{j\gamma,im}^{2}]\right) ^{2}\right]  \notag \\
	& \leq KM_{j}+\max_{i\leq n}(\mathbb{E}[A_{j,i}^{\top}A_{j,i}])^{2}\leq
	KM_{j}^{2}.  \label{P_C1_MGCLT_L2_5}
\end{align}
By (\ref{P_C1_MGCLT_L1_6}), we also have%
\begin{equation}
	\max_{g\in\mathcal{G}_{1}}\max_{i\in I_{g}}\frac{\lambda_{\max}(D_{1,i})}{%
		n_{g}}=\max_{g\in\mathcal{G}_{1}}\max_{i\in I_{g}}\frac{\sum_{i^{\prime
			}=i+1}^{N_{g}+n_{g}}\lambda_{\max}(\mathbb{E}[A_{1,i^{\prime}}A_{1,i^{%
				\prime}}^{\top}])}{n_{g}}\leq K.  \label{P_C1_MGCLT_L2_6}
\end{equation}
Getting back to (\ref{P_C1_MGCLT_L2_1a}), we note that since $A_{1,i}$ are
independent across $i$,%
\begin{align}
	\mathrm{Var}\left( \sum_{g\in\mathcal{G}_{1}}\sum_{i=N_{g}+1}^{N_{g}+n_{g}-1}%
	\frac{A_{1,i}^{\top}D_{1,i}A_{1,i}}{n_{g}^{2}T}\right) & =\sum _{g\in%
		\mathcal{G}_{1}}\sum_{i=N_{g}+1}^{N_{g}+n_{g}-1}\frac{\mathrm{Var}%
		(A_{1,i}^{\top}D_{1,i}A_{1,i})}{\left( n_{g}^{2}T\right) ^{2}}  \notag \\
	& \leq\sum_{g\in\mathcal{G}_{1}}\sum_{i=N_{g}+1}^{N_{g}+n_{g}-1}\frac{%
		\mathbb{E}[(A_{1,i}^{\top}D_{1,i}A_{1,i})^{2}]}{\left( n_{g}^{2}T\right) ^{2}%
	}  \notag \\
	& \leq\sum_{g\in\mathcal{G}_{1}}\sum_{i=N_{g}+1}^{N_{g}+n_{g}-1}\frac{%
		(\lambda_{\max}(D_{1,i}))^{2}\mathbb{E}[(A_{1,i}^{\top}A_{1,i})^{2}]}{\left(
		n_{g}^{2}T\right) ^{2}}  \notag \\
	& \leq K\sum_{g\in\mathcal{G}_{1}}%
	\sum_{i=N_{g}+1}^{N_{g}+n_{g}-1}(n_{g}T)^{-2}M_{1}^{2}\leq
	KM_{1}^{2}T^{-2}\sum_{g\in\mathcal{G}_{1}}n_{g}^{-1}  \notag \\
	& \leq KM_{1}^{2}G_{1}T^{-2}\max_{g\in\mathcal{G}_{1}}n_{g}^{-1}\leq
	KG_{1}^{-1}\max_{g\in\mathcal{G}_{1}}n_{g}^{-1}  \label{P_C1_MGCLT_L2_7}
\end{align}
where the third inequality is by (\ref{P_C1_MGCLT_L2_5}) and (\ref%
{P_C1_MGCLT_L2_6}), and the last inequality is by (\ref{P_MGCLT_V_6b}). This
establishes the desired result in (\ref{P_C1_MGCLT_L2_1a}).

Similarly,%
\begin{align}
	& \mathrm{Var}\left( \sum_{g\in\mathcal{G}_{1}}%
	\sum_{i=N_{g}+2}^{N_{g}+n_{g}-1}\sum_{i^{\prime}=N_{g}+1}^{i-1}\frac{%
		A_{1,i}^{\top}D_{1,i}A_{1,i^{\prime}}}{n_{g}^{2}T}\right)  \notag \\
	& =\sum_{g\in\mathcal{G}_{1}}\sum_{i=N_{g}+2}^{N_{g}+n_{g}-1}\sum_{i^{\prime
		}=N_{g}+1}^{i-1}\frac{\mathrm{Var}(A_{1,i}^{\top}D_{1,i}A_{1,i^{\prime}})}{%
		(n_{g}^{2}T)^{2}}=\sum_{g\in\mathcal{G}_{1}}\sum_{i=N_{g}+2}^{N_{g}+n_{g}-1}%
	\sum_{i^{\prime}=N_{g}+1}^{i-1}\frac{\mathbb{E}[(A_{1,i}^{\top
		}D_{1,i}A_{1,i^{\prime}})^{2}]}{(n_{g}^{2}T)^{2}}  \notag \\
	& \leq\sum_{g\in\mathcal{G}_{1}}\sum_{i=N_{g}+2}^{N_{g}+n_{g}-1}\sum_{i^{%
			\prime}=N_{g}+1}^{i-1}\frac{(\lambda_{\max}(D_{1,i}))^{2}\lambda_{\max}(%
		\mathbb{E}[A_{1,i^{\prime}}A_{1,i^{\prime}}^{\top}])\mathbb{E}%
		[A_{1,i}^{\top}A_{1,i}]}{(n_{g}^{2}T)^{2}}  \notag \\
	& \leq K\sum_{g\in\mathcal{G}_{1}}\sum_{i=N_{g}+2}^{N_{g}+n_{g}-1}\frac {%
		M_{1}}{n_{g}T^{2}}\leq KM_{1}G_{1}T^{-2}\leq KT^{-1},
	\label{P_C1_MGCLT_L2_8}
\end{align}
where the second inequality is by (\ref{P_C1_MGCLT_L1_6}) and (\ref%
{P_C1_MGCLT_L2_6}), and the fourth inequality is by Assumptions \ref{A1}(i)
and \ref{A5}(ii). This shows (\ref{P_C1_MGCLT_L2_1b}).

To derive (\ref{P_C1_MGCLT_L2_2}), we use the same notation for the
structure in $\mathcal{G}_{1}$ employed in verifying (\ref{P_C1_MGCLT_L2_1})
for $j=1$ above. That is for any $g\in\mathcal{G}_{1}$, $I_{g}=\{N_{g}+1,%
\ldots ,N_{g}+n_{g}\}$ where $N_{g}\equiv\sum_{g^{\prime}<g}n_{g^{\prime}}$.
Moreover, for each $g\in\mathcal{G}_{1}$, we order $I_{g}$ according to the
group identity $g_{2}(i)$ such that $g_{2}(i)$ is non-decreasing. Based on
the values of $g_{2}(i)$, we can divide $I_{g}$ to $M_{g}$ subsets ($1\leq
M_{g}\leq G_{2}$): $I_{g,l}\equiv\{i_{g,l-1}+1,\ldots,i_{g,l}\}$ ($%
l=1,\ldots,M_{g}$)\ with $i_{g,0}\equiv N_{g}$, $i_{g,M_{g}}\equiv
N_{g}+n_{g}$ and $i_{g,j}\in I_{g}$ for $j=1,\ldots,M_{g}$ such that $%
g_{2}(i)=g_{2}(i^{\prime})$ for any $i,i^{\prime}\in I_{g,l}$. For each $i$,
let%
\begin{equation*}
	Q_{1,i}\equiv\{i^{\prime}\in I_{g_{2}(i)}\text{: }i^{\prime}\leq
	N_{g_{1}(i)}\} \text{ \ and \ }Q_{2,i}\equiv\{i^{\prime}\in I_{g_{2}(i)}%
	\text{: }N_{g_{1}(i)}+1\leq i^{\prime}\leq i-1\}.
\end{equation*}
By definition, $Q_{1,i}$ includes the individuals who share the same group
identity with individual $i$ under model 2 but are in groups $I_{g^{\prime}}$
with $g^{\prime}<g_{1}(i)$ under model 1. On the other hand, $Q_{2,i}$
includes the individuals who share the same group identity with individual $%
i $ under model 2, and are in the same group $g_{1}(i)$ of individual $i$
under model 1 but have their index strictly less than $i$. Since $%
N_{g_{1}(i)}\leq i-1$, it is clear that $Q_{1,i}$ and $Q_{2,i}$ are mutually
exclusive, and $\{i^{\prime}\in I_{g_{2}(i)}$: \ $i^{\prime}<i\}=Q_{1,i}\cup
Q_{2,i}$.

Using the above notations, for any $i\in I_{g}$ and any $g\in \mathcal{G}%
_{1} $ we can write%
\begin{align}
	\tilde{U}_{1,i}\tilde{U}_{2,i}& =(n_{g}n_{g_{2}(i)}T)^{-1}\left(
	\sum_{i^{\prime }=N_{g}+1}^{i-1}A_{1,i}^{\top }A_{1,i^{\prime }}\right)
	\left( \sum_{\{i^{\prime }\in I_{g_{2}(i)}:\text{ }i^{\prime
		}<i\}}A_{2,i}^{\top }A_{2,i^{\prime }}\right)  \notag \\
	& =\sum_{i_{1}=N_{g}+1}^{i-1}\sum_{i_{2}\in Q_{1,i}}\frac{A_{1,i_{1}}^{\top
		}A_{1,i}A_{2,i}^{\top }A_{2,i_{2}}}{n_{g}n_{g_{2}(i)}T}%
	+\sum_{i_{1}=N_{g}+1}^{i-1}\sum_{i_{2}\in Q_{2,i}}\frac{A_{2,i_{2}}^{\top
		}A_{2,i}A_{1,i}^{\top }A_{1,i_{1}}}{n_{g}n_{g_{2}(i)}T}.
	\label{P_C1_MGCLT_L2_9}
\end{align}%
Let $d_{12,i}\equiv \mathbb{E}[A_{1,i}A_{2,i}^{\top }]$ and $d_{21,i}\equiv
d_{12,i}^{\top }$. Then by the independence of $A_{j,i}$ across $i$, we
obtain 
\begin{align}
	\sum_{i\leq n}\mathbb{E}[\tilde{U}_{1,i}\tilde{U}_{2,i}|\mathcal{F}%
	_{nT,i-1}]& =\sum_{g\in \mathcal{G}_{1}}\sum_{i\in I_{g}}\sum_{i_{2}\in
		Q_{1,i}}\sum_{i_{1}=N_{g}+1}^{i-1}\frac{A_{2,i_{2}}^{\top
		}d_{21,i}A_{1,i_{1}}}{n_{g}n_{g_{2}(i)}T}  \notag \\
	& \text{ \ \ \ \ \ }+\sum_{g\in \mathcal{G}_{1}}\sum_{i\in
		I_{g}}\sum_{i_{2}\in Q_{2,i}}\sum_{i_{1}=N_{g}+1}^{i-1}\frac{%
		A_{2,i_{2}}^{\top }d_{21,i}A_{1,i_{1}}}{n_{g}n_{g_{2}(i)}T}.
	\label{P_C1_MGCLT_L2_10}
\end{align}%
To study the two terms in the RHS of (\ref{P_C1_MGCLT_L2_10}), we first
investigate $d_{12,i}$. Applying Lemma \ref{L00} (with $X_{1}=A_{1,i}$ and $%
X_{2}=A_{2,i}$) yields%
\begin{equation}
	\max_{i\leq n}\lambda _{\max }(d_{12,i}d_{21,i})\leq \max_{i\leq n}\lambda
	_{\max }(\mathbb{E}[A_{1,i}A_{1,i}^{\top }])\lambda _{\max }(\mathbb{E}%
	[A_{2,i}A_{2,i}^{\top }])\leq K,  \label{P_C1_MGCLT_L2_11}
\end{equation}%
where the second inequality is by (\ref{P_C1_MGCLT_L1_6}).

Consider any $g\in\mathcal{G}_{1}$. Since $I_{g}=\cup_{l\leq M_{g}}I_{g,l}$,
and\ for any\ $i\in I_{g,l}$, $Q_{1,i}=Q_{1,i_{g,l}}$ and $%
n_{g_{2}(i)}=n_{g_{2}(i_{g,l})}$, we can write%
\begin{align}
	\sum_{i\in I_{g}}\sum_{i_{2}\in Q_{1,i}}\sum_{i_{1}=N_{g}+1}^{i-1}\frac{%
		A_{2,i_{2}}^{\top}d_{21,i}A_{1,i_{1}}}{n_{g}n_{g_{2}(i)}T} & =\sum_{l\leq
		M_{g}}\sum_{i\in I_{g,l}}\sum_{i_{2}\in Q_{1,i_{g,l}}}\sum
	_{i_{1}=N_{g}+1}^{i-1}\frac{A_{2,i_{2}}^{\top}d_{21,i}A_{1,i_{1}}}{%
		n_{g}n_{g_{2}(i_{g,l})}T}  \notag \\
	& =\sum_{l\leq M_{g}}\sum_{i_{2}\in Q_{1,i_{g,l}}}\frac{A_{2,i_{2}}^{\top}}{%
		n_{g}n_{g_{2}(i_{g,l})}T}\sum_{i\in
		I_{g,l}}\sum_{i_{1}=N_{g}+1}^{i-1}d_{21,i}A_{1,i_{1}}.
	\label{P_C1_MGCLT_L2_12}
\end{align}
By the definition of $I_{g,l}$, 
\begin{equation*}
	\sum_{i\in
		I_{g,l}}\sum_{i_{1}=N_{g}+1}^{i-1}d_{21,i}A_{1,i_{1}}=%
	\sum_{i=i_{g,l-1}+1}^{i_{g,l}}d_{21,i}%
	\sum_{i_{1}=N_{g}+1}^{i-1}A_{1,i_{1}}=D_{1,g}^{l}%
	\sum_{i_{1}=N_{g}+1}^{i_{g,l-1}}A_{1,i_{1}}+%
	\sum_{i_{1}=i_{g,l-1}+1}^{i_{g,l}-1}D_{2,gi_{1}}^{l}A_{1,i_{1}},
\end{equation*}
where $D_{1,g}^{l}\equiv\sum_{i=i_{g,l-1}+1}^{i_{g,l}}d_{21,i}$ and $%
D_{2,gi}^{l}\equiv\sum_{i^{\prime}=i+1}^{i_{g,l}}d_{21,i^{\prime}}$. Hence
the summation in the RHS of the second equality of (\ref{P_C1_MGCLT_L2_12})
can be further decomposed into two terms which leads to%
\begin{align}
	\sum_{g\in\mathcal{G}_{1}}\sum_{i\in I_{g}}\sum_{i_{2}\in Q_{1,i}}\sum
	_{i_{1}=N_{g}+1}^{i-1}\frac{A_{2,i_{2}}^{\top}d_{21,i}A_{1,i_{1}}}{%
		n_{g}n_{g_{2}(i)}T} & =\sum_{g\in\mathcal{G}_{1}}\sum_{l\leq
		M_{g}}\sum_{i_{2}\in Q_{1,i_{g,l}}}\sum_{i_{1}=N_{g}+1}^{i_{g,l-1}}\frac{%
		A_{2,i_{2}}^{\top}D_{1,g}^{l}A_{1,i_{1}}}{n_{g}n_{g_{2}(i_{g,l})}T}  \notag
	\\
	& \text{ \ \ \ \ \ }+\sum_{g\in\mathcal{G}_{1}}\sum_{l\leq M_{g}}\sum
	_{i_{2}\in Q_{1,i_{g,l}}}\sum_{i_{1}=i_{g,l-1}+1}^{i_{g,l}-1}\frac{%
		A_{2,i_{2}}^{\top}D_{2,gi_{1}}^{l}A_{1,i_{1}}}{n_{g}n_{g_{2}(i_{g,l})}T}.
	\label{P_C1_MGCLT_L2_14}
\end{align}
For any $T\times1$ real vector $a$, (\ref{P_C1_MGCLT_L2_11}) and the
Cauchy-Schwarz inequality,%
\begin{align*}
	a^{\top}(D_{1,g}^{l})^{\top}D_{1,g}^{l}a &
	=\sum_{i_{1}^{\prime},i_{2}^{\prime}=i_{g,l-1}+1}^{i_{g,l}}a^{%
		\top}d_{12,i_{1}^{\prime}}d_{21,i_{2}^{\prime}}a\leq\left(
	\sum_{i^{\prime}=i_{g,l-1}+1}^{i_{g,l}}\left\Vert
	a^{\top}d_{12,i^{\prime}}\right\Vert \right) ^{2} \\
	& \leq\max_{i\leq n}\lambda_{\max}(d_{12,i}d_{21,i})\left( \sum_{i^{\prime
		}=i_{g,l-1}+1}^{i_{g,l}}(a^{\top}a)^{1/2}\right) ^{2}\leq K(a^{\top
	}a)(i_{g,l}-i_{g,l-1})^{2},
\end{align*}
which implies that%
\begin{equation}
	\max_{g\in\mathcal{G}_{1}}\max_{l\leq M_{g}}\max_{i\in I_{g,l}}\frac {%
		\lambda_{\max}((D_{1,g}^{l})^{\top}D_{1,g}^{l})}{(i_{g,l}-i_{g,l-1})^{2}}%
	\leq K.  \label{P_C1_MGCLT_L2_13a}
\end{equation}
Similarly, we can show that%
\begin{equation}
	\max_{g\in\mathcal{G}_{1}}\max_{l\leq M_{g}}\max_{i\in I_{g,l}}\frac {%
		\lambda_{\max}((D_{2,gi}^{l})^{\top}D_{2,gi}^{l})}{(i_{g,l}-i_{g,l-1})^{2}}%
	\leq K.  \label{P_C1_MGCLT_L2_13b}
\end{equation}
By the independence of $A_{1,i}$ and $A_{2,i}$ across $i$,%
\begin{align}
	& \mathrm{Var}\left( \sum_{g\in\mathcal{G}_{1}}\sum_{l\leq M_{g}}\sum
	_{i_{2}\in Q_{1,i_{g,l}}}\sum_{i_{1}=N_{g}+1}^{i_{g,l-1}}\frac{%
		A_{2,i_{2}}^{\top}D_{1,g}^{l}A_{1,i_{1}}}{n_{g}n_{g_{2}(i_{g,l})}T}\right) 
	\notag \\
	& =\sum_{g\in\mathcal{G}_{1}}\sum_{l\leq M_{g}}\sum_{i_{2}\in
		Q_{1,i_{g,l}}}\sum_{i_{1}=N_{g}+1}^{i_{g,l-1}}\frac{\mathbb{E}%
		[A_{1,i_{1}}^{\top}(D_{1,g}^{l})^{\top}\mathbb{E}[A_{2,i_{2}}A_{2,i_{2}}^{%
			\top}]D_{1,g}^{l}A_{1,i_{1}}]}{(n_{g}n_{g_{2}(i_{g,l})}T)^{2}}  \notag \\
	& \leq\max_{i\leq n}(\lambda_{\max}(\mathbb{E}[A_{j,i}A_{j,i}^{\top}]))^{2}%
	\sum_{g\in\mathcal{G}_{1}}\sum_{l\leq M_{g}}\sum_{i_{2}\in
		Q_{1,i_{g,l}}}\sum_{i_{1}=N_{g}+1}^{i_{g,l-1}}\frac{\lambda_{%
			\max}((D_{1,g}^{l})^{\top}D_{1,g}^{l})M_{1}}{(n_{g}n_{g_{2}(i_{g,l})}T)^{2}}
	\notag \\
	& \leq KM_{1}T^{-2}\sum_{g\in\mathcal{G}_{1}}\sum_{l\leq M_{g}}\frac {%
		(i_{g,l}-i_{g,l-1})^{2}(i_{g,l-1}-N_{g})(\#Q_{1,i_{g,l}})}{%
		(n_{g}n_{g_{2}(i_{g,l})})^{2}}  \notag \\
	& \leq KM_{1}T^{-2}\sum_{g\in\mathcal{G}_{1}}\sum_{l\leq M_{g}}\frac {%
		i_{g,l}-i_{g,l-1}}{n_{g}}\leq KG_{1}M_{1}T^{-2}\leq KT^{-1}
	\label{P_C1_MGCLT_L2_15}
\end{align}
where the $\#Q_{1,i_{g,l}}$ denotes the cardinality of $Q_{1,i_{g,l}}$, the
second inequality is by (\ref{P_C1_MGCLT_L1_6}) and (\ref{P_C1_MGCLT_L2_13a}%
), and the third inequality is by $n_{g}\geq i_{g,l-1}-N_{g}$ and $%
n_{g_{2}(i_{g,l})}\geq\#Q_{1,i_{g,l}}+i_{g,l}-i_{g,l-1}$ such that $%
(i_{g,l}-i_{g,l-1})(\#Q_{1,i_{g,l}})n_{g_{2}(i_{g,l})}^{-2}\leq1$, and the
fourth inequality is due to $\sum_{l\leq M_{g}}(i_{g,l}-i_{g,l-1})=n_{g}$,
and the last inequality is by $%
G_{1}M_{1}T^{-1}=G_{1}M_{1}(nT)^{-1/2}(n/T)^{1/2}\leq K$ where the
inequality holds by Assumptions \ref{A1}(i) and \ref{A5}(ii). Similarly, 
\begin{align}
	& \mathrm{Var}\left( \sum_{g\in\mathcal{G}_{1}}\sum_{l\leq M_{g}}\sum
	_{i_{2}\in Q_{1,i_{g,l}}}\sum_{i_{1}=i_{g,l-1}+1}^{i_{g,l}-1}\frac{%
		A_{2,i_{2}}^{\top}D_{2,gi_{1}}^{l}A_{1,i_{1}}}{n_{g}n_{g_{2}(i_{g,l})}T}%
	\right)  \notag \\
	& =\sum_{g\in\mathcal{G}_{1}}\sum_{l\leq M_{g}}\sum_{i_{2}\in
		Q_{1,i_{g,l}}}\sum_{i_{1}=i_{g,l-1}+1}^{i_{g,l}-1}\frac{\mathbb{E}%
		[A_{1,i_{1}}^{\top }(D_{2,gi_{1}}^{l})^{\top}\mathbb{E}%
		[A_{2,i_{2}}A_{2,i_{2}}^{\top}]D_{2,gi_{1}}^{l}A_{1,i_{1}}]}{%
		(n_{g}n_{g_{2}(i_{g,l})}T)^{2}}  \notag \\
	& \leq\max_{i\leq n}(\lambda_{\max}(\mathbb{E}[A_{j,i}A_{j,i}^{\top}]))^{2}%
	\sum_{g\in\mathcal{G}_{1}}\sum_{l\leq M_{g}}\sum_{i_{2}\in
		Q_{1,i_{g,l}}}\sum_{i_{1}=i_{g,l-1}+1}^{i_{g,l}-1}\frac{\lambda_{\max
		}((D_{2,gi_{1}}^{l})^{\top}D_{2,gi_{1}}^{l})M_{1}}{%
		(n_{g}n_{g_{2}(i_{g,l})}T)^{2}}  \notag \\
	& \leq KM_{1}T^{-2}\sum_{g\in\mathcal{G}_{1}}\sum_{l\leq M_{g}}\frac {%
		(i_{g,l}-i_{g,l-1})^{2}(i_{g,l-1}-N_{g})(\#Q_{1,i_{g,l}})}{%
		(n_{g}n_{g_{2}(i_{g,l})})^{2}}\leq KG_{1}M_{1}T^{-2}\leq KT^{-1}.
	\label{P_C1_MGCLT_L2_16}
\end{align}
Since $\sum_{g\in\mathcal{G}_{1}}\sum_{i\in I_{g}}\sum_{i_{2}\in
	Q_{1,i}}\sum_{i_{1}=N_{g}+1}^{i-1}\frac{\mathbb{E}[A_{2,i_{2}}^{%
		\top}d_{21,i}A_{1,i_{1}}]}{n_{g}n_{g_{2}(i)}T}=0$, by (\ref{P_C1_MGCLT_L2_14}%
), (\ref{P_C1_MGCLT_L2_15}), (\ref{P_C1_MGCLT_L2_16}) and Markov's
inequality, we have%
\begin{equation}
	\sum_{g\in\mathcal{G}_{1}}\sum_{i\in I_{g}}\sum_{i_{2}\in Q_{1,i}}\sum
	_{i_{1}=N_{g}+1}^{i-1}\frac{A_{2,i_{2}}^{\top}d_{21,i}A_{1,i_{1}}}{%
		n_{g}n_{g_{2}(i)}T}=O_{p}(T^{-1/2}).  \label{P_C1_MGCLT_L2_17}
\end{equation}

We next study the second term in the RHS of (\ref{P_C1_MGCLT_L2_10}). For
any\ $i\in I_{g,l}$, by the arrangement in $I_{g}$, it is clear that $%
Q_{2,i}=\{i_{g,l-1}+1,\ldots ,i-1\}$ and hence 
\begin{equation}
	\sum_{i_{2}\in Q_{2,i}}\sum_{i_{1}=N_{g}+1}^{i-1}\frac{A_{2,i_{2}}^{\top
		}d_{21,i}A_{1,i_{1}}}{n_{g}n_{g_{2}(i)}T}=\sum_{i_{2}=i_{g,l-1}+1}^{i-1}%
	\sum_{i_{1}=N_{g}+1}^{i_{g,l-1}}\frac{A_{2,i_{2}}^{\top }d_{21,i}A_{1,i_{1}}%
	}{n_{g}n_{g_{2}(i_{g,l})}T}+\sum_{i_{1},i_{2}=i_{g,l-1}+1}^{i-1}\frac{%
		A_{2,i_{2}}^{\top }d_{21,i}A_{1,i_{1}}}{n_{g}n_{g_{2}(i_{g,l})}T}.
	\label{P_C1_MGCLT_L2_18}
\end{equation}%
Since%
\begin{align}
	\sum_{i\in
		I_{g,l}}\sum_{i_{2}=i_{g,l-1}+1}^{i-1}\sum_{i_{1}=N_{g}+1}^{i_{g,l-1}}\frac{%
		A_{1,i_{1}}^{\top }d_{12,i}A_{2,i_{2}}}{n_{g}n_{g_{2}(i_{g,l})}T}&
	=\sum_{i_{1}=N_{g}+1}^{i_{g,l-1}}\frac{A_{1,i_{1}}^{\top }}{%
		n_{g}n_{g_{2}(i_{g,l})}T}\sum_{i=i_{g,l-1}+2}^{i_{g,l}}d_{12,i}%
	\sum_{i_{2}=i_{g,l-1}+1}^{i-1}A_{2,i_{2}}  \notag \\
	& =\sum_{i_{1}=N_{g}+1}^{i_{g,l-1}}\sum_{i_{2}=i_{g,l-1}+1}^{i_{g,l}-1}\frac{%
		A_{1,i_{1}}^{\top }(D_{2,gi_{2}}^{l})^{\top }A_{2,i_{2}}}{%
		n_{g}n_{g_{2}(i_{g,l})}T},  \label{P_C1_MGCLT_L2_19}
\end{align}%
by the similar arguments for showing (\ref{P_C1_MGCLT_L2_15}), we get%
\begin{align}
	& \mathrm{Var}\left( \sum_{g\in \mathcal{G}_{1}}\sum_{l\leq M_{g}}\sum_{i\in
		I_{g,l}}\sum_{i_{2}=i_{g,l-1}+1}^{i-1}\sum_{i_{1}=N_{g}+1}^{i_{g,l-1}}\frac{%
		A_{1,i_{1}}^{\top }d_{12,i}A_{2,i_{2}}}{n_{g}n_{g_{2}(i_{g,l})}T}\right) 
	\notag \\
	& =\sum_{g\in \mathcal{G}_{1}}\sum_{l\leq
		M_{g}}\sum_{i_{1}=N_{g}+1}^{i_{g,l-1}}\sum_{i_{2}=i_{g,l-1}+1}^{i_{g,l}-1}%
	\frac{\mathbb{E}[A_{1,i_{1}}^{\top }(D_{2,gi_{2}}^{l})^{\top }\mathbb{E}%
		[A_{2,i_{2}}A_{2,i_{2}}^{\top }]D_{2,gi_{2}}^{l}A_{1,i_{1}}]}{%
		(n_{g}n_{g_{2}(i_{g,l})}T)^{2}}  \notag \\
	& \leq \max_{i\leq n}(\lambda _{\max }(\mathbb{E}[A_{j,i}A_{j,i}^{\top
	}]))^{2}\sum_{g\in \mathcal{G}_{1}}\sum_{l\leq
		M_{g}}\sum_{i_{1}=N_{g}+1}^{i_{g,l-1}}\sum_{i_{2}=i_{g,l-1}+1}^{i_{g,l}-1}%
	\frac{\lambda _{\max }((D_{2,gi_{2}}^{l})^{\top }D_{2,gi_{2}}^{l})M_{1}}{%
		(n_{g}n_{g_{2}(i_{g,l})}T)^{2}}  \notag \\
	& \leq KM_{1}T^{-2}\sum_{g\in \mathcal{G}_{1}}\sum_{l\leq M_{g}}\frac{%
		(i_{g,l}-i_{g,l-1})^{3}(i_{g,l-1}-N_{g})}{(n_{g}n_{g_{2}(i_{g,l})})^{2}} 
	\notag \\
	& \leq KM_{1}T^{-2}\sum_{g\in \mathcal{G}_{1}}\sum_{l\leq M_{g}}\frac{%
		i_{g,l}-i_{g,l-1}}{n_{g}}\leq KG_{1}M_{1}T^{-2}\leq KT^{-1}.
	\label{P_C1_MGCLT_L2_20}
\end{align}%
Next note that%
\begin{align}
	& \sum_{i\in I_{g,l}}\sum_{i_{1},i_{2}=i_{g,l-1}+1}^{i-1}\frac{%
		A_{2,i_{2}}^{\top }d_{21,i}A_{1,i_{1}}}{n_{g}n_{g_{2}(i)}T}  \notag \\
	& =\sum_{i\in I_{g,l}}\sum_{i_{1}=i_{g,l-1}+1}^{i-1}\frac{A_{2,i_{1}}^{\top
		}d_{21,i}A_{1,i_{1}}}{n_{g}n_{g_{2}(i_{g,l})}T}+2\sum_{i\in
		I_{g,l}}\sum_{i_{1}=i_{g,l-1}+2}^{i-1}\sum_{i_{2}=i_{g,l-1}+1}^{i_{1}-1}%
	\frac{A_{2,i_{2}}^{\top }d_{21,i}A_{1,i_{1}}}{n_{g}n_{g_{2}(i_{g,l})}T} 
	\notag \\
	& =\sum_{i=i_{g,l-1}+1}^{i_{g,l}-1}\frac{A_{2,i}^{\top }D_{2,gi}^{l}A_{1,i}}{%
		n_{g}n_{g_{2}(i_{g,l})}T}+2\sum_{i_{1}=i_{g,l-1}+2}^{i_{g,l}-1}%
	\sum_{i_{2}=i_{g,l-1}+1}^{i_{1}-1}\frac{A_{2,i_{2}}^{\top
		}D_{2,gi_{1}}^{l}A_{1,i_{1}}}{n_{g}n_{g_{2}(i_{g,l})}T}.
	\label{P_C1_MGCLT_L2_21}
\end{align}%
By the independence of $A_{1\gamma ,i}$ and $A_{2\gamma ,i}$ across $i$,%
\begin{align}
	& \mathrm{Var}\left( \sum_{g\in \mathcal{G}_{1}}\sum_{l\leq
		M_{g}}\sum_{i=i_{g,l-1}+1}^{i_{g,l}-1}\frac{A_{2,i}^{\top
		}D_{2,gi}^{l}A_{1,i}}{n_{g}n_{g_{2}(i_{g,l})}T}\right)  \notag \\
	& =\sum_{g\in \mathcal{G}_{1}}\sum_{l\leq
		M_{g}}\sum_{i=i_{g,l-1}+1}^{i_{g,l}-1}\frac{\mathrm{Var}(A_{2,i}^{\top
		}D_{2,gi}^{l}A_{1,i})}{(n_{g}n_{g_{2}(i_{g,l})}T)^{2}}\leq \sum_{g\in 
		\mathcal{G}_{1}}\sum_{l\leq M_{g}}\sum_{i=i_{g,l-1}+1}^{i_{g,l}-1}\frac{%
		\mathbb{E}[(A_{2,i}^{\top }D_{2,gi}^{l}A_{1,i})^{2}]}{%
		(n_{g}n_{g_{2}(i_{g,l})}T)^{2}}  \notag \\
	& \leq \sum_{g\in \mathcal{G}_{1}}\sum_{l\leq
		M_{g}}\sum_{i=i_{g,l-1}+1}^{i_{g,l}-1}\frac{\mathbb{E}[(A_{1,i}^{\top
		}(D_{2,gi}^{l})^{\top }D_{2,gi}^{l}A_{1,i})(A_{2,i}^{\top }A_{2,i})]}{%
		(n_{g}n_{g_{2}(i_{g,l})}T)^{2}}  \notag \\
	& \leq \sum_{g\in \mathcal{G}_{1}}\sum_{l\leq
		M_{g}}\sum_{i=i_{g,l-1}+1}^{i_{g,l}-1}\frac{\lambda _{\max
		}((D_{2,gi}^{l})^{\top }D_{2,gi}^{l})(\mathbb{E}[||A_{1,i}||^{4}]\mathbb{E}%
		[||A_{2,i}||^{4}])^{1/2}}{(n_{g}n_{g_{2}(i_{g,l})}T)^{2}}  \notag \\
	& \leq K\sum_{g\in \mathcal{G}_{1}}\sum_{l\leq M_{g}}\frac{%
		(i_{g,l}-i_{g,l-1})^{3}M_{1}M_{2}}{(n_{g}n_{g_{2}(i_{g,l})}T)^{2}}\leq
	KM_{1}M_{2}T^{-2}\sum_{g\in \mathcal{G}_{1}}n_{g}^{-1}\leq
	KG_{2}^{-1}\max_{g\in \mathcal{G}_{1}}n_{g}^{-1},  \label{P_C1_MGCLT_L2_22}
\end{align}%
where the second inequality is by the Cauchy-Schwarz inequality, the third
inequality is by H\"{o}lder's inequality, the fourth inequality is by\ (\ref%
{P_C1_MGCLT_L2_5}) and\ (\ref{P_C1_MGCLT_L2_13b}), and the last inequality
is by Assumptions \ref{A1}(i) and \ref{A5}(ii). By the similar arguments for
showing (\ref{P_C1_MGCLT_L2_15}), we also have 
\begin{align}
	& \mathrm{Var}\left( \sum_{g\in \mathcal{G}_{1}}\sum_{l\leq
		M_{g}}\sum_{i_{1}=i_{g,l-1}+2}^{i_{g,l}-1}\sum_{i_{2}=i_{g,l-1}+1}^{i_{1}-1}%
	\frac{A_{2,i_{2}}^{\top }D_{2,gi_{1}}^{l}A_{1,i_{1}}}{%
		n_{g}n_{g_{2}(i_{g,l})}T}\right)  \notag \\
	& =\sum_{g\in \mathcal{G}_{1}}\sum_{l\leq
		M_{g}}\sum_{i_{1}=i_{g,l-1}+2}^{i_{g,l}-1}\sum_{i_{2}=i_{g,l-1}+1}^{i_{1}-1}%
	\frac{\mathrm{Var}(A_{2,i_{2}}^{\top }D_{2,gi_{1}}^{l}A_{1,i_{1}})}{%
		(n_{g}n_{g_{2}(i_{g,l})}T)^{2}}  \notag \\
	& =\sum_{g\in \mathcal{G}_{1}}\sum_{l\leq
		M_{g}}\sum_{i_{1}=i_{g,l-1}+2}^{i_{g,l}-1}\sum_{i_{2}=i_{g,l-1}+1}^{i_{1}-1}%
	\frac{\mathbb{E}[A_{1,i_{1}}^{\top }(D_{2,gi_{1}}^{l})^{\top }\mathbb{E}%
		[A_{2,i_{2}}A_{2,i_{2}}^{\top }]D_{2,gi_{1}}^{l}A_{1,i_{1}}]}{%
		(n_{g}n_{g_{2}(i_{g,l})}T)^{2}}  \notag \\
	& \leq \max_{i\leq n}(\lambda _{\max }(\mathbb{E}[A_{j,i}A_{j,i}^{\top
	}]))^{2}\sum_{g\in \mathcal{G}_{1}}\sum_{l\leq
		M_{g}}\sum_{i_{1}=i_{g,l-1}+2}^{i_{g,l}-1}\sum_{i_{2}=i_{g,l-1}+1}^{i_{1}-1}%
	\frac{\lambda _{\max }((D_{2,gi_{1}}^{l})^{\top }D_{2,gi_{1}}^{l})M_{1}}{%
		(n_{g}n_{g_{2}(i_{g,l})}T)^{2}}  \notag \\
	& \leq KM_{1}T^{-2}\sum_{g\in \mathcal{G}_{1}}\sum_{l\leq M_{g}}\frac{%
		(i_{g,l}-i_{g,l-1})^{4}}{(n_{g}n_{g_{2}(i_{g,l})})^{2}}\leq
	KG_{1}M_{1}T^{-2}\leq KT^{-1}.  \label{P_C1_MGCLT_L2_23}
\end{align}%
Collecting the results in\ (\ref{P_C1_MGCLT_L2_18})-(\ref{P_C1_MGCLT_L2_23}%
), we deduce that%
\begin{align*}
	\sum_{g\in \mathcal{G}_{1}}\sum_{i\in I_{g}}\sum_{i_{2}\in
		Q_{2,i}}\sum_{i_{1}=N_{g}+1}^{i-1}\frac{A_{2,i_{2}}^{\top
		}d_{21,i}A_{1,i_{1}}}{n_{g}n_{g_{2}(i)}T}& =\sum_{g\in \mathcal{G}%
		_{1}}\sum_{l\leq M_{g}}\sum_{i=i_{g,l-1}+1}^{i_{g,l}-1}\frac{\mathbb{E}%
		[A_{2,i}^{\top }D_{2,gi}^{l}A_{1,i}]}{n_{g}n_{g_{2}(i_{g,l})}T} \\
	& \text{ \ \ \ \ \ }+O_{p}(T^{-1/2}+G_{2}^{-1/2}\max_{g\in \mathcal{G}%
		_{1}}n_{g}^{-1/2}),
\end{align*}%
which together with (\ref{P_C1_MGCLT_L2_10}) and (\ref{P_C1_MGCLT_L2_17})
implies%
\begin{equation}
	\sum_{i\leq n}(\mathbb{E}[\tilde{U}_{1,i}\tilde{U}_{2,i}|\mathcal{F}%
	_{nT,i-1}]-\mathbb{E}[\tilde{U}_{1,i}\tilde{U}%
	_{2,i}])=O_{p}(T^{-1/2}+G_{2}^{-1/2}\max_{g\in \mathcal{G}_{1}}n_{g}^{-1/2}).
	\label{P_C1_MGCLT_L2_24}
\end{equation}%
We can switch the role of $\mathcal{G}_{1}$ and $\mathcal{G}_{2}$, and use
the same argument above to show that 
\begin{equation}
	\sum_{i\leq n}(\mathbb{E}[\tilde{U}_{1,i}\tilde{U}_{2,i}|\mathcal{F}%
	_{nT,i-1}]-\mathbb{E}[\tilde{U}_{1,i}\tilde{U}%
	_{2,i}])=O_{p}(T^{-1/2}+G_{1}^{-1/2}\max_{g\in \mathcal{G}_{2}}n_{g}^{-1/2}).
	\label{P_C1_MGCLT_L2_25}
\end{equation}%
Since 
\begin{align}
	& \min \left\{ G_{1}^{-1}\max_{g\in \mathcal{G}_{2}}n_{g}^{-1},\text{\ }%
	G_{2}^{-1}\max_{g\in \mathcal{G}_{1}}n_{g}^{-1}\right\} \leq \left(
	G_{1}^{-1/2}\max_{g\in \mathcal{G}_{2}}n_{g}^{-1/2}\right) \left(
	G_{2}^{-1/2}\max_{g\in \mathcal{G}_{1}}n_{g}^{-1/2}\right)  \notag \\
	& \leq \frac{G_{1}^{-1}\max_{g\in \mathcal{G}_{1}}n_{g}^{-1}+G_{2}^{-1}%
		\max_{g\in \mathcal{G}_{2}}n_{g}^{-1}}{2}\leq \max_{j=1,2}\left(
	G_{j}^{-1}\max_{g\in \mathcal{G}_{j}}n_{g}^{-1}\right) ,
	\label{P_C1_MGCLT_L2_26}
\end{align}%
the desired result in (\ref{P_C1_MGCLT_L2_2}) follows from (\ref%
{P_C1_MGCLT_L2_24}), (\ref{P_C1_MGCLT_L2_25}) and (\ref{P_C1_MGCLT_L2_26}%
).\hfill $Q.E.D.$

\bigskip

\begin{lemma}
	\textit{\label{C2_MGCLT_L1}\ Under the conditions of Theorem }\ref{MGCLT_V},
	we have for $j=1,2$%
	\begin{equation}
		\max_{g\in\mathcal{G}_{j}}\max_{i\in I_{g}}\mathbb{E}\left[ \left\vert
		M_{j}^{-1/2}(n_{g}T)^{1/2}\tilde{U}_{j,i}\right\vert ^{4}\right] \leq K.
		\label{C2_MGCLT_L1_1}
	\end{equation}
\end{lemma}

\noindent\textsc{Proof of Lemma \ref{C2_MGCLT_L1}}.\ We only prove (\ref%
{C2_MGCLT_L1_1}) for $j=1$, since the proof for $j=2$ follows the same
arguments. We shall use the structure of $\mathcal{G}_{1}$ employed in
verifying (\ref{P_C1_MGCLT_L2_1}) for $j=1$ in the proof of Lemma \ref%
{C1_MGCLT_L2}. That is for any $g\in\mathcal{G}_{1}$, $I_{g}=\{N_{g}+1,%
\ldots,N_{g}+n_{g}\}$ where $N_{g}\equiv\sum_{g^{\prime}<g}n_{g^{\prime}}$.
Following these notations, we can write%
\begin{equation*}
	\tilde{U}_{1,i}=(n_{g}T^{1/2})^{-1}\sum_{i^{\prime}=N_{g}+1}^{i-1}A_{1,i}^{%
		\top}A_{1,i^{\prime}}
\end{equation*}
for any $i\in I_{g}$ and any $g\in\mathcal{G}_{1}$. Since $\{A_{j,i}^{\top
}A_{j,i^{\prime}}\}_{i^{\prime}=1}^{i-1}$ are independent conditioning on $%
A_{j,i}$, we can use Rosenthal's inequality of independent random variables
to obtain%
\begin{align}
	& \mathbb{E}[(n_{g}T^{1/2}\tilde{U}_{1,i})^{4}|A_{1,i}]=\mathbb{E}\left[
	\left. \left(
	\sum_{i^{\prime}=N_{g}+1}^{i-1}A_{1,i}^{\top}A_{1,i^{\prime}}\right)
	^{4}\right\vert A_{1,i}\right]  \notag \\
	& \text{ \ \ \ \ \ \ \ \ \ \ \ \ \ \ \ \ \ \ \ }\leq K\left( \sum
	_{i^{\prime}=N_{g}+1}^{i-1}\mathbb{E}\left[ \left. (A_{1,i}^{\top
	}A_{1,i^{\prime}})^{2}\right\vert A_{1,i}\right] \right) ^{2}+K\sum
	_{i^{\prime}=N_{g}+1}^{i-1}\mathbb{E}\left[ \left. (A_{1,i}^{\top
	}A_{1,i^{\prime}})^{4}\right\vert A_{1,i}\right] .  \label{P_C2_MGCLT_L1_1}
\end{align}
By the independence of $A_{j,i}$ and $A_{j,i^{\prime}}$ for any $i\neq
i^{\prime}$,%
\begin{align}
	& \sum_{i^{\prime}=N_{g}+1}^{i-1}\mathbb{E}\left[ \left. (A_{1,i}^{\top
	}A_{1,i^{\prime}})^{2}\right\vert A_{1,i}\right] \overset{}{=}A_{1,i}^{\top
	}\left( \sum_{i^{\prime}=N_{g}+1}^{i-1}\mathbb{E}[A_{1,i^{\prime}}A_{1,i^{%
			\prime}}^{\top}]\right) A_{1,i}  \notag \\
	& \text{ \ \ \ \ \ \ \ \ \ \ \ \ \ \ \ \ \ \ \ \ \ \ \ \ \ \ \ \ \ \ \ \ \ \ 
	}\overset{}{\leq}(A_{1,i}^{\top}A_{1,i})\sum_{i^{\prime}=N_{g}+1}^{i-1}%
	\lambda_{\max}(\mathbb{E}[A_{1,i^{\prime}}A_{1,i^{\prime}}^{\top}])\text{ }%
	\overset{}{\leq}Kn_{g}(A_{1,i}^{\top}A_{1,i}),  \label{P_C2_MGCLT_L1_2}
\end{align}
where the second inequality is by\ (\ref{P_C1_MGCLT_L1_6}). Therefore, 
\begin{equation}
	\max_{g\in\mathcal{G}_{1}}\max_{i\in I_{g}}\mathbb{E}\left[ \left(
	n_{g}^{-1}\sum_{i^{\prime}=N_{g}+1}^{i-1}\mathbb{E}\left[ \left.
	(A_{1,i}^{\top}A_{1,i^{\prime}})^{2}\right\vert A_{1,i}\right] \right) ^{2}%
	\right] \leq K\max_{i\leq n}\mathbb{E}[(A_{1,i}^{\top}A_{1,i})^{2}]\leq
	KM_{1}^{2},  \label{P_C2_MGCLT_L1_3}
\end{equation}
where the second inequality is by\ (\ref{P_C1_MGCLT_L2_5}).

For any $i,i^{\prime}=1,\ldots,n$ with $i\neq i^{\prime}$, by the
independence between $A_{j,i}$ and $A_{j,i^{\prime}}$, and the triangle
inequality%
\begin{align}
	\mathbb{E}\left[ (A_{1,i}^{\top}A_{1,i^{\prime}})^{4}\right] & =\mathbb{E}%
	\left[ \left\vert \sum_{m\in\mathcal{M}_{1}}\tilde{\Psi}_{1\gamma,im}\tilde{%
		\Psi}_{1\gamma,i^{\prime}m}\right\vert ^{4}\right]  \notag \\
	& \leq\sum_{m_{1},m_{2},m_{3},m_{4}}|\mathbb{E}[\tilde{\Psi}%
	_{1\gamma,im_{1}}\cdots\tilde{\Psi}_{1\gamma,im_{4}}\tilde{\Psi}%
	_{1\gamma,i^{\prime}m_{1}}\cdots\tilde{\Psi}_{1\gamma,i^{\prime}m_{4}}]| 
	\notag \\
	& \leq\max_{m_{1},m_{2},m_{3},m_{4}}|\mathbb{E}[\tilde{\Psi}_{1\gamma
		,i^{\prime}m_{1}}\cdots\tilde{\Psi}_{1\gamma,i^{\prime}m_{4}}]|\sum
	_{m_{1},m_{2},m_{3},m_{4}}|\mathbb{E}[\tilde{\Psi}_{1\gamma,im_{1}}\cdots%
	\tilde{\Psi}_{1\gamma,im_{4}}]|  \notag \\
	& \leq\max_{i\leq n}\max_{m\in\mathcal{M}_{1}}\mathbb{E}[|\tilde{\Psi }%
	_{1\gamma,im}|^{4}]\sum_{m_{1},m_{2},m_{3},m_{4}}|\mathbb{E}[\tilde{\Psi }%
	_{1\gamma,im_{1}}\tilde{\Psi}_{1\gamma,im_{2}}\tilde{\Psi}_{1\gamma,im_{3}}%
	\tilde{\Psi}_{1\gamma,im_{4}}]|  \notag \\
	& \leq K\sum_{m_{1},m_{2},m_{3},m_{4}}|\mathbb{E}[\tilde{\Psi}_{1\gamma
		,im_{1}}\tilde{\Psi}_{1\gamma,im_{2}}\tilde{\Psi}_{1\gamma,im_{3}}\tilde{%
		\Psi }_{1\gamma,im_{4}}]|,  \label{P_C2_MGCLT_L1_4}
\end{align}
where the third inequality is by H\"{o}lder's inequality and the last
inequality is by\ (\ref{P_C1_MGCLT_L1_4}). Consider any $%
m_{1},m_{2},m_{3},m_{4}$ from $\mathcal{M}_{1}$ with $m_{1}\leq\ldots\leq
m_{4}$. Let $r\equiv\max_{j=1,2,3}\{m_{j+1}-m_{j}\}$ and $%
j_{r}\equiv\min\{j=1,2,3:m_{j+1}-m_{j}=r\}$. It is clear that $r$ ranges
from $0$ to $M_{1}-1$. When $r=0$, we have $m_{1}=\ldots=m_{4}=m$ for some $%
m\in\mathcal{M}_{1}$\ and in this case%
\begin{equation}
	|\mathbb{E}[\tilde{\Psi}_{1\gamma,im_{1}}\tilde{\Psi}_{1\gamma,im_{2}}\tilde{%
		\Psi}_{1\gamma,im_{3}}\tilde{\Psi}_{1\gamma,im_{4}}]|=\mathbb{E}[\tilde{\Psi}%
	_{1\gamma,im}^{4}]\leq K,  \label{P_C2_MGCLT_L1_5}
\end{equation}
where the inequality is by (\ref{P_C1_MGCLT_L1_4}). We next consider the
case that $r=1,\ldots,M_{1}-1$. By H\"{o}lder's inequality and Corollary
A.2. in \cite{HallHeyde1980}, 
\begin{align}
	& \left\vert \mathbb{E}[\tilde{\Psi}_{1\gamma,im_{1}}\tilde{\Psi}%
	_{1\gamma,im_{2}}\tilde{\Psi}_{1\gamma,im_{3}}\tilde{\Psi}_{1\gamma,im_{4}}]-%
	\mathbb{E}[\tilde{\Psi}_{1\gamma,im_{1}}\cdots\tilde{\Psi}_{1\gamma
		,im_{j_{r}}}]\mathbb{E}[\tilde{\Psi}_{1\gamma,im_{j_{r}+1}}\cdots\tilde{\Psi 
	}_{1\gamma,im_{4}}]\right\vert  \notag \\
	& \leq(\alpha_{i}(r))^{\delta/(4+\delta)}\left\Vert \tilde{\Psi}%
	_{1\gamma,im_{1}}\cdots\tilde{\Psi}_{1\gamma,im_{j_{r}}}\right\Vert
	_{(4+\delta)/j_{r}}\left\Vert \tilde{\Psi}_{1\gamma,im_{j_{r}+1}}\cdots 
	\tilde{\Psi}_{1\gamma,im_{4}}\right\Vert _{(4+\delta)/(4-j_{r})}  \notag \\
	& \leq(\alpha_{i}(r))^{\delta/(4+\delta)}\max_{m\in\mathcal{M}%
		_{1}}\left\Vert \tilde{\Psi}_{1\gamma,im}\right\Vert _{(4+\delta)}^{4}\leq
	K(\alpha _{i}(r))^{\delta/(4+\delta)},  \label{P_C2_MGCLT_L1_6}
\end{align}
where the last inequality is by (\ref{P_C1_MGCLT_L1_4}). Since $\mathbb{E}[%
\tilde{\Psi}_{1\gamma,im}]=0$, by (\ref{P_C2_MGCLT_L1_6}) we have 
\begin{equation}
	\left\vert \mathbb{E}[\tilde{\Psi}_{1\gamma,im_{1}}\tilde{\Psi}_{1\gamma
		,im_{2}}\tilde{\Psi}_{1\gamma,im_{3}}\tilde{\Psi}_{1\gamma,im_{4}}]\right%
	\vert \leq K(\alpha_{i}(r))^{\delta/(4+\delta)}  \label{P_C2_MGCLT_L1_7}
\end{equation}
if $j_{r}=1$ or $3$. On the other hand if $j_{r}=2$, then 
\begin{equation}
	\left\vert \mathbb{E}[\tilde{\Psi}_{1\gamma,im_{1}}\cdots\tilde{\Psi}%
	_{1\gamma,im_{4}}]\right\vert \leq K(\alpha_{i}(r))^{\delta/(4+\delta
		)}+\left\vert \mathbb{E}[\tilde{\Psi}_{1\gamma,im_{1}}\tilde{\Psi}%
	_{1\gamma,im_{2}}]\mathbb{E}[\tilde{\Psi}_{1\gamma,im_{3}}\tilde{\Psi }%
	_{1\gamma,im_{4}}]\right\vert .  \label{P_C2_MGCLT_L1_8}
\end{equation}
Since for each $r_{0}=1,\ldots,M_{1}-1$, there are at most $%
3(M_{1}-r_{0})(1+r_{0})^{2}$ of indices $m_{1},m_{2},m_{3},m_{4}$ satisfying 
$1\leq m_{1}\leq\ldots\leq m_{4}\leq M_{1}$ such that $\max_{j=1,2,3}%
\{m_{j+1}-m_{j}\}=r_{0}$. From (\ref{P_C2_MGCLT_L1_5}), (\ref%
{P_C2_MGCLT_L1_6}) and (\ref{P_C2_MGCLT_L1_7}) we deduce that%
\begin{align}
	& \sum_{m_{1}\leq\ldots\leq m_{4}}\left\vert \mathbb{E}[\tilde{\Psi}%
	_{1\gamma,im_{1}}\tilde{\Psi}_{1\gamma,im_{2}}\tilde{\Psi}_{1\gamma,im_{3}}%
	\tilde{\Psi}_{1\gamma,im_{4}}]\right\vert  \notag \\
	& \leq\sum_{m\in\mathcal{M}_{1}}\mathbb{E}[\tilde{\Psi}_{1\gamma,im}^{4}]+K%
	\sum_{r_{0}=1}^{M_{1}-1}(\alpha_{i}(r_{0}))^{\delta/(4+%
		\delta)}(M_{1}-r_{0})(1+r_{0})^{2}  \notag \\
	& \text{ \ \ \ \ \ \ }+\sum_{m_{1}\leq\ldots\leq m_{4}}\left\vert \mathbb{E}[%
	\tilde{\Psi}_{1\gamma,im_{1}}\tilde{\Psi}_{1\gamma,im_{2}}]\mathbb{E}[\tilde{%
		\Psi}_{1\gamma,im_{3}}\tilde{\Psi}_{1\gamma,im_{4}}]\right\vert  \notag \\
	& \leq KM_{1}+KM_{1}\sum_{r_{0}=1}^{M_{1}-1}(\alpha_{i}(r_{0}))^{\delta
		/(4+\delta)}r_{0}^{2}+\left( \sum_{m_{1},m_{2}\in\mathcal{M}_{1}}\left\vert 
	\mathbb{E}[\tilde{\Psi}_{1\gamma,im_{1}}\tilde{\Psi}_{1\gamma,im_{2}}]\right%
	\vert \right) ^{2}  \notag \\
	& \leq KM_{1}+KM_{1}^{2}\leq KM_{1}^{2},  \label{P_C2_MGCLT_L1_9}
\end{align}
where the third inequality is by Assumption \ref{A1}(iv) and (\ref%
{P_C1_MGCLT_L1_5}). Since for any $m_{1},m_{2},m_{3},m_{4}\in \mathcal{M}%
_{1} $ there are at most 24 ways to order them from the smallest to the
largest, by (\ref{P_C2_MGCLT_L1_4}) and (\ref{P_C2_MGCLT_L1_9}), we deduce
that $\mathbb{E}[(A_{1,i}^{\top}A_{1,i^{\prime}})^{4}]\leq KM_{1}^{2}$ for
any $i\neq i^{\prime}$, which implies that%
\begin{equation}
	\max_{i\leq n}n_{g}^{-1}\sum_{i^{\prime}=N_{g}+1}^{i-1}\mathbb{E}\left[
	(M_{1}^{-1/2}A_{1,i}^{\top}A_{1,i^{\prime}})^{4}\right] \leq K.
	\label{P_C2_MGCLT_L1_10}
\end{equation}
Collecting the results in (\ref{P_C2_MGCLT_L1_1}), (\ref{P_C2_MGCLT_L1_3}), (%
\ref{P_C2_MGCLT_L1_10}), we obtain%
\begin{align*}
	\max_{i\in I_{g}}\mathbb{E}[|M_{1}^{-1/2}(n_{g}T)^{1/2}\tilde{U}_{1,i}|^{4}]
	& =\max_{g\in\mathcal{G}_{1}}\max_{i\in I_{g}}\mathbb{E}\left[ n_{g}^{-2}%
	\mathbb{E}[(M_{1}^{-1/2}n_{g}T^{1/2}\tilde{U}_{1,i})^{4}|A_{1,i}]\right] \\
	& \leq K\max_{g\in\mathcal{G}_{1}}\max_{i\in I_{g}}\mathbb{E}\left[ \left(
	n_{g}^{-1}\sum_{i^{\prime}=N_{g}+1}^{i-1}\mathbb{E}\left[ \left.
	(M_{1}^{-1/2}A_{1,i}^{\top}A_{1,i^{\prime}})^{2}\right\vert A_{1,i}\right]
	\right) ^{2}\right] \\
	& \text{ \ \ \ \ \ \ \ }+K\max_{g\in\mathcal{G}_{1}}\max_{i\in I_{g}}\left(
	n_{g}^{-2}\sum_{i^{\prime}=N_{g}+1}^{i-1}\mathbb{E}\left[
	(M_{1}^{-1/2}A_{1,i}^{\top}A_{1,i^{\prime}})^{4}\right] \right) \overset{}{%
		\leq}K,
\end{align*}
which shows (\ref{C2_MGCLT_L1_1}) for $j=1$.\hfill$Q.E.D.$

\section{Further Examples of Linear Models \label{sec:linear-section}}

In this section, we provide several examples from the linear panel model%
\begin{equation*}
	y_{i,t}=x_{i,t}^{\top }\beta +u_{i,t}+\varepsilon _{i,t},
\end{equation*}%
where $y_{i,t}$ is the dependent variable, $x_{i,t}$ are the observed
regressors, $u_{i,t}$ is a unobservable term which may include individual
and/or time fixed effects, and $\varepsilon _{i,t}$ is i.i.d. across $i$ and 
$t$ and normally distributed with mean zero and variance $\sigma ^{2}$. The
data are $\{z_{i,t}\}_{i\leq n,t\leq T}$, where $z_{it}=(y_{i,t},x_{i,t}^{%
	\top })^{\top }$. As illustrated in the examples below, $u_{i,t}$ takes
different forms under various model specifications.

\paragraph{Example 1. (Time Invariant Fixed Effect)}

In this model, $u_{i,t}$ includes a time invariant fixed effect, i.e., $%
u_{i,t}=\gamma _{i}$. Since there is a group structure in $t$ and no group
structure in $i$, we have $\mathcal{G}=\{1,\ldots ,n\}$, $I_{g}=\{g\}$ for
any $g\in \mathcal{G}$, $\mathcal{M}=\{1\}$ and $I_{m}=\{1,\ldots ,T\}$ for $%
m\in \mathcal{M}$.\ Therefore $\gamma _{g,1}=\gamma _{g}\ $for any $g\in 
\mathcal{G}$, and the log-likelihood function for any $i=g$ and any $t=m$ is%
\begin{equation*}
	\log \psi \left( z_{it};\theta ,\gamma _{g,m}\right) =-\frac{\log (\sigma
		^{2})}{2}-\frac{1}{2\sigma ^{2}}\left( y_{i,t}-x_{i,t}^{\top }\beta -\gamma
	_{g}\right) ^{2},
\end{equation*}%
which leads to the joint-likelihood function%
\begin{equation*}
	\sum_{g\in \mathcal{G}}\sum_{i\in I_{g}}\sum_{m\in \mathcal{M}}\sum_{t\in
		I_{m}}\psi \left( z_{it};\theta ,\gamma _{g,m}\right) =-\frac{nT\log (\sigma
		^{2})}{2}-\frac{1}{2\sigma ^{2}}\sum_{t\leq T}\sum_{i\leq n}\left(
	y_{i,t}-x_{i,t}^{\top }\theta -\gamma _{i}\right) ^{2},
\end{equation*}%
where $\theta =(\beta ^{\top },\sigma ^{2})^{\top }$.

\paragraph{Example 2. (Time Varying Fixed Effect)}

Consider the time variate fixed effect model where $u_{i,t}=\gamma _{i,1}$
for $t=1,\ldots ,T_{0}$ and $u_{i,t}=\gamma _{i,2}$ for $t=T_{0}+1,\ldots ,T$%
. In this case, there is no group structure along $i$ but a group structure
on $t$. Hence we have $\mathcal{G}=\{1,\ldots ,n\}$, $I_{g}=\{g\}$ for any $%
g\in \mathcal{G}$, $\mathcal{M}=\{1,2\}$ and%
\begin{equation*}
	I_{m}=\left\{ 
	\begin{array}{cc}
		\{1,\ldots ,T_{0}\}, & \text{if }m=1 \\ 
		\{T_{0}+1,\ldots ,T\}, & \text{if }m=2%
	\end{array}%
	\right. .
\end{equation*}%
Therefore the log-likelihood function for any $i=g$ and any $t\in I_{m}$ is%
\begin{equation*}
	\log f\left( z_{it};\theta ,\gamma _{g,m}\right) =-\frac{\log (\sigma ^{2})}{%
		2}-\frac{1}{2\sigma ^{2}}\left( y_{i,t}-x_{i,t}^{\top }\beta -\gamma
	_{g,m}\right) ^{2}
\end{equation*}%
which leads to the joint-likelihood function%
\begin{equation*}
	\sum_{g\in \mathcal{G}}\sum_{i\in I_{g}}\sum_{m\in \mathcal{M}}\sum_{t\in
		I_{m}}\psi \left( z_{it};\theta ,\gamma _{g,m}\right) =-\frac{nT\log (\sigma
		^{2})}{2}-\frac{1}{2\sigma ^{2}}\sum_{i\leq n}\sum_{m\in \{1,2\}}\sum_{t\in
		I_{m}}\left( y_{i,t}-x_{i,t}^{\top }\beta -\gamma _{im}\right) ^{2},
\end{equation*}%
where $\theta =(\beta ^{\top },\sigma ^{2})^{\top }$.

\paragraph{Example 3. (Time Fixed Effect)}

In this example, only the time fixed effect shows up in $u_{i,t}$, i.e., $%
u_{i,t}=\zeta _{t}$ for any $i$ and any $t$. Since there is no grouping
structure in $t$ and a group structure in $i$, we have $\mathcal{G}=\{1\}$, $%
I_{g}=\{1,\ldots ,n\}$ for any $g\in \mathcal{G}$, $\mathcal{M}=\{1,\ldots
,T\}$ and $I_{m}=\{m\}$ for $m\in \mathcal{M}$.\ Therefore $\gamma
_{1,m}=\zeta _{m}\ $for any $m\in \mathcal{M}$, and the log-likelihood
function for any $i=g$ and any $t=m$ is%
\begin{equation*}
	\log \psi \left( z_{it};\theta ,\gamma _{g,m}\right) =-\frac{\log (\sigma
		^{2})}{2}-\frac{1}{2\sigma ^{2}}\left( y_{i,t}-x_{i,t}^{\top }\beta -\zeta
	_{m}\right) ^{2},
\end{equation*}%
which leads to the joint-likelihood function%
\begin{equation*}
	\sum_{g\in \mathcal{G}}\sum_{i\in I_{g}}\sum_{m\in \mathcal{M}}\sum_{t\in
		I_{m}}\psi \left( z_{it};\theta ,\gamma _{g,m}\right) =-\frac{nT\log (\sigma
		^{2})}{2}-\frac{1}{2\sigma ^{2}}\sum_{t\leq T}\sum_{i\leq n}\left(
	y_{i,t}-x_{i,t}^{\top }\theta -\zeta _{t}\right) ^{2},
\end{equation*}%
where $\theta =(\beta ^{\top },\sigma ^{2})^{\top }$.

\paragraph{Example 4.\ (Grouped Time Fixed Effects)}

In this example, there is a group structure in $i$ but no group structure in 
$t$, i.e., $u_{i,t}=\gamma _{g\left( i\right) ,t}$, where $g\left( \cdot
\right) $ is a known function with range $\{1,\ldots ,G\}$. Hence in this
case, $\mathcal{G}=\{1,\ldots ,G\}$, $I_{g}=\{g(i)=g:i=1,\ldots ,n\}$ for
any $g\in \mathcal{G}$, $\mathcal{M}=\{1,\ldots ,T\}$ and $I_{m}=\{m\}$ for $%
m\in \mathcal{M}$. \cite{bonhomme2015grouped} investigated this model where
the function $g\left( \cdot \right) $ was considered unknown, and one of
their goals\ was to estimate the latent group structure that is implied by $%
g\left( \cdot \right) $.\ The log-likelihood function for any $i\in I_{g}$
and any $t=m\ $is%
\begin{equation*}
	\log \psi \left( z_{it};\theta ,\gamma _{g,m}\right) =-\frac{\log (\sigma
		^{2})}{2}-\frac{1}{2\sigma ^{2}}\left( y_{i,t}-x_{i,t}^{\top }\beta -\gamma
	_{g,t}\right) ^{2}
\end{equation*}%
which leads to the joint-likelihood function%
\begin{equation*}
	\sum_{g\in \mathcal{G}}\sum_{i\in I_{g}}\sum_{m\in \mathcal{M}}\sum_{t\in
		I_{m}}\psi \left( z_{it};\theta ,\gamma _{g,m}\right) =-\frac{nT\log (\sigma
		^{2})}{2}-\frac{1}{2\sigma ^{2}}\sum_{g\in \mathcal{G}}\sum_{i\in
		I_{g}}\sum_{t\leq T}\left( y_{i,t}-x_{i,t}^{\top }\beta -\gamma
	_{g,t}\right) ^{2},
\end{equation*}%
where $\theta =(\beta ^{\top },\sigma ^{2})^{\top }$.

\bigskip

Below, we present an expansion of the likelihood function and discuss
Assumption \ref{A5}.

\bigskip

\noindent \textbf{Example 1. (Time Invariant Fixed Effect ctd.)}. In this
example, the expansion of the likelihood function can be written as%
\begin{equation*}
	\sum_{i\leq n}\hat{\Psi}_{i}(\hat{\phi}_{i})=\sum_{i\leq n}\hat{\Psi}%
	_{i}(\phi _{i}^{\ast })-2^{-1}\sum_{i\leq n}\hat{\Psi}_{\gamma ,i}^{2}/\Psi
	_{\gamma \gamma ,i}+o_{p}\left( n^{1/2}T^{-1}\right) ,
\end{equation*}%
where $\phi _{i}\equiv (\theta ^{\top },\gamma _{i})^{\top }$, $\hat{\Psi}%
_{i}(\phi _{i})\equiv T^{-1}\sum_{t\leq T}\psi (z_{i,t};\phi _{i})$, $\hat{%
	\Psi}_{\gamma ,i}\equiv T^{-1}\sum_{t\leq T}\partial \psi (z_{i,t};\phi
_{i}^{\ast })/\partial \gamma $ and $\Psi _{\gamma \gamma ,i}\equiv
T^{-1}\sum_{t\leq T}\mathbb{E}[\partial ^{2}\psi (z_{i,t};\phi _{i}^{\ast
})/\partial \gamma ^{2}]$. Moreover we have $G=n$, $M=1$, $n_{g}=1$, $%
T_{m}=T $ so Assumption \ref{A5} boils down to $n^{4/p}/T=o\left( 1\right) $
and $n/\left( nT\right) ^{1/2}=\left( n/T\right) ^{1/2}=O\left( 1\right) $.

\bigskip

\noindent \textbf{Example 2. (Time Varying Fixed Effect ctd.)}. In this
example, the expansion of the likelihood function can be written as%
\begin{align*}
	\sum_{i\leq n}\sum_{m\in \mathcal{M}}T_{m}\hat{\Psi}_{im}(\hat{\phi}_{im})=&
	\sum_{i\leq n}\sum_{m\in \mathcal{M}}T_{m}\hat{\Psi}_{im}(\phi _{im}^{\ast })
	\\
	& -2^{-1}\sum_{i\leq n}\sum_{m\in \mathcal{M}}T_{m}\hat{\Psi}_{\gamma
		,im}^{\top }\Psi _{\gamma \gamma ,im}^{-1}\hat{\Psi}_{\gamma
		,im}+o_{p}\left( (nM)^{1/2}\right) ,
\end{align*}%
where $\phi _{im}\equiv (\theta ^{\top },\gamma _{im})^{\top }$, $\hat{\Psi}%
_{im}\equiv T_{m}^{-1}\sum_{t\in I_{m}}\psi (z_{i,t};\phi _{im})$, $\hat{\Psi%
}_{\gamma ,im}\equiv T_{m}^{-1}\sum_{t\in I_{m}}\partial \psi (z_{i,t};\phi
_{im}^{\ast })/\partial \gamma $ and $\Psi _{\gamma \gamma ,im}\equiv
T_{m}^{-1}\sum_{t\in I_{m}}\mathbb{E}[\partial ^{2}\psi (z_{i,t};\phi
_{im}^{\ast })/\partial \gamma ^{2}]$. Since $G=n$\textbf{, }$M=2$\textbf{, }%
$n_{g}=1$\textbf{, }$T_{1}=T_{0}$ and $T_{2}=T-T_{0}$, Assumption \ref{A5}
holds in this example if $n^{2/p}(T_{0}^{-1/2}+(T-T_{0})^{-1/2})=o(1)$ and $%
n/\left( nT\right) ^{1/2}=\left( n/T\right) ^{1/2}=O\left( 1\right) $\textbf{%
	.}

\bigskip

\noindent \textbf{Example 3. (Time Fixed Effect ctd.)}.\ In this example,
the expansion of the likelihood function can be written as%
\begin{equation*}
	\sum_{t\leq T}\hat{\Psi}_{t}(\hat{\phi}_{t})=\sum_{t\leq T}\hat{\Psi}%
	_{t}(\phi _{t}^{\ast })-2^{-1}\sum_{t\leq T}\hat{\Psi}_{\gamma ,t}^{2}/\Psi
	_{\gamma \gamma ,t}+o_{p}\left( T^{1/2}n^{-1}\right) ,
\end{equation*}%
where $\phi _{t}\equiv (\theta ^{\top },\gamma _{t})^{\top }$, $\hat{\Psi}%
_{t}(\phi _{t})\equiv n^{-1}\sum_{i\leq n}\psi (z_{i,t};\phi _{t})$, $\hat{%
	\Psi}_{\gamma ,t}\equiv n^{-1}\sum_{i\leq n}\partial \psi (z_{i,t};\phi
_{t}^{\ast })/\partial \gamma $ and $\Psi _{\gamma \gamma ,t}\equiv
n^{-1}\sum_{i\leq n}\mathbb{E}[\partial ^{2}\psi (z_{i,t};\phi _{t}^{\ast
})/\partial \gamma ^{2}]$. In this case, $G=1$\textbf{, }$M=T$\textbf{, }$%
n_{g}=n$\textbf{, }$T_{m}=1$. Hence Assumption \ref{A5} becomes $%
T^{4/p}/n=o\left( 1\right) $ and $T/\left( nT\right) ^{1/2}=\left(
T/n\right) ^{1/2}=O\left( 1\right) $.

\bigskip

\noindent \textbf{Example 4.\ (Grouped Time Fixed Effects ctd.)}. In this
example, the expansion of the likelihood function can be written as%
\begin{equation*}
	\sum_{t\leq T}\sum_{g\in \mathcal{G}}n_{g}\hat{\Psi}_{gt}(\hat{\phi}%
	_{gt})=\sum_{t\leq T}\sum_{g\in \mathcal{G}}n_{g}\hat{\Psi}_{gt}(\phi
	_{gt}^{\ast })-2^{-1}\sum_{t\leq T}\sum_{g\in \mathcal{G}}n_{g}\hat{\Psi}%
	_{\gamma ,gt}^{\top }\Psi _{\gamma \gamma ,gt}^{-1}\hat{\Psi}_{\gamma
		,gt}+o_{p}\left( (GT)^{1/2}\right)
\end{equation*}%
where $\phi _{gt}\equiv (\theta ^{\top },\gamma _{g(i)t})^{\top }$, $\hat{%
	\Psi}_{gt}\equiv n_{g}^{-1}\sum_{i\in I_{g}}\psi (z_{i,t};\phi _{gt}^{\ast
}) $, $\hat{\Psi}_{\gamma ,gt}\equiv n_{g}^{-1}\sum_{i\in I_{g}}\partial
\psi (z_{i,t};\phi _{gt}^{\ast })/\partial \gamma $ and $\Psi _{\gamma
	\gamma ,im}\equiv n_{g}^{-1}\sum_{i\in I_{g}}\mathbb{E}[\partial ^{2}\psi
(z_{i,t};\phi _{gt}^{\ast })/\partial \gamma ^{2}]$. Since $M=T$\textbf{, }$%
n_{g}=\#I_{g}$ where $\#I_{g}$ denotes the cardinality of set $I_{g}$, and $%
T_{m}=1$, Assumption \ref{A5} holds in this example if $(GT)^{2/p}\max_{g%
	\leq G}n_{g}^{-1/2}=o(1)$ and $(GT)(nT)^{-1/2}=GT^{1/2}n^{-1/2}=O(1)$.

\subsection{Time Invariant vs. Time Varying Individual Effects\label%
	{V_Test_1}}

Suppose that model 1 specifies a time invariant individual fixed effect with
joint likelihood function $\sum_{i\leq n}\sum_{t\leq T}\psi _{1}\left(
z_{i,t};\theta _{1},\gamma _{1,i}\right) $, and model 2 specifies a time
varying individual fixed effect with joint likelihood function $\sum_{i\leq
	n}\sum_{m=1}^{M_{2}}\sum_{t\in I_{m}}\psi _{2}\left( z_{i,t};\theta
_{2},\gamma _{2,im}\right) $ where $M_{2}>1$ is a fixed integer.\ In this
example, we have $\mathcal{G}_{1}=\{1,\ldots ,n\}$ and $\mathcal{M}%
_{1}=\{1\} $ for model 1, and $\mathcal{G}_{2}=\{1,\ldots ,n\}$ and $%
\mathcal{M}_{2}=\{1,\ldots ,M_{2}\}$ for model 2.\footnote{%
	Since $G_{2}=n$ in this example, we let $M_{2}$ be a fixed integer such that
	Assumption \ref{A5}(ii) is satisfied.}

We first provide the specific forms of $\tilde{V}_{j,i}$ and $\tilde{U}%
_{j,i} $ in this example. It is clear that 
\begin{equation*}
	\tilde{\Psi}_{1\gamma ,i}=\frac{T^{-1/2}\sum_{t\leq T}\psi _{1\gamma
		}(z_{i,t};\phi _{1,i}^{\ast })}{\left( T^{-1}\sum_{t\leq T}\mathbb{E}\left[
		-\psi _{1\gamma \gamma }(z_{i,t};\phi _{1,i}^{\ast })\right] \right) ^{1/2}}
\end{equation*}%
and 
\begin{equation*}
	\tilde{\Psi}_{2\gamma ,im}=\frac{T_{m}^{-1/2}\sum_{t\in I_{m}}\psi _{2\gamma
		}(z_{i,t};\phi _{2,im}^{\ast })}{\left( T_{m}^{-1}\sum_{t\in I_{m}}\mathbb{E}%
		\left[ -\psi _{2\gamma \gamma }(z_{i,t};\phi _{2,im}^{\ast })\right] \right)
		^{1/2}},
\end{equation*}%
and we suppress the dependence of $\tilde{\Psi}_{1\gamma ,im}$ on $m$ since $%
\mathcal{M}_{1}=\{1\}$. Therefore,%
\begin{equation*}
	\tilde{V}_{1,i}=(2T^{1/2})^{-1}\text{\ }\tilde{\Psi}_{1\gamma ,i}^{2}\text{
		, \ }\tilde{V}_{2,i}=(2T^{1/2})^{-1}\sum_{m\in \mathcal{M}_{2}}\tilde{\Psi}%
	_{2\gamma ,im}^{2}\text{ \ and \ }\tilde{U}_{j,i}=0\text{ for }j=1,2\text{.}
\end{equation*}%
The expansion of the QLR statistic in (\ref{R_QLR_2}) thus implies that%
\begin{equation}
	\widehat{QLR}_{n,T}=n^{-1/2}\sum_{i\leq n}\left( \tilde{\Psi}_{1,i}-\tilde{%
		\Psi}_{2,i}+\frac{\tilde{\Psi}_{1\gamma ,i}^{2}-\sum_{m\in \mathcal{M}_{2}}%
		\tilde{\Psi}_{2\gamma ,im}^{2}}{2T^{1/2}}\right) +o_{p}(n^{-1/2}).
	\label{QLR_Case1}
\end{equation}%
Applying Theorem \ref{QLR_Stat} obtains the asymptotic normality for $%
\widehat{QLR}_{n,T}$ with variance 
\begin{equation*}
	\omega _{n,T}^{2}=n^{-1}\sum_{i\leq n}\mathrm{Var}\left( \tilde{\Psi}_{1,i}-%
	\tilde{\Psi}_{2,i}+\frac{\tilde{\Psi}_{1\gamma ,i}^{2}-\sum_{m\in \mathcal{M}%
			_{2}}\tilde{\Psi}_{2\gamma ,im}^{2}}{2T^{1/2}}\right) .
\end{equation*}

In the case that $\psi _{1}\left( \cdot \right) $ and $\psi _{2}\left( \cdot
\right) $ take the same function form (say $\psi \left( \cdot \right) $),
model 2 nests model 1 under the null, since $\phi _{1,i}^{\ast }=\phi
_{2,im}^{\ast }$ by the definitions of $\phi _{1,i}^{\ast }$ and $\phi
_{2,im}^{\ast }$ and their identification conditions. Therefore, the
expansion of the QLR statistic in (\ref{QLR_Case1}) is reduced to%
\begin{equation}
	\widehat{QLR}_{n,T}=(4nT)^{-1/2}\sum_{i\leq n}\left( \tilde{\Psi}_{1\gamma
		,i}^{2}-\sum_{m\in \mathcal{M}_{2}}\tilde{\Psi}_{2\gamma ,im}^{2}\right)
	+o_{p}(n^{-1/2}).  \label{case1_2}
\end{equation}%
Therefore in this scenario%
\begin{equation}
	\omega _{n,T}^{2}=(4nT)^{-1}\sum_{i\leq n}\mathrm{Var}\left( \tilde{\Psi}%
	_{1\gamma ,i}^{2}-\sum_{m\in \mathcal{M}_{2}}\tilde{\Psi}_{2\gamma
		,im}^{2}\right) ,  \label{case1_2b}
\end{equation}%
and Assumption \ref{A6}(i) holds as long as $n^{-1}\sum_{i\leq n}\mathrm{Var}%
(\tilde{\Psi}_{1\gamma ,i}^{2}-\sum_{m\in \mathcal{M}_{2}}\tilde{\Psi}%
_{2\gamma ,im}^{2})$ is bounded away from zero. We next provides the formula
of $\omega _{n,T}^{2}$ in the simplified case that $\mathbb{E}[\partial
^{2}\psi (z_{i,t};\phi _{i,t}^{\ast })/\partial \gamma ^{2}]$ is constant
for any $i$ and any $t$, and\ $\partial \psi (z_{i,t};\phi _{i,t}^{\ast
})/\partial \gamma $ is independent with constant variance across $t$.

\begin{lemma}
	\textit{\label{case_1_iid}\ }Consider two models with $\mathcal{G}%
	_{1}=\{1,\ldots ,n\}$ and $\mathcal{M}_{1}=\{1\}$ for model 1, and $\mathcal{%
		G}_{2}=\{1,\ldots ,n\}$ and $\mathcal{M}_{2}=\{1,\ldots ,M_{2}\}$ for model
	2 where $M_{2}$ is a fixed positive integer. \textit{Suppose that: (i) }$%
	\psi _{1}\left( \cdot \right) $ and $\psi _{2}\left( \cdot \right) $ take
	the same form $\psi \left( \cdot \right) $\textit{; (ii) }$\psi _{\gamma
		,i,t}\equiv \partial \psi (z_{i,t};\phi _{i,t}^{\ast })/\partial \gamma $ is
	independent across $t$ with $\mathbb{E}[\psi _{\gamma ,i,t}^{4}]<K$; (iii) $%
	\mathbb{E}[\partial ^{2}\psi (z_{i,t};\phi _{i,t}^{\ast })/\partial \gamma
	^{2}]=\Psi _{\gamma \gamma }$ where $\Psi _{\gamma \gamma }$ is a nonzero
	constant; and (iv) $\min_{m\in \mathcal{M}_{2}}(T_{m}/T)\geq K^{-1}$. Then%
	\begin{align}
		\omega _{n,T}^{2}=& \frac{n^{-1}\sum_{i\leq n}\bar{\sigma}_{\gamma
				,i}^{4}+\sum_{m=1}^{M_{2}}(1-2w_{m})n^{-1}\sum_{i\leq n}\bar{\sigma}_{\gamma
				,im}^{4}}{2T\Psi _{\gamma \gamma }^{2}}  \notag \\
		& +O\left( T^{-1}\max_{m\in \mathcal{M}_{2}}T_{m}^{-1}\right) ,
		\label{case_1_iid_1}
	\end{align}%
	where $w_{m}\equiv T_{m}T^{-1}$, $\bar{\sigma}_{\gamma ,i}\equiv \sqrt{%
		T^{-1}\sum_{t\leq T}\sigma _{\gamma ,i,t}^{2}}$, $\bar{\sigma}_{\gamma
		,im}\equiv \sqrt{T_{m}^{-1}\sum_{t\in I_{m}}\sigma _{\gamma ,i,t}^{2}}$ and $%
	\sigma _{\gamma ,i,t}^{2}\equiv \mathrm{Var}(\psi _{\gamma ,i,t})$.
\end{lemma}

When $\psi _{\gamma ,i,t}$ has constant variance across $t$, i.e., $\sigma
_{\gamma ,i,t}^{2}=\sigma _{\gamma ,i}^{2}$ for some finite constant $\sigma
_{\gamma ,i}^{2}$, we have 
\begin{equation*}
	n^{-1}\sum_{i\leq n}\bar{\sigma}_{\gamma ,i}^{4}=n^{-1}\sum_{i\leq n}\bar{%
		\sigma}_{\gamma ,im}^{4}=\sigma _{\gamma ,i}^{4},
\end{equation*}%
and the approximated expression of $\omega _{n,T}^{2}$ in (\ref{case_1_iid_1}%
) becomes 
\begin{equation}
	\omega _{n,T}^{2}=\frac{M_{2}-1}{2}\frac{n^{-1}\sum_{i\leq n}\sigma _{\gamma
			,i}^{4}}{T\Psi _{\gamma \gamma }^{2}}+O\left( T^{-1}\max_{m\in \mathcal{M}%
		_{2}}T_{m}^{-1}\right) ,  \label{case_1_iid_2}
\end{equation}%
where the second term in the RHS of (\ref{case_1_iid_2}) is asymptotically
negligible if $\max_{m\in \mathcal{M}_{2}}T_{m}^{-1}=o(1)$.

\subsection{Homogeneous vs. Heterogeneous Time Effects \label{V_Test_2}}

Suppose that model 1 specifies a group heterogeneous time effect with joint
likelihood function $\sum_{g=1}^{G_{1}}\sum_{i\in I_{g}}\sum_{t\leq T}\psi
_{1}\left( z_{i,t};\theta _{1},\gamma _{1,gt}\right) $ where $G_{1}>1$ is a
fixed integer, and model 2 specifies a homogeneous time fixed effect with
joint likelihood function $\sum_{i\leq n}\sum_{t\leq T}\psi _{2}\left(
z_{i,t};\theta _{2},\gamma _{2,t}\right) $. In this example, we have $%
\mathcal{G}_{1}=\{1,\ldots ,G_{1}\}$ and $\mathcal{M}_{1}=\{1\ldots ,T\}$
for model 1, and $\mathcal{G}_{2}=\{1\}$ and $\mathcal{M}_{2}=\{1,\ldots
,T\} $ for model 2.\footnote{%
	Since $M_{1}=T$ in this example, we let $G_{1}$ be a fixed integer such that
	Assumption \ref{A5}(ii) is satisfied.}

Under the specifications of model 1 and model 2,%
\begin{equation*}
	\tilde{\Psi}_{1\gamma ,i,t}=\frac{\psi _{1\gamma }(z_{i,t};\phi
		_{1,i,t}^{\ast })}{\left( n_{g}^{-1}\sum_{i\in I_{g}}\mathbb{E}\left[ -\psi
		_{1\gamma \gamma }(z_{i,t};\phi _{1,i,t}^{\ast })\right] \right) ^{1/2}}
\end{equation*}%
and 
\begin{equation*}
	\tilde{\Psi}_{2\gamma ,i,t}=\frac{\psi _{2\gamma }(z_{i,t};\phi _{2,t}^{\ast
		})}{\left( n^{-1}\sum_{i\leq n}\mathbb{E}\left[ -\psi _{2\gamma \gamma
		}(z_{i,t};\phi _{2,t}^{\ast })\right] \right) ^{1/2}}
\end{equation*}%
where we write $\tilde{\Psi}_{j\gamma ,im}$ as $\tilde{\Psi}_{j\gamma ,i,t}$
since $\mathcal{M}_{j}=\{1\ldots ,T\}$. Without loss of generality, we
suppose that for any $g\in \mathcal{G}_{1}$, $I_{g}=\{N_{g}+1,\ldots
,N_{g}+n_{g}\}$ where $N_{g}\equiv \sum_{g^{\prime }<g}n_{g^{\prime }}$.
Then for any $i\in I_{g}$ and any $g\in \mathcal{G}_{1}$ 
\begin{align*}
	\tilde{V}_{1,i}& =\frac{A_{1,i}^{\top }A_{1,i}}{2n_{g}T^{1/2}}\text{, \ }%
	\tilde{U}_{1,i}=\frac{\sum_{i^{\prime }=N_{g}+1}^{i-1}A_{1,i}^{\top
		}A_{1,i^{\prime }}}{n_{g}T^{1/2}}\text{,} \\
	\tilde{V}_{2,i}& =\frac{A_{2,i}^{\top }A_{2,i}}{2nT^{1/2}}\text{ \ and \ }%
	\tilde{U}_{2,i}=\frac{\sum_{i^{\prime }=1}^{i-1}A_{2,i}^{\top
		}A_{2,i^{\prime }}}{nT^{1/2}},
\end{align*}%
where $A_{j,i}=(\tilde{\Psi}_{j\gamma ,i,t})_{t\leq T}$ for $j=1,2$.%
\footnote{%
	Following the trandition that the summaion over an empty set is zero, we
	have $\tilde{U}_{1,N_{g}+1}=0$ and $\tilde{U}_{2,1}=0$ in the above display.}
Applying Theorem \ref{QLR_Stat}, we obtain the asymptotic normality for $%
\widehat{QLR}_{n,T}$ with variance 
\begin{equation*}
	\omega _{n,T}=\sqrt{n^{-1}\sum_{i\leq n}\mathrm{Var}\left( \tilde{\Psi}%
		_{1,i}-\tilde{\Psi}_{2,i}+\tilde{V}_{1,i}-\tilde{V}_{2,i}+\tilde{U}_{1,i}-%
		\tilde{U}_{2,i}\right) }.
\end{equation*}

Moreover, the lemma below shows that $\tilde{V}_{j,i}$ ($j=1,2$) are
asymptotically negligible, and hence the asymptotic normality of $\widehat{%
	QLR}_{n,T}$ is driven by $n^{-1}\sum_{i\leq n}(\tilde{\Psi}_{1,i}-\tilde{\Psi%
}_{2,i}+\tilde{U}_{1,i}-\tilde{U}_{2,i})$. Moreover, the variance of $%
\widehat{QLR}_{n,T}$ is given by%
\begin{equation*}
	\omega _{n,T}^{\ast }=\sqrt{n^{-1}\sum_{i\leq n}\left( \mathrm{Var}(\tilde{%
			\Psi}_{1,i}-\tilde{\Psi}_{2,i})+\mathrm{Var}(\tilde{U}_{1,i}-\tilde{U}%
		_{2,i})\right) }.
\end{equation*}%
Hence Assumption \ref{A6}(i) holds as long as $\sum_{i\leq n}\mathrm{Var}(%
\tilde{U}_{1,i}-\tilde{U}_{2,i})$ is bounded away from zero.

\begin{lemma}
	\textit{\label{case_2} }Consider two models with $\mathcal{G}_{1}=\{1,\ldots
	,G_{1}\}$ and $\mathcal{M}_{1}=\{1\ldots ,T\}$ for model 1 where $G_{1}$ is
	a fixed positive integer, and $\mathcal{G}_{2}=\{1\}$ and $\mathcal{M}%
	_{2}=\{1,\ldots ,T\}$ for model 2. Suppose that Assumptions \ref{A1}, \ref%
	{A2}, \ref{A3}, \ref{A4}, \ref{A5}(i) and \ref{A7} hold for models 1 and 2,
	and $\sum_{i\leq n}\mathrm{Var}(\tilde{U}_{1,i}-\tilde{U}_{2,i})>K^{-1}$.
	Then 
	\begin{equation}
		\frac{\widehat{QLR}_{n,T}-\mathbb{E}[S_{n,T}]-QLR_{n,T}^{\ast }}{\omega
			_{n,T}^{\ast }}\rightarrow _{d}N(0,1),  \label{case_2_0}
	\end{equation}%
	where $\mathbb{E}[S_{n,T}]=(2nT)^{-1/2}\sum_{g\in \mathcal{G}_{1}}\sum_{i\in
		I_{g}}\left( n_{g}^{-1}\mathbb{E}[A_{1,i}^{\top }A_{1,i}]-n^{-1}\mathbb{E}%
	[A_{2,i}^{\top }A_{2,i}]\right) $, and%
	\begin{align}
		(\omega _{n,T}^{\ast })^{2}& =n^{-1}\sum_{i\leq n}\mathrm{Var}(\tilde{\Psi}%
		_{1,i}-\tilde{\Psi}_{2,i})  \notag \\
		& \text{ \ \ \ \ }+\frac{\sum_{g\in \mathcal{G}_{1}}(\left\Vert \Sigma
			_{1\gamma ,g}\right\Vert ^{2}-2w_{g}\left\Vert \Sigma _{12\gamma
				,g}\right\Vert ^{2})+\left\Vert \sum_{g\in \mathcal{G}_{1}}w_{g}\Sigma
			_{2\gamma ,g}\right\Vert ^{2}}{2nT}  \notag \\
		& \text{ \ \ \ \ }+O\left( n^{-1}\max_{g\in \mathcal{G}_{1}}n_{g}^{-1}%
		\right) .  \label{case_2_1}
	\end{align}%
	where $w_{g}\equiv n_{g}n^{-1}$, $\Sigma _{j\gamma ,g}\equiv
	n_{g}^{-1}\sum_{i\in I_{g}}\mathbb{E}[A_{j\gamma ,i}A_{j\gamma ,i}^{\top }]$
	and $\Sigma _{12\gamma ,g}\equiv n_{g}^{-1}\sum_{i\in I_{g}}\mathbb{E}%
	[A_{1,i}A_{2,i}^{\top }]$.
\end{lemma}

To get a more explicit form for $(\omega _{n,T}^{\ast })^{2}$, we consider
the scenario that the conditions of Lemma \ref{case_1_iid} hold. In this
case, model 1 nests model 1 under the null hypothesis and $\mathrm{Var}(%
\tilde{\Psi}_{1,i}-\tilde{\Psi}_{2,i})=0$. Moreover, $\Sigma _{j\gamma
	,g}=\Sigma _{12\gamma ,g}$ with 
\begin{equation*}
	\left\Vert \Sigma _{1\gamma ,g}\right\Vert ^{2}=\left\Vert \Sigma _{12\gamma
		,g}\right\Vert ^{2}=\Psi _{\gamma \gamma }^{-2}\sum_{t\leq T}\bar{\sigma}%
	_{\gamma ,gt}^{4}\text{ \ \ and \ \ }\left\Vert \sum_{g\in \mathcal{G}%
		_{1}}w_{g}\Sigma _{2\gamma ,g}\right\Vert ^{2}=\Psi _{\gamma \gamma
	}^{-2}\sum_{t\leq T}\bar{\sigma}_{\gamma ,t}^{4}
\end{equation*}%
where $\bar{\sigma}_{\gamma ,gt}\equiv \sqrt{n_{g}^{-1}\sum_{i\in
		I_{g}}\sigma _{\gamma ,i,t}^{2}}$ and $\bar{\sigma}_{\gamma ,t}^{2}\equiv 
\sqrt{n^{-1}\sum_{i\leq n}\sigma _{\gamma ,i,t}^{2}}$. Hence the expression
in (\ref{case_2_1}) leads to 
\begin{equation*}
	(\omega _{n,T}^{\ast })^{2}=\frac{T^{-1}\sum_{t\leq T}\bar{\sigma}_{\gamma
			,t}^{4}+\sum_{g\in \mathcal{G}_{1}}(1-2w_{g})T^{-1}\sum_{t\leq T}\bar{\sigma}%
		_{\gamma ,gt}^{4}}{2n\Psi _{\gamma \gamma }^{2}}+O\left( n^{-1}\max_{g\in 
		\mathcal{G}_{1}}n_{g}^{-1}\right) .
\end{equation*}

\subsection{Individual vs. Time Fixed Effect Models \label{V_Test_3}}

Suppose that model 1 specifies individual fixed effects\ with joint
likelihood $\sum_{i\leq n}\sum_{t\leq T}\psi_{1}\left(
z_{i,t};\theta_{1},\gamma _{1,i}\right) $, while model 2 specifies\ time
fixed effects with joint likelihood $\sum_{i\leq n}\sum_{t\leq
	T}\psi_{2}\left( z_{i,t};\theta _{1},\gamma_{2,t}\right) $.\ In this case,
we have $\mathcal{G}_{1}=\{1,\ldots,n\}$ and $\mathcal{M}_{1}=\{1\}$ for
model 1, and $\mathcal{G}_{2}=\{1\}$ and $\mathcal{M}_{2}=\{1,\ldots,T\}$
for model 2.

In this example, it is clear that 
\begin{equation*}
	\tilde{\Psi}_{1\gamma ,i}=\frac{T^{-1/2}\sum_{t\leq T}\psi _{1\gamma
		}(z_{i,t};\phi _{1,i}^{\ast })}{\left( T^{-1}\sum_{t\leq T}\mathbb{E}\left[
		-\psi _{1\gamma \gamma }(z_{i,t};\phi _{1,i}^{\ast })\right] \right) ^{1/2}}%
	\text{\ \ and \ \ }\tilde{\Psi}_{2\gamma ,i,t}=\frac{\psi _{2\gamma
		}(z_{i,t};\phi _{2,t}^{\ast })}{\left( n^{-1}\sum_{i\leq n}\mathbb{E}\left[
		-\psi _{2\gamma \gamma }(z_{i,t};\phi _{2,t}^{\ast })\right] \right) ^{1/2}}
\end{equation*}%
where we suppress the dependence of $\tilde{\Psi}_{1\gamma ,im}$ on $m$
since $\mathcal{M}_{1}=\{1\}$, and write $\tilde{\Psi}_{2\gamma ,im}$ as $%
\tilde{\Psi}_{2\gamma ,i,t}$ since $\mathcal{M}_{2}=\{1,\ldots ,T\}$.
Therefore,%
\begin{equation*}
	\tilde{V}_{1,i}=\frac{\tilde{\Psi}_{1\gamma ,i}^{2}}{2T^{1/2}}\text{, \ }%
	\tilde{U}_{1,i}=0\text{, \ }\tilde{V}_{2,i}=\frac{A_{2,i}^{\top }A_{2,i}}{%
		2nT^{1/2}}\text{ \ and \ }\tilde{U}_{2,i}=\frac{\sum_{i^{\prime
			}=1}^{i-1}A_{2,i}^{\top }A_{2,i^{\prime }}}{nT^{1/2}},
\end{equation*}%
where $A_{2,i}=(\tilde{\Psi}_{2\gamma ,i,t})_{t\leq T}$.\footnote{%
	Following the trandition that the summaion over an empty set is zero, we
	have $\tilde{U}_{2,1}=0$ in the above display.} Applying Theorem \ref%
{QLR_Stat}, we obtain the asymptotic normality for $\widehat{QLR}_{n,T}$
with variance 
\begin{equation*}
	\omega _{n,T}=\sqrt{n^{-1}\sum_{i\leq n}\mathrm{Var}\left( \tilde{\Psi}%
		_{1,i}-\tilde{\Psi}_{2,i}+\tilde{V}_{1,i}-\tilde{V}_{2,i}-\tilde{U}%
		_{2,i}\right) }.
\end{equation*}%
Since $\tilde{V}_{2,i}$ is the same as the one in the Example 2, we can show
that it is asymptotically negligible and the asymptotic normality of $%
\widehat{QLR}_{n,T}$ is driven by $n^{-1}\sum_{i\leq n}(\tilde{\Psi}_{1,i}-%
\tilde{\Psi}_{2,i}+\tilde{V}_{1,i}-\tilde{U}_{2,i})$. As we shall prove in
the lemma below, the variance of $\widehat{QLR}_{n,T}$ in this example is
given by%
\begin{equation*}
	\omega _{n,T}^{\ast }=\sqrt{n^{-1}\sum_{i\leq n}\left( \mathrm{Var}(\tilde{%
			\Psi}_{1,i}-\tilde{\Psi}_{2,i}+\tilde{V}_{1,i})+\mathrm{Var}(\tilde{U}%
		_{2,i})\right) }.
\end{equation*}%
Hence Assumption \ref{A6}(i) holds as long as $\sum_{i\leq n}\mathrm{Var}(%
\tilde{U}_{2,i})$ is bounded away from zero.

\begin{lemma}
	\textit{\label{case_3}\ }Consider two models with $\mathcal{G}%
	_{1}=\{1,\ldots ,n\}$ and $\mathcal{M}_{1}=\{1\}$ for model 1, and $\mathcal{%
		G}_{2}=\{1\}$ and $\mathcal{M}_{2}=\{1,\ldots ,T\}$ for model 2. Suppose
	that Assumptions \ref{A1}, \ref{A2}, \ref{A3}, \ref{A4}, \ref{A5}(i) and \ref%
	{A7} hold for models 1 and 2, and $\sum_{i\leq n}\mathrm{Var}(\tilde{U}%
	_{2,i})>K^{-1}$. Then%
	\begin{equation}
		\frac{\widehat{QLR}_{n,T}-\mathbb{E}[S_{n,T}]-QLR_{n,T}^{\ast }}{\omega
			_{n,T}^{\ast }}\rightarrow _{d}N(0,1),  \label{case_3_0}
	\end{equation}%
	where $\mathbb{E}[S_{n,T}]=(2nT)^{-1/2}\sum_{i\leq n}(\mathbb{E}[\tilde{\Psi}%
	_{1\gamma ,i}^{2}]-n^{-1}\mathbb{E}[A_{2,i}^{\top }A_{2,i}])$, and%
	\begin{equation}
		(\omega _{n,T}^{\ast })^{2}=n^{-1}\sum_{i\leq n}\mathrm{Var}(\tilde{\Psi}%
		_{1,i}-\tilde{\Psi}_{2,i}+\tilde{V}_{1,i})+\frac{\left\Vert
			n^{-1}\sum_{i\leq n}\mathbb{E}[A_{2,i}A_{2,i}^{\top }]\right\Vert ^{2}}{2nT}%
		+O(n^{-2}).  \label{case_3_1}
	\end{equation}
\end{lemma}

\subsection{Proofs for Lemmas \protect\ref{case_1_iid}, \protect\ref{case_2}
	and \protect\ref{case_3}}

\noindent \textsc{Proof of Lemma \ref{case_1_iid}}. Recall that in the two
models investigated in this lemma, we have $\mathcal{G}_{1}=\{1,\ldots ,n\}$
and $\mathcal{M}_{1}=\{1\}$ for model 1, and $\mathcal{G}_{2}=\{1,\ldots
,n\} $ and $\mathcal{M}_{2}=\{1,\ldots ,M_{2}\}$ for model 2 where $M_{2}$
is a fixed positive integer. Under Conditions (i, iii) of the Lemma, we can
write%
\begin{equation*}
	\tilde{\Psi}_{1\gamma ,i}^{2}=\frac{\left( \sum_{m\in \mathcal{M}%
			_{2}}(T_{m}/T)^{1/2}\tilde{\Psi}_{\gamma ,im}\right) ^{2}}{-\Psi _{\gamma
			\gamma }}=\sum_{m=1}^{M_{2}}\frac{T_{m}\tilde{\Psi}_{\gamma ,im}^{2}}{-T\Psi
		_{\gamma \gamma }}+2\sum_{m=2}^{M_{2}}\frac{\sum_{m^{\prime
			}=1}^{m-1}T_{m}^{1/2}\tilde{\Psi}_{\gamma ,im}T_{m^{\prime }}^{1/2}\tilde{%
			\Psi}_{\gamma ,im^{\prime }}}{-T\Psi _{\gamma \gamma }}
\end{equation*}%
and $\tilde{\Psi}_{2\gamma ,im}^{2}=\tilde{\Psi}_{\gamma ,im}^{2}/(-\Psi
_{\gamma \gamma })$ for any $m\in \mathcal{M}_{2}$ and any $i=1,\ldots ,n$.
Therefore,%
\begin{equation}
	\tilde{\Psi}_{1\gamma ,i}^{2}-\sum_{m\in \mathcal{M}_{2}}\tilde{\Psi}%
	_{2\gamma ,im}^{2}=\sum_{m=1}^{M_{2}}(1-T_{m}/T)\frac{\tilde{\Psi}_{\gamma
			,im}^{2}}{\Psi _{\gamma \gamma }}-2\sum_{m=2}^{M_{2}}\sum_{m^{\prime
		}=1}^{m-1}\frac{T_{m}^{1/2}\tilde{\Psi}_{\gamma ,im}T_{m^{\prime }}^{1/2}%
		\tilde{\Psi}_{\gamma ,im^{\prime }}}{T\Psi _{\gamma \gamma }}.
	\label{case1_3}
\end{equation}%
Under Condition (ii) of the lemma, $\tilde{\Psi}_{\gamma ,im}$ are
independent across $m$ for any $i$, which together with the expression in\ (%
\ref{case1_3}) implies that%
\begin{align}
	& \mathrm{Var}\left( \tilde{\Psi}_{1\gamma ,i}^{2}-\sum_{m\in \mathcal{M}%
		_{2}}\tilde{\Psi}_{2\gamma ,im}^{2}\right) =\sum_{m=1}^{M_{2}}(1-T_{m}/T)^{2}%
	\frac{\mathrm{Var}(\tilde{\Psi}_{\gamma ,im}^{2})}{\Psi _{\gamma \gamma }^{2}%
	}  \notag \\
	& \text{ \ \ \ \ \ \ \ \ \ \ \ \ \ \ \ \ \ \ \ \ \ \ \ \ \ \ \ \ \ \ \ \ \ \
		\ \ \ \ \ \ \ \ \ \ \ \ \ \ \ \ \ }+4\sum_{m=2}^{M_{2}}\sum_{m^{\prime
		}=1}^{m-1}\frac{\mathrm{Var}(T_{m}^{1/2}\tilde{\Psi}_{\gamma ,im})\mathrm{Var%
		}(T_{m^{\prime }}^{1/2}\tilde{\Psi}_{\gamma ,im^{\prime }})}{T^{2}\Psi
		_{\gamma \gamma }^{2}}.  \label{case1_4}
\end{align}%
By Condition (ii) of the lemma, $\mathrm{Var}(T_{m}^{1/2}\tilde{\Psi}%
_{\gamma ,im})=\sum_{t\in I\,_{m}}\sigma _{\gamma ,i,t}^{2}$ for any $m$,
and hence%
\begin{align}
	& 4T^{-2}\sum_{m=2}^{M_{2}}\sum_{m^{\prime }=1}^{m-1}\mathrm{Var}(T_{m}^{1/2}%
	\tilde{\Psi}_{\gamma ,im})\mathrm{Var}(T_{m^{\prime }}^{1/2}\tilde{\Psi}%
	_{\gamma ,im^{\prime }})  \notag \\
	& =4T^{-2}\sum_{m=2}^{M_{2}}\sum_{m^{\prime }=1}^{m-1}\left( \sum_{t\in
		I\,_{m}}\sigma _{\gamma ,i,t}^{2}\right) \left( \sum_{t\in I\,_{m^{\prime
	}}}\sigma _{\gamma ,i,t}^{2}\right)  \notag \\
	& =2\left( T^{-1}\sum_{t\leq T}\sigma _{\gamma ,i,t}^{2}\right)
	^{2}-2\sum_{m=1}^{M_{2}}w_{m}^{2}\left( T_{m}^{-1}\sum_{t\in I\,_{m}}\sigma
	_{\gamma ,i,t}^{2}\right) ^{2}.  \label{case1_5}
\end{align}%
Without loss of generality for any $m\in \mathcal{M}_{2}$, let $%
I_{m}=\{t_{m,1},\ldots ,t_{m,T_{m}}\}$. Then 
\begin{equation*}
	\tilde{\Psi}_{\gamma ,im}^{2}=T_{m}^{-1}\sum_{t\in I_{m}}\psi _{\gamma
		,i,t}^{2}+2T_{m}^{-1}\sum_{j=2}^{T_{m}}\sum_{j^{\prime }=1}^{j-1}\psi
	_{\gamma ,i,t_{m,j}}\psi _{\gamma ,i,t_{m,j^{\prime }}}.
\end{equation*}%
Let $\kappa _{\gamma ,i,t}^{4}\equiv \mathrm{Var}(\psi _{\gamma ,i,t}^{2})$.
By Condition (ii) of the lemma%
\begin{align*}
	\mathrm{Var}(\tilde{\Psi}_{\gamma ,im}^{2})& =T_{m}^{-2}\sum_{t\in I_{m}}%
	\mathrm{Var}(\psi _{\gamma
		,i,t}^{2})+4T_{m}^{-2}\sum_{j=2}^{T_{m}}\sum_{j^{\prime }=1}^{j}\mathrm{Var}%
	(\psi _{\gamma ,i,t_{m,j}})\mathrm{Var}(\psi _{\gamma ,i,t_{m,j^{\prime }}})
	\\
	& =T_{m}^{-2}\sum_{t\in I_{m}}\kappa _{\gamma ,i,t}^{4}+2T_{m}^{-2}\left[
	\left( \sum_{t\in I_{m}}\sigma _{\gamma ,i,t}^{2}\right) ^{2}-\sum_{t\in
		I_{m}}\sigma _{\gamma ,i,t}^{4}\right] ,
\end{align*}%
which implies that%
\begin{align}
	\sum_{m=1}^{M_{2}}(1-T_{m}/T)^{2}\mathrm{Var}(\tilde{\Psi}_{\gamma
		,im}^{2})& =\sum_{m=1}^{M_{2}}(1-w_{m})^{2}T_{m}^{-2}\left( \sum_{t\in
		I_{m}}\kappa _{\gamma ,i,t}^{4}-2\sum_{t\in I_{m}}\sigma _{\gamma
		,i,t}^{4}\right)  \notag \\
	& \text{ \ \ \ \ \ }+2\sum_{m=1}^{M_{2}}(1-w_{m})^{2}\left(
	T_{m}^{-1}\sum_{t\in I_{m}}\sigma _{\gamma ,i,t}^{2}\right) ^{2}.
	\label{case1_6}
\end{align}%
Combining the results in (\ref{case1_4}), (\ref{case1_5}) and (\ref{case1_6}%
), we have

\begin{align}
	\mathrm{Var}\left( \tilde{\Psi}_{1\gamma ,i}^{2}-\sum_{m\in \mathcal{M}_{2}}%
	\tilde{\Psi}_{2\gamma ,im}^{2}\right) & =\frac{2\left[ \left(
		T^{-1}\sum_{t\leq T}\sigma _{\gamma ,i,t}^{2}\right)
		^{2}+\sum_{m=1}^{M_{2}}(1-2w_{m})\left( T_{m}^{-1}\sum_{t\in I_{m}}\sigma
		_{\gamma ,i,t}^{2}\right) ^{2}\right] }{\Psi _{\gamma \gamma }^{2}}  \notag
	\\
	& +\frac{\sum_{m=1}^{M_{2}}(1-w_{m})^{2}T_{m}^{-2}\left( \sum_{t\in
			I_{m}}\kappa _{\gamma ,i,t}^{4}-2\sum_{t\in I_{m}}\sigma _{\gamma
			,i,t}^{4}\right) }{\Psi _{\gamma \gamma }^{2}}.  \label{case1_7}
\end{align}%
Since $M_{2}$ is fixed, by $\mathbb{E}[\psi _{\gamma ,i,t}^{4}]<K$,%
\begin{equation*}
	\sum_{m=1}^{M_{2}}(1-w_{m})^{2}T_{m}^{-2}\left( \sum_{t\in I_{m}}\kappa
	_{\gamma ,i,t}^{4}+2\sum_{t\in I_{m}}\sigma _{\gamma ,i,t}^{4}\right) \leq
	KM_{2}\max_{m\in \mathcal{M}_{2}}T_{m}^{-1}=O\left( \max_{m\in \mathcal{M}%
		_{2}}T_{m}^{-1}\right) ,
\end{equation*}%
which together with (\ref{case1_7})\ and the definitions of $\bar{\sigma}%
_{\gamma ,i}$ and $\bar{\sigma}_{\gamma ,im}$ yields 
\begin{equation*}
	\sum_{i\leq n}\mathrm{Var}\left( \tilde{\Psi}_{1\gamma ,i}^{2}-\sum_{m\in 
		\mathcal{M}_{2}}\tilde{\Psi}_{2\gamma ,im}^{2}\right) =\frac{2\left(
		\sum_{i\leq n}\bar{\sigma}_{\gamma
			,i}^{4}+\sum_{m=1}^{M_{2}}(1-2w_{m})\sum_{i\leq n}\bar{\sigma}_{\gamma
			,im}^{4}\right) }{\Psi _{\gamma \gamma }^{2}}+O\left( \max_{m\in \mathcal{M}%
		_{2}}T_{m}^{-1}\right) .
\end{equation*}%
The claim in (\ref{case_1_iid_1}) now follows from (\ref{case1_2b}) and the
above equation.\hfill $Q.E.D.$

\bigskip

\noindent \textsc{Proof of Lemma \ref{case_2}}. Recall that in the two
models investigated in this lemma, we have $\mathcal{G}_{1}=\{1,\ldots
,G_{1}\}$ and $\mathcal{M}_{1}=\{1\ldots ,T\}$ for model 1 where $G_{1}$ is
a fixed positive integer, and $\mathcal{G}_{2}=\{1\}$ and $\mathcal{M}%
_{2}=\{1,\ldots ,T\}$ for model 2. First notice that by $G_{1}<K$, $G_{2}=1$
and $M_{1}=M_{2}=T$, Assumption \ref{A5}(ii) holds. By the independence of $%
\{z_{i,t}\}$ across $i$ for all $t$, we $\tilde{\Psi}_{1,i}-\tilde{\Psi}%
_{2,i}$ is uncorrelated with $\tilde{U}_{1,i}-\tilde{U}_{2,i}$ for any $i$.
Therefore, 
\begin{equation}
	\mathrm{Var}(\tilde{\Psi}_{1,i}-\tilde{\Psi}_{2,i}+\tilde{U}_{1,i}-\tilde{U}%
	_{2,i})=\mathrm{Var}(\tilde{\Psi}_{1,i}-\tilde{\Psi}_{2,i})+\mathrm{Var}(%
	\tilde{U}_{1,i}-\tilde{U}_{2,i}).  \label{P_case_2_1a}
\end{equation}%
The condition $\sum_{i\leq n}\mathrm{Var}(\tilde{U}_{1,i}-\tilde{U}%
_{2,i})>K^{-1}$ maintained in the lemma thus implies that 
\begin{equation}
	n(\omega _{n,T}^{\ast })^{2}>K^{-1}.  \label{P_case_2_1b}
\end{equation}

We first show (\ref{case_2_0}). Since $\tilde{V}_{j,i}$ are independent
across $i$, 
\begin{align}
	& \mathrm{Var}\left( n^{-1/2}\sum_{i\leq n}\left( \tilde{V}_{1,i}-\tilde{V}%
	_{2,i}\right) \right) =n^{-1}\sum_{i\leq n}\mathrm{Var}\left( \tilde{V}%
	_{1,i}-\tilde{V}_{2,i}\right)  \notag \\
	& \leq n^{-1}\sum_{g\in \mathcal{G}_{1}}\sum_{i\in I_{g}}n_{g}^{-2}\mathrm{%
		Var}\left( T^{-1/2}A_{1,i}^{\top }A_{1,i}\right) +n^{-3}\sum_{i\leq n}%
	\mathrm{Var}\left( T^{-1/2}A_{2,i}^{\top }A_{2,i}\right)  \notag \\
	& \leq K\left( n^{-2}+n^{-1}\sum_{g\in \mathcal{G}_{1}}n_{g}^{-1}\right)
	\leq Kn^{-1}\max_{g\in \mathcal{G}_{1}}n_{g}^{-1}  \label{P_case_2_1}
\end{align}%
where the second inequality is by (\ref{P_C1_MGCLT_L2_5}) (where we show
that $\mathrm{Var}(M_{j}^{-1/2}A_{j,i}^{\top }A_{j,i})\leq K$ with $M_{j}=T$
here). Next note that 
\begin{align}
	\omega _{n,T}^{2}& =n^{-1}\sum_{i\leq n}\mathrm{Var}(\tilde{\Psi}_{1,i}-%
	\tilde{\Psi}_{2,i}+\tilde{U}_{1,i}-\tilde{U}_{2,i})+n^{-1}\sum_{i\leq n}%
	\mathrm{Var}(\tilde{V}_{1,i}-\tilde{V}_{2,i})  \notag \\
	& +2n^{-1}\sum_{i\leq n}\mathrm{Cov}(\tilde{\Psi}_{1,i}-\tilde{\Psi}_{2,i}+%
	\tilde{U}_{1,i}-\tilde{U}_{2,i},\tilde{V}_{1,i}-\tilde{V}_{2,i})  \notag \\
	& =(\omega _{n,T}^{\ast })^{2}+2n^{-1}\sum_{i\leq n}\mathrm{Cov}(\tilde{\Psi}%
	_{1,i}-\tilde{\Psi}_{2,i}+\tilde{U}_{1,i}-\tilde{U}_{2,i},\tilde{V}_{1,i}-%
	\tilde{V}_{2,i})+O\left( n^{-1}\max_{g\in \mathcal{G}_{1}}n_{g}^{-1}\right) ,
	\label{P_case_2_2}
\end{align}%
where the second equality is by (\ref{P_case_2_1}). By H\"{o}lder's
inequality,%
\begin{align*}
	& \left\vert n^{-1}\sum_{i\leq n}\mathrm{Cov}(\tilde{\Psi}_{1,i}-\tilde{\Psi}%
	_{2,i}+\tilde{U}_{1,i}-\tilde{U}_{2,i},\tilde{V}_{1,i}-\tilde{V}%
	_{2,i})\right\vert \\
	& \leq n^{-1}\sum_{i\leq n}(\mathrm{Var}(\tilde{\Psi}_{1,i}-\tilde{\Psi}%
	_{2,i}+\tilde{U}_{1,i}-\tilde{U}_{2,i})\mathrm{Var}(\tilde{V}_{1,i}-\tilde{V}%
	_{2,i}))^{1/2} \\
	& \leq \left( n^{-1}\sum_{i\leq n}\mathrm{Var}(\tilde{\Psi}_{1,i}-\tilde{\Psi%
	}_{2,i}+\tilde{U}_{1,i}-\tilde{U}_{2,i})\right) ^{1/2}\left(
	n^{-1}\sum_{i\leq n}\mathrm{Var}(\tilde{V}_{1,i}-\tilde{V}_{2,i})\right)
	^{1/2},
\end{align*}%
which together with the definition of $\omega _{n,T}^{\ast }$, (\ref%
{P_case_2_1a}) and (\ref{P_case_2_1}) implies that 
\begin{equation}
	\left\vert n^{-1}\sum_{i\leq n}\mathrm{Cov}(\tilde{\Psi}_{1,i}-\tilde{\Psi}%
	_{2,i}+\tilde{U}_{1,i}-\tilde{U}_{2,i},\tilde{V}_{1,i}-\tilde{V}%
	_{2,i})\right\vert =O\left( \omega _{n,T}^{\ast }n^{-1/2}\max_{g\in \mathcal{%
			G}_{1}}n_{g}^{-1/2}\right) .  \label{P_case_2_3}
\end{equation}%
By (\ref{P_case_2_1b}), (\ref{P_case_2_2}) and (\ref{P_case_2_3}), we can
deduce that%
\begin{equation}
	\frac{\omega _{n,T}^{2}}{(\omega _{n,T}^{\ast })^{2}}=1+O\left( \frac{%
		\max_{g\in \mathcal{G}_{1}}n_{g}^{-1/2}}{n^{1/2}\omega _{n,T}^{\ast }}%
	\right) +O\left( \frac{\max_{g\in \mathcal{G}_{1}}n_{g}^{-1}}{n(\omega
		_{n,T}^{\ast })^{2}}\right) =1+O\left( \max_{g\in \mathcal{G}%
		_{1}}n_{g}^{-1/2}\right) .  \label{P_case_2_4}
\end{equation}%
The claim in (\ref{case_2_0}) thus follows from Theorem \ref{QLR_Stat} and
Theorem \ref{MGCLT_V}, Assumption \ref{A7}(iii) and Slutsky's theorem.

We next turn to the proof of (\ref{case_2_1}). Since $\mathbb{E}[A_{j,i}]=0$
and $A_{j,i}$ are independent across $i$, for any $i\in I_{g}$ and any $g\in 
\mathcal{G}_{1}$ 
\begin{align*}
	\sum_{i\in I_{g}}\mathrm{Var}(\tilde{U}_{1,i})&
	=\sum_{i=N_{g}+2}^{N_{g}+n_{g}}\sum_{i^{\prime }=N_{g}+1}^{i-1}\frac{\mathrm{%
			Var}(A_{1,i}^{\top }A_{1,i^{\prime }})}{n_{g}^{2}T}%
	=\sum_{i=N_{g}+2}^{N_{g}+n_{g}}\sum_{i^{\prime }=N_{g}+1}^{i-1}\frac{\mathrm{%
			Tr}(\mathbb{E}[A_{1,i}A_{1,i}^{\top }]\mathbb{E}[A_{1,i^{\prime
		}}A_{1,i^{\prime }}^{\top }])}{n_{g}^{2}T} \\
	& =(2T)^{-1}\mathrm{Tr}\left( \left( n_{g}^{-1}\sum_{i\in I_{g}}\mathbb{E}%
	[A_{1,i}A_{1,i}^{\top }]\right) ^{2}-n_{g}^{-2}\sum_{i\in I_{g}}(\mathbb{E}%
	[A_{1,i}A_{1,i}^{\top }])^{2}\right) \\
	& =(2T)^{-1}\left( \left\Vert \Sigma _{1\gamma ,g}\right\Vert
	^{2}-n_{g}^{-2}\sum_{i\in I_{g}}\left\Vert \mathbb{E}[A_{1,i}A_{1,i}^{\top
	}]\right\Vert ^{2}\right) ,
\end{align*}%
which together with (\ref{P_C1_MGCLT_L1_6}) in the proof of Lemma \ref%
{C1_MGCLT_L1} implies that%
\begin{equation}
	\sum_{g\in \mathcal{G}_{1}}\sum_{i\in I_{g}}\mathrm{Var}(\tilde{U}%
	_{1,i})=(2T)^{-1}\sum_{g\in \mathcal{G}_{1}}\left\Vert \Sigma _{1\gamma
		,g}\right\Vert ^{2}+O\left( \max_{g\in \mathcal{G}_{1}}n_{g}^{-1}\right) .
	\label{P_case_2_var_2}
\end{equation}%
Similarly,%
\begin{align}
	\sum_{i\leq n}\mathrm{Var}(\tilde{U}_{2,i})& =(2n^{2}T)^{-1}\mathrm{Tr}%
	\left( \left( \sum_{i\leq n}\mathbb{E}[A_{2,i}A_{2,i}^{\top }]\right)
	^{2}-\sum_{i\leq n}(\mathbb{E}[A_{2,i}A_{2,i}^{\top }])^{2}\right)  \notag \\
	& =(2T)^{-1}\left\Vert \sum_{g\in \mathcal{G}_{1}}w_{g}\Sigma _{2\gamma
		,g}\right\Vert ^{2}+O\left( n^{-1}\right) ,  \label{P_case_2_var_3}
\end{align}%
and%
\begin{align}
	2\sum_{i\leq n}\mathrm{Cov}(\tilde{U}_{1,i},\tilde{U}_{2,i})&
	=2(nT)^{-1}\sum_{g\in \mathcal{G}_{1}}n_{g}^{-1}\sum_{i\in I_{g}}\mathrm{Cov}%
	\left( \sum_{i^{\prime }=N_{g}+1}^{i-1}A_{1,i}^{\top }A_{1,i^{\prime
	}},\sum_{i^{\prime }=1}^{i-1}A_{2,i}^{\top }A_{2,i^{\prime }}\right)  \notag
	\\
	& =2(nT)^{-1}\sum_{g\in \mathcal{G}_{1}}n_{g}^{-1}%
	\sum_{i=N_{g}+2}^{N_{g}+n_{g}}\sum_{i^{\prime }=N_{g}+1}^{i-1}\mathrm{Tr}(%
	\mathbb{E}[A_{1,i^{\prime }}A_{2,i^{\prime }}^{\top }]\mathbb{E}%
	[A_{2,i}A_{1,i}^{\top }])  \notag \\
	& =(nT)^{-1}\sum_{g\in \mathcal{G}_{1}}n_{g}^{-1}\mathrm{Tr}\left(
	\sum_{i,i^{\prime }\in I_{g}}\mathbb{E}[A_{1,i}A_{2,i}^{\top }]\mathbb{E}%
	[A_{2,i^{\prime }}A_{1,i^{\prime }}^{\top }]-\sum_{i\in I_{g}}\mathbb{E}%
	[A_{1,i}A_{2,i}^{\top }]\mathbb{E}[A_{2,i}A_{1,i}^{\top }]\right)  \notag \\
	& =T^{-1}\sum_{g\in \mathcal{G}_{1}}w_{g}\left\Vert \Sigma _{12\gamma
		,g}\right\Vert ^{2}-(nT)^{-1}\sum_{g\in \mathcal{G}_{1}}n_{g}^{-1}\sum_{i\in
		I_{g}}\left\Vert \mathbb{E}[A_{1,i}A_{2,i}^{\top }]\right\Vert ^{2}  \notag
	\\
	& =T^{-1}\sum_{g\in \mathcal{G}_{1}}w_{g}\left\Vert \Sigma _{12\gamma
		,g}\right\Vert ^{2}+O\left( n^{-1}\right) ,  \label{P_case_2_var_4}
\end{align}%
where the last equality is by (\ref{P_C1_MGCLT_L2_11})\ in the proof of
Lemma \ref{C1_MGCLT_L2}. Collecting the results in (\ref{P_case_2_var_2}), (%
\ref{P_case_2_var_3}) and (\ref{P_case_2_var_4}), we obtain 
\begin{equation*}
	\sum_{i\leq n}\mathrm{Var}(\tilde{U}_{1\gamma ,i}^{\ast }-\tilde{U}_{2\gamma
		,i}^{\ast })=\frac{\sum_{g\in \mathcal{G}_{1}}(\left\Vert \Sigma _{1\gamma
			,g}\right\Vert ^{2}-2w_{g}\left\Vert \Sigma _{12\gamma ,g}\right\Vert
		^{2})+\left\Vert \sum_{g\in \mathcal{G}_{1}}w_{g}\Sigma _{2\gamma
			,g}\right\Vert ^{2}}{2T}+O\left( \max_{g\in \mathcal{G}_{1}}n_{g}^{-1}\right)
\end{equation*}%
which establishes (\ref{case_2_1}).\hfill $Q.E.D.$

\bigskip

\noindent \textsc{Proof of Lemma \ref{case_3}}. Recall that in the two
models investigated in this lemma, we have $\mathcal{G}_{1}=\{1,\ldots ,n\}$
and $\mathcal{M}_{1}=\{1\}$ for model 1, and $\mathcal{G}_{2}=\{1\}$ and $%
\mathcal{M}_{2}=\{1,\ldots ,T\}$ for model 2. First notice that Assumption %
\ref{A5}(ii) holds with $G_{1}=n$, $M_{1}=1$, $G_{2}=1$\ and $M_{2}=T$. In
the proof of Lemma \ref{case_2}, we have shown that%
\begin{equation}
	\mathrm{Var}\left( n^{-1/2}\sum_{i\leq n}\tilde{V}_{2,i}\right)
	=n^{-3}\sum_{i\leq n}\mathrm{Var}\left( T^{-1/2}A_{2,i}^{\top
	}A_{2,i}\right) \leq Kn^{-2}.  \label{P_case_3_1}
\end{equation}%
Therefore, using the same arguments for deriving (\ref{P_case_2_4}) obtains%
\begin{equation*}
	\frac{\omega _{n,T}^{2}}{(\omega _{n,T}^{\ast })^{2}}=1+O\left(
	n^{-1}\right) ,
\end{equation*}%
and the claim in (\ref{case_3_0}) thus follows from Theorem \ref{QLR_Stat}
and Theorem \ref{MGCLT_V} and Slutsky's theorem.

By the independence of $z_{i,t}$ across $i$ for all $t$%
\begin{equation}
	\mathrm{Var}\left( n^{-1/2}\sum_{i\leq n}(\tilde{\Psi}_{1,i}-\tilde{\Psi}%
	_{2,i}+\tilde{V}_{1,i}-\tilde{U}_{2,i})\right) =n^{-1}\sum_{i\leq n}\mathrm{%
		Var}(\tilde{\Psi}_{1,i}-\tilde{\Psi}_{2,i}+\tilde{V}_{1,i})+n^{-1}\sum_{i%
		\leq n}\mathrm{Var}(\tilde{U}_{2,i}).  \label{P_case_3_2}
\end{equation}%
The expression in (\ref{case_3_1}) is can be derived directly from (\ref%
{P_case_3_2}) and (\ref{P_case_2_var_3}) in the proof of Lemma \ref{case_2}%
.\hfill $Q.E.D.$

\end{document}